\documentclass[a4paper,titlepage, 11pt,twoside,openright]{book}
\usepackage[utf8]{inputenc}
\usepackage[T1]{fontenc}
\usepackage[english]{babel}

\usepackage{amsmath,amsfonts,amssymb,amsthm,amscd}
\usepackage{appendix}
\usepackage{array}
\usepackage{bbm}
\usepackage{bm}
\usepackage{blkarray}
\usepackage{booktabs}
\usepackage{braket}
\usepackage{calc}
\usepackage{cancel}
\usepackage{caption}
\usepackage{cases}
\usepackage{cite}
\usepackage{color}
\usepackage{comment}
\usepackage[autostyle,italian=guillemets]{csquotes}
\usepackage{dsfont}
\usepackage{empheq}
\usepackage{enumitem}
\usepackage{fancyhdr}
\usepackage{float}
\usepackage[Sonny]{fncychap}
\usepackage[hang,flushmargin]{footmisc}
\usepackage[a4paper,inner=2.6cm,outer=2.6cm,top=3.5cm,bottom=3cm]{geometry}
\usepackage{graphicx}
\usepackage[unicode,implicit]{hyperref}
\usepackage{latexsym}
\usepackage{mathtools}
\usepackage{multicol}
\usepackage{multirow}
\usepackage{nccmath}
\usepackage{nicefrac}
\usepackage{pdfpages}
\usepackage{physics}
\usepackage{slashed}
\usepackage{subcaption}
\usepackage{tcolorbox}
\usepackage{tensor}
\usepackage{textcomp}
\usepackage{tikz} 
\usepackage{tikz-cd}
\usepackage{tikzsymbols}
\usepackage{todonotes}
\usepackage{transparent}
\usepackage[normalem]{ulem}
\usepackage{url}
\usepackage{verbatim}
\usepackage{wasysym}
\usepackage[all]{xy}
\usepackage{yfonts} 
\usepackage{yhmath}

\usetikzlibrary{decorations.markings,arrows.meta,shapes.misc,decorations.pathmorphing,calc,bending}
\tikzset{
	branch cut/.style={
		decorate,decoration=snake,
		to path={
			(\tikztostart) -- (\tikztotarget) \tikztonodes
		}
	}
}

\newcommand{\fncyblank }{\fancyhf{}}

{\cleardoublepage\fncyblank\null\vfill\begin{center}%
	\bfseries \abstractname \end{center}}%
{\vfill \null}

\definecolor{MycolorBlue}{HTML}{3882DB}

\hypersetup{%
	pdftitle    = {Exploring String Theory Solutions: BH thermodynamics with alpha prime corrections and type II compactifications}
	pdfkeywords = {black hole, string theory, entropy, temperature,
		thermodynamics, Wald, Hawking, Smarr,  branes, supergravity, supersymmetry, higher order, corrections,  heterotic superstring}
	pdfauthor   = {Matteo Zatti},
	plainpages  = true,
	colorlinks  = true,
	citecolor   = blue,
	urlcolor    = blue,
	linkcolor   = black
}

\newcommand{\bmat}{\left(\begin{array}}
\newcommand{\emat}{\end{array}\right)}

\newcommand{\pr}{\mathbbm{R}}
\newcommand{\pz}{\mathbbm{Z}}

\def\CK {{\cal K}}
\def\a {\alpha}
\def\b {\beta}

\def\1{{\bf 1}}
\def\2{{\bf 2}}
\def\3{{\bf 3}}
\def\4{{\bf 4}}
\def\6{{\bf 6}}

\def\targ#1#2{\genfrac{[}{]}{0pt}{}{#1}{#2}}
\def\targ2#1#2{\genfrac{}{}{0pt}{}{#1}{#2}}

\definecolor{mygr}{rgb}{0,0.6,0}
\definecolor{mygrey}{rgb}{0,0.1,0.2}
\definecolor{myblue}{rgb}{0,0.5,0.9}
\definecolor{myblue2}{rgb}{0,0.5,0.5}
\definecolor{myblue3}{rgb}{0,0.7,0.9}
\definecolor{myblue4}{rgb}{0,0.6,0.6}
\definecolor{myorange}{rgb}{1,0.5,0}
\definecolor{mypurple}{rgb}{0.6,0,1}
\definecolor{mygolden}{rgb}{1,0.8,0.2}
\definecolor{mycyan}{rgb}{0,1,1}
\definecolor{mymagenta}{rgb}{1,0,1}
\definecolor{mykiwi}{rgb}{0.8,1,0.5}
\definecolor{mybrown}{cmyk}{0.14, 0.42, 0.56, 0.2}
\definecolor{myturq}{cmyk}{0.99, 0, 0.2, 0.4}
\definecolor{myaubergine2}{cmyk}{0.4, 0.5, 0, 0.1}
\definecolor{myaubergine}{cmyk}{0.6,0.85,0,0}
\definecolor{CycleGreen}{cmyk}{0.52,0,1,0}
\definecolor{CycleBrown}{cmyk}{0, 0.4, 0.9, 0.2}
\DeclareFontFamily{U}{rcjhbltx}{}
\DeclareFontShape{U}{rcjhbltx}{m}{n}{<->rcjhbltx}{}
\DeclareSymbolFont{hebrewletters}{U}{rcjhbltx}{m}{n}
\DeclareMathSymbol{\lamed}{\mathord}{hebrewletters}{108}
\DeclareMathSymbol{\mem}{\mathord}{hebrewletters}{109}
\DeclareMathSymbol{\ayin}{\mathord}{hebrewletters}{96}
\DeclareMathSymbol{\tsadi}{\mathord}{hebrewletters}{118}
\DeclareMathSymbol{\qof}{\mathord}{hebrewletters}{113}
\DeclareMathSymbol{\resh}{\mathord}{hebrewletters}{114}
\DeclareMathSymbol{\pe}{\mathord}{hebrewletters}{112}
\DeclareMathSymbol{\pesofit}{\mathord}{hebrewletters}{80}
\DeclareMathSymbol{\samekh}{\mathord}{hebrewletters}{115}
\DeclareMathSymbol{\tav}{\mathord}{hebrewletters}{116}
\DeclareMathSymbol{\vav}{\mathord}{hebrewletters}{119}
\DeclareMathSymbol{\het}{\mathord}{hebrewletters}{120}
\DeclareMathSymbol{\yod}{\mathord}{hebrewletters}{121}
\DeclareMathSymbol{\zayin}{\mathord}{hebrewletters}{122}
\DeclareMathSymbol{\alephdot}{\mathord}{hebrewletters}{128}
\DeclareMathSymbol{\tsadisofit}{\mathord}{hebrewletters}{90}
\DeclareMathSymbol{\shin}{\mathord}{hebrewletters}{152}

\newcommand{\bk}[1]{{\color{black}  #1}}

\newtheorem{conjecture}{Conjecture}
\def\CN {{\cal N}}

\def\CK {{\cal K}}

\def\d {{\rm d}}
\def\be{\begin{equation}}
	\def\ee{\end{equation}}
\def\bea{\begin{eqnarray}}
	\def\eea{\end{eqnarray}}
\def\bes{\begin{subequations}}
	\def\ees{\end{subequations}}

\def\eps{{\epsilon}}
\def\oh{\frac{1}{2}}

\def\re{\mbox{Re}\, }
\def\im{\mbox{Im}\, }

\def\om{\omega}
\def\Om{\Omega}
\def\p {{\partial}}

\def\g {{\gamma}}

\def\th {{\theta}}

\newcommand{\cF}{\mathcal{F}}

\newcommand{\cK}{\mathcal{K}}

\newcommand{\cN}{\mathcal{N}}
\newcommand{\cO}{\mathcal{O}}

\newcommand{\IR}{\mathbb{R}}
\newcommand{\IZ}{\mathbb{Z}}

\newenvironment{eqn*}{\begin{equation*}\begin{aligned}}{\end{aligned}\end{equation*}\noindent}
\makeatletter
\newsavebox\myboxA
\newsavebox\myboxB
\newlength\mylenA

\newcommand*\xoverline[2][0.75]{%
	\sbox{\myboxA}{$\m@th#2$}%
	\setbox\myboxB\null
	\ht\myboxB=\ht\myboxA%
	\dp\myboxB=\dp\myboxA%
	\wd\myboxB=#1\wd\myboxA
	\sbox\myboxB{$\m@th\overline{\copy\myboxB}$}
	\setlength\mylenA{\the\wd\myboxA}
	\addtolength\mylenA{-\the\wd\myboxB}%
	\ifdim\wd\myboxB<\wd\myboxA%
	\rlap{\hskip 0.5\mylenA\usebox\myboxB}{\usebox\myboxA}%
	\else
	\hskip -0.5\mylenA\rlap{\usebox\myboxA}{\hskip 0.5\mylenA\usebox\myboxB}%
	\fi}
\makeatother

\def\be{\begin{equation}}
	\def\ee{\end{equation}}

\newcommand{\diff}{\mathrm{d}}

\newcommand{\zp}{{f}_{p}}
\newcommand{\zm}{{f}_{w}}
\newcommand{\zz}{{g}}
\definecolor{applegreen}{rgb}{0.55, 0.71, 0.0}

\begin{document}

	\allowdisplaybreaks

	\frontmatter
	
	\pagestyle{fancy}
	\frontmatter 
	\fancyhf{}
	\fancyhf[EFC,OFC]{\thepage}
	\pagenumbering{roman}
	
	\begin{titlepage}
		\vspace*{-0.25in}
		\begin{center}

			\noindent\makebox[\linewidth]{\rule{\textwidth}{2pt}}\vspace{0.15cm} \\
			{ \LARGE \bfseries Exploring String Theory Solutions: \\ Black hole thermodynamics with $\alpha'$ corrections \\ and type II compactifications \\[0.3cm] }
			\noindent\makebox[\linewidth]{\rule{\textwidth}{2pt}}

			\vspace{1cm}
			\normalsize{Memoria de Tesis Doctoral realizada por
				\vspace{0.15cm}\\
				\textbf{Matteo Zatti\vspace{0.15cm}}\\
				presentada ante el Departamento de Física Teórica\\ de la Universidad Autónoma de Madrid\\ para optar al Título de Doctor en Física Teórica\\}
			\vspace{1cm}
			\normalsize{Tesis Doctoral dirigida por}\vspace{0.15cm}\\
			\vspace{0.5cm}
			\normalsize{\textbf{Fernando G. Marchesano Buznego}\vspace{0.15cm}}\\
			\normalsize{Investigador Científico del CSIC}\\
			\vspace{0.5cm}
			\normalsize{\textbf{Tomás Ortín Miguel}\vspace{0.15cm}}\\
			\normalsize{Profesor de Investigación del CSIC}\\
			\vspace{1cm}
			\normalsize{Departamento de Física Teórica\\ Universidad Autónoma de Madrid}\\
			\vspace{0.35cm}
			\normalsize{Instituto de Física Teórica UAM-CSIC}\\
			\vspace{0.3cm}

			\begin{figure}[H]
				\centering\hspace{1.25cm}
				\begin{subfigure}[c]{0.3\textwidth}
					\centering
					\includegraphics[width=0.9\textwidth]{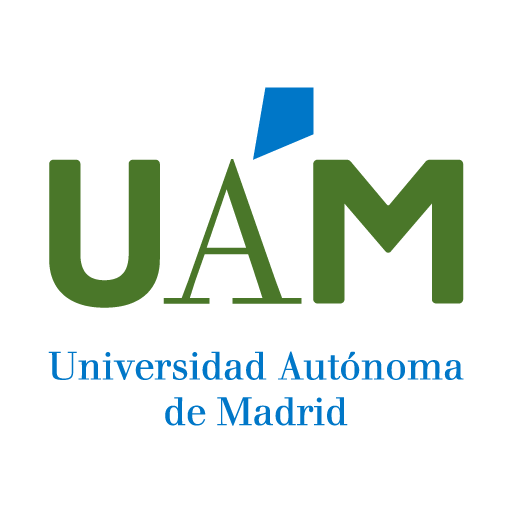}\hfill
				\end{subfigure}\hspace{0.5cm}
				\begin{subfigure}[c]{0.5\textwidth}
					\centering
					\includegraphics[width=0.9\textwidth]{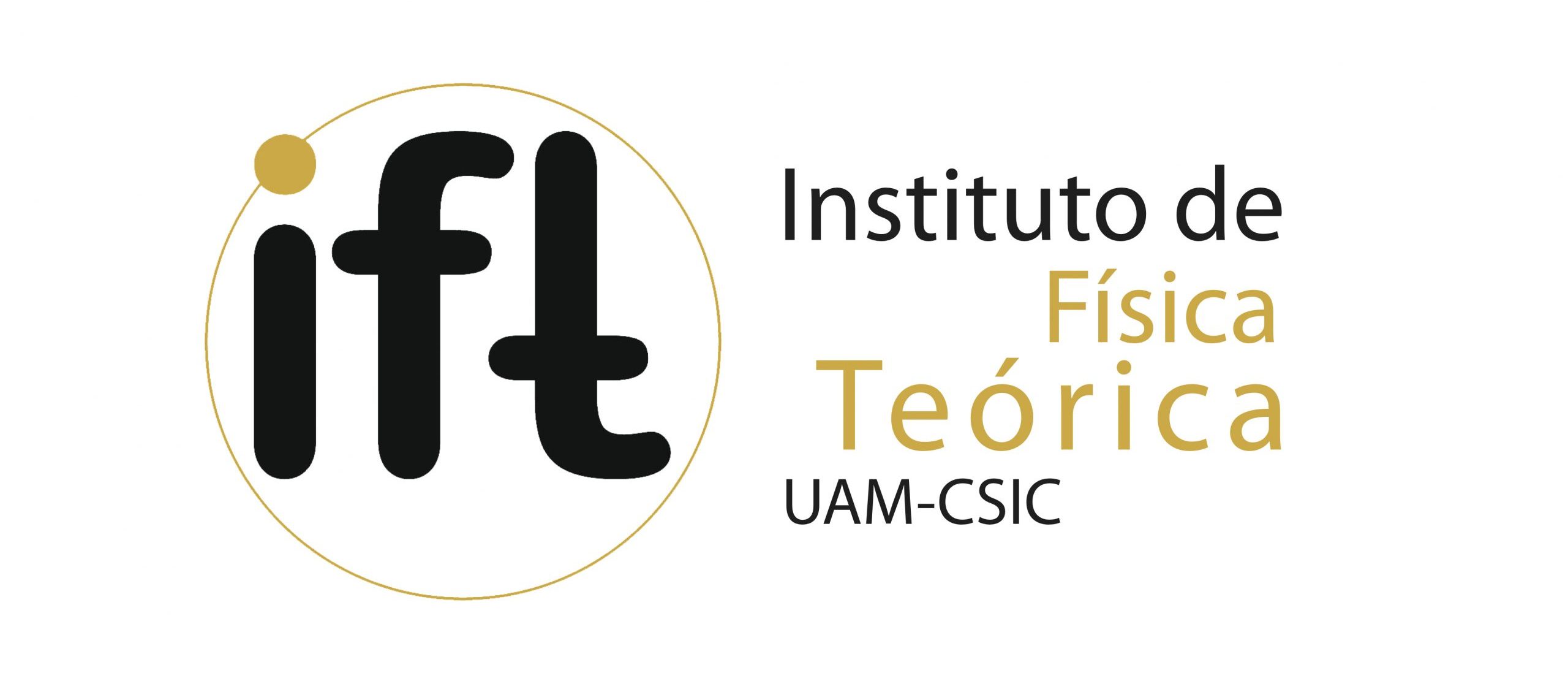}\hfill
				\end{subfigure}\vspace{-0.75cm}\\
				\begin{subfigure}[b]{0.45\textwidth}
					\centering
					\includegraphics[width=0.9\textwidth]{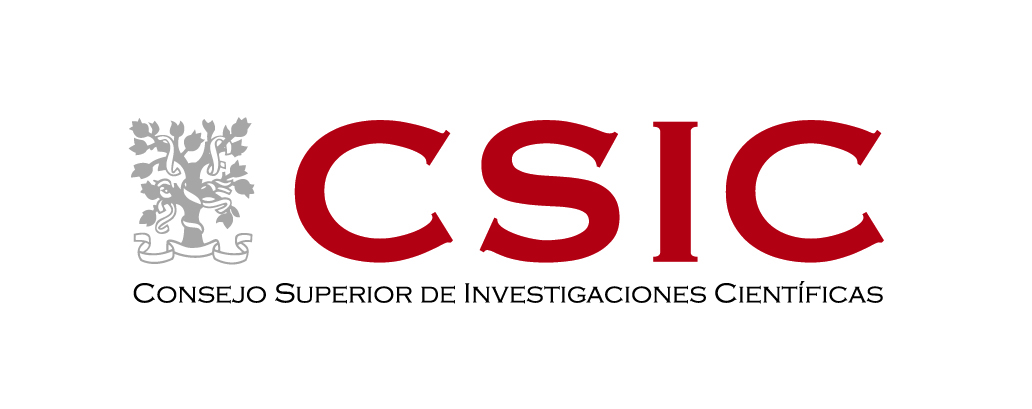}\hfill
				\end{subfigure}
				\begin{subfigure}[h]{0.45\textwidth}
					\vspace{-2cm}
					\includegraphics[width=0.9\textwidth]{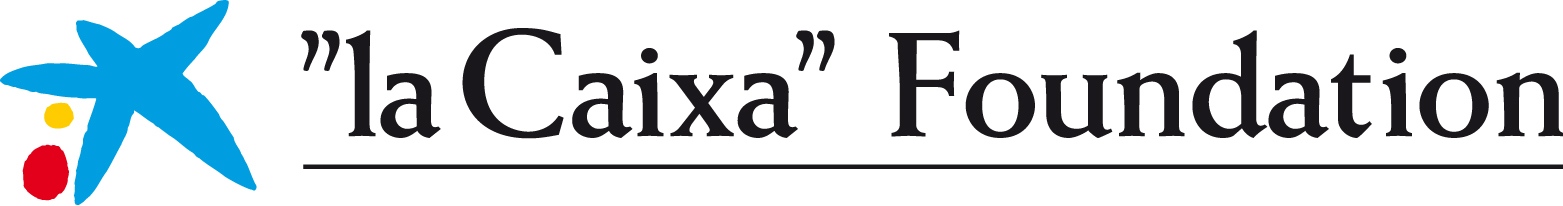}\hfill
				\end{subfigure} \\
			\end{figure}

			\vspace{0.25cm}
			\normalsize Madrid, Septiembre de 2024
		\end{center}
	\end{titlepage}
	
	\clearpage{\pagestyle{empty}\cleardoublepage}

	
	\thispagestyle{empty}
	\phantom{a}\vspace{5cm} 
	\begin{flushright} 
		\textit{To Miriam and my family.}
	\end{flushright} 
	
	\clearpage{\pagestyle{empty}\cleardoublepage}
	
	
	\ChRuleWidth{0.0pt}
	\renewcommand{\headrulewidth}{0pt}
	\ChTitleVar{\Large\sf}

	\chapter*{}
	
	\vspace*{-6cm}
	\textbf{\Large   Lista de publicaciones}

	\thispagestyle{empty}
	\vspace*{1.5cm}
	\noindent
	\textbf{Esta tesis doctoral está basada en los siguientes artículos:}
	\vspace*{0.5cm}

	\noindent [1]\, \textit{Non-supersymmetric black holes with $\alpha'$ corrections,}\\
	\phantom{[1]\, }P. A. Cano, T. Ortin, A. Ruiperez, \textbf{M. Zatti} \\
	\phantom{[1]\, }\href{https://doi.org/10.1007/JHEP03(2022)103}{JHEP 03 (2022) 103}  \href{https://arxiv.org/abs/2111.15579}{[2111.15579]}
	
	\noindent [2]\, \textit{Extremal stringy black holes in equilibrium at first order in $\alpha'$,}\\
	\phantom{[2]\, }T. Ortin, A. Ruiperez, \textbf{M. Zatti}\\
	\phantom{[2]\, }\href{https://doi.org/10.21468/SciPostPhysCore.6.4.072}{SciPost Phys. Core 6 (2023) 072}  \href{https://arxiv.org/abs/2112.12764}{[2112.12764]}
	
	\noindent [3]\, \textit{New instabilities for non-supersymmetric AdS$_4$ orientifold vacua,}\\
	\phantom{[3]\, } F. Marchesano, J. Quirant, \textbf{M. Zatti}\\
	\phantom{[3]\, }\href{https://doi.org/10.1007/JHEP10(2022)026}{JHEP 10 (2022) 026}  \href{https://arxiv.org/abs/2207.14285}{[2207.14285]}
	
	\noindent [4]\, \textit{Non-extremal, $\alpha'$-corrected black holes in 5-dimensional heterotic superstring theory,}\\
	\phantom{[4]\, }P. A. Cano, T. Ortin, A. Ruiperez,\textbf{M. Zatti}\\
	\phantom{[4]\, }\href{https://doi.org/10.1007/JHEP12(2022)150}{JHEP 12 (2022) 150} \href{https://arxiv.org/abs/2210.01861}{[2210.01861]}
	
	\noindent [5]\, \textit{On scalar charges and black hole thermodynamics,}\\
	\phantom{[5]\, } R. Ballesteros, C. Gomez-Fayren, T. Ortin,\textbf{M. Zatti}\\
	\phantom{[5]\, }\href{https://doi.org/10.1007/JHEP05(2023)158}{JHEP 05 (2023) 158} \href{https://arxiv.org/abs/2302.11630}{[2302.11630]}

	\noindent [6]\, \textit{Wald entropy in Kaluza-Klein black holes,}\\
	\phantom{[6]\, } C. Gomez-Fayren, P. Meessen, T. Ortin,\textbf{M. Zatti}\\
	\phantom{[6]\, }\href{https://doi.org/10.1007/JHEP08(2023)039}{JHEP 08 (2023) 039} \href{https://arxiv.org/abs/2305.01742}{[2305.01742]}
	
	\noindent [7]\, \textit{Torsion in cohomology and dimensional reduction,}\\
	\phantom{[7]\, } G. F. Casas, F. Marchesano, \textbf{M. Zatti}\\
	\phantom{[7]\, }\href{https://doi.org/10.1007/JHEP09(2023)061}{JHEP 09 (2023) 061} \href{https://arxiv.org/abs/2306.14959}{[2306.14959]}
	
	\noindent [8]\, \textit{$\alpha'$ corrections to 4-dimensional non-extremal black holes,}\\
	\phantom{[8]\, } \textbf{M. Zatti}\\
	\phantom{[8]\, }\href{https://doi.org/10.1007/JHEP11(2023)185}{JHEP 11 (2023) 185} \href{https://arxiv.org/abs/2308.12879}{[2308.12879]}
	
	\noindent [9]\, \textit{Revisiting $\alpha'$ corrections to heterotic two-charge black holes,}\\
	\phantom{[9]\, } S. Massai, A. Ruiperez, \textbf{M. Zatti}\\
	\phantom{[9]\, } \href{https://doi.org/10.1007/JHEP04(2024)150}{JHEP 04 (2024) 150}
	\href{https://arxiv.org/abs/2311.03308}{[2311.03308]}

	\phantom{\cite{Cano:2021nzo,Ortin:2021win,Marchesano:2022rpr,Cano:2022tmn,Ballesteros:2023iqb,Gomez-Fayren:2023wxk,Casas:2023wlo,Zatti:2023oiq,Massai:2023cis}}
	
	\noindent 	\textbf{Otros artículos escritos durante la realización del doctorado y que no se incluyen en esta tesis son}
	\vspace*{0.5cm}
	
	\noindent [10]\, \textit{Democratic actions with scalar fields: symmetric
		$\sigma$-models, supergravity actions}\\
		\phantom{[10]\, } \textit{and the effective theory of the type~IIB superstring,} \\
	\phantom{[10]\, } J. J. Fernández-Melgarejo, G. Giorgi, C. Gomez-Fayren, T. Ortin, \textbf{M. Zatti}\\
	\phantom{[10]\, } 
	\href{https://doi.org/10.21468/scipostphyscore.7.4.068}{SciPost Phys. Core 7 (2024) 068}
	\href{https://arxiv.org/abs/2401.00549}{[2401.00549]}

	\noindent [11]\, \textit{Gravitational higher-form symmetries and the origin of hidden symmetries}\\
	\phantom{[11]\, } \textit{in Kaluza-Klein compactifications,} \\
	\phantom{[11]\, }  C. Gomez-Fayren, T. Ortin, \textbf{M. Zatti}\\
	\phantom{[11]\, } 
	\href{https://doi.org/10.21468/SciPostPhysCore.8.1.010}{SciPost Phys. Core 8 (2025) 010}
	\href{https://arxiv.org/abs/2405.16706}{[2405.16706]}
	
	\phantom{\cite{Fernandez-Melgarejo:2023kwk,Gomez-Fayren:2024cpl}}
	
	\clearpage{\pagestyle{empty}\cleardoublepage}
	
	
	\ChRuleWidth{0.0pt}
	\renewcommand{\headrulewidth}{0pt}
	\ChTitleVar{\Large\sf}

	\chapter*{}
	
	\vspace*{-5cm}
	\textbf{\Large  Acknowledgments}

	\vspace*{1.5cm}
	\noindent	
	I would like to express my deepest gratitude to my supervisors, Fernando and Tomás, for their invaluable guidance, unwavering support, and profound insight throughout the course of my doctoral research. Both of you are examples of the kind of researcher, and more importantly, the kind of person I aspire to be. \\

	\noindent I am also grateful to my colleagues and collaborators for the interesting conversations and their friendship. Their perspectives have been invaluable in advancing my research and broadening my understanding of the field, allowing me to feel part of a global community.\\
	
	\noindent I extend my heartfelt appreciation to Miriam and my family for their support, encouragement, and understanding throughout this journey. Their love, patience, and encouragement have been a constant source of strength and motivation for me. \\
	
	\noindent During the realization of this thesis, I have been supported by the fellowship LCF/BQ/DI20/ 11780035 from ``La Caixa'' Foundation (ID100010434), by the MCI, AEI, FEDER (UE) grants PGC2018-095205-B-10 (“Gravity, Supergravity and Superstrings” (GRASS)), PID2021-125700 NBC21 (GRASS), PGC2018-095976-B-C21 and PID2021-123017NB-I00, and by the grant IFT Centro de Excelencia Severo Ochoa CEX2020-001007-S.
	
	\clearpage{\pagestyle{empty}\cleardoublepage}

	
	\ChRuleWidth{0.0pt}
	\renewcommand{\headrulewidth}{0pt}
	\ChTitleVar{\Large\sf}
	
	\chapter*{}
	
	\vspace*{-5cm}
	\textbf{\Large Introduction}
	
	\vspace*{1.5cm}

	\noindent String theory is the best example we have of a quantum theory of gravity. It is the only theory we know whose amplitudes describe graviton scatterings and are perturbatively finite to all orders. The goal of this thesis is to study some phenomenological aspects of certain low-energy, effective limits of string theory. More precisely, we analyze the classical solutions of certain effective superstring actions.
	
	This thesis is divided in two main parts. In part \ref{part:TO} we study black hole solutions of the effective action of heterotic string theory (HST). This is done at first order in $\alpha'$, i.e. incorporating in the effective theory part of the worldsheet first order quantum corrections. Since $\alpha'$ is dimensionful, such corrections enter in the action in the form of higher derivative terms. All the black holes we consider are well-known solutions of the theory at zeroth order in $\alpha'$. In this work we present the explicit, analytical, $\alpha'$ corrections of such solutions and we completely characterize their thermodynamics. Among the possible methods available to describe black hole thermodynamics in theories with higher derivative corrections, we have Wald's formalism and the Euclidean on-shell action method. We use both of them. However, Wald's formalism in its original formulation has some limitations which prevent us from applying it safely to theories containing gauge symmetries, such as HST. The realization of such a fact started a program whose goal was revisiting Wald's formalism. In this part we also present some of the milestones of such a program. The extended Wald's formalism is the one we use to characterize HST black hole's thermodynamics. 
	
	The content of part \ref{part:TO} is organized in the following way: in chapter \ref{ch:2} we review the original formulation of Wald's formalism and we explain what is the extended Wald's formalism. In this chapter we summarize the results of \cite{Ballesteros:2023iqb, Gomez-Fayren:2023wxk}. In particular, we explain how one can define scalar charges in a coordinate independent way and how the mathematical tools used to extend Wald's formalism (the gauge-covariant Lie derivatives) naturally emerge in Kaluza--Klein (KK) dimensional reductions.  In chapter \ref{ch:3} we shortly review HST at first order in $\alpha'$. We then explain how to solve the EOMs of HST in order the compute the corrections to several families of black hole solutions. We consider 4-dimensional, 4-charge and 5-dimensional, 3-charge black holes. We evaluate the corrections in both the extremal (supersymmetric and non-supersymmetric) and non-extremal cases. In the supersymmetric case these black holes are the HST version of the Strominger--Vafa black holes. We further generalize the extremal solutions obtaining the corrections for a configuration with an arbitrary number of extremal BHs in equilibrium. We fully characterize the thermodynamics of all the solutions using the extended Wald's formalism. This chapter summarizes the results of \cite{Cano:2021nzo,Ortin:2021win,Cano:2022tmn,Zatti:2023oiq}. In chapter \ref{ch:4} we further investigate the thermodynamics of 2-charge black holes comparing the results obtained with the extended Wald's formalism to those obtained using the Euclidean on-shell action method and duality arguments. This chapter is based on \cite{Massai:2023cis}.

	In part \ref{part:FM} we study the properties of certain type II compactifications of the form $X_4 \times X_6$. We first consider massive type IIA orientifold compactifications of the form AdS$_4 \times X_6$, where $X_6$ admits a Calabi-Yau metric and is threaded by background fluxes. In the literature, these vacua are described in terms of smeared sources, which is unsatisfactory, especially because of the presence of the orientifold planes. We check that a 10d description with localized sources exists and we study the its non-perturbative stability.
	We then focus on compactifications where the internal manifold $X_6$ has non-trivial integer (co)homology. Conventional wisdom dictates that the $\mathbb{Z}_N$ factors in the integral cohomology group of a compact manifold do not affect the lower dimensional theory because they cannot be detected via smooth p-forms. We revisit this lore in light of the dimensional reduction of string theory on G-structure metric that leads to a supersymmetric EFT.

	The content of part \ref{part:FM} is organized in the following way: in chapter \ref{ch:5} we present the results of \cite{Marchesano:2022rpr} regarding some families of AdS$_4 \times X_6$ compactifications of massive type IIA. We start with a brief review of the democratic formulation of type IIA effective action and its compactifications. We then show  explicitly that within the regime in which we can describe localized sources via smeared delta functions with a small correction, we can still solve the equations of motion. From the perspective of the non-perturbative stability, we find that non-supersymmetric vacua admit superextremal branes which can trigger decays.
	In chapter \ref{ch:6} we discuss type II compactifications with torsion factors in its singular cohomology groups. Our main result is the proposal contained in \cite{Casas:2023wlo}. We propose that, if some massive p-form eigenmodes of the Laplacian are much lighter than the Kaluza-Klein scale and enter the EFT, whenever torsion cycles are calibrated, is possible to extract the associated topological information using their smeared delta forms. More precisely, it is possible to compute their linking number.

	\clearpage{\pagestyle{empty}\cleardoublepage}

	\ChRuleWidth{0.0pt}
	\renewcommand{\headrulewidth}{0pt}
	\ChTitleVar{\Large\sf}
	
	\chapter*{}
	
	\vspace*{-5cm}
	\textbf{\Large  Introducción}
	
	\vspace*{1.5cm}

	\noindent La teoría de cuerdas es el mejor ejemplo que tenemos de una teoría cuántica de la gravedad. Es la única teoría que conocemos cuyas amplitudes describen dispersiones de gravitones y son perturbativamente finitas a todos los órdenes. El objetivo de esta tesis es estudiar algunos aspectos fenomenológicos de ciertos límites efectivos de baja energía de la teoría de cuerdas. En concreto, analizamos las soluciones clásicas de ciertas acciones efectivas de supercuerdas.	
	
	Esta tesis se divide en dos partes principales. En la parte \ref{part:TO} estudiamos soluciones de agujeros negros de la acción efectiva de la teoría de cuerdas heterótica (HST). Esto se realiza a primer orden en $\alpha'$, es decir, incorporando en la parte efectiva de la teoría correcciones cuánticas de primer orden en la acción de \textit{worldsheet}. Dado que $\alpha'$ tiene dimensiones, tales correcciones entran en la acción en forma de términos de orden superior. Todos los agujeros negros que consideramos son soluciones bien conocidas de la teoría a orden cero en $\alpha'$. En este trabajo presentamos las correcciones explícitas, analíticas, en $\alpha'$, de tales soluciones y caracterizamos completamente su termodinámica. Entre los posibles métodos disponibles para describir la termodinámica de los agujeros negros en teorías con correcciones derivadas de orden superior, tenemos el formalismo de Wald y el método de acción Euclidiea \textit{on-shell}. Utilizamos ambos. Sin embargo, el formalismo de Wald en su formulacione original tiene algunas limitaciones que nos impiden aplicarlo a teorías que contienen simetrías de gauge, como la HST. La comprensión de este hecho dio inicio a un programa cuyo objetivo era revisar el formalismo de Wald. En esta parte también presentamos algunos de los hitos de dicho programa. El formalismo extendido de Wald es el que utilizamos para caracterizar la termodinámica de los agujeros negros de HST.
	
	La organización de la parte \ref{part:TO} es la siguiente: en el capítulo \ref{ch:1} hacemos una breve revisión de la teoría de cuerdas. Este capítulo se basa en material de texto. En el capítulo \ref{ch:2} revisamos la formulación original del formalismo de Wald y explicamos qué es el formalismo extendido de Wald. En este capítulo resumimos los resultados de \cite{Ballesteros:2023iqb, Gomez-Fayren:2023wxk}. En particular, explicamos cómo se pueden definir las cargas escalares de manera independiente de las coordenadas y cómo las herramientas matemáticas utilizadas para extender el formalismo de Wald (las derivadas de Lie covariantes gauge) emergen en forma natural en las reducciones dimensionales de Kaluza--Klein (KK). En el capítulo \ref{ch:3} revisamos brevemente la HST a primer orden en $\alpha'$. Despues explicamos cómo resolver las ecuaciones de movimiento de la HST para calcular las correcciones a varias familias de soluciones de agujeros negros. Consideramos agujeros negros en 4 dimensiones con 4 cargas y en 5 dimensiones con 3 cargas. Evaluamos las correcciones tanto en los casos extremos (supersimétricos y no supersimétricos) como en los no extremales. En el caso supersimétrico, estos agujeros negros son la versión de la HST de los agujeros negros de Strominger--Vafa. Además, generalizamos las soluciones extremales obteniendo las correcciones para una configuración con un número arbitrario de agujeros negros extremales en equilibrio. Caracterizamos completamente la termodinámica de todas las soluciones utilizando el formalismo extendido de Wald. Este capítulo resume los resultados de \cite{Cano:2021nzo,Ortin:2021win,Cano:2022tmn,Zatti:2023oiq}. En el capítulo \ref{ch:4} investigamos más a fondo la termodinámica de los agujeros negros de 2 cargas comparando los resultados obtenidos con el formalismo extendido de Wald con aquellos obtenidos usando el método de acción Euclidea \textit{on-shell} y argumentos de dualidad. Este capítulo se basa en \cite{Massai:2023cis}.
	
	En la parte \ref{part:FM} estudiamos las propiedades de ciertas compactificaciones de las teorias de tipo II de la forma $X_4 \times X_6$. En primer lugar, consideramos compactificaciones \textit{orientifold} de la teoría IIA masiva de la forma AdS$_4 \times X_6$, donde $X_6$ admite una métrica de Calabi-Yau y está atravesada por flujos de \textit{background}. En la literatura, estos vacíos tienen una descripción en términos de fuentes difuminadas, lo cual es insatisfactorio, especialmente debido a la presencia de los planos \textit{orientifold}. Verificamos que existe una descripción de 10 dimensiones con fuentes localizadas y estudiamos su estabilidad no-perturbativa.
	A continuación, nos centramos en compactificaciones donde la variedad interna $X_6$ tiene homología entera no-trivial. La tradición dice que los factores $\mathbb{Z}_N$ en el grupo de cohomología integral de una variedad compacta no afectan a la teoría dimensional inferior porque no pueden detectarse a través de formas suaves $p$. Revisemos esta creencia a la luz de la reducción dimensional de la teoría de cuerdas en una métrica con G-estructura que conduce a una EFT supersimétrica.
	
	La organización de la parte \ref{part:FM} es la siguiente: en el capítulo \ref{ch:5} presentamos los resultados de \cite{Marchesano:2022rpr} sobre algunas familias de compactificaciones AdS$_4 \times X_6$ de tipo IIA masivas. Comenzamos con una breve revisión de la formulación democrática de la acción efectiva de tipo IIA y sus compactificaciones. Despues mostramos explícitamente que dentro del régimen en el cual podemos describir fuentes localizadas a través de funciones delta difuminadas con una pequeña corrección, aún podemos resolver las ecuaciones de movimiento. Desde la perspectiva de la estabilidad no-perturbativa, encontramos que los vacíos no supersimétricos admiten branas superextremales que pueden desencadenar decaimientos del vacío.
	En el capítulo \ref{ch:6} discutimos compactificaciones de tipo II con cohomología entera que contiene elementos de torsión. Nuestro resultado principal es la propuesta contenida en \cite{Casas:2023wlo}: proponemos que, si algunos autovectores masivos del laplaciano son mucho más ligeros que la escala de Kaluza--Klein y entran en la EFT, siempre que los ciclos de torsión estén calibrados, es posible extraer la información topológica asociada utilizando sus formas delta difuminadas. En concreto, es posible calcular su número de enlace.

	\clearpage{\pagestyle{empty}\cleardoublepage}

	\clearpage{\pagestyle{empty}\cleardoublepage}

	
	\ChRuleWidth{0.5pt}
	\renewcommand{\headrulewidth}{0pt}
	\ChTitleVar{\Large\sf}

	\renewcommand{\contentsname}{Contents}
	\tableofcontents

	\clearpage{\pagestyle{empty}\cleardoublepage}


	\mainmatter	
	
	\ChRuleWidth{0.5pt}
	\renewcommand{\headrulewidth}{0.5pt}
	\ChTitleVar{\Large\sf}	
	\pagestyle{fancy}
	\fancyhead{}
	\fancyhead[LE]{\nouppercase{\hfill\leftmark}}
	\fancyhead[RO]{\nouppercase{\rightmark\hfill}}

	\part{Review} \label{part:review}
	\chapter{Introduction to String Theory} \label{ch:1}

In this chapter we briefly introduce string theory. The content presented is based on textbook material \cite{Ortin:2015hya, Blumenhagen:2013fgp, Ibanez:2012zz, Hull1987, Green:2012oqa, Green:2012pqa, Polchinski:1998rq, Polchinski:1998rr}.

\section{Worldsheet formulation}

\subsection{Closed bosonic strings}
The core of string theory is the idea that fundamental objects may not be point-like particles but some extended objects. If this is the case, their physics in the perturbative regime and in absence of interactions should be governed by the classical  world-volume Nambu--Goto action 
\begin{equation}
	S_{NG} = - T \int d^{p+1} \xi \sqrt{|\gamma|} \,, 
\end{equation}
with $T = 1/(2 \pi \alpha')$ where $\alpha'$ is a constant with dimensions of length squared and
\begin{equation}
	\gamma = \det \gamma_{\alpha\beta} \,, \qquad \gamma_{\alpha\beta} = \partial_\alpha X^\mu \partial_\beta X^\nu g_{\mu\nu}\,, \qquad \alpha,\beta = 1,\dots p \,, \quad \mu,\nu = 1, \dots d \,.
\end{equation}
Here the $\xi^\alpha$ parametrize the $p+1$ dimensional world-volume, $X^\mu$ are scalars fields defined on the world-volume which can be interpreted as coordinates of a $d$-dimensional target-space; $g_{\mu\nu}$ is a metric on the target space and $\gamma_{\alpha\beta}$ is the induced metric on the world-sheet. In this chapter we use mostly plus signature.  It is easy to verify that the classical equations of motion for $X^\mu$ given by the Nambu--Goto action are equivalent to those obtained from the Polyakov action
\begin{equation}\label{eq:polyakovaction}
	S_P = -\frac{T}{2} \int d^{p+1}\xi \sqrt{|\gamma|} \gamma^{\alpha\beta} \partial_\alpha X^\mu \partial_\beta X^\nu g_{\mu\nu} \,,
\end{equation}
if we treat $\gamma_{\alpha\beta}$ as an independent field and we impose its equations of motion. The case $p=1$, i.e. when the extended object is a string, is special. 

The gauge symmetries of the action with $p=1$ allow us to trivialize $\gamma_{\alpha\beta}$ setting it to the flat metric $\eta_{\alpha\beta}$. This gauge is called in the literature the conformal gauge. Choosing also the metric $g_{\mu\nu}$ to be the Minkowski one $\eta_{\mu\nu}$, we can study explicitly the spectrum of the theory on flat background. Classically the equations of motion are $\partial_\alpha \partial^\alpha X^\mu = 0$. Assuming closed boundary conditions $X^\mu (\xi^0, \xi^1) \sim  X^\mu (\xi^0, \xi^1 + \ell )$ we obtain that $X^\mu$ must be the sum of two contributions: a wave $X^\mu_L$ propagating to the left depending on $\xi^+ = \xi^0 + \xi^1$ and a wave $X^\mu_R$ propagating to the right depending on $\xi^- = \xi^0 - \xi^1$. Explicitly we have 
\begin{subequations}
	\begin{align}
		\partial_- X^\mu & = \frac{2\pi}{\ell} \sqrt{\frac{\alpha'}{2}} \sum_{n\in \mathbb{Z}} \alpha^\mu_n \, e^{-\frac{2\pi i}{\ell}n\xi^-} \,, \\[1mm]
		\partial_+ X^\mu & = \frac{2\pi}{\ell} \sqrt{\frac{\alpha'}{2}} \sum_{n\in \mathbb{Z}} \bar{\alpha}^\mu_n \, e^{-\frac{2\pi i}{\ell}n\xi^+} \,, 
	\end{align}
\end{subequations} 
where $\alpha^\mu_n$ and $\bar{\alpha}^\mu_n$ represent the weights of the different oscillatory modes. Not all the weights are independent and we have some constraints coming from the requirement that the functions are real, the relation of $\alpha^\mu_0$ and $\bar{\alpha}^\mu_0$ with the total momentum of the closed string and the so-called level-matching condition.\footnote{It is a consistency constraint which is equivalent to asking that the left and right waves carry the same energy.} On top of that, we have extra constraints coming from the equations of motion of $\gamma_{ij}$. Even if we gauge fixed it to be the flat metric, its equations of motion could not be trivial. This last set of constraints can be implemented in a simple way provided that we break explicit SO$(1,d-1)$ Lorentz covariance. Finally, we have some residual gauge symmetry. Combining everything, we obtain that we lose two towers of oscillatory modes: one is completely fixed; the other one contains a single independent mode, its zero mode. We are left therefore with manifest SO$(d-2)$ covariance. Such a gauge is called in the literature the light-cone gauge. 

We want to study the quantum spectrum in the light-cone gauge. We use this gauge because it is the simplest to handle.\footnote{It can be shown that even if the covariance is not manifest, it is still implicitly preserved (i.e. the Lorentz group is not broken).} Following the standard quantization procedure, we obtain that the surviving $\alpha^i_n$ and $\bar{\alpha}^i_n$ with $n\ne 0$ and $i = 1, \dots d-2$ represent two independent sets of creation and annihilation operators for the oscillatory modes
\begin{equation}\label{eq:bosonicOscllators}
	[\alpha^i_n, \alpha^j_m] = [\bar{\alpha}^i_n,\bar{\alpha}^j_m] = n \,\delta^{ij} \delta_{n+m,0} \,,\qquad [{\alpha}^i_n,\bar{\alpha}^j_m] = 0 \,.
\end{equation} 
The other operators coming from the independent dynamical fields in the light-cone gauge represent positions and momentum operators. In particular, $\alpha^i_0 = \bar{\alpha}^i_0$ and they are proportional to the momentum $p^i$ carried in the direction $X^i$. In this setup the mass-squared operator $m^2$ has the explicit form
\begin{equation}
	\alpha' m^2  = 2  \left( N_\alpha + \bar{N}_{\bar{\alpha}} + 2 a \right) \,,
\end{equation}	
where $a$ is a parameter related to the number $d$ of bosons $X^\mu$ and $N$ is the number operator.\footnote{Notice that with our normalization of the $\alpha$ we are weighting the modes with a factor proportional to the mode level.} The consistency of the spectrum impose us to set $a = -1$ and $d = 26$. We can easily see, then, that the ground state of the spectrum $\ket{0}$ has negative mass and it is tachyonic. The level-matching condition is $N = \bar{N}$ and the first exited state is massless and has the form $ \alpha_{-1}^{i} \bar{\alpha}_{-1}^j \ket{0}$. This object is a tensor of SO$(d-2)$, which is the little group of the massless $d$-dimensional Lorentz representations. Decomposing $ \alpha_{-1}^{i} \bar{\alpha}_{-1}^j \ket{0}$ into the traceless and symmetric part, antisymmetric part and trace part, we find the little group representations of the degrees of freedom of a spin 2 massless particle, an antisymmetric tensor and a scalar field. Embedding the degrees of freedom in $d$-dimensional covariant objects we have
\begin{equation}
	(g_{\mu\nu} \,, B_{\mu\nu} \,,  \phi) \,.
\end{equation}
The 2-form $B$ is the so-called Kalb--Ramond (KR) field and the scalar field $\phi$ is the dilaton.

\subsection{Open bosonic strings}

Apart from closed strings, we have open strings. We obtain them by solving the classical equations of motion with a different set of boundary conditions. Instead of periodicity, we require special boundary conditions at $\xi^1 = 0, \ell$. Such conditions naturally emerge from the request that the variational principle is well defined. They are
\begin{equation}
	(D) :\quad \partial_0 X^\mu \big\vert_{\xi^1 = 0 , \ell} = 0\,, \qquad \quad (N): \quad \partial_1 X^\mu  \big\vert_{\xi^1 = 0 , \ell} = 0 \,.
\end{equation}
Dirichlet (D) boundary conditions fix the position of one of the endpoints along a direction. The physical interpretation of Neumann (N) boundary conditions is that the endpoint propagates with zero longitudinal velocity.
For the time direction $X^0$ we can only choose N boundary conditions. For all the other directions, we can choose independently the conditions on the two endpoints. We have the four cases NN, DD, ND and DN.

If we impose $p+1$ Neumann boundary conditions and $d-p-1$  Dirichlet boundary conditions we are defining a $p+1$-dimensional surface on which the endpoint of the string moves. Given that the string momentum in the Dirichlet directions is not conserved, total momentum conservation requires that these surfaces are (non-perturbative) dynamical objects which exchange momentum with the string. In particular, we refer to the $p+1$-dimensional surface as the worldvolume of a $p$-dimensional D$p$-brane.

Proceeding as in the closed string case performing light-cone quantization, we find that now we have a single set of creation operators $\alpha^i_n$ with $i = 1,\dots d-2$ (the left and right modes are not independent). $n$ is integer for NN and DD boundary conditions and semi-integer for mixed boundary conditions. The associated mass operator is 
\begin{equation}
	\alpha' m^2 = N_{\alpha} + a + \frac{\nu}{16} + \alpha' T^2 (\Delta X)^2 \,,
\end{equation}
where $a = -1$, $N$ is the number operator, $\nu$ is the number of directions with mixed boundary conditions and $(\Delta X)^2 $ is the sum of the squared distances of the endpoints in the DD directions. The ground state is tachyonic. The first excited states are built with no mixed boundary conditions and with no separation between the endpoints. This is equivalent to requiring that the open strings starts and ends on the locus of a D$p$-brane. The DD boundary conditions break the Lorentz group to $SO(1,p)$, whose little group is $SO(p-1)$. Indicating with $i = 1,\dots p-1$ the NN directions (we removed the light-cone ones) and $a = 1, \dots d - p -1 $ the DD directions, the massless states are $\alpha_{-1}^i \ket{0} $ and $\alpha_{-1}^a \ket{0}$. The former encodes the degrees of freedom of a massless vector; the latter $d - p -1$ scalars. If we now consider $N$ coincident D$p$ branes, we increase the degeneracy of the ground state. For oriented strings, we can have a total of $N^2$ configurations, one for each pair of branes on which the string endpoints lie. We obtain $\alpha_{-1}^i \ket{m,n,0}$  and $\alpha_{-1}^a \ket{m,n,0}$ where $m, n = 1 \dots N$ indicates the starting and ending brane. We have therefore the  massless, $(p+1)$-dimensional world-volume fields 
\begin{equation}
	A^I \,, \qquad \phi^{a\,I} \,,
\end{equation}
with $ I = 1,\dots N^2 $. The $I$ index can be interpreted as an adjoint index of $U(N)$. Finally, we can consider configurations with generic intersecting branes with $|\Delta X| = 0$. All the non-scalar states are massive. The ground state and the scalars built with the creation operators corresponding to mixed boundary conditions can be tachyonic, massless or massive depending on the particular configuration.

\subsection{Closed superstrings}

The theory described by the Polyakov action (\ref{eq:polyakovaction}) has a problem. The ground state has negative mass, signaling that we are not expanding around a stable vacuum. We want to look for an tachyon-free alternative to bosonic string theory. We consider the 2-dimensional supergravity action
\begin{equation}
	\begin{split}
		S = & - \frac{T}{2} \int d^2 \xi \sqrt{|\gamma|} \bigg[\gamma^{\alpha\beta} \partial_\alpha X^\mu \partial_\beta X^\nu g_{\mu\nu} + i \bar{\psi}^\mu \slashed{D} \psi_\mu \\ & \hspace{1cm}- 2 i \bar{\chi}_\alpha\rho^\beta\rho^\alpha \psi^\mu \partial_\beta X_\mu  + \frac{1}{2}(\bar{\chi}_\alpha\rho^\beta \rho^\alpha \chi_\beta)(\bar{\psi}^\mu \psi_\mu)\bigg] \,,
	\end{split}
\end{equation}
where we have $d$ 2-dimensional scalar superfields $(X^\mu,\psi^\mu)$ coupled to the 2-dimensional supergravity multiplet $(e{}^\alpha{}_a, \chi_\alpha)$. In particular, $\psi^\mu$ and $\chi_\alpha$ are real 2-component spinors. $\rho^\alpha = e^\alpha{}_a \rho^a$ where the $\rho^a$ are 2-dimensional gamma matrices. Using the gauge symmetries of the theory we can decouple the supergravity multiplet. Such gauge fixing is called the superconformal gauge and we obtain the action
\begin{equation}
	S = - \frac{T}{2} \int d^2 \xi \bigg[\eta^{\alpha\beta} \partial_\alpha X^\mu \partial_\beta X^\nu g_{\mu\nu} + i \bar{\psi}^\mu \slashed{\partial}\, \psi_\mu \bigg] \,.
\end{equation}

We study again the closed strings with flat spacetime metric $g_{\mu\nu} = \eta_{\mu\nu}$. The classical solutions for the bosons $X^\mu$ are the same of the bosonic string case. Thus, we focus on the fermions. Their equations of motion are $\slashed{\partial} \psi^\mu = 0$. Again we have to impose proper boundary conditions, but now we can choose between periodic (Ramond) or anti-periodic (Neveu--Schwarz) boundary conditions. The choice can be made independently on the left and right moving modes and we obtain the classical solutions $\psi^\mu = \psi^\mu_L + \psi^\mu_R $
\begin{subequations}
	\begin{align}
		\psi^\mu_L & = \sqrt{\frac{2 \pi}{\ell}} 	\sum_{k \in \mathbb{Z} + s} \bar{b}^\mu_k e^{-\frac{2\pi i}{\ell} k\,\xi^+} \,, \\[2mm]
		\psi^\mu_R & = \sqrt{\frac{2 \pi}{\ell}} 	\sum_{k \in \mathbb{Z} + s} {b}^\mu_k e^{-\frac{2\pi i}{\ell} k\,\xi^-} \,,
	\end{align}
\end{subequations}	
with $s = 0$ for Ramond (R) boundary conditions and $s = 1/2$ for Neveu--Schwarz (NS) boundary conditions. Again, not all the ${b}^\mu_k$ and $\bar{b}^\mu_k$ are independent and we still have to impose the equations of motions of the supergravity multiplet, exploit the residual gauge symmetry we have, require the spinors to be real and impose the level-matching condition. In the light-cone gauge this fixes completely two of the spinors.

We want to study now the quantized theory spectrum (with Minkowski target space metric). The standard quantization procedure produces, in addition to (\ref{eq:bosonicOscllators}), the two sets of anticommutators
\begin{equation}
	\{b^i_n,b^j_m \} = \{ \bar{b}^i_n,\bar{b}^j_m \} = \delta^{ij} \delta_{m+n,0} \,, \qquad \{b^i_n,\bar{b}^j_m \}  = 0 \,.
\end{equation}
We identify the  $b^i_n$ and $\bar{b}^i_n$ for $n \ne 0 $ as fermionic creation and annihilation operators. The zero modes $b^i_0$ and $\bar{b}^i_0$ exist only for Ramond boundary conditions. They can still be interpreted as raising and lowering operators, but they commute with the mass squared operator. Therefore, they are interpreted as operators generating a ground state degeneracy. In particular, in even dimensions, we have a degeneracy of $2^{(d-1)/2}$. The mass-squared operator is
\begin{equation}
	\alpha' m^2  = 2  \left( N_\alpha + \bar{N}_{\bar{\alpha}} + N_b + \bar{N}_{\bar{b}} + a + \bar{a} \right) \,,
\end{equation}	
where $a$ and $\bar{a}$ are now parameters whose dependence on $d$ depends on the choice of boundary conditions we made for the right modes and for the left modes. Therefore, we have a total of 4 sectors depending on the independent conditions we picked: (NS,NS), (R,R), (NS,R) and (R,NS). We restrict now to the right modes (the discussion of the left modes is identical). The consistency of the spectrum requires $d=10$ and $a = -1/2$ for NS boundary conditions and $a = 0$ for R boundary conditions. In $d = 10$ dimensions the Ramond ground state is massless and has 16 degenerate states which can be divided in two groups of 8 elements, depending on their chirality. Each of the 8 states can be seen as the propagating degrees of freedom of a 10-dimensional Majorana--Weyl spinor. We indicate them with $\ket{+}_R$ and $\ket{-}_R$. The ground state of the NS sector is instead tachyonic and we indicate it with $\ket{0}_{NS}$.  The ground states of the four sectors are built by taking the relevant tensor products of R and  NS ground states.   

\subsubsection{The type II spectrum}
Instead of picking a particular sector we can consider all of them at once and project out some of the states. This procedure is called the GSO projection. Concretely, we quotient the spectrum selecting certain eigenstates of the $(-1)^F$ and $(-1)^{\bar{F}}$ operators, where $F$ counts the number of fermionic right modes creation operators applied on the vacua and $\bar{F}$ those of the left modes. We define them so that $\bar{F} = F=1$ for $\ket{0}_{NS}$ and $\ket{-}_R$ and $\bar{F} = F= 0$ for $\ket{+}_R$. This choice can be justified either by demanding spacetime supersymmetry or by imposing modular invariance of the torus partition function, i.e. demanding the cancellation of the anomalies of the global part of the world-sheet diffeomorphism group. In the NS sector we always want to eliminate the tachyonic vacuum and we set $(-1)^F = (-1)^{\bar{F}} = 1$. We have, then, 4 possible choices for the Ramond sector. For $(-1)^F = (-1)^{\bar{F}} = \pm 1$ we obtain the type 2B$_+$ and  type 2B$_-$ theories. For  $(-1)^F = -(-1)^{\bar{F}} = \pm 1$ we obtain the type 2A$_{+-}$ and  type 2A$_{-+}$ theories. The difference between the two type 2B theories is the chirality of the fermions and the presence of a self-dual 4-form in 2B$_+$ and an antiself-dual 4-form in 2B$_-$. The two type 2A theories have the same spectrum but the supergravity theories they form differ in the signs of the Chern-Simons terms. 

Explicitely, the closed string massless states of the type 2A$_{+-}$ theory which satisfy the level-matching condition are
\begin{equation}
	\begin{split}
		& \bar{b}^i_{-1/2} \ket{0}_{NS} \otimes b^j_{-1/2} \ket{0}_{NS} \,, \quad\quad  \ket{+}_R \otimes \ket{-}_R \,, \\[2mm] &  \ket{+}_R \otimes b_{-1/2}^i \ket{0}_{NS}\,, \quad \hspace{1.5cm} \bar{b}^i_{-1/2} \ket{0}_{NS} \otimes \ket{-}_R \,.
	\end{split}
\end{equation}	
Organizing the states into 10-dimensional covariant objects we find
\begin{equation}
	\text{IIA} \hspace{-1.5cm}\begin{split}
		(\text{NS},\text{NS}):&  \quad g_{\mu\nu} \,, B_{\mu\nu} \,,  \phi \,, \hspace{1cm}	(\text{R},\text{R}): \quad   C^{(1)}_\mu\,, C^{(3)}_{\mu\nu\rho} \,, \\[2mm]
		(\text{R},\text{NS}):&  \quad \lambda^1 \,, \psi^1_\mu  \hspace{1.9cm}
		(\text{NS},\text{R}): \quad \lambda^2 \,, \psi^2_\mu  \,,
	\end{split}	
\end{equation}
where $g_{\mu\nu}$ is a graviton, $C^{(3)}$ a 3-form, $B$ is a rank-2 form called the Kalb--Ramond (KR) field, $C^{(1)}$ is a 1-form, $\phi$ is a real scalar called the dilaton, $\lambda^i$ are two spin 1/2 fermions called dilatinos and $\psi_\mu^i$ are two spin 3/2 fermions called gravitinos.	The two dilatinos and the two gravitinos have opposite chiralities. For the type 2B$_+$ we have instead 
\begin{equation}
	\begin{split}
		& \bar{b}^i_{-1/2} \ket{0}_{NS} \otimes b^j_{-1/2} \ket{0}_{NS} \,, \quad\quad  \ket{+}_R \otimes \ket{+}_R \,, \\[2mm] &  \ket{+}_R \otimes b_{-1/2}^i \ket{0}_{NS}\,, \quad \hspace{1.5cm} \bar{b}^i_{-1/2} \ket{0}_{NS} \otimes \ket{+}_R \,.
	\end{split}
\end{equation}		
Organizing the states into 10-dimensional covariant objects we find
\begin{equation}
	\text{IIB} \hspace{-1.2cm}\begin{split}
		(\text{NS},\text{NS}):&  \quad g_{\mu\nu} \,, B_{\mu\nu} \,,  \phi \,, \hspace{1cm}	(\text{R},\text{R}): \quad   C^{(0)}\,, C^{(2)}_{\mu\nu} \,, C^{(4)}_{\mu\nu\rho\sigma}  \,, \\[2mm]
		(\text{R},\text{NS}):&  \quad \lambda^1 \,, \psi^1_\mu  \hspace{1.9cm}
		(\text{NS},\text{R}): \quad \lambda^2 \,, \psi^2_\mu  \,,
	\end{split}	
\end{equation}
where $g_{\mu\nu}$ is a graviton, $C^{(4)}$ is a selfdual 4-form, $B$ and $C^{(2)}$ are two 2-forms, $\phi$ and $C^{(0)}$ are two real scalars, $\lambda^i$ are two dilatinos and $\psi_\mu^i$ are two gravitinos. The two dilatinos and the two gravitions have the same chirality.

\subsubsection{The heterotic spectrum}

The idea is now to build the spectrum considering a right-moving sector governed by $d=10$ superstring theory and a left-moving sector governed by $d=26$ bosonic string theory. In order to glue the two modes and obtain a $d=10$ spacetime theory we have to compactify the 16 extra dimensions. We consider therefore a tours T$^{16} = \mathbb{R}^{16}/\Gamma$, where $\Gamma$ is a lattice. The consistency of the torus partition function constrains the lattice to be a Euclidean, even and self-dual lattice.  If we interpret the lattice as the root lattice of a group $G$ with rank 16 we have a natural group action of $G$ and we can organize the spectrum into representations of $G$.  The only two inequivalent possibilities are the groups SO$(32)$ (we indicate the theory with HO) and E$_8 \times $E$_8$ (we indicate the theory with HE). 

We construct now the spectrum explicitly. The mass squared operator is now  $ m^2 = m_L^2 + m_R^2$ with
\begin{subequations}
	\begin{align}
		\alpha' m^2_L & = 2  \left( \bar{N}_\alpha + \bar{a} \right) + p^I_L \, p^I_L\,, \\[2mm]
		\alpha' m^2_R & = 2  \left( N_\alpha + N_\beta + a \right) \,,
	\end{align}
\end{subequations}	
where $\bar{a} = -1$ and $a = -1/2, 0$ for respectively NS and R boundary conditions. $p^I_L$ with $I = 1,\dots 16$ is the momentum propagating in the internal directions. $\bar{N}_\alpha$ is the number operator built with the external $\bar{\alpha}^i_n$ and the internal $\bar{\alpha}^I_n$.  ${N}_a$ and ${N}_b$ are the number operators built with $\alpha^i_n$ and $b^i_n$. The massless states which satisfy the level-matching condition $m_L^2 = m_R^2 = 0$ we obtain after we perform a GSO projection of the right modes spectrum are (we indicate with $\ket{0}$ the bosonic left-moving vacuum) 
\begin{equation}
	\begin{split}
		& \bar{\alpha}^i_{-1} \ket{0} \otimes b^j_{-1/2} \ket{0}_{NS} \,, \hspace{1cm} \bar{\alpha}^i_{-1} \ket{0} \otimes \ket{+}_R \,, \\[2mm]
		&  \bar{\alpha}^I_{-1} \ket{0} \otimes b^j_{-1/2} \ket{0}_{NS}  \,, \hspace{1cm}  \ket{p^I_L p^I_L = 2} \otimes b^j_{-1/2} \ket{0}_{NS} \,, \\[2mm]
		&  \bar{\alpha}^I_{-1} \ket{0} \otimes \ket{+}_{R}  \,, \hspace{2.1cm}  \ket{p^I_L p^I_L = 2} \otimes \ket{+}_{R} \,.
	\end{split}
\end{equation}		
We can identify the states with the degrees of freedom of the fields (we indicate with $i$ the states generated with external bosonic operators and $I$ those generated with internal operators)
\begin{equation}
	\text{HE/HO} \hspace{-1.2cm}\begin{split}
		(i,\text{NS}):&  \quad g_{\mu\nu} \,, B_{\mu\nu} \,,  \phi \,, \hspace{1cm}	(i,\text{R}): \quad  \lambda \,, \psi_\mu \,, \\[2mm]
		(I,\text{NS}):&  \quad V^M_\mu   \hspace{2.6cm}
		(I,\text{R}): \quad  \eta^M  \,,
	\end{split}	
\end{equation}
where $g_{\mu\nu}$ is a graviton, $B$ is the Kalb--Ramond (KR) field, $\phi$ is the dilaton, $\lambda$ is a dilatino, $\psi_\mu$ is a gravitino, $V_\mu^M$ are gauge vectors and $\eta^M$ are the spin 1/2 superpartners of te gauge vectors called gauginos.	The index $M$ labels the 496 generators of E$_8\times$E$_8$ for HE and SO$(32)$ for HO. In particular, the states built with $\ket{p_L^Ip_L^I = 2}$ represent the non Abelian part of the algebra. Those built with $\bar{\alpha}^I_{-1}$ represent the U$(1)^{16}$ Cartan subalgebra. 

\subsubsection{Other 10-dimensional superstring theories}

For completeness, let us mention that the type II and Heterotic string theories are not the only 10-dimensional supersymmetric string theories available. There exist also the type I theories, which are built by taking a proper orientifold quotient of the type IIB theory. Such a quotient makes the string non orientable. Moreover, it introduces a non-dynamical object called O9-plane. It acts as a source for the $C_{10}$ RR field. For consistency, we need to compensate its contribution inserting 32 D$9$ branes. The cancellation removes 1-point functions called RR tadpoles which generate divergences already in the Klein--bottle amplitude. This implies that the closed sector of the type I is consistent only if it is coupled to the open string sector. Finally, if we drop the requirement of having a supersymmetric theory, we can build other tachyon-free string theories. 

\subsection{Open superstrings}

As we have done in the case of the bosons $X^\mu$, we can describe the open strings by considering different boundary conditions for the fermions $\psi^\mu$. These must be of the form
\begin{equation} \label{eq:boundarycond}
	\psi_L(0) = \pm \psi_R(0) \,, \qquad \psi_L(\ell) = \pm \, \eta \psi_R(\ell) \,.
\end{equation}
$\eta$ is also a sign and we have a total of four different combinations of signs. The configurations with $\eta = 1$ are in the R sector and the configurations with $\eta = -1$ are in the NS sector. Dirichlet boundary conditions correspond to picking the minus in (\ref{eq:boundarycond}) and Neumann boundary conditions correspond to picking the plus. Solving the equations of motion we obtain a single set of oscillation modes $b_k^\mu$. 
Depending on the boundary conditions $k$ is either integer or half-integer
\begin{equation}
	\text{NN}/\text{DD}: \quad \begin{cases}  k \in \mathbb{Z}  & \text{R} \\
		 k \in \mathbb{Z}/2 & \text{NS}  
	\end{cases} \,, \qquad \text{ND}/\text{DN}: \quad \begin{cases}  k \in \mathbb{Z}  & \text{NS} \\
	k \in \mathbb{Z}/2 & \text{R}
	\end{cases}
\end{equation}
Moreover, it is important to notice that for DD and DN boundary conditions, the weights of the right-moving modes are minus the weights of the left-moving modes, i.e. we have in the expansion $-b_k^\mu$. Performing the light-cone quantization we end up with the creation operators $b^i_k$, $i = 1,\dots d-2$. The mass-squared operator is
\begin{equation}
	\alpha' m^2  =  N_\alpha +  N_b +  a + \frac{\nu_{NS}}{8} +  \alpha' \, T^2 (\Delta X)^2 \,,
\end{equation}
where $N_\alpha$ and $N_\beta$ are the number operators for the transverse modes and $(\Delta X)^2 $ is the sum of the squared distance of the string endpoints in the DD directions. In the R sector $a = \nu_{NS} = 0$; in the NS sector $a = -1/2$ and $\nu_{NS}$ is equal to the number of directions with mixed boundary conditions.

In the heterotic string theory we cannot separate the left- and right-moving modes and we do not have open strings. We consider then type II theories with strings ending on a single D$p$ brane. We introduce the index $i = 1\,\dots p-1$ and $a = 1\,\dots 9-p$. Using GSO projections, in the NS sector we can always remove the ground state. Then, we have $b^i_{-1/2} \ket{0}_{NS}$ and $b^a_{-1/2} \ket{0}_{NS}$ which are a massless vector and $9-p$ scalars under the little group $SO(p-1)$.\footnote{Left- and right-moving modes are not independent. Now we use $\ket{0}_{NS}$ to indicate the tensor product of left and right NS vacuum.} The lowest state one can build in the R sector is always the vacuum. In particular, for an even number of DD directions we can construct $\ket{\pm}_R \otimes \ket{\pm}_R $ and for an odd number of DD directions we can construct $\ket{\pm}_R \otimes \ket{\mp}_R $.\footnote{Essentially, every sign difference in front of the mode weights in the Ramond sector with DD boundary conditions introduce a sign into the way the $(-1)^F$ operator acts on the right vacua. This depends on the details of the construction of the $F$ operator.} The ground state massless spinor organizes into spinorial representations of the little group $SO(p-1)$. We obtain the fields living on the $(p+1)$-dimensional worldvolume
\begin{equation}
	A \,, \qquad \phi^{a} \,, \qquad \lambda^j \,,
\end{equation}
for $p$ even in the type IIA and $p$ odd in the type IIB theory. These are exactly the states of a $U(1)$ supermultiplet with 16 supercharges in $p+1$ dimensions. 

\section{Construction of effective actions}

In this section we explain how to obtain a spacetime effective action for string theory. More precisely, we want an action whose dynamical field content is given by the massless spacetime fields contained in the string spectrum and whose amplitudes reproduce those of the world-sheet action up to a certain order in the perturbative expansion. We will see that the construction of the effective action is not straightforward and some assumptions are necessary. 

In order to construct an effective field theory (EFT) we first need to understand how to compute string theory scattering amplitudes. We focus on the the simplest case of closed bosonic string theory. The starting point is the assumption that the conformal symmetry of the classical theory is not anomalous, i.e. it is not broken by quantum effects. Then, exploiting the correspondence of states and vertex operators of CFTs, we can compute amplitudes considering an Euclidean path integral with the insertion of proper vertex operators. For the scattering of $n$ particles of type $\{\Lambda_i\}$ and momentum $\{k_i\}$ with $i = 1, \dots n$ whose interactions are governed by the Polyakov action (\ref{eq:polyakovaction}) with flat spacetime metric $g_{\mu\nu} = \eta_{\mu\nu}$, we can write
\begin{equation}
	\mathcal{A}(\Lambda_1, k_1,\dots ,\Lambda_n, k_n) = \int \mathcal{D}  \gamma_{\alpha\beta} \, \mathcal{D}X^\mu \, e^{-S_{P,E}} \, \prod_{i=1}^n  V_{\Lambda_i}(k_i) \,,
\end{equation}
where $S_{P,E}$ is the Euclidean Polyakov action and the $V_{\Lambda_i}$ are the $n$ vertex operators representing the absorption or the emission of the scattered string states. In order to match spacetime amplitudes with the world-sheet ones, we have to identify the vertex operators corresponding to the massless string states. In the case of the bosonic string theory we have the graviton $g_{\mu\nu}$, the KR 2-form $B_{\mu\nu}$ and the dilaton $\phi$. The explicit form of the graviton vertex operator is
\begin{equation}
	V = -\frac{1}{4\pi \alpha'} \int d^2 \xi  \sqrt{|\gamma|} \,\gamma^{\alpha\beta} \partial_\alpha X^\mu \partial_\beta X^\nu h_{\mu\nu} \,,
\end{equation}
with $h_{\mu\nu}$ a symmetric matrix. Notice that these operators are in general non-local. If we exponentiate the graviton operator we expect to describe a coherent state of gravitons. And indeed we can write
\begin{equation}
	\int \mathcal{D}  \gamma_{\alpha\beta} \, \mathcal{D}X^\mu \, e^V e^{-S_{P,E}}  = \int \mathcal{D}  \gamma_{\alpha\beta} \, \mathcal{D}X^\mu \, e^{-S_{P,E}'}  \,,
\end{equation}
where $S_{P,E}'$ is the Polyakov action (\ref{eq:polyakovaction}) in Euclidean signature with spacetime metric $g_{\mu\nu} = \eta_{\mu\nu} + h_{\mu\nu}$. The path integral is now describing string states propagating in curved spacetime. With the insertion of proper operators describing dilatons and KR field coherent states, we eventually end up with (in Minkowski signature)
\begin{equation} \label{eq:poliakovCurved}
	S_P'' = -\frac{1}{4\pi \alpha'} \int d^2 \xi \sqrt{|\gamma|} \left[\gamma^{\alpha\beta} \partial_\alpha X^\mu \partial_\beta X^\nu g_{\mu\nu} + \epsilon^{\alpha \beta} \partial_\alpha X^\mu \partial_\beta X^\nu B_{\mu\nu} + \alpha' \phi \, R \right] \,,
\end{equation}
where $\epsilon^{\alpha\beta}$ is the world-sheet Levi-Civita tensor and $R$ is the world-sheet Ricci scalar. This action is the most general renormalizable action one can consider for a non-linear sigma model. It is referred to as the Polyakov action on curved background. 

We want now to analyze the structure of the amplitudes perturbative expansion. In the Polyakov action we have the coupling $\alpha'$. As usual, its powers organize the amplitudes loop expansion. Since $\alpha'$ is dimensionful, in the induced EFTs $\alpha'$ corrections will correspond to the insertion in the action of higher derivative operators. However, $\alpha'$ is not the unique parameter we have. To show this explicitly we restrict to the Polyakov action with vanishing KR field and constant dilaton $\phi = \phi_0$. The action can be written (up to total derivatives) as the sum of the Polyakov action (\ref{eq:polyakovaction}) and a term with form $\phi_0 \chi$ where
\begin{equation}
	\chi(\Sigma) = \frac{1}{4\pi}\int_\Sigma d^2\xi  \, R + \frac{1}{2\pi} \int_{\partial \Sigma} ds \, k \,,
\end{equation}
is the Euler constant of the surface $\Sigma$ describing the closed string states scattering. $k$ is the world-sheet metric extrinsic curvature. Expanding the world-sheet metric measure using the fixed topology measure element $\mathcal{D}\gamma_{\alpha\beta}^{\,g}$, we can write the path integral as 
\begin{equation}
	\int \mathcal{D}\gamma_{\alpha\beta} \, \mathcal{D}X^\mu \, e^{-S_{P,E} - \phi_0 \chi} = \sum_{g} g_s^{-\chi(\Sigma_g)} \int \mathcal{D} \gamma_{\alpha\beta}^{\,g} \, \mathcal{D} X^\mu \, e^{-S_{P,E}}
\end{equation}
where $g_s = e^{\phi_0}$ is the \textit{string coupling constant} and $g$ is a parameter labeling the different topologies. In scatterings of $n$ oriented and closed strings,  the surface $\Sigma_g$ can be always conformally mapped to a compact genus $g$ surface with $n$ punctures whose Euler characteristic is given by $\chi(\Sigma_g) = 2 - 2 g - n$. We find that $g_s$ powers are also associated with loop counting.  Although both $g_s$ and $\alpha'$ can be used as expansion parameters, their interpretation is different. $\alpha'$ is a truly independent physical parameter defining the theory. $g_s$ instead labels different vacua of the same theory.

The loops of $\alpha'$ and $g_s$ have a different nature. $g_s$ corrections are sensitive to the loops in the world-sheet of the scattering of closed string states. $\alpha'$ corrections are sensitive to the loop corrections of the scattering of classical fields states on a local patch of the world-sheet. To see the latter explicitly, we consider the normal coordinate expansion of the world-sheet scalars around a classical solution
\begin{equation}
	X^\mu = X^\mu_{\text{cl}} + \pi^\mu + \Gamma^\mu_{\rho\sigma} \pi^\rho \pi^\sigma + \dots \,.
\end{equation} 
and we replace it in the Polyakov action $S_{P,E}$ in conformal gauge, with non trivial target space metric $g_{\mu\nu}$. We obtain
\begin{equation}
	S_{P,E}[X] = S_{P,E}[X_{\text{cl}}] + \frac{1}{4\pi \alpha'} \int d^2 \xi  \left[ \partial \pi \cdot \partial \pi + A_{\mu\nu}  \pi^\mu \pi^\nu + A_{\mu\nu i} \partial^i \pi^\mu \pi^\nu + \dots \right] \,,
\end{equation}
where the dots indicate terms with a higher number of $\pi^\mu$s, and the coefficients $A$ are combinations of derivatives of $X^\mu_{\text{cl}}$ and of the background metric $g_{\mu\nu}$ evaluated on the classical solution $X_{\text{cl}}^\mu$. If we rescale the $\pi^\mu$s to absorb the $\alpha'$ factor, we obtain that a term with $n$ copies of $\pi$ carries a factor $(\alpha')^{n/2}$. Such terms can be seen as interaction terms for the classical fields contained into the A coefficients. The $\pi^\mu$s represent the degrees of freedom propagating into an irreducible scattering amplitude of classical fields.

The program of precisely matching the EFT and the world-sheet amplitudes is not straightforward. If we impose specifically the cancellation of the Weyl anomaly for the Polyakov action in curved spacetime (\ref{eq:poliakovCurved})  we obtain constraints over $(g_{\mu\nu}, B_{\mu\nu}, \phi)$. In particular, Weyl symmetry is not broken as long as the trace of the stress energy tensor is vanishing. The quantum-corrected trace of the stress energy tensor can be written as
\begin{equation} \label{eq:traceStressenergy}
	2 \alpha' T^\alpha{}_\alpha = \alpha' \beta^\phi R + \beta_{\mu\nu}^g \gamma^{\alpha\beta} \partial_\alpha X^\mu \partial_\beta X^\nu + \beta_{\mu\nu}^B \epsilon^{\alpha \beta} \partial_\alpha X^\mu \partial_\beta X^\nu  = 0 \,.
\end{equation} 
The relation (\ref{eq:traceStressenergy}) is exact to all orders in the perturbative expansion parameter $\alpha'$ for each choice of topology. The absence of anomalies then is equivalent to requiring $\beta^\phi = \beta_{\mu\nu}^g =  \beta_{\mu\nu}^B = 0$. The $\beta$s are quantities which only involve the spacetime fields  $(g_{\mu\nu}, B_{\mu\nu}, \phi)$ and, at leading order in $\alpha'$, their explicit form in the critical dimension $d=26$ is
\begin{subequations}
	\begin{align}
		\beta_{\mu\nu}^g & = \alpha' \left(R_{\mu\nu} - \frac{1}{4} H_{\mu}{}^{\rho\sigma} H_{\nu \rho \sigma} + 2 \nabla_\mu \nabla_\nu \phi \right) + \mathcal{O}(\alpha'{}^2) \,, \\[2mm]
		\beta_{\mu\nu}^B & = 	\alpha'\left(-\frac{1}{2}\nabla_\lambda H^\lambda{}_{\mu\nu} + H^\lambda{}_{\mu\nu}\nabla_\lambda \phi\right) + \mathcal{O}(\alpha'{}^2) \,,\\[2mm]
		\beta^\phi & = \alpha' \left((\nabla\phi)^2 - \frac{1}{2}\nabla^2\phi - \frac{1}{24}H^2\right) + \mathcal{O}(\alpha'{}^2) \,,
	\end{align}
\end{subequations}
where curvature tensors and covariant derivatives are built with the spacetime metric $g_{\mu\nu}$. The relations we obtain by imposing the vanishing of the $\beta$s can be interpreted as equations of motion of an effective theory. It is pretty simple to verify that such a theory is given by
\begin{equation}
	S = \frac{g_s^2}{2\kappa^2_{26}}\int d^{26}x \sqrt{|g|} e^{-2\phi} \left[R+4(\partial \phi)^2 - \frac{1}{12} H^2 + \mathcal{O}(\alpha')\right] \,.
\end{equation}
where $\kappa_{26}$ is the gravitational coupling constant. We want to see how $\alpha'$ and $g_s$ combine to produce such a constant in generic dimension $d$. In the scattering of $n$ gravitons we have $n-2$ vertices, each of them carrying a factor of $\kappa_d$. 
If we have no loops, the coupling constant of the process is then $\kappa_d^{n- 2}$, which reduces to $\kappa_d^{-2}$ if we normalize the external legs. From the world-sheet perspective, with $g = 0$ we have a factor $g_s^{-2}$. The rescaling of the $\pi$s we performed induces a rescaling of the path integral measure and we obtain an overall factor of $\alpha'{}^{-d/2}$. Finally, even if we are at tree-level in world-sheet loops, in order to realize the tree-level scattering of $n$ classical gravitons with need $n-2$ vertices and at least 1 loop in the $\pi^\mu$s. The vertices with two $\pi^\mu$s do not add factors of $\alpha'$,  but the loop gives us an extra $\alpha'$. We obtain that the EFT amplitudes match at tree-level the world-sheet ones provided that 
\begin{equation}
	\kappa^2_d \sim g_s^2 \alpha'{}^{\frac{d-2}{2}} \,.
\end{equation}
Notice that $\kappa_d$ has dimension of $[L]^{(d-2)/2}$ and the expression is consistent.

The procedure described so far, based on beta functions, is useful to extract the leading order spacetime action of bosonic string theory but its extension to more general cases is much more complicated. Moreover, it is not clear in this perspective how to implement stringy effects which are sensitive to the world-sheet topology. Other successful methods used in the literature exploit, for instance, the assumption of the existence of quantum-protected symmetry and dualities, such as supersymmetry and T-duality. Instead of determining the whole action with a top-down computation, one determines just a few terms with a top-down approach and then reconstructs the missing ones restoring the broken symmetries or dualities.  

\section{Non perturbative aspects of string theory}

\subsection{Dualities}

The five supersymmetric 10-dimensional string theories we presented are not independent theories. They are nodes of a web of dualities. In this section we briefly describe one of the dualities involved, namely T-duality. We will use it extensively in the main text. Then, we briefly present the relations among closed 10-dimensional superstring theories. 

\subsubsection{T-duality}

T-duality is a map between two theories compactified on a circle S$^1$. It is a perturbative duality. We will have to apply it only to the common bosonic sector, therefore we restrict ourselves to study T-duality in bosonic string theory.

T-duality can be understood in several ways. The first description is from the point of view of the spectrum. On a compact direction we must impose a new type of boundary condition. Assuming that $X^{d-1} \sim X^{d-1} + 2\pi R $, with $R$ the radius of the compact dimension, we impose  $X^{d-1}(\xi^0 ,\xi^1 ) = X^{d-1}(\xi^0 ,\xi^1 + \ell ) + 2 \pi R n $, where $n\in\mathbb{Z}$ is the string winding number. The compactness of $X^{d-1}$ also implies the quantization of the momentum along such direction. We have $p^{d-1} = m/R$, with $m\in\mathbb{Z}$ the quanta of momentum carried by the string. The mass-squared operator for closed strings becomes
\begin{equation}
	\alpha' m^2  = 2  \left( N_\alpha + \bar{N}_{\bar{\alpha}} + 2 a \right)  + \frac{R^2}{\alpha'} m^2 + \frac{\alpha'}{R^2} n^2 \,,
\end{equation}
with level-matching condition $N_\alpha - \bar{N}_{\bar{\alpha}} = m n $. T-duality is the map  
\begin{equation}
	m \leftrightarrow n \,, \qquad R \leftrightarrow \alpha'/R \,,
\end{equation}
which exchanges winding and momentum quanta and inverts the radius of the compact direction. At the level of the oscillators it can be seen as the map which flip the sign of the right modes 
\begin{equation}
	(X_L^\mu, \psi_L^\mu) \rightarrow (X_L^\mu, \psi_L^\mu) \,, \qquad (X_R^\mu, \psi_R^\mu) \rightarrow (- X_R^\mu, - \psi_R^\mu)  \,.
\end{equation}
Notice that for open strings T-duality exchanges D and N boundary conditions.

A second description is in term of transformations of world-sheet background fields. Assuming that the compact direction is an isometry direction of the target space fields and using adapted coordinates, the transformations of the common bosonic sector are encoded in the Buscher rules \cite{Buscher:1987sk,Buscher:1987qj} (we split the indices as $\mu = (i,z),$ where $z$ corresponds to the compact direction)
\begin{equation} \label{eq:busch}
	\begin{split}
		& g_{zz}' = 1/g_{zz} \,,  \hspace{5.35cm}  B_{iz}' = g_{iz}/g_{zz} \,, \\[2mm]
		& g_{iz}' = B_{\mu z} /g_{zz} \,,  \hspace{5cm} B'_{ij} = B_{ij} + 2 g_{[i|z} B_{j]z}/g_{zz} \,, \\[1mm]
		& g_{ij}' = g_{ij} -  (g_{iz}g_{jz}- B_{iz}B_{jz})/g_{zz} \,, \hspace{2cm} \phi' = \phi - \frac{1}{2} \log |g_{zz}| \,.
	\end{split}
\end{equation}
The procedure to derive the transformation properties of the metric and the KR field is the following. We start by gauging the translations along the isometry direction. We then impose that the introduced connection is pure gauge via the insertion of a proper Lagrange multiplier $\tilde{X}^z$. We remove now the connection replacing its equations of motion and we set to zero $X^z$ given that now its equations of motion are trivially satisfied. Imposing that the new action has the same form of the starting one with $X^z$ replaced by $\tilde{X}^z$ we recover the relations (\ref{eq:busch}). In order to derive the transformation rule of the dilaton it is necessary to take into account quantum effects and consider the path integral. The transformation is necessary to compensate the Jacobian of the change of coordinates of the path integral measure. Notice that the transformations receive $\alpha'$ corrections when applied to actual solutions of the spacetime EFTs, unless the higher order $\beta$ functions vanish exactly. In terms of lower-dimensional fields and using the dictionary of appendix \ref{sec-dictionary}, the Buscher rules at leading order in $\alpha'$ are simply 
\begin{equation}
	A \leftrightarrow C \,, \qquad k \leftrightarrow 1/k \,,
\end{equation}
where $A$ is the Kaluza--Klein (KK) vector, $C$ is the winding vector and $k$ is the KK scalar. The lower dimensional perspective is particularly efficient to study the transformations properties of the other fields. 

Notice that T-duality can be generalized to an $O(n,n)$ rotation for compactifications on $T^n$ and also admit non-Abelian formulations.

\subsubsection{Relations among supersymmetric string theories}

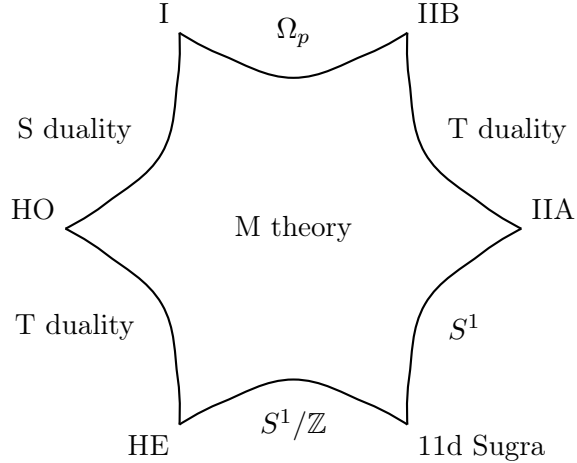
\begin{figure}[h]
	\centering
	\begin{tikzpicture}
		\coordinate[label=above right: IIA] (A) at (0:3);
		\coordinate[label=above right: IIB] (B) at (60:3);
		\coordinate[label=above left: I] (C) at (120:3);
		\coordinate[label=above left: HO] (D) at (180:3);
		\coordinate[label= below left: HE] (E) at (240:3);
		\coordinate[label= below right: 11d Sugra] (F) at (300:3);
		
		\coordinate (center) at (barycentric cs:A=1,B=1,C=1,D=1,E=1,F=1);
		\node at (center) {M theory};
		
		\node at ($(A)!0.5!(B)$) {\hspace{1cm} T duality};
		\node at ($(B)!0.5!(C)$) {$\Omega_p$};
		\node at ($(C)!0.5!(D)$) {S duality \hspace{1cm} $ $};
		\node at ($(D)!0.5!(E)$) {T duality \hspace{1cm} $ $};
		\node at ($(E)!0.5!(F)$) {$S^1/\mathbb{Z}$};
		\node at ($(F)!0.5!(A)$) {$S^1$};
		
		\coordinate (A') at (7.5:2.5);
		\coordinate (B') at (67.5:2.5);
		\coordinate (C') at (127.5:2.5);
		\coordinate (D') at (187.5:2.5);
		\coordinate (E') at (247.5:2.5);
		\coordinate (F') at (307.5:2.5);
		
		\coordinate (A''') at (-7.5:2.5);
		\coordinate (B''') at (52.5:2.5);
		\coordinate (C''') at (112.5:2.5);
		\coordinate (D''') at (172.5:2.5);
		\coordinate (E''') at (232.5:2.5);
		\coordinate (F''') at (292.5:2.5);
		
		\coordinate (A'') at (30:2);
		\coordinate (B'') at (90:2);
		\coordinate (C'') at (150:2);
		\coordinate (D'') at (210:2);
		\coordinate (E'') at (270:2);
		\coordinate (F'') at (330:2);

		
		\draw[black, thick, smooth] plot [smooth,tension=0.7] coordinates { (A) (A') (A'') (B''') (B) };
		\draw[black, thick, smooth] plot [smooth,tension=0.7] coordinates { (B) (B') (B'') (C''') (C) };
		\draw[black, thick, smooth] plot [smooth,tension=0.7] coordinates { (C) (C') (C'') (D''') (D) };
		\draw[black, thick, smooth] plot [smooth,tension=0.7] coordinates { (D) (D') (D'') (E''') (E) };
		\draw[black, thick, smooth] plot [smooth,tension=0.7] coordinates { (E) (E') (E'') (F''') (F) };
		\draw[black, thick, smooth] plot [smooth,tension=0.7] coordinates { (F) (F') (F'') (A''') (A) };
	\end{tikzpicture}
	\caption{M theory and its 10- and 11- dimensional supersymmetric limits and dualities.} \label{fig:dualities}
\end{figure}

10-dimensional superstring theories are related by a web of dualities (see fig. \ref{fig:dualities}). Type IIA compactified on a circle with radius $R_{10}$ is T-dual to type IIB compactified on the circle with radius $\alpha'/R_{10}$. At strong $g_s$ coupling the massive modes of type IIA cannot be ignored. In particular, the solitonic modes become light and the mass scaling of the lightest tower (represented by D0 branes) is compatible with the one of the KK modes of an 11-dimensional theory compactified on a circle with vev\footnote{Vacuum expectation value.} $R_{11} = \alpha'{}^{1/2} g_s$, signaling a decompactification limit. The resulting 11-dimensional theory is called M theory and it is interpreted as the non-perturbative completion of type IIA. M-theory at low-energy is described by the unique $\mathcal{N} =1$, 11-dimensional supergravity.\footnote{M theory contains extended objects which arise as non-perturbative, solitonic configurations in 11-dimensional supergravity. Therefore, 11-dimensional supergravity is an effective description of M theory.} M theory compactified on an interval produces the non-perturbative completion of HE.\footnote{M theory on a circle is described by the Horava–Witten theory. The effective supergravity action of this theory is determined by its matter content and it is the same one obtains for HE. Horava–Witten theory is therefore conjectured to be the non-perturbative completion of HE.} From the quotient of type IIB  with respect to the world-sheet parity operator $\Omega_p$ we obtain type I. The heterotic string theories compactified on a circle are related by a T-duality transformation. HO and type I are related by and S-duality, i.e. type I can be regarded as the strong coupling limit of HO. In particular, under S-duality $g_s$ is inverted and the fundamental string of HO is exchanged with the D1 brane of type I. Finally, type IIB has a global SL$(2,\mathbb{Z})$ symmetry which acts as S-duality. Due to this web of dualities (see fig.\ref{fig:dualities}), M theory is conjectured to be the unique theory underlying string theory. The 10-dimensional superstring theories are interpreted as its different perturbative limits.

\subsection{Extended objects}
D$p$ branes and fundamental strings (F1) are not the only extended objects of string theory. Solving  the equations of motion of the EFTs we can find extended objects with no simple world-sheet description. Moreover, exploiting the dualities of superstring theories we can predict the existence of new objects as the duals of the known ones. The goal of this section is to review the extended objects we are going to refer to in the main text, without giving a complete list. We start by recalling that, so far, we know that in all superstring theories except type I we have F1 strings. Then, in type IIA we have D$p$ branes with $p$ even, in type IIB D$p$ branes with $p$ odd, in type I, D1, D5 and D9 and in HE/HO no D$p$ branes.

Other than these extended objects, some string theories contain NS5 branes. The NS5 brane can be found as an explicit solution of the supergravity theory acting as a magnetic source for the KR field. Its existence can be predicted equivalently as the SL$(2,\mathbb{Z})$ partner of the D5 brane in type IIB. The dual of the NS5 brane should correspond to an extended object in type IIA. Given that T-duality tells us that such an object must source the magnetic charge of the KK vector we call it KK6A monopole. There is no obstruction then to add an NS5 brane also in type IIA and consider the dual state KK6B in type IIB. Notice that by construction KK6A/B can exists only in the compactified theory. If we quotient type IIB to obtain type I, we get that NS5 is projected out. Now, with an S-duality, we can map the D5 and D1 branes of type I to the NS5 and F1 of HO. If we compactify HO on a circle we can conclude that HE must contain a KK6. Introducing an NS5 in HE we obtain that also HO must have a KK6.

With an analogous reasoning we can conclude that in type IIB we need extra solitonic S7 and S9 branes. If we compactify type IIA on the circle, T-duality implies that we need an extra KK8A and an extra KK9A. Many more extended extended objects are allowed if we compactify further the superstring theories.

\section{The Swampland Program}

String theory is the best example we have of a quantum theory of gravity. It is indeed the only theory we know whose amplitudes are finite and describe graviton scatterings.\footnote{In the case of maximal supergravity, we have not encountered any divergent amplitude so far and there are arguments guaranteeing finiteness up to a certain number of loops. However, there is no proof that divergences are absent at all orders, as in  string theory \cite{Polchinski:1998rq}.} However, in order to connect with experiments we need to reduce the number of spacetime dimensions to 4, compactifying the extra ones. This process produces as many different IR limits as possible vacuum geometries. The collection of all the EFTs which arise as an IR limit of string theory is called the \textit{string landscape}. Despite that the string landscape is huge on its own, there are a lot of field theories which cannot be derived from string theory or, more generally, from a theory of quantum gravity (i.e. a UV-complete theory containing gravity). The collection of the field theories which are not EFTs of theory of quantum gravity is called the \textit{swampland}. In this framework takes place the \textit{swampland program} which aims to clarify the criteria that allow to identify which field theories cannot be consistently obtained from a theory of quantum gravity. A common practice in the field is to call EFTs all the theories which are not UV-complete. Thus, an EFT is in the swampland if it is a theory which is not UV-complete and cannot be uplifted to a theory of quantum gravity.

Due to their heuristic nature, swampland's criteria are often formulated as conjectures, supported by a wide collection of arguments and examples but without a rigorous proof.  Such arguments are usually of three kinds: conditions derived from microscopic models, common characteristics expressed by string theory vacua and constraints derived from EFTs. Moreover, it turns out that the more rigorous the arguments, the less the conjectures constrain EFTs. Looking at the criteria individually it is therefore not evident that the swampland program is a reliable approach to study quantum gravity. However, its relevance becomes clear once the conjectures are considered all together. They are often strictly related and point in a common direction. A single conjecture should be therefore regarded as a node of a web which is constantly expanding and collecting more and more arguments.	In the following we are going to briefly review some conjectures, including specifically those we are going to refer to in the main text. For a more comprehensive review of the subject the interested reader can check \cite{Palti:2019pca, vanBeest:2021lhn} and references therein.

\subsubsection{The main conjectures}

We can identify three main conjectures in the swampland program: the no global symmetries conjecture (NGSC), the weak gravity conjecture (WGC) and the swampland distance conjecture (SDC).
\vspace{0.2cm}
\begin{tcolorbox} 
	$\textbf{No Global Symmetry Conjecture}$ \\
	A theory coupled to gravity can-not admit global symmetries.
\end{tcolorbox}
\vspace{0.1cm}
\noindent We have evidence for the NGSC derived with different and independent approaches based on perturbative string theory (all global symmetries on the world-sheet are gauged in target space), AdS/CFT (absence of charged operator on the boundary corresponding to charged operators in the bulk\footnote{This has been proven for symmetries which are splittable on the boundary \cite{Harlow:2018jwu}.}) and black holes physics (violation of entropy bounds\footnote{A reference for this \cite{Susskind:1995da} However, in the argument proposed there might be a loophole. The modification of the Bekenstein--Hawking entropy due to quantum corrections is not discussed.}). Note that the conjecture applies at the UV scale. Therefore, we can admit global symmetries of the EFTs if they are not exact, i.e. they break in the UV.

\vspace{0.1cm}
\vspace{0.2cm}
\begin{tcolorbox}
	$ \textbf{Weak Gravity Conjecture ($d\ge4$)} $ \\
	Given a theory in $d\ge4$ spacetime dimensions coupled to gravity with a $U(1)$ gauge symmetry, with the gauge coupling g, 	
	
	\begin{itemize} 
		\item (Electric WGC) There exists a state with mass $M$ and charge $q$ satisfying 
		\begin{equation} 
			M \le \sqrt{\frac{d-2}{d-3}}\,gq \,\left(M_P^{(d)}\right)^{\frac{d-2}{2}} \,.
		\end{equation} 
		\item (Magnetic WGC) There exists a cutoff scale $\Lambda$ such that
		\begin{equation}
			\Lambda \lesssim g\left(M_P^{(d)}\right)^{\frac{d-2}{2}} \,.
		\end{equation}
	\end{itemize}	
\end{tcolorbox}	
\vspace{0.2cm}
\noindent The conjecture was originally proposed in \cite{Arkani-Hamed:2006emk}. The physical idea behind the electric WGC is that extremal charged black holes must be able to discharge\footnote{Notice that despite this assumption is reasonable, it is not clear if it is inconsistent to have black holes which are not able to discharge. A problem that can emerge from the impossibility for the black holes to discharge, is the existence of stable remnants. Such stable remnants may affect the pair production rate of the black hole \cite{Banks:1992is,Giddings:1993km} and drive the renormalized Newton constant to
	zero \cite{Susskind:1995da}.} and the precise form of the bound has been obtained considering Reissner--Nordstr\"om black hole solutions of Einstein--Maxwell theory in $d$-dimensions. The idea behind the magnetic WGC is that a black hole should be able to get rid of its magnetic charge too. The precise relation has been obtained assuming that magnetic monopoles are generated at a cutoff scale $\Lambda$ as solitonic configurations. Notice that the magnetic WGC can be interpreted as a sharpened version of the NGSC: in the limit $g = 0$ the gauge symmetry becomes global and the cutoff scale goes to zero. The statements are supported by a variety of string theory constructions. 
\vspace{0.1cm}
\begin{tcolorbox} 
	$\textbf{Swampland Distance Conjecture }$ \\
	Consider a theory of gravity with moduli space $\mathcal{M}$ which is parameterized by the expectation value of some free field $\phi^i$. Then
	\begin{itemize}
		\item $\forall$ $P\in\mathcal{M}$, $s>0$, $\exists$ $Q_s \in \mathcal{M}$ such that $d(P,Q_s) > s$, i.e. every configuration admits a boundary at infinite geodesic distance.
		\item There exist an infinite tower of states with mass scale $M(Q_\infty)$ such that
		\begin{equation}
			M(Q_{\infty})\sim M(P)e^{-\beta d(P,Q_{\infty})}\,, \quad \quad \beta > 0\,.
		\end{equation}  
	\end{itemize}
\end{tcolorbox}	
\vspace{0.1cm}
\noindent The conjecture was originally proposed in \cite{Ooguri:2006in}. A priori, we do not have an argument to say that a consistent theory of gravity cannot be formulated in $d=4$ dimensions without an infinite tower of states. However, inspired by the consistency of string theory, it is conjectured that UV-complete theories must contain an infinite tower of states. From explicit examples we know that the exponential scaling is typical of towers which arise from compactifications, but the existence of a light tower for every asymptotic region specifically requires string theory. Recently some bottom-up argument have been proposed which allow us to recover the exponential behavior.\footnote{They exploit the covariant entropy bounds \cite{Calderon-Infante:2023ler}, the species scale properties \cite{vandeHeisteeg:2023ubh} and the thermodynamic properties of the so-called minimal black holes \cite{Basile:2023blg}.}

\subsubsection{Refinements}

The WGC admits several refinements. If we consider theories with $U(1)^N$ gauge group we have to introduce the concept of a vector of charge-to-mass ratios  $\vec{Z} = \vec{Q}/M$. Then, to ensure the discharge of an extremal black hole, we have to require that there exist states with charge-to-mass ratio vectors $\vec{z}_i$ such that their convex hull contains $\vec{Z}$. If the theory has a higher-dimensional origin, we expect the WGC to be realized in the higher-dimensional setup. We find that an EFT must have not only a single state satisfying the WGC, but an entire tower. In the case of the gauge group $U(1)^N$, instead of a tower we need a lattice and the conjecture has the name of \textit{lattice weak gravity conjecture}. If the states which satisfy the WGC are smaller extremal black holes we say that the conjecture is satisfied in a \textit{mild} form.
If we now consider theories with scalar interactions we expect further refinements of the WGC. In particular, we have to take into account that the mass $M$ of the emitted states may depend on the value of the scalar fields $\phi^i$. This may generate non trivial differences in the results one obtains with extended states and probe particles. Given a black hole with scalar hair, a probe particle can feel a Yukawa force arising from the variation of its mass $M(\phi)$. Instead, in a multi-center black hole solution, the value of the mass of the black hole might be independent of the profile of $\phi$, due to the fact that regularity at the event horizon fixes the value of the scalar fields in term of the gauge charges of the system. The statement that there must always exist at least one  state whose charge-to-mass ratio makes it self-repulsive (taking into account the Yukawa interactions) is called \textit{repulsive force conjecture}. Finally, we can generalize the WGC and its refinements to the case of a black brane. Again, we demand the existence of extended objects which allow the extremal branes to discharge. The relevant ratio is now $Q/T$, where $T$ is the extended object tension. The WGC has a strong version called \textit{sharpened weak gravity conjecture}. It  consists in the requirement that the bound is saturated only by supersymmetric states. 

Combining the strong version of the WGC and its formulation for branes one can conclude that AdS vacuum must be non-perturbatively unstable. 
\vspace{0.1cm}
\begin{tcolorbox} 
	$\textbf{AdS Instability Conjecture }$ \\
	Any non-supersymmetric AdS vacuum is at best metastable and has  eventually to decay.
\end{tcolorbox}	
\noindent The conjecture was originally proposed in \cite{Ooguri:2016pdq}. The idea is that in non-susy AdS vacua it is possible to nucleate bubbles whose boundaries are non-BPS branes which satisfy the WGC but do no saturate the bound. For such branes it is energetically favorable to increase the bubble radius. The initial vacuum is then metastable because it will be replaced by the vacuum contained inside the bubble. 

The last conjecture we want to review is the \textit{AdS distance conjecture}. It states that 
\vspace{0.1cm}
\begin{tcolorbox} 
	$\textbf{AdS Distance Conjecture }$ \\
	In AdS spacetime there is always a tower of states whose mass scale $m$ is related to the cosmological constant $\Lambda$ by
	\begin{equation}
		m \sim |\Lambda|^\alpha \,,
	\end{equation}
	with $\alpha$ a positive order-one number.
\end{tcolorbox}	
\noindent
The conjecture was originally proposed in \cite{Lust:2019zwm}. This statement is based mainly on string theory models. However, it may also be interpreted as a refinement of the SDC. Indeed, assuming that $\Lambda$ is a function of the moduli (as it happens in string theory examples), we can associate the limit $\Lambda \rightarrow 0$ to  an infinite distance limit in the moduli space. Then, a tower of states becomes light with vanishing $|\Lambda|$. The \textit{AdS distance conjecture} admits a \textit{strong} version saying that $\alpha = 1/2$ for susy vacua, not allowing for scales separation. 

The absence of scale separation signals that an EFT has no predictive power at best; at worst, it is inconsistent. Indeed, the cosmological constant scale corresponds to the minimal mass a particle can have in order to have Compton length shorter than the size of the observable universe. At the same time, given that EFTs works in regimes where the infinite tower of states can be integrated out, they cannot be valid above the mass scale of the tower. In order to be consistent and compatible with our universe, the cosmological constant scale and the mass scale of tower must be separated.\footnote{See \cite{Coudarchet:2023mfs} for a recent review on the state of the art of scale separation in string theory.}	
	 
	\clearpage{\pagestyle{empty}\cleardoublepage}

	\part{Black Hole Solutions} \label{part:TO}
	
		\chapter{Revisiting Wald's Thermodynamics} \label{ch:2}

We review Wald's formalism for black hole thermodynamics, pointing out some limitations of the standard formulation and the proposals adopted to extend it. One of the main problems is that the entropy obtained by  Iyer and Wald is not gauge invariant in presence of matter fields with gauge freedoms. A possible way to solve this issue is via the introduction of gauge-covariant Lie derivatives. We put on solid ground such proposal showing that the gauge-covariant Lie derivative naturally emerges in the context of Kaluza--Klein dimensional reduction. 

A second problem is the lack of a prescription to identify certain terms which should appear in the first law of thermodynamics. The most elusive ones are the scalar charges, because, for a long time, they were missing a coordinate-independent definition. We propose a possible definition for them as the integral of a closed form built out of the isometries of the scalar target space and spacetime. As a byproduct, we prove that the scalar charges are not independent quantities and are fixed by the values of the gauge charges (if present) at the black hole horizon. 

We illustrate the details and subtleties of the extended version of Wald's formalism studying the thermodynamics of a symmetric $\sigma$-model coupled to gravity and Abelian gauge fields in four dimensions. In such a context we prove explicitly some properties of the electromagnetic potentials which are relevant for rotating black holes. 

\section{Review of black hole thermodynamics} \label{sec-reviewBHthermody}

In this section we review some basic facts about black hole solutions and their physics. After recalling the notion of a black hole and the solutions available in General Relativity, we will review the laws of black hole mechanics and their thermodynamic interpretation. Then, we will briefly describe Wald's formalism and black hole thermodynamics in theories with higher derivatives. We conclude pointing out some limitations of Wald's formalism and how it is possible to overcome them properly modifying Wald's method.

\subsection{Basic properties of black holes}

A black hole is a solution of an effective theory of gravity in $d$-dimensions which represents a configuration of the gravitational field (the metric) with certain properties. A first precise characterization of such properties for asymptotically flat configurations is due to Penrose \cite{Penrose:1968ar}, which defined a black hole as the set of points of a spacetime which cannot be connected to the future null infinity through an outgoing null geodesic. The boundary $\mathcal{H}$ of a black hole is a null $(d-1)$-dimensional hypersurface called event horizon. 

This formal characterization might be obscure at first glance, but we can easily extract some insights of why a black hole has such a name. The fact that null geodesics passing through a black hole cannot reach the future null infinity means that the black hole represents a configuration whose gravitational fields is so strong that everything crossing the event horizon $\mathcal{H}$ is trapped and not even light can escape.

If we restrict ourselves to General Relativity in $d=4$ dimensions and we consider stationary spacetimes we have several results which further characterize the allowed black hole solutions. Collectively, such results can be summarized in a statement on the structure of the event horizon $\mathcal{H}$ and in a uniqueness theorem for the black hole solutions. Starting with the characterization of the horizon, the request of a stationary spacetime is equivalent to having an isometry whose generator is asymptotically timelike. The rigidity theorem \cite{Hawking:1971vc,Hawking:1973uf} proves that a stationary black hole spacetime which solves Einstein equations in vacuum must be either static or axisymmetric; $\mathcal{H}$ must be a Killing horizon and the associated Killing vector is a linear combination of two commuting Killing vectors: a spacelike Killing vector generating a $\text{U}(1)$ isometry and a timelike one. Regarding the uniqueness of the solutions, the first solution discovered is the Schwarzschild black hole \cite{Schwarzschild:1916uq}. The solution has been proven to be the unique spherically-symmetric black hole solution of General Relativity \cite{Birkhoff:1923} and the unique static black hole solution \cite{Israel:1967wq}. The rotating generalization of the the Schwarzschild solution is Kerr black hole \cite{Kerr:1963ud} and it is the unique stationary, rotating black hole solution of General Relativity \cite{Carter:1971zc,Robinson:1975bv}. Coupling General Relativity to Maxwell theory some of the previous theorems have a natural extension \cite{Israel:1967za}. The most general static black hole solution is the Reissner--Nordstr\"om black hole \cite{Reissner:1916cle,Nordstrom:1918}, which is a charged generalization of the Schwarzschild solution, and the Majumdar-Papapetrou solutions \cite{Papaetrou:1947ib}, which represent a configuration of several extremal Reissner--Nordstr\"om black holes in equilibrium. The most general axisymmetric black hole solution is the Kerr-Newmann black hole \cite{Kerr:1963ud}, which represents a charged and rotating configuration. Considering other kind of matter coupled to General Relativity, one may expect that it is always possible to find black hole solutions. This expectation is in general satisfied. However, the solutions may not represent a non trivial generalization of those we just described. It is conjectured that asymptotically-flat black hole solutions are uniquely specified once mass, angular momenta, electric charges and magnetic charges are fixed.\footnote{This implies in particular, the black hole solutions of a theory with scalar fields are not characterized by a new independent charge associated to the scalar fields profiles. See section \ref{sec-simpleexample} for further details.} Such conjecture is often refereed to as "no-hair theorem" \cite{Ruffini:1971bza}.

An interesting property of black hole solutions of General Relativity is that they satisfy a set of relations involving variations of geometrical quantities. In the literature they are called the laws of black hole mechanics. The first law is an identity originally proved for symmetric perturbations\footnote{The perturbed solutions has the same isometries of the unperturbed one.} of Kerr black holes \cite{Bardeen:1973gs}. The variation of the area of a black hole $\delta A$, the surface gravity of its Killing horizon $\kappa$, the variation of the mass $\delta M$, the variation of the angular momentum $\delta J$ and the horizon angular velocity $\Omega$ combine into 
\begin{equation}
	\delta M = \frac{\kappa}{8 \pi G_N } \delta A + \Omega \, \delta J \,,
\end{equation}
where $G_N$ is the Newton constant. Such relation has been further proven to be valid for accretion processes and generalized to setups with matter\footnote{We are adding to General Relativity a matter content in the form of an action for other fundamental fields.} \cite{Hawking:1972hy, Sudarsky:1992ty}. The second law \cite{Hawking:1971tu} says that if Einstein equations hold and matter satisfies the null energy condition, the area of the spacelike sections of the event horizon cannot decrease moving towards future null infinity. It can be written as $\delta A \ge 0  $. The zeroth and the third law, both involve the surface gravity $\kappa$. The former says that $\kappa$ is constant along the event horizon and it has been proven with some kinds of matter satisfying the dominant energy condition \cite{Bardeen:1973gs}. The latter says that a process that reduces $\kappa$ to zero requires an infinite amount of time. The third law was proposed in \cite{Bardeen:1973gs} and a proof assuming the weak energy condition was given in \cite{Israel:1986gqz}. However, some counterexamples have been found (see for instance \cite{Kehle:2022uvc}).

The four laws of black hole mechanics resemble the laws of thermodynamics and for this reason Bekenstein \cite{Bekenstein:1973ur} proposed to identify the area $A$ of a black hole with its entropy (up to a proportionality constants). Thus, $T \sim \kappa$. With the revolutionary work of Hawking  \cite{Hawking:1974rv,Hawking:1975vcx} such proposals have been put on solid ground. Hawking indeed proved in the semi-classical approximation\footnote{We are considering quantum matter in a purely classical spacetime.} that a black hole should emit black body radiation at the Hawking temperature $T_H$
\begin{equation}
	T_H = \frac{\kappa}{2\pi}\,,
\end{equation}
which implies that the entropy, known in the literature as Bekenstein-Hawking entropy, should be 
\begin{equation}
	S_{BH} = \frac{A}{4 G_N} \,.
\end{equation}

The results discussed so far are valid in General Relativity in $d=4$ dimension. If we consider generic EFTs of gravity in generic dimensions most of them are no longer valid or they require non-trivial modifications to be valid. Working with General Relativity but choosing $d>4$ we obtain that rigidity and uniqueness theorems no longer apply \cite{Emparan:2008eg, Horowitz:2012nnc}. In particular, the hypothesis of having stationary or static solutions do not fix anymore the event horizon topology and several extended configurations with non-trivial horizons are allowed. The number of spacetime dimensions is not affecting the thermodynamic relations and the Bekenstein--Hawking results still apply in General Relativity in $d>4$. However, if we consider EFTs of gravity containing higher-derivative terms, the first, second and third laws do not hold anymore. This can be partially overcome using Wald's formalism. With such a formalism one can show that properly adapting some definitions it is still possible to prove a first law of thermodynamics \cite{Wald:1993nt, Iyer:1994ys}. A proof of the second law in EFTs of gravity\footnote{The higher derivative corrections are treated as perturbative corrections. The second law is satisfied order by order in the perturbative expansion} is given in \cite{Wall:2015raa, Hollands:2022fkn}.

\subsection{Wald's Formalism}

In this section we want to briefly recall the basic ideas of Wald's formalism \cite{Wald:1993nt, Iyer:1994ys}. This method is general and applies to every diffeomorphism-invariant theory of gravity. In order to obtain a first law of thermodynamics during the proof we will have to restrict ourselves to stationary configurations admitting a bifurcation surface.\footnote{Wald's results are extended for an arbitrary cross-section of the event horizon in \cite{Jacobson:1993vj}.} Finally, a third implicit hypothesis is assumed by Wald, i.e. the absence of fields transforming non-trivially under gauge transformations. The consequences of dropping such hypothesis will be discussed in the next section.

The first step of Wald's formalism is the construction of a rank-$(d-1)$ conserved current exploiting the diffeomorphism invariance of the action.
We start by considering the variation of the action $S[\varphi]$ (we indicate with $\varphi$ a generic dynamical field). One obtains a term proportional to the equations of motion $\mathbf{E}_\varphi$ plus a total derivative 
\begin{equation}
	\delta S = \int \mathbf{E}_\varphi \wedge \delta \varphi + d \mathbf{\Theta}(\varphi, \delta\varphi) \,,
\end{equation}
where $\mathbf{\Theta}$ is linear in $\delta \varphi$. If we restrict ourselves to a diffeomorphism transformation $\delta_\xi x^\mu  = \xi^\mu$ we can use Noether second theorem to write (off-shell) the first term  as a total derivative which vanishes on-shell. Introducing  $\mathbf{E}_\varphi \wedge \delta_\xi \varphi = d \mathbf{S}_\xi$ we get then
\begin{equation} \label{eq_waldLHS}
	\delta_\xi S = \int d \mathbf{\Theta}'(\varphi, \delta_\xi \varphi) \,,
\end{equation}
with $\mathbf{\Theta}' = \mathbf{S}_\xi + \mathbf{\Theta}$. At the same time, acting with $\delta_\xi$ on the Lagrangian form $\mathbf{L}$ and using the Cartan magic formula we get 
\begin{equation} \label{eq_waldRHS}
	\delta_\xi S =  - \int d \iota_\xi \mathbf{L} \,.
\end{equation} 
Combining (\ref{eq_waldLHS}) and (\ref{eq_waldRHS}) we get the off-shell relation 
\begin{equation}
	\int d \left[\mathbf{\Theta}'(\varphi,\delta_\xi \varphi) + \iota_\xi \mathbf{L} \right] = 0 \,,
\end{equation} 
which implies the existence of a rank-$(d-1)$ closed current $\mathbf{J}$ and rank-$(d-2)$ current $\mathbf{Q}_\xi$ called Noether--Wald current
\begin{equation}
	\mathbf{J} = d\mathbf{Q}_\xi = \left[\mathbf{\Theta}'(\varphi,\delta_\xi \varphi) + \iota_\xi \mathbf{L} \right]  \,.
\end{equation}

The second step is the construction of two conserved rank-$(d-2)$ currents out of the Noether--Wald current. For the first one, we assume that $\xi = k$ is a Killing vector which leaves invariant all the fields, i.e. $\delta_k \varphi = 0$. We obtain on-shell
\begin{equation}
	d \mathbf{Q}_k \doteq \iota_k \mathbf{L} \,.
\end{equation}
Given that $\delta_k \mathbf{L} = -d \iota_k \mathbf{L} = 0$ we know that there exists a rank-$(d-2)$ form $\mathbf{\omega}_k$ such that $\iota_k \mathbf{L} = d \mathbf{\omega}_k $ which allows us to define the generalized Komar current \cite{Komar:1958wp}
\begin{equation}
	\mathbf{K}_k =  \mathbf{Q}_k - \mathbf{\omega}_k \,, \qquad d \mathbf{K}_k = 0 \,.
\end{equation}
For the second current we start considering the symplectic form 
\begin{equation} \label{eq_step1}
	\mathbf{\omega}(\varphi,\delta \varphi, \delta_k \varphi) = \delta\mathbf{\Theta}(\varphi,\delta_k \varphi)- \delta_k \mathbf{\Theta}(\varphi, \delta \varphi)\,.
\end{equation}
Assuming that $k$ is a Killing vector which leaves invariant all the fields we obtain that $\mathbf{\omega}(\varphi,\delta \varphi, \delta_k \varphi) = 0$. Moreover, assuming $k$ is not modified by the perturbations we are considering, i.e. $\delta k = 0$, and that $\delta\varphi$ satisfies the linearized equations of motion, we obtain\footnote{This last hypothesis is necessary to conclude that not only $\mathbf{\Theta} \doteq \mathbf{\Theta}'$, but also  $\delta\mathbf{\Theta} \doteq \delta\mathbf{\Theta}'$. This is due to the fact that $\delta \mathbf{S}_\xi$ is proportional to the linearized equations of motion \cite{PereniguezRodriguez:2022eal}.}
\begin{equation} \label{eq_step2}
	\begin{split}
		\omega(\phi,\delta \phi, \delta_k \phi) & \doteq (\delta \mathbf{J} - \iota_k \delta \mathbf{L}) + \mathcal{L}_k \mathbf{\Theta}(\phi,\delta\phi)  \\[2mm]
		& \doteq d \left[\delta \mathbf{Q}_k + \iota_k \mathbf{\Theta}(\phi,\delta \phi)\right] \,.
	\end{split}
\end{equation}
Therefore, we can define the closed rank-$(d-2)$ form 
\begin{equation}\label{eq_OmegaNaive}
	\mathbf{W}_k = \delta \mathbf{Q}_k + \iota_k \mathbf{\Theta}(\phi,\delta \phi) \,.
\end{equation}

The final step is the integration of the rank-$(d-2)$ closed forms over suitable regions of spacetime. Integrating their exterior derivative on the $(d-1)$-dimensional spacelike surface $\Sigma$ whose boundary is the union of the spatial infinity and the event horizon bifurcation surface\footnote{It is defined as the intersection of the future event horizon $\mathcal{H}^+$ and the past event horizon $\mathcal{H}^-$.} $\mathcal{BH}$ and using Stokes theorem we obtain the non-trivial relations
\begin{align}
	\int_\infty \mathbf{K}_k & = \int_{\mathcal{BH}} \mathbf{K}_k \,, \\[2mm]
	\int_\infty \mathbf{W}_k  & = \int_{\mathcal{BH}} \mathbf{W}_k \label{eq_wald3}\,.	
\end{align}
We will see explicitly with an example that the second identity is nothing but the first-law of thermodynamics. The first identity is producing instead the Smarr formula, which can be understood as the integrated version of the first law. To conclude, we present the explicit result for theories of pure gravity with a stationary black hole. In this case the Killing vector is $k = \partial_t - \Omega \partial_\phi$, where $\Omega$ is a constant which correspond to the angular velocity of the event horizon. Exploiting the linearity of $\mathbf{W}_k$ in $k$ we can separate in the LHS of (\ref{eq_wald3}) the temporal and angular parts. Then \cite{Iyer:1994ys}
\begin{subequations} \label{eq_defdeltaMdeltaJ}
	\begin{align} 
		& \delta M = \int_\infty \delta \mathbf{Q}_t + \iota_t \mathbf{\Theta}(\phi,\delta \phi) \,, \\[1mm] 
		& \delta J = \int_\infty \delta \mathbf{Q}_\phi + \iota_\phi \mathbf{\Theta}(\phi,\delta \phi) \,.
	\end{align}
\end{subequations}
For the RHS, we use the fact that $k$ vanishes on $\mathcal{BH}$ to drop $\iota_k \mathbf{\Theta}$. Then, repeating all the algorithm described distinguishing the metric from the other fields we get \cite{Iyer:1994ys}
\begin{equation}  \label{eq_defintbh}
	\delta \int_{\mathcal{BH}} \mathbf{Q}_k = \frac{\kappa}{2\pi} \delta S_W \,,
\end{equation} 
where $S_W$ is given by
\begin{equation}
	S_W = 2 \pi \int_{\mathcal{BH}} \frac{\delta \mathbf{L}}{\delta R^{ab}} n^{ab}
\end{equation}
and $n^{ab}$ is the binormal to the bifurcation surface $\mathcal{BH}$.

\subsection{Extended Wald's formalism}  \label{sec:extendedWaldformalism}

The original formulation of Wald's thermodynamics presented in the previous sections has some limitations:
\begin{itemize}
	\item It is not clear how the other physical charges relevant to describe a black hole configuration should appear within the first law (electric charges, magnetic charges, scalar charges, supercharges, \dots).
	\item It is not clear how to apply the Wald's procedure to theories which contain fields which transform non trivially under gauge transformations. 
\end{itemize}
In a series of recent works \cite{Elgood:2020svt,Elgood:2020mdx,Elgood:2020nls,Meessen:2022hcg, Mitsios:2021zrn, Ortin:2022uxa,Ballesteros:2023iqb,Gomez-Fayren:2023wxk,Bandos:2023zbs, Ballestaros:2023ipa} both problems have been solved. Starting with the second issue, the reason why we need to assume that all the fields transform trivially under gauge transformations is to be able to impose unambiguously the relation 
\begin{equation} \label{eq_naivevariation}
	\delta_k \varphi = 0 \,.
\end{equation}
Indeed, if we describe $\delta_k$ with the standard Lie derivative and $\varphi$ is transforming under gauge transformations, the condition $\delta_k \varphi = 0$ is not gauge invariant because $\mathcal{L}_k \varphi = 0 $ is not preserved under gauge transformations. The solution to the problem is finding a gauge covariant definition for $\delta_k  \varphi = 0$. The most natural possibility from the mathematical perspective is the notion of gauge-covariant Lie derivative. The idea is that the transformation generated by $k$ should be a combination of a diffeomorphism and a gauge transformations. Indicating with $\mathbb{L}_k$ the gauge-covariant Lie derivative, with $\Lambda_k$ the parameter of the induced gauge transformations and with $\delta_{\Lambda_k}$ the operator implementing the transformation, we have
\begin{equation} \label{eq_defgaugecovariantderivative}
	\mathbb{L}_k \varphi = \mathcal{L}_k \varphi - \delta_{\Lambda_k} \varphi \,.
\end{equation}
We should then replace (\ref{eq_naivevariation}) with
\begin{equation} \label{eq_correctvariation}
	\delta_k \varphi = -\mathbb{L}_k \varphi = 0 \,.
\end{equation}
We will dedicate the section \ref{sec-gaugecovariantLiederivative} to justify the use of the gauge-covariant Lie derivative $\mathbb{L}_k$. With this change we are now able to apply consistently Wald's formalism to a larger class of theories, involving fields which transforms non-trivially under gauge transformations. 

We want to explore the consequences of using the gauge-covariant Lie derivative. If we repeat the steps of the previous section we can check that the only explicit use we made of $\delta_k = -\mathcal{L}_k$ is between equations (\ref{eq_step1}) and (\ref{eq_step2}). Using (\ref{eq_correctvariation}) one gets 
\begin{equation} \label{eq_modified1stlaw}
	\mathbf{\omega}(\varphi,\delta \varphi, \delta_k \varphi) = d \left[\delta \mathbf{Q}_k + \iota_k \mathbf{\Theta}(\varphi,\delta \varphi)\right] - \delta_{\Lambda_k} \mathbf{\Theta}(\varphi,\delta \varphi) \,.
\end{equation}
Writing the last term as a total derivative $\delta_{\Lambda_k} \mathbf{\Theta}(\varphi,\delta \varphi) = d \mathbf{\varpi}_{k}$ we obtain a closed form $\mathbf{W}_k$ which differs from (\ref{eq_OmegaNaive}) for the last term\footnote{Equation (\ref{eq_modified1stlaw}) with the vanishing of $\mathbf{\omega}(\phi,\delta \phi, \delta_k \phi)$ imply that $\delta_{\Lambda_k} \mathbf{\Theta}'(\phi,\delta \phi)$ is exact. However, at the best of our knowledge, there is no general procedure to build explicitly $\varpi_{k}$ and we have to do it case by case.}
\begin{equation}
	\mathbf{W}_k = \delta \mathbf{Q}_k + \iota_k \mathbf{\Theta}(\varphi,\delta \varphi) - \mathbf{\varpi}_{k} \,.
\end{equation}
Integrating $\mathbf{W}_k$ we obtain a non trivial modification of the first law of thermodynamics. In section \ref{sec-simpleexample} studying an example we will see that the extra terms naturally account for the variations of some of the missing charges. More precisely, we obtain the terms proportional to the variations of the gauge charges associated to the gauge transformations which enter in the definition of the gauge-covariant Lie derivative. In particular, $U(1)$ electric charges have been addressed for the first time in \cite{Elgood:2020svt}, $U(1)$ magnetic charges in \cite{Ortin:2022uxa} and supercharges in \cite{Bandos:2023zbs}. Notice that the non-trivial modifications may also modify the entropy formula \cite{Elgood:2020nls}. However, a general closed formula for the modified entropy is lacking so far and the modifications must be studied case by case.

The use of gauge-covariant Lie derivatives almost solves all our issues. However, there is still a kind of term which is expected to contribute to the first law which is not yet included: the term associated to the scalar charges \cite{Gibbons:1996af}. How to define them in theories with the structure of symmetric non-linear sigma coupled to Abelian gauge fields and how they appear in the first law of thermodynamics is studied in \cite{Ballesteros:2023iqb} and is the topic of section \ref{sec-simpleexample}. Setups with non-trivial scalar potentials are instead studied in \cite{Ballestaros:2023ipa}. For the sake of the consistency of this section, we anticipate some of the ideas we will discuss later. To implement scalar charges in the first law two things are required: a coordinate-independent definition of the scalar charges and the proper identification of them within the first law. The first one is achieved integrating a proper closed rank-$(d-2)$ form built out of the global symmetries (or dualities) of the scalar fields on a $(d-2)$-dimensional surface. This definition is reviewed in section \ref{sec-definitioofScalarCharge}. For the second one, we will see that we have to modify the interpretation of the integrals of $\mathbf{W}_k$ in \cite{Iyer:1994ys}. More precisely (\ref{eq_defdeltaMdeltaJ}) and (\ref{eq_defintbh}) do not hold in general. We will provide more details in section \ref{sec-simpleexample}.

\section{The gauge covariant Lie derivative} \label{sec-gaugecovariantLiederivative}

The spacetime symmetries of gauge field configurations cannot be treated independently of the gauge transformations. The main observation is that one has to search for symmetries in the complete bundle and that those symmetries, when seen from (or projected to) the base	space, are combinations of a diffeomorphism and a ``induced'' or ``compensating'' gauge transformation. This gauge transformation depends on the
diffeomorphism and cannot be ignored or separated from it. As a consequence, most fields cannot be treated as simple tensors under diffeomorphisms as in \cite{Iyer:1994ys}. This is a fundamental fact that we are going to prove using the KK framework,\footnote{A purely principal-bundle-based approach can	be found in \cite{Prabhu:2015vua}.} but one can arrive at this conclusion by considering spinors in curved spacetime.
Spinors (and Lorentz tensors) are defined in appropriate bundles connected to the tangent space on which local Lorentz transformations act. Usually, they are treated as scalars under diffeomorphisms but it is not difficult to see	that this description is incorrect: let us consider spinor fields in Minkowski spacetime and let us consider the effect of an infinitesimal global Lorentz spacetime (\textit{i.e.}~not tangent space) transformation on the spinors with parameter $\sigma^{ab}$. If they are treated as scalars they will transform as such, that is\footnote{In this simple example we are working in Cartesian coordinates and we are not distinguishing between spacetime and tangent	space indices.}
\begin{equation}
	\delta_{\sigma}\psi
	=
	-\mathcal{L}_{k_{\sigma}}\psi
	= -\iota_{k_{\sigma}}d\psi\,,
	\hspace{1cm}
	k_{\sigma}
	\equiv
	\sigma^{\mu}{}_{\nu}x^{\nu}\partial_{\mu}\,.
\end{equation}

\noindent
Thus, they will not transform as spinors under that spacetime transformation
as they certainly should under a Lorentz transformation.

The solution to the above problem comes from the following observation: the
spacetime diffeomorphism generated by the Killing vector $k_{(\sigma)}$ induces
a tangent space Lorentz transformation with a parameter that is minus the
(automatically antisymmetric) derivative of the Killing vector, also known as
\textit{Killing bivector} or \textit{Lorentz momentum map}. In this case

\begin{equation}
	-\partial_{\mu}k_{(\sigma) \nu}= \sigma_{\mu\nu}\,,
\end{equation}

\noindent
as expected. In more general settings the parameter of the induced local
Lorentz transformation includes a term proportional to the spin connection
\cite{Ortin:2002qb} and is, indeed, local.

The combination of the Lie derivative and the compensating Lorentz
transformation for the infinitesimal diffeomorphism generated by an arbitrary
Killing vector field $k$ is known as the \textit{spinorial Lie derivative}
$\mathbb{L}_{k}$ and was first introduced by Lichnerowicz and Kosmann in
Refs.~\cite{kn:Lich,kn:Kos,kn:Kos2} and later studied and extended in
Refs.~\cite{Hurley:1994cfa,Vandyck:1988ei,Vandyck:1988gc,Ortin:2002qb} also as
the \textit{Lorentz-covariant Lie derivative} or as the \textit{Lie-Lorentz
	derivative}. One of its main properties is that it transforms covariantly
under further diffeomorphisms and local Lorentz transformations. Thus, the
invariance of the spinor field $\psi$ under the infinitesimal diffeomorphism
generated by $k$ reads

\begin{equation}
	\mathbb{L}_{k}\psi
	=
	0\,,  
\end{equation}

\noindent
and it is an invariant statement.\footnote{This and similar equations can be seen as
	equations determining the values of the vector fields and gauge parameters
	that, combined, leave invariant the fields, known as \textit{reducibility
		(or Killing) parameters} \cite{Barnich:2001jy}. Our approach stresses the
	invariance of the equations which is an important ingredient in the gauge
	invariance of the final results.} 

This section will be organized in the following way: first we will review the approach used by  \cite{Ortin:2015hya,Elgood:2020svt,Elgood:2020mdx,Elgood:2020nls}
to construct more general Lie covariant	derivatives with analogous properties. Then we will test
the simplest of these constructions (the Lie-Maxwell derivative for U$(1)$ gauge fields) using the KK framework.

\subsection{Bottom-up construction of  the gauge-covariant Lie derivative}

Following \cite{Ortin:2015hya,Elgood:2020svt,Elgood:2020mdx,Elgood:2020nls} we want to construct the gauge-covariant Lie derivative determining the compensating gauge transformations necessary for the covariance.

We start by considering local Lorentz transformations. We want to determine which is the compensating gauge transformation we have to consider to make the Lie derivative gauge covariant in the simplest case of the Vielbein $e^a$. Under a local Lorentz transformation with parameter $\sigma^a{}_b \in \text{SO}(1,d-1)$ the Vielbein transforms as $e^a{}' = \sigma^a{}_b e^b$. We want to impose then that
\begin{equation} \label{eq_liederivativevierbein}
	(\mathbb{L}_\xi e^a{} )' = \sigma^a{}_b \, \mathbb{L}_\xi e^b
\end{equation}
Applying definition (\ref{eq_defgaugecovariantderivative}) we obtain that (\ref{eq_liederivativevierbein}) is satisfied provided that the compensating local Lorentz transformations with parameter $\Lambda_\xi{}^a{}_b \in \text{SO}(1,d-1)$ satisfies
\begin{equation} \label{eq_defLambda}
	(\Lambda_\xi{}^a{}_b)' = \sigma^a{}_c \, \Lambda_\xi{}^c{}_d \, (\sigma^{-1})^d{}_b + \iota_\xi d (\sigma^a{}_c ) (\sigma^{-1})^c{}_b \,.
\end{equation}
Equation (\ref{eq_defLambda}) has a very well known structure. It is closely related to the transformation law of the connection of $\text{SO}(1,d-1)$ (see for instance \cite{Ortin:2015hya}). We can conclude then that the compensating gauge transformations must have the form
\begin{equation}
	\Lambda_\xi{}^a{}_b = \iota_\xi \omega{}^a{}_b - P_\xi{}^a{}_b \,,
\end{equation}
where $\omega^a{}_b$ is the spin connection 1-form and $P_\xi{}_{ab}$ is a generic antisymmetric tangent-space tensor. If we now enforce the property that, when $\xi = k$ is a Killing vector, $\mathbb{L}_k e^a$ must vanish we obtain 
\begin{equation}
	\mathbb{L}_k e^a = \mathcal{D} k^a + P_k{}^a{}_b \, e^b = 0\,,
\end{equation}
which is solved by $P_k{}_{ab} = \mathcal{D}_{[a} k_{b]}$.
For future reference, let us apply now the Lorentz-covariant Lie derivative to the spin connection 1-form $\omega^{ab}$ and to the curvature 2-form $R^{ab}$. We obtain \cite{Elgood:2020svt}
\begin{subequations}
	\begin{align}
		& \mathbb{L}_\xi \omega^{ab} = \mathcal{D} P^{ab}_\xi + \iota_\xi R^{ab} \,, \\[2mm] 
		& \mathbb{L}_\xi R^{ab} = \mathcal{D} \iota_\xi R^{ab} + 2 P_\xi{}^{[a}{}_c R^{c|b]} \,.
	\end{align}
\end{subequations}
If we now impose the $\xi = k$ is a Killing vector it is not difficult to verify that both the Lorentz-covariant Lie derivatives vanishes. In particular we obtain the so called Lorentz momentum map equation
\begin{equation} \label{eq_momentummaplorentz}
	\mathcal{D} P^{ab}_k + \iota_k R^{ab} = 0 \,.
\end{equation}
Notice that this last relation can be used as an alternative definition of $P^{ab}_k$. It is satisfied by Killing bivectors.

We consider now the Abelian, rank-1, $U(1)$ gauge field $A$, whose gauge transformations are $A' = A + d \lambda$. We want to determine which is the form of the compensating gauge transformations $\Lambda_\xi$ we have to introduce in $\mathbb{L}_\xi$ in such a way that
\begin{equation}
	(\mathbb{L}_\xi A)' = \mathbb{L}_\xi A \,.
\end{equation}
The compensating gauge transformation $\Lambda_\xi$ now must satisfy
\begin{equation}
	d \Lambda_\xi ' = d \Lambda_\xi + d\iota_\xi d \lambda \,,
\end{equation}
which is solved by
\begin{equation}\label{eq_compensating }
	\Lambda_\xi  = \iota_\xi A - P_\xi 
\end{equation}
where $P_\xi $ is a gauge-invariant object defined up to an additive constant. Now we want to determine $P_\xi$ imposing that $\mathbb{L}_\xi A$ must vanish for a vector $\xi = k$ whenever $\mathbb{L}_k F$ vanishes, with $F = dA$. For a generic $\xi$ we have
\begin{equation}
	\mathbb{L}_\xi F = \mathcal{L}_\xi F = d \iota_\xi F \,.
\end{equation}
Imposing the vanishing of $\mathbb{L}_k F$ we obtain the condition $d \iota_k F = 0$ which implies that there exist an object $\bar{P}_k$ such that $\iota_k F + d \bar{P}_k = 0$. If we now impose the vanishing of $\mathbb{L}_k A $ we obtain the condition
\begin{equation} \label{eq_KillingderA}
	\mathbb{L}_k A = \iota_k F + dP_k = 0\,.
\end{equation}
Equation (\ref{eq_KillingderA}) is called momentum map equation and can be solved requiring that $P_{k} = \bar{P}_k$. Notice that equation (\ref{eq_KillingderA}) has the same structure and origin of (\ref{eq_momentummaplorentz}): they are both obtained applying the gauge-covariant Lie derivative to the connection associated to the gauge transformation. However, now we do not have a field which is more fundamental than the connection $A$, as it is the Vielbein $e^a$ for the spin connection $\omega^a{}_b$. Therefore, the momentum map equation (\ref{eq_KillingderA}) is all what we have to define $P_k$ and we are not able to provide an explicit gauge-invariant expression for it as for the case of the Lorentz momentum map. Anyway, we will see in some examples that this is not an issue for the purpose of evaluating thermodynamic quantities.

In a similar way more general gauge covariant Lie derivatives can be built \cite{Elgood:2020nls,Bandos:2023zbs}.

\subsection{Emergence of gauge-covariant Lie derivative in Kaluza--Klein theory}\label{sec-emergenceKK}

In this section we will study the simplest example of Kaluza--Klein theory, which is 5-dimensional Einstein gravity compactified on a circle $(\text{S}^1)$. We restrict ourselves to the case in which the circle is an isometry direction. Comparing the structure of the null geodesics in 4- and 5-dimension we conclude that if the 4-dimensional metric has an event horizon, so does the 5-dimensional one. Moreover, if the 4-dimensional horizon is a Killing horizon generated by the 4-dimensional Killing vector $l$, there exist an embedding of $l$ into a 5-dimensional Killing vector $\hat{l}$ such that $\hat{l}$ is generating the 5-dimensional event horizon. From the 4-dimensional perspective, the embedding is not unique because it is defined up to gauge transformations of the Kaluza--Klein vector. Such ambiguity in the 5-dimensional perspective is absorbed in the possibility of performing certain change of coordinates. We will show that imposing the Killing equations for $\hat{l}$ we obtain gauge-covariant relations for the 4-dimensional fields. Such relations have the structure of gauge-covariant Lie derivative generated by $l$. Therefore, the gauge-covariant Lie derivative is a natural object in principal bundles.

\subsubsection{Basic Kaluza--Klein theory}

Consider pure Einstein gravity in 5 dimensions parametrized by the coordinates
$\hat{x}^{\hat{\mu}}$.\footnote{We write hats over all 5-dimensional objects
	to distinguish them from the 4-dimensional ones. The 5\textsuperscript{th}
	coordinate will be denoted as $x^{4} = z$ and the corresponding
	(\textit{world}) index will be $\underline{z}$ to distinguish it from the
	corresponding, not underlined, 5\textsuperscript{th} tangent space
	direction. Thus, $(\hat{\mu}) = (\mu,\underline{z})$, $(\hat{a})=(a,z)$,
	etc. We use a mostly minus signature and the rest of the conventions are
	those used in Ref.~\cite{Ortin:2015hya}.} The only dynamical field is the
5-dimensional metric $\hat{g}_{\hat{\mu}\hat{\nu}}$ and the 5-dimensional line
element is
\begin{equation}
	ds_{(5)}^{2}
	=
	\hat{g}_{\hat{\mu}\hat{\nu}}dx^{\hat{\mu}}x^{\hat{\nu}}\,.
\end{equation}	
In this theory, the dynamics of the metric field is dictated by the Einstein-Hilbert action
\begin{equation}
	\label{eq:5dEHaction}
	S[\hat{g}]
	=
	\frac{1}{16\pi G_{N}^{(5)}}\int d^{5}x\,\sqrt{|\hat{g}|}\, \hat{R}\,,
\end{equation}
where $G_{N}^{(5)}$ is the 5-dimensional analog of the Newton constant. If the 5\textsuperscript{th} coordinate is periodic $z\sim z+2\pi \ell$,
where $\ell$ is some length scale, all the components in the metric can be	expanded in Fourier series.\footnote{$\ell$ will be related to the asymptotic radius of the compact dimension $R$ and the value at infinity of the Kaluza--Klein scalar by $R = \ell \, k_\infty$. Notice that $R$ is the only quantity which is independent from rescalings of the $z$ coordinate, cfr. equation (\ref{eq_rescalingz}).} Since the higher modes correspond to fields which
appear as massive from the non-compact 4-dimensional world perspective and
since their masses can be made arbitrarily high by choosing the size of the
5\textsuperscript{th} direction small enough, we can safely ignore them at low
energies and work with the zero modes only, which are the components of a
$z$-independent metric
\begin{equation}
	\label{eq:adaptedcoordinates}
	\partial_{\underline{z}} \hat{g}_{\hat{\mu}\hat{\nu}}=0\,.  
\end{equation}	
Thus, in this scenario the metric admits an isometry generated by a spacelike Killing vector $	\hat{k} = \hat{k}^{\hat{\mu}}\partial_{\hat{\mu}}
=	\partial_{\underline{z}}$
and the coordinates we are using $(x^{\hat{\mu}})=(x^{\mu},x^{4}\equiv z)$ are	coordinates adapted to the isometry. 

The 5-dimensional metric can be decomposed in terms of fields which transform as 4-dimensional fields under 5-dimensional reparametrizations that
respect the gauge choice in Eq.~(\ref{eq:adaptedcoordinates}) (coordinates
adapted to the isometry). There is a scalar $k$ (the Kaluza-Klein
(KK) scalar field), a vector $A = A_\mu dx^\mu$ (the KK vector) and a metric $ds_{(4)}^{2}
=
g_{\mu\nu}dx^{\mu}dx^{\nu}$ with components given by
\begin{subequations}
	\begin{align}
		k^{2} &= -\hat{g}_{\underline{z}\underline{z}}\,, \\[2mm]
		A_{\mu} & =
		\hat{g}_{\mu\underline{z}}/\hat{g}_{\underline{z}\underline{z}}\,, \\[2mm]
		g_{\mu\nu} & =
		\hat{g}_{\mu\nu}
		-\hat{g}_{\mu\underline{z}}\hat{g}_{\nu\underline{z}}/\hat{g}_{\underline{z}\underline{z}}\,.
	\end{align}
\end{subequations}
The 5-dimensional line element can be rewritten in terms of the 4-dimensional KK fields we have just defined as
\begin{equation}
	ds_{(5)}^{2}
	=
	ds_{(4)}^{2} -k^{2}\left(dz+A\right)^{2}\,.
\end{equation}	
The 5-dimensional reparametrizations are generated by $z$-independent 5-dimensional vectors $\hat{\xi}^{\hat{\mu}}$ which act on the 5-dimensional
metric according to	
\begin{equation}
	\delta_{\hat{\xi}}\hat{g}_{\hat{\mu}\hat{\nu}}
	=
	-\mathcal{L}_{\hat{\xi}}\hat{g}_{\hat{\mu}\hat{\nu}}
	=
	-\left(\hat{\xi}^{\hat{\rho}}\partial_{\hat{\rho}}\hat{g}_{\hat{\mu}\hat{\nu}}
	+2\partial_{(\hat{\mu}}\hat{\xi}^{\hat{\rho}}\hat{g}_{\hat{\nu})\hat{\rho}}
	\right)\,.
\end{equation}
It follows that their action on the 4-dimensional fields is 
\begin{subequations}
	\begin{align}
		\delta_{\hat{\xi}} k
		& =
		-\hat{\xi}^{\rho}\partial_{\rho}k\,,
		\\[2mm]
		\delta_{\hat{\xi}}A_{\mu}
		& =
		-\left(\hat{\xi}^{\rho}\partial_{\rho} A_{\mu}
		+\partial_{\mu}\hat{\xi}^{\rho} A_{\rho}\right)
		-\partial_{\mu}\hat{\xi}^{\underline{z}}\,,
		\\[2mm]
		\delta_{\hat{\xi}}g_{\mu\nu}
		& =
		-\left(\hat{\xi}^{\rho}\partial_{\rho}g_{\mu\nu}
		+2\partial_{(\mu}\hat{\xi}^{\rho}g_{\nu)\rho}
		\right)\,.
	\end{align}
\end{subequations}
These transformations can be interpreted as 4-dimensional general coordinate transformations generated by the 4-dimensional vector $ 	\xi^{\mu}\equiv \hat{\xi}^{\mu}$ plus standard gauge transformations $\delta_{\chi}A = d\Lambda$ generated by the gauge parameter $\Lambda \equiv -\hat{\xi}^{\underline{z}}$. Therefore, $A$ plays the role of a 1-form connection with gauge-invariant field strength $F = dA$.

There is only one $z$-dependent 5-dimensional general coordinate
transformation that preserves the gauge Eq.~(\ref{eq:adaptedcoordinates}). It is generated by the vector field $	\hat{\eta} = z\partial_{\underline{z}} $ and it only acts on the $z$ coordinate as a rescaling. If $z' = e^{\alpha} z$ then 
\begin{equation} \label{eq_rescalingz}
	k' = e^{-\alpha} k\,, \qquad A'  = e^{\alpha} A\,, \qquad g_{\mu\nu}' = g_{\mu\nu}\,.
\end{equation}
Observe that the vector field that generates these rescalings does not commute with the Killing vector that generates translations in the internal dimension \cite{Gomez-Fayren:2023wxk}.

Following Scherk and Schwarz \cite{Scherk:1979zr}, in order to find the
equations of motion that govern the dynamics of the 4-dimensional fields it is
convenient to use the Vielbein formalism, making a particular choice for the
decomposition of the 5-dimensional one $\hat{e}{}^{\hat{a}}{}_{\hat{\mu}}$ in
terms of the 4-dimensional fields $e^{a}{}_{\mu},A_{\mu},k$ that breaks the
5-dimensional Lorentz group down to the 4-dimensional one:
\begin{equation}
	\label{eq:standardVielbeinAnsatz}
	\left( \hat{e}^{\hat{a}}{}_{\hat{\mu}} \right) = 
	\left(
	\begin{array}{c@{\quad}c}
		e^{a}{}_{\mu} & kA_{\mu} \\
		&\\[-3pt]
		0       & k    \\
	\end{array}
	\right)\!, 
	\hspace{1cm}
	\left(\hat{e}_{\hat{a}}{}^{\hat{\mu}} \right) =
	\left(
	\begin{array}{c@{\quad}c}
		e_{a}{}^{\mu} & 0  \\
		& \\[-3pt]
		-A_{a}       & k^{-1} \\
	\end{array}
	\right)\,.
\end{equation}
Here $A_{a}= e_{a}{}^{\mu} A_{\mu}$ and we will assume that all 4-dimensional fields with Lorentz indices $a,b,c,\ldots$ have been contracted with the 4-dimensional Vielbein. The above expressions can also be written in the form
\begin{subequations}
	\begin{align}
		\hat{e}^{a}
		& =
		e^{a}\,, \hspace{1cm}
		&
		\hat{e}_{a}
		& =
		e_{a}-\iota_{a}A \partial_{\underline{z}}\,,
		\\[2mm]
		\hat{e}^{z}
		& =
		k(dz+A)\,,
		&
		\hat{e}_{z}
		& =
		k^{-1}\partial_{\underline{z}}\,,
	\end{align}
\end{subequations}
where $\iota_{a}$ indicates the inner product  with $e_{a}$, that is, $\iota_{a}A = e_{a}{}^{\mu}A_{\mu}$. With this decomposition, the non-vanishing components of the spin
connection\footnote{Our spin connection satisfies
	$\mathcal{D}e^{a}=de^{a}-\omega^{a}{}_{b}\wedge e^{b}=0$ in 5 and 4
	dimensions.}  are
\begin{equation}
	\label{eq:standardspinconnectionreduction}
	\begin{array}{rclcrcl}
		\hat{\omega}_{abc} & = & \omega_{abc}\,, & \hspace{2cm}&
		\hat{\omega}_{abz} & = & \frac{1}{2} k F_{ab}\,,
		\\[2mm]
		\hat{\omega}_{zbc} & = & -\frac{1}{2} k F_{bc}\,,
		& \hspace{2cm} & 
		\hat{\omega}_{zbz} & = & -\partial_{b} \ln{k}\,.\\
	\end{array}
\end{equation}
Carefully decomposing the 5-dimensional Einstein equations we obtain the following 3 equations involving 4-dimensional fields \cite{Gomez-Fayren:2023wxk}	
\begin{subequations}
	\label{eq:4deomKKframe}
	\begin{align}
		R^{a}+k^{-1}\mathcal{D}\iota^{a}dk +\tfrac{1}{2}k^{2}F^{ab}\iota_{b}F
		& =
		0\,,
		\\[2mm]
		\mathcal{D}_{b}\left(k^{3} F^{ba}\right)
		& =
		0\,,
		\\[2mm]
		\mathcal{D}^{2}k\,
		+\tfrac{1}{4}k^{3}F^{2}
		& =
		0\,.
	\end{align}
\end{subequations}
The 5-dimensional Einstein-Hilbert action
Eq.~(\ref{eq:5dEHaction}) in the Vielbein formalism, takes the form
\begin{equation}
	\label{eq:5dEHactionVielbein}
	S[\hat{e}]
	=
	\frac{1}{16\pi G_{N}^{(5)}}
	\int \hat{\star}(\hat{e}^{\hat{a}}\wedge \hat{e}^{\hat{b}})
	\wedge \hat{R}_{\hat{a}\hat{b}}\,.
\end{equation}
The action from which the 4-dimensional equations (\ref{eq:4deomKKframe}) can be derived can be obtained by substituting the above decompositions of the 5-dimensional Vielbein and curvature in terms of the 4-dimensional fields. We get
\begin{equation}
	\label{eq:4dEHactionVielbein0}
	S[e,A,k]
	=
	\frac{1}{16\pi G_{N}^{(5)}}
	\int \left\{ k\left[ -\star(e^{a}\wedge e^{b})
	\wedge R_{ab} +\tfrac{1}{2}k^{2}F\wedge \star F \right]
	+d\left[2\star dk\right]\right\}\wedge dz\,.
\end{equation}
Integrating over the internal coordinate $z\in [0,2\pi \ell]$ and using the
$z$-dependence of the 4-form, we get up to total derivatives
\begin{equation}
	\label{eq:4dEHactionVielbein}
	S[e,A,k]
	=
	\frac{2\pi\ell}{16\pi G_{N}^{(5)}}
	\int  k\left[ -\star(e^{a}\wedge e^{b})
	\wedge R_{ab} +\tfrac{1}{2}k^{2}F\wedge \star F \right]\,.
\end{equation}	
It is not too difficult \cite{Gomez-Fayren:2023wxk} to see that the equations that one gets from this action are combinations of Eqs.~(\ref{eq:4deomKKframe}) and, therefore,	equivalent to them.\footnote{Notice that this is a non-trivial check. It is not guaranteed that replacing the ansatz directly in the action we obtain a consistent truncation.} The factor of $k$ in front of the Einstein-Hilbert term in Eq.~(\ref{eq:4dEHactionVielbein0}) indicates that the 4-dimensional metric $g_{\mu\nu}$ is not in the (conformal) Einstein frame, in which, by definition, the Einstein-Hilbert term has no additional scalar factors. The Einstein-frame metric is clearly related to $g_{\mu\nu}$ by a Weyl rescaling
with some power of the KK scalar $k$. If we want the rescaling to preserve the normalization of the metric at spatial infinity in the non-compact directions, we must use a power	of $k/k_{\infty}$ and not just of $k$ to rescale it. Thus, we define the 4-dimensional Einstein-frame metric $g_{E\, \mu\nu}$ and Vielbein	$e_{E}{}^{a}{}_{\mu}$ and the Einstein-frame KK vector field $A_{E\, \mu}$ by	
\begin{equation}
	g_{\mu\nu}= \left(k/k_{\infty}\right)^{-1}g_{E\, \mu\nu}\,,
	\hspace{.5cm}
	e^{a}{}_{\mu} = \left(k/k_{\infty}\right)^{-1/2}e_{E}{}^{a}{}_{\mu}\,,
	\hspace{.5cm}
	A_{\mu} = k_{\infty}^{1/2}A_{E\, \mu}\,.
\end{equation}
Neglecting total derivatives, we arrive \cite{Gomez-Fayren:2023wxk} to the Einstein-frame action 
\begin{equation}
	\label{eq:4dEHactionVielbeinEframe}
	\begin{aligned}
		S[e_{E},A_{E},k]
		& =
		\frac{1}{16\pi G_{N}^{(4)}}
		\int \bigg\{ -\star_{E}(e_{E}{}^{a}\wedge e_{E}{}^{b})
		\wedge R_{E\, ab}
		+\tfrac{3}{2}d\log{k}\wedge \star_{E} d\log{k} \\[1mm]
		& \hspace{1.5cm}	+\tfrac{1}{2}k^{3}F_{E}\wedge \star F_{E} \bigg\}\,,
		\\
	\end{aligned}
\end{equation}
with the 4-dimensional Newton constant given by 
\begin{equation}
	\label{eq:4-5Newtonconstant}
	G_{N}^{(4)}
	=
	\frac{G_{N}^{(5)}}{2\pi R}\,.
\end{equation}	
Finally, we redefine $k$ in terms of an unconstrained scalar field $\phi$ which can take any real value
\begin{equation}
	k
	=
	e^{\phi/\sqrt{3}}\,,
\end{equation}
and we arrive to the final form of our action
\begin{equation}
	\label{eq:4dEHactionVielbeinEframestandard}
	\begin{split}
		S[e_{E},A_{E},\phi]
		& =
		\frac{1}{16\pi G_{N}^{(4)}}
		\int \bigg\{ -\star_{E}(e_{E}{}^{a}\wedge e_{E}{}^{b})
		\wedge R_{E\, ab}
		+\tfrac{1}{2}d\phi\wedge \star_{E} d\phi \\[1mm]
		& \hspace{1.5cm}	+\tfrac{1}{2}e^{\sqrt{3}\phi}F_{E}\wedge \star_{E} F_{E} \bigg\} \,.
	\end{split}
\end{equation}
This is a particular Einstein-Maxwell-dilaton (EMD) model with $a=-\sqrt{3}$ in the parametrization used in Ref.~\cite{Ortin:2015hya}. After all these redefinitions, the relation between the 5-dimensional line element and the 4-dimensional Einstein-frame line element $	ds^{2}_{E\,(4)}	= 	g_{E\, \mu\nu}dx^{\mu}dx^{\nu}$
and other Einstein-frame fields is
\begin{equation}
	\label{eq:5dimensionalmetric}
	ds_{(5)}^{2}
	=
	e^{-(\phi-\phi_{\infty})/\sqrt{3}}ds^{2}_{E\,(4)}
	- e^{2\phi/\sqrt{3}}
	\left[dz+e^{\frac{\phi_{\infty}}{2\sqrt{3}}}A_{E}\right]^{2}\,,
\end{equation}
where $e^{\phi_{\infty}/\sqrt{3}} =	k_{\infty}$.

\subsubsection{Dimensional reduction of the Killing horizon}

An important question is whether the presence of event horizons in	the 4-dimensional metric implies their presence in the 5-dimensional one. We need to study the behaviour of null geodesics in the two setups. This is done in the next section. We find that the behavior of null geodesics in the 5-dimensional spacetime is determined by Eqs.~(\ref{eq:d5geodesicequations}), which we have shown to be equivalent to the 4-dimensional Eqs.~(\ref{eq:d4geodesicequations2})
plus the equation of conservation of $P_{z}$.  The second of
Eqs.~(\ref{eq:d4geodesicequations2}) is particularly interesting because it tells us that the lightcones of the 5-dimensional metric are equal to those of the 4-dimensional one, times a circle.\footnote{The conformal rescaling that brings us to the Einstein metric leaves the lightcones invariant.}
5-dimensional, massless, $P_{z}\neq 0$ particles which move over the
5-dimensional lightcone are seen to move inside the 4-dimensional one. In particular, this means that, if the 4-dimensional metric has event horizons, so does the 5-dimensional one at the same place in the 4-dimensional coordinates. The 5-dimensional horizon simply has one more dimension, parametrized by $z$, fibered over the 4-dimensional one and we will denote both the 4- and 5-dimensional event horizons by $\mathcal{H}$.

The main feature of the 4-dimensional geometries we are considering is the fact that they all admit a Killing vector $l=l^\mu \partial_\mu$ which is the generator of the event horizon $\mathcal{H}$. It can be characterized by the property
\begin{equation}
	l^{2}=l^{\mu}g_{\mu\nu}l^{\nu} \stackrel{\mathcal{H}}{=}0\,.  
\end{equation}	
If we trivially uplift $l$ to 5d and we compute its norm on the 5-dimensional horizon $\mathcal{H}$ we find that it is not a null vector anymore
\begin{equation}
	l^{\mu}\hat{g}_{\mu\nu}l^{\nu} \stackrel{\mathcal{H}}{=} -k^{2}(\iota_{l}A)^{2}\,.
\end{equation}
Thus, from the 5-dimensional point of view, the event horizon is not the	Killing horizon of $l$.	It is natural to search for a 5-dimensional extension of $l$, that we will denote by $\hat{l}$, whose Killing horizon coincides with the event horizon,
that is,	
\begin{equation}
	\hat{l}^{2}=\hat{l}^{\hat{\mu}}\hat{g}_{\hat{\mu}\hat{\nu}}\hat{l}^{\hat{\nu}}
	\stackrel{\mathcal{H}}{=}0\,.   
\end{equation}
Assuming that $\hat{l}$ has the form $\hat{l}=l+f\hat{k}$ we have
\begin{equation}
	\hat{l}^{2}
	\stackrel{\mathcal{H}}{=}
	-k^{2}(f+\iota_{l}A)^{2}\,.    
\end{equation}
The RHS vanishes if we assume that
\begin{equation}
	\label{eq:fg}
	f=-\iota_{l}A +\gamma\,,
	\,\,\,\,
	\text{where}
	\,\,\,\,
	\left. \gamma\right|_{\mathcal{H}}
	=
	0\,.
\end{equation}
Now, we want to impose that $\hat{l}$ is a Killing vector of the
5-dimensional metric. If $\hat{l}$ does not depend on $z$ we obtain,
\begin{subequations}
	\begin{align}
		&	\mathcal{L}_{\hat{l}}\hat{g}_{\underline{z}\underline{z}}
		=
		-2k\mathcal{L}_{l}k =0 \,, \\[1mm]
		&	\mathcal{L}_{\hat{l}}\hat{g}_{\mu\underline{z}}
		=
		-2kA_{\mu}\mathcal{L}_{l}k -k^{2}\left(\mathcal{L}_{l}A_{\mu}+\partial_{\mu}f\right) =0\,, \\[1mm]
		& 	\mathcal{L}_{\hat{l}}\hat{g}_{\mu\nu}
		=	\mathcal{L}_{l}g_{\mu\nu}
		-2kA_{\mu}A_{\nu}\mathcal{L}_{l}k
		-2k^{2}\left(\mathcal{L}_{l}A_{(\mu}+\partial_{(\mu}f\right)A_{\nu)}
		=
		0\,.
	\end{align}
\end{subequations}
The condition $\mathcal{L}_{l}k = 0$ in typical black hole setups is always satisfied.\footnote{If we consider static metrics, $l = \partial_t$ and the condition $\mathcal{L}_t k = 0$ is the requirement that the scalar $k$ is time-independent.} The 5-dimensional Killing equations are satisfied provided that 
\begin{equation} \label{eq_condizione}
	\mathcal{L}_{l}A_{\mu}+\partial_{\mu}f=0\,,
\end{equation}
Using Eq.~(\ref{eq:fg}) and	differential-form language, and rescaling the equation with $k^{-1/2}_{\infty}$, this condition takes the form
\begin{equation}
	\iota_{l}F_{E}+d(k^{-1/2}_{\infty}\gamma)=0\,.  
\end{equation}
This is nothing but the Maxwell momentum map equation introduced in
(\ref{eq_KillingderA}) with $k^{-1/2}_{\infty}\gamma$ playing the role of	momentum map $ P_{E\, l}$, and, taking into account that
$ \left. \gamma\right|_{\mathcal{H}} = 0$ and that the momentum map is defined	only up to an additive constant, we conclude that
\begin{equation}
	k^{-1/2}_{\infty}\gamma
	=
	P_{E\, l}-\left.P_{E\,l}\right|_{\mathcal{H}}
	\equiv
	\overline{P}_{E\, l}\,,
\end{equation}
and we arrive at 
\begin{equation}
	\label{eq:fg2}
	f=-k_{\infty}^{1/2}\left(\iota_{l}A_{E} -\overline{P}_{E\, l}\right)\,.
\end{equation}
Comparing (\ref{eq_condizione}) and (\ref{eq:fg2}) with the results of the previous section, we can conclude that the condition of invariance of the gauge field $A_{E}$ under the isometry generated by $l$ is nothing but 
\begin{equation}
	\mathbb{L}_{l}A_{E} =  \iota_{l}F_{E}+dP_{E\, l} = 0\,,
\end{equation}
where the ``compensating gauge transformation'' parameter $\Lambda_{l}$ is $\Lambda_{l} \equiv	\iota_{l}A_{E}-P_{E\, l} $ 	and $\mathbb{L}_{l}A_{E}$ is the gauge-covariant Lie (or Lie-Maxwell) derivative	of $A_{E}$ with respect to $l$, cfr.$\,\,$equation (\ref{eq_KillingderA}).\footnote{See also
	Ref.~\cite{Heusler:1993cj,Ortin:2015hya,Elgood:2020svt}.} The emergence of this formula in the KK	framework is one of our main results. Thus, we have constructed a 5-dimensional extension of $l$ (the
\textit{uplift} of $l$), namely	
\begin{equation}
	\label{eq:hatldef}
	\hat{l}
	=
	l-k_{\infty}^{1/2}\left(\iota_{l}A_{E} -\overline{P}_{E\, l}\right)\hat{k}\,,  
\end{equation}
which is a Killing vector of the 5-dimensional metric and whose Killing
horizon is a S$^{1}$ fibration over the Killing horizon of $l$. On the Killing horizon itself we can write
\begin{equation}
	\hat{l}
	\stackrel{\mathcal{H}}{=}
	l -k_{\infty}^{1/2}\Omega \hat{k}\,,  
\end{equation}
where the constant $\Omega$ is given by
\begin{equation}
	\Omega
	=
	\left.\iota_{l}A_{E}\right|_{\mathcal{H}}\,.
\end{equation}

If we restrict to 4-dimensional static solutions, $l = \partial_t$ and $\Omega$ can be identified with the	electrostatic potential evaluated over the horizon
\begin{equation}
	\Omega = \Phi_\mathcal{H}\,,  
\end{equation}
which is a gauge-dependent quantity. Since the gauge transformations of the 4-dimensional KK vector field are 5-dimensional diffeomorphisms which are not 5-dimensional isometries, this result is not surprising. However, the ambiguity in the value of $\Omega$ can be eliminated by demanding the 5-dimensional metric to be asymptotically flat with the following normalization\footnote{We assume the 4-dimensional metric to be asymptotically-flat as well.}
\begin{equation}
	ds^{2}_{(5)}
	\longrightarrow 
	\eta_{\mu\nu}dx^{\mu}dx^{\nu} -k^{2}_{\infty}dz^{2}\,,
\end{equation}
or, equivalently, that the KK vector field vanishes at spatial
infinity. Then,
\begin{equation}
	\Omega = \bar{\Phi}\,,  
\end{equation}
where $\Phi$ is the (gauge-invariant) difference of electrostatic potential between the horizon and spatial infinity. Without this condition, the coordinates $t$ and $z$ are entangled at infinity in electrically-charged black holes, for instance.	There is another interpretation for the constant $\Omega$: the linear momentum
of free-falling observers in the direction $z$, given by
\begin{equation}
	P_{z} \equiv \hat{g}_{\underline{z}\hat{\mu}}\dot{x}^{\hat{\mu}}\,,  
\end{equation}
is a conserved quantity. When the KK vector is electric,
$\iota_{l}A= A_{t}\neq 0$, observers with $P_{z}=0$, however, are
moving in the $z$ direction with velocity
\begin{equation}
	\frac{dz}{dt}
	=
	-\iota_{l}A\,.
\end{equation}
This fact can be interpreted as the dragging of inertial frames by the spacetime, which has momentum in the direction $z$. A particle that starts at infinity with zero velocity in the compact direction and falls	radially towards the horizon will acquire a non-vanishing velocity in the
internal direction that will equal $k_{\infty}^{1/2}\Omega$ at the horizon. This is very similar to what happens in the Kerr spacetime to zero angular momentum observers (\textit{ZAMO}s) and, geometrically, it has to do with the fact that the 5-dimensional vector $\partial_{t}$ is not hypersurface-orthogonal. A difference, however, is that in these spacetimes	there may not be a static limit where $\hat{g}_{tt}\neq 0$. A similar phenomenon happens in the magnetic case in which $\iota_{\partial_{\varphi}}A\neq 0$. For vanishing $P_{z}$, either $\dot{z}=\dot{\varphi}=0$ or	
\begin{equation}
	\frac{dz}{d\varphi}
	=
	-\iota_{\partial_{\varphi}}A\,.
\end{equation}

To end this section, we can prove that the surface gravity of the
5-dimensional Killing horizon coincides with that of the 4-dimensional one. Again we restrict to static 4-dimensional metric. First, observe that the standard definition of the 4-dimensional surface	gravity is invariant under Weyl rescalings of the metric when we write it in	the form
\begin{equation}
	\nabla_{\mu}l^{2}
	\stackrel{\mathcal{H}}{=}
	-2\kappa l_{\mu}\,,
\end{equation}
and, therefore, we can use this definition in the Einstein or KK frames. The 1-form dual to the Killing vector $\hat{l}$ if given by
\begin{equation}
	\hat{l}_{\hat{\mu}}dx^{\hat{\mu}}
	=
	\left[g_{t\mu} -k_{\infty}k^{2}\overline{P}_{E\, l}A_{E\, \mu}\right]dx^{\mu}
	-k_{\infty}^{1/2}k^{2}\overline{P}_{E\, l}dz\,.
\end{equation}
It follows that 
\begin{equation}
	\hat{l}_{\hat{\mu}}dx^{\hat{\mu}}
	\stackrel{\mathcal{H}}{=}
	l_{\mu}dx^{\mu} \,,
\end{equation}
so the pullbacks of the 1-forms $\hat{l}_{\hat{\mu}}dx^{\hat{\mu}}$ and
$l_{\mu}dx^{\mu}$ are identical over the horizon even if the dual vectors are	not. Then, on $\mathcal{H}$ only, using the vanishing of
$\overline{P}_{E\, l}$ and $g_{tt}$ there, we find
\begin{equation}
	\begin{aligned}
		\hat{\nabla}_{\mu}\hat{l}^{2}
		& 
		=
		\nabla_{\mu}g_{tt}
		=
		\nabla_{\mu}l^{2}
		=
		-2\kappa l_{\mu}
		=
		-2\kappa \hat{l}_{\mu} \,,
		\\[2mm]
		\hat{\nabla}_{\underline{z}}\hat{l}^{2}
		& =
		0
		=
		-2\kappa \hat{l}_{\underline{z}}\,,
	\end{aligned}
\end{equation}
thus showing that the 4- and 5-dimensional surface gravities are the same.

\section{The scalar charge}  \label{sec-definitioofScalarCharge}

It is widely believed that one of the defining characteristics of classical black holes is that they have no ``hair''. The concept of black hole hair is a very broad one but, for stationary black holes it can be defined as any parameter that enters the metric and which cannot be eliminated through a coordinate transformation which is not a function of the charges of the theory which are conserved by virtue of a local
symmetry (mass, angular momenta, electric charges) or a topological property (magnetic charges) or the asymptotic values of the scalars (\textit{moduli}). 

Scalar charges, typically defined through the asymptotic behavior at spatial infinity of the scalars in the black hole spacetime, are not protected by any conservation law. In ungauged theories the only local symmetries scalar fields transform under are diffeomorphisms but the conserved charges associated to	them are the gravitational ones: the mass and linear and angular momenta. Scalar	fields only transform under global symmetries of the action or of the equations of motion to which we will refer to as \textit{dualities}. However, the	charges associated to those symmetries in stationary black hole spacetimes vanish identically. They seem to have nothing to do with the conventionally-defined black hole scalar charges. Gauging the global	symmetries does not help because the gauge symmetry would be associated to some 1-form gauge fields and the conserved charges would have the interpretation of electric and magnetic charges. 

Therefore, according to our definition of hair, scalar charges are understood as hair and, according to the \textit{no-hair conjecture}, no black hole	solutions with regular horizons (henceforth to be referred to as ``regular	black holes'') carrying scalar charges should be expected. Any scalar charges	possessed by gravitationally collapsing matter should be radiated away in the	black hole formation. However, there are many regular black hole solutions
carrying non-vanishing scalar charges such as dilaton black holes and their generalizations.\footnote{For a review with many references, see
	Ref.~\cite{Ortin:2015hya}.}

The solution to this apparent counterexample of the no-hair conjecture lies in the distinction between primary and secondary hair \cite{Coleman:1991ku}:	in all the regular black hole solutions with non-vanishing scalar charges, those charges are not independent parameters but very specific functions of the independent conserved charges which are allowed by the no-hair conjecture	and they are (by definition) secondary hair. In the solutions in which the
scalar charges are truly independent parameters, such as the
Janis-Newman-Winicour solution \cite{Janis:1968zz} or the Agnese-La Camera	solutions \cite{Agnese:1994zx} and their generalizations \cite{Ortin:2015hya}, there are no regular horizons but naked singularities unless the scalar charge	takes the value of the specific function of the conserved charges we mentioned
above (simply zero in the JNW solution). This kind of scalar hair is, by	definition, primary hair and it is the one which would actually be forbidden by the conjecture.

The scalar charges which are allowed by the no-hair conjecture remain,
nevertheless, quite mysterious: What are the values of the scalar charges allowed in a given theory?  Why are those values allowed and no others? And,	even more basic: Is there a coordinate-independent definition of scalar charge? 

In this section we are going to show how we can define coordinate-independent scalar charges defining them as the integrals of closed $(d-2)$-forms. These charges are manifestly coordinate and gauge independent and satisfy a Gauss law in stationary black hole spacetimes. This definition relies on the existence of conserved charges associated to global symmetries and on the existence of a
timelike Killing vector whose Killing horizon coincides with the black hole's event horizon and whose action leaves invariant all the physical fields. Therefore, there is a scalar charge associated to each global symmetry and the number of charges may or may not coincide with the number of scalar fields.

\subsection{A possible definition}

In static, spherically symmetric black holes the scalar charge $\Sigma^x$ associated to a scalar field $\phi^x$ in $d$ spacetime dimensions is conventionally defined through the asymptotic behavior of the field at spatial infinity 
\begin{equation}\label{eq:conventionaldefinition}
	\phi^x \sim \phi_\infty^x + \frac{4 \pi G_N^{(d)}}{\omega_{(d-2)}(d-3)} \frac{\Sigma^x}{r^{d-3}} + \mathcal{O}\left(\frac{1}{r^{2(d-3)}}\right)\,,
\end{equation}
where $G_N^{(d)}$ is the Newton constant in $d$ dimensions and $\omega_{(d-2)}$ is the volume of the unit $(d-2)$-sphere. As explained previously, this definition is not really satisfactory and we would like to obtain the scalar charge as the integral of a closed form. In this way the charge would be manifestly independent of the coordinates used and the integration surface considered.

A simple idea would be looking for a global symmetry (or a duality, i.e. a symmetry of the equations of motion) acting non trivially on the scalar fields and use the associate closed Noether current $J_A$ (or the associated Noether--Gaillard--Zumino current in the case of a duality \cite{Gaillard:1981rj}). However, it is not difficult to verify that in examples representing static black hole solutions, the charge defined as the integral of this current on a $(d-1)$-dimensional surface not only does not reproduce the charges $\Sigma^{x}$, but vanishes identically \cite{Ballesteros:2023iqb}.

In stationary black hole spacetimes, though, we can extract a rank $d-2$ form out of the rank-$(d-1)$ current $J_A$. Let us assume that all the fields (denoted collectively with $\varphi$) are invariant under the isometry generated by the spacetime vector $k$,	$\delta_{k}\varphi=0$. For the scalar fields $\phi^{x}$ it means that their Lie derivatives with respect to that vector vanish $\mathcal{L}_k  \phi^x = 0$. For fields which admit gauge transformations we have to impose instead that their gauge-covariant Lie derivatives vanish $\mathbb{L}_k \varphi = 0$. Then, if all the fields are invariant under $\delta_{k}$, so must the $J_A$. Therefore we obtain
\begin{equation}
	\begin{aligned}
		\delta_{k}J_{A}
		& =
		-\mathbb{L}_k J_{A}
		\\[1mm]
		& = -\mathcal{L}_k  J_{A} + \delta_{\Lambda_k}  J_{A} \\[1mm]
		& \doteq
		d \left[-\iota_{k} J_{A} + \chi_A \right]
		\\[1mm]
		& = 0\,,
	\end{aligned}
\end{equation}
where we used the on-shell closure of $J_A$ and we introduced $\chi_A$ as an object such that $d\chi_A = \delta_{\Lambda_k} J_A$. The expression in brackets is a closed $(d-2)$-form
\begin{equation}
	\mathbf{Q}_{A}[k] = \iota_{k} J_{A} - \chi_A \,, 
\end{equation}
which can be integrated over $(d-2)$-dimensional, spacelike, closed surfaces to obtain a charge	
\begin{equation}
	Q_{A,k}
	=
	\int_{\Sigma^{d-2}} \mathbf{Q}_{A}[k]  \,.
\end{equation}
In \cite{Ballesteros:2023iqb} there are several examples corresponding to	static dilaton and axidilaton black holes and the charges defined with this procedure reproduce the values of the conventionally-defined scalar charges
Eq.~(\ref{eq:conventionaldefinition}). It is also worth stressing that there might be more symmetries than scalar	fields as in the case we are going to analyze in the next section. Not all of them will be independent and the conventionally-defined scalar charges $\Sigma^{x}$ can
be expressed in terms of the charges $Q_{A,k}$ that we have just defined.

This definition has been introduced in \cite{Ballesteros:2023iqb}. However, it is worth mentioning that there is a slightly different procedure that allows us to obtain an equivalent result in the particular case of dilaton black holes in Ref.~\cite{Pacilio:2018gom}. 

\section{A simple example} \label{sec-simpleexample}

We are now going to apply all the machinery we introduced in a simple example proving the first law of black hole thermodynamics. We will show that thanks to the introduction of the gauge-covariant Lie derivative, electric and magnetic charges naturally arise. Furthermore, taking into account hitherto ignored contributions to the integrals at spatial infinity we are able to recover the scalar charges as thermodynamical potentials conjugate to the variations of the moduli.

We study 4-dimensional theories whose scalar kinetic terms are described by symmetric sigma models in which the scalar fields map spacetime into a target space which is a symmetric	Riemannian homogeneous space G/H. Furthermore, our theories include Abelian 1-forms and	we are going to assume that the couplings of the scalars to those 1-forms are such that the equations of motion, enhanced with the Bianchi identities satisfied by the 2-form field strengths, are invariant under the duality group G.\footnote{The extension to higher dimensions and higher-rank forms is straightforward using the results	of Ref.~\cite{Bandos:2016smv} for the Noether-Gaillard-Zumino currents.} 

In these theories we can associate a conserved scalar charge $\mathcal{Q}_{A,k}$ to each of the generators $T_A$ of G, even if	some of the transformations (the electric-magnetic duality rotations in	particular) do not leave the action invariant. As a result, according to our definition, there are always more scalar charges than scalars $\phi^x$. Nevertheless, one can verify in concrete examples \cite{Ballesteros:2023iqb} that the conventional scalar charges can be recovered as	combinations of the ones we have defined, matching the result obtained by Gibbons, Kallosh and Kol in Ref.~\cite{Gibbons:1996af} (see also Ref.~\cite{Astefanesei:2018vga})\footnote{The general form of the theories that we consider is identically to that of the theories considered by GKK in Ref.~\cite{Gibbons:1996af} but in our approach it is crucial to know the global symmetries of the theory.} 	
\begin{equation}
	\mathcal{Q}_{A\,k}\, \delta^{A}_{\infty}
	=
	\tfrac{1}{4}\Sigma^{x}g_{xy\, \infty}\delta \phi^{y}_{\infty}\,,
\end{equation}
where $\Sigma^x$ is the scalar charge defined via the asymptotic expansions of the scalar fields $\phi^x$, $g_{xy}$ is the target space metric and $\delta^A_\infty$ is built using the variation of the generators of G. As a byproduct, we are going to	find a general expression for the scalar charges in terms of the conserved gauge charges $q^M$ and the electromagnetic potentials $\Phi^M_\mathcal{H}$ evaluated at the event horizon
\begin{equation}
	\mathcal{Q}_{A\, k}
	=
	-\Omega_{MP}T_{A}{}^{P}{}_{N}\Phi^{M}_{\mathcal{H}}q^{N}\,,
\end{equation}
proving the scalar hair is \textit{secondary hair}.
The main and final result of this section is the explicit derivation of the first law of thermodynamics from first principles using the extended Wald's formalism for stationary black holes
\begin{equation*}
	\delta M
	=
	T \delta S
	+\Omega\delta J
	-\left(\Omega_{MN}\bar{\Phi}^M \right)\delta q^{N}
	-\left(\Omega_{MP}T_{A}{}^{P}{}_{N} \, \bar{\Phi}^M q^{N} \right)
	\delta^{A}_{\infty}\,,
\end{equation*}
where $S$ is the Bekenstein--Hawking entropy, $M$ is the ADM mass, $T$ is the Hawking temperature, $J$ the angular momentum, $\Omega$ the horizon angular velocity and $\bar{\Phi}^M  = \Phi^M_\mathcal{H} - \Phi^M_\infty$ is a symplectic vector built with the electrostatic and magnetostatic potentials taking the difference of their values at the horizon $\mathcal{H}$ and at spatial infinity. With a proper gauge fixing we can always set $\Phi^M_\infty = 0$ and the coefficient of $\delta^A_\infty$ is exactly given by the scalar charges $\mathcal{Q}_{A\,k}$. Notice that the right-hand side of this expression only contains the variations of quantities which are independent physical parameters of the black hole solutions. The variations of the scalar charges cannot and do not appear, as predicted by the no-hair theorem.

\subsection{Symmetric $\sigma$-models coupled to Abelian gauge fields}	
\label{sec-theory}
We are going to consider 4-dimensional ungauged
supergravity-inspired theories containing $n_{S}$ scalar fields $\phi^{x}$	that parametrize a symmetric coset space $G/H$, $n_{V}$ 1-form fields $A^{\Lambda}=A^{\Lambda}{}_{\mu}dx^{\mu}$ with 2-form field strengths $	F^{\Lambda} = dA^{\Lambda}$, and gravity which we will describe through the Vierbein $e^{a}=e^{a}{}_{\mu}dx^{\mu}$. Up to two derivatives, they can be described by
the generic action

\begin{equation}
	\label{eq:action-05}
	\begin{split}
		S
		= & 
		\frac{1}{16\pi G_{N}^{(4)}}\int \bigg[
		-\star (e^{a}\wedge e^{b}) \wedge R_{ab}
		+\tfrac{1}{2}g_{xy}d\phi^{x}\wedge \star d\phi^{y}
		-\tfrac{1}{2}I_{\Lambda\Sigma}F^{\Lambda}\wedge \star F^{\Sigma} \\[1mm]
		& \hspace{1.5cm}-\tfrac{1}{2}R_{\Lambda\Sigma}F^{\Lambda}\wedge F^{\Sigma} \bigg]\,,
	\end{split}
\end{equation}
where the kinetic matrix $I=\left(I_{\Lambda\Sigma}\right)$ is
negative-definite and we are going to assume that the positive-definite
$\sigma$-model metric $g_{xy}(\phi)$ is invariant under the action of $G$ (the
duality group) which also leaves invariant the set of all equations of motion
plus the Bianchi identities of the theory. This assumption will be translated
into conditions for the scalar-dependent matrices
$I=\left(I_{\Lambda\Sigma}\right)$ and $R= \left(R_{\Lambda\Sigma}\right)$	shortly. The equations of motion are defined by (here $\varphi$ stands for all the fields of the theory)
\begin{equation} \label{eq:variationsigmamodel}
	\delta S
	=
	\int\left\{
	\mathbf{E}_{a}\wedge \delta e^{a} + \mathbf{E}_{x}\delta\phi^{x}
	+\mathbf{E}_{\Lambda}\wedge \delta A^{\Lambda}  
	+d\mathbf{\Theta}(\varphi,\delta\varphi)
	\right\}\,,
\end{equation}
and given by (we ignore the overall factor)
\begin{subequations}
	\begin{align}
		\begin{split}
			\label{eq:Ea}
			\mathbf{E}_{a}
			& =
			\iota_{a}\star(e^{b}\wedge e^{c})\wedge R_{bc}
			+\tfrac{1}{2}g_{xy}\left(\iota_{a}d\phi^{x} \star d\phi^{y}
			+d\phi^{x}\wedge \iota_{a}\star d\phi^{y}\right) \\[1mm]
			& \hspace{.5cm}
			-\tfrac{1}{2}I_{\Lambda\Sigma}\left(\iota_{a}F^{\Lambda}\wedge\star
			F^{\Sigma} -F^{\Lambda}\wedge \iota_{a}\star F^{\Sigma}\right)\,,
		\end{split} \\[2mm]
		\begin{split}
			\mathbf{E}_{x}
			& =
			-g_{xy}\left\{d\star d\phi^{y}
			+\Gamma_{zw}{}^{y}d\phi^{z}\wedge\star d\phi^{w} \right\}
			-\tfrac{1}{2}\partial_{x}I_{\Lambda\Sigma} F^{\Lambda}\wedge\star
			F^{\Sigma} \\[1mm]
			& \hspace{.5cm} -\tfrac{1}{2}\partial_{x}R_{\Lambda\Sigma} F^{\Lambda}\wedge F^{\Sigma}\,,
		\end{split}
		\\[2mm]
		\mathbf{E}_{\Lambda}
		& =
		d F_{\Lambda}\,, \\[2mm]
		\mathbf{\Theta}(\varphi,\delta\varphi)
		& =
		-\star (e^{a}\wedge e^{b})\wedge \delta \omega_{ab}
		+g_{xy}\star d\phi^{x}\delta\phi^{y}
		-F_{\Lambda}\wedge \delta A^{\Lambda}\,. 	\label{eq:Theta}
	\end{align}
\end{subequations}
where we have defined the dual 2-form field strength
\begin{equation}
	\label{eq:dualfieldstrengthsdef}
	F_{\Lambda}
	\equiv
	I_{\Lambda\Sigma}\star F^{\Sigma}+R_{\Lambda\Sigma}F^{\Sigma}\,.
\end{equation}
The original and dual 2-forms can be combined into a symplectic vector of
2-forms\footnote{The symplectic nature of this vector will be proven shortly.}
\begin{equation}
	\left(F^{M}\right)
	\equiv
	\left(
	\begin{array}{c}
		F^{\Lambda} \\ F_{\Lambda} \\
	\end{array}
	\right)\,.
\end{equation}
The Bianchi identities of the original 2-form field strength $F^{\Lambda}$ 	and  the Maxwell equations $\mathbf{E}_{\Lambda}=0$ can be written as
\begin{equation}
	\label{eq:equations}
	dF^{M}=0\,.  
\end{equation}
These equations can be interpreted as Bianchi identities implying the local
existence of 1-form potentials $F^{M}=dA^{M}$. 

The set of equations (\ref{eq:equations}) is invariant under arbitrary
GL$(2n_{V},\mathbb{R})$ transformations
\begin{equation}
	F^{M\,\prime}  =S^{M}{}_{N}F^{N}\,, 
\end{equation}
but we have to take into account the rest of the equations and an important
constraint: the components of $F^{M}$ are not independent and, therefore,
$F^{M}$ satisfies the following \textit{twisted self-duality constraint}
\begin{equation}
	\label{eq:twisted}
	\star F^{M}= -\Omega^{MN}\mathcal{M}_{NP}F^{P}\,,  
\end{equation}
\noindent
where $\mathcal{M}_{MN}$ is the $2n_{V}\times 2n_{V}$ symmetric symplectic\footnote{$\mathcal{M}$ satisfies $\mathcal{M}^T \Omega \mathcal{M} = \Omega$\,.} matrix
\begin{equation}
	\begin{aligned}
		\left(\mathcal{M}_{MN}\right)
		=
		\left(
		\begin{array}{lr}
			I +RI^{-1}R
			&
			-RI^{-1}
			\\
			& \\
			-I^{-1}R
			&
			I^{-1}
			\\
		\end{array}
		\right)\,,
		\\
	\end{aligned}
\end{equation}
and	
\begin{equation}
	\left(\Omega_{MN}\right)
	= 
	\left(
	\begin{array}{lr}
		0 & \mathbbm{1}_{n_{V}\times n_{V}}   \\
		& \\
		-\mathbbm{1}_{n_{V}\times n_{V}}      & 0 \\
	\end{array}
	\right)\,.
\end{equation}
As a consequence, the set of Maxwell equations and Bianchi identities will
only be invariant under the subset of GL$(2n_{V},\mathbb{R})$ transformations
that preserve this constraint, which is possible provided that $\mathcal{M}$
transforms as

\begin{equation}
	\label{eq:Mtransform}
	\mathcal{M}'
	=
	\left(\Omega^{-1}S\Omega\right) \mathcal{M} S^{-1}\,,
	\hspace{1cm}
	S = \left(S^{M}{}_{N}\right)\,.
\end{equation}
The invariance of the Einstein equations (in particular, of the stress energy tensor) implies that $S$ must satisfy	
\begin{equation}
	\label{eq:symplecticdef}
	S^{T}\Omega S =\Omega\,,  
\end{equation}
which means that $S \in \text{Sp}(2n_{V},\mathbb{R})$
\cite{Gaillard:1981rj}. Going back to Eq.~(\ref{eq:Mtransform}), we find that 
\begin{equation}
	\label{eq:Mtransform2}
	\mathcal{M}^{-1\, \prime}
	=
	S\mathcal{M}^{-1} S^{T}\,.
\end{equation}
It is clear that these transformations are associated to
transformations of the scalars. The infinitesimal transformations of the scalars that leave the	equations of motion invariant must necessarily be generated by the Killing	vectors of the $\sigma$-model metric $g_{xy}$, which we are going to denote by $\{k_{A}{}^{x}(\phi)\}$.\footnote{These transformations leave exactly
	invariant the energy-momentum tensor of the scalars, which is the only piece
	of the Einstein equations that we had not studied, and transform covariantly
	the first two terms of the scalar equations of motion.
} 	
The infinitesimal transformations of the 1-form fields are
\begin{equation}
	\begin{aligned}
		S \sim	\mathbbm{1}_{2n_{V}\times 2n_{V}} +\alpha^{A}T_{A}\,,
		\qquad 	T_{A}  =
		\left(T_{A}{}^{M}{}_{N}\right) \,.
	\end{aligned}
\end{equation}
$S$ is symplectic if $\left(\Omega T_{A}\right)^{T}
=
\Omega T_{A}$. Then, it can be easily seen \cite{Ballesteros:2023iqb} that the whole scalar equations of motion
transform as	
\begin{equation}
	\delta_{A}\mathbf{E}_{x}
	= -\partial_{x}k_{A}{}^{y}\mathbf{E}_{y}\,,
\end{equation}
under the transformations
\begin{equation}
	\delta_{A}\phi^{x}= k_{A}{}^{x}\,,
	\hspace{1cm}
	\delta_{A}F^{M} = T_{A}{}^{M}{}_{N}F^{N}\,,
\end{equation}
provided that the $n_{V}\times n_{V}$, symmetric, \textit{period matrix} $\mathcal{N} = R+iI$ satisfies the equivariance condition 
\begin{equation}
	\label{eq:equivariancecondition}
	k_{A}{}^{x}\partial_{x}\mathcal{N} = \delta_{A}\mathcal{N}\,,  
\end{equation}
with
\begin{equation}
	\label{eq:infinitesimalfractionallinearN}
	\delta_{A}\mathcal{N}_{\Lambda\Sigma}
	=
	T_{A\, \Lambda\Sigma}
	+T_{A}{}_{\Lambda}{}^{\Omega}\mathcal{N}_{\Omega\Sigma}
	-\mathcal{N}_{\Lambda\Omega}T_{A}{}^{\Omega}{}_{\Sigma}
	-\mathcal{N}_{\Lambda\Gamma} T_{A}{}^{\Gamma\Omega}\mathcal{N}_{\Omega\Sigma}\,.
\end{equation}

\subsection{Electric, magnetic and scalar charges} \label{sec-charges}

\subsubsection{Electric charges}
There are several equivalent definitions of electric charge in the literature. Most of them are integrals of certain forms over asymptotic surfaces. In our case we will use 
\begin{equation}
	q_\Sigma = \frac{1}{16 \pi G_N^{(4)}}\int_{S^2_\infty}	F_\Sigma \,.
\end{equation} 
Notice that the condition $dF_\Sigma = 0$ is nothing but the Maxwell equations of motion. 

For completeness, we explain how to extract the electric charge in more general cases.	We start considering $U(1)$ gauge transformations within theories which depend only on the rank 2 field strengths $F^\Sigma = d A^\Sigma$. Ignoring total derivatives,  the equations of motion of $A^\Sigma$ have the form (from now on we ignore the overall factor of the action)
\begin{equation}
	\mathbf{E}_\Sigma = d\, \left[ - \frac{\delta \mathbf{L}}{\delta F^\Sigma}\right] \equiv d \, F_\Sigma \,.
\end{equation}
Therefore, we have a closed rank-$(d-2)$ form which is closed on shell which is nothing but the dual field strength. Integrating on a closed $(d-2)$-dimensional surface $\Sigma^{d-2}$ we obtain
\begin{equation}
	q_\Sigma =  \int_{\Sigma^{d-2}} F_\Sigma \,,
\end{equation}
which represents the amount of electric charge contained in the region of space bounded by $\Sigma^{d-2}$. 

We consider now the case of gauge symmetries generated by more general $p$-forms $\Lambda^\Sigma$. This is the method used in \cite{Elgood:2020nls}. The general idea is that we are always able to associate to the gauge transformations through Noether theorem a closed rank-$(d-2)$ form which can be integrate on a $(d-2)$-dimensional surface. For concreteness, suppose that $A^\Sigma$ are now rank-$(p+1)$ forms, that we have the gauge transformations $A^\Sigma{}' = A^\Sigma + d \Lambda^\Sigma$ and that we are in $d$ spacetime dimensions. We also assume that the Lagrangian depends on $A^\Sigma$ only via their field strengths $F^\Sigma = dA^\Sigma$. The variation of the action is (we indicate with $\varphi$ the fields of the theory which are not $A^\Sigma$)
\begin{equation}
	\delta S = \int \mathbf{E}_\Sigma \,\wedge\delta A^\Sigma + \mathbf{E}_\varphi \wedge \delta \varphi + d \mathbf{\Theta} (A^\Sigma, \varphi, \delta A^\Sigma,\delta \varphi) \,.
\end{equation}
Specifying the formula for gauge transformations generated by $\Lambda^\Sigma$, assuming that $\varphi$ is not transforming under such transformations and integrating by parts making use of the Noether identities, we obtain the off-shell identity 
\begin{equation}\label{eq:actionTOtalderivative}
	0 = \delta_\Lambda S = \int d \left[(-1)^{d-p - 1}\,  \mathbf{E}_\Sigma \wedge \Lambda^\Sigma + (-1)^{d-p-2} \frac{\delta \mathbf{L}}{\delta F^\Sigma} \wedge d \Lambda^\Sigma  \right] \,.
\end{equation}
We conclude that we have a rank-$(d-1)$, off-shell closed current $\mathbf{J}_\Lambda$
\begin{equation}\label{eq:defofJofMatteo}
	\mathbf{J}_\Lambda = (-1)^{d-p - 1}\,  \mathbf{E}_\Sigma \wedge \Lambda^\Sigma + (-1)^{d-p-2} \frac{\delta \mathbf{L}}{\delta F^\Sigma} \wedge d \Lambda^\Sigma \,.
\end{equation} 
We would like to write it as $ \mathbf{J}_\Lambda = d \mathbf{Q}_\Lambda $. Recalling that 
\begin{equation}
	\mathbf{E}_\Sigma =  (-1)^{d-p - 1} d \frac{\delta \mathbf{L}}{\delta F^\Sigma} \equiv d F_\Sigma\,,
\end{equation}
where we introduced the dual field strength $F_\Lambda$,	 we easily obtain the rank-$(d-2)$ form
\begin{equation}
	\mathbf{Q}_\Lambda =  F_\Sigma \wedge \Lambda^\Sigma \,.
\end{equation} 
$\mathbf{Q}_\Lambda$ is closed whenever $\Lambda^\Sigma$ is closed. Integrating on a closed $(d-2)$-dimensional surface $\Sigma^{d-2}$ we obtain as many charges as allowed harmonic p-forms supported over  $\Sigma^{d-2}$. Indeed, let's consider the integral  
\begin{equation}
	\int_{\Sigma^{d-2}}  F_\Sigma \wedge \Lambda^\Sigma \,.
\end{equation}
This integrals on-shell is vanishing unless $\Lambda^\Sigma$ is harmonic. Expanding $\Lambda^\Sigma$ in a basis on harmonic p-forms $\{h^i\}$
\begin{equation}
	\Lambda^\Sigma = \Lambda^\Sigma{}_{i} \, h^i  \,,
\end{equation}
we can associate a charge $q_\Sigma{}^i$ to every parameter $\Lambda^\Sigma{}_{i}$ and we obtain 
\begin{equation}
	q_\Sigma{}^i =  \int_{\Sigma^{d-2}}  F_\Sigma \wedge h^i \,.
\end{equation}

Let us finally comment on the cases in which there is an explicit dependence on the gauge connections $A^\Sigma$. If the Lagrangian is still exactly gauge invariant, then there must be some other field transforming under the gauge transformation and compensating the variation.\footnote{This is the case, for instance, of minimally coupled theories. Another example is a theory where the field strengths definitions contain Chern-Simons terms.} Then the procedure to follow is the same: we build an off-shell-closed Noether current and extract a rank-$(d-2)$ current out of it. The main difference is that now there will be contributions from $\mathbf{E}_\varphi$ and the form of $\mathbf{\Theta}$ will change. An explicit example of this is contained in \cite{Elgood:2020nls}. If the Lagrangian is instead gauge invariant up to total derivative, one simply has to subtract from (\ref{eq:actionTOtalderivative}) that total derivative and follow the same steps.

\subsubsection{Magnetic charges}

A simple way to introduce the magnetic charge is defining it as the electric charge of the electromagnetic dual gauge field $\tilde{A}$. In our case we obtain the charges $p^\Sigma$
\begin{equation}
	p^\Sigma = \frac{1}{16 \pi G_N^{(4)}}\int_{S^2_\infty}	 F^{\Sigma}\,.
\end{equation} 
The condition $dF^\Sigma = 0$ is nothing but the Bianchi identity.

In the more general case of $p$-forms in $d$ dimensions we obtain (we ignore the overall factors of the action)
\begin{equation}
	p^\Sigma{}_j = \int_{\Sigma^{d-2}}	 F^{\Sigma} \wedge \tilde{h}_j \,,
\end{equation} 
where $\{\tilde{h}_j\}$ is a basis of harmonic $(d-p-2)$-forms. In the cases in which the field strength is not closed\footnote{This is typical when Chern-Simons terms are involved.} one has to replace $F^\Sigma$ with the closed current $d A^\Sigma$. Notice that the integral is not vanishing because $A^\Sigma$ in general will not be globally defined and $d A^\Sigma$ will be a harmonic form.

\subsubsection{Scalar charges}	\label{sec-scalarcharge}
Not all the symmetries of the equations of motion that we have studied are	symmetries of the action. However, as shown in Ref.~\cite{Gaillard:1981rj}, there is an on-shell conserved current for each of them, the so-called \textit{Noether-Gaillard-Zumino (NGZ) current}. The	simplest way to construct them is by contracting the scalar equations of motion with the Killing vectors that generate them. Using the Killing vector equation and the equivariance condition we get \cite{Bandos:2016smv}
\begin{equation}
	\label{eq:kAxEx}
	\begin{aligned}
		k_{A}{}^{x}\mathbf{E}_{x}
		& =
		-d \star \hat{k}_{A} -\tfrac{1}{2}\Omega_{MP}T_{A}{}^{P}{}_{N}F^{M}\wedge F^{N}
		\\
		& \\
		& =
		-d\left[ \star \hat{k}_{A}
		+\tfrac{1}{2}\Omega_{MP}T_{A}{}^{P}{}_{N}A^{M}\wedge F^{N}\right]
		+\tfrac{1}{2}\Omega_{MP}T_{A}{}^{P}{}_{N}A^{M}\wedge \mathbf{E}^{N}\,,
	\end{aligned}
\end{equation}

\noindent
where we have collected in a symplectic vector of 3-forms the Maxwell equations
and Bianchi identities:
\begin{equation}
	\left(\mathbf{E}^{M}\right)
	\equiv
	\left(
	\begin{array}{c}
		\mathbf{E}^{\Lambda} \\ \mathbf{E}_{\Lambda} \\
	\end{array}
	\right)\,,
\end{equation}
and where we have denoted by $\hat{k}_{A} = k_{A}{}^{x}g_{xy}d\phi^{y}$ the pullback of the 1-form dual to the target space Killing vector $k_{A}$. Therefore, we find that the NGZ currents 
\begin{equation}
	\star j_{A}
	\equiv
	-\star \hat{k}_{A} -\tfrac{1}{2}\Omega_{MP}T_{A}{}^{P}{}_{N}A^{M}\wedge F^{N}\,.
\end{equation}
Following the logic of the previous section, we require that all the fields are invariant under the isometry generated by a spacetime Killing vector $k$ and we obtain
\begin{equation}
	\delta_{k}\star j_{A} \doteq
	d\left\{-\iota_{k}\star j_{A}
	-\tfrac{1}{2}\Omega_{MP}T_{A}{}^{P}{}_{N}\Lambda_{k}{}^{M} F^{N}\right\} = 0\,,
\end{equation}
where $\Lambda_{k}{}^M$ is the compensating gauge transformation associated to $A^M$. Exploiting the explicit definition of $\Lambda^M$ and the momentum map equation $\mathbb{L}_k A^M = 0$
\begin{equation}
	\chi_{k}
	=
	\iota_{k}A-P_{k}\,, \qquad \iota_k F + dP_k{}^M = 0\,,
\end{equation}
we eventually obtain	
\begin{equation}
	\label{eq:charge2form}
	\mathbf{Q}_{A}[k]
	=
	\iota_{k}\star \hat{k}_{A}
	+\Omega_{MP}T_{A}{}^{P}{}_{N}P_{k}{}^{M}F^{N}\,.
\end{equation}
Now, integrating over 2-dimensional, spacelike, closed surfaces (and restoring
the normalization) we get the charges associated to the NGZ currents:
\begin{equation}
	\mathcal{Q}_{A,k}
	=
	\frac{1}{16\pi G_{N}^{(4)}}
	\int_{\Sigma^{2}} \left\{\iota_{k}\star \hat{k}_{A}
	+\Omega_{MP}T_{A}{}^{P}{}_{N}P_{k}{}^{M}F^{N}\right\}\,.
\end{equation}

Notice that this definition depends on the value of the momentum map over the
integration surface. The Maxwell momentum map is defined only up to an
additive constant. This constant can be chosen so that
$\left.P_{k}{}^{M}\right|_{\infty}=0$. That is the choice that allows us to
recover the values of the conventionally-defined scalar charges
Eq.~(\ref{eq:conventionaldefinition}). However, other choices are
possible. The form of the first law that we are going to find includes an
additional term that takes into account that possibility so that the first law
is invariant under a change of asymptotic value of the Maxwell momentum maps.

It is also worth stressing that in the case we are considering (a
symmetric $\sigma$-model) there are always more symmetries than scalar
fields. Therefore, there are more scalar charges than scalars. However, one can verify in examples \cite{Ballesteros:2023iqb} that the conventionally-defined scalar charges $\Sigma^{x}$ can be expressed in terms of the charges $Q_{A,k}$ that we have just defined and not all of them will be independent.

Finally, notice that on the bifurcation surface	we have
\begin{equation}
	\mathbf{Q}_{A}[k]
	\stackrel{\mathcal{BH}}{=}
	\Omega_{MP}T_{A}{}^{P}{}_{N}P_{k}{}^{M}_{\mathcal{H}}F^{N}\,,
\end{equation}
and, therefore
\begin{equation} \label{eq:scalarChargeExp}
	\mathcal{Q}_{A\, k}
	=
	-\Omega_{MP}T_{A}{}^{P}{}_{N}\Phi^{M}_{\mathcal{H}}q^{N}\,.  
\end{equation}
This formula gives a universal relation	between the scalar charges of a black hole and the electric and magnetic	charges and potentials evaluated on the horizon generalizing the result found
in Ref.~\cite{Pacilio:2018gom} in a gauge-invariant way. Observe that the existence of a bifurcate Killing horizon is crucial to prove that the scalar hair is  secondary hair\cite{Coleman:1991ku}.\footnote{Static, spherically-symmetric solutions of pure gravity and dilaton gravity with primary scalar hair
	(\textit{i.e.}~scalar fields with charges which are independent parameters
	of the solutions) can be found in Refs.~\cite{Janis:1968zz,Agnese:1994zx}
	(see also the higher-dimensional generalizations in Chapter~16 of
	Ref.~\cite{Ortin:2015hya}) and are singular.}

\subsection{Electromagnetic potentials} \label{sec-potentials}

Once we determine the magnetic and electric charges, we can identify the electromagnetic potentials with the coefficients of the variation of the charges in the first law. According to  \cite{Elgood:2020mdx, Ortin:2022uxa, Gomez-Fayren:2023wxk}, for configurations which admit a Killing vector field $k$ which generates the BH horizon and leaves invariant all the fields, such coefficients are strictly related to the momentum maps $P_k^\Sigma$. In the simple case of a 2-form $F^\Sigma$, the associated momentum map $P_k^\Sigma$ defined by the equation
\begin{equation}
	\iota_k F^\Sigma + d P_k^\Sigma = 0 \,,
\end{equation}
is a function. At the bifurcation surface $\mathcal{BH}$, the Killing vector $k$ vanishes. We obtain that $ P_k^\Sigma$ is closed there, which implies that it must have a constant value over the bifurcation surface. We identify that value with the electrostatic potential 
\begin{equation}
	P_k^\Sigma  	\stackrel{\mathcal{BH}}{=} \Phi^\Sigma_\mathcal{\mathcal{BH}} \,.
\end{equation}
Using the fact that the event horizon is generated by $k$, one can easily prove that $P_k^\Sigma$ is constant over the whole event horizon and not only on the bifurcation surface. It is not clear how to extend this argument for higher-rank momentum maps, but the issue is not relevant for the cases studied in this thesis.\footnote{For higher-rank forms, the differential equations describing the flow of $P_k{}^\Sigma$ along the event horizon do not admit a unique solution if we fix the boundary conditions at the bifurcation surface.} For the magnetostatic potential, the same logic applies with the dual momentum map $P_k{}_\Sigma$
\begin{equation}
	P_k{}_{\, \Sigma} 	\stackrel{\mathcal{BH}}{=} \Phi_\Sigma{}_{\,\mathcal{\mathcal{BH}}} \,.
\end{equation}  

For generic $(p+2)$-forms $F^\Sigma$ in $d$-dimensions, the momentum map $P{}_k{}^\Sigma$ will be a $p$-form and the dual momentum map $P_k{}_{\,\Sigma}$ a $(d-p-4)$-form. At the bifurcation surface the momentum maps are closed forms, which implies that their pullback on the the $(d-2)$-dimensional compact bifurcation surface admits an expansion in harmonic forms. We have
\begin{subequations} 
	\begin{align}
		P_k^\Sigma  & \stackrel{\mathcal{BH}}{=} \left(\Phi^\Sigma_{\mathcal{BH}}\right)_i \, h^i + d \, e^\Sigma \,,  \\[2mm]   P_k{}_{\,\Sigma} &  \stackrel{\mathcal{BH}}{=} \left(\Phi_\Sigma{}_{\,\mathcal{BH}}\right)^j \, \tilde{h}_j + d \,e_{\,\Sigma} \,,
	\end{align}
\end{subequations}
where $\{h^i\}$ and $\{\tilde{h}_j\}$ are bases of harmonic $p$- and $(d-p-4)$-forms supported on the bifurcation surface. Therefore, there are as many electromagnetic potentials as charges. In particular, we obtain
\begin{subequations} \label{eq:momentummapIntegral}
	\begin{align}
		\int_{\Sigma^{d-2}} F_\Sigma \wedge P_k^\Sigma  \,\,\, & \stackrel{\mathcal{BH}}{=}   \,\,\,  \Phi^\Sigma_{\mathcal{BH}}{}_{\,i} \, \, q_\Sigma{}^i \,, \\[2mm]
		\int_{\Sigma^{d-2}} F^\Sigma \wedge P_k{}_{\,\Sigma}  \,\,\, & \stackrel{\mathcal{BH}}{=}   \,\,\,  \Phi_\Sigma{\,}_{\mathcal{BH}}{}^j \, \, q^\Sigma{}_j \,.
	\end{align}
\end{subequations}

In most cases, with adapted coordinates and imposing proper asymptotic conditions on the fields, the integrals (\ref{eq:momentummapIntegral}) vanish asymptotically. However, there are examples for which it is not the case (cfr. the KK black hole in the previous section). We are going to explain what happens in those cases. 

If a $p+2$-form $F^\Sigma$ is vanishing asymptotically then $dP_k^\Sigma$ vanishes too and $P_k^\Sigma$ is closed. Then, we can expand
\begin{equation}
	P_k^\Sigma  \stackrel{\infty}{=} \left(\Phi^\Sigma_{\infty}\right)_i \, h^i + d \, e^\Sigma \,,
\end{equation}
with $\{h^i\}$ a basis of harmonic $p$-forms supported on $\Sigma_\infty^{d-2}$ and we obtain
\begin{equation} \label{eq:Integral}
	\int_{\Sigma^{d-2}} F_\Sigma \wedge P_k^\Sigma  \,\,\,  \stackrel{\infty}{=}  \,\,\,  \Phi^\Sigma_{\infty}{}_{\,i} \, \, q_\Sigma{}^i \,.
\end{equation}
The same applies for $F_\Sigma$
\begin{equation} \label{eq:Integral2}
	\int_{\Sigma^{d-2}} F^\Sigma \wedge P_k{\,}_\Sigma  \,\,\,  \stackrel{\infty}{=}  \,\,\,  \Phi_\Sigma{\,}_{\infty}{}^{\,j} \, \, q^\Sigma{}_j \,.
\end{equation}
If $F^\Sigma$ is not vanishing asymptotically,  $P_k^\Sigma$ is not closed anymore and it has a coexact component. However, one can prove that only the closed part of $P_k^\Sigma$ can contribute to the integral (\ref{eq:Integral}). We prove this in appendix \ref{sec-potentialsProve}.

\subsection{First law of thermodynamics}

\label{sec-firstlaw}

Taking into account the results obtained in
Refs.~\cite{Elgood:2020svt,Elgood:2020mdx,Elgood:2020nls,Ortin:2022uxa} for the inclusion of matter fields in Wald's formalism
\cite{Lee:1990nz,Wald:1993nt,Iyer:1994ys}, which we have summarized in section \ref{sec-reviewBHthermody}, the first law of black hole
thermodynamics for a non-extremal black hole whose bifurcate horizon coincides	with the Killing horizon of the Killing vector field
$k=\partial_{t}-\Omega\partial_{\varphi}$, can be derived by integrating the
on-shell identity
\begin{equation}
	d\mathbf{W}_k
	\doteq
	0\,,
\end{equation}
over a spacelike hypersurface with boundaries at spatial infinity
(S$^{2}_{\infty}$ in $d=4$) and at the bifurcation sphere $\mathcal{BH}$ and applying
the Stokes theorem. We recall that 
\begin{equation}
	\mathbf{W}[k]
	\equiv
	\delta\mathbf{Q}[k] +\iota_{k}\mathbf{\Theta}(\varphi,\delta\varphi)-\varpi_{k}\,,
\end{equation}
where $\mathbf{Q}[k]$, $\mathbf{\Theta}(\varphi,\delta\varphi)$ and
$\varpi_{k}$ are defined in section \ref{sec-reviewBHthermody}.

Let us start computing the Noether--Wald charge $\mathbf{Q}[k]$. The gauge-invariant variations of the fields under the action of a diffeomorphism generated by the vector $\xi$ are
\begin{subequations}
	\begin{align}
		& \delta_\xi e^a = - \mathbb{L}_\xi e^a = - [\mathcal{D} \xi^a + P_\xi{}^a{}_b ] \,, \\[2mm]
		& \delta_\xi A^\Sigma = - \mathbb{L}_\xi A^\Sigma = - [\iota_\xi F^\Sigma + d P_\xi{}^\Sigma] \,,\\[2mm]
		& \delta_\xi \phi^x = - \mathbb{L}_\xi \phi^x = - \iota_\xi d \phi^x \,.
	\end{align}
\end{subequations}
Replacing these variations into (\ref{eq:variationsigmamodel}) we obtain
\begin{equation}\label{eq:variationsigmamodel2}
	\begin{split}
		\delta_\xi S = & - \int \bigg\{D \mathbf{E}_a \xi^a + \mathbf{E}^{[a} \wedge e^{b]} P_\xi{}_{ab} + \mathbf{E}_\phi \, \iota_\xi d\phi + \mathbf{E}_\Sigma \wedge \iota_\xi F^\Sigma+ d \mathbf{E}_\Sigma \, P_\xi{}^\Sigma   \\
		& \hspace{1.5cm} - d \mathbf{\Theta}'(\varphi,\delta \varphi) \bigg\} \,,
	\end{split}
\end{equation}
with 
\begin{equation}
	\mathbf{\Theta}'(\varphi,\delta_\xi\varphi) = \mathbf{\Theta}(\varphi,\delta_\xi\varphi) + \mathbf{E}_a \xi^a + \mathbf{E}_\Sigma P^\Sigma\,.
\end{equation}	
The second and the fifth terms of (\ref{eq:variationsigmamodel2}) vanish identically. The former because of the symmetry of Einstein equations, the latter because $\mathbf{E}_\Sigma$ is closed.\footnote{Notice that two cancellations can be obtained as a consequence of the Noether identities associated to the theory invariance under local Lorentz transformations and gauge transformations.} Replacing the explicit expressions of $\mathbf{E}_a$, $\mathbf{E}_\phi$ and $\mathbf{E}_\Sigma$ one can verify that the first, third and fourth terms combine into an expression which is identically zero (off-shell). Therefore, we have
\begin{equation}
	\delta_\xi S = \int d \mathbf{\Theta}' \,,
\end{equation}
and we can identify $\mathbf{\Theta}'$ with the one introduced in section \ref{sec-reviewBHthermody}.	After some algebra, $\mathbf{\Theta}'$ has the explicit expression
\begin{equation}
	\mathbf{\Theta}'(\varphi,\delta_\xi \varphi) = -\iota_\xi \mathbf{L} + d \left[\star (e^a \wedge e^b) P_\xi{}_{ab} -  F_\Sigma P_\xi{}^\Sigma \right] \,,
\end{equation}
and we obtain the Noether--charge
\begin{equation}
	\mathbf{Q}[k]
	=
	\star (e^{a}\wedge e^{b})P_{k\, ab} -P_{k}{}^{\Lambda}F_{\Lambda}\,.
\end{equation}
A quick calculation gives	
\begin{equation}
	\label{eq:deltaQ}
	\delta\mathbf{Q}[k]
	=
	P_{k\, ab}\, \delta \star (e^{a}\wedge e^{b})
	+\star (e^{a}\wedge e^{b})\delta P_{k\, ab}
	-F_{\Lambda}\delta P_{k}{}^{\Lambda}
	-P_{k}{}^{\Lambda}\delta F_{\Lambda}\,.
\end{equation}
Let us move now to the presymplectic 3-form $\mathbf{\Theta}$. Its expression is given in Eq.~(\ref{eq:Theta}) and another short calculation gives	
\begin{equation}
	\label{eq:ikTheta}
	\begin{aligned}
		\iota_{k}\mathbf{\Theta}(\varphi,\delta\varphi) 
		& =
		-\iota_{k}\star (e^{a}\wedge e^{b})\wedge
		\delta \omega_{ab} -\star (e^{a}\wedge e^{b})\wedge \delta\iota_{k}
		\omega_{ab} +g_{xy}\iota_{k}\star d\phi^{x}\delta\phi^{y}
		\\[2mm]
		& \hspace{.5cm}
		-\tfrac{1}{2}\iota_{k}F_{\Lambda}\wedge \delta A^{\Lambda}
		-\tfrac{1}{2}F_{\Lambda}\wedge \delta\iota_{k} A^{\Lambda}\,.
	\end{aligned}
\end{equation}
By assumption $k$ leaves invariant all the fields, \textit{i.e.} $\delta_k \varphi = 0$. Exploiting the properties of the Lie derivative,\footnote{In particular if $k$ is a Killing vector then $[\mathcal{L}_k, \star ] = 0$ } it is simple to verify that $\delta_k F_\Lambda = 0$. Since, on-shell, the dual 1-forms obey the same equations as the original ones, we can define the \textit{dual (magnetic) momentum maps} $P_{k\,\Lambda}$ through the equation	
\begin{equation}
	\delta_k A_\Sigma = - \mathbb{L}_k A_\Sigma = -[\iota_{k}F_{\Lambda}+dP_{k\,\Lambda}] =0\,,
\end{equation}

\noindent
and, substituting this definition in the above expression and integrating by
parts, we get

\begin{equation}
	\label{eq:ikTheta2}
	\begin{aligned}
		\iota_{k}\mathbf{\Theta}
		& =
		-\iota_{k}\star (e^{a}\wedge e^{b})\wedge
		\delta \omega_{ab} -\star (e^{a}\wedge e^{b})\wedge \delta\iota_{k}
		\omega_{ab} +g_{xy}\iota_{k}\star d\phi^{x}\delta\phi^{y}
		\\[2mm]
		& \hspace{.5cm}
		+P_{k\,\Lambda}\wedge \delta F^{\Lambda}
		-F_{\Lambda}\wedge \delta\iota_{k} A^{\Lambda}\,,
	\end{aligned}
\end{equation}
up to an irrelevant total derivative. Finally, we move to $\mathbf{\varpi}_k$. In the case at hands we have to deal with $U(1)$ gauge transformations and local Lorentz transformations. We indicate with $\delta_{\Lambda_k}$ the compensating gauge transformations. A simple calculation gives \cite{Ortin:2022uxa}
\begin{equation}
	\begin{aligned}
		\delta_{\Lambda_{k}} \mathbf{\Theta}(\varphi,\delta\varphi)	& =
		-\delta_{\Lambda_{k}}\left[\star (e^{a}\wedge e^{b})\wedge \delta
		\omega_{ab}\right]
		-F_{\Lambda}\wedge \delta_{\Lambda_{k}}\delta A^{\Lambda}
		\\[2mm]
		& =
		-\star (e^{a}\wedge e^{b})\wedge \mathcal{D} \, \delta \Lambda_{k\, ab}
		-F_{\Sigma}\wedge d \, \delta \Lambda_{k}{}^\Sigma
		\\[2mm]
		& =
		d\left\{ -\star (e^{a}\wedge e^{b})\wedge\delta \Lambda_{k\, ab}
		-F_{\Lambda}\delta \Lambda_{k}{}^{\Lambda}
		\right\}\,.
	\end{aligned}
\end{equation}
Therefore,
\begin{equation}
	\varpi_{k}
	=
	-\star (e^{a}\wedge e^{b})\wedge\delta \Lambda_{k\, ab}
	-F_{\Lambda}\delta \Lambda_{k}{}^{\Lambda}\,.
\end{equation}
Replacing the explicit expressions of the parameters of the induced Lorentz and Maxwell gauge transformations
\begin{subequations}
	\begin{align}
		\Lambda_{k}{}^{ab}
		& =
		\iota_{k}\omega^{ab}-P_{k}{}^{ab}\,,
		\\[2mm]
		\Lambda_{k}{}^{\Lambda}
		& =
		\iota_{k}A^{\Lambda}-P_{k}{}^{\Lambda}\,,
	\end{align}
\end{subequations}
and combining all these partial results, we arrive at	
\begin{equation}
	\begin{aligned}
		\mathbf{W}[k]
		& =	P_{k\, ab}\delta \star (e^{a}\wedge e^{b})
		-\iota_{k}\star (e^{a}\wedge e^{b})\wedge\delta \omega_{ab}
		\\[1mm]
		& \hspace{.5cm}
		-P_{k}{}^{\Lambda}\delta F_{\Lambda}
		+P_{k\,\Lambda}\delta F^{\Lambda}
		+g_{xy}\iota_{k}\star d\phi^{x}\delta\phi^{y}\,.
	\end{aligned}
\end{equation}

Let us consider the integral of $\mathbf{W}[k]$ at spatial infinity first, restoring the global factor $1/(16\pi G_{N}^{(4)})$. The first two terms give the gravitational contribution\footnote{Cfr. with the standard procedure presented in section \ref{sec-reviewBHthermody}.}
\begin{equation}
	\frac{1}{16\pi G_{N}^{(4)}}  \int_{S^{2}_{\infty}}\left\{
	P_{k\, ab}\delta \star (e^{a}\wedge e^{b})
	-\iota_{k}\star (e^{a}\wedge e^{b})\wedge\delta \omega_{ab}
	\right\}
	=
	\delta M-\Omega\delta J\,,
\end{equation}
while the third and fourth give
\begin{equation}
	\frac{1}{16\pi G_{N}^{(4)}}  \int_{S^{2}_{\infty}}\left\{
	-P_{k}{}^{\Lambda}\delta F_{\Lambda}
	+P_{k\,\Lambda}\delta F^{\Lambda}
	\right\}
	=
	-\Phi^{\Lambda}_{\infty}\delta q_{\Lambda}
	+\Phi_{\Lambda\,\infty} \delta p^{\Lambda}\,,
\end{equation}
where $\Phi^{\Lambda}_{\infty}$ and $\Phi_{\Lambda\,\infty}$ are the values of the electrostatic and magnetostatic potentials at spatial infinity and $q_\Lambda$ and $q^\Lambda$ are, respectively, the electric and magnetic charges. Let us consider the last term. Using the definition of scalar charges we have proposed and the identity
$g_{xy}=g^{AB}k_{A\, x} k_{B\, y}$ we can write
\begin{equation}
	\int_{S^{2}_{\infty}} g_{xy}\iota_{k}\star d\phi^{x}\delta\phi^{y}
	= \left(\mathcal{Q}_{A}
	-\Omega_{MP}T_{A}{}^{P}{}_{N}\Phi^{M}_{\infty}q^{N}\right)\delta^{A}_{\infty}
\end{equation}
where we have defined $\delta^{A} \equiv g^{AB}k_{B\, y}\delta\phi^{y}$. The bifurcation surface is defined by the property $k=0$ and, on it,
\begin{equation}
	P_{k\, ab}
	\stackrel{\mathcal{BH}}{=}
	\kappa n_{ab}\,,
\end{equation}
where $n^{ab}$ is the binormal to the horizon with the normalization
$n^{ab}n_{ab}=-2$ and $\kappa$ is the surface gravity. Therefore,
\begin{equation}
	\int_{\mathcal{BH}}\mathbf{W}[k] =
	\frac{\kappa \delta A_{\mathcal{H}}}{8\pi G_{N}^{(4)}}
	-\Phi^{\Lambda}_{\mathcal{H}}\delta q_{\Lambda}
	+\Phi_{\Lambda\, \mathcal{H}}\delta p^{\Lambda}\,,
\end{equation}

\noindent
where $A_{\mathcal{H}}$ is the area of the horizon and
$\Phi^{\Lambda}_{\mathcal{H}}$ and $\Phi_{\Lambda\,\mathcal{H}}$ are the
values of the electrostatic and magnetostatic potentials over the horizon (constant according to the generalized zeroth law).
Putting all the pieces together we arrive at our main result:\footnote{The overall sign of the electric and
	magnetic terms is unconventional. It is due to the definition of
	$F_{\Lambda}$ with a negative-definite kinetic matrix
	$I_{\Lambda\Sigma}$. It can be easily be changed, but the relative sign
	between the electric and magnetic terms can only be changed at the expense
	of losing explicit symplectic invariance.}

\begin{equation}
	\label{eq:firstlaw1}
	\delta M
	=
	\frac{\kappa \delta A_{\mathcal{H}}}{8\pi G_{N}^{(4)}}
	+\Omega\delta J
	-\Omega_{MN}\left(\Phi^{M}_{\mathcal{H}}-\Phi^{M}_{\infty}\right)\delta q^{N}
	-\left(\mathcal{Q}_{A\, k}
	-\Omega_{MP}T_{A}{}^{P}{}_{N}\Phi^{M}_{\infty}q^{N}\right)\delta^{A}_{\infty}\,.
\end{equation}
In this expression the object $\delta^{A}_{\infty}$ is unusual, but it
just reflects the different forms in which the dualities of the theory can	modify the values of the moduli at infinity, which are also naturally associated to the charges that we have defined.The last term involving $\Phi^{M}_{\infty}$ is also unusual, but it has to be
there if we are going to allow for potentials which do not vanish at infinity. In the examples analyzed in \cite{Ballesteros:2023iqb} we have $\Phi^{M}_{\infty}=0$ and the scalar charges take the expected value. Furthermore, in that case, the scalar term can be brought to the form found in Ref.~\cite{Gibbons:1996af} (up to the normalization of the charges):
\begin{equation}
	-\mathcal{Q}_{A\,k}\delta^{A}_{\infty}
	=
	-\tfrac{1}{4}\Sigma^{x}g_{xy\, \infty}\delta \phi^{y}_{\infty}\,,
\end{equation}
where the scalar charges $\Sigma^x$ are defined through the asymptotic expansions of the scalar fields. Finally, if we plug expression (\ref{eq:scalarChargeExp}) into the first law we arrive at	
\begin{equation}
	\label{eq:firstlaw2}
	\delta M
	=
	T \delta S
	+\Omega\delta J
	-\left(\Omega_{MN}\bar{\Phi}^M \right)\delta q^{N}
	-\left(\Omega_{MP}T_{A}{}^{P}{}_{N} \, \bar{\Phi}^M q^{N} \right)
	\delta^{A}_{\infty}\,,
\end{equation}
where $S$ is the Bekenstein--Hawking entropy, $T$ is the Hawking temperature, and $\bar{\Phi}^M  = \Phi^M_\mathcal{H} - \Phi^M_\infty$ is a symplectic vector built with the electrostatic and magnetostatic potentials taking the difference of their values at the horizon $\mathcal{H}$ and at spatial infinity.
	\clearpage{\pagestyle{empty}\cleardoublepage}

		\chapter{Heterotic String Theory black holes with 	$\alpha'$ corrections} \label{ch:3}

We study 5- and 4-dimensional black hole solutions of the Heterotic Superstring effective action at first order in	$\alpha'$ with, respectively, 3 and 4 charges of arbitrary signs. For a particular choice of the relative signs of these charges the solutions are supersymmetric in the extremal limit, but other signs give rise to extremal, non-supersymmetric black holes. We provide fully	analytic $\mathcal{O}(\alpha')$ solutions and we completely characterize their thermodynamics computing their Hawking temperatures, Wald entropies, masses, gauge charges and their dual thermodynamic potentials. We verify that all these quantities are related by the first law of extended black hole mechanics and by the Smarr formula once we include a potential associated to the dimensionful parameter $\alpha'$ and the scalar charges. We check that applying T-duality, the solutions transform as expected. In the extremal limit, the masses of some of these black holes, once expressed in terms of the physical charges, are corrected in a complicated way. We show that the shift	is always negative, in agreement with the Weak Gravity Conjecture. The solutions with no corrections to the mass are precisely those we generalize to multicenter solutions. We obtain an explicit example of cancellation of forces among non-supersymmetric extremal BHs with $\alpha'$ corrections. We study the non-perturbative stability of these solutions, finding that charge conservation is not compatible with fragmentation processes. We study two families of 5-dimensional solutions and four families of 4-dimensional solutions which differs by the relative signs of the charges and are summarized in the tables \ref{tab:5dBhs} and \ref{tab:4dBhs}.
\begin{table}[h]
	\renewcommand{\arraystretch}{1.5}
	\begin{center}
		\begin{tabular}{c|c c c|c}
			5d/3 charges & EXT &   Multi-Center & SUSY & Non EXT \\
			\hline
			$+$ & \checkmark & \checkmark & \checkmark & \checkmark \\
			\hline
			$-$ & \checkmark & \checkmark & $\times$ & \checkmark  \\
		\end{tabular}
		\caption{\textit{The check mark indicates the solutions whose corrections have been successfully computed. The SUSY column indicates whether the extremal solution is supersymmetric or not. The 3 charges are related to amount of winding, momentum carried and NS5 branes contained in the configuration. The labels $\pm$ describe the relative signs of the momentum and winding charges.}}
		\label{tab:5dBhs}
	\end{center}
\end{table}
\begin{table}[h]
	\renewcommand{\arraystretch}{1.5}
	\begin{center}
		\begin{tabular}{c|c c c|c}
			4d/4 charges & EXT &   Multi-Center & SUSY & Non EXT (KK6 = NS5) \\
			\hline
			$+ +$ & \checkmark & \checkmark & \checkmark & \checkmark  \\
			\hline
			$+ -$ & \checkmark & $-$  & $\times$ & \checkmark \\
			\hline
			$- +$ & \checkmark & \checkmark & $\times$ & \checkmark \\
			\hline
			$- -$ & \checkmark & $-$ & $\times$ & \checkmark \\
		\end{tabular}
		\caption{\textit{The \checkmark indicates the solutions whose corrections have been successfully computed. The $-$ indicates the solutions whose corrections have not been computed. The SUSY column indicates whether the extremal solution is supersymmetric or not. The 4 charges are related to amount of winding, momentum carried, NS5 branes and Kaluza--Klein monopoles (KK6) contained in the configuration. The labels $\pm$ describe the relative signs of the momentum and winding charges and of the NS5 and KK6 charges. The non-extremal solutions have 4 charges, but the NS5 and the KK6 charges are not independent.}}
		\label{tab:4dBhs}
	\end{center}
\end{table}

\section{Introduction}
Black hole (BH) solutions are an excellent setup to test string theory as a theory of quantum gravity. The matching between the Bekenstein-Hawking entropy and the microscopic states counting for the 5-dimensional BPS black hole solution of the string effective action considered by Strominger and Vafa \cite{Strominger:1996sh} is still one of the main successes of Superstring Theory. Soon after, the non-extremal version of such BH was studied and, using U-duality arguments, some limits of the Bekenstein-Hawking entropy have been matched with the microscopic state counting \cite{Horowitz:1996ay}. Since then, a lot of work has been done to extend these results to other solutions and to higher order in $\alpha'$. From the macroscopic side, a major development is due to the introduction of the entropy function formalism \cite{Sen:2007qy}, whose ideas have been used to compute the corrections to the entropy of asymptotically-flat extremal BHs using only their near-horizon limit \cite{Castro:2007hc, Castro:2007ci, Castro:2008ne, DominisPrester:2008ynb}. From the microscopic side, the corrections to the Cardy formula have been computed in \cite{Castro:2008ys, Kutasov:1998zh, Kraus:2005vz}, finding a perfect match.

Despite this success, a proof of the existence of a regular BH connecting the computed near horizon metric and an asymptotically-flat region was lacking. A fully analytical extremal supersymmetric solution with near-horizon metric $AdS_2 \times S^3$ was found for the first time in \cite{Cano:2018qev}, allowing to evaluate independently the corrections to the thermodynamic quantities which require the knowledge of the asymptotic fall-off.\footnote{It is important to remark that there are methods to define such quantities which circumvent this necessity and require only the knowledge of the near-horizon geometry, provided that some extra assumptions hold. For instance, assuming the existence and the regularity of the solution one may use a generalized Komar integral (see for example \cite{Kastor:2008xb}, \cite{Ortin:2021ade} and a recent application \cite{Cano:2023dyg}) to compute the charges at the horizon.} In a similar fashion, the $\alpha'$ corrections to more general families of charged static extremal BH solutions and stationary BH solutions were computed in \cite{Chimento:2018kop,Cano:2018brq, Cano:2019ycn, Cano:2021rey, Cano:2021nzo, Ortin:2021win}, increasing the landscape of the corrections already known \cite{Campbell:1991kz,Natsuume:1994hd,Giveon:2009da}.
Only very recently the corrections to the non-extremal 5- and 4- dimensional BHs of Strominger and Vafa have been obtained \cite{Cano:2022tmn, Zatti:2023oiq}. Another recent development in the computation of $\alpha'$ corrections to the thermodynamics is the method described in \cite{Reall:2019sah} to determine higher derivative corrections. The method essentially allows to compute the first-order higher derivative corrections to BH thermodynamics using the knowledge of the zeroth-order solution only. Such advance makes no longer necessary to solve the corrected equations of motion (EOMs), but it does not spoil the relevance of obtaining an analytical correction. Indeed, the extra information contained in the analytical solution at first-order in $\alpha'$ is still relevant because it can be used to obtain the second-order corrections as described in \cite{Ma:2023qqj} and explicitly applied in \cite{Cano:2023dyg}. Moreover, the method of \cite{Reall:2019sah} has never been applied and tested with the 10-dimensional Heterotic String Theory (HST) effective action at first order in $\alpha'$. 
The goal of this chapter is to summarize the results of \cite{Cano:2021nzo, Ortin:2021win, Cano:2022tmn, Zatti:2023oiq}, where the corrections to several 5- and 4-dimensional BH solutions of HST have been computed. 

The setup we are working with is that of the 10-dimensional HST effective action in the Bergshoeff-de Roo formulation \cite{Bergshoeff:1989de}, with fermions and Yang-Mills fields consistently truncated. Among the string theory effective theories, that of the HST has two properties which make it special and particularly suitable to obtain explicit solutions. On the one hand, it is the only 10-dimensional effective action which has $\alpha'$ corrections already at first order. On the other hand, most of the higher-derivative contributions to the EOMs are proportional to the zeroth-order EOMs.\footnote{See the Lemma proven in \cite{Bergshoeff:1989de}.} Therefore, the first-order EOMs take a much simpler form when evaluated for a correction of a solution of the zeroth order EOMs. On top of that, we have full control of the supersymmetry transformations.

In order to solve the 10-dimensional EOMs we start by making a spherically-symmetric ansatz well suited to perform a dimensional reduction over a torus. Knowing a priori the number of independent unknown functions we need is not a simple task. However, following the logic of \cite{Cano:2021nzo}, we can obtain constraints among them by  using the duality transformations of the HST effective action. With those constraints we can solve some of the EOMs with standard methods. Some of them, however, reduce once combined to higher-order differential equations  that we can only solve with the help of a symbolic manipulation program using the techniques of \cite{Cano:2022tmn}. 

The duality transformation we use is T-duality. T-duality arises because of the presence of toroidal compact directions, and it takes its simplest form when expressed in term of the fields obtained performing the dimensional reduction. Such a representation \cite{Bergshoeff:1994dg} is equivalent to the well-known Buscher rules \cite{Buscher:1987sk,Buscher:1987qj} and has been used to extend them to type II theories \cite{Bergshoeff:1995as,Meessen:1998qm} and to the HST effective action at first order in $\alpha'$ \cite{Bergshoeff:1995cg,Elgood:2020xwu}.
The reason why we can use T-duality transformations to constrain the unknown functions of the ansatz is that all the lower-dimensional fields descending from the dimensional reduction of the Kalb-Ramond (KR) field receive explicit $\alpha'$ corrections. These explicit corrections are interchanged and mixed with the implicit corrections contained in the unknown functions. The explicit corrections can be exactly evaluated because they only require the knowledge of the zeroth order solutions. They can then be used to constrain non-trivially the implicit corrections of the unknown functions.  


In order to compute the corrections to the macroscopic entropy we compute the Wald entropy. However, we cannot directly apply Iyer and Wald's entropy formula \cite{Lee:1990nz,Wald:1993nt, Iyer:1994ys} because of the presence of Chern-Simons terms in the KR field strength, as it is well understood \cite{Elgood:2020nls,Cano:2022tmn}. A first strategy to deal with Chern-Simons terms was proposed in \cite{Sahoo:2006pm} and successfully applied in \cite{DominisPrester:2008ynb,Faedo:2019xii}. Recently, an extension of Wald's algorithm has been proposed \cite{Elgood:2020svt,Elgood:2020mdx,Elgood:2020nls} in order to obtain an entropy formula explicitly gauge invariant and frame independent. The entropy formula proposed in \cite{Elgood:2020nls} has been successfully tested in several examples \cite{Cano:2019ycn,Cano:2022tmn, Zatti:2023oiq, Massai:2023cis}. 

The program of revisiting the Wald's formalism which started with the research of a gauge-invariant and frame-independent entropy formula has recently developed further \cite{Ortin:2021ade,Mitsios:2021zrn,Meessen:2022hcg,Ortin:2022uxa,Ballesteros:2023iqb,Gomez-Fayren:2023wxk,Bandos:2023zbs,Ballestaros:2023ipa}. The main advance is related to the understanding of the role of the chemical potentials associated to the magnetic charges \cite{Ortin:2022uxa}, the role of the chemical potentials associated to the dimensionful parameters appearing in the effective action \cite{Meessen:2022hcg} (see \cite{Kastor:2009wy} for the seminal work on this topic) and the role of the scalar charges \cite{Ballesteros:2023iqb} (first studied in \cite{Gibbons:1996af}). These works produced proposals not specific for HST which can be tested with our analytical solution. For instance, a highly non-trivial test for the proposed entropy formula is the matching between the Hawking temperature $T_H$ and the temperature obtained from the thermodynamic relation $\delta S/\delta M = 1/T$.

This chapter is organized in the following way: in section \ref{sec-HST}, we review the HST effective action with $\alpha'$ corrections. In section \ref{sec-ansatz}, we describe the black hole solutions whose corrections we are going to compute and we present the ansatz we will use. In section \ref{sec-SolutionConstruction}, we explain the different techniques and approaches we used to solve the equations of motions. Finally in section \ref{sec-thermod}, we describe the thermodynamics of the black holes.

\section{HST with $\alpha'$ corrections}\label{sec-HST}

In this chapter we study some classes of Heterotic Superstring (HST) black-hole solutions. In all the cases considered we have vanishing fermionic fields. For the sake of self-consistency, we give a short description of the bosonic sector of the HST effective action and the fermionic supersymmetry transformation rules to first order in $\alpha'$. The bosonic action is enough to extract the equations of motion of the bosonic fields. Studying the fermionic supersymmetry transformations is fundamental to determine whether our solutions are supersymmetric or not.	

\subsubsection{The action}	
The first-order $\alpha'$ corrections in the effective action of the HST were studied in \cite{Gross:1986mw, Metsaev:1987zx, Bergshoeff:1989de}. While different approaches were used, it was later shown in \cite{Chemissany:2007he} that the resulting effective actions are equivalent up to field redefinitions. In this thesis we are going to use the scheme of \cite{Bergshoeff:1989de} with the conventions of	Ref.~\cite{Ortin:2015hya}.\footnote{The relation between the fields used here and those in Ref.~\cite{Bergshoeff:1989de} can be found in	Ref.~\cite{Fontanella:2019avn}.} 

The bosonic sector of the HST effective action describes the massless bosonic degrees of freedom of the HST:	the (string-frame) Zehnbein $e^{a}=e^{a}{}_{\mu}dx^{\mu}$, the Kalb-Ramond 2-form $B=\tfrac{1}{2}B_{\mu\nu}dx^{\mu}\wedge dx^{\nu}$, the dilaton $\phi$	and the Yang-Mills field $A^{A}=A^{A}{}_{\mu}dx^{\mu}$ (where $A,B,C,\ldots$ take	values in the Lie algebra of the gauge group). In order to conveniently describe the HST effective action and its EOMs we introduce some objects. Given the (torsionless, metric-compatible) Levi-Civita spin connection 1-form $\omega^{a}{}_{b}=\omega_{\mu}{}^{a}{}_{b}dx^{\mu}$ which in our convention satisfies the Cartan structure equation
\begin{equation}
	\mathcal{D}e^{a}\equiv de^{a}-\omega^{a}{}_{b}\wedge e^{b}=0\,,
\end{equation}
and has curvature 2-form
\begin{equation}
	{R}{}^{{a}}{}_{{b}} = 
	d \omega{}^{{a}}{}_{{b}}
	- {\omega}{}^{{a}}{}_{{c}}
	\wedge  
	{\omega}{}^{{c}}{}_{{b}}\,,
\end{equation}
we define two torsionful spin connections
\begin{equation}
	\Omega_{(\pm)}^{(0)}{}^a{}_b =	{\omega}^{{a}}{}_{{b}}
	\pm
	\tfrac{1}{2}{H}^{(0)}_{{\mu}}{}^{{a}}{}_{{b}}dx^{{\mu}}\,,
\end{equation}
where ${H}^{(0)}$ is the zeroth-order field strength of the KR 2-form $B$
\begin{equation}
	{H}^{(0)} = d B \,.
\end{equation}
From the torsionful spin connection $\Omega_{(-)}^{(0)}{}^a{}_b$ we can build the associated curvature 2-form and the Lorentz-Chern-Simons 3-form
\begin{subequations}
	\begin{align}
		{R}^{(0)}_{(-)}{}^{{a}}{}_{{b}}
		& =
		d {\Omega}^{(0)}_{(-)}{}^{{a}}{}_{{b}}
		- {\Omega}^{(0)}_{(-)}{}^{{a}}{}_{{c}}
		\wedge  
		{\Omega}^{(0)}_{(-)}{}^{{c}}{}_{{b}}\,, \label{eq:Rminus}
		\\[2mm]
		{\omega}^{{\rm L}\, (0)}_{(-)} 
		& =   
		d{\Omega}^{ (0)}_{(-)}{}^{{a}}{}_{{b}} \wedge 
		{\Omega}^{ (0)}_{(-)}{}^{{b}}{}_{{a}} 
		-\tfrac{2}{3}
		{\Omega}^{ (0)}_{(-)}{}^{{a}}{}_{{b}} \wedge 
		{\Omega}^{ (0)}_{(-)}{}^{{b}}{}_{{c}} \wedge
		{\Omega}^{ (0)}_{(-)}{}^{{c}}{}_{{a}}\,.  
	\end{align}
\end{subequations}
Analogously, we can define the curvature 2-form and the Chern-Simons 3-form of the Yang-Mills fields 
\begin{subequations}
	\begin{align}
		{F}^{A}	& =	d{A}^{A}+\tfrac{1}{2} f_{BC}{}^{A}{A}^{B}\wedge{A}^{C}\,, \\[2mm]
		{\omega}^{\rm YM}	& = dA_{A}\wedge {A}^{A}+\tfrac{1}{3}f_{ABC}{A}^{A}\wedge{A}^{B}\wedge{A}^{C}\,,
	\end{align}
\end{subequations}
where we have used the Killing metric of the gauge group's
Lie algebra in the relevant representation to lower the indices. We define now the first-order field strength of the KR 2-form
\begin{equation}
	\label{eq:H1def-08}
	H^{(1)}
	= 
	d{B}
	+\frac{\alpha'}{4}\left({\omega}^{\rm YM}+{\omega}^{{\rm L}\, (0)}_{(-)}\right)\,.    
\end{equation}
Notice that the $H^{(1)}$ field strength is invariant under both Yang-Mills and local-Lorentz gauge transformations because they induce a compensating Nicolai-Townsend transformations of $B$. We finally introduce  the so-called ``$T$-tensors'', which encode the explicit $\alpha'$ corrections in the action, in the equations of motion and in the Bianchi identity of the Kalb-Ramond 2-form
\begin{equation}
	\label{eq:Ttensors-08}
	\begin{array}{rcl}
		{T}^{(4)}
		& \equiv &
		\dfrac{\alpha'}{4}\left[
		{F}_{A}\wedge{F}^{A}
		+
		{R}_{(-)}{}^{{a}}{}_{{b}}\wedge {R}_{(-)}{}^{{b}}{}_{{a}}
		\right]\,,
		\\
		& & \\ 
		{T}^{(2)}{}_{{\mu}{\nu}}
		& \equiv &
		\dfrac{\alpha'}{4}\left[
		{F}_{A}{}_{{\mu}{\rho}}{F}^{A}{}_{{\nu}}{}^{{\rho}} 
		+
		{R}_{(-)\, {\mu}{\rho}}{}^{{a}}{}_{{b}}{R}_{(-)\, {\nu}}{}^{{\rho}\,  {b}}{}_{{a}}
		\right]\,,
		\\
		& & \\    
		{T}^{(0)}
		& \equiv &
		{T}^{(2)\,\mu}{}_{{\mu}}\,.
		\\
	\end{array}
\end{equation}
The  string-frame HST effective action is then, to first order in $\alpha'$,
\begin{equation}
	\label{heterotic-08}
	S_{\text{HST}}
	=
	\frac{g_{s}^{2}}{16\pi G_{N}^{(10)}}
	\int d^{10}x\sqrt{|{g}|}\, 
	e^{-2{\phi}}\, 
	\left\{
	{R} 
	-4(\partial{\phi})^{2}
	+\tfrac{1}{12}{H}^{(1)\, 2}
	-\tfrac{1}{2}T^{(0)}
	\right\}\,,
\end{equation}
where $R$ is the Ricci scalar of the string-frame metric
$g_{\mu\nu}=\eta_{ab}e^{a}{}_{\mu}e^{b}{}_{\nu}$, $G_{N}^{(10)}$ is the
10-dimensional Newton constant, $g_{s}$ is the HST coupling constant (the
vacuum expectation value of the dilaton $e^{<\phi>}$ which we will identify
with the asymptotic value of the dilaton $e^{\phi_{\infty}}$ in
asymptotically-flat black-hole solutions). The 10-dimensional Newton constant, the string length $\ell_s = \sqrt{\alpha'}$ and the string coupling
constant $g_s$ are related by
\begin{equation}
	\label{eq:10dNewtonconstant-08}
	G_{N}^{(10)} = 8\pi^{6}g_{s}^{2}\ell_{s}^{8}\,.  
\end{equation}
Notice that the expression of the action we are considering contains terms of order $\mathcal{O}(\alpha'{}^2)$ that we must drop in the computation of the EOMs. However, this expression of the action has the advantage of being manifestly invariant under gauge transformations. Dropping the $\mathcal{O}(\alpha'{}^2)$ terms, it would only be invariant up to terms of order $\mathcal{O}(\alpha'{}^2)$.

\subsubsection{Equations of motion}	
Now we want to compute the EOMs. The naive variation of the action (\ref{heterotic-08}) leads to very complicated equations of motion which contain terms with higher derivatives. However, it can be shown that all of them come from the variation of the torsionful spin connection and they are proportional to the zeroth-order EOMs (see the lemma proven in Ref.~\cite{Bergshoeff:1989de}). 
Therefore, in order to compute the corrections to a solution of the zeroth order EOMs, we can consistently ignore those terms. The variation of the HST action with $\delta \Omega_{(-)}{}^a{}_b = 0$ leads, then, to a set of EOMs that can be written in the form
\begin{subequations}
	\begin{align}
		R_{\mu\nu} -2\nabla_{\mu}\partial_{\nu}\phi
		+\tfrac{1}{4}{H}_{\mu}{}^{\rho\sigma}{H}_{\nu\rho\sigma}
		-T^{(2)\,}_{\mu\nu} = 0 	\label{eq:eq1-08} \,, \\[2mm]
		(\partial \phi)^{2} -\tfrac{1}{2}\nabla^{2}\phi
		-\tfrac{1}{4\cdot 3!}{H}^{2}
		+\tfrac{1}{8}T^{(0)} = 0 	\label{eq:eq2-08}  \,, \\[2mm]
		\nabla_{\mu}\left(e^{-2\phi}H^{\mu\nu\rho}\right) = 0 	\label{eq:eq3-08} \,, \\[2mm]
		\alpha' e^{2\phi}\nabla_{(+)\, \mu}\left(e^{-2\phi}F^{A\, \mu\nu}\right) = 
		0 	\label{eq:eq4-08} \,,
	\end{align}
\end{subequations}
where $\nabla_{(+)\, \mu}$ is the covariant derivative which is covariant with respect to $\Omega_{(+)}^{(0)}{}^a{}_b$ and the YM gauge transformation. Notice that the YM fields can be consistently truncated. Finally, the KR field strength satisfies the Bianchi identity
\begin{equation}\label{eq:bianchi-04}
	d H^{(1)} - \frac{1}{3}T^{(4)} = 0 \,.
\end{equation}

\subsubsection{Supersymmetry transformations}	
To first order in $\alpha'$ and for vanishing fermions, the supersymmetry
transformation rules of the gravitino $\psi_{\mu}$, dilatino $\lambda$ and
gaugini $\chi^{A}$ (all of them 32-component Majorana-Weyl spinors,
$\psi_{\mu},\chi^{A}$ and $\epsilon$ with positive chirality and $\lambda$ with negative
chirality) are

\begin{eqnarray}
	\label{eq:gravitino}
	\delta_{\epsilon} \psi_{a}
	& = &
	\nabla^{(+)}{}_{a}\,  \epsilon
	\equiv
	\left(\partial_{a}-\tfrac{1}{4} \Omega^{(+)}{}_{a \, bc} \Gamma^{bc}\right)
	\epsilon \,,
	\\[2mm]
	\label{eq:dilatino}
	\delta_{\epsilon} \lambda
	& = &
	\bigg( \partial_{a} \phi \Gamma^{a}
	-\tfrac{1}{12} H_{abc} \Gamma^{abc} \bigg) \epsilon \,,
	\\[2mm]
	\label{eq:gaugino}
	\alpha'\delta_{\epsilon} \chi^{A}
	& = &
	- \tfrac{1}{4} \alpha' F^{A}{}_{ab} \Gamma^{ab} \epsilon\,. 
\end{eqnarray}

\section{The Ansatz} \label{sec-ansatz}
In this section we present the ansatz we use to solve HST equations of motion. All the ansatz are given in terms of the  the graviton, the dilaton and the Kalb--Ramond 2-form; all the other 10-dimensional fields are consistently truncated. The solutions represent black holes in a number of dimensions $d$ lower than 10 and are built considering proper T$^{10-d}$ torus compactifications. The coordinates chosen for the ansatz are nothing but those adapted to the internal manifold isometries. In each case considered we describe the known leading-order solutions of the $0^{\rm th}$-order equations of motion of the HST effective action and the ansatz we will use to obtain the corrected solution of the $1^{\rm st}$-order equations of motion of the 10-dimensional HST effective action.

\subsection{5-dimensional, 3-charge BHs}

A well-known family of solutions of the 10-dimensional HST effective action at zeroth order in $\alpha'$ is 5-dimensional, 3-charge, black holes \cite{Horowitz:1996ay}. These solutions are built compactifying HST on a T$^5$ torus. 

\subsubsection{Non-extremal leading-order solution} 
An explicit ansatz for the non-vanishing fields is  (we indicate with a hat the 10-dimensional fields)
\begin{subequations}
	\label{eq:3-charge10dsolutionzerothorder}
	\begin{align}
		d\hat{s}^{2}
		& =
		\frac{1}{\mathcal{Z}_{+}\mathcal{Z}_{-}}Wdt^{2}
		-\mathcal{Z}_{0}(W^{-1}dr^{2}+r^{2}d\Omega_{(3)}^{2})
		\nonumber \\[1mm]
		& 
		-\frac{k_{\infty}^{2}\mathcal{Z}_{+}}{\mathcal{Z}_{-}}
		\left[dz+\beta_{+}k_{\infty}^{-1}
		\left(\mathcal{Z}^{-1}_{+}-1\right)dt\right]^{2}
		-dy^{\tilde{m}}dy^{\tilde{m}} \,,
		\label{eq:d10metriczerothorder}
		\\[2mm]
		\hat{H}^{(0)}
		& = 
		\beta_{-}d\left[ k_{\infty}\left(\mathcal{Z}^{-1}_{-}-1\right)
		dt \wedge dz\right]
		+\beta_{0}r^{3}\mathcal{Z}'_{0}\omega_{(3)}\,,
		\\[2mm]
		e^{-2\hat{\phi}}
		& =
		e^{-2\hat{\phi}_{\infty}}
		\mathcal{Z}_{-}/\mathcal{Z}_{0}\,,
	\end{align}
\end{subequations}
where $z$ and $y^{\tilde{m}}$ with $\tilde{m} = 1,\dots 4$ are coordinates of the T$^5$ torus with periodicity $2\pi \ell_s$. The prime indicates derivation with respect to the radial coordinate $r$ and 
\begin{subequations}
	\begin{align}
		d\Omega^{2}_{(3)}
		& =
		\frac{1}{4}\left[ (d\psi+\cos{\theta}d\varphi)^{2}
		+ d\Omega^{2}_{(2)} \right]\,,
		\\[2mm]
		d\Omega^{2}_{(2)}
		& =
		d\theta^{2}+\sin^{2}{\theta}d\varphi^{2}\,,\\[2mm]
		\omega_{(3)} &  = \frac{1}{8}d\cos{\theta}\wedge d\varphi\wedge d\psi\,,
	\end{align}
\end{subequations}
are, respectively, the metrics of the round 3- and 2-spheres of unit radii and the volume 3-form of the former. This ansatz depends on 4 functions	
\begin{equation}
	\mathcal{Z}_+ \,, \quad \mathcal{Z}_-\,, \quad \mathcal{Z}_0 \,, \quad W \,,
\end{equation}
and reduces to the one for extremal black holes (in particular, for the Strominger-Vafa black hole \cite{Strominger:1996sh}) when the so-called ``blackening factor'' $W$ is absent or, equivalently, $W=1$. In the non-extremal case the equations of motion are solved at $0^{\rm th}$ order in $\alpha'$ for
\cite{Horowitz:1996ay}			
\begin{equation}
	\label{eq:Zs3-chargezerothorder-04}
	\mathcal{Z}_i = 1+ \frac{q_i}{r^2} \,, \qquad
	W
	=
	1+\frac{\omega}{r^{2}} \,,
	\quad
	i=0,+,-\,,
\end{equation}
where asymptotic flatness and the standard normalization of the metric at
spatial infinity have already been imposed, leaving just 4 integration	constants.
\begin{equation}
	q_{0}\,,\,\,q_{+}\,,\,\,q_{-}\,,\,\,\omega\,.
\end{equation}
These are related to the other constants appearing in the solution by the following 3 relations
\begin{equation}
	\label{eq:omegaqbetarelation}  
	\beta_{i} = s_{i}\sqrt{1-\frac{\omega}{q_{i}}}\,,
	\hspace{1cm}
	s_{i}^{2}=1\,.
\end{equation}
Finally, $\hat{\phi}_\infty$ and $k_{\infty}$ are moduli corresponding to the  asymptotic values of the dilaton and of the scalar field describing the radii of non trivial internal circles in string units 
\begin{equation}
	\text{vol}(\text{S}^1_{\infty,z})/2\pi = R_z \equiv k_{\infty} \ell_s \,.
\end{equation}
We end up having a solution with 6 independent parameters. Notice that this is the exact amount of parameters that, according to the no-hair conjectures, we expect in a non-extremal solution with three Abelian charges and two scalar fields.\footnote{We are ignoring the KK fields with a trivial profile.} $q_i$, $\omega$, $\beta_i$ can be indeed expressed in terms of the physical quantities characterizing the solution: the mass, the three Abelian charges and the two moduli.

Without loss of generality we can always choose $\omega <0$. Indeed, the $\omega>0$ case with radial coordinate $r$ and parameters $q_i$ can be mapped to the case with $\tilde{\omega}<0$, radial coordinate $\tilde{r}$ and parameters $\tilde{q}_i$ by the coordinate transformation and redefinition of the parameters\footnote{This statement is easy to verify at zeroth order. We verified it at first order in $\alpha'$ by building solutions for both cases and checking that they are mapped into each other \cite{Cano:2022tmn}.}
\begin{equation}
	\label{eq:omegatominusomegacoordinatetransformation}
	\tilde{r}^{2} = r^{2}+{\omega}\,, \qquad 	 \tilde{\omega} = -\omega\,, \qquad \tilde{q}_{i}
	=
	{q}_{i}-\omega  \,, 
\end{equation}
Negative values of the parameters $q_{0\,\pm}$ can also be related to positive ones by similar
transformations. However, these transformations shift $\omega$ by positive
quantities and we may end up violating the assumed negativity of
$\omega$. Moreover, it is well known that they have to be strictly positive if
we want to obtain regular black hole in the extremal limit. Thus, we will assume $q_i >0$.

\subsubsection{Non-extremal, 10-dimensional ansatz at first order in $\alpha'$}

Based on our knowledge of the $0^{\rm th}$-order non-extremal solution, we propose an educated ansatz
for the 3-charge 5-dimensional black-hole solution with 7 independent functions\footnote{Solving the equations of motion we will see that only 6 of them are truly independent. Different choice of the 7th function are equivalent to performing a change of coordinates. This freedom turns out to be useful to write the final solutions in a simpler way.} of the radial coordinate
$r$

\begin{equation}
	\label{eq:7functions}
	\mathcal{Z}_{0}\,,\,\,\mathcal{Z}_{+}\,,\,\,\mathcal{Z}_{-}\,,\,\,
	\mathcal{Z}_{h0}\,,\,\,\mathcal{Z}_{h-}\,,\,\,W_{tt}\,,\,\,W_{rr}\,.
\end{equation}

\noindent
The new functions $\mathcal{Z}_{h0}$ and $\mathcal{Z}_{h-}$ are, respectively,
identical to $\mathcal{Z}_{0}$ and $\mathcal{Z}_{-}$ at $0^{\rm th}$ order and they
have to be introduced because these functions get different corrections when
they occur in different components of the fields of the solutions. The
functions $W_{tt}$ and $W_{rr}$ reduce to $W$ at $0^{\rm th}$ order and are
needed because $W$ gets different corrections when it is part of the $tt$ or
the $rr$ components of the metric. The functions in Eq.~(\ref{eq:7functions}) are assumed to have the following
form ($i=0,+,-$, $j=tt, rr$):
\begin{equation}
	\begin{aligned}
		\mathcal{Z}_{i}
		& =
		1 + \frac{q_{i}}{r^{2}} + \alpha' \delta \mathcal{Z}_{i}\,,
		\\[2mm]
		\mathcal{Z}_{hi}
		& =
		1 + \frac{q_{i}}{r^{2}} + \alpha' \delta \mathcal{Z}_{hi}\,,
		\\[2mm]
		W_{j}
		& =
		1+\frac{\omega}{r^{2}} + \alpha' \delta W_{j}\,.
	\end{aligned}
\end{equation}

\noindent
Thus, they become the functions of the $0^{\rm th}$-order ansatz
Eqs.~(\ref{eq:Zs3-chargezerothorder-04}) when $\alpha'=0$
(note that $\mathcal{Z}_{h0}=\mathcal{Z}_{0}$, $\mathcal{Z}_{h-}=\mathcal{Z}_{-}$ and $W_{tt}=W_{rr}=W$).	In terms of these functions and constants, the 10-dimensional fields are
assumed to be given by			
\begin{subequations}
	\label{eq:3-charge10dsolution-04}
	\begin{align}
		d\hat{s}^{2}
		& =
		\frac{1}{\mathcal{Z}_{+}\mathcal{Z}_{-}}W_{tt}dt^{2}
		-\mathcal{Z}_{0}(W_{rr}^{-1}dr^{2}+r^{2}d\Omega_{(3)}^{2})
		\nonumber \\[1mm]
		& 
		-\frac{k_{\infty}^{2}\mathcal{Z}_{+}}{\mathcal{Z}_{-}}
		\left[dz+\beta_{+}k_{\infty}^{-1}
		\left(\mathcal{Z}^{-1}_{+}-1\right)dt\right]^{2}
		-dy^{m}dy^{m}\,,
		\label{eq:d10metric-04}
		\\[2mm]
		\hat{H}^{(1)}
		& = 
		\beta_{-}d\left[ k_{\infty}\left(\mathcal{Z}^{-1}_{h-}-1\right)
		dt \wedge dz\right]
		+\beta_{0}r^{3}\mathcal{Z}'_{h0}\omega_{(3)}\,,
		\\[2mm]
		e^{-2\hat{\phi}}
		& =
		-\frac{2 c_{\hat{\phi}}}{r^{3} \mathcal{Z}_{h-}'}  \left(\frac{\mathcal{Z}_{h-}}{\mathcal{Z}_-}\right)^2 \left(\frac{W_{tt}}{W_{rr}}\right)^{1/2} \frac{\mathcal{Z}_-}{ \mathcal{Z}_{0}}   \,,
	\end{align}
\end{subequations}
where $c_{\hat{\phi}}$ is a constant that is determined by setting the asymptotic value of the dilaton to $\hat{\phi}_\infty$. The form of the ansatz for the dilaton has been chosen to identically satisfy the KR field equation of motion and reduces to the $0^{\rm th}$-order one,		Eq.~(\ref{eq:3-charge10dsolutionzerothorder}), when
$\mathcal{Z}_{h0}=\mathcal{Z}_{0}$, $\mathcal{Z}_{h-}=\mathcal{Z}_{-}$ and
$W_{tt}=W_{rr}=W$. We expect the same number of independent physical parameters as in the $0^{\rm th}$-order solutions, namely 6. We are going to assume that the three $0^{\rm th}$-order relations		Eqs.~(\ref{eq:omegaqbetarelation}) are satisfied at first order as well, without corrections 
\begin{equation}
	\beta_{i} = s_{i}\sqrt{1-\frac{\omega}{q_{i}}}\,,
	\hspace{1cm}
	s_{i}^{2}=1\,.
\end{equation}

\subsubsection{5-dimensional form}
We can write the ansatz in terms of 5-dimensional fields by exploiting the dictionary between higher and lower dimensional fields of appendix \ref{sec-dictionary}. We obtain in the string frame\footnote{In order to perform the dimensional reduction we used the zeroth-order solution to verify that $\omega^{(L)}_{trz}$ vanishes. No further on-shell relations have been used. To get an explicit expression for $\hat{B}_{\hat{\mu}\hat{\nu}}$ we fixed some integration constants imposing the absence of $\alpha'$ corrections to the asymptotic value of the fields and to the charge associated to $C_w^{(1)}$. See \cite{Cano:2022tmn} for further details.}
\begin{subequations}
	\label{eq:5dsolution1storder}
	\begin{align}
		ds^{2}
		& =
		\frac{W_{tt}}{\mathcal{Z}_{+}\mathcal{Z}_{-}}dt^{2}
		-\mathcal{Z}_{0}
		\left({W}^{-1}_{rr}dr^{2}+r^{2}d\Omega^{2}_{(3)}\right)\,,
		\\[2mm]
		\label{eqM015}
		H^{(1)}
		& =
		\beta_{0} r^{3} \mathcal{Z}_{h0}'\,\omega_{(3)}\,, 
		\\[2mm]
		\label{eqM010}
		k
		& =
		k_{\infty} \sqrt{\mathcal{Z}_{+}/\mathcal{Z}_{-}}\,,
		\\[2mm]
		\label{eqM011}
		A
		& =
		k_{\infty}^{-1}\beta_{+}\left[-1+\frac{1}{\mathcal{Z}_{+}}\right]dt \,,
		\\[2mm]
		\label{eqM013}
		C^{(1)}
		& =
		k_{\infty}\beta_{-}\left[-1+\frac{1}{\mathcal{Z}_{h-}}
		\left(1+\alpha'\frac{\Delta_{C}}{\beta_{-}}\right)\right]
		dt \,,
		\\[2mm]
		\label{eqM005}
		e^{-2\phi}
		& =
		- \frac{2\,c_{\phi}}{r^{3} \mathcal{Z}_{h-}'}
		\left(\frac{\mathcal{Z}_{h-}}{\mathcal{Z}_-}\right)^2 \left(\frac{W_{tt}}{W_{rr}}\right)^{1/2}
		\frac{\sqrt{\mathcal{Z}_{+}\mathcal{Z}_{-}}}{\mathcal{Z}_{0}} \,,
	\end{align}
\end{subequations}
with $c_{\phi} = c_{\hat{ \phi}}\, k_{\infty}$ and	
\begin{equation}
	\Delta_{C}
	=
	\frac{-\left(\beta_{-}\mathcal{Z}_{+}\mathcal{Z}_{-}'+\beta_{+}\mathcal{Z}_{+}'\mathcal{Z}_{-}\right)W'+2\left(\beta_{+}+\beta_{-}\right)\mathcal{Z}_{+}'W\mathcal{Z}_{-}'}{8\,\mathcal{Z}_{0}\mathcal{Z}_{-}\mathcal{Z}_{+}}\,.
\end{equation}
The metric in the \textit{modified
	Einstein-frame} is given by\footnote{It is the unique Einstein frame in which the metric is asymptotically flat with the standard normalization. It has been introduced in \cite{Maldacena:1996ky}. See appendix \ref{sec-Einsteinnormalization} for further details}			
\begin{equation}
	\label{eq:modifiedEinsteinmetriccorrected}
	ds^{2}_{E} = F
	\left[f^{2}W_{tt}dt^{2}
	-f^{-1}\left(W_{rr}^{-1}dr^{2}+r^{2}d\Omega_{(3)}^{2}\right)\right]\,,
\end{equation}

\noindent
with

\begin{equation}
	F = \left[ -\frac{2c_{\phi} e^{2\phi_{\infty}}
		\mathcal{Z}_{h-}^{2}}{r^{3}\mathcal{Z}_{h-}'\mathcal{Z}_{-}^{2}}
	\right]^{2/3} \left(\frac{W_{tt}}{W_{rr}}\right)^{1/3} \,, \qquad \quad f^{-3}
	=
	\mathcal{Z}_{+}\mathcal{Z}_{-}\mathcal{Z}_{0} \,.
\end{equation}

\noindent 
The gauge fields in the \textit{modified Einstein normalization}\footnote{It is the global rescaling of the 5d fields that absorb all the explicit occurrence of the moduli in the action. Such normalization has been introduced in \cite{Gomez-Fayren:2023wxk} without a specific name. See appendix \ref{sec-Einsteinnormalization} for further details} are
\begin{subequations}
	\begin{align}
		A_E & = A \, e^{\frac{2}{3}\phi_\infty} \,, \\[2mm]
		C^{(1)}_E & =  C^{(1)}  e^{\frac{2}{3}\phi_\infty} \,, \\[2mm]
		B_E & =  B  \, e^{\frac{4}{3}\phi_\infty} \,,
	\end{align}
\end{subequations}

\noindent
Finally, the auxiliary combination $k^{(1)}$ involved in T-duality transformations that we obtain using the dictionary of appendix \ref{sec-dictionary} is
\begin{equation}
	k^{(1)}
	=
	k_{\infty} \sqrt{\mathcal{Z}_{+}/\mathcal{Z}_{-}}
	\left(1+\alpha' \Delta_{k}\right)\,,
	\label{eqM012}
\end{equation}
with
\begin{equation}
	\label{eq:Deltak}
	\Delta_{k}
	=
	\frac{-W\left(\mathcal{Z}_{+}\mathcal{Z}_{-}'-\mathcal{Z}_{-}\mathcal{Z}_{+}'\right)^{2}+\left(\beta_{-}\mathcal{Z}_{+}\mathcal{Z}_{-}'+\beta_{+}\mathcal{Z}_{-}\mathcal{Z}_{+}'\right)^{2}}{8\,\mathcal{Z}_{0}\mathcal{Z}_{-}^{2}\mathcal{Z}_{+}^{2}}\,.
\end{equation}

\subsubsection{Multi-center leading-order solution}

In the extremal case we have trivial $W$ and we can write the ansatz in the form 
\begin{subequations}\label{eq:ansatz5dEqui}
	\begin{align}
		d\hat{s}^{2}
		& =
		\frac{1}{\mathcal{Z}_{+}\mathcal{Z}_{-}}Wdt^{2}
		-\mathcal{Z}_{0} d\sigma^2
		\nonumber \\[1mm]
		& 
		-\frac{k_{\infty}^{2}\mathcal{Z}_{+}}{\mathcal{Z}_{-}}
		\left[dz+\beta_{+}k_{\infty}^{-1}
		\left(\mathcal{Z}^{-1}_{+}-1\right)dt\right]^{2}
		-dy^{\tilde{m}}dy^{\tilde{m}} \,,
		\\[2mm]
		\hat{H}^{(0)}
		& = 
		\beta_{-}d\left[ k_{\infty}\left(\mathcal{Z}^{-1}_{-}-1\right)
		dt \wedge dz\right]
		+\beta_{0}\star_\sigma d \mathcal{Z}_0 \,,
		\\[2mm]
		e^{-2\hat{\phi}}
		& =
		e^{-2\hat{\phi}_{\infty}}
		\mathcal{Z}_{-}/\mathcal{Z}_{0}\,,
	\end{align}
\end{subequations}
with $d\sigma^2$ the metric of the flat 4-dimensional Euclidean space $\mathbb{E}^4$ and $\star_\sigma$ the Hodge star operator defined on it. The equations of motion are solved provided that $\mathcal{Z}_i$ are harmonic functions in $\mathbb{E}^4$. Imposing asymptotic flatness, we have the general solution 
\begin{equation}\label{eq:harmonics5d}
	\mathcal{Z}_i = 1 + \sum_{a = 1}^{n_c} \frac{q_i^a}{r^2_a} \,.
\end{equation}
where $n_c$ is the number of poles of the harmonic function and $r_a$ is the distance in $\mathbb{E}^4$ from the $a$th pole located at $x^k_a$
\begin{equation}
	r_a^2 = (x^k_a - x^k)^2 \,.
\end{equation}
Expanding the ansatz (\ref{eq:ansatz5dEqui}) close to the poles' locations we can see that they are nothing but the position of 3-charge, 5-dimensional, extremal black holes described by the ansatz (\ref{eq:3-charge10dsolutionzerothorder}). Therefore, the integration constants appearing in the solution have the same interpretation that they had in the non-extremal case with the difference that now $\beta_i^2 = 1$ and that we have three charges $q_+^a$, $q_-^a$, $q_0^a$ for every pole of the harmonic functions. 

In order to solve the EOMs it turns out to be not necessary to modify the 10-dimensional ansatz introducing new functions and we will use (\ref{eq:ansatz5dEqui}).

\subsection{4-dimensional, 4-charge BHs}

Another set of solutions of the 10-dimensional two-derivative theory is obtained considering a simple dimensional reduction over a torus T$^6$. These solutions represent 4-dimensional, 4-charge black holes\cite{Horowitz:1996ay}.

\subsubsection{Non-extremal leading-order solution}
An explicit ansatz is  (we indicate with an hat the 10-dimensional fields)
\begin{subequations} \label{eq:anatz4d}
	\begin{align}
		\begin{split}
			d\hat{s}^2 \, = & \;\; \frac{W}{\mathcal{Z}_+\mathcal{Z}_-}dt^2-\mathcal{Z}_0\mathcal{Z}_\mathcal{H}(W^{-1} dr^2 + r^2d\Omega_{(2)}^2) \\[2mm] & \, -\ell_{\infty }^2\frac{\mathcal{Z}_0}{\mathcal{Z}_\mathcal{H}} \bigg[dw + \ell_{\infty }^{-1}\beta_{\mathcal{H}} \, q_{\mathcal{H}} \cos\theta d\varphi \bigg]^2 \\[2mm] & \, -k_{\infty}^2\frac{\mathcal{Z}_+}{\mathcal{Z}_-} \left[\,dz + k_{\infty}^{-1}\beta_+(\mathcal{Z}^{-1}_+-1)\,dt\right]^2 - dy^{\tilde{m}} dy^{\tilde{m}}\,,
		\end{split} \\[2mm]
		\hat{H} \, = & \;\; k_{\infty}\beta_- \,d\left[(\mathcal{Z}^{-1}_{-}-1)\,dt\wedge dz\right] + \ell_{\infty }\beta_{0} \, r^2 \mathcal{Z}_{0}'\,\omega_{(2)} \wedge d w\,, \\[2mm]
		e^{-2\hat{\phi}} \, = & \; \; 	e^{-2\hat{\phi}_\infty} {\mathcal{Z}_-}/{\mathcal{Z}_0}\,, 
	\end{align}
\end{subequations}
where $z$, $w$ and $y^{\tilde{m}}$ with $\tilde{m} = 1,\dots 4$ are coordinates of the T$^6$ torus with periodicity $2\pi \ell_s$. The prime indicates derivation with respect to the radial coordinate $r$ and
\begin{subequations}
	\begin{align}
		d\Omega_{(2)}^2 & = d\theta^2+\sin^2(\theta) d\varphi^2 \,, \\[2mm]
		\omega_{(2)} & = \sin \theta \, d\theta \wedge d\varphi \,.
	\end{align}
\end{subequations}
$k_{\infty}$ and $\ell_{\infty }$ are moduli corresponding to the  asymptotic values of the scalar fields describing the radii of the two non-trivial internal circles in string units 
\begin{equation}
	\text{vol}(\text{S}^1_{\infty,z})/2\pi = R_z \equiv k_{\infty} \ell_s \,, \qquad 	\text{vol}(\text{S}^1_{\infty,w})/2\pi = R_w \equiv \ell_{\infty} \ell_s \,,
\end{equation}
and we have assumed that the asymptotic value of the scalars associated to the $\text{T}^4$ is 1, in such a way that $\text{vol}(\text{T}^4) = (2\pi \ell_s)^4 $.
$\hat{\phi}_\infty$ is the asymptotic value of the dilaton. 
The ansatz depends on 5 functions 
\begin{equation}
	\mathcal{Z}_+ \,,\quad \mathcal{Z}_- \,,\quad \mathcal{Z}_0 \,,\quad \mathcal{Z}_\mathcal{H} \,,\quad W \,,
\end{equation}
and reduces to the one for extremal black holes when $W=1$. The equations of motion are solved at $0^{\rm th}$ order in $\alpha'$ for
\cite{Horowitz:1996ay}			
\begin{equation}
	\mathcal{Z}_i = 1+ \frac{q_i}{r} \,, \qquad
	W
	=
	1+\frac{\omega}{r} \,,
	\quad
	i=0,+,-,\mathcal{H}\,,
\end{equation}

\noindent
where asymptotic flatness and the standard normalization of the metric at spatial infinity have already been imposed.  The ansatz solves the EOMs if
\begin{equation}
	\omega = q_i \left(1-\beta_i^2\right) \,.
\end{equation}  
The integration constants $q_i$, $\omega$ are then related to the $\beta_i$ by the following 4 relations
\begin{equation}
	\beta_{i} = s_{i}\sqrt{1-\frac{\omega}{q_{i}}}\,,
	\hspace{1cm}
	s_{i}^{2}=1\,.
\end{equation}
Therefore, we have a total of 8 independent integration constants, namely: $q_i$, $\omega$, $k_\infty$, $\ell_{\infty}$ and $\phi_\infty$. They as many as we expect for a non-extremal black hole with 4 charges and 3 scalar fields. In particular, one can express $\omega$, $q_i$ and $\beta_i$ in terms of the physical mass, the 4 gauge charges and the 3 moduli. 

Finally, we assume that  $q_i > 0$ and $\omega < 0$. The first condition is necessary to obtain regular solutions at zeroth order. The second condition can always be satisfied. Indeed, a solution with $\omega > 0$ can be mapped into a solution with $\tilde{\omega} < 0 $ with the change of coordinates and the dictionary between the $q_i$
\begin{equation}
	\tilde{r} = r + \omega\,, \qquad \tilde{\omega} = - \omega \,, \qquad \tilde{q}_i = q_i - \omega \,.
\end{equation}

\subsubsection{Non-extremal, 10-dimensional ansatz at first order in $\alpha'$}

In order to describe a $1^{\rm st}$-order non-extremal solution carrying 4 Abelian gauge charges, we propose an educated ansatz with 8 independent functions
\begin{equation}
	\mathcal{Z}_{0}\,,\,\,\mathcal{Z}_{+}\,,\,\,\mathcal{Z}_{-}\,,\,\,\mathcal{Z}_{\mathcal{H}}\,,\,\,
	\mathcal{Z}_{h0}\,,\,\,\mathcal{Z}_{h-}\,,\,\,W_{tt}\,,\,\,W_{rr}\,,
\end{equation}
which takes the form
\begin{subequations}\label{eq021}
	\begin{align}
		\begin{split}
			d\hat{s}^2 \, = & \;\; \frac{W_{tt}}{\mathcal{Z}_+\mathcal{Z}_-}dt^2-\mathcal{Z}_0\mathcal{Z}_\mathcal{H}(W_{rr}^{-1} dr^2 + r^2d\Omega_{(2)}^2) \\[2mm] & \, -\ell_{\infty }^2\frac{\mathcal{Z}_0}{\mathcal{Z}_\mathcal{H}} \bigg[dw + \ell_{\infty }^{-1}\beta_{\mathcal{H}} \, q_{\mathcal{H}} \cos\theta d\varphi \bigg]^2 \\[2mm] & \, -k_{\infty}^2\frac{\mathcal{Z}_+}{\mathcal{Z}_-} \left[\,dz + k_{\infty}^{-1}\beta_+(\mathcal{Z}^{-1}_+-1)\,dt\right]^2 - dy^{\tilde{m}} dy^{\tilde{m}}\,,
		\end{split} \\[2mm]
		\hat{H} \, = & \;\; k_{\infty}\beta_- \,d\left[(\mathcal{Z}^{-1}_{h-}-1)\,dt\wedge dz\right] + \ell_{\infty }\beta_{0} \, r^2 \mathcal{Z}_{h0}'\,\omega_{(2)} \wedge d w\,, \\[2mm]
		e^{-2\hat{\phi}} \, = & \; \; -\frac{c_{\hat{ \phi}}}{r^2 \mathcal{Z}_{h-}'} \sqrt{\frac{W_{tt}}{W_{rr}}} \left(\frac{\mathcal{Z}_{h-}}{\mathcal{Z}_{-}}\right)^2 \frac{\mathcal{Z}_-}{\mathcal{Z}_0}\,. \label{eq:dilaton10d4charge}
	\end{align}
\end{subequations}
$c_{\hat{ \phi}}$ is a constant that we will fix later imposing that the asymptotic value of the dilaton is $\hat{\phi}_\infty$. The ansatz for the dilaton has been chosen in such a way that the Kalb-Ramond EOM is automatically satisfied. In order to recover the $0^{\rm th}$-order configuration for $\alpha'=0$ we assume that the unknown functions have the form
\begin{subequations}
	\begin{align}
		& 	\mathcal{Z}_i = 1 + \frac{q_i}{r} + \alpha' \delta \mathcal{Z}_i \,, \hspace{1.3cm} i = 0, \pm \,, \mathcal{H} \\[2mm]
		& 	\mathcal{Z}_{hi} = 1 + \frac{q_i}{r} + \alpha' \delta \mathcal{Z}_{hi} \,, \hspace{1cm}  i = 0, - \,, \\[2mm]
		&	W_{j} = 1 + \frac{\omega}{r} +\alpha' \delta {W}_j \,,\hspace{1.2cm}  j = tt, rr \,.
	\end{align}
\end{subequations}
We assume an ansatz for the $\beta$s such that this relation is not modified at first order in $\alpha'$, i.e. we consider
\begin{equation}
	\beta_i = s_i\sqrt{1-\frac{\omega}{q_i}}\,,
\end{equation}		

\subsubsection{4-dimensional form}
Using the relation between 10d and 4d fields summarized in appendix \ref{sec-dictionary}, our ansatz in the 4d string frame takes the form\footnote{In order to perform the dimensional reduction we used the zeroth order solution to verify that $\omega^{(L)}_{trz}$ vanishes. No further on-shell relations have been used. To get an explicit expression for $\hat{B}_{\hat{\mu}\hat{\nu}}$ we fixed some integration constants imposing the absence of $\alpha'$ corrections to the asymptotic value of the fields and to the charge associated to $C_w^{(1)}$. See \cite{Zatti:2023oiq} for further details.} (we omit the indices over the trivial $\text{T}^4$)
\begin{subequations} \label{ansatz4d}
	\begin{align}
		ds^2 & =
		\frac{W_{tt}}{\mathcal{Z}_{+}\mathcal{Z}_{-}}dt^{2}
		-\mathcal{Z}_{0}\mathcal{H}\left({W}^{-1}_{rr}dr^{2}+r^{2}d\Omega^{2}_{(2)}\right)\,, \\[4mm]
		G_{mn} & \equiv \begin{pmatrix} \ell^2 & 0 \\ 0 & k^2   \end{pmatrix} = \begin{pmatrix} \ell_\infty^2{\mathcal{Z}_0}/{\mathcal{Z}_\mathcal{H}} & 0 \\ 0 & k_\infty^2 {\mathcal{Z}_+}/{\mathcal{Z}_-}  \end{pmatrix} \,, \label{eqscalars4d} \qquad m,n \in \{w, z\} \,, \\[4mm]
		A^m & = \begin{pmatrix}\ell_\infty^{-1}\beta_\mathcal{H} \, q \cos \theta \, d\varphi 		
			,& k_\infty^{-1}\beta_+\left[-1+\mathcal{Z}_+^{-1}\right]dt  
		\end{pmatrix}\,, \label{eqvecA4d} \\[4mm]
		C^{(1)}_m & = \begin{pmatrix}
			\ell_\infty \beta_0 \, q_0 \cos \theta \, d\varphi  , & 	k_\infty \beta_-\left[-1+\mathcal{Z}_{h-}^{-1}  \left(1+\alpha' \beta_-^{-1} \Delta_C \right)\right] dt
		\end{pmatrix} \,,  \label{eqvecC4d} \\[4mm] 
		e^{-2\phi} & =  -\frac{c_{{ \phi}}}{r^2 \mathcal{Z}_{h-}'} \sqrt{\frac{W_{tt}}{W_{rr}}} \left(\frac{\mathcal{Z}_{h-}}{\mathcal{Z}_{-}}\right)^2 \sqrt{\frac{\mathcal{Z}_+\mathcal{Z}_-}{\mathcal{Z}_\mathcal{H}\mathcal{Z}_0}}  \label{eqdilaton4d} \,,
	\end{align}
\end{subequations}
with $c_{\phi} = c_{\hat{\phi}} k_\infty l_\infty$ and
\begin{equation}
	\Delta_C = \frac{-W'\left(\beta_{-}\mathcal{Z}_+\mathcal{Z}_{-}'+\beta_+\mathcal{Z}_+'\mathcal{Z}_-\right)+2\left(\beta_{+}+\beta_{-}\right)\mathcal{Z}_+'W\mathcal{Z}_-'}{8\,\mathcal{Z}_0\mathcal{Z}_-\mathcal{Z}_+} + \mathcal{O}(\alpha) \,.
\end{equation}
The metric in the \textit{modified Einstein frame} takes the form
\begin{equation} \label{eq:modifiedEInstein4d}
	ds^2_E = ds^2 e^{-2(\phi-\phi_\infty)} = F \left[\frac{W_{tt}}{f} dt^2 - f\left(W_{rr}^{-1}dr^2 + r^2 d\Omega_{(2)} \right)\right] \,,
\end{equation}
where 
\begin{equation}
	F = -\frac{c_{{ \phi}}\,e^{2\phi_\infty}}{r^2 \mathcal{Z}_{h-}'} \sqrt{\frac{W_{tt}}{W_{rr}}} \left(\frac{\mathcal{Z}_{h-}}{\mathcal{Z}_{-}}\right)^2 \,, \qquad f = \sqrt{\mathcal{Z}_+\mathcal{Z}_-\mathcal{Z}_0\mathcal{Z}_\mathcal{H}} \,.
\end{equation}
The 4d gauge fields in the \textit{modified Einstein normalization} are
\begin{subequations}
	\begin{align}
		A^m_E & = A^m e^{\phi_\infty} \,, \\[2mm]
		C^{(1)}_m{}_E & =  C^{(1)}_m  e^{\phi_\infty} \,.
	\end{align}
\end{subequations}
Finally, the combinations $k^{(1)}$ and $\ell^{(1)}$ defined in appendix \ref{sec-dictionary} and involved in T-duality transformations take the form
\begin{equation} \label{eq:scalarcombination}
	\qquad k^{(1)}  =  k_\infty\sqrt{\frac{\mathcal{Z}_+}{\mathcal{Z}_-}} \left(1+\alpha' \Delta_k\right)\,,	 
	\qquad \ell^{(1)}  = \ell_\infty \sqrt{\frac{\mathcal{Z}_0}{\mathcal{Z}_\mathcal{H}}}\left(1+\alpha' \Delta_\ell\right)\,,
\end{equation}
with
\begin{subequations}
	\begin{align}
		& \Delta_k = \frac{-W\left(\mathcal{Z}_+\mathcal{Z}_-'-\mathcal{Z}_-\mathcal{Z}_+'\right)^2+\left(\beta_{-}\mathcal{Z}_+\mathcal{Z}_{-}'+\beta_+\mathcal{Z}_{-}\mathcal{Z}_+'\right)^2}{8\,\mathcal{Z}_0\mathcal{Z}_-^2\mathcal{Z}_+^2} + \mathcal{O}(\alpha)\,, \\[2mm]
		& \Delta_\ell = \frac{-W\left(\mathcal{Z}_0\mathcal{Z}_\mathcal{H}'-\mathcal{Z}_\mathcal{H}\mathcal{Z}_0'\right)^2-\left(\beta_\mathcal{H} \mathcal{Z}_0\mathcal{Z}_\mathcal{H}'+\beta_0\mathcal{Z}_\mathcal{H}\mathcal{Z}_0'\right)^2}{8\,\mathcal{Z}_0^3 \mathcal{Z}_\mathcal{H}^3} + \mathcal{O}(\alpha) \,.
	\end{align}
\end{subequations}

\subsubsection{Multi-center leading-order solution}

In the extremal case we have $W = 1$ and we can write the ansatz in the form 
\begin{subequations}\label{eq:4dansatzLeading}
	\begin{align}
		d\hat{s}^{2}
		& =
		\frac{1}{\mathcal{Z}_{+}\mathcal{Z}_{-}}Wdt^{2}
		-\mathcal{Z}_{0} d\sigma^2
		\nonumber \\[1mm]
		& 
		-\frac{k_{\infty}^{2}\mathcal{Z}_{+}}{\mathcal{Z}_{-}}
		\left[dz+\beta_{+}k_{\infty}^{-1}
		\left(\mathcal{Z}^{-1}_{+}-1\right)dt\right]^{2}
		-dy^{\tilde{m}}dy^{\tilde{m}} \,,
		\\[2mm]
		\hat{H}^{(0)}
		& = 
		\beta_{-}d\left[ k_{\infty}\left(\mathcal{Z}^{-1}_{-}-1\right)
		dt \wedge dz\right]
		+\beta_{0}\star_\sigma d \mathcal{Z}_0 \,,
		\\[2mm]
		e^{-2\hat{\phi}}
		& =
		e^{-2\hat{\phi}_{\infty}}
		\mathcal{Z}_{-}/\mathcal{Z}_{0}\,,
	\end{align}
\end{subequations}
with $d\sigma^2$ the metric of a 4-dimensional metric
\begin{equation}
	d \sigma^2 = \mathcal{Z}_\mathcal{H} d\vec{x}_{(3)}^2 + \ell_{\infty }^2 \mathcal{Z}_\mathcal{H}^{-1} \bigg[dw + \ell_{\infty }^{-1}\beta_{\mathcal{H}} \, \chi \bigg]^2 \,,
\end{equation}
where $\vec{x}_{(3)}$ are coordinates of a flat 3-dimensional Euclidean space $\mathbb{E}^3$ and $\chi$ is a 1-form satisfying
\begin{equation}
	d \chi = \star_{(3)} d \mathcal{Z}_\mathcal{H}\,,
\end{equation}
with $\star_{(3)}$ the Hodge star operator defined on $\mathbb{E}^3$.
The equations of motion are solved provided that $\mathcal{Z}_i$ are harmonic functions in $\mathbb{E}^3$. Imposing asymptotic flatness, we have the general solution 
\begin{equation}\label{eq:harmonics}
	\mathcal{Z}_i = 1 + \sum_{a = 1}^{n_c} \frac{q_i^a}{r_a} \,,
\end{equation}
where $n_c$ is the number of poles of the harmonic functions and $r_a$ is the distance in $\mathbb{E}^3$ from the $a$th pole located at $x^k_a$ to $x^k$
\begin{equation}
	r_a^2 = (x^k_a - x^k)^2 \,.
\end{equation}
Locally, $\chi$ has the form
\begin{equation}
	\chi
	=
	\sum_{a} q^{a}_{\mathcal{H}} \cos{\theta_{a}} d\phi_{a}\,.
\end{equation}
Expanding the ansatz (\ref{eq:4dansatzLeading}) close to the poles' locations we can see that they are nothing but the position of 4-charge, 4-dimensional, extremal black holes described by the ansatz (\ref{eq:anatz4d}). Therefore, the integration constants appearing in the solution have the same interpretation that they had in the non-extremal case with the difference that now $\beta_i^2 = 1$ and that we have four charges $q_+^a$, $q_-^a$, $q_0^a$, $q^a_{\mathcal{H}}$ for every pole of the harmonic functions. 

We will solve the EOMs in the case $\beta_0 = \beta_\mathcal{H}$.\footnote{This condition is necessary but not sufficient for supersymmetry (see later). Moreover, we know that in the single-center case the structure of the corrections is much simpler.}  In this case the leading-order ansatz (\ref{eq:4dansatzLeading}) is enough and we do not have to introduce new independent functions.

\section{Solution construction} \label{sec-SolutionConstruction}

Our goal is to find an explicit and analytic expression for the unknown functions we introduced in our ansatz. To do so, we have to replace our expressions for the 10-dimensional metric, dilaton and KR field into the 10 dimensional HST EOMs and solve them. Given that we are working at first order in $\alpha'$, within the perturbative regime of validity of theory we can expand the EOMs in series of $\alpha'$  and drop higher order terms. As a result, we obtain linear differential equations for the unknown deformations. In most cases this system of equations will be highly coupled. There is no general procedure to solve it, but the tools and the techniques we can exploit are based on a few simple ideas
\begin{itemize}
	\item The \textit{dualities} of the HST effective action may give us non-trivial relations among the unknown functions.
	\item Instead of solving the system directly, we can  expand the differential equations in powers of the radial coordinate and determine first the asymptotic behavior of the unknown functions. Then, we can try to reconstruct the full analytical solution using a symbolic manipulation program that is capable to make a good guess for the \textit{generating function} given a very large number of terms of the series.
\end{itemize}
Independently of the procedure we follow to solve the differential equations we will end up with several integration constants. According to the no-hair theorem, we expect to be able to determine all of them in terms of physical conserved charges. This is indeed what happens in all the cases we have analyzed \cite{Cano:2021nzo,Ortin:2021win,Cano:2022tmn,Zatti:2023oiq,Massai:2023cis}. The full list of conditions we use is the following
\begin{itemize}
	\item We require that, asymptotically, the fields have the same normalization they have at $0^{\rm th}$-order in $\alpha'$ \,.
	\item We require that the event horizon of the black hole is regular, i.e. we impose that the unknown functions are not divergent when evaluated at the horizon.
	\item We perform changes of coordinates such as shifts of the radial coordinate  to absorb nonphysical integration constants.  
	\item We choose a thermodynamic ensemble. More precisely, we pick some thermodynamic quantities and we require that the relations among them and the $0^{\rm th}$-order parameters appearing in the  $0^{\rm th}$-order solution do not receive $\alpha'$ corrections. Notice that this is completely equivalent to absorb some non-physical integration constants in a redefinition of the dummy parameters appearing in our ansatz. 
\end{itemize}
In the rest of this section, we provide further details on the resolution procedure case by case and we present the explicit form of  the black hole solutions.

\subsection{5-dimensional, non-extremal, 3-charge BHs}

\subsubsection{Solving the EOMs} \label{sec-5d3csolving}
We start by the equations which are the simplest to solve: the Kalb-Ramond
(KR) equation (\ref{eq:eq3-08}) and Bianchi identity Eq.~(\ref{eq:bianchi-04}).\footnote{Our ansatz is written in terms of the KR
	field strength and, therefore, we must impose the KR Bianchi identity. } As a matter of fact, the ansatz for the dilaton has been chosen in such a way that the KR equation (\ref{eq:eq3-08}) is automatically solved.				
On the other hand, the only non-trivial component of the Bianchi identity
Eq.~(\ref{eq:bianchi-04}) to first order in $\alpha'$ is the $r\theta\phi\psi$ one
and, using the $0^{\rm th}$-order solution, it becomes a differential equation for
$\delta\mathcal{Z}_{h0}$ which is solved by

\begin{equation}
	\label{eq:deltaZh0}
	\delta\mathcal{Z}_{h0}
	=   d^{(0)}_{h0} +\frac{d^{(2)}_{h0}}{r^{2}}
	+\frac{2 q_{0}^{3} + \omega \left(q_{0}^{2} + 9 q_{0} r^{2} + 6
		r^{4}\right)}{2 q_{0} r^{2} (q_{0} + r^{2})^{2}}
	-\frac{3 \omega}{q_{0}^{2}}\log\left(1+\frac{q_{0}}{r^{2}}\right)\,,
\end{equation}

\noindent
where $d^{(0)}_{h0}$ and $d^{(2)}_{h0}$ are integration constants. Imposing
that $\mathcal{Z}_{h0}$ does not receive $\alpha'$ corrections at infinity we
obtain $d^{(0)}_{h0} = 0$.

The remaining equations of motion for the 6 remaining functions form a highly coupled system of 5
independent differential equations. Note that we have more unknown functions than equations, and the reason is that our ansatz contains some gauge freedom: one of the functions can be chosen at will by means of a transformation of the radial coordinate. This freedom will prove to be very useful to express the solution in a simple form.
Despite the complexity of the equations, they can solved by the
following procedure: first, we use as an ansatz for the $\delta \mathcal{Z}$s
and $\delta W$s the following series with arbitrary coefficients
\begin{equation}
	\label{eq:seriesdef}
	\delta \mathcal{Z}_{i} = \sum_{n = 1} \frac{d^{(2n)}_{i}}{r^{2n}}\,,
	\quad
	\delta \mathcal{Z}_{hi} = \sum_{n = 1} \frac{d^{(2n)}_{hi}}{r^{2n}}\,,
	\quad
	\delta W_{j} = \sum_{n = 1} \frac{d^{(2n)}_{wj}}{r^{2n}}\,.
\end{equation} 
Notice that we have assumed that all the powers in $1/r$ are even and that
there is no correction to the asymptotic values of $\mathcal{Z}$s and
$W$s. Furthermore, it follows that the coefficient $c_{\hat{\phi}}$ that sets $\lim_{r\rightarrow\infty} \hat\phi =\hat\phi_{\infty}$ is given by 

\begin{equation}
	c_{\hat{\phi}}
	=
	\left(q_{-} + \alpha' d^{(2)}_{h-}\right) e^{-2\hat{\phi}_{\infty}}\,.
\end{equation}

Plugging this ansatz into the Einstein equations (\ref{eq:eq1-08})
and the dilaton equation (\ref{eq:eq2-08}) and demanding that
they are solved order by order in powers of of $1/r$ one obtains algebraic
equations for the $d^{(2n)}$ coefficients that one can solve for large
values of $n$.  It turns out that not all of these coefficients are determined
by these equations: the equations of motion are solved for arbitrary values of	the coefficients $d^{(2)}_{i}$, $d^{(2)}_{hi}$, $d_{wj}^{(2)}$ and
$d^{(2n)}_{h-}$ with $n\ge 3$. Having determined the coefficients for a large
enough value of $n$ we can determine the functions associated to those
expansions resuming the series.\footnote{We have used a symbolic
	manipulation program that makes a good guess for those functions based on a very large number of terms of the series.} Finally, we have to check that	those functions do solve exactly the equations of motion to first order in $\alpha'$.

\subsubsection{Regularity conditions}

We want to determine the integration constants. We set $d^{(2n)}_{h-} = 0 $ with $n\ge 3$ (different choices of these coefficients can be absorbed in proper change of coordinates). We now impose regularity of the solution at the horizon. At $0^{\rm th}$-order the horizon is located at $r = \sqrt{-\omega}$. At first order it could be shifted by $\alpha' \delta r$. Therefore, we
consider an expansion of the 5-dimensional fields in $z = (r-r_{H})$ with
$r_{H} = \sqrt{-\omega}+\alpha' \delta r$. We obtain

\begin{subequations}
	\begin{align}
		e^{-2\phi}
		& =
		y_{\phi}^{(0)} + \alpha' y_{\phi}^{(0,\log)} \log{z} +\mathcal{O}(z)\,,
		\\[2mm]
		k
		& =
		y_{k}^{(0)} +  \alpha' y_{k}^{(0,\log)} \log{z} +\mathcal{O}(z)\,,
		\\[2mm]
		g_{E\,tt}
		& =
		\alpha'y_{tt}^{(0)}+  y_{tt}^{(1)}z + \alpha'y_{tt}^{(1,\text{log})} z
		\log{z}
		+\mathcal{O}(z^{2})\,,
		\\[2mm]
		\begin{split}
			g_{E\,rr}
			& =
			e^{-\frac{4}{3}\phi} \bigg[ \alpha'\frac{y_{rr}^{(-2)}}{z^{2}}
			+\frac{y_{rr}^{(-1)}}{z} + y_{rr}^{(0)}
			+\alpha\frac{'y_{rr}^{(-1,\log)}}{z}\log{z} \\[1mm]
			& \quad +
			\alpha'y_{rr}^{(0,\log)}\log{z} +  \mathcal{O}(z) \bigg]\,,
		\end{split}
		\\[2mm]
		g_{E\,\theta\theta}
		& =
		y_{\theta\theta}^{(0)}  + \mathcal{O}(z)\,,
	\end{align}
\end{subequations}

\noindent
and

\begin{subequations}
	\begin{align}
		F_{E \, r t}
		& =
		y^{(0)}_{F} +\mathcal{O}(z)\,,
		\\[2mm]
		G^{(1)}_E{}_{r t}
		& =
		y^{(0)}_{G} +\mathcal{O}(z)\,,
		\\[2mm]
		H^{(1)}_E{}_{\psi\theta\varphi}
		& =
		y^{(0)}_{H} + \mathcal{O}\left(z\right)\,,
	\end{align}
\end{subequations}

\noindent
where the constants $y_{i}$ are combinations of $d_{i}$, $d_{hi}$, $d_{wj}$
and $\delta r$. The $\alpha'$ factors indicate the terms that are purely
first-order corrections. In order to have a regular horizon we ask that the
scalars have a finite near-horizon limit. This leads to the conditions

\begin{equation}
	y_{\phi}^{(0,\log)} = 0 \,,
	\qquad
	y_{k}^{(0,\log)} = 0 \,.
\end{equation} 
The requirement that there is an event horizon at $z = 0$ implies the
vanishing of the constant part of the $g_{E\,tt}$ component of the metric
\begin{equation}
	y_{tt}^{(0)} = 0 \,.
\end{equation}
Demanding that $g_{E\,rr}$ approaches the horizon at most as $1/z$, we obtain
the conditions
\begin{equation}
	y_{rr}^{(-2)} = 0\,,
	\qquad y_{rr}^{(-1,\log)} = 0\,.
\end{equation}
Finally, in order to have a finite Hawking temperature, we have to impose
\begin{equation}
	y_{tt}^{(1,\log)} = 0 \,.
\end{equation}
Combining all these conditions we obtain expressions for $\delta r$,
$d_{0}^{(2)}$ and $d_{-}^{(2)}$ in terms of the 5 undetermined parameters
$d_{tt}^{(2)}$, $d_{rr}^{(2)}$, $d_{h0}^{(2)}$, $d_{h-}^{(2)}$ and
$d_{+}^{(2)}$. 4 integration constants can be determined by demanding that the mass and the 3 asymptotic charges do not get $\alpha'$ corrections. The remaining integration constant can be interpreted as the freedom of choosing the position of the horizon, \textit{i.e.}~the value of $\delta r$, and it can be eliminated through a change of coordinates.  We finally obtain
\begin{subequations}
	\begin{align}
		d_{0}^{(2)}
		& =
		\frac{\omega \left(8 q_{0}^{2}-18 q_{0} \omega+13 \omega^{2}\right)}{8
			(q_{0}-\omega)^{2} (2 q_{0}-\omega)} +
		\frac{(q_{0}-\omega)(2q_{-}-\omega)(2q_{+} - \omega)}{\omega (q_{-} -
			q_{+})(2 q_{0} - \omega)}\, d_{-}^{(2)}\,,
		\\[2mm]
		d_{tt}^{(2)}
		& =
		\frac{2q_{-}-\omega}{q_{-} - q_{+}}\, d_{-}^{(2)} \,,
		\\[2mm]
		d_{rr}^{(2)}
		& = 
		\frac{\omega}{q_{0}-\omega} + \frac{(2 q_{-} - \omega)(2q_{+} -
			\omega)}{\omega(q_{-}-q_{+})}\,d_{-}^{(2)} \,,
		\\[2mm]
		d_{-}^{(2)}
		& =
		-\frac{3\omega^{2}(q_{-} -
			q_{+})(2q_{0}-3\omega)}{4(q_{0}-\omega)^{2}\left[4q_{-}q_{+}
			+4q_{0}(q_{-}+q_{+}-\omega)-4\omega\left(q_{-}+q_{+}\right)+3\omega^{2}\right]}\,,
		\\[4mm]
		\delta r
		& =
		d_{h0}^{(2)} = d_{h-}^{(2)} = d_{+}^{(2)} = 0 \,.
	\end{align}
\end{subequations}

\subsubsection{Regular solutions}

The explicit expression of the $\delta \mathcal{Z}$s and $\delta W$s are 			
\begin{subequations}\label{eq:sol5d3charge}
	\begin{align}
		\delta \mathcal{Z}_{h0}
		& =
		\frac{2 q_{0}^{3} + \omega \left(q_{0}^{2} + 9 q_{0} r^{2} + 6
			r^{4}\right)}{2 q_{0} r^{2} (q_{0} + r^{2})^{2}}
		-\frac{3\omega}{q_{0}^{2}}
		\log{\mathcal{Z}_{0}}\,,
		\\[2mm]
		\begin{split}
			\delta \mathcal{Z}_{0}
			& =
			\frac{8q_{0}^{6} -24q_{0}^{5}\omega -r^{4}\omega^{3}
				\left(r^{2} +2\omega\right) +q_{0}^{3}\omega
				\left(4r^{4} -26r^{2}\omega
				-7\omega^{2}\right)}{4q_{0}r^{2}
				\left(q_{0} +r^{2}\right)^{2} (q_{0} -\omega)^{2} (2 q_{0}
				-\omega)}
			\\[1mm]
			& \hspace{.5cm}
			+\frac{ \omega q_{0}^{3} \left(8r^{2} +22\omega\right)
				+r^{2}\omega^{2} \left(2r^{4} +11r^{2}\omega
				-4\omega^{2}\right)
				+q_{0}\omega^{2}
				\left(-10r^{4} +25r^{2}\omega +2\omega^{2}\right)}{4r^{2}
				\left(q_{0} +r^{2}\right)^{2} (q_{0} -\omega)^{2} (2q_{0}
				-\omega)}
			\\[1mm]
			& \hspace{.5cm}
			+\frac{(q_{0} -\omega)(2q_{-} -\omega)(2q_{+} -\omega)}{\omega(q_{-}
				-q_{+}) (2q_{0} -\omega)r^{2}}d_{-}^{(2)}
			+\frac{\omega^{2}q_{0}\left(2r^{2} +\omega\right)
				-\omega^{2}r^{2}\left(r^{2}
				+2\omega\right)}{4q_{0}^{2}r^{2}
				(q_{0} -\omega)^{2}} \log{\mathcal{Z}_{0}}\,, 
		\end{split}
		\\[2mm]
		\delta\mathcal{Z}_{h-}
		& =
		-\frac{\omega q_{-}}{2 \left(q_{0} -\omega\right) r^{4}}
		-\frac{q_{-}(q_{-} -\omega)(2q_{+} -\omega)}{\omega (q_{-} -q_{+})
			r^{4}}\,d_{-}^{(2)}  \,,  
		\\[2mm]
		\delta\mathcal{Z}_{-}
		& = 
		\delta Z_{h-} + \mathcal{Z}_{-}\left[ \Delta_{k}
		-\frac{\Delta_{C}}{\beta_{+}}
		+\frac{r^{3}}{4}
		\bigg( \frac{\delta Z_{h-}'}{q_{-}}
		-\frac{\text{T}\left[\delta Z_{h-}'\right]}{q_{+}}\bigg)\right]\,, 
		\\[2mm]
		\delta\mathcal{Z}_{+}
		& =
		-\mathcal{Z}_{+}\frac{\Delta_{C}}{\beta_{+}}
		+\text{T}\left[\delta Z_{h-}\right] \,, 
		\\[2mm]
		\begin{split}
			\delta W_{tt}
			& =
			-\frac{\omega^{2}}{2 \left(q_{0}-\omega\right) r^{4}}  -
			\frac{\beta_{-}(W+\beta_{+}\beta_{-})+ W(\beta_{+} + \beta_{-})}{8
				\beta_{+} \mathcal{Z}_{0} \mathcal{Z}_{-}} \mathcal{Z}_{-}' W'
			\\[1mm]
			& \hspace{.5cm}
			+\frac{(2q_{-} -\omega)(r^{2} +\omega)}{(q_{-} -q_{+})r^{4}}\,
			d_{-}^{(2)} \,,
		\end{split}
		\\[2mm]
		\begin{split}
			\delta W_{rr}
			& =
			-\frac{\omega \left(r^{2} +\omega\right)\left[-4q_{0}^{3}
				+q_{0}\omega \left(5r^{2} -2\omega\right) +r^{2}\omega
				(2r^{2} +\omega) + q_{0}^{2} (-4 r^{2} +6
				\omega)\right]}{4q_{0}r^{4}\left(q_{0} +r^{2}\right)
				(q_{0} -\omega)^{2}}
			\\[1mm]
			& \hspace{.5cm}
			+\frac{(2q_{-} -\omega)(2q_{+} -\omega)(r^{2}
				+\omega)}{\omega(q_{-} -q_{+})r^{4}}d_{-}^{(2)}
			+\frac{\omega^{2} \left(2 r^{4} +3 r^{2}\omega
				+\omega^{2}\right)}{4 q_{0}^{2} r^{2} (q_{0} -\omega)^{2}}
			\log{\mathcal{Z}_{0}} \,,
		\end{split} 
	\end{align}
\end{subequations}

\noindent
where T is an operator implementing the transformation of the parameters\footnote{This transformation is nothing but a T-duality in the S$^{1}_z$ direction. See the 4-dimensional case for more details or \cite{Cano:2022tmn}.}			
\begin{equation}
	\label{eq:Tdualityparameters}
	q_{\pm} \leftrightarrow q_{\mp} \,,
	\qquad
	\beta_{\pm} \leftrightarrow \beta_{\mp} \,,
	\qquad
	k_{\infty} \leftrightarrow 1/k_{\infty}\,.
\end{equation}
It is easy to verify that the solution is self-dual under the action of T. 

\begin{figure}[h]
	\centering
	\includegraphics[width=0.6\textwidth]{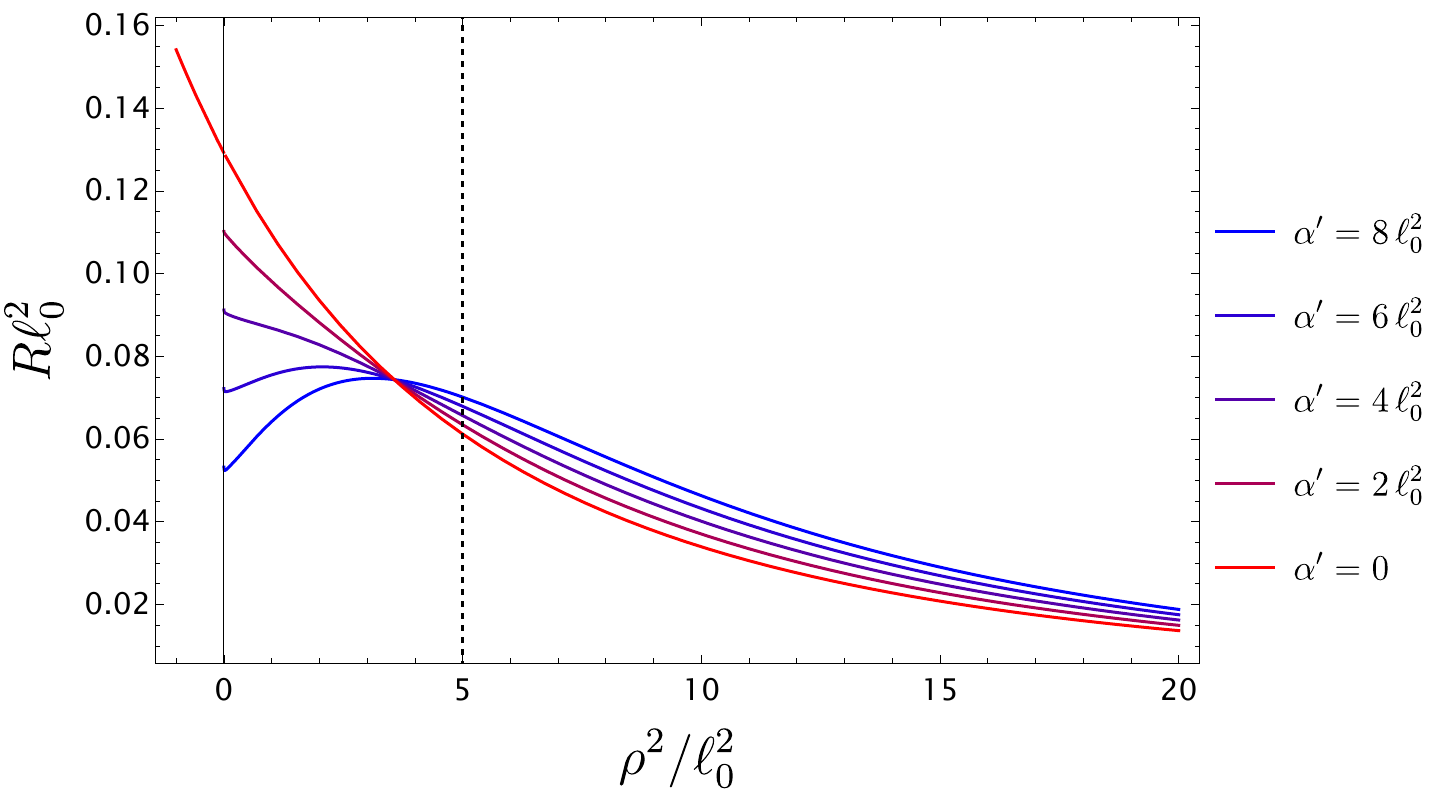}
	\caption{\textit{The Ricci scalar as a function of the radial coordinate for $q_{+} = 40 \,\ell_0^2$, $q_{-} = 20\, \ell_0^2 $, $q_{0} = 10\, \ell_0^2$,
			$\omega = -5\,\ell_0^2$, $s_{+}s_{-} = -1$ for different values of
			$\alpha'$. We normalized the units setting $\ell_0 = 1$.}}
	\label{fig:P1}
\end{figure}

\begin{figure}[h]
	\centering 
	\includegraphics[width=0.6\textwidth]{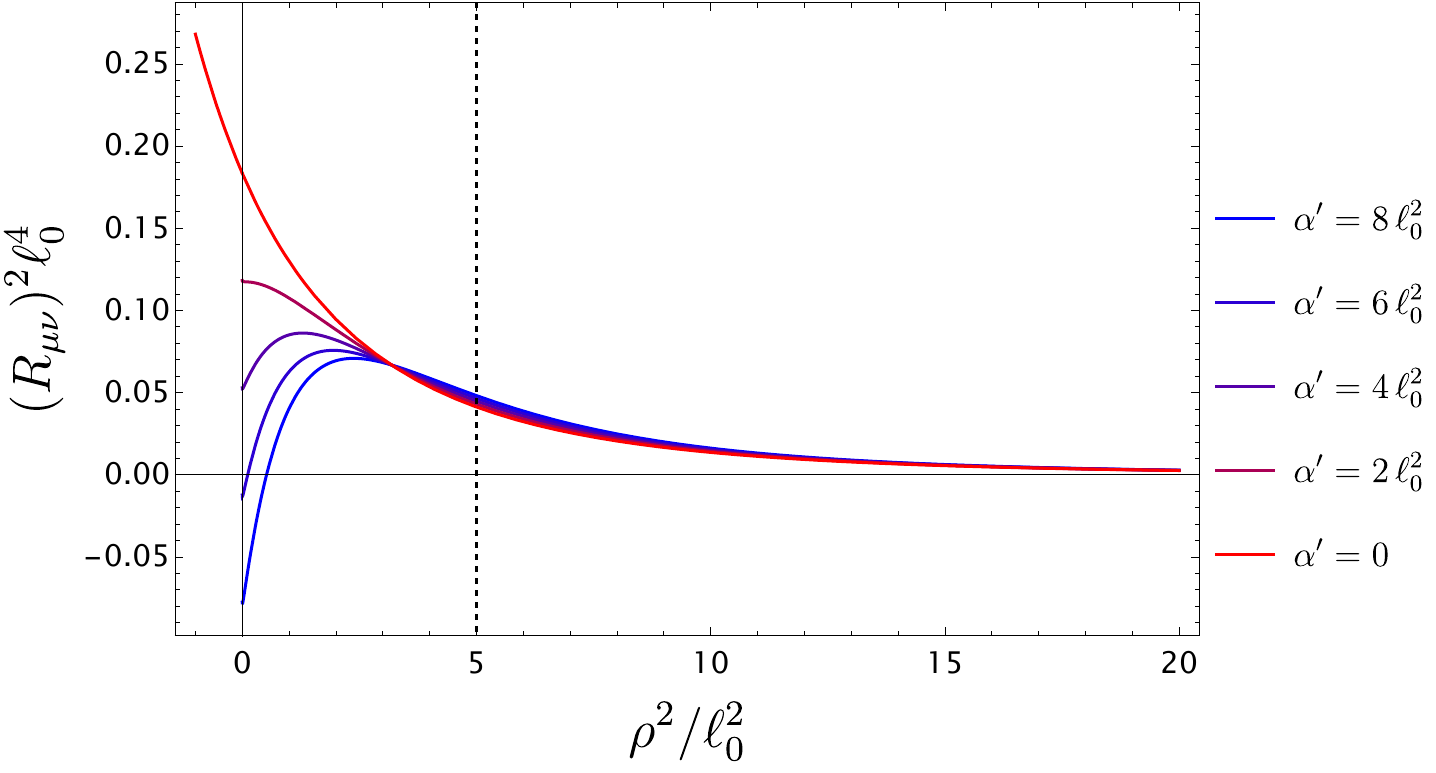}
	\caption{\textit{The $R_{\mu\nu}R^{\mu\nu}$ invariant as a function of the
			radial coordinate for $q_{+} = 40 \,\ell_0^2$, $q_{-} = 20\, \ell_0^2 $, $q_{0} = 10\, \ell_0^2$,
			$\omega = -5\,\ell_0^2$, $s_{+}s_{-} = -1$ for different values of
			$\alpha'$. We normalized the units setting $\ell_0 = 1$.}}
	\label{fig:P2}
\end{figure}

\begin{figure}[h]
	\centering 
	\includegraphics[width=0.6\textwidth]{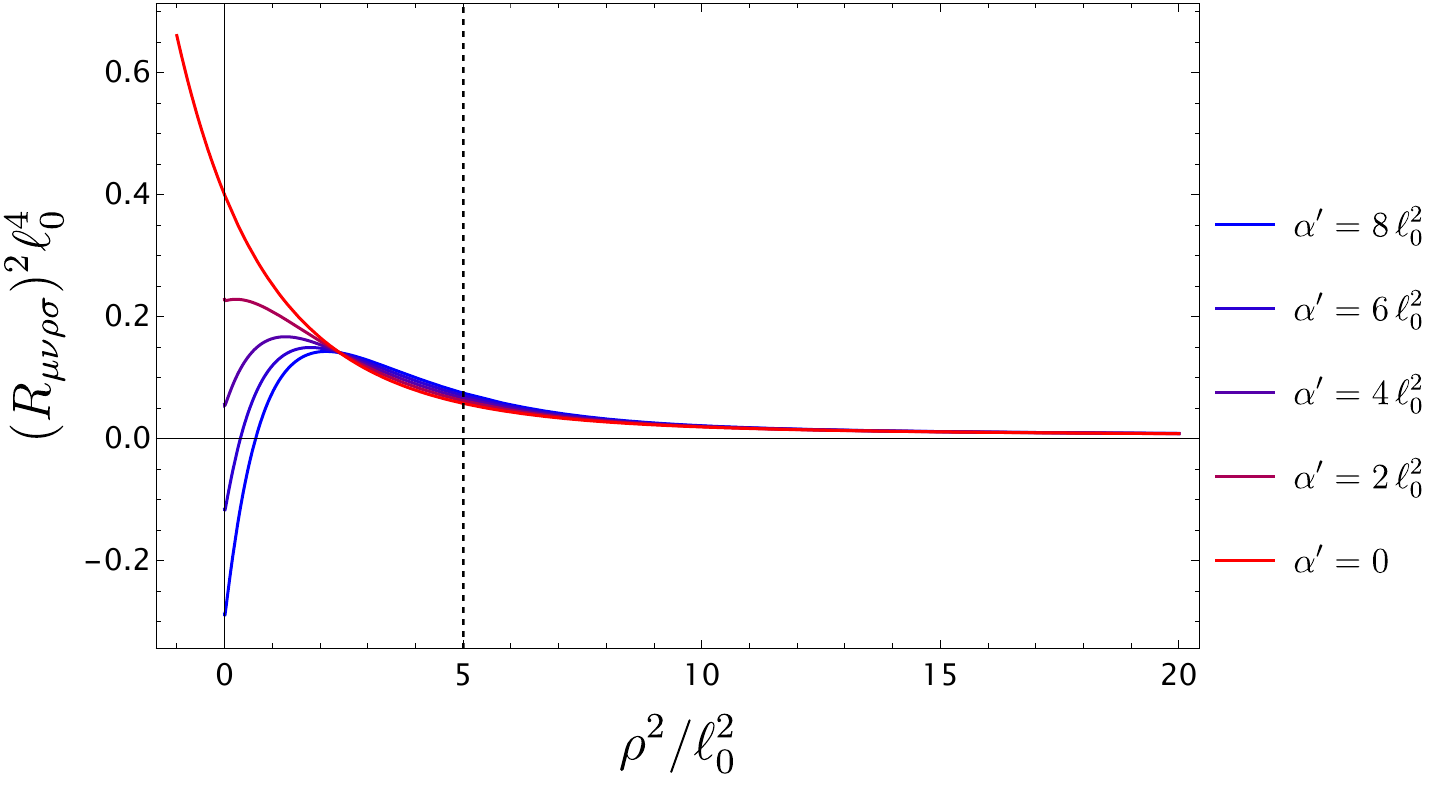}
	\caption{\textit{The Kretschmann invariant
			$R_{\mu\nu\rho\sigma}R^{\mu\nu\rho\sigma}$ as a function of the radial coordinate for $q_{+} = 40 \,\ell_0^2$, $q_{-} = 20\, \ell_0^2 $, $q_{0} = 10\, \ell_0^2$,
			$\omega = -5\,\ell_0^2$, $s_{+}s_{-} = -1$ for different values of
			$\alpha'$. We normalized the units setting $\ell_0 = 1$.}}
	\label{fig:P3}
\end{figure}

To end this section, we present Figs.~\ref{fig:P1}, \ref{fig:P2} and \ref{fig:P3} in which we plot several curvature invariants for a typical choice of integration constants $q_{+},q_{-},q_{0},\omega$ for several values of $\alpha'$ as a way to visualize the effect of those	corrections which have to be small anyway. The plots do not extend beyond $\rho^{2}=0$ for $\alpha'\neq 0$ because there is a logarithmic singularity at	that point. There seem to be no other curvature singularities for larger values of $\rho^{2}$, including the position of the inner horizon, which is slightly displaced to the right of $\rho^{2}=0$ by the $\alpha'$ corrections.


\subsection{5-dimensional, extremal, 3-charge BHs}

\subsubsection{Solving the EOMs}

The solutions can be easily obtained as a limit of (\ref{eq:sol5d3charge}) for $\omega = 0$. However, notice that it is possible to avoid the process of reconstructing the generating functions. In this case, only 3 unknown functions are necessary \cite{Cano:2021nzo}. We have indeed
\begin{equation}
	\mathcal{Z}_0 = \mathcal{Z}_{h0} \,, \quad \mathcal{Z}_+ \,,  \quad \mathcal{Z}_- = \mathcal{Z}_{h-} \,, \quad W_{tt} = W_{rr} = 1 \,.
\end{equation}
The Bianchi identity can be used to determine $\mathcal{Z}_0$. The KR equation is not automatically satisfied, but can be used to determine $\mathcal{Z}_-$. Finally, the Einstein equations and the dilaton equation form a system of linear differential equations for $ \delta \mathcal{Z}_+$ that can be solved directly.

The integration constants are determined demanding that the gauge charges do not receive $\alpha'$ corrections, namely that the numerator of the $1/r^2$ term of the $\mathcal{Z}_i$ is not modified, and that the asymptotic normalization of the fields is the same they have at $0^{\rm th}$-order.

\subsubsection{Regular, extremal solutions}

Explicitly, the three functions $\mathcal{Z}_0$, $\mathcal{Z}_+$, and $\mathcal{Z}_-$ are
\begin{subequations}
	\begin{align}
		\mathcal{Z}_{0}
		& = 1 + \frac{q_0}{r^2} + \alpha'
		\frac{q_{0}^{2}}{r^{2}(q_{0}+r^{2})^{2}}\,,
		\\[2mm]
		\mathcal{Z}_{+} 
		& =
		1 + \frac{q_+}{r^2} - \alpha' (1 + \beta_{+}\beta_{-})\frac{q_{+}q_{-}}{r^{2}(q_{0}+r^{2})(q_{-}+r^{2})} \,,
		\\[2mm]
		\mathcal{Z}_{-} & = 1 + \frac{q_-}{r^2}\,.
	\end{align}
\end{subequations}
where we used that, for $\omega = 0$ 
\begin{equation}
	\beta_+ = s_+ \,, \quad \beta_- = s_- \,, \quad \beta_0 = s_0 \,.
\end{equation}

\subsubsection{Supersymmetry}
\label{sec-d5-susy}

The ansatz (\ref{eq:3-charge10dsolution-04}) takes the form (\ref{eq:ansatz5dEqui}) in the extremal case. Let us determine the unbroken supersymmetries of the field configurations for arbitrary choices of the functions $\mathcal{Z}_{0},\mathcal{Z}_{+},\mathcal{Z}_{-}$.\footnote{See also
	Ref.~\cite{Chimento:2018kop}.}  We use the Zehnbein basis
\begin{equation}
	\begin{aligned}
		{\hat e}^{0} & = \frac{1}{\sqrt{\mathcal{Z}_{+}\mathcal{Z}_{-}}}dt\,,
		\hspace{.7cm}
		{\hat e}^{1} = \sqrt{\mathcal{Z}_{0}}dr\,,
		\hspace{.7cm}
		{\hat e}^{i+1} = \sqrt{\mathcal{Z}_{0}}r/2 v^{i}\,,
		\\[4mm]
		{\hat e}^{5}&  =
		k_{\infty}\sqrt{\frac{\mathcal{Z}_{+}}{\mathcal{Z}_{-}}}
		\left[dz+\beta_{+}k_{\infty}^{-1}
		\left(\mathcal{Z}^{-1}_{+}-1\right)dt\right]\,,
		\hspace{.7cm}
		{\hat e}^{m} = dy^{m}\,,
	\end{aligned}
\end{equation}

\noindent
where the $v^{i}$, $i=1,2,3$ are the SU$(2)$ left-invariant Maurer-Cartan
1-forms, satisfying the Maurer-Cartan equation

\begin{equation}
	\label{eq:SU2MC}
	dv^{i} = -\tfrac{1}{2}\epsilon^{ijk}v^{j}\wedge v^{k}\,.
	\hspace{1cm}
	\epsilon^{123}=+1\,,
\end{equation}

Plugging this configuration into the Killing spinor equations
$\delta_{\epsilon} \psi_{a}=0$, $\delta_{\epsilon} \lambda=0$ and
$\delta_{\epsilon} \chi^{A}=0$ with the supersymmetry variations in
Eqs.~(\ref{eq:gravitino})-(\ref{eq:gaugino}), we immediately see that the
third of them (the gaugini's) is automatically satisfied. It is not hard to
see that the second (the dilatini) is satisfied for supersymmetry parameters
satisfying the two (compatible) conditions

\begin{subequations}
	\begin{align}
		\label{eq:susyprojector1}
		\tfrac{1}{2}\left(1-\beta_{-}\Gamma^{05} \right)\epsilon
		& =
		0\,,
		\\[2mm]
		\label{eq:susyprojector2}
		\tfrac{1}{2}\left(1+\beta_{0}\Gamma^{1234} \right)\epsilon
		& =
		0\,.
	\end{align}
\end{subequations}

Solving $\delta_{\epsilon} \psi_{0}=0$ with a time-independent spinor, though,
demands $\beta_{-}=\beta_{+}$, and, using this condition in
$\delta_{\epsilon} \psi_{1}=0$, we find that

\begin{equation}
	\epsilon = (\mathcal{Z}_{+}\mathcal{Z}_{-})^{-1/4}\epsilon_{0}\,,  
\end{equation}

\noindent
where $\epsilon_{0}$ is an $r-$independent spinor that satisfies the above conditions.	

The equations $\delta_{\epsilon} \psi_{i}=0$ with $i=2,3,4$ (the directions in
the 3-sphere) take the form

\begin{equation}
	\label{eq:KSEi1}
	\left(v_{i}-\tfrac{1}{4}(1+\beta_{0})\Gamma^{i1}\right)\epsilon_{0}=0\,,
\end{equation}

\noindent
where $v_{i}$ are the vectors dual to the left-invariant Maurer-Cartan 1-forms
in SU$(2)$ (S$^{3}$), $v^{i}$, defined above in Eq.~(\ref{eq:SU2MC}).
When $\beta_{0}=-1$, $\epsilon_{0}$ is just a constant spinor. When
$\beta_{0}=+1$, we can rewrite the equation in
the form

\begin{equation}
	\label{eq:KSEi2}
	\left(d-v^{i}T_{i}\right)\epsilon_{0}=0\,,
\end{equation}

\noindent
where we have defined the SU$(2)$ generators

\begin{equation}
	T_{i} \equiv \tfrac{1}{2}\Gamma^{i1}\,.
\end{equation}

\noindent
Indeed, it can be checked that they satisfy the commutation relations

\begin{equation}
	[T_{i},T_{j}] = -\varepsilon_{ijk}T_{k}\,,
\end{equation}

\noindent
in the subspace of spinors satisfying Eq.~(\ref{eq:susyprojector2}) with
$\beta_{0}=+1$. Since, by definition,\footnote{Here we are using the
	conventions and results of Ref.~\cite{Alonso-Alberca:2002wsh}.}
$v^{i}T_{i}=-u^{-1}du$, where $u$ is a generic element of SU$(2)$,
Eq.~(\ref{eq:KSEi2}) is equivalent to

\begin{equation}
	\label{eq:KSEi3}
	d(u\epsilon_{0})=0\,,
	\,\,\,\,\,
	\Rightarrow
	\,\,\,\,\,
	\epsilon_{0} = u^{-1}\epsilon_{00}\,,
\end{equation}

\noindent
where $\epsilon_{00}$ may, at most, depend on $z$. However,
$\delta_{\epsilon} \psi_{5}=0$ is solved for $z$-independent $\epsilon_{00}$
upon use of the projector Eq.~(\ref{eq:susyprojector1}) with
$\beta_{-}=\beta_{+}$, and the rest of the Killing spinor equations,
$\delta_{\epsilon} \psi_{m}=0$, are trivially solved for $y^{m}$-independent
(\textit{i.e.}~constant) $\epsilon_{00}$.

The conclusion is that the field configurations with $\beta_{-}=\beta_{+}$
(and arbitrary $\beta_{0}=\pm 1$) are the only supersymmetric ones, although the Killing
spinors are quite different for $\beta_{0}=+1$ and $\beta_{0}=-1$ cases. This
result is true regardless of the values of the functions
$\mathcal{Z}_{0},\mathcal{Z}_{+},\mathcal{Z}_{-}$ which means that the
$\alpha'$ corrections preserve the unbroken
supersymmetries of the zeroth-order solution.

\subsection{5-dimensional, multi-center, 3-charge BHs}

\subsubsection{Solving the EOMs}

We are going to solve the EOMs replacing explicitly the ansatz (\ref{eq:ansatz5dEqui}). The details of the computations can be found in \cite{Chimento:2018kop, Ortin:2021win}. Some intermediate steps are collected in appendix \ref{sec-curvature}. We assume that the $\mathcal{Z}_i$ functions  depend only on the coordinates of the 4-dimensional Euclidean space $\mathbb{E}^4$ and that the $\beta_i$ are signs. After some algebra we obtain that the Bianchi identity (\ref{eq:bianchi-04}) is equivalent to the condition 
\begin{equation}\label{eq:zeta0}
	\Delta_{(4)} \left\{\mathcal{Z}_0 - \frac{\alpha'}{4}\left(\frac{\partial_m \mathcal{Z}_0 \, \partial^m \mathcal{Z}_0 }{\mathcal{Z}_0^2}\right)\right\} = 0 \,,
\end{equation}
where $\Delta_{(4)}$ represents the Laplacian operator on $\mathbb{E}^4$ and $\partial_m$ are the derivatives dual to the vierbeins $v^m$. It is solved by
\begin{equation}
	\mathcal{Z}_0  = \mathcal{Z}_0^{(0)} + \frac{\alpha'}{4}\left(\partial \log \mathcal{Z}_0^{(0)}\right)^2  \,, 
\end{equation}
where $\mathcal{Z}_0^{(0)} $ is any harmonic of $\mathbb{E}^4$. The KR equation of motion (\ref{eq:eq3-08}) is in its turn equivalent to
\begin{equation}\label{eq:zeta-}
	\Delta_{(4)} \mathcal{Z}_- = 0 \,.
\end{equation} 
We conclude that $\mathcal{Z}_-$ must be a harmonic function on $\mathbb{E}^4$. Finally, we have a single independent EOM among the Einstein equations (\ref{eq:eq1-08}) and the dilaton EOM (\ref{eq:eq2-08}). In particular, using the vielbein basis (\ref{eq:vierbeinBasis}) introduced in appendix \ref{sec-curvature}, the $++$ component of Einstein equations reduces to
\begin{equation} \label{eq:zeta+}
	\Delta_{(4)} \left\{\mathcal{Z}_+ + \alpha'\frac{(1+\beta_+ \beta_-)}{4}\left(\frac{\partial_m \mathcal{Z}_+ \, \partial^m \mathcal{Z}_- }{\mathcal{Z}_0 \mathcal{Z}_-}\right) \right\} = 0 \,,
\end{equation}
which is solved by 
\begin{equation}
	\mathcal{Z}_+ = \mathcal{Z}_+^{(0)} - \alpha'\frac{(1+\beta_+ \beta_-)}{4}\left(\frac{\partial_m \mathcal{Z}_+^{(0)} \, \partial^m \mathcal{Z}_- }{\mathcal{Z}_0^{(0)} \mathcal{Z}_-}\right) \,,
\end{equation}
where $\mathcal{Z}_+^{(0)}$ is any harmonic of $\mathbb{E}^4$.

\subsubsection{Regular solutions}

Notice that the equations (\ref{eq:zeta0}), (\ref{eq:zeta-}) and (\ref{eq:zeta+}) determine the $\mathcal{Z}_i$ up to harmonic functions. These harmonic functions can be determined requiring regularity at the horizon, finding the proper asymptotic normalization of the fields and selecting the microcanonical ensemble. With the ansatz for the $0^{\rm th}$-order part of  $\mathcal{Z}_i$ given by (\ref{eq:harmonics5d}) we obtain explicitly 
\begin{subequations}
	\begin{align}
		\mathcal{Z}_{+}
		& = 
		\mathcal{Z}_{+}^{(0)}
		-\alpha'\left[
		(1+\beta_+ \beta_- ) \,\mathcal{Z}_{-}^{-1}\mathcal{Z}_{0}^{(0)\,-1}
		\sum_{a,b}\frac{q^{a}_{+}q^{b}_{-}n^{m}_{a}n^{m}_{b}}{r^{3}_{a}r^{3}_{b}}
		\right]\,, \\[2mm]
		\mathcal{Z}_{0}
		& =
		\mathcal{Z}_{0}^{(0)}
		+\alpha'\left[
		\mathcal{Z}_{0}^{(0)\,-2}\sum_{a,b}\frac{q^{a}_{0}q^{b}_{0}n^{m}_{a}n^{m}_{b}}{r^{3}_{a}r^{3}_{b}}
		\right]\,, \\[2mm]
		\mathcal{Z}_{-}
		& =
		\mathcal{Z}_{-}^{(0)} \,, 
	\end{align}
\end{subequations}
where we have defined the unit radial vectors
\begin{equation}
	n^{m}_{a}\equiv (x^{m}-x^{m}_{a})/r_{a}\,.
\end{equation}

\subsection{4-dimensional, non-extremal, 4-charge BHs}

\subsubsection{Solving the EOMs}

A possible approach to solve the EOMs would be the method of generating function reconstruction presented in section \ref{sec-5d3csolving}. Despite there being no obstruction to its application, all attempts to obtain an explicit solution with 4 independent parameters $q_i$ failed.  We focus therefore to the particular case with 3 independent charges. Therefore, we have
\begin{equation}
	q_0 = q_\mathcal{H} = q \,,  \qquad q_- \,, \qquad q_+ \,.
\end{equation}

Replacing the ansatz (\ref{eq021}) into the EOMs and the Bianchi identity of HST we can easily determine some of the unknown functions. The KR equation (\ref{eq:eq3-08}) is automatically satisfied and has been used to determine the expression (\ref{eq:dilaton10d4charge}) of the dilaton $\hat{\phi}$. Expanding the Bianchi identity (\ref{eq:bianchi-04}) in $\alpha'$ and dropping $\mathcal{O}(\alpha'{}^2)$ terms we obtain a second-order differential equation for $\delta \mathcal{Z}_{h0}$. It can easily be solved providing
\begin{equation}
	\begin{split}
		\delta \mathcal{Z}_{h0} = & (1+s_0 s_{\mathcal{H}})\left[\frac{q^4+q^3 \omega+11 q^2 r \omega+15 q r^2 \omega+6 r^3 \omega}{4 q^2 r (q+r)^3}-  \frac{3 \omega }{2q^3}\log \left(\frac{q+r}{r}\right)\right]+ \frac{d_{h0}^{(1)}}{r}\,,
	\end{split}
\end{equation}
where $d_{h0}^{(1)}$ is an integration constant (the second integration constant has been already fixed asking that the asymptotic value of $\mathcal{Z}_{h0}$ is not modified). Imposing that the charge associated with $C^{(1)}_w$ is not renormalized we obtain $d_{h0}^{(1)} = 0$. The dilaton and Einstein equations form instead a complicate system of coupled differential equations. Once we replace our ansatz together with the expressions for $\hat{\phi}$ and $\delta\mathcal{Z}_{h0}$ and we drop $\mathcal{O}(\alpha'{}^2)$ terms we obtain a total of 9 non-trivial equations (we indicate with $\mathbb{E}_{\hat{\mu}\hat{\nu}}$ the components of the Einstein equations (\ref{eq:eq1-08}) and with $\mathbb{E}_\phi$ the dilaton EOM (\ref{eq:eq2-08}))
\begin{equation}
	\{\mathbb{E}_{tt}, \mathbb{E}_{rr},\mathbb{E}_{\theta\theta},\mathbb{E}_{\varphi\varphi},\mathbb{E}_{ww},\mathbb{E}_{zz},\mathbb{E}_{tz},\mathbb{E}_{\varphi w},\mathbb{E}_{\phi}\}\,.
\end{equation}
However, not all of them are independent. For instance, we can drop $\mathbb{E}_{\varphi\varphi}$ and $\mathbb{E}_{\varphi w}$ because they are combinations of the other EOMs. $\mathbb{E}_{ww}$ turns out to be a second-order differential equation for the combination $\delta \mathcal{Z}_{\mathcal{H}} - \delta \mathcal{Z}_0$. Solving it, we obtain
\begin{equation}
	\begin{split}
		\delta \mathcal{Z}_{\mathcal{H}}  = &  \; \delta \mathcal{Z}_0  + (d_0^{(1)}-d_\mathcal{H}^{(1)})\left[\frac{4 q (q-\omega)}{r \omega^2}- \frac{(2 q - \omega)(2 q r + q w - r \omega)}{r \omega^3}\log\left(1+\frac{\omega}{r}\right)\right] \\& 
		- \frac{q(1+s_0s_{\mathcal{H}})(q-\omega)}{4 r (q + r)^3} \,,
	\end{split}
\end{equation} 
where we have imposed that both $\mathcal{Z}_{\mathcal{H}}$ and $\mathcal{Z}_{0}$ vanish asymptotically. $d_0^{(1)}$ and $d_\mathcal{H}^{(1)}$ are integration constants and represent the poles of the $1/r$ terms of $\delta \mathcal{Z}_{0} $ and $\delta \mathcal{Z}_{\mathcal{H}} $. $\mathbb{E}_{\theta\theta}$ is an algebraic constraint for $\delta W_{rr}$. We can use it to determine  $\delta W_{rr}$ as a function of the other unknown functions and their derivatives (we omit at this stage the actual expression because of its lengthiness)
\begin{equation}
	\delta W_{rr} = f \left(r,\delta W_{tt},\delta W_{tt}',\delta\mathcal{Z}_0,\delta\mathcal{Z}_0',\delta\mathcal{Z}_0'',\delta\mathcal{Z}_-,\delta\mathcal{Z}_-',\delta\mathcal{Z}_{h-},\delta\mathcal{Z}_{h-}',\delta\mathcal{Z}_{h-}''\right)\,.
\end{equation}
We are left with 5 equations and 5 unknown functions. Despite the complexity of the system is possible to solve it with the procedure of \cite{Cano:2022tmn}. First, we consider an ansatz for the unknown functions with arbitrary coefficients
\begin{equation}
	\delta \mathcal{Z}_i = \sum_{k>0} \frac{d_i^{(k)}}{r^k} \,, \qquad \delta \mathcal{Z}_{h-} = \sum_{k>0} \frac{d_{h-}^{(k)}}{r^k} \,, \qquad
	\delta W_{j} = \sum_{k>0} \frac{d_{wj}^{(k)}}{r^k}\,,
\end{equation}
where we have only assumed that the $\alpha'$ corrections do not modify the asymptotic value of the $\mathcal{Z}$ and $W$ functions. Then, we replace the series expansions into the EOMs and we demand that they are solved order by order in powers of of $1/r$. In this way, we obtain a set of algebraic equations for the coefficients $d^{(k)}$ for arbitrarily large values of  $k$. Solving such equations we obtain the asymptotic expansion of the unknown functions in powers of $1/r$. The coefficients of one of these functions are not determined, signaling that only 4 of the 5 EOMs left are truly independent. More precisely, we find a family of solutions which depend on the coefficients
\begin{equation}
	\left\{d_{h-}^{(1)}, d_-^{(1)},d_+^{(1)},d_\mathcal{H}^{(1)},d_{0}^{(1)},d_{wt}^{(1)},d_{wr}^{(1)}, d_{h-}^{(k)}  \right\} \,, \qquad k \ge 3 \,.
\end{equation}
Imposing that the charges associated with $A^z$ and $C_z^{(1)}$ are not renormalized (i.e. they do not receive $\alpha'$ corrections) we fix $d_{h-}^{(1)} = d_+^{(1)} = 0$. Then, we set to zero the coefficients $d_{h-}^{(k)}$ with $k\ge3$. This can always be done without loss of generality because it is equivalent to performing a proper change of coordinates. We obtain
\begin{equation}
	\delta \mathcal{Z}_{h-} = \frac{1}{2}\left(d_{wt}^{(1)}-d_{wr}^{(1)}-2d_{-}^{(1)}\right)\frac{q_-}{r^2} \,.
\end{equation}
With this expression we can easily determine some other quantities. First of all we notice that expanding the dilaton in series we get
\begin{equation}
	e^{-2\phi} \sim \frac{c_\phi}{q_-} + \mathcal{O}(1/r) \,,
\end{equation}
which fixes $c_\phi = e^{-2\phi_\infty} q_-$. Second, we can use the compatibility with the T-duality constraints to extract $\mathcal{Z}_{\pm}$ (see appendix \ref{sec-tduality} for more details). The action of T-duality along the $z$ direction on the lower dimensional fields is 
\begin{equation}
	T_z: \qquad 	C^{(1)}_z \leftrightarrow A^z \,, \qquad k  \leftrightarrow 1/k^{(1)} \,, \qquad	ds^2_{E} \leftrightarrow ds^2_{E} \,, \qquad e^{-2\phi}\leftrightarrow e^{-2\phi} \,. \\
\end{equation}
Assuming that $T_z$ can be implemented by
\begin{equation} 
	T_z: \qquad 	q_+ \leftrightarrow q_- \,, \qquad \beta_+ \leftrightarrow \beta_- \,, \qquad k_\infty \leftrightarrow 1/k_\infty \,,
\end{equation}
we obtain (\ref{eqtdualzmbis}), which gives
\begin{equation}\label{eqdeltazm}
	\delta \mathcal{Z}_- = \delta \mathcal{Z}_{h-} + \mathcal{Z}_- \left[\Delta_k - \frac{\Delta_C}{\beta_+}+\frac{r^2}{2}\left(\frac{\delta \mathcal{Z}_{h-}'}{q_-} - \frac{T_z[\delta\mathcal{Z}_{h-}']}{q_+}\right)\right] \,,
\end{equation} 
where $T_z$ is the operator implementing the T-duality transformation. This assumption imposes non-trivial constraints on some of the integration constants appearing in the series describing the expansion of $\delta \mathcal{Z}_-$. Indeed, the series $\{d^{(k)}_-\}$ satisfies (\ref{eqdeltazm}) provided that
\begin{equation}
	T_z \left[d_{h-}^{(2)}\right] = \frac{1}{2} q_+ \left(d_{wt}^{(1)}- d_{wr}^{(1)}\right) \,, \qquad d_-^{(1)} = \frac{(q_+-q_-)}{2q_- - w}d_{wt}^{(1)} \,.
\end{equation}
Expanding equation (\ref{eqtdualzmh}) we obtain an expression for $\delta \mathcal{Z}_+$ which matches the series $\{d_+^{(k)}\}$ without further constraints
\begin{equation}
	\delta \mathcal{Z}_+ = T_z \left[\delta \mathcal{Z}_{h-}\right] - \mathcal{Z}_+ \frac{\Delta_C}{\beta_+}\,.
\end{equation}
Once we replace the expressions obtained for $\delta \mathcal{Z}_-$, $\delta \mathcal{Z}_+$, $\delta\mathcal{Z}_\mathcal{H}$, $\delta\mathcal{Z}_{h-}$ and $\delta W_{rr}$ into the EOMs we obtain a set of differential equations for $\delta W_{tt}$ and $\delta \mathcal{Z}_0$. In particular, $\mathbb{E}_{zz}$ is a first-order differential equations which involves only $\delta W_{tt}$. Solving it we obtain
\begin{equation}
	\begin{split}
		\delta W_{tt} = \quad &\frac{ r \omega^2-q_- \omega (2 r+\omega)}{8 r^2 (q+r)^2 (q_-+r)}-\frac{\beta_- \beta_+ q_- q_+ \omega (r+\omega)}{4 r^2 (q+r)^2 (q_-+r) (q_+-\omega)} \\
		&  + \frac{d_{wt}^{(1)} (2 q_++2 r+\omega)}{2 r^2}-\frac{d_{wr}^{(1)} \omega}{2 r^2} \,.
	\end{split}
\end{equation}
Replacing this expression for $\delta W_{tt}$ into the EOMs, we are left with a fourth-order differential equation for $\delta \mathcal{Z}_0$. Instead of solving it directly, we focus on finding the generating function of the coefficients $\{d_0^{(k)}\}$. Replacing the guessed generating function into the fourth-order order differential equation we can verify that it is exactly solved and we actually find the expression of $\delta \mathcal{Z}_0$ (we omit again the actual expression because of its lengthiness). With the explicit expression of $\delta \mathcal{Z}_0$ one can finally reconstruct the explicit expressions of all the $\delta \mathcal{Z}$s and $\delta W$s. It is then possible to verify that the expressions obtained solve exactly all the EOMs of the HST effective action at first-order in $\alpha'$.

\subsubsection{Regularity conditions}

At zeroth order the horizon lies at $r = -\omega$. At first order it may be shifted and placed at $r_H =-\omega + \alpha' \delta r$. Therefore, we expand the 4d fields around $z = (r-r_H)$. We obtain
\begin{subequations}
	\begin{align}
		e^{-2 \phi } & \quad  = \quad  y^{(0)}_\phi + y^{(0,\log)}_\phi \log z + \mathcal{O}(z \log z) \,, \\
		k & \quad  = \quad y^{0}_k + \mathcal{O}(z) \,, \\
		\ell & \quad  = \quad y^{(0)}_\ell + y^{(0,\log)}_\ell \log z + \mathcal{O}(z \log z)\,, \\
		g_{tt,E} & \quad  = \quad y^{(0)}_{tt} +  y^{(1,\log)}_{tt} z \log z + \mathcal{O}(z) \,, \\
		g_{rr,E} & \quad  = \quad \frac{y^{(-2)}_{rr} }{z^2} +\frac{y^{(-1)}_{rr} }{z} +\frac{y^{(-1,\log)}_{rr} }{z} \log z + y^{(0,\log)}_{rr} \log z + \mathcal{O}(1) \,, \\
		g_{\theta\theta,E} & \quad  = \quad 	y^{(0)}_{\theta\theta} +  \mathcal{O}(z \log z) \,, \\
		F^w_{\theta \varphi} & \quad  =  \quad y^{(0)}_{Fw}  \,, \\
		F^z_{tr}  & \quad  = \quad  y^{(0)}_{Fz} + \mathcal{O}(z)\,, \\
		G_w & \quad  = \quad y^{(0)}_{Gw}  \,, \\
		G_z & \quad  = \quad   y^{(0)}_{Gz} + \mathcal{O}(z) \,,
	\end{align}
\end{subequations}
where the $y_i^{(k)}$s are combinations of $q_i$, $\omega$, $\beta_i$, $d_\mathcal{H}^{(1)}$, $d_{0}^{(1)}$, $d_{wt}^{(1)}$,  $d_{wr}^{(1)}$, $\delta r$, $\phi_\infty$, $k_\infty$, $\ell_\infty$  and $\alpha'$. Imposing that the BH horizon is placed at $z = 0$, we obtain the condition 
\begin{equation} \label{eqcondreg1}
	y^{(0)}_{tt} = 0 \,.
\end{equation}
Imposing that the scalars have a finite value on the BH horizon, we get
\begin{equation}\label{eqcondreg2}
	y^{(0,\log)}_\phi  = y^{(0,\log)}_\ell = 0 \,.
\end{equation}
Demanding that the Hawking temperature is finite, we obtain 
\begin{equation}\label{eqcondreg3}
	y^{(1,\log)}_{tt} = y^{(-2)}_{rr} = y^{(-1,\log)}_{rr} = 0 \,.
\end{equation}
The conditions (\ref{eqcondreg1}),(\ref{eqcondreg2}),(\ref{eqcondreg3}) together with the requirement that the BH horizon is not shifted, \textit{i.e.} $\delta r = 0$, lead to
\begin{subequations}
	\begin{align}
		d_0^{(1)} & = d_\mathcal{H}^{(1)} \,, \\[2mm]
		\begin{split}
			d_{wt}^{(1)} & = \frac{ \omega (\omega-2 q)}{(q-\omega) (\omega-2 q_+)}d_\mathcal{H}^{(1)} +\frac{ s_0 s_\mathcal{H} \, q \, \omega \left(2 q^2-6 q \omega+5 \omega^2\right)}{20 (q-\omega)^4 (\omega-2 q_+)} \\
			& \quad +\frac{\omega \left(-4 q^3+22 q^2 \omega-35 q \omega^2+25 \omega^3\right)}{40 (q-\omega)^4 (\omega-2 q_+)} \,, 
		\end{split} \\[4mm]
		\begin{split}
			d_{wr}^{(1)} & = \frac{ (2 q-\omega)}{q-\omega}d_\mathcal{H}^{(1)}-\frac{s_0 s_{\mathcal{H}} \,q  \left(2 q^2-6 q \omega+5 \omega^2\right)}{20 (q-\omega)^4} \\
			& \quad +\frac{4 q^3-12 q^2 \omega+15 q \omega^2-15 \omega^3}{40 (q-\omega)^4} \,.
		\end{split}
	\end{align}
\end{subequations}

\subsubsection{Regular solutions}

\begin{figure}[h]
	\centering
	\includegraphics[width=0.6\linewidth]{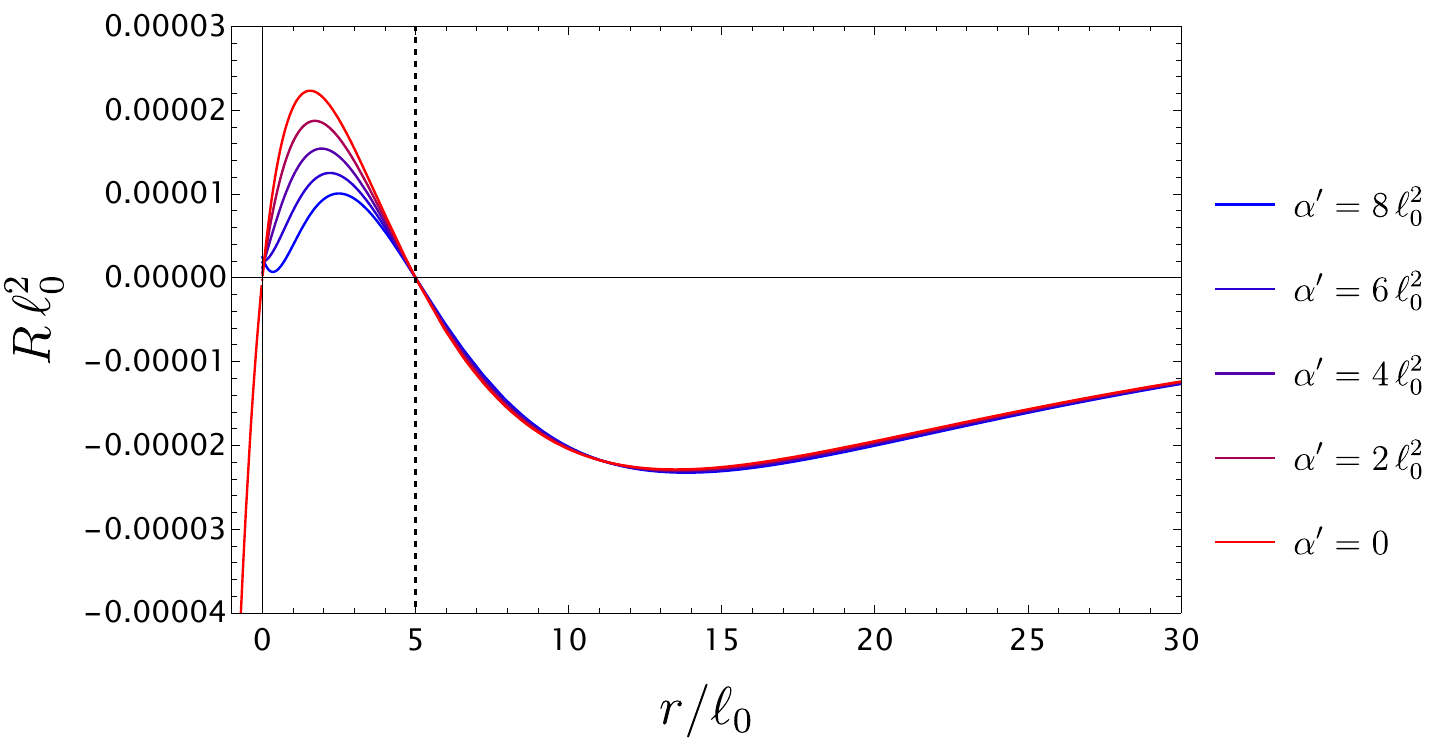}
	\caption{\textit{The Ricci scalar as a function of the radial coordinate for $q_+ = 40 \,\ell_0 $, $q_- = 20 \, \ell_0 $, $q = 10 \, \ell_0 $, $\omega = -5 \, \ell_0 $, $s_+ s_- = s_0 s_{\mathcal{H}} = 1$, for different values of $\alpha'$. We normalized the units setting $\ell_0 = 1$.}}
	\label{figricci}
\end{figure}

\begin{figure}[h]
	\centering
	\includegraphics[width=0.6\linewidth]{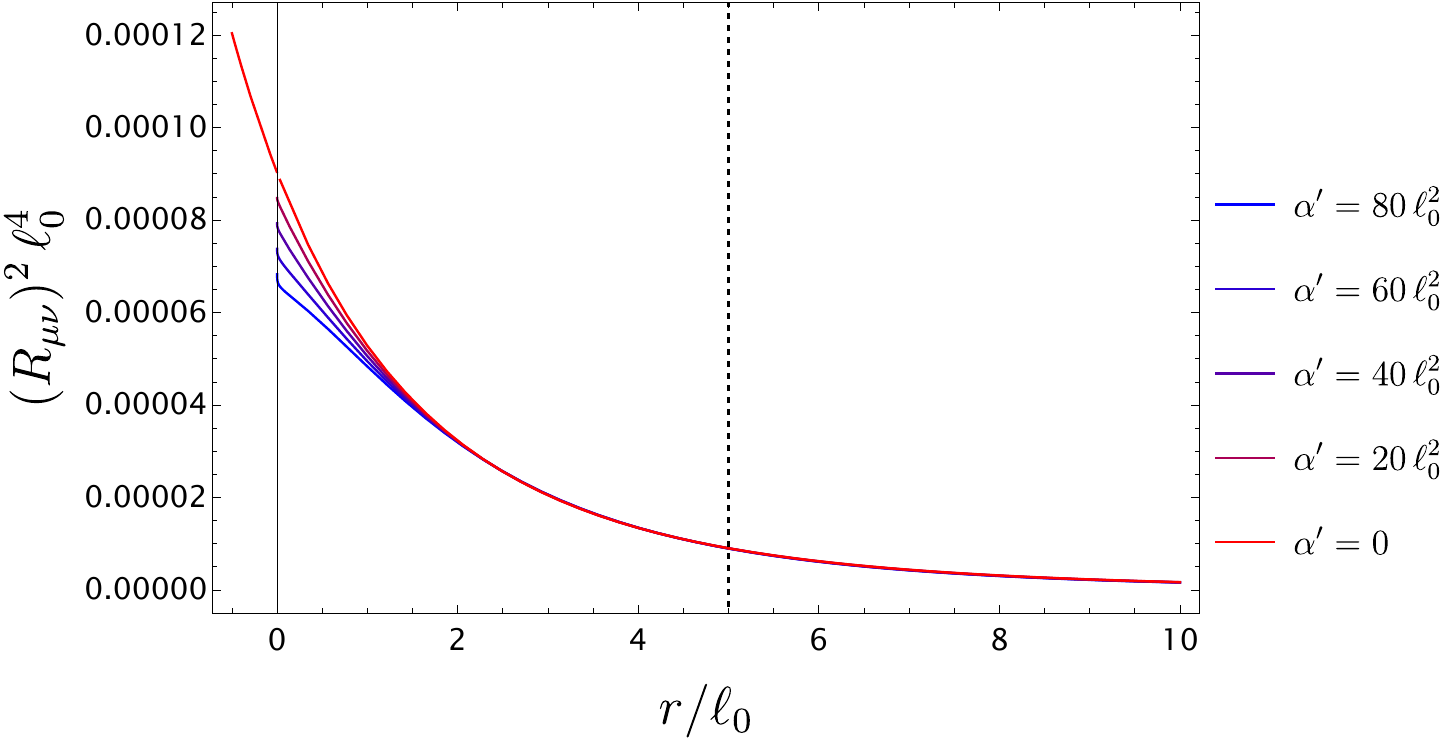}
	\caption{\textit{The $R_{\mu\nu}R^{\mu\nu}$ invariant as a function of the radial coordinate for $q_+ = 40 \,\ell_0 $, $q_- = 20 \, \ell_0 $, $q = 10 \, \ell_0 $, $\omega = -5 \, \ell_0 $, $s_+ s_- = s_0 s_{\mathcal{H}} = 1$ for different values of $\alpha'$. We normalized the units setting $\ell_0 = 1$.}}
	\label{figriccitensor}
\end{figure}

\begin{figure}[h]
	\centering
	\includegraphics[width=0.6\linewidth]{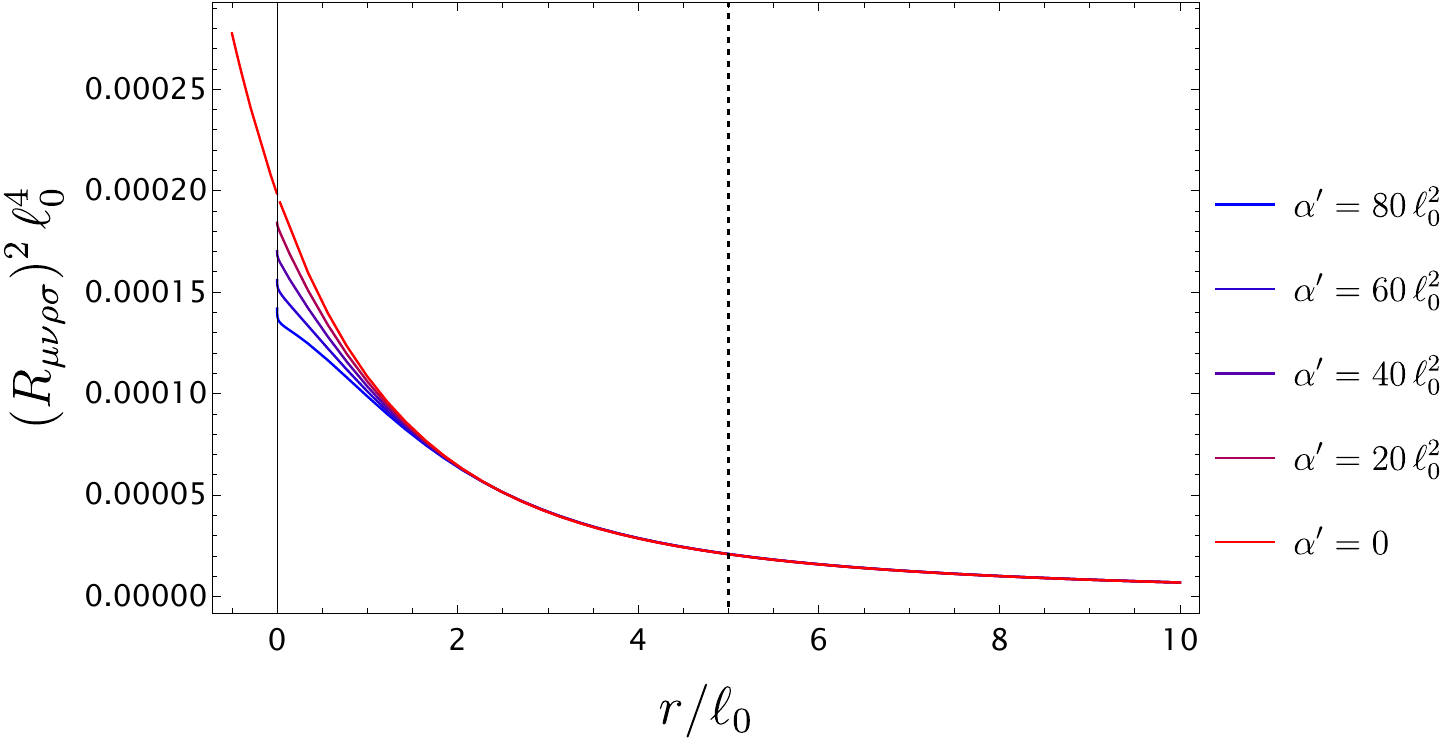}
	\caption{\textit{The Kretschmann invariant $R_{\mu\nu\rho\sigma}R^{\mu\nu\rho\sigma}$ as a function of the radial coordinate for $q_+ = 40 \,\ell_0 $, $q_- = 20 \, \ell_0 $, $q = 10 \, \ell_0 $, $\omega = -5 \, \ell_0 $, $s_+ s_- = s_0 s_{\mathcal{H}} = 1$, for different values of $\alpha'$. We normalized the units setting $\ell_0 = 1$.}}
	\label{figriemann}
\end{figure}

Identifying the physical mass\footnote{In this case we are not picking the microcanonical ensemble, but we are imposing that in the extremal limit we recover the extremal BHs mass. See \cite{Zatti:2023oiq} for more details.} we determine $d_\mathcal{H}^{(1)}$ and we obtain the regular solution 
\begin{subequations}
	\begin{align}
		\begin{split}
			\delta \mathcal{Z}_{h0} = & \; (1+s_0 s_{\mathcal{H}})\left[\frac{q^4+q^3 \omega+11 q^2 r \omega+15 q r^2 \omega+6 r^3 \omega}{4 q^2 r (q+r)^3}-  \frac{3 \omega }{2q^3}\log \mathcal{Z}_0 \right]\,,
		\end{split} \\[4mm]
		\begin{split}
			\delta \mathcal{Z}_{0} = & \; \frac{d_{\mathcal{H}}^{(1)}}{r}+\frac{\omega^2 (2 q-\omega) (s_0 s_{\mathcal{H}}+4) (\omega  q-3 \omega  r + 2 rq-2 r^2) }{40 q^3 r (q-\omega)^3}\log \mathcal{Z}_0 \\
			& +\frac{1}{120 q^2 r (q+r)^3 (q-\omega)^3} \bigg[-q r^2 (q-3 \omega) \left(5 q^3+57 q \omega^2-26 \omega^3\right) \\
			& +q^2 r \left(-16 q^4+51 q^3 \omega-165 q^2 \omega^2+205 q \omega^3-51 \omega^4\right) \\
			& +2 q^3 (q-\omega) \left(18 q^3-58 q^2 \omega+41 q \omega^2-19 \omega^3\right) \\
			& +24 r^4 \omega^2 (2 q-\omega)  +36 r^3 \omega^2 (2 q-\omega) (q+\omega) \bigg]  \\
			& + \frac{s_0 s_\mathcal{H}}{
				240 q^2 r (q+r)^3 (q-\omega)^3} \bigg[q^3 (q-\omega) \left(48 q^3-128 q^2 \omega+91 q \omega^2-29 \omega^3\right)\\
			&+q r^2 \left(10 q^4-30 q^3 \omega+9 q^2 \omega^2+86 q \omega^3-39 \omega^4\right)\\
			&+2 q^2 r \left(16 q^4-51 q^3 \omega+30 q^2 \omega^2+20 q \omega^3-9 \omega^4\right)\\
			& +12 r^4 \omega^2 (2 q-\omega)+18 r^3 \omega^2 (2 q-\omega) (q+\omega)\bigg] \,,
		\end{split} \\[4mm]
		\begin{split}
			\delta \mathcal{Z}_{\mathcal{H}} = & \; \delta \mathcal{Z}_0 
			- \frac{q(1+s_0s_{\mathcal{H}})(q-\omega)}{4 r (q + r)^3} \,,
		\end{split} \\[4mm]
		\begin{split}
			\delta \mathcal{Z}_{h-} = & \; \frac{ q_- (2 q-\omega) (q_--\omega)}{r^2 (q-\omega) (\omega-2 q_-)}d_{\mathcal{H}}^{(1)}-\frac{q q_- \left(2 q^2-6 q \omega+5 \omega^2\right) (q_--\omega) (s_{0} s_{\mathcal{H}}-1)}{20 r^2 (q-\omega)^4 (\omega-2 q_-)} \\
			& +\frac{q_- \omega^2 \left[q^2+q (q_--3 \omega)+\omega (4 \omega-3 q_-)\right]}{8 r^2 (q-\omega)^4 (\omega-2 q_-)} \,,
		\end{split} \\[4mm]
		\begin{split}
			\delta \mathcal{Z}_- = & \; \delta \mathcal{Z}_{h-} + \mathcal{Z}_- \left[\Delta_k - \frac{\Delta_C}{\beta_+}+\frac{r^2}{2}\left(\frac{\delta \mathcal{Z}_{h-}'}{q_-} - \frac{T_z[\delta\mathcal{Z}_{h-}']}{q_+}\right)\right] \,,
		\end{split} \\[4mm]
		\begin{split}
			\delta \mathcal{Z}_+ = & \; T_z \left[\delta \mathcal{Z}_{h-}\right] - \mathcal{Z}_+ \frac{\Delta_C}{\beta_+} \,,
		\end{split} \\[4mm]
		\begin{split}
			\delta W_{tt} = & \; \frac{q \omega \left(2 q^2-6 q \omega+5 \omega^2\right) (r+\omega) (s_0 s_{\mathcal{H}}-1)}{20 r^2 (q-\omega)^4 (\omega-2 q_+)} -\frac{\omega (2 q-\omega) (r+\omega)}{r^2 (q-\omega) (\omega-2 q_+)}d_{\mathcal{H}}^{(1)} \\
			& \frac{q^2 \omega^2 (2 q_++2 r+\omega)}{8 r^2 (q-\omega)^4 (\omega-2 q_+)}  +\frac{\omega^2}{8 r (q+r)^2 (q_-+r)} \\
			& +\frac{q_- \omega^2 (2 r+\omega)-q_- q_+ \omega [2 r (\beta_- \beta_++1)+2 \beta_- \beta_+ \omega+\omega]}{8 r^2 (q+r)^2 (q_-+r) (q_+-\omega)} \\
			&+\frac{5 \omega^5}{16 r^2 (q-\omega)^4 (\omega-2 q_+)} +\frac{5 q_+ \omega^4}{8 r^2 (q-\omega)^4 (\omega-2 q_+)} \\
			& -\frac{q \omega^3 (4 q_++5 r+3 \omega)}{8 r^2 (q-\omega)^4 (\omega-2 q_+)} +\frac{5 \omega^4}{8 r (q-\omega)^4 (\omega-2 q_+)}+\frac{3 \omega^4}{16 r^2 (q-\omega)^4} \,,
		\end{split} \\[4mm]
		\begin{split}
			\delta W_{rr} = & \; \frac{(2 q-\omega) (r+\omega)}{r^2 (q-\omega)}d_{\mathcal{H}}^{(1)}+\frac{\omega^2 (2 q-\omega) (r+\omega) (2 r+\omega) (s_0 s_{\mathcal{H}}+4)}{20 q^3 r (q-\omega)^3} \log \mathcal{Z}_0 \\
			& + \frac{(s_0 s_{\mathcal{H}}-1)}{120 q^2 r^2 (q+r)^4 (q-\omega)^4}\bigg[-12 r^6 \omega^2 \left(2 q^2-3 q \omega+\omega^2\right) \\
			& -6 q^7 \omega \left(2 q^2-6 q \omega+5 \omega^2\right) \\
			& -6 r^5 \left(2 q^5-6 q^4 \omega+19 q^3 \omega^2-15 q^2 \omega^3-2 q \omega^4+3 \omega^5\right) \\
			& +r^4 \left(-46 q^6+124 q^5 \omega-177 q^4 \omega^2-5 q^3 \omega^3+125 q^2 \omega^4-45 q \omega^5-6 \omega^6\right)\\
			& -q r^3 \left(64 q^6-136 q^5 \omega+42 q^4 \omega^2+221 q^3 \omega^3-167 q^2 \omega^4+15 q \omega^5+21 \omega^6\right) \\
			& +q^2 r^2 \left(-48 q^6+84 q^5 \omega+37 q^4 \omega^2-157 q^3 \omega^3-3 q^2 \omega^4+53 q \omega^5-26 \omega^6\right) \\
			& +q^3 r \left(-12 q^6-12 q^5 \omega+118 q^4 \omega^2-145 q^3 \omega^3+21 q^2 \omega^4+5 q \omega^5-5 \omega^6\right)	\bigg] \\
			& +\frac{\omega^2}{8 q^2 r^2 (q+r)^2 (q-\omega)^4} \bigg[-4 r^4 \left(2 q^2-3 q \omega+\omega^2\right) \\
			& -q r \left(q^4+9 q^2 \omega^2-7 q \omega^3+3 \omega^4\right)+r^3 \left(-11 q^3+3 q^2 \omega+12 q \omega^2-6 \omega^3\right)\\
			& +q^2 \omega \left(-q^3+2 q^2 \omega-4 q \omega^2+\omega^3\right) \\
			& +r^2 \left(-2 q^4-16 q^3 \omega+17 q^2 \omega^2-3 q \omega^3-2 \omega^4\right)\bigg] \,.
		\end{split} 
	\end{align}
\end{subequations}
with
\begin{align}
	\begin{split}
		d_{\mathcal{H}}^{(1)} = \; &  \frac{1}{  40 (q-\omega)^3 (2 q-\omega) D}\bigg\{-8 q^3 (s_0 s_{\mathcal{H}}-1) (q (q_-+q_+)+2 q_- q_+) \\
		& +\omega^3 [(q^2 (48 s_0 s_{\mathcal{H}}-98)+q (q_-+q_+) (54 s_0 s_{\mathcal{H}}-59)+4 q_- q_+ (8 s_0 s_{\mathcal{H}}+7)] \\
		& -4 q \omega^2 [q^2 (8 s_0 s_{\mathcal{H}}-13)+2 q (q_-+q_+) (8 s_0 s_{\mathcal{H}}-13)+q_- q_+ (22 s_0 s_{\mathcal{H}}+13)] \\
		& +4 q^2 \omega [2 q^2 (s_0 s_{\mathcal{H}}-1)+q (q_-+q_+) (9 s_0 s_{\mathcal{H}}-14)+2 q_- q_+ (8 s_0 s_{\mathcal{H}}-3)]\\
		& +\omega^4 (-16 s_0 s_{\mathcal{H}} (2 q+q_-+q_+)+72 q+11 (q_-+q_+))+2 \omega^5 (4 s_0 s_{\mathcal{H}}-9)\bigg\} \,,
	\end{split} \\[2mm]
	D = & \;-2 q (q_-+q_+-\omega)+3 \omega (q_-+q_+)-4 q_- q_+-2 \omega^2 \,, \label{eqdefD}
\end{align}
and $T_z$, $T_w$ are the T-duality operators
\begin{subequations} \label{eqtdualoperator}
	\begin{align}
		&T_z: \qquad 	q_+ \leftrightarrow q_- \,, \qquad \beta_+ \leftrightarrow \beta_- \,, \qquad k_\infty \leftrightarrow 1/k_\infty \,,\\
		&T_w: \qquad 	q\leftrightarrow q \,,  \qquad \beta_{\mathcal{H}} \leftrightarrow \beta_0 \,, \qquad \ell_\infty \leftrightarrow 1/\ell_\infty  \,.
	\end{align}
\end{subequations}
It can be verified that, applying (\ref{eqtdualoperator}) to this solution, the ansatz (\ref{eq021}) obeys the transformation rules 
\begin{subequations}
	\begin{align}
		&T_z: \qquad 	C^{(1)}_z \leftrightarrow A^z \,, \qquad k  \leftrightarrow 1/k^{(1)} \,, \qquad	ds^2_{E} \leftrightarrow ds^2_{E} \,, \qquad e^{-2\phi}\leftrightarrow e^{-2\phi} \,, \\
		&T_w: \qquad 	C^{(1)}_w \leftrightarrow A^w \,, \qquad \ell \leftrightarrow 1/\ell^{(1)} \,, \qquad	ds^2_{E} \leftrightarrow ds^2_{E} \,, \qquad e^{-2\phi}\leftrightarrow e^{-2\phi}  \,.
	\end{align}
\end{subequations}
We conclude by plotting some of the curvature invariants (see figures \ref{figricci}, \ref{figriccitensor}, \ref{figriemann}). They show explicitly that the solution obtained for a typical choice of charges and mass within the range of validity of the perturbative regime does not present curvature singularities outside the BH horizon. The plots do not extend beyond $r = 0$ because there is a logarithmic singularity at that point.


\subsection{4-dimensional, extremal, 4-charge BHs}

\subsubsection{Solving the EOMs}

We can easily obtain an extremal solution by taking the $\omega = 0$ limit of the regular solution presented in the previous section. However, this is not the most general result available. In the extremal limit it is possible to compute the corrections to the solutions with 4 independent charges. 

According to \cite{Cano:2021nzo} in the extremal case we can simplify ansatz (\ref{eq021}). We only need 5 independent functions
\begin{equation}
	\mathcal{Z}_H \,, \qquad \mathcal{Z}_0\,, \qquad \mathcal{Z}_{h0} \,, \qquad \mathcal{Z}_+\,, \qquad \mathcal{Z}_- = \mathcal{Z}_{h-} \,, \qquad W_{rr} = W_{tt} = 1 \,.
\end{equation}
The Bianchi identity can be used to determine $\mathcal{Z}_{h0}$. The KR equation is automatically solved by our ansatz for the dilaton (\ref{eq:dilaton10d4charge}). Finally, the dilaton and the Einstein equations form them a system of coupled linear differential equations for $\delta\mathcal{Z}_\mathcal{H}$, $\delta\mathcal{Z}_0$, $\delta\mathcal{Z}_+$ and $\delta \mathcal{Z}_-$. This system can be solved directly, without using the method of the generating function reconstruction.

In order to determine the integration constants we demand that the 4 gauge charges maintain their $0^{\rm th}$-order expressions, which is equivalent to asking for the absence of the $1/r$ term in the asymptotic expansion of $\delta\mathcal{Z}_{h0}$, $\delta\mathcal{Z}_+$ and $\delta\mathcal{Z}_-$. We also demand that all the fields keep their asymptotic values. Finally we determine the corrections to $\delta \mathcal{Z}_\mathcal{H}$ and $\delta \mathcal{Z}_0$ by demanding that their $1/r$ poles at $r=0$ vanish. One can verify that this last two conditions are compatible with regularity at the horizon and they are equivalent to determine the corrections to the mass and the position of the event horizon.

\subsubsection{Regular, extremal solutions}

We can divide the solutions in two families \cite{Cano:2021nzo}, depending on the value of $\beta_\mathcal{H} \beta_0$. In the two cases the form of the corrections changes significantly. In the case $\beta_\mathcal{H} \beta_0 = 1$, we get
\begin{subequations}
	\begin{align}
		\mathcal{Z}_{+}
		& =
		1+\frac{{q}_{+}}{r}
		-\frac{\alpha'}{4}(\beta_{+}\beta_{-}+1)
		\frac{{q}_{+}{q}_{-}}{r(r+{q}_{0})(r+{q})(r+{q}_{-})}\,,
		\\[2mm]
		\mathcal{Z}_{-}
		& =
		1+\frac{{q}_{-}}{r}\,,
		\\[2mm]
		\mathcal{Z}_{0}
		& =
		{\cal Z}_{0h}= 1+\frac{{q}_{0}}{r}+\frac{\alpha'}{4}\frac{{q}^{\,2}(r+{q}_{0})^{2}+{q}_{0}^{2}(r+{q})^{2}}{r(r+{q}_{0})^{2}(r+{q})^{3}}\,,
		\\[2mm]
		\mathcal{Z}_\mathcal{H}
		& =
		1+\frac{{q}}{r}\,.
	\end{align}
\end{subequations}
In the case with $\beta_0 \beta_\mathcal{H} = -1$, we obtain, instead,

\begin{subequations}
	\label{eq:deltaZs4-chargecase2.2}
	\begin{align}
		\begin{split}
			\delta \mathcal{Z}_{0}
			& =
			\frac{1}{4(q_{0}-q)^{5}} \Bigg\{
			-\frac{q(q_{0}-q)^{5}}{(r+q)^{3}}
			+\frac{(q_{0}-q)^{4}(q+q_{0})}{(r+q)^{2}}
			+\frac{\left(q-q_{0}\right)^{4}q_{0}}{(r+q_{0})^{2}} \\[1mm]
			&
			+\frac{(2 q_{0}-q)(q_{0}-q)^{3}}{r+q_{0}}
			-\frac{(q_{0}-q)^{3} (8q_{0}+q)}{r+q}
			-\frac{2 q_{0}(2 q^{3}+29 q^{2} q_{0}+12 q q_{0}^{2}-3q_{0}^{3})}{r} \\[1mm]
			&
			-\frac{4 q q_{0}^{2} \left(q{}^{2}+3 q_{0} \left(2
				q+q_{0}\right)\right)}{r^{2}}
			+\frac{2 q q_{0}}{(q_{0}-q)r^{3}}
			\bigg\{q q_{0} \log{\left(\frac{q_{0}}{q}\right)}\bigg[3 q r (q+12r) \\[1mm]
			&
			+q_{0} \left[2 \left(q{}^{2}+18 q r+12 r^{2}\right)
			+3 q_{0} \left(4 q+7 r+2q_{0}\right)\right]
			\bigg] \\[1mm]
			&
			-\log{\left(\frac{r+q_{0}}{r+q}\right)}
			\Big[9 q{}^{2}
			r^{3}+ 3 q_{0} q r \left(q{}^{2}+12 q r+4 r^{2}\right) \\[1mm]
			&
			+q_{0}^{2} \big[2
			q{}^{3}+36 q{}^{2} r+24 q r^{2}-r^{3}+3 q q_{0} \left(4 q+7 r+2
			q_{0}\right)\big]\Big]\bigg\}\Bigg\}\,,
		\end{split} \\[4mm]
		\delta \mathcal{Z}_{h0}
		& =
		\frac{(q_{0}-q)[q_{0}(r+q) +q(r+q_{0})]}{4(r+q)^{3}(r+q_{0})^{2}}\,, 
		\\[4mm]
		\begin{split}
			\delta \mathcal{H}
			& =
			\frac{1}{2(q_{0}-q)^{5}}\Bigg\{\frac{q}{r} \left(12 q_{0} q{}^{2}+29
			q_{0}^{2} q+2 q_{0}^{3}-3 q^{3}\right) \\[1mm]
			&
			+\frac{2 q_{0} q^{2}}{r^{2}} \left(6 q_{0} q+q_{0}^{2}+3
			q{}^{2}\right)
			-\frac{q_{0}(q_{0}-q)^{3}}{r+q}
			-\frac{3 q(q_{0}-q)^{3}}{r+q_{0}} \\[1mm]
			&
			+\frac{q_{0} q}{r^{3}(q_{0}-q)}
			\bigg[r^{3} \left(12 q_{0} q+9q_{0}^{2}-q^{2}\right)
			\log{\left(\frac{r+q_{0}}{r+q}\right)} \\[1mm]
			&
			+q_{0} q \Big[12 q_{0} \left(3 r^{2}+3 r q+q{}^{2}\right)+q_{0}^{2} (3
			r+2 q)+3 q \left(8 r^{2}+7 r q+2 q{}^{2}\right)\Big]\times \\[1mm]
			&
			\times\left[
			\log{\left(\frac{r+q_{0}}{r+q}\right)}-\log{\left(\frac{q_{0}}{q}\right)}\right]
			\bigg]\Bigg\}\,,
		\end{split} \\[4mm]
		\delta \mathcal{Z}_{+}
		& =
		\frac{q_{+}}{q_{-}}\delta \mathcal{Z}_{-}
		-(\beta_{+}\beta_{-}+1)\frac{q_+ q_-}{4 r (r+q_-)(r+q_0)(r+q)}\,,
		\\[4mm]
		\begin{split}
			\delta \mathcal{Z}_{-}
			& =
			\frac{q_{-}}{4r^{3}(q_{0}-q)^{5}}
			\Bigg\{
			8 q_{0}q r^{3}
			\log{\left(\frac{r+q_{0}}{r+q}\right)} \\[1mm]
			&
			+(q_{0}+q) \bigg[2 q_{0} q (q_{0} q-3 r^{2})
			\log{\left(\frac{1+r/q}{1+r/q_{0}}\right)}+r (q_{0}-q)
			\left[r(q_{0}+q) -2q_{0}q\right]\bigg]
			\Bigg\}\,.
		\end{split}
	\end{align}
\end{subequations}
The explicit form of the constant $\hat{c}_\phi$ appearing in (\ref{eq:dilaton10d4charge}) is
\begin{equation}
	\hat{c}_{\phi}
	  =
	e^{-2\hat{\phi}_{\infty}} q_- \left\{1-\frac{\alpha'}{4}\Upsilon\right\}\,,
\end{equation}
where we have defined\footnote{This quantity is proportional to the correction to the attractor value of the dilaton.}
\begin{equation}
	\label{eq:Upsilon}
	\Upsilon
	=
	\frac{1}{(q_{0}-q)^{5}}\left[6q_{0}q(q_{0}+q)\log{\left(\frac{q_{0}}{q}\right)}
	-(q_{0}-q)(q_{0}^{2}+q^{2}+10q_{0}q)\right]\,.
\end{equation}

\subsubsection{Supersymmetry}
\label{sec-d4-susy}

The ansatz (\ref{eq021}) in the extremal case takes the form 
\begin{subequations}\label{eq:4-charge10dsolution}
	\begin{align}
		\begin{split}
			d\hat{s}^2 \, = & \;\; \frac{1}{\mathcal{Z}_+\mathcal{Z}_-}dt^2-\mathcal{Z}_0\mathcal{Z}_\mathcal{H}(dr^2 + r^2d\Omega_{(2)}^2) \\[2mm] & \, -\ell_{\infty }^2\frac{\mathcal{Z}_0}{\mathcal{Z}_\mathcal{H}} \bigg[dw + \ell_{\infty }^{-1}\beta_{\mathcal{H}} \, q_{\mathcal{H}} \cos\theta d\varphi \bigg]^2 \\[2mm] & \, -k_{\infty}^2\frac{\mathcal{Z}_+}{\mathcal{Z}_-} \left[\,dz + k_{\infty}^{-1}\beta_+(\mathcal{Z}^{-1}_+-1)\,dt\right]^2 - dy^{\tilde{m}} dy^{\tilde{m}}\,,
		\end{split} \\[2mm]
		\hat{H} \, = & \;\; k_{\infty}\beta_- \,d\left[(\mathcal{Z}^{-1}_{-}-1)\,dt\wedge dz\right] + \ell_{\infty }\beta_{0} \, r^2 \mathcal{Z}_{h0}'\,\omega_{(2)} \wedge d w\,, \\[2mm]
		e^{-2\hat{\phi}} \, = & \; \; -\frac{c_{\hat{ \phi}}}{r^2 \mathcal{Z}_{-}'} \frac{\mathcal{Z}_-}{\mathcal{Z}_0}\,.
	\end{align}
\end{subequations}
We 	use the Zehnbein basis
\begin{equation}\label{eq:basis4d}
	\begin{aligned}
		{\hat e}^{0} & = \frac{1}{\sqrt{\mathcal{Z}_{+}\mathcal{Z}_{-}}}dt\,,
		\hspace{.7cm}
		{\hat e}^{1} = \sqrt{\mathcal{Z}_{0}\mathcal{H}}\, dr\,,
		\hspace{.7cm}
		{\hat e}^{2} = \sqrt{\mathcal{Z}_{0}\mathcal{H}}\, r v^{2}\,,
		\hspace{.7cm}
		{\hat e}^{3} = \sqrt{\mathcal{Z}_{0}\mathcal{H}}\, r v^{1}\,,
		\\[2mm]
		{\hat e}^{4}&  =
		\ell_{\infty}\sqrt{\frac{\mathcal{Z}_{0}}{\mathcal{H}}}
		\left[dw+\beta \ell_{\infty}^{-1}q\cos{\theta}d\varphi \right]\,,
		\\[2mm]
		{\hat e}^{5}&  =
		k_{\infty}\sqrt{\frac{\mathcal{Z}_{+}}{\mathcal{Z}_{-}}}
		\left[dz+\beta_{+}k_{\infty}^{-1}
		\left(\mathcal{Z}^{-1}_{+}-1\right)dt\right]\,,
		\hspace{.7cm}
		{\hat e}^{m} = dy^{m}\,,
	\end{aligned}
\end{equation}

\noindent
where

\begin{equation}
	v^{1} =\sin{\theta} d\varphi\,,
	\hspace{1cm}
	v^{2} = d\theta\,,
\end{equation}

\noindent
are the horizontal components of the left-invariant Maurer-Cartan 1-form of the
SU$(2)/$U$(1)$ coset space.\footnote{The details of this construction are
	given in appendix \ref{app:coset}.}

The dilatino Killing spinor equation (KSE) is $\delta_{\epsilon} \lambda = 0$,
where the supersymmetry variation of the dilatino with vanishing fermions is
given in Eq.~(\ref{eq:dilatino}). Substituting the values of the fields, it
can be brought to the form

\begin{equation}
	\left\{
	\frac{\mathcal{Z}_{0}'}{\mathcal{Z}_{0}}
	\left[1+\beta_{0}\frac{\mathcal{Z}_{h0}'}{\mathcal{Z}_{0}'}\Gamma^{1234}\right]
	-\frac{\mathcal{Z}_{-}'}{\mathcal{Z}_{-}}
	\left[1-\beta_{-}\Gamma^{05}\right]
	+\left(\frac{\mathcal{Z}_{-}''}{\mathcal{Z}_{-}'}+\frac{2}{r}\right)
	\right\}\epsilon =0\,.
\end{equation}
We can solve this equation without demanding any relations between
$\mathcal{Z}_{0}$ and $\mathcal{Z}_{-}$ if we demand the following two
conditions on the functions:

\begin{subequations}
	\begin{align}
		\mathcal{Z}_{h0}'
		=
		\mathcal{Z}_{0}'\,, \qquad 
		\mathcal{Z}_{-}'
		\propto
		1/r^{2}\,,
	\end{align}
\end{subequations}

\noindent
which are satisfied by the zeroth-order solutions, and the following
conditions on the Killing spinors:

\begin{subequations}
	\begin{align}
		\label{eq:4d5braneprojector}
		\left[1+\beta_{0}\Gamma^{1234}\right]\epsilon
		& =
		0\,,
		\\
		& \nonumber \\
		\label{eq:4dwaveprojector}
		\left[1-\beta_{-}\Gamma^{05}\right] \epsilon
		& =
		0\,,
	\end{align}
\end{subequations}

\noindent
which are compatible and reduce the number of independent components of the
spinors to a $1/4$ of the total 16 real components.

The $a = 0$ component of the gravitino KSE $\delta_{\epsilon} \psi_{a}=0$,
where the supersymmetry variation of the gravitino with vanishing fermions
is given in Eq.~(\ref{eq:gravitino}), can be brought to the form 

\begin{equation}
	\left\{\partial_{0}
	-\frac{\left[\log{(\mathcal{Z}_{+}\mathcal{Z}_{-})}\right]'}{4(\mathcal{Z}_{0}\mathcal{H})^{1/2}}\left[1-\beta_{-}\frac{\left[\log{(\mathcal{Z}_{+}^{\beta_{+}\beta_{-}}\mathcal{Z}_{-})}\right]'}{\left[\log{(\mathcal{Z}_{+}\mathcal{Z}_{-})}\right]'}\Gamma^{05}\right]\right\}\epsilon
	=
	0\,,
\end{equation}

\noindent
and can be solved by a spinor satisfying $\partial_{0}\epsilon=0$ and the condition (\ref{eq:4dwaveprojector}) if
\begin{equation}
	\label{eq:b+b-=1}
	\beta_{+}\beta_{-}=+1\,.  
\end{equation}
The $a=1$ component of the gravitino KSE takes the form
\begin{equation}
	\left\{
	\partial_{r}
	+\tfrac{1}{4}\beta_{-}
	\left[\log{(\mathcal{Z}_{+}^{\beta_{+}\beta_{-}}\mathcal{Z}_{-})}\right]' \Gamma^{05}
	\right\}\epsilon
	=
	0\,,
\end{equation}

\noindent
and, after use of the conditions (\ref{eq:4dwaveprojector}) and (\ref{eq:b+b-=1}) it can be
solved by

\begin{equation}
	\epsilon = (\mathcal{Z}_{+}\mathcal{Z}_{-})^{-1/4}\epsilon_{0}\,,  
\end{equation}

\noindent
where $\epsilon_{0}$ is an $r$-independent spinor that satisfies the conditions
Eqs.~(\ref{eq:4d5braneprojector}) and (\ref{eq:4dwaveprojector}). The $a=4$ component takes the form

\begin{equation}
	\left\{
	\partial_{4}
	-\tfrac{1}{2}\Gamma^{14}\Omega_{(+)\, 414}
	\left[1+\beta_{0}
	\frac{\left[\log{(\mathcal{Z}_{0}/\mathcal{H})}\right]'}
	{\left[\frac{\mathcal{Z}_{h0}'}{\mathcal{Z}_{0}}+\beta\beta_{0}\frac{q/r^{2}}{\mathcal{H}}\right]}\Gamma^{1234}\right]
	\right\}\epsilon
	=
	0\,,
\end{equation}

\noindent
and, if we demand

\begin{subequations}
	\begin{align}
		\beta\beta_{0}
		=
		+1\,, \qquad 
		\mathcal{H}'
		=
		-q/r^{2}\,,  
	\end{align}
\end{subequations}

\noindent
it is solved by a spinor satisfying Eq.~(\ref{eq:4d5braneprojector}) and $ \partial_{4}\epsilon=0$. The $a=5$ component can be written in the form

\begin{equation}
	\left\{
	\partial_{5}
	-\tfrac{1}{2}\Gamma^{01}\Omega_{(+)\, 501}
	\left[1-\beta_{-}
	\frac{\left[\log{(\mathcal{Z}_{+}/\mathcal{Z}_{-})}\right]'}
	{\left[\log{(\mathcal{Z}_{+}^{\beta_{+}\beta_{-}}/\mathcal{Z}_{-})}\right]'}
	\Gamma^{05}
	\right]
	\right\}\epsilon
	=
	0\,,
\end{equation}

\noindent
and, if the conditions (\ref{eq:b+b-=1}) and
(\ref{eq:4dwaveprojector}) are satisfied, it is solved if, in addition, $ \partial_{5}\epsilon=0$.
Finally, using all the conditions derived so far, the $a=2,3$ equations can be
combined into a single differential equation for the spinor
$\epsilon_{0}=\epsilon_{0}(\theta,\phi)$\footnote{The last 4 components of the
	gravitino KSE are trivially solved by $y$-independent spinors and the
	conditions $\partial_{0,4,5}\epsilon=0$ imply that $\epsilon_{0}$ is
	independent of the coordinates $r$, $t$, $w$, $z$ and $y^{i}$.}

\begin{equation}
	\label{eq:angularKSE}
	\left\{d +\tfrac{1}{2}\Gamma^{13}\sin{\theta} d\varphi
	+\tfrac{1}{2}\Gamma^{12}d\theta
	+\tfrac{1}{2}\Gamma^{23}\cos{\theta} d\varphi\right\}\epsilon_{0}
	=
	0\,.
\end{equation}

\noindent
We can make the following identifications with the generators of the su$(2)$
algebra

\begin{equation}
	P_{1} = \tfrac{1}{2}\Gamma^{13}\,,
	\hspace{1cm}
	P_{2} = \tfrac{1}{2}\Gamma^{12}\,,
	\hspace{1cm}
	M = \tfrac{1}{2}\Gamma^{23}\,,  
\end{equation}

\noindent
because they satisfy the commutation relations Eq.~(\ref{eq:su2algebrasplit}).
Then, using the components of the Maurer-Cartan 1-form in
Eq.~(\ref{eq:componentsMC1-form}), the KSE (\ref{eq:angularKSE}) can be
rewritten in the form

\begin{equation}
	\label{eq:angularKSE2}
	\left\{d +P_{a}v^{a}+M\vartheta \right\}\epsilon_{0}
	=
	0\,.
\end{equation}

\noindent
The 1-forms in this equation can be seen to add up to the left-invariant
Maurer-Cartan 1-form $V=-u^{-1}du$ (Eq.~(\ref{eq:MC1-form})), and the equation
can be rewritten in the form

\begin{equation}
	\label{eq:angularKSE3}
	\left\{d -u^{-1}du\right\}\epsilon_{0}
	=
	-u^{-1}d(u\epsilon_{0})=0\,.
\end{equation}

\noindent
Thus, it is solved by

\begin{equation}
	\epsilon_{0}
	=
	u^{-1}\epsilon_{00}
	=
	e^{(\theta-\pi/2)\tfrac{1}{2}\Gamma^{12}}   e^{\varphi\tfrac{1}{2}\Gamma^{13}}\epsilon_{00}\,,
\end{equation}

\noindent
where $\epsilon_{00}$ is a constant spinor that satisfies the conditions
(\ref{eq:4d5braneprojector}) and (\ref{eq:4dwaveprojector}).

Taking into account that we are considering configurations that, at zeroth
order in $\alpha'$ are determined by the functions 
\begin{equation}
	\mathcal{Z}_i = 1 + \frac{q_i}{r} \,, \qquad \mathcal{Z}_{h0} = 1 + \frac{q_0}{r} \,, \qquad i = 0,\pm, \mathcal{H}
\end{equation}
which may have additional corrections at the next
order, we can summarize our results as follows: only the configurations of the
form Eq.~(\ref{eq:4-charge10dsolution}) which satisfy all the conditions

\begin{equation}
	\beta_{+}=\beta_{-}\,,
	\hspace{.5cm}
	\beta_{0}=\beta\,,
	\hspace{.5cm}
	\mathcal{Z}_{h0} = \mathcal{Z}_{0}\,,
	\hspace{.5cm}
	\mathcal{Z}_{-} = 1+\frac{q_{-}}{r}\,,
	\hspace{.5cm}
	\mathcal{H} = 1+\frac{q}{r}\,,
\end{equation}

\noindent
are supersymmetric, and their Killing spinors take the form

\begin{equation}
	\epsilon
	=
	(\mathcal{Z}_{+}\mathcal{Z}_{-})^{-1/4}
	e^{(\theta-\pi/2)\tfrac{1}{2}\Gamma^{12}}
	e^{\varphi\tfrac{1}{2}\Gamma^{13}}
	\epsilon_{00}\,,  
\end{equation}

\noindent
where the constant spinor $\epsilon_{00}$ satisfies

\begin{equation}
	\left[1+\beta_{0}\Gamma^{1234}\right]\epsilon_{00} = 0\,,
	\hspace{1cm}
	\left[1-\beta_{-}\Gamma^{05}\right] \epsilon_{00} = 0 \,.
\end{equation}

\subsection{4-dimensional, multi-center, 4-charge BHs}

\subsubsection{Solving the EOMs}

We are going solve explicitly the EOMs replacing the ansatz (\ref{eq:4dansatzLeading}) and assuming that $\mathcal{Z}_0$, $\mathcal{Z}_+$ and $\mathcal{Z}_-$ are functions in the 4-dimensional Gibbons-Hawking space whose metric is
\begin{equation}
	d \sigma^2 = \mathcal{Z}_\mathcal{H} d\vec{x}_{(3)}^2 + \ell_{\infty }^2 \mathcal{Z}_\mathcal{H}^{-1} \bigg[dw + \ell_{\infty }^{-1}\beta_{\mathcal{H}} \, \chi \bigg]^2 \,.
\end{equation}
where $\vec{x}_{(3)}$ are coordinates of a flat 3-dimensional Euclidean space $\mathbb{E}^3$ and $\chi$ is a 1-form satisfying
\begin{equation} \label{eq:Constr2}
	d \chi = \star_{(3)} d \mathcal{Z}_\mathcal{H}\,.
\end{equation}
Notice that (\ref{eq:Constr2}) implies that $ \mathcal{Z}_\mathcal{H}$ is a harmonic function of $\mathbb{E}^3$ and $\chi$ is a 1-form of $\mathbb{E}^3$.	
The details of the computations can be found in \cite{Chimento:2018kop, Ortin:2021win}. Some intermediate steps are collected in appendix \ref{sec-curvature}. We recall that the $\beta_i$ are signs. We restrict ourselves to the case $\beta_{0} \beta_{\mathcal{H}} = 1$. The Bianchi identity (\ref{eq:bianchi-04}) is equivalent to 
\begin{equation}
	\Delta_{(4)} \left\{\mathcal{Z}_0 - \frac{\alpha'}{4}\left(\frac{\partial_m \mathcal{Z}_0 \, \partial^m \mathcal{Z}_0 }{\mathcal{Z}_0^2}+\frac{\partial_m \mathcal{Z}_\mathcal{H} \, \partial^m \mathcal{Z}_\mathcal{H} }{\mathcal{Z}_\mathcal{H}^2}\right)\right\} = 0 \,,
\end{equation}
where $\Delta_{(4)}$ represents the Laplacian operator on the 4-dimensional Gibbons--Hawking space and $\partial_m$ are the derivatives dual to the vierbeins $v^m$. It is solved by
\begin{equation}
	\mathcal{Z}_0  = \mathcal{Z}_0^{(0)} + \frac{\alpha'}{4}\left[\left(\partial \log \mathcal{Z}_0^{(0)}\right)^2 + \left(\partial \log \mathcal{Z}_\mathcal{H}^{(0)}\right)^2\right]  \,, 
\end{equation}
where $\mathcal{Z}_0^{(0)} $ is any harmonic function of the Gibbons--Hawking space. Notice that if we further assume that $\mathcal{Z}_0^{(0)}$ is independent of $w$, then $\mathcal{Z}_0^{(0)}$ is a harmonic function with respect to the metric $d\sigma^2$ iff it is a harmonic function in $\mathbb{E}^3$. In this way we recover the leading-order solution (\ref{eq:harmonics}). The KR equation of motion (\ref{eq:eq3-08}) is equivalent to
\begin{equation}
	\Delta_{(4)} \mathcal{Z}_- = 0 \,.
\end{equation} 
Again, restricting to $\mathcal{Z}_-$ independent of $w$ we obtain
\begin{equation}
	\Delta_{(3)} \mathcal{Z}_- = 0 \,,
\end{equation} 
where $\Delta_{(3)}$ is the Laplacian operator on $\mathbb{E}^3$.
We conclude that $\mathcal{Z}_-$ must be a harmonic function of $\mathbb{E}^3$. Finally, we have a single independent EOM among the Einstein equations (\ref{eq:eq1-08}) and the dilaton EOM (\ref{eq:eq2-08}). The result is the same we got in the case of the 5-dimensional BH. Using the vielbein basis (\ref{eq:vierbeinBasis}) introduced in appendix \ref{sec-curvature}, the $++$ component of Einstein equations reduces to
\begin{equation} 
	\Delta_{(4)} \left\{\mathcal{Z}_+ + \alpha'\frac{(1+\beta_+ \beta_-)}{4}\left(\frac{\partial_m \mathcal{Z}_+ \, \partial^m \mathcal{Z}_- }{\mathcal{Z}_0 \mathcal{Z}_-}\right) \right\} = 0 \,,
\end{equation}
which is solved by 
\begin{equation}
	\mathcal{Z}_+ = \mathcal{Z}_+^{(0)} - \alpha'\frac{(1+\beta_+ \beta_-)}{4}\left(\frac{\partial_m \mathcal{Z}_+^{(0)} \, \partial^m \mathcal{Z}_- }{\mathcal{Z}_0^{(0)} \mathcal{Z}_-}\right) \,,
\end{equation}
where $\mathcal{Z}_+^{(0)}$ is a harmonic function of the Gibbons--Hawking space. Assuming that $\mathcal{Z}_+^{(0)}$ is independent of $w$, we restrict ourselves to the case where  $\mathcal{Z}_+^{(0)}$ is a harmonic function of $\mathbb{E}^3$ and we connect with the leading-order solution (\ref{eq:harmonics}).

\subsubsection{Regular solutions}

The $\mathcal{Z}_i$ are determined up to harmonic functions in $\mathbb{E}^3$. This harmonic functions can be fixed requiring regularity at the horizon, the proper asymptotic normalization of the fields and selecting the microcanonical ensemble. With the ansatz for the $0^{\rm th}$-order part of  $\mathcal{Z}_i$ given by (\ref{eq:harmonics}), we obtain explicitly 
\begin{subequations}
	\begin{align}
		\mathcal{Z}_{+}
		& =
		\mathcal{Z}_{+}^{(0)}
		-\frac{\alpha'}{4} (1+\beta_+ \beta_- )
		\left[
		\mathcal{Z}_{0}^{(0)\,-1}\mathcal{Z}_{-}^{-1}\mathcal{Z}_\mathcal{H}^{-1}
		\sum_{a,b}\frac{
			q^{a}_{+}q^{b}_{-}n_{a}^{m}n_{b}^{m}}{r^{2}_{a}r^{2}_{b}}
		\right] \,,
		\\[2mm]
		\mathcal{Z}_{0}
		& =
		\mathcal{Z}_{0}^{(0)}
		+\frac{\alpha'}{4}\left[
		\mathcal{Z}_{0}^{(0)\,-2}\mathcal{Z}_\mathcal{H}^{-1}
		\sum_{a,b}\frac{q^{a}_{0}q^{b}_{0}n_{a}^{m}n_{b}^{m}}{r^{2}_{a}r^{2}_{b}}
		+\mathcal{Z}_\mathcal{H}^{-3}\sum_{a,b}\frac{q^{a}_{\mathcal{H}}q^{b}_{\mathcal{H}}n_{a}^{m}n_{b}^{m}}{r^{2}_{a}r^{2}_{b}}
		\right] \,, \\[2mm]
		\mathcal{Z}_- & = \mathcal{Z}_-^{(0)} \,, \\[2mm]
		\mathcal{Z}_\mathcal{H} & = \mathcal{Z}_\mathcal{H}^{(0)} \,,
	\end{align}
\end{subequations}
where we have defined the unit radial vectors
\begin{equation}
	n^{m}_{a}\equiv (x^{m}-x^{m}_{a})/r_{a}\,.
\end{equation}			

\section{Thermodynamics with $\alpha'$ corrections}	 \label{sec-thermod}

In this section we compute the thermodynamic quantities of the $\alpha'$-corrected solutions found in the previous section. We start by explaining how to compute the mass, the temperature, the entropy, the gauge charges, the scalar charges and the chemical potentials. We then apply the formulae to the 5-dimensional, 3-charge and 4-dimensional, 4-charge solutions. 

\subsection{General overview} \label{sec-thermodynamics}

\subsubsection{Mass and scalar charges} 

The expression for the mass $M$ can be obtained by applying the ADM formula. In practice, we can just identify $M$ by looking at the asymptotic behavior of the $tt$ component of the metric in the modified Einstein frame \cite{Maldacena:1996ky},
\begin{equation}
	g_{E}{}_{tt}=e^{-\frac{4}{d-2}(\phi-\phi_{\infty})} g_{tt}\approx 1- \frac{16\pi G_N^{(d)} }{(d-2)\omega_{(d-2)}}\frac{M}{r^{d-3}}+\dots \,,
\end{equation}
where $\omega_{(d-2)}$ is the volume of the unit ${\mathbb S}^{(d-2)}$ sphere. We recall that 
\begin{equation}
	g_s^{(d)}\,{}^2 = g_s^2\, \text{Vol}_{10-d}/(2\pi\ell_s)^n \,, \qquad G_N^{(d)}  = G_N^{(10)} \, \text{Vol}_{10-d} \,,
\end{equation} 
with 
\begin{equation}
	\text{Vol}(\text{T}^4) = (2\pi \ell_s)^4\,, \qquad \text{Vol}(\text{S}_w^1) = 2\pi\ell_{\infty} \ell_s \,, \qquad \text{Vol}(\text{S}_z^1) = 2\pi k_{\infty} \ell_s \,.
\end{equation}
For the scalar fields $\varphi^x$ normalized in such a way that their kinetic terms have form
\begin{equation}
	\begin{split}
		\frac{1}{16 \pi G_N^{(d)}} \int dx^{d} \sqrt{|g_E|} \left[\frac{1}{2} g_{xy} \partial_\mu \varphi^x \partial^\mu \varphi^y \right]   \,,
	\end{split}
\end{equation}
where $g_{xy}$ is the scalar metric, we can compute the scalar charges $\Sigma^x$ via the asymptotic expansion
\begin{equation}
	\phi^x \sim \phi_\infty^x + \frac{4 \pi G_N^{(d)}}{\omega_{(d-2)}(d-3)} \frac{\Sigma^x}{r^{d-3}} + \dots
\end{equation}

\subsubsection{Temperature} 

The metric of static spherically-symmetric black holes can always be written as
\begin{equation}
	ds^2 = g_{tt}(r) dt^2 - g_{rr}(r) dr^2 - R(r)^2 d\Omega_{(d-2)}^2 \,.
\end{equation}
The Hawking temperature is 
\begin{equation}
	T_H = \frac{1}{4\pi} \frac{\partial_r g_{tt}}{\sqrt{|g_{tt}g_{rr}}|}\bigg\rvert_{r = r_H} \,,
\end{equation}
where $r_H$ is the position of the outer horizon. It corresponds to the (largest, finite) root of the metric function $g_{tt}$,
\begin{equation}
	g_{tt}(r_H)=0\, .
\end{equation} 
Equivalently, we can compute the inverse temperature $\beta$ demanding regularity in the Euclidean section \cite{Gibbons:1976ue}. For the Einstein-frame metrics (\ref{eq:modifiedEinsteinmetriccorrected}) and (\ref{eq:modifiedEInstein4d}), a simple computation produces
\begin{equation}\label{eq:beta}
	\beta= 4\pi \frac{W_{tt}'}{\sqrt{ h \, W_{tt}W_{rr}^{-1} }} \bigg\rvert_{r = r_H} \,, \qquad h = \begin{cases}
		\mathcal{Z}_+ \mathcal{Z}_- \mathcal{Z}_0  \,, & \quad d = 5 \,, \\
		\mathcal{Z}_+ \mathcal{Z}_- \mathcal{Z}_0 \, \mathcal{Z}_\mathcal{H} \,, & \quad d = 4  \,.
	\end{cases}
\end{equation}
Notice that, depending of our choice of boundary conditions, the position of the horizon $r_H$ may shifted by the $\alpha'$ corrections.

\subsubsection{Gauge charges and dual potentials.}

In 5 dimensions we are going to deal with two gauge vectors (KK and winding vectors) carrying electric charges and one rank-2 gauge potential (the 5-dimensional KR field) carrying magnetic charge. Dualizing the KR field we can equivalently work with three electrically charged gauge vectors. In 4 dimensions we have instead four gauge vectors, two of them coming from the dimensional reduction of the metric and two of them coming from the dimensional reduction of the KR field. Two of the four vectors carry electric charge and 2 of them magnetic charge. Again, dualizing the two vectors carrying magnetic charges, we can work with four electrically-charged gauge vectors.

For the computation of the electric and magnetic charges we are going to apply the definitions of section \ref{sec-charges}. Given that we are only interested in the value of the charges and not in the actual construction of the closed currents, we will compute them integrating on a sphere at infinity. In this a way, the only contribution is coming from the currents one would get with the 2-derivative theory. 

For the computation of the potentials we follow \ref{sec-potentials}. In all cases we have just to determine the electrostatic potential of a gauge vector. Therefore, we have to integrate the momentum-map equation 
\begin{equation}
	\iota_k F + dP_k = 0 \,,
\end{equation}
with $F = dA $ a rank 2 form and $P_k$ a scalar function. To do so, we can profit of the fact that we are considering static configurations with spherical symmetry. In particular, the gauge fields can be written in the static gauge. Then the momentum map is simply
\begin{equation}
	P_k = -A_t \,,
\end{equation}  
and the electrostatic potentials are
\begin{equation}
	\Phi_{\mathcal{BH}} = -A_t \bigg \vert_{\mathcal{BH}} \,, \qquad \Phi_{\infty} = -A_t \bigg \vert_{\infty} \,,
\end{equation}  
In the cases at hand we further have that $A_t$ vanishes asymptotically. Thus, the potential $\Phi$ appearing in the first law is just
\begin{equation}
	\Phi \equiv \Phi_{\mathcal{BH}} - \Phi_\infty = \Phi_{\mathcal{BH}} \,.
\end{equation}

\subsubsection{Black hole entropy.}  

In higher-derivative theories the entropy can be computed with Wald's formula \cite{Wald:1993nt, Iyer:1994ys}. However, one of the key assumptions in its derivation does not hold in presence of gravitational Chern-Simons terms, such as the ones present in the heterotic theory.\footnote{Namely, that the transformation of the $d$-form Lagrangian $\mathbf L$ under diffemorphisms is $\delta_{\xi}{\mathbf L}={\cal L}_{\xi}{\mathbf L}$, being ${\cal L}_{\xi}$ the Lie derivative with respect to $\xi$.} As a consequence, different strategies have been proposed in the literature in order to circumvent this issue (see e.g.~\cite{Sahoo:2006vz, Sahoo:2006pm, Faedo:2019xii, Elgood:2020xwu, Ortin:2020xdm, Ma:2022nwq} for a limited list of references), which mainly involve a convenient rewriting of the action. Nevertheless, it is also possible to extend Wald's formalism to properly account for gravitational Chern-Simons terms. Doing so, general expressions for the black hole entropy were obtained in \cite{Tachikawa:2006sz} and more recently in \cite{Elgood:2020nls}. It is more convenient for us to  make use of the entropy formula given in \cite{Elgood:2020nls}, as it has been derived precisely in the context of the heterotic effective action. We report it here for completeness,
\begin{equation}
	\label{eq:Waldentropyformula-09}
	S
	=
	\frac{(-1)^{d+1} {\hat g}_{s}^{2}}{8\hat{G}_{N}}
	\int_{\mathcal{BH}}
	e^{-2\hat\phi}
	\left\{
	\left[
	\hat \star ({\hat e}^{\hat a}\wedge {\hat e}^{\hat b})
	+\frac{\alpha'}{2}\star \hat R_{(-)}{}^{\hat a \hat b}
	\right]{\hat n}_{\hat a\hat b}
	+(-1)^{d}\frac{\alpha'}{2}\Pi_{n}\wedge \hat\star {\hat H}
	\right\}\,,
\end{equation}
where $\mathcal{BH}$ stands for the bifurcation surface of the event horizon and $\hat R_{(-)}{}^{\hat a \hat b}$ is the curvature two-form defined in eq.~(\ref{eq:Rminus}). $\Pi_{n}$ is the vertical Lorentz momentum map associated to the binormal to the Killing horizon, $\hat n^{\hat a\hat b}$, and it is defined by the property
\begin{equation}
	d\Pi_{n}
	\stackrel{\mathcal{BH}}{=}
	{\hat R}_{(-)}{}^{\hat a\hat b}{\hat n}_{\hat a\hat b}\,.
\end{equation}
The formula for the entropy is gauge invariant and frame independent. Performing a local Lorentz transformation we can always put the Vielbein components $\hat{e}_{\hat{\mu}}{}^{\hat{a}}$ in an upper triangular form. In such a frame and with our ansatz, $\Pi_{n}$ has the explicit expression
\begin{equation}
	\Pi_{n}
	\stackrel{\mathcal{BH}}{=}
	{\hat \Omega}_{(-)}{}^{\hat a\hat b}{\hat n}_{\hat a\hat b}\,,
\end{equation}
and \eqref{eq:Waldentropyformula-09} can be easily evaluated.

\subsection{Non-extremal, 5-dimensional BHs}	

\subsubsection{Gauge charges, dual potentials and brane sources}

Our 5d solution represents a 3-charge BH. More precisely, the BH is electrically charged with respect to the two gauge vectors, $A_E$ and $C^{(1)}_{E}$, and magnetically charged with respect the Kalb--Ramond field $B_E$.  Explicitly, we have (all the fields are 5d and in the modified Einstein normalization)
\begin{subequations}
	\begin{align}
		& Q_+ \equiv \frac{1}{16 \pi G_N^{(5)}} \int_{S^3_\infty}e^{-4/3 \phi}k^2 \star_E F_E = \frac{\pi}{4 G_N^{(5)}}g_s^{(5)}{}^{-2/3} k_\infty q_+ \beta_+ \,,\\[2mm]
		& Q_- \equiv \frac{1}{16 \pi G_N^{(5)}} \int_{S^3_\infty}e^{-4/3 \phi}k^{-2} \star_E G_E = \frac{\pi}{4 G_N^{(5)}}g_s^{(5)}{}^{-2/3} k_\infty^{-1} q_- \beta_- \,,\\[2mm]
		& Q_0 \equiv -\frac{1}{16 \pi G_N^{(5)}} \int_{S^3_\infty} H_E = \frac{\pi}{4 G_N^{(5)}}g_s^{(5)}{}^{4/3}  q_0 \beta_0 \,, 
	\end{align}
\end{subequations}
The normalization of the charges is chosen in such a way that the $Q_i$ are quantized in units of $1/(\ell_s g_s^{(5)}{}^{2/3})$. The associated gauge potentials are
\begin{subequations}
	\begin{align}
		& \Phi_+ = \frac{k_\infty^{-1}}{\beta_+}g_s^{2/3}\left(1+\alpha'\Delta_{\Phi+}\right) \,, \\[2mm]
		& \Phi_- = \frac{k_\infty}{\beta_-}g_s^{2/3}\left(1+\alpha'\Delta_{\Phi-}\right) \,, \\[2mm]
		& \Phi_0 = \frac{1}{\beta_0}g_s^{-4/3}\left(1+\alpha'\Delta_{\Phi0}\right) \,,
	\end{align}
\end{subequations}
with 
\begin{subequations}
	\label{eq:potentialsforomeganegative}
	\begin{align}
		\Delta_{\Phi+}
		& =
		\frac{\omega\beta_{+}}{2 (q_{0}-\omega) (q_{+}-\omega) \beta_{-} }
		+\frac{3 \omega (2q_{0}-3 \omega )(2q_{-} - \omega)}{4
			(q_{0}-\omega)^{2}{D}} \,,
		\\[2mm]
		\Delta_{\Phi-}
		& =
		\frac{\omega\beta_{-}}{2 (q_{0}-\omega) (q_{-}-\omega) \beta_{+} }
		+\frac{3 \omega (2q_{0}-3 \omega )(2q_{+} - \omega)}{4
			(q_{0}-\omega)^{2}{D}} \,,
		\\[2mm]
		\Delta_{\Phi0}
		& =
		\frac{4q_{-}q_{+}\omega + 16 q_{0} \omega (q_{-}+q_{+}-\omega)- 22\omega^{2}(q_{-}+q_{+})+21 \omega^{3}}{4 (q_{0} -\omega)^{2}D}\,,
	\end{align}
\end{subequations}
where $D$ is given by 
\begin{equation}
	\label{eq:D}
	D
	\equiv
	4q_{-}q_{+}    -4\omega(q_{0}+q_{-}+q_{+})    +4q_{0}(q_{-}+q_{+})+3 \omega^{2}\,.
\end{equation}

In order to verify the quantization of $Q_{0,-}$ we work in 10d HST. We notice that if we have a total of $N_{-} \in \mathbb{Z}$ fundamental strings wrapped along the $z$ direction and we couple the associated current to the 10d HST effective action, the KR equation takes the form 
\begin{equation}
	\frac{g_s^2}{16 \pi G_N^{(10)}} \int_{V_8} d \left[e^{-2 \hat{\phi}} \hat{\star} \hat{H} + \mathcal{O}(\alpha'{}^2)\right] = T_{F1} N_{-} \,,
\end{equation}
where $V_8$ is such that $\partial{V_8} = \text{T}^4 \times \text{S}^3_\infty $ and $T_{F1} = 1/2 \pi \alpha'$. Evaluating the LHS we obtain
\begin{equation}
	\frac{\pi k_\infty^{-1} \ell_s}{4 G_N^{(5)}} \frac{1}{2 \pi \ell_s^2} \beta_- q_- = Q_- g_s^{(5)}{}^{2/3} \ell_s T_{F1} \equiv T_{F1} N_- \,.
\end{equation}
If we have a total of $N_{0} \in \mathbb{Z}$ NS5 branes wrapped along the $z$ directions and the internal torus $\text{T}^4$  and we couple the associated current to the effective action, the Bianchi identity of the KR field takes the form
\begin{equation}
	-\frac{1}{16 \pi G_N^{(10)}}\int_{V_4} d H - \frac{1}{3} T^{(4)} =  T_{\text{NS5}} \, N_0 \,,
\end{equation}
with $T_{\text{NS5}}^{-1} = (2\pi\ell_s)^5 \ell_s g_s^2$ and $V_4$ the region of spacetime contained in $\partial V_4 = S_\infty^3$. We have
\begin{equation}
	\frac{\pi \ell_s g_s^{(5)}{}^2}{4 G_N^{(5)}} \frac{1}{(2\pi\ell_s)^5 \ell_s g_s^2} = Q_0 g_s^{(5)}{}^{2/3} \ell_s T_{\text{NS5}} \equiv T_{\text{NS5}} N_0 \,.
\end{equation}
Finally, the quantization of $\mathcal{Q}_+$ follows using T-duality
\begin{equation}
	Q_-   \quad \xrightarrow{T_z} \quad  \frac{k_\infty g_s^{(5)}{}^{-2/3}}{4 G_N^{(5)}} \, \beta_+\,q_+ = \mathcal{Q}_+ = N_+ \frac{1}{\ell_s  g_s^{(5)}{}^{2/3}}\,, \qquad N_+ \in \mathbb{Z}\,.
\end{equation}

From the point of view of string theory, the solution is a superposition of solitonic 5-branes (NS5) wrapped around the directions parametrized by the coordinates $y^{1},\cdots,y^{4},z$, fundamental strings (F1) wound around the circle parametrized by $z$ and waves (W) carrying momentum propagating along the same circle (see table \ref{diagram5d-08}). In the extremal case, the integers $N_i$ have a clear interpretation. Their norm counts the number of sources and their signs distinguish between brane and antibranes. In the non-extremal case, the situation is more subtle and the parameters do not have a clear interpretation. However, it has been proposed in similar settings that they may correspond to the difference between the number of branes and the number of antibranes \cite{Horowitz:1996ay}. 
\begin{table}[h]
	\begin{center}
		\begin{tabular}{c|cccccccccc|c}
			&$t$&$z$&$y^{1}$&$y^{2}$&$y^{3}$&$y^{4}$&$x^{1}$&$x^{2}$&$x^{3}$& $x^{4}$ & $\#$\\
			\hline
			F1&$\times$ &$\times$&$\sim$&$\sim$&$\sim$&$\sim$&$-$&$-$&$-$&$-$ & $N_-$\\
			\hline
			W&$\times$&$\times$&$\sim$&$\sim$&$\sim$&$\sim$&$-$&$-$&$-$&$-$ & $N_+$ \\
			\hline
			NS5&$\times$&$\times$&$\times$&$\times$&$\times$&$\times$&$-$&$-$&$-$&$-$ & $N_0$
		\end{tabular}
		\caption{\textit{Sources associated to the five-dimensional black holes. The symbol
				$\times$ stands for the worldvolume directions and $-$ for the transverse
				directions. The symbol $\sim$ denotes a transverse direction over
				which the corresponding object has been smeared.}}
		\label{diagram5d-08}
	\end{center}
\end{table}
\subsubsection{Mass, entropy and temperature}

The mass of the BH is
\begin{equation}
	M = \frac{3 \pi}{8 G_N^{(5)}} \left[\frac{2}{3}(q_++ q_- + q_0) - \omega\right]
\end{equation}
The horizon radius, the Hawking temperature, and the Bekenstein-Hawking and Wald entropies
are given by
\begin{subequations}
	\begin{align}
		R_{H}^{(1)}
		& =
		R_{H}^{(0)}
		\left\{1 +\alpha'\,\left[\frac{\omega}{8 (q_{0} -\omega)^{2}}
		-\frac{1}{6(q_{0} -\omega)\beta_{+}\beta_{-}}\right] \right\} \,,
		\\[2mm]
		\begin{split}
			T^{(1)}_{H}
			& =
			T^{(0)}_{H}
			\bigg\{1+\alpha'\bigg[\frac{1}{2(q_{0} -\omega)\beta_{-}\beta_{+}}
			-\frac{8q_{0}^{2}(q_{-}+q_{+}-\omega)+9q_{-}q_{+}\omega}{2(q_{0}-\omega)^{2}D}
			\\[1mm]
			& \hspace{.5cm}
			+ \frac{4q_{0}\left(4q_{-}q_{+} +11q_{-}\omega +11q_{+}\omega
				-12\omega^{2}\right) +9\omega^{3}}{8(q_{0}-\omega)^{2}D} \bigg] \bigg\}\,,
		\end{split}
		\\[2mm]
		S_{BH}^{(1)}
		& =
		S_{BH}^{(0)}
		\left\{1+\alpha'\,\left[\frac{3\omega}{8 (q_{0} -\omega)^{2}}
		-\frac{1}{2(q_{0} -\omega) \beta_{+}\beta_{-}}\right] \right\}\,,
		\\[2mm]
		S^{(1)}_{W} 
		& =
		S^{(0)}_{BH}
		\left\{1+\alpha'\,\left[\frac{8q_{0} - 9\omega}{8(q_{0} -\omega)^{2}}
		+ \frac{1}{2(q_{0} -\omega)\beta_{+}\beta_{-}}\right] \right\} \,, \label{eq:waldentropy}
	\end{align}
\end{subequations}
where $D$ is still given by Eq.~(\ref{eq:D}) and we have the zeroth order quantities
\begin{subequations}
	\begin{align}
		\label{eq:RH0omeganegative}
		R_{H}^{(0)}
		& =
		\left[ (q_{+}-\omega)(q_{-}-\omega)(q_{0}-\omega)\right]^{1/6}\,,
		\\[2mm]
		\label{eq:zerothorderHtemperatureomeganegative}
		T_{H}^{(0)}
		& =
		\frac{1}{2\pi} \frac{-\omega}{\sqrt{ (q_{+}-\omega)(q_{-}-\omega)(q_{0}-\omega)}}\,,
		\\[2mm]
		\label{eq:zerothorderBHentropyomeganegative}
		S_{BH}^{(0)}
		& =
		\frac{\pi^{2}}{2G_{N}^{(5)}}\sqrt{ (q_{+}-\omega)(q_{-}-\omega)(q_{0}-\omega)}\,.
	\end{align}
\end{subequations}

\subsubsection{First law, smarr formula and scalar charge}

We can test if the computed quantities satisfy the first law computing the variations of the gauge charges $Q_i$, the entropy $S_W$ and $S_{BH}$, the mass $M$ and the moduli $k_\infty$, $\phi_\infty$. We assume that $G_N^{(5)}$ is independent of such variations. We find that only with the Wald entropy (\ref{eq:waldentropy}) precisely satisfies
\begin{equation} 
	\delta S^{(1)}_{W}
	=
	\frac{1}{T^{(1)}_{H}} \bigg[ \delta M^{(1)}
	-\Phi_{i} \delta Q_{i} - {\mathcal{Q}_k} \delta k_\infty - {\mathcal{Q}_\ell} \delta \ell_\infty - {\mathcal{Q}_\phi} \delta \phi_\infty
	-\Phi^{\alpha'}\delta\alpha'\bigg]\,.
\end{equation}

The coefficients $\mathcal{Q}_i$ are related to the numerators of the $1/r^2$ term in the asymptotic expansion of the scalar fields. In particular, given a 2-derivative theory containing Abelian vector fields and scalars coupled to gravity with the scalar kinetic sector
\begin{equation}
	\begin{split}
		\frac{1}{16 \pi G_N^{(5)}} \int dx^{5} \sqrt{|g_E|} \left[\frac{1}{2} g_{xy} \partial_\mu \varphi^x \partial^\mu \varphi^y \right]   \,,
	\end{split}
\end{equation}
we obtain (see \cite{Gibbons:1996af})
\begin{equation}
	\mathcal{Q}_x = -\frac{1}{4} {g}_{x y} \,\Sigma^y \,,
\end{equation}
where $\Sigma^x$ are the scalar charges defined by the asymptotic expansion of the $\varphi^x$.	In the modified Einstein frame the kinetic term of the scalar sector of the HST action has the form (see appendix \ref{sec-Einsteinnormalization})
\begin{equation}
	\frac{1}{16 \pi G_N^{(5)}} \int d^5x \sqrt{g_E} \bigg[\frac{4}{3}\, (\partial \phi)^2 + (\partial \log k )^2\bigg] \,.
\end{equation} 
At zeroth order we precisely recover
\begin{subequations}
	\begin{align}
		& \mathcal{Q}_\phi \, \delta \phi_\infty = - \frac{2}{3}\Sigma_{\phi} \, \delta \phi_\infty + \mathcal{O}(\alpha') \,, \label{eqChargeD}\\
		& \mathcal{Q}_k \,  \delta k_\infty = - \frac{1}{2 }\Sigma_{\log k} \,  \delta \log k_\infty  + \mathcal{O}(\alpha') \label{eqChargeK}\,.
	\end{align}
\end{subequations}
At first order in $\alpha'$ we should apply the technique of \cite{Ballesteros:2023iqb} to HST. This will be the goal of a future work and is beyond the scope of this one \cite{Ortin:2024emt}. The explicit forms of the $\mathcal{Q}$s we get are 
\begin{subequations}
	\begin{align}
		\mathcal{Q}_\phi & = \frac{2}{3}\Phi_+ \, Q_+  + \frac{2}{3}\Phi_- \, Q_- - \frac{4}{3}\Phi_0 \, Q_0  \,, \\[2mm]
		\mathcal{Q}_k & = k_\infty^{-1} \left[-\Phi_+ \, Q_+  + \Phi_- \, Q_- \right] \,, 
	\end{align}
\end{subequations}
It is possible to verify that equation (\ref{eqChargeK}) is still valid at first order in $\alpha'$, but (\ref{eqChargeD}) is not satisfied anymore. If we expand $\delta G_N^{(5)} = 0 $ we obtain
\begin{equation}
	\delta \alpha' = - \frac{4}{3} \alpha' \delta \phi_\infty \,.
\end{equation}
We can then see that the tern of the first law proportional to $\delta \alpha'$ is actually contributing to the dilaton charge. We retrieve 
\begin{equation}
	\mathcal{Q}_\phi \, \delta \phi_{\infty} + \Phi_{\alpha'} \delta \alpha' = - \frac{2}{3} {\Sigma}_{\phi} \, \delta \phi_{\infty} \,.
\end{equation}		

The explicit form of the potentials $\Phi_{\alpha'}$ is 	
\begin{equation}
	\label{eq:Phialphaomeganegative}
	\Phi^{\alpha'}
	=
	-\frac{\pi \omega}{4G_{N}^{(5)}}
	\frac{(8q_{0}-9\omega)\beta_{+}\beta_{-}
		+4(q_{0}-\omega)}{8(q_{0}-\omega)^{2}\beta_{+}\beta_{-}}\,.
\end{equation}
Finally, all the quantities satisfy the Smarr formula Eq.~(\ref{eq:firstSmarrformula})
\begin{equation}
	\label{eq:firstSmarrformula}
	M^{(1)}
	=
	\tfrac{3}{2}S_{W}^{(1)}T_{H}^{(1)} +\Phi^{+}Q_{+} +\Phi^{-}Q_{-}
	+\Phi^{0}Q_{0}  +\Phi^{\alpha'}\alpha'\,,
\end{equation}

\subsection{Non-extremal, 4-dimensional BHs} \label{secTherm}

\subsubsection{Gauge Charges and Brane Sources}
Our 4d solution represents a 4-charge BH. More precisely, the BH is electrically charged with respect to two of the gauge vectors, $A^z_E$ and $C^{(1)}_{z\,E}$, and magnetically charged with respect to $A^w_E$ and $C^{(1)}_{w\,E}$.  Explicitly, we have (all the fields are 4d and in the modified Einstein normalization)\footnote{The expression for the electric charge has been obtained integrating the current  $ J = \delta (\star \mathcal{L})/\delta F$ (which is closed on shell) on the surface at infinity and dropping the terms which are not contributing.}
\begin{subequations} \label{eqdefCharges}
	\begin{align}
		Q_+ & \equiv \frac{1}{16 \pi G_N^{(4)}} \int_{S^2_\infty} e^{-2\phi}k^2\star_E F^z_E = \frac{k_\infty}{4 G_N^{(4)}g_s^{(4)}} \beta_{+}q_+ \,, \\
		Q_- & \equiv \frac{1}{16 \pi G_N^{(4)}} \int_{S^2_\infty} e^{-2\phi}k^{-2}\star_E G_{z\, E}^{(1)} =  \frac{k_\infty^{-1}}{4 G_N^{(4)} g_s^{(4)}} \beta_{-}q_-  \,, \\
		Q_\mathcal{H} & \equiv -\frac{1}{16 \pi G_N^{(4)}} \int_{S^2_\infty}  F^w_E =  \frac{\ell_\infty^{-1}g_s^{(4)}}{4 G_N^{(4)}} \, \beta_\mathcal{H}\,q  \,, \\
		Q_0 & \equiv -\frac{1}{16 \pi G_N^{(4)}} \int_{S^2_\infty}  G_{w\, E}^{(1)} =  \frac{\ell_\infty g_s^{(4)}}{4 G_N^{(4)}} \, \beta_0\,q \,. 
	\end{align}
\end{subequations}

The normalization of the charges is chosen so that the $Q_i$ have the correct Dirac quantization. With our normalization of the action such condition takes the form (see \cite{Duff:1994an} or \cite{Ortin:2015hya} for a more recent reference)
\begin{equation} \label{eqDirac}
	Q_i Q_j \in \frac{1}{16 \pi G_N^{(4)}} 2 \pi \mathbb{Z} \,.
\end{equation}
One can verify that the charges (\ref{eqdefCharges}) satisfy (\ref{eqDirac}) checking that they are quantized in units of $1/(\ell_s g_s^{(4)})$ and recalling that, in our conventions,
\begin{equation}
	\left(\ell_s g_s^{(4)}\right)^{-2} = \frac{1}{16\pi G_N^{(4)}} 2\pi \,.
\end{equation}
The quantization of $\mathcal{Q}_\mathcal{H}$ is the simplest to prove. The absence of Dirac-Misner singularities in the 10d ansatz for the metric implies that
\begin{equation}
	N_\mathcal{H} = \frac{2 \, q \beta_{\mathcal{H}}}{R_w}  \in \mathbb{Z} \,,
\end{equation}
and with a straightforward manipulation we can write the quantized quantity as
\begin{equation}
	N_\mathcal{H}  =  Q_{\mathcal{H}} \ell_s  g_s^{(4)} \,.
\end{equation}
The quantization of $Q_0$ easily follows from the fact that it is the quantity T-dual to $Q_\mathcal{H}$. We have indeed 
\begin{equation}
	Q_\mathcal{H}  =  \frac{\ell_\infty^{-1}g_s^{(4)}}{4 G_N^{(4)}} \, \beta_\mathcal{H}\,q \quad \xrightarrow{T_w} \quad \frac{\ell_\infty g_s^{(4)}}{4 G_N^{(4)}} \, \beta_0\,q = \mathcal{Q}_0 = N_0 \frac{1}{\ell_s  g_s^{(4)}} \,, \qquad N_0 \in \mathbb{Z}\,.
\end{equation}
In order to verify the quantization of $Q_-$ we work in 10d HST. We notice that, if we have a total of $N_{-} \in \mathbb{Z}$ fundamental strings wrapped along the $z$ direction and we couple the associated current to the 10d HST effective action, the KR equation takes the form 
\begin{equation}
	\frac{g_s^2}{16 \pi G_N^{(10)}} \int_{V_8} d \left[e^{-2 \hat{\phi}} \hat{\star} \hat{H} + \mathcal{O}(\alpha'{}^2)\right] = T_{F1} N_{-} \,,
\end{equation}
where $V_8$ is such that $\partial{V_8} = \text{T}^4 \times \text{S}^2_\infty \times \text{S}^1_w$ and $T_{F1} = 1/2 \pi \alpha'$. Evaluating the LHS we obtain
\begin{equation}
	\frac{k_\infty^{-1} \ell_s}{4 G_N^{(4)}} \frac{1}{2 \pi \ell_s^2} \beta_- q_- = Q_- g_s^{(4)} \ell_s T_{F1} \equiv T_{F1} N_- \,.
\end{equation}
The quantization of $\mathcal{Q}_+$ follows using T-duality
\begin{equation}
	Q_-   \quad \xrightarrow{T_z} \quad  \frac{k_\infty}{4 G_N^{(4)}g_s^{(4)}} \, \beta_+\,q_+ = \mathcal{Q}_+ = N_+ \frac{1}{\ell_s  g_s^{(4)}}\,, \qquad N_+ \in \mathbb{Z}\,.
\end{equation}

From the point of view of string theory, the solution is a superposition of solitonic 5-branes (NS5) and Kaluza-Klein monopoles (KK6) wrapped around the directions parametrized by the coordinates $y^{1},\cdots,y^{4},z$, fundamental strings (F1) wound around the circle parametrized by $z$ and waves (W) carrying momentum propagating along the same circle (see table \ref{diagram4d-08}). In the extremal case the integers $N_i$ have a clear interpretation. Their norm counts the number of sources and their signs distinguish between brane and antibranes. In the non-extremal case the situation is more subtle and the parameters do not have a clear interpretation. However, in similar settings it has been proposed that they may correspond to the difference between the number of branes and the number of antibranes \cite{Horowitz:1996ay}. 
\begin{table}[h]
	\begin{center}
		\begin{tabular}{c|cccccccccc|c}
			&$t$&$z$&$y^{1}$&$y^{2}$&$y^{3}$&$y^{4}$&$w$&$x^{1}$&$x^{2}$&$x^{3}$ & $\#$\\
			\hline
			F1&$\times$ &$\times$&$\sim$&$\sim$&$\sim$&$\sim$&$\sim$&$-$&$-$&$-$ & $N_-$\\
			\hline
			W&$\times$&$\times$&$\sim$&$\sim$&$\sim$&$\sim$&$\sim$&$-$&$-$&$-$ & $N_+$ \\
			\hline
			NS5&$\times$&$\times$&$\times$&$\times$&$\times$&$\times$&$\sim$&$-$&$-$&$-$ & $N_0$\\
			\hline
			KK6&$\times$&$\times$&$\times$&$\times$&$\times$&$\times$&$\sim$&$-$&$-$&$-$ & $N_\mathcal{H}$\\
		\end{tabular}
		\caption{\textit{Sources associated to the four-dimensional black holes. The symbol
				$\times$ stands for the worldvolume directions and $-$ for the transverse
				directions. The symbol $\sim$ denotes a transverse direction over
				which the corresponding object has been smeared.}}
		\label{diagram4d-08}
	\end{center}
\end{table}

\subsubsection{Mass, temperature and entropy}
The mass of the BH is
\begin{equation}
	M = \frac{1}{4 G_N^{(4)}} \left[2 q + q_- + q_+ -2\omega -  (1-s_0 s_\mathcal{H})\frac{\alpha'}{5 (2q-\omega)} \right]\,.
\end{equation}
The Hawking temperature is 
\begin{equation} \label{eqthawking}
	T_H = \frac{-\omega}{4 \pi\sqrt{ (q-\omega)^2(q_+-\omega)(q_--\omega)}} (1+ \alpha' \Delta_T)\,,
\end{equation}
with
\begin{equation}
	\begin{split}
		\Delta_T = & \frac{q_- q_+ (\beta_- \beta_+-1)}{8 (q-\omega)^2 (\omega-q_-) (\omega-q_+)} + \bigg[4 q \omega \left(-5 q^2  -13 qq_--13 qq_++q_- q_+\right) \\
		& +4 q^2 ( 5 qq_-+5 qq_++2 q_- q_+)+\omega^3 (-34 q-8 q_--8q_+) \\
		& +q \omega^2 (50 q+33 q_-+ 33q_+)+8 \omega^4\bigg] \frac{(s_0 s_{\mathcal{H}}-1)}{40 (q-\omega)^3 (2 q-\omega) D} \\
		& + \bigg[\omega^3 \left(-6 q^2 -19 qq_--19 qq_+ +4 q_-^2+26 q_- q_++4 q_+^2\right) \\
		& +\omega^2 \left(12 q^2 q_-+12 q^2 q_+  + 9 qq_-^2+20 q q_- q_++9 q q_+^2-23 q_-^2 q_+ -23 q_- q_+^2 \right)\\
		& +\omega^4 (10 q-4 q_--4 q_+) +\omega \left(20 q_-^2 q_+^2-6 q^2 q_-^2-16 q^2q_- q_+-6 q^2 q_+^2\right)   \\
		& +4 q q_- q_+ (qq_-+qq_+-2 q_- q_+)\bigg] \frac{1}{8 (q-\omega)^3 (\omega-q_-) (\omega-q_+) D} \,,
	\end{split}
\end{equation}
where $D$ is the quantity introduced in eq (\ref{eqdefD}). Finally, the entropy is 
\begin{equation}
	S_W = \frac{\pi}{G_N^{(4)}} \sqrt{(q_+ -w)(q_--w)\left[(q-w)^2 + \alpha' \Delta_S\right]} \,,
\end{equation}
with 
\begin{equation}
	\Delta_S = \frac{5 q^2 - 9 q \omega + 3 \omega^2}{10q^2 - 15 q \omega + 5 \omega^2 } + \frac{q \beta_0 \beta_{\mathcal{H}} (10 q^2 - 19 q \omega + 8 \omega^2)}{20(q-\omega)^2(2q-\omega)}+ \frac{q_- q_+ \beta_+ \beta_-}{4 (q_+ - \omega)(q_- - \omega)} \,.
\end{equation}

\subsubsection{First law, smarr formula and scalar charges}

We want to verify that the thermodynamic quantities we computed satisfy the first law. In order to express the variation of the entropy in term of the physical quantities, we first determine the variation of the mass and charges with respect to the variation of the parameters $q_i$, $w$, $k_\infty$, $\ell_\infty$, $\phi_\infty$ and $\alpha'$, assuming that $ G_N^{(4)} $ is a fixed constant
\begin{subequations}
	\begin{align}
		\delta Q_0 = \; & - \frac{g_s^{(4)} \ell_\infty}{4 G_N^{(4)}} \left[\frac{2 q - \omega}{2 q \, \beta_0}\delta q - \frac{1}{2 \beta_0} \delta \omega \right] + Q_0 \, \delta \phi_\infty + Q_0 \, \frac{\delta \ell_\infty }{\ell_\infty}\,, \\
		\delta Q_\mathcal{H} = \; & - \frac{g_s^{(4)} \ell_\infty^{-1}}{4 G_N^{(4)}} \left[\frac{2 q - \omega}{2 q\, \beta_\mathcal{H}}\delta q - \frac{1}{2 \beta_\mathcal{H}} \delta \omega \right]  + Q_\mathcal{H} \, \delta \phi_\infty - Q_\mathcal{H} \, \frac{\delta \ell_\infty }{\ell_\infty} \,, \\
		\delta Q_- = \; & \frac{k_\infty^{-1}}{4 G_N^{(4)} g_s^{(4)}} \left[\frac{2 q_- - \omega}{2 q_- \beta_-}\delta q_- - \frac{1}{2 \beta_-} \delta \omega \right] - Q_- \, \delta \phi_\infty - Q_- \, \frac{\delta k_\infty}{k_\infty} \,, \\
		\delta Q_+ = \; & \frac{k_\infty}{4 G_N^{(4)} g_s^{(4)}} \left[\frac{2 q_+ - \omega}{2 q_+ \beta_i}\delta q_+ - \frac{1}{2 \beta_+} \delta \omega \right] - Q_+ \, \delta \phi_\infty + Q_+ \, \frac{\delta k_\infty}{k_\infty} \,, \\
		\begin{split}
			\delta M = \; &  \frac{1}{4 G_N^{(4)}} \bigg\{\left[2 + \alpha'\frac{2(1-s_0 s_\mathcal{H}) }{5(2q - \omega)^2}\right]\delta q + \delta q_- + \delta q_+  \\
			& + \left[-2 + \alpha'\frac{(-1 + s_0 s_\mathcal{H})}{5(2q - \omega)^2}\right] \delta \omega + \frac{(-1+s_0 s_\mathcal{H})}{10 q - 5 \omega} \delta\alpha' \bigg\} \,.
		\end{split}
	\end{align}
\end{subequations}
Then, we express the variations $\delta q_i $ and $\delta \omega$ in term of the variations of the physical charges and mass. Finally, we replace them in the variation of the entropy expressed in terms of $\delta q_i $, $\delta \omega$ and $\delta \alpha'$. We obtain
\begin{equation}
	\delta S = \frac{1}{T} \left(  \delta M -{\Phi_i} \,\delta Q_i - {\Phi_{\alpha'}} \, \delta \alpha' - {\mathcal{Q}_k} \delta k_\infty - {\mathcal{Q}_\ell} \delta \ell_\infty - {\mathcal{Q}_\phi} \delta \phi_\infty \right) \,.
\end{equation} 

The temperature $T$ appearing in the first law matches the Hawking temperature (\ref{eqthawking}), providing an highly non-trivial check that the gauge invariant entropy formula proposed by \cite{Elgood:2020nls} is the correct one to use (repeating the same computation with the standard Iyer--Wald prescription we do not recover the Hawking temperature). 

The coefficients $\Phi_\pm$ match the electrostatic potentials of \cite{Elgood:2020nls}, with the subtlety that they must be computed with the fields in the modified Einstein normalization (it will be relevant in order to have  a canonically normalized scalar charge). They are defined by
\begin{equation}
	\Phi_+ 	\stackrel{\mathcal{BH}}{=}  -\iota_t A^z_E \,, \qquad  \Phi_-	\stackrel{\mathcal{BH}}{=}  -\iota_t C^{(1)}_{z \, E} \,.
\end{equation}	
The coefficients $\Phi_{0,\mathcal{H}}$ match the magnetic potentials of \cite{Ortin:2022uxa}. They are defined as the electrostatic potential of the dual gauge fields. Again, they must be computed with the fields in the modified Einstein normalization
\begin{equation}
	\Phi_{\mathcal{H}} 		\stackrel{\mathcal{BH}}{=}  -\iota_t A^{w}_E{}^D \,, \qquad  \Phi_0		\stackrel{\mathcal{BH}}{=}  -\iota_t C^{(1)}{}^D_{\omega \, E} \,.
\end{equation}	
The simplest way to compute them is evaluating first $\Phi_0$ and then obtaining $\Phi_{\mathcal{H}}$ performing a T-duality transformation. In order to compute  $\Phi_0$ we could dualize directly $C^{(1)}{}^D_\omega$. However, it is simpler to perform the dualization in 10d and then perform a dimensional reduction. Therefore, we start dualizing the KR field directly in in 10d 
\begin{equation}
	\hat{H}^{(7)} = e^{-2 \hat{\phi}} \hat{\star} \, \hat{H} \,.
\end{equation}
We have, then\footnote{The dimensional reduction is straightforward because now the Bianchi identity is just $d \mathcal{H}^{(7)} = 0$ and the only non-vanishing components are those with the form $\mathcal{H}^{(7)}_{\mu \nu m y^1 \dots y^4}$. In particular, in the relation between higher- and lower-dimensional fields there are no explicit $\alpha'$ corrections.}
\begin{equation}
	G^{(1)}{}^D_\omega = \frac{1 }{2} \hat{H}^{(7)}_{\mu \nu \bar{z} y^1\dots y^4} \, dx^\mu \wedge dx^\nu \,,
\end{equation}
and the modified Einstein frame field strength 
\begin{equation}
	G^{(1)}_{w \, E}{}^D = G^{(1)}_{w}{}^D \, e^{\phi_\infty} \,.
\end{equation}
The explicit expressions of the potentials are
\begin{subequations}
	\begin{align}
		\Phi_+ & = \frac{k_\infty^{-1}g_s^{(4)}}{\beta_+}\left(1+\alpha' \Delta_{\Phi+}\right) \,, \\
		\Phi_- & =  \frac{k_\infty g_s^{(4)}}{\beta_-}\left(1+\alpha' \Delta_{\Phi-}\right) \,, \\
		\Phi_0 & =  \frac{\ell_\infty^{-1}}{g_s^{(4)} \, \beta_0}\left(1+\alpha' \Delta_{\Phi0}\right) \,, \\
		\Phi_\mathcal{H} & = \frac{\ell_\infty}{g_s^{(4)} \,\beta_\mathcal{H}}\left(1+\alpha' \Delta_{\Phi\mathcal{H}}\right) \,,
	\end{align}	
\end{subequations} 
with
\begin{subequations}
	\begin{align}
		\begin{split}
			\Delta_{\Phi+} = &  \; \frac{  \omega (q-2 \omega) (\omega-2 q_-) (s_0 s_\mathcal{H}+4)}{10 (q-\omega)^3 D} +\frac{\beta_- \beta_+ q_- \omega}{8 (q-\omega)^2 (q_--\omega) (q_+-\omega)} \,,
		\end{split} \\
		\begin{split}
			\Delta_{\Phi-} = & \; \frac{  \omega (q-2 \omega) (\omega-2 q_+) (s_0 s_\mathcal{H}+4)}{10 (q-\omega)^3 D} +\frac{\beta_- \beta_+ q_+ \omega}{8 (q-\omega)^2 (q_--\omega) (q_+-\omega)} \,,
		\end{split} \\
		\begin{split}
			\Delta_{\Phi0} = & \; s_0 s_\mathcal{H} \bigg[4 \omega^2 \left(3 q^2 +qq_-+q q_++3 q_- q_+\right) \\
			&-8 q \omega \left(q^2+2 q q_-+2 qq_++3 q_- q_+\right)+8 q^2 (q q_-+ qq_++2 q_- q_+) \\
			& +\omega^3 (2 q-q_--q_+)-2 \omega^4\bigg] \frac{1}{40 (q-\omega)^3 (2 q-\omega) D}\\
			&  + \bigg[ -8 \omega^2 \left(9 q^2+18 q q_-+18 q q_++4 q_- q_+\right) \\
			& +4 q \omega \left(2 q^2+19 q q_-+19 qq_++16 q_- q_+\right)-8 q^2 (q q_-+qq_++2 q_- q_+) \\
			& +8 \omega^3 (16 q+7 q_-+7q_+)-48 \omega^4 \bigg]  \frac{1}{40 (q-\omega)^3 (2 q-\omega) D}	 \,,
		\end{split} \\
		\begin{split}
			\Delta_{\Phi\mathcal{H}} = &  \; \Delta_{\Phi0}\,,
		\end{split} 
	\end{align}	
\end{subequations} 
where $D$ is the quantity defined in equation (\ref{eqdefD}). 

Let us consider now the coefficients of the variations of the moduli. The explicit form of the $\mathcal{Q}$s we get is
\begin{subequations}
	\begin{align}
		\mathcal{Q}_\phi & = \Phi_+ \, Q_+  + \Phi_- \, Q_- - \Phi_0 \, Q_0 - \Phi_\mathcal{H} \, Q_\mathcal{H} \,, \\
		\mathcal{Q}_k & = k_\infty^{-1} \left[-\Phi_+ \, Q_+  + \Phi_- \, Q_- \right] \,, \\
		\mathcal{Q}_\ell & = \ell_\infty^{-1} \left[-\Phi_0 \, Q_0  + \Phi_\mathcal{H} \, Q_\mathcal{H}\right] \equiv  0 \,.
	\end{align}
\end{subequations}
If we expand $\delta G_N^{(4)} = 0 $, we obtain
\begin{equation}
	\delta \alpha' = - 2 \alpha' \delta \phi_\infty \,.
\end{equation}
The potential $\Phi_\alpha$ contributes to the dilaton charge. Indeed, we recover 
\begin{subequations}
	\begin{align}
		& 	\mathcal{Q}_\phi \, \delta \phi_{\infty}  = -  {\Sigma}_{\phi} \, \delta \phi_{\infty} - \Phi_{\alpha'} \delta \alpha'\,, \\
		& \mathcal{Q}_k \,  \delta k_\infty = - \frac{1}{2 }\Sigma_{\log k} \,  \delta \log k_\infty   \,, \\
		& \mathcal{Q}_\ell \, \delta \ell_\infty = - \frac{1}{2}\Sigma_{\log \ell} \,  \delta \log \ell_\infty  \,.
	\end{align}
\end{subequations}
Finally, the explicit expression of the potential $\Phi_{\alpha'}$ is
\begin{equation}
	\begin{split}
		\Phi_{\alpha'} = & \; \frac{1}{G_N^{(4)}} \bigg[\frac{4 q^2(-1+s_0 s_\mathcal{H})-5q(-4+s_0 s_\mathcal{H})\omega - 20\omega^2}{160(q-\omega)^3}  \\ 
		& + \frac{5 q_- q_+ (q-\omega) \omega \beta_- \beta_+}{160 (q-\omega)^3(q_- - \omega)(q_+ - \omega)}\bigg]\,,
	\end{split}
\end{equation}
and it is exactly the one which allows the Smarr formula to be satisfied\footnote{The presence of a potential in the Smarr formula is expected for every independent dimensionful parameter. See for instance \cite{Ortin:2021win,Mitsios:2021zrn,Meessen:2022hcg}.}
\begin{equation}
	M = 2 S_W T_H + \Phi_i Q_i + 2 \Phi_{\alpha'} \,\alpha' \,.
\end{equation}

\subsection{Extremal BHs and Weak Gravity Conjecture}

\subsubsection{Thermodynamics}
The thermodynamic quantities can be easily obtained with the extremal limit $\omega = 0$. The temperature vanishes
\begin{equation}
	T_H = 0 \,.
\end{equation}
The gauge charges are in 5d
\begin{subequations}
	\begin{align}
		& Q_+ = \frac{\pi}{4 G_N^{(5)}}g_s^{(5)}{}^{-2/3} k_\infty q_+ \beta_+  = N_+/\ell_q  \,,\\[2mm]
		& Q_-  = \frac{\pi}{4 G_N^{(5)}}g_s^{(5)}{}^{-2/3} k_\infty^{-1} q_- \beta_-  = N_-/\ell_q  \,,\\[2mm]
		& Q_0  = \frac{\pi}{4 G_N^{(5)}}g_s^{(5)}{}^{4/3}  q_0 \beta_0  = N_0/\ell_q  \,, 
	\end{align}
\end{subequations}
and in 4d
\begin{subequations}
	\begin{align}
		Q_+ &  = \frac{k_\infty}{4 G_N^{(4)}}\left(g_s^{(4)}\right)^{-1} \beta_{+}q_+ = N_+/\ell_q \,, \\
		Q_- &  =  \frac{k_\infty^{-1}}{4 G_N^{(4)} } \left(g_s^{(4)}\right)^{-1}  \beta_{-}q_- = N_-/\ell_q \,, \\
		Q_\mathcal{H} & =  \frac{\ell_\infty^{-1}}{4 G_N^{(4)}} g_s^{(4)}  \, \beta_\mathcal{H}\,q_\mathcal{H} = N_\mathcal{H}/\ell_q \,, \\
		Q_0 &  =  \frac{\ell_\infty}{4 G_N^{(4)}} g_s^{(4)}   \, \beta_0\,q_0 = N_0/\ell_q\,,
	\end{align}
\end{subequations}
with $N_i \in \mathbb{Z}$ and $\beta_i^2 = 1$. The parameter $\ell_q$ is defined as
\begin{equation} \label{eq:ellq}
	\ell_q = \left(g_s^{(d)}\right)^{\frac{2}{d-2}} \ell_s \,.
\end{equation}
Once we replace in the entropy the physical charges we obtain
\begin{equation}
	S_W = 2\pi \sqrt{|N_+ N_- | (k + 2 + \beta_{+} \beta_{-})} \,,
\end{equation}
with
\begin{equation}
	k = \begin{cases}
		|N_0| \,, & \quad \text{if} \, d = 5 \,. \\[1mm] 
		|N_0 N_\mathcal{H}| + \beta_0 \beta_\mathcal{H} \,, & \quad \text{if} \, d = 4  \,.
	\end{cases}
\end{equation}
The mass is
\begin{equation}
	M
	=
	\frac{k_{\infty}}{\ell_s}|N_-|
	+\frac{1}{\ell_{s}k_{\infty}}|N_+|
	+\frac{1}{\ell_{s} \ell_\infty g_{s}^{(d)}{}^{2}}|N_{0}|
	+\frac{\ell_{\infty}}{\ell_{s}g_{s}^{(d)}{}^{2}}|N_\mathcal{H}|+\delta M \,,
\end{equation}

\noindent
with $N_\mathcal{H} = 0$ and $\ell_\infty = 1 $ in $d = 5$. The mass shift $\delta M$ reads
\begin{equation}
	\label{eq:shiftM4d}
	\delta M
	= 
	(d-5)\frac{(1-\beta_0 \beta_\mathcal{H})}{2}\frac{16\ell_{\infty} }{\ell_{s}g_{s}^{(d)}{}^{2} |N_{0}|}f(v)\,,
\end{equation}

\noindent
Notice that $\delta M \ne 0$ only in $d=4$ for non-supersymmetric solutions such that $\beta_0 \beta_\mathcal{H} = -1$. $f(v)$ is defined as

\begin{equation}\label{eq:fv}
	f(v)
	=
	\frac{v^{4}-8 v^{3}+8 v-1+12 v^{2} \log{\left(v\right)}}{(v-1)^{5}}\,,
	\quad
	v
	=
	\frac{|N_\mathcal{H}|}{|N_{0}|}\ell_{\infty}^{2}\,.
\end{equation}
Notice that despite
the apparent singularity of $f(v)$ for $v=1$, this function is actually smooth
for every $v\ge 0$, as illustrated in Fig.~\ref{fig:fv}. In fact, we have $\lim_{v\rightarrow 1} f(v)=2/5$.

\subsubsection{Weak Gravity Conjecture}
\begin{figure}[t!]
	\begin{center}
		\includegraphics[width=0.55\textwidth]{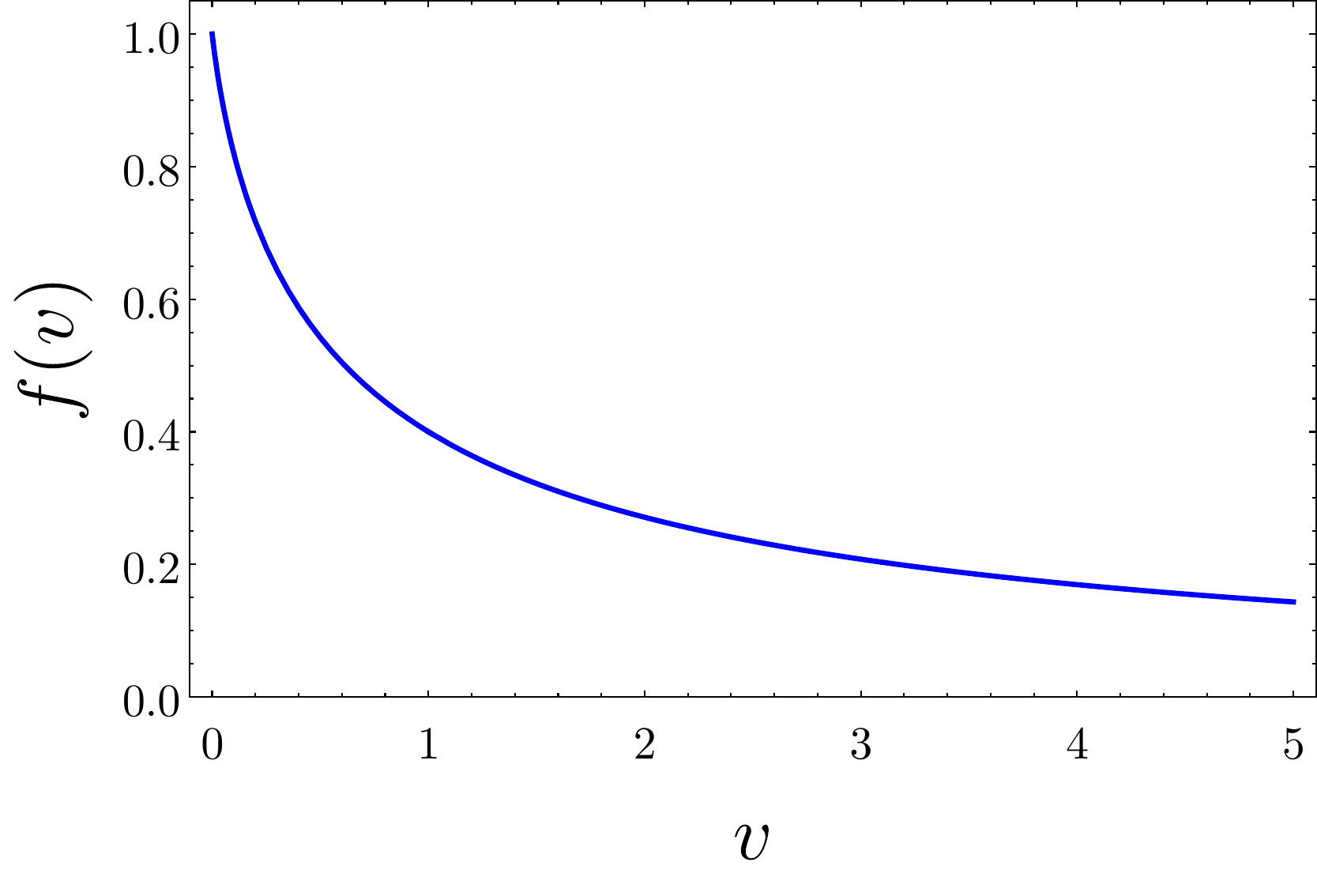} 
		\caption{Function $f(v)$ defined in (\ref{eq:fv}) that
			controls the shift to the mass (\ref{eq:shiftM4d}). Observe
			that it is smooth everywhere and that it is positive,
			meaning that the mass is corrected negatively, in agreement
			with the mild form of the WGC.}
		\label{fig:fv}
	\end{center}
\end{figure}

Higher-derivative corrections to the mass of extremal black holes are a	matter of interest in the context of the Weak Gravity Conjecture
\cite{Cheung:2018cwt,Hamada:2018dde,Bellazzini:2019xts,Charles:2019qqt,Loges:2019jzs,Cano:2019oma,Cano:2019ycn,Andriolo:2020lul,Loges:2020trf,Cano:2020qhy,Cano:2021tfs,Arkani-Hamed:2021ajd}. Originally,
this conjecture has been formulated for black holes charged under a single
$U(1)$ field, and it states that, in a consistent theory of Quantum Gravity,
the corrections to the extremal charge-to-mass ratio $Q/M$ should be
positive. The logic of this statement lies in the fact that, in this way, the
decay of extremal black holes is possible in terms of energy and charge
conservation.  The extension of this conjecture to the case of black holes
with multiple charges is subtle \cite{Jones:2019nev}, but, as a general rule,
one can see that the corrections to the mass should be negative in order to
allow for the decay of extremal black holes. Now, since our black holes are an
explicit solution of string theory, they should satisfy the WGC, assuming it
is correct. We check that, indeed, $f(v)>0$  (see
Fig.~\ref{fig:fv}), which implies that $\delta M<0$ for all the values of the
charges.  This is a quite non-trivial test of the validity of the WGC in
string theory, that adds up to the ones already found in
Refs.~\cite{Cano:2019oma,Cano:2019ycn}.

\subsection{Multicenter black holes and forces cancellation}

\subsubsection{Thermodynamics}
In a configurations with $n_c$ black holes the zeroth-order harmonic functions $\mathcal{Z}_i$ have $n_c$ poles. Such poles are located at $x^k_a $ with $a = 1, \dots n_c$ and $x^k$ are coordinates of $\mathbb{E}^{d-1}$. $\alpha'$ corrections do not modify the positions of the BHs and the horizon of the $a$th black hole is located at $r_a = 0$, which we recall is defined by	
\begin{equation}
	r_a^2 = (x^k_a - x^k)^2 \,.
\end{equation}
The near-horizon metric of the $a$th BH located at $r_a = 0$ is at leading-order (in the $r_a$ expansion) the metric of an extremal BH. We can easily evaluate the associated Hawking temperature $T^a$ and we obtain that, as expected, it vanishes 
\begin{equation}
	T^a_H = 0 \,.
\end{equation}
We can compute the gauge charges $Q_i^a$ associated to the $a$th black hole applying the definitions of section \ref{sec-charges} considering a surface enclosing only the $a$th center. We obtain in 5d
\begin{subequations}
	\begin{align}
		& Q_+^a = \frac{\pi}{4 G_N^{(5)}}g_s^{(5)}{}^{-2/3} k_\infty q_+^a \beta_+ \,,\\[2mm]
		& Q_-^a  = \frac{\pi}{4 G_N^{(5)}}g_s^{(5)}{}^{-2/3} k_\infty^{-1} q_- \beta_-   \,,\\[2mm]
		& Q_0^a  = \frac{\pi}{4 G_N^{(5)}}g_s^{(5)}{}^{4/3}  q_0^a \beta_0   \,, 
	\end{align}
\end{subequations}
and in 4d
\begin{subequations}
	\begin{align}
		Q_+^a &  = \frac{k_\infty}{4 G_N^{(4)}}\left(g_s^{(4)}\right)^{-1} \beta_{+}q_+^a  \,, \\
		Q_-^a &  =  \frac{k_\infty^{-1}}{4 G_N^{(4)} } \left(g_s^{(4)}\right)^{-1}  \beta_{-}q_-^a  \,, \\
		Q_\mathcal{H}^a & =  \frac{\ell_\infty^{-1}}{4 G_N^{(4)}} g_s^{(4)}  \, \beta_\mathcal{H}\,q_\mathcal{H}^a  \,, \\
		Q_0^a &  =  \frac{\ell_\infty}{4 G_N^{(4)}} g_s^{(4)}   \, \beta_0\,q_0^a \,,
	\end{align}
\end{subequations}
with $\beta_i^2 = 1$. All the charges $Q_i^a$ are quantized according to
\begin{equation}
	Q_i^a  \ell_q = N^a_i \in \mathbb{Z} \,,
\end{equation}
with $\ell_q$ defined in (\ref{eq:ellq}). Applying the definition of the charges with an asymptotic surface enclosing all the BHs we obtain the total charges
\begin{equation}
	Q_i = \sum_{a = 1}^{n_c} Q_i^a \,, \qquad N_i = \sum_{a = 1}^{n_c} N_i^a
\end{equation}  
The total mass is
\begin{equation}
	M
	=
	\frac{k_{\infty}}{\ell_s}|N_-|
	+\frac{1}{\ell_{s}k_{\infty}}|N_+|
	+\frac{1}{\ell_{s} \ell_\infty g_{s}^{(d)}{}^{2}}|N_{0}|
	+\frac{\ell_{\infty}}{\ell_{s}g_{s}^{(d)}{}^{2}}|N_\mathcal{H}| \,,
\end{equation} 
with $N_\mathcal{H} = 0$ and $\ell_\infty = 1 $ in $d = 5$. Given that we are considering configurations with BHs whose charges $Q_i$ have the same signs $\beta_i$, 
\begin{equation}
	|N_i| =  \sum_{a = 1}^{n_c} |N_i^a| \,,
\end{equation}
and the mass of the configuration is just the sum of the masses of isolated BHs with charges $Q_i^a$
\begin{equation}
	M = \sum_{a = 1}^{n_c} M^a \,, 
\end{equation}
where
\begin{equation}
	M_a = \frac{k_{\infty}}{\ell_s}|N_-^a|
	+\frac{1}{\ell_{s}k_{\infty}}|N_+^a|
	+\frac{1}{\ell_{s} \ell_\infty g_{s}^{(d)}{}^{2}}|N_{0}^a|
	+\frac{\ell_{\infty}}{\ell_{s}g_{s}^{(d)}{}^{2}}|N_\mathcal{H}^a| \,,
\end{equation}
with $N_\mathcal{H}^a = 0$ and $\ell_\infty = 1 $ in $d = 5$. Finally, we can compute the contribution of the $a$th BH to the total entropy evaluating (\ref{eq:Waldentropyformula-09}) at the surface $r_a = 0$. We find that the contribution is just the one of an isolated BH with charges $Q_i^a$. We have indeed
\begin{equation}
	S_W = \sum_{a = 1}^{n_c}2\pi \sqrt{|N_+^a N_-^a | (k^a + 2 + \beta_{+} \beta_{-})} \,, \qquad 
\end{equation}
with\footnote{Recall that in $d =4$ we have only solutions with $\beta_0 \beta_\mathcal{H} = 1$.}
\begin{equation}
	k^a = \begin{cases}
		|N_0^a| \,, & \quad \text{if} \, d = 5 \,. \\[1mm] 
		N_0^a N_\mathcal{H}^a + 1 \,, & \quad \text{if} \, d = 4  \,.
	\end{cases}
\end{equation}

\subsubsection{Cancellation of forces}

\begin{figure}[ht!]
	\begin{minipage}{0.48\textwidth}	
		\begin{center}
			\includegraphics[scale=0.4,trim=40 0 0 0,clip]{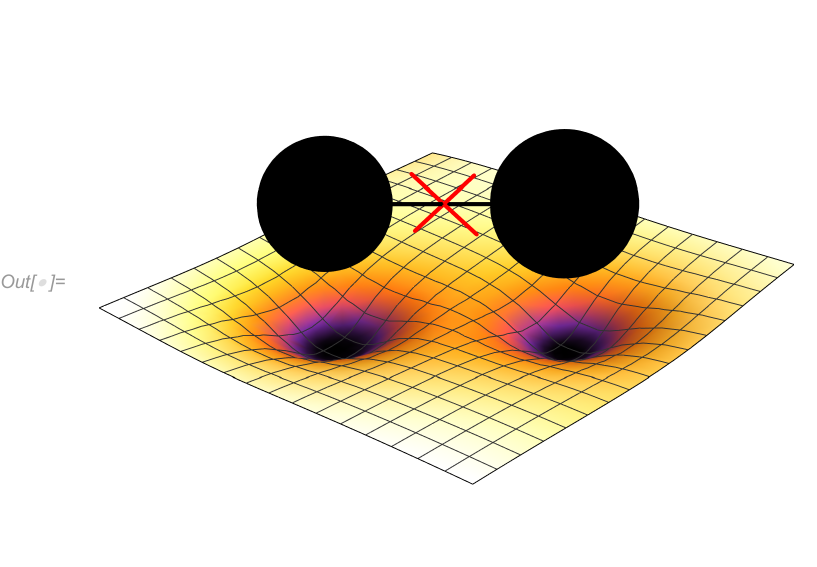}
			\caption{The absence of struts (conical singularities) joining the centers is
				a necessary condition to interpret the solution as black holes in
				equilibrium without the help of external forces..}
			\label{fig:nost1}
		\end{center}
	\end{minipage}
	\begin{minipage}{0.04\textwidth}	
		$$ $$
	\end{minipage}
	\begin{minipage}{0.48\textwidth}	
		\begin{center}
			\includegraphics[scale=0.4,trim=40 0 0 0,clip]{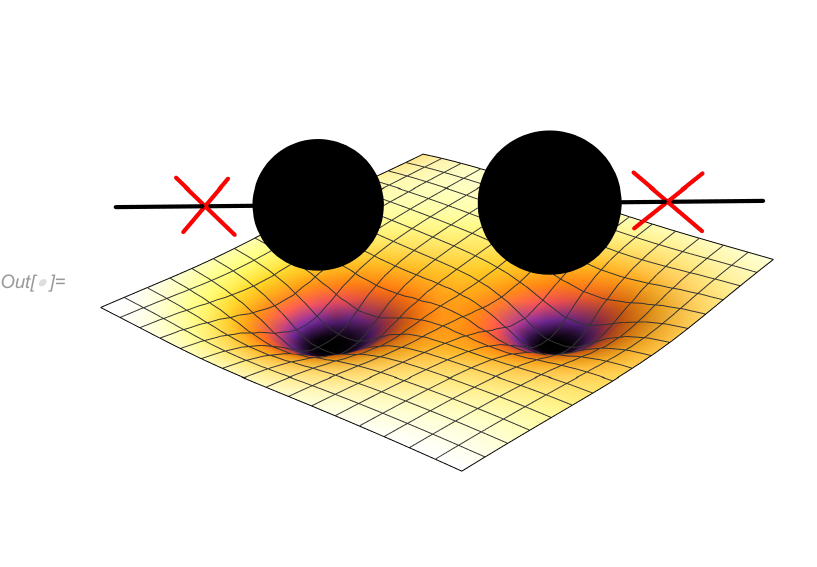}
			\caption{The absence of struts (conical singularities) extending from the
				centers to infinity is another necessary condition to interpret the solution
				as black holes in equilibrium without the help of external forces..}
			\label{fig:nost2}
		\end{center}
	\end{minipage}
\end{figure}

The mere existence of a static multi-center solution is sufficient to conclude that there is forces cancellation among the black holes of the configuration. However, it is not enough to prove that the charge-to-mass ratios of the BHs involved have the critical value to compensate attraction and repulsion. Indeed, it	has to be taken into account that exists solutions describing collinear Schwarzschild
black holes in static equilibrium 	\cite{Israel-Kahn,Costa:2000kf}. These solutions, however, have conical	singularities in the line joining the centers or extending from the centers to
infinity, known as \textit{struts}, associated to the external forces	necessary to hold the system in equilibrium. Therefore, in order to determine the critical charge-to-mass ratio it is necessary to ensure that the solutions do not present these singularities and they have regular horizons \cite{Brill:1963yv,Hartle:1972ya} (see
Figs.~\ref{fig:nost1} and \ref{fig:nost2}).\footnote{For an extended discussion of this problem we refer to the Introduction of \cite{Meessen:2017rwm}
	and references therein.} Our solutions are fully regular and do not presents \textit{struts}. We can safely conclude that the charge-to-mass ratios of the extremal BHs studied are those ensuring cancellation of the forces. Interestingly, we find that $\alpha'$ corrections do not spoil the equilibrium even in the case non-supersymmetric case. However, this cannot be considered a counter example of the strong form of the Weak Gravity Conjecture, which is essentially saying that the cancellation of forces is realized only in supersymmetric contests. Indeed, higher order corrections in $\alpha'$ or $g_s$ may still spoil the equilibrium.

\subsubsection{Fragmentation}
Clearly, a necessary condition for the fragmentation process to be allowed is
that the conserved charges of the initial and final configurations are
identical. At the two-derivative level, the conserved charges are proportional
to the total numbers of the different fundamental objects, which means that
these numbers must not change. The number of centers $n_{c}$ can, in
principle, change, but it does not appear explicitly in the entropy formula at
this order. Then, at this order, the fragmentation is, in principle, allowed,
but entropically disfavored. 

In presence of $\alpha'$ corrections (which introduce
Chern-Simons terms), one can define several notions of charge and not all of
them are necessarily conserved \cite{Marolf:2000cb}. Therefore, the first
thing we have to do is to figure out which notions of charge are conserved and
which are not.  A thorough analysis of all the possible notions of charge,
their physical interpretation and their properties in this context requires
much more work and will be carried out elsewhere \cite{kn:ORZ}. Thus, here we
will just focus on one of them, the solitonic 5-brane charge. The presence of
NS5 branes modifies the Bianchi identity of the Kalb-Ramond 2-form $\hat{B}$ as
follows:
\begin{equation}\label{eq:Bianchi}
	\frac{1}{16 \pi {G}^{(10)}_{\rm N}}
	\left[d\hat{H}-\frac{\alpha'}{4}\hat{R}_{(-)}{}^{\hat{a}}{}_{\hat{b}}\wedge
	\hat{R}_{(-)}{}^{\hat{b}}{}_{\hat{a}}\right]
	=
	\hat{\star} \hat{J}_{NS5}\,.
\end{equation}
The current $J_{NS5}$ describes the coupling of external sources (NS5 branes) to
the magnetic dual of the KR 2-form. Following \cite{Marolf:2000cb}, we refer
to it as the brane-source current. By definition, it is localized, which in
this context means that it vanishes whenever the sourceless (supergravity)
equations of motion are satisfied. For instance, in the five-dimensional
configurations we have studied, the brane-source current associated to NS5
branes is given by 
\begin{equation}
	\hat{\star}\hat{J}_{NS5}
	=
	-T_{NS5}\sum_{a} N^{a}_0 \star_{4} \delta^{(4)}(x-x_{a})\,,
\end{equation}

\noindent
since it is precisely at the centers where
equations
\begin{equation}
	d \star_\sigma d \mathcal{Z}_{\pm,0} = 0
\end{equation}
are not satisfied. Therefore, the
brane-source charge, defined as the integral of $\hat\star \hat{J}_{NS5}$, is
proportional to (minus) the total number of NS5 branes,

\begin{equation}
	\int_{\mathbb{E}^{4}} \hat{\star}\hat{J}_{NS5}
	=
	- T_{NS5}\sum_{a}N^{a}_0
	= 
	- T_{NS5}{N_0} \,.
\end{equation}
As explained in \cite{Marolf:2000cb}, brane-source charges are not conserved quantities in presence of Chern-Simons terms. The heterotic case is,	however, a bit peculiar because, when taking a exterior derivative in	\eqref{eq:Bianchi}, we arrive to
\begin{equation}
	d\hat\star\hat{J}_{NS5}
	=
	\frac{\hat{g}_{s}^{2} \alpha'}{32 \pi G^{(10)}_{\rm N}}
	\hat{\mathcal{D}}_{(-)} \hat{R}_{(-)\, \hat{a} \hat{b}}\wedge \hat{R}_{(-)}{}^{\hat{a}\hat{b}}=0\,,
\end{equation}

\noindent
if the Bianchi identity of the curvature tensor of the torsionful spin
connection $\hat{R}_{(-)}{}^{\hat{a} \hat{b}}$, defined in
Eq.~(\ref{eq:Rminus}), is not modified by the
presence of sources, that is

\begin{equation}
	\hat{\cal D}_{(-)}\hat{R}_{(-)}{}_{\hat{a} \hat{b}}
	=
	\hat {\cal D}\hat{R}_{\hat{a}\hat{b}}=0\,.
\end{equation}
This implies that the total number of 5-branes, $N_0$, must remain constant in
the fragmentation process. 

Another charge we can define is the Maxwell solitonic 5-brane
charge, which in the five-dimensional case is given by
\begin{equation}
	-\frac{1}{16 \pi {G}^{(10)}_{\rm N}}
	\int_{\mathbb{E}^{4}} d\hat{H} = - T_{NS5} (N_0 + n_c)
\end{equation}
with $n_c$ the number of centers. An important difference among the brane-source and the Maxwell charge is that only the first one is localized (\textit{i.e.} it satisfies a Gauss law in $\mathbb{E}^4$). However, both are conserved (\textit{i.e.} do not vary in time. The integrals are indeed independent of the particular space-like surface chosen). Hence, it is evident that the fragmentation is forbidden if both Maxwell and brane-source charges are conserved. An analogous analysis in the
four-dimensional case yields the same conclusion.

	\clearpage{\pagestyle{empty}\cleardoublepage} 
	
	\chapter{Tests of 2-charge black holes thermodynamics with $\alpha'$ corrections}\label{ch:4}

We find solutions of the heterotic string effective action describing the first-order $\alpha'$ corrections to two-charge black holes at finite temperature. Making explicit use of these solutions, we compute the corrections to the thermodynamic quantities: temperature, chemical potentials, mass, charges and entropy. We check that the first law of black hole mechanics is satisfied and that the thermodynamics agrees with the one extracted from the Euclidean on-shell action. Finally, we show that our results are in agreement with the corrections for the thermodynamics recently predicted by Chen, Maldacena and Witten.

\section{Introduction}

The study of two-charge black holes has attracted much attention since the first investigations of black holes in string theory. This is mainly due to the fact that they are supposed to describe perhaps the simplest configuration in string theory which has a non-vanishing degeneracy of BPS states. This microscopic system consists of a fundamental heterotic string with winding $Q_{w}$ and momentum $Q_p$ charges along a compact direction ${\mathbb S}^{1}_{y}$. The degeneracy of BPS states of this system was computed by Dabholkar and Harvey  in \cite{Dabholkar:1989jt, Dabholkar:1990yf}, and it is given by 
\begin{equation}
	S_{\rm{micro}}(Q_p, Q_w)=\log d\left(Q_p, Q_w\right)=4\pi \sqrt{Q_p Q_w}\,.
\end{equation}
Being a BPS degeneracy, it must be protected when extrapolating it to the finite string-coupling regime where an effective black hole description is expected to exist (a priori). In other words, it should be possible to match this BPS degeneracy with the Bekenstein-Hawking entropy of the corresponding black hole. However, when trying to do so one finds a puzzle: even though there is a supergravity solution with the same charges and preserving the same supersymmetries as the Dabholkar-Harvey states \cite{Sen:1994eb, Cvetic:1995uj, Dabholkar:1995nc, Callan:1995hn}, it describes a singular black hole with vanishing horizon area. Hence, the naive macroscopic entropy that can be associated to the two-charge system vanishes. 

In order to explain this mismatch, Sen proposed in \cite{Sen:1995in} that two-charge black holes have a small horizon of string size, which, therefore, cannot be resolved by supergravity unless the latter is supplemented with higher-derivative terms capturing stringy $\alpha'$ corrections.\footnote{For this reason, these black holes are often referred to as \emph{small} black holes.} Almost ten years after this proposal, it was claimed in \cite{Dabholkar:2004yr, Dabholkar:2004dq} that four-derivative corrections in the context of type IIA on ${\mathbb K}_{3}\times {\mathbb T}^2$ (which is dual to heterotic on ${\mathbb T}^6$) stretch the horizon of two-charge black holes (hiding the singularity behind) and, what is even more remarkable, also give the precise contribution to the black hole entropy so that it reproduces the microstate counting of the two-charge system. 

These results, however, have been recently questioned in a series of papers \cite{Cano:2018hut, Ruiperez:2020qda, Cano:2021dyy} in which, working directly within the heterotic theory, it has been shown that $\alpha'$ corrections do not remove the singularity of BPS two-charge black holes. Furthermore, it has been argued that the configuration studied in \cite{Dabholkar:2004yr, Dabholkar:2004dq} should correspond to a regular four-dimensional black hole whose entropy accidentally matches the microscopic degeneracy of the two-charge system, but which carries different charges and preserves less supersymmetry. The fact that it preserves less supersymmetry is indeed the smoking gun of the presence of additional sources (NS5 branes and Kaluza-Klein monopoles), which would be the ultimate reason explaining why this four-dimensional black hole has a regular horizon. 

The fact that the two-charge system does not seem to admit a black hole description in the BPS limit is something that appears rather natural from the point of view of the correspondence between black holes and fundamental strings \cite{Susskind:1993ws, Horowitz:1996nw, Horowitz:1997jc, Damour:1999aw} (see also \cite{Chen:2021emg, Chen:2021dsw, Brustein:2021cza, Matsuo:2022kvx, Balthazar:2022hno, Ceplak:2023afb} for recent discussions). According to this proposal, black holes should turn into highly-excited strings when their sizes are of the order of the string scale. This has been recently discussed by Chen, Maldacena and Witten in \cite{Chen:2021dsw} precisely in the context of the two-charge system. Let us consider a two-charge black hole at finite temperature. It can be described in supergravity by a solution with a large (macroscopic) horizon. However, if the black hole starts losing its mass it will reach the string size before reaching extremality, which would imply that the right description of the system near extremality should be a sort of self-gravitating string solution \cite{Horowitz:1997jc, Damour:1999aw} rather than a solution with a horizon \cite{Chen:2021dsw}.\footnote{See also \cite{Mathur:2018tib} for a complementary point of view on this.}

In this chapter we will mainly focus on two-charge black holes at finite temperature.
More concretely, we consider two-charge black holes in heterotic string theory and we study how the first-order $\alpha'$ corrections modify the solutions and their thermodynamic properties. The corrections to the thermodynamics have been recently studied in \cite{Chen:2021dsw}, exploiting the fact that the two-charge solutions can be obtained by perfoming suitable $O(2, 2)$ transformations to the Schwarzschild-Tangherlini solution, whose $\alpha'$ corrections had been already studied in \cite{Callan:1988hs}. In principle this method can be used not only to obtain the corrections to the thermodynamics but also the corrected solutions themselves, which were not provided in \cite{Chen:2021dsw}. This is just technically more involved, as one would have to take into account that the  $O(2, 2)$ transformations receive $\alpha'$ corrections \cite{Bergshoeff:1995cg, Kaloper:1997ux, Bedoya:2014pma, Ortin:2020xdm, Eloy:2020dko}.\footnote{As explained in \cite{Chen:2021dsw}, one can ignore the explicit corrections to the $O(2, 2)$ transformations if the goal is just to obtain the corrected thermodynamics.} This was precisely the strategy followed in \cite{Giveon:2009da}. However, as pointed out in \cite{Chen:2021dsw}, the corrected thermodynamics obtained in these two references do not agree within each other. Our main motivation here is to perform an independent “first principles” computation of the corrected solutions and their thermodynamics; we are going to find the corrected solutions explicitly by solving the $\alpha'$-corrected equations of motion and then compute the thermodynamic quantities with standard methods.

Anticipating our results, we are going to show that the $\alpha'$ corrections to the thermodynamics that we compute fully agree with those of \cite{Chen:2021dsw}. This is a strong consistency check of both approaches, as well as of the methods employed and of the results obtained in previous related works by two of the authors and collaborators, see e.g.~\cite{Cano:2018hut, Ruiperez:2020qda, Cano:2021dyy, Cano:2018qev, Chimento:2018kop, Cano:2018brq, RuiperezVicente:2020qfw, Elgood:2020nls, Elgood:2020xwu, Cano:2021nzo, Ortin:2021win, Cano:2022tmn, Zatti:2023oiq} and references therein. In particular, we want to emphasize that the (singular) solutions found in \cite{Cano:2018hut, Ruiperez:2020qda, Cano:2021dyy} are properly recovered from the non-extremal ones we have found in this chapter after taking the extremal limit in a suitable form. This is discussed in subsection~\ref{sec:BPS_limit} and  further confirms the conclusions of \cite{Cano:2018hut, Ruiperez:2020qda, Cano:2021dyy}, yet from a different perspective.

The organization of the rest of the chapter is the following. In section~\ref{sec:solutions} we review the two-derivative solutions describing heterotic two-charge black holes in arbitrary dimension ($4\le d\le 9$) and then provide the details about the $\alpha'$-corrected solutions, focusing on the four- and five-dimensional cases. In section~\ref{sec:BH_thermodynamics} we compute the thermodynamic quantities of the solutions and express them using two-different parametrizations: fixing the value of the mass and charges (micro-canonical ensemble) and fixing the inverse temperature and the chemical potentials (grand-canonical ensemble). We show that the results we get are consistent with the first law of black hole mechanics. Then in section~\ref{sec:chargesfromI} we corroborate the results of section~\ref{sec:BH_thermodynamics} by employing an alternative method to compute the corrected thermodynamics, namely from the Euclidean on-shell action. Finally, in section~\ref{sec:2QsBHfromST} we compare our results with those of \cite{Chen:2021dsw} finding that they are in perfect agreement. The appendices contain additional information on the procedure followed to find the corrected solutions in app.~\ref{app:correctedsol}.

\section{$\alpha'$ corrections to heterotic two-charge black holes}\label{sec:solutions}

\subsection{Two-charge black holes at leading order in $\alpha'$}

Let us begin by reviewing the two-derivative solution describing non-extremal two-charge black holes in $d$ dimensions \cite{Horowitz:1996nw}. Given that in subsection~\ref{sec:alpha_corrected_BHs} we will solve the corrected ten-dimensional equations of motion, here we directly present the solution in its ten-dimensional form. However, since the solutions have a ${\mathbb T}^{(9-d)}$ torus playing a trivial role, we feel free to ignore these torus directions from now on.\footnote{Taking them into account just amounts to add the flat metric on the torus $-d{\vec z}^2_{(9-d)}$ to the $(d+1)$-dimensional metric \eqref{eq:metric}.} Doing so, the resulting $(d+1)$-dimensional solution is given by 
\begin{eqnarray}
	\label{eq:metric}
	{\diff}{\hat s}^2&=&\frac{f}{\zp \zm}\,{\diff}t^2-f^{-1}{\diff}\rho^2-\rho^2 {\diff}\Omega^2_{(d-2)}-k^2_{\infty}\frac{\zp}{\zm}\left({\diff}y+\beta_pk^{-1}_{\infty}\left(\zp^{-1}-1\right){\diff}t\right)^2\,,\\
	{\hat B}&=&\beta_w k_{\infty}\left(\zm^{-1}-1\right) {\diff}t \wedge {\diff}y\,, \\
	e^{2\hat \phi}&=&e^{2\hat \phi_{\infty}}\zm^{-1}\,,
\end{eqnarray}
where $\diff{\hat s}$ represents the line element  in the string frame and
\begin{equation}
	\zp=1+\frac{q_{p}}{\rho^{d-3}}\, , \hspace{1cm}\zm=1+\frac{q_{w}}{\rho^{d-3}}\, , \hspace{1cm} f=1-\frac{\rho_s^{d-3}}{\rho^{d-3}}\, .
\end{equation} 
The parameters $q_p$, $q_w$ and $\rho_s$ are related to the charges and mass of the solutions. Together with the moduli ${\hat \phi}_{\infty}$ and $k_{\infty}$  (representing the asymptotic values of the dilaton and the Kaluza-Klein scalar), they constitute the set of independent parameters of the solutions since $\beta_{p}$ and $\beta_{w}$ are subject to the following constraints,
\begin{equation}
	\rho_s^{(d-3)}=q_{p}\left(\beta^2_{p}-1\right)=q_{w}\left(\beta^2_{w}-1\right)\, ,
\end{equation}
implying that
\begin{equation}
	\beta_{i}=\epsilon_{i}\sqrt{1+ \frac{\rho_s^{d-3}}{q_{i}}}\, , \hspace{1cm} i=\left\{p, w\right\}\, .
\end{equation}
where $\epsilon^2_{i}=1$.  These correspond to the signs of the winding and momentum charges, respectively. In the BPS limit ($\rho_s\to 0$), the solution with $\epsilon_{w}=\epsilon_{p}$ is supersymmetric, while the one with $\epsilon_{w}=-\epsilon_{p}$ does not preserve any supersymmetry. The analysis of the Killing spinor equations for these configurations can be found for instance in \cite{Ruiperez:2020qda, Cano:2021dyy, Cano:2021nzo}. Notice that this solution in $d=5$ is a particular case of (\ref{eq:3-charge10dsolutionzerothorder}) with the identification
\begin{equation}\label{eq:map1}
	q_p = q_+ \,, \quad q_- = q_w \,, \quad q_0 = 0 \,, \quad \omega = - \rho_s^2 \,,
\end{equation}
instead in $d=4$ we obtain a particular case of (\ref{eq:anatz4d}) with the identification \begin{equation}\label{eq:map2}
	q_p = q_+ \,, \quad q_- = q_w \,, \quad q_0 = q_{\mathcal{H}} =  0 \,, \quad \omega = - \rho_s \,.
\end{equation}

\subsection{$\alpha'$-corrected solutions}
\label{sec:alpha_corrected_BHs}

Our aim now is to compute the first-order $\alpha'$ corrections to these two-charge black holes. As usual, we treat the $\alpha'$ corrections in a perturbative fashion and ignore ${\cal O}(\alpha'^2)$ terms. The first-order $\alpha'$ corrections in the effective action of the heterotic superstring were studied in \cite{Gross:1986mw, Metsaev:1987zx, Bergshoeff:1989de}. While different approaches were used, it was later shown in \cite{Chemissany:2007he} that the resulting effective actions are equivalent up to field redefinitions. Here we choose to work in the Bergshoeff-de Roo scheme, \cite{Bergshoeff:1989de}. Our conventions are reviewed in section \ref{sec-HST}.

Before entering into the details of the corrected solutions,  the strategy we have followed in order to find the corrected solutions is the same of presented in section \ref{sec-SolutionConstruction}. The interested reader is referred to appendix~\ref{app:correctedsol} form more details.  It turns out that an educated ansatz to solve the corrected equations of motion is the following,
\begin{subequations} \label{eq:ansatzAlej}
	\begin{align}
		\label{eq:ansatz_metric}
		\begin{split}
			{\diff}{\hat s}^2&=\frac{f}{\zp  {\tilde f}_{w}}{\diff}t^2-\zz\left(f^{-1}{\diff}\rho^2+\rho^2 {\diff}\Omega^2_{(d-2)}\right)  \\[1mm]				
			& \quad -k^2_{\infty}\frac{\zp}{ {\tilde f}_{w}}\left[{\diff}z+\beta_{p}k^{-1}_{\infty}\left(\zp^{-1}-1\right){\diff}t\right]^2\,, 
		\end{split} \\[4mm]
		\label{eq:ansatz_B}
		{\hat B}&=\beta_w k_{\infty}\left({f}_{w}^{-1}-1\right) {\diff}t \wedge {\diff}z\,, \\[4mm]
		e^{-2\hat{\phi} }&= - \frac{(d-3) c_{\hat{\phi}}}{\rho^{d-2} f'_w} \left(\frac{f_w}{\tilde{f}_w}\right)^2 \tilde{f}_w \, g^{-(d-3)/{2}}\,,
	\end{align}
\end{subequations}
where the functions  $f, f_p, f_w, {\tilde f}_w, g$ and the dilaton $\hat \phi$ are assumed to depend only on the radial coordinate $\rho$. The ansatz for the dilaton has been obtained solving the KR field equations of motion. For consistency with the perturbative approach, they must be of the form
\begin{equation}
	\begin{aligned}
		{f}_{p}=\,&1+\frac{q_p}{\rho^{d-3}}+\alpha' {\delta f_p}\, , \hspace{5mm} {\tilde f}_{w}=\,1+\frac{q_w}{\rho^{d-3}}+\alpha' {\delta {\tilde f}_{w}}\, ,\hspace{5mm}\zz=\,1+\alpha' {\delta \zz}\, ,\\[1mm]
		{f}=&\,1-\frac{\rho_s^{d-3}}{\rho^{d-3}}+\alpha' {\delta f}\, ,\hspace{5mm} {f}_{w}=\,1+\frac{q_w}{\rho^{d-3}}+\alpha' {\delta {f}_{w}}\, .
	\end{aligned}
\end{equation}
Notice that the structure of the ansatz is slightly different from that of (\ref{eq:3-charge10dsolution-04}) and (\ref{eq021}). Therefore, there is no simple match among the unknown functions corrections. 

After linearization in $\alpha'$, the equations of motion boil down to a linear system of inhomogeneous second-order ODEs for the unknown functions $\delta f_p, \delta {\tilde f}_w, \delta g, \delta f, \delta f_w$ and $\hat \phi$. The strategy we are going to follow to solve them is the same as in \cite{Cano:2022tmn}, which consists of performing an asymptotic expansion (large $\rho$) of the unknown functions and solving the equations of motion order by order. Following this procedure, we can determine all the coefficients of the asymptotic expansion except for a few of them which remain free: the integration constants.  Once the form of the asymptotic solution has been found, we resum the asymptotic series with the help of \texttt{Mathematica}. The final step is to fix the integration constants by imposing regularity at the horizon and suitable boundary conditions. Our choice here will be such that we keep the asymptotic charges and the mass fixed: \textit{i.e.}, we are going to give the form of the corrected solution in the micro-canonical ensemble.

In what follows we give the corrected solutions in $d=5$ and $d=4$, as well as its dimensional reduction on ${\mathbb S}^{1}_{y}$. Finally, we study their BPS limits and check that they agree with the corrected solutions found in \cite{Cano:2018hut, Ruiperez:2020qda, Cano:2021dyy}.

\subsubsection{Five-dimensional black holes}
Let us first consider the $d=5$ case. Imposing that the asymptotic value of the string coupling is not renormalized, namely
\begin{equation}
	\lim_{\rho\to\infty}{\hat \phi}={\hat \phi}_{\infty}\, .
\end{equation}
we obtain $c_{\hat{\phi}} = q_w e^{-2 \hat{\phi}_\infty}$. After fixing the integration constants in the way we have explained, we find the following solution:
\begin{eqnarray}
	\delta \zp&=&-\frac{q_p \rho _s^2 \left(1+\frac{\beta_w}{\beta_p}\right) \log \left(1+\frac{q_w}{\rho ^2}\right)}{2 q_w^2 \rho ^2  }\nonumber\\
	&+&\frac{1}{32 q_w \left(\rho ^2+q_w\right) \rho ^6 }\left\{16 q_p \rho_s^2 \rho ^4-q_w q_p \rho ^2 \left(9 \rho _s^2+32 q_w\right)+7 q_w^2 q_p \rho_s^2 \right.\nonumber\\
	&+&\left.\frac{\beta_w}{\beta_p}\left[16 q_p \rho_s^2 \rho ^4+8 q_w \rho ^2 \left(q_p \rho_s^2-2 q_w \left(\rho_s^2+2 q_p\right)\right)+8 q_w^2 q_p \rho_s^2\right]\right\}\,,\\[4mm]
	\delta {\tilde f}_w&=&-\frac{\rho_s^2 \left(1+\frac{\beta_w}{\beta_p}\right) \log \left(1+\frac{q_w}{\rho ^2}\right)}{2 q_w\rho ^2 }+\frac{\beta_w\rho_s^2 \left(q_w+2 \rho^2\right) }{4\beta_p\rho^6}+\frac{7 q_w \rho_s^2}{32\rho^6}+\frac{\rho_s^2}{2 \rho^4}\nonumber\\
	&+& \frac{9\rho_s^2 (q_p-q_w)}{4 \rho^2 \left(4 \rho_s^2 (q_w+q_p)+4 q_wq_p+3 \rho_s^4\right)}\,,\\[4mm]
	\delta \zz&=&\frac{\rho_s^2 \left(1+\frac{\beta_w}{\beta_p}\right) \log \left(1+\frac{q_w}{\rho ^2}\right)}{2 q_w^2}+\frac{\beta_w\rho _s^2 \left(q_w-2 \rho ^2\right)}{4\beta_p q_w \rho ^4 }-\frac{7\rho_s^2}{32 \rho^4}\nonumber\\
	&+&\frac{1}{8 \rho^2} \left(-\frac{4 \rho_s^2}{q_w}+\frac{18 q_p \left(\rho_s^2+2 q_w\right)}{4 \left(q_w+q_p\right) \rho_s^2+3 \rho_s^4+4 q_w q_p}-9\right)\,,\\[4mm]
	\delta f&=&\frac{\rho_s^4 \left(1+\frac{\beta_w}{\beta_p}\right) \log \left(1+\frac{q_w}{\rho ^2}\right)}{2 q_w^2\rho ^2 }-\frac{\beta_w \rho_s^2 \left(2 \rho_s^2 \rho ^4+q_w \rho_s^2 \rho ^2+q_w^2 \left(3 \rho_s^2-4 \rho ^2\right)\right)}{4q_w \beta_p\rho^6\left(\rho ^2+q_w\right) }\nonumber\\
	&+&\frac{\rho_s^2q_w }{\left(\rho ^2+q_w\right)\rho^4}-\frac{3 \left(\rho ^2+q_p\right)}{4 \rho^4}+\frac{\rho_s^4 \left(-16 \rho ^4+9 q_w \rho ^2-7 q_w^2\right)}{32q_w \rho^6\left(\rho ^2+q_w\right)}\nonumber\\
	&+&\frac{ \left(\rho ^2+q_p\right) \left(2 q_w q_p-\left(q_w-2 q_p\right) \rho_s^2\right)}{2\rho^4\left(4 \left(q_w+q_p\right) \rho_s^2+3 \rho_s^4+4 q_w q_p\right)}\, ,\\[4mm]
	\delta {f}_{w}&=&-\frac{\rho_s^2 \left(1+\frac{\beta_w}{\beta_p}\right) \log \left(1+\frac{q_w}{\rho ^2}\right)}{2 q_w\rho ^2 }-\frac{\beta_w\rho_s^2 \left(q_w-2 \rho ^2\right)}{4\beta_p  \rho ^6} -\frac{q_w\rho_s^2}{32 \rho^6}\nonumber\\
	&+&\frac{\rho_s^2 \left(4+\frac{18 q_w \left(q_w-q_p\right)}{4 \left(q_w+q_p\right) \rho_s^2+3 \rho_s^4+4 q_w q_p} \right)}{8 \rho^4}\, ,
\end{eqnarray}

\subsubsection{Four-dimensional black holes}

We proceed as in the five-dimensional case. We still have  $c_{\hat{\phi}} = q_w e^{-2 \hat{\phi}_\infty} $. The solutions is then: 
\begin{eqnarray}
	\delta \zp &=& \frac{\beta_p\beta_wq_p \left[q_p \left(4 q_w^3 (\rho_s-3 \rho)+\rho_s q_w^2 \rho-3 \rho_s q_w \rho^2-6 \rho_s \rho^3\right)-6 \rho_s q_w^3 \rho\right]}{48 q_w^2 \rho^4 (q_p+\rho_s) (q_w+\rho)} \nonumber\\
	&-& \frac{q_p \left[q_w^3 (39 \rho-10 \rho_s)+q_w^2 \rho (17 \rho_s+3 \rho)+9 \rho_s q_w \rho^2+18 \rho_s \rho^3\right]}{144 q_w^2 \rho^4 (q_w+\rho)}\nonumber\\
	&+& \log \left(1+\frac{q_w}{\rho}\right) \left(\frac{\beta_p\beta_wq_p^2 \rho_s}{8 q_w^3 \rho (q_p+\rho_s)}+\frac{q_p \rho_s}{8 q_w^3 \rho}\right) \, ,\\[4mm]
	\delta {\tilde f}_w &=& \frac{\beta_p\beta_wq_p \rho_s \left(4 q_w^2+3 q_w \rho-6 \rho^2\right)}{48 q_w \rho^4 (q_p+\rho_s)} + \frac{q_w^2 (10 \rho_s-3 \rho)+9 \rho_s q_w \rho-18 \rho_s \rho^2}{144 q_w \rho^4} \nonumber\\
	&+& \frac{q_p-q_w}{\rho \left(4 q_p q_w +3 q_p \rho_s +3 \rho_s q_w +2 \rho_s^2 \right)} \nonumber\\
	&+& \log \left(1+\frac{q_w}{\rho}\right)\left(\frac{\beta_p\beta_wq_p \rho_s}{8 q_w^2 \rho (q_p+\rho_s)}+\frac{\rho_s}{8 q_w^2 \rho}\right) \,,\\[4mm]
	\delta \zz &=& \frac{\beta_p\beta_wq_p \rho_s \left(2 q_w^2-3 q_w \rho+6 \rho^2\right)}{24 q_w^2 \rho^3 (q_p+\rho_s)} -\frac{\rho_s+3 q_w}{8 q_w \rho^2}-\frac{5 \rho_s}{36 \rho^3} \nonumber\\
	&-& \frac{(\rho_s+2 q_w) \left(-3 q_p \rho_s+2 q_p q_w-2 \rho_s^2+6 q_w^2+\rho_s q_w\right)}{4 q_w^2 \rho \left(3 q_p \rho_s+4 q_p q_w+2 \rho_s^2+3 \rho_s q_w\right)}  \nonumber\\
	&-& \log \left(1+\frac{q_w}{\rho}\right) \left(\frac{\beta_p\beta_wq_p \rho_s}{4 q_w^3 (q_p+\rho_s)}+\frac{\rho_s}{4 q_w^3}\right) \,,\\[4mm]
	\delta f &=& \frac{\beta_p\beta_wq_p \rho_s \left(2 q_w^3 (6 \rho-5 \rho_s)-\rho_s q_w^2 \rho+3 \rho_s q_w \rho^2+6 \rho_s \rho^3\right)}{48 q_w^2 \rho^4 (q_p+\rho_s) (q_w+\rho)} \nonumber\\
	&-& \frac{40 q_p \rho_s^2 q_w^4+30 \rho_s^3 q_w^3 (q_p+q_w)+20 \rho_s^4 q_w^3+\rho^4 \left(288 q_w^3+144 \rho_s q_w^2\right)}{144 q_w^2 \rho^4 (q_w+\rho) \left[3 \rho_s (q_p+q_w)+4 q_p q_w+2 \rho_s^2\right]} \nonumber\\
	&+& \frac{\rho^3 \left(-288 q_w^3 (q_p+q_w)-144 \rho_s q_w^2 (q_p+q_w)+54 \rho_s^3 (q_p+q_w)+72 q_p \rho_s^2 q_w+36 \rho_s^4\right)}{144 q_w^2 \rho^4 (q_w+\rho) \left[3 \rho_s (q_p+q_w)+4 q_p q_w+2 \rho_s^2\right]} \nonumber\\
	&+& \frac{\rho^2 \left[9 \rho_s^2 q_w^2 (5 q_p+q_w)+3 \rho_s^3 q_w (9 q_p+11 q_w)+18 \rho_s^4 q_w -288 q_p q_w^4-132 q_p \rho_s q_w^3 \right]}{144 q_w^2 \rho^4 (q_w+\rho) \left[3 \rho_s (q_p+q_w)+4 q_p q_w+2 \rho_s^2\right]} \nonumber\\
	&+&  \frac{\rho \left[156 q_p \rho_s q_w^4+\rho_s^2 q_w^3 (185 q_p+117 q_w)+3 \rho_s^3 q_w^2 (17 q_p+43 q_w)+34 \rho_s^4 q_w^2\right]}{144 q_w^2 \rho^4 (q_w+\rho) \left[3 \rho_s (q_p+q_w)+4 q_p q_w+2 \rho_s^2\right]} \nonumber\\
	&-& \log \left(1+\frac{q_w}{\rho}\right) \left(\frac{\beta_p\beta_wq_p \rho_s^2}{8 q_w^3 \rho (q_p+\rho_s)}+\frac{\rho_s^2}{8 q_w^3 \rho}\right) \, ,\\[4mm]
	\delta {f}_{w} &=& -\frac{\beta_p\beta_wq_p \rho_s \left(2 q_w^2-3 q_w \rho+6 \rho^2\right)}{48 q_w \rho^4 (q_p+\rho_s)}+\frac{q_w (q_w-q_p)}{\rho^2 \left(3 \rho_s q_p+ 3 \rho_s q_w +4 q_p q_w+2 \rho_s^2\right)} \nonumber\\
	&-& \frac{\rho_s \left(8 q_w^2-9 q_w \rho+18 \rho^2\right)}{144 q_w \rho^4}+\frac{q_w}{48 \rho^3}  \nonumber\\
	&+& \log \left(1+\frac{q_w}{\rho}\right) \left(\frac{\beta_p\beta_wq_p \rho_s}{8 q_w^2 \rho (q_p+\rho_s)}+\frac{\rho_s}{8 q_w^2 \rho}\right) \,. 
\end{eqnarray}

\subsection{Dimensional reduction on ${\mathbb S}^1_y$}

The dimensional reduction of this solution to five dimensions can be carried out using the formulae obtained in \cite{Elgood:2020xwu}, which are collected in appendix~\ref{sec-dictionary}. Applying them to the configuration at hands, we get
\begin{eqnarray}
	\diff s^2&=\,&\frac{f}{\zp  {\tilde f}_{w}}{\diff}t^2-\zz\left(f^{-1}{\diff}\rho^2+\rho^2 {\diff}\Omega^2_{(d-2)}\right)\, ,\\[1mm]
	B&=\,&0\,,\\[1mm]
	A&=\,&\beta_{p}\, k^{-1}_{\infty}\left(\zp^{-1}-1\right){\diff}t\,,\hspace{1cm} C=\,\beta_{w}\, k_{\infty}\left[\zm^{-1}\left(1+\alpha' \beta_w^{-1}\Delta_C \right)-1\right]{\diff}t\,,\\[1mm]
	e^{2\phi}&=&e^{2\hat\phi}\,k^{-1}_{\infty} \left(\frac{{\tilde f}_w}{f_p}\right)^{1/2}   \,, \hspace{1cm} k=k_{\infty}\left(\frac{f_p}{{\tilde f}_w}\right)^{1/2}\, ,
\end{eqnarray}
where $\diff s$ represents the line element in the string frame and 
\begin{equation}\label{eq:DeltaC}
	\Delta_{C}=\frac{2\left(\beta_p+\beta_w\right)f f'_p f'_w-f'\left(\beta_w f_p f'_w+\beta_p f'_p f_w\right) }{8f_p f_w} .
\end{equation}

\subsection{Extremal limit}\label{sec:BPS_limit}

The extremal limit is implemented by setting $\rho_s\to 0$ while keeping the charge parameters $q_p$ and $q_w$ fixed. The $\alpha'$ corrections in this limit have been already studied in the recent literature \cite{Cano:2018hut, Ruiperez:2020qda, Cano:2021dyy}. The corrected solution in arbitrary dimension is given by \cite{Ruiperez:2020qda, Cano:2021dyy}:\footnote{Here we are focusing on the supersymmetric case $\epsilon_{p}=\epsilon_{w}$, which was the case analyzed in \cite{Cano:2018hut, Ruiperez:2020qda, Cano:2021dyy}. Surprisingly, in the non-supersymmetric case $\epsilon_{p}=-\epsilon_{w}$ the corrections simply vanish as the first-order correction in \eqref{eq:fp} is multiplied by $1+\epsilon_p \epsilon_w$.}
\begin{eqnarray}
	f&=&g=1\,,\\[1mm]
	\label{eq:fp}
	f_{p}&=&1+\frac{q_p}{\rho^{d-3}}-\frac{(d-3)^2\alpha'}{2} \frac{q_p q_w}{\rho^{d-1}\left(\rho^{d-3}+q_w\right)}\,,\\[1mm]
	{\tilde f}_{w}&=&f_{w}=1+\frac{q_w}{\rho^{d-3}}\,.
\end{eqnarray}
We have checked that this solution is precisely recovered from the non-extremal ones we have presented presented in subsection~\ref{sec:alpha_corrected_BHs} upon taking $\rho_s\to 0$.\footnote{While the limit is smooth in the five-dimensional case, in the four-dimensional one it must be taken before fixing the integration constants, as the expressions for the latter (which we have not provided explicitly) diverge when $\rho_s\to0$.} This is an interesting consistency check of our solutions.

\section{Black hole thermodynamics}\label{sec:BH_thermodynamics}

In this section we compute the thermodynamic quantities of the $\alpha'$-corrected solutions found in the previous section. We use the same methods of \ref{sec-thermodynamics}. Then, we apply the corresponding formulae to the five- and four-dimensional solutions. 

\subsection{Generalities}

In order to facilitate the comparison with the previous literature \cite{Giveon:2009da, Chen:2021dsw} (which is something that we will do later in section~\ref{sec:2QsBHfromST}), we introduce the notation which is used in the aforementioned references:
\begin{equation}
	q_{i}=\rho_s^{d-3} \sinh^2\gamma_i\,, \hspace{1cm} i=\left\{p, w\right\}\, .
\end{equation}
In addition, we will write down the different expressions both in the micro- and grand-canonical ensembles. By definition, the first is the one in which the expressions for the mass $E$ and charges, $Q_{p}$ and $Q_{w}$, take the same form as in the two-derivative solution (the solution in section~\ref{sec:alpha_corrected_BHs} is given using this parametrization). In turn, what is fixed in the latter ensemble are the inverse temperature $\beta$ and the chemical potentials, $\Phi_{p}$ and $\Phi_{w}$.

The inverse temperature $\beta$ is given by
\begin{equation}
	\beta= 4\pi \frac{\sqrt{g f_p {\tilde f}_{w}}}{f'}\Bigg|_{\rho=\rho_h}\, ,
\end{equation}
where $\rho_h$ is the position of the outer horizon. The latter corresponds to the (largest, finite) root of the metric function $f$, $f(\rho_h)=0$. As a consequence of our choice of boundary conditions the position of the horizon $\rho_h$ is shifted by the $\alpha'$ corrections.	We have two gauge vectors electrically charged. The chemical potentials associated to the Kaluza-Klein and winding vectors are
\begin{subequations}\label{eq:chem_pot}
	\begin{align}
		\Phi_{p} & =\xi^{\mu}A_{\mu}|_{\infty}-\xi^{\mu}A_{\mu}|_{\rho=\rho_h}\,, \\[2mm] \Phi_{w} & =\xi^{\mu}C_{\mu}|_{\infty}-\xi^{\mu}C_{\mu}|_{\rho=\rho_h}\,,
	\end{align}
\end{subequations}
where $\xi=\partial_t$ is the Killing vector that generates the horizon. The associated charges are 
\begin{subequations}
	\begin{align}
		Q_{p} & = \frac{1}{16\pi G_N}\int_{{\mathbb S}^{(d-2)}_{\infty}}e^{-2\left(\phi-\phi_{\infty}\right)}k_{(1)}^2 \star F \,, \\[2mm]
		Q_{w} & =\frac{1}{16\pi G_N}\int_{{\mathbb S}^{(d-2)}_{\infty}}e^{-2\left(\phi-\phi_{\infty}\right)}k^{-2} \star G \, ,
	\end{align}
\end{subequations}
where $F=\diff A$, $G=\diff C$, $k_{(1)}$ is the scalar combination given in \eqref{eq:scalarcombination} and $G_N$ is the $d$-dimensional Newton constant,
\begin{equation}
	G_N=\frac{\hat G_N}{2\pi R_{y}}=\frac{\hat G_N}{2\pi k_{\infty}\ell_s}\, ,
\end{equation}
being $\hat G_N$ the $(d+1)$-dimensional one. The rest of the thermodynamic quantities are computed exactly as described in sections (\ref{sec-thermodynamics}). 

\subsection{Thermodynamic quantities in the micro-canonical ensemble}

\subsubsection{Five-dimensional black holes.}
By definition, the expressions for the charges are the same as in the two-derivative solution, namely
\begin{eqnarray}
	Q_{p}&=&\frac{\epsilon_{p} k_{\infty} \pi}{8G_N} \rho_s^2 \sinh \left(2\gamma_p\right)\,, \\[1mm]
	Q_{w}&=&\frac{\epsilon_{w} \pi}{8G_N k_{\infty}} \rho_s^2 \sinh \left(2\gamma_w\right)\,, \\[1mm]
	E&=&\frac{\pi \rho _s^2}{8G_N} \left[1+\cosh \left(2 \gamma _p\right)+\cosh \left(2 \gamma
	_w\right)\right]\, .
\end{eqnarray}
Contrarily to the charges, the inverse temperature \eqref{eq:beta} and the chemical potentials \eqref{eq:chem_pot} receive $\alpha'$ corrections. Parametrizing them as follows,
\begin{eqnarray}
	\beta&=&2\pi  \cosh \gamma_p \cosh\gamma_w \rho_s \left(1+\frac{\alpha' \Delta\beta}{\rho_s^2}\,\right)\,,\\[1mm]
	\Phi_{p}&=&\frac{\epsilon_{p}\tanh\gamma_p}{k_{\infty}}\left(1+\frac{\alpha' \Delta \Phi_{p}}{\rho_s^2}\right)\,,\\[1mm]
	\Phi_{w}&=&\epsilon_{w}k_{\infty}\tanh\gamma_w\left(1+\frac{\alpha' \Delta \Phi_{w}}{\rho_s^2}\right)\,,
\end{eqnarray}
we get 
\begin{eqnarray}
	\Delta \beta &=\,&-\frac{\epsilon_{p}\epsilon_w}{2}\tanh\gamma_p \tanh\gamma_w-\,\frac{9\left(4\sinh^2\gamma_p \sinh^2\gamma_w-1\right)}{8\left(4\cosh^2\gamma_p \cosh^2\gamma_w-1\right)}\,, \\[1mm]
	\Delta\Phi_{p}&=\,&-\frac{\epsilon_{p}\epsilon_w \tanh \gamma_w}{\sinh \left(2\gamma_p\right)}-\frac{9\cosh\left(2\gamma_w\right)}{4\left(4\cosh^2\gamma_p \cosh^2\gamma_w-1\right)}\,, \\[1mm]
	\Delta\Phi_{w}&=\,& -\frac{\epsilon_{p}\epsilon_w \tanh \gamma_p}{\sinh \left(2\gamma_w\right)}-\frac{9\cosh\left(2\gamma_p\right)}{4\left(4\cosh^2\gamma_p \cosh^2\gamma_w-1\right)}\,. 
\end{eqnarray}
Finally, the result that we obtain for the black hole entropy is
\begin{equation}
	S=\frac{\pi^2 \rho_s^3\cosh\gamma_p \cosh\gamma_w }{2G_N}\left[1+\frac{\alpha'}{8\rho_s^2}\left(9+4\epsilon_{p}\epsilon_{w}\tanh \gamma_p \tanh \gamma_w\right)\right]\, .
\end{equation}
These expressions pass several consistency checks. First, one can verify that the first law of black-hole mechanics,
\begin{equation}\label{eq:first_law}
	\diff E=\beta^{-1} \diff S+ \Phi_p \,\diff Q_p + \Phi_w \,\diff Q_w\, ,
\end{equation}
is obeyed. Second, the corrections agree with those of \cite{Cano:2022tmn}, where three-charge black holes were considered, in the limit in which the third charge, associated to the presence of NS5 branes, goes  to zero. Finally, the expressions are consistent with T-duality, which exchanges $\gamma_p \leftrightarrow \gamma_w$ and sends $k_{\infty}\to 1/k_{\infty}$. One can see that the mass, entropy and temperature are left invariant, whereas the chemical potentials and charges are interchanged, as expected. 

\subsubsection{Four-dimensional black holes.}
The expressions for the charges and mass read,
\begin{eqnarray}
	Q_{p}&=& \frac{\epsilon_p k_{\infty} }{8G_N} \rho_s \sinh \left(2\gamma_p\right)\,, \\[1mm]
	Q_{w}&=& \frac{\epsilon_{w} k_{\infty}^{-1} }{8G_N } \rho_s \sinh \left(2\gamma_w\right)\,, \\[1mm]
	E&=&\frac{\rho _s}{8G_N} \left[2+\cosh \left(2 \gamma _p\right)+\cosh \left(2 \gamma
	_w\right)\right]\, .
\end{eqnarray}
The inverse temperature and the chemical potential receive the following $\alpha'$ corrections,
\begin{eqnarray}
	\beta&=&4\pi  \cosh \gamma_p \cosh\gamma_w \rho_s \left(1+\frac{\alpha' \Delta\beta}{\rho_s^2}\right)\,,\\[1mm]
	\Phi_{p}&=& \epsilon_p k_\infty^{-1} \tanh\gamma_p\left(1+\frac{\alpha' \Delta \Phi_{p}}{\rho_s^2}\right)
	\,,\\[1mm]
	\Phi_{w}&=& \epsilon_w k_{\infty}\tanh\gamma_w\left(1+\frac{\alpha' \Delta \Phi_{w}}{\rho_s^2}\right)\,,
\end{eqnarray}
where
\begin{eqnarray}
	\Delta \beta &=\,& \frac{\cosh (2\gamma_p)\left[1-2\cosh(2\gamma_w)\right]+\cosh (2\gamma_w)}{2\cosh (2\gamma_p)\left[1+2\cosh(2\gamma_w)\right]+2 \cosh (2\gamma_w)} - \frac{\epsilon_p \epsilon_w  \tanh \gamma_p \tanh \gamma_w}{8}   \,, \\[1mm]
	\Delta\Phi_{p}&=\,& -\frac{2 \cosh (2\gamma_w)}{\cosh (2\gamma_p)\left[1+2\cosh(2\gamma_w)\right]+ \cosh (2\gamma_w)}- \frac{\epsilon_p \epsilon_w\tanh (\gamma_w)}{4\sinh (2\gamma_p)} \,, \\[1mm]
	\Delta\Phi_{w}&=\,&  -\frac{2 \cosh (2\gamma_p)}{\cosh (2\gamma_p)\left[1+2\cosh(2\gamma_w)\right]+2 \cosh (2\gamma_w)}-\frac{\epsilon_p \epsilon_w\tanh (\gamma_p)}{4\sinh (2\gamma_w)} \,.
\end{eqnarray}
Finally, the expression for the entropy is
\begin{equation}
	S = \frac{\pi \rho_s^2 \cosh \gamma_p \cosh \gamma_w}{G_N}  \left[1+ \frac{\alpha'}{2\rho_s^2}\left(1+ \frac{\epsilon_p \epsilon_{w}\tanh \gamma_p \tanh \gamma_w}{4} \right) \right]\,.
\end{equation}
These corrections agree with those of \cite{Zatti:2023oiq}, where the corrections to a family of four-charge black holes have been computed. As in the five-dimensional case, the thermodynamic quantities we have obtained transform as expected under T-duality and obey the first law of black-hole mechanics \eqref{eq:first_law}.

\subsection{Thermodynamic quantities in the grand-canonical ensemble} 

In order to obtain the thermodynamics in the grand-canonical ensemble, we must consider a different choice of boundary conditions. This can be simply implemented by considering a different parametrization of the solution,
\begin{equation} \label{eqChangeEns}
	\rho_s\to \rho_s+ \alpha' \delta \rho_s\left(\rho_s, \gamma_i\right)\,, \hspace{1cm} \gamma_{i}\to \gamma_{i}+\frac{\alpha' \delta \gamma_{i}\left(\rho_s, \gamma_j\right)}{\rho_s} \,, 
\end{equation}
and fixing $\delta \rho_s$ and $\delta \gamma_i$ by imposing the vanishing of the corrections to $\beta$ and the chemical potentials $\Phi_{i}$. The resulting expressions for the thermodynamic quantities associated to the five- and four-dimensional solutions are given below.

\subsubsection{Five-dimensional black holes.}
\begin{eqnarray}
	\label{eq:chemicalpot_GCensemble}
	\beta&=\,&2\pi \rho_s\cosh \gamma_p\, \cosh \gamma_w\,,\hspace{5mm}\Phi_{p}=\,\frac{\epsilon_{p} \tanh \gamma_p}{k_{\infty}}\,,\hspace{5mm}\Phi_{w}=\,\epsilon_{w}k_{\infty}\tanh \gamma_w\,,\\[1mm]
	\label{eq:Qp}
	Q_{p}&=\,&\frac{\epsilon_{p} k_{\infty}\pi}{8G_N}\left[\rho_s^2\sinh\left(2\gamma_p\right)+\alpha' \epsilon_{p}\epsilon_{w}\tanh \gamma_w\right]\,,\\[1mm]
	\label{eq:Qw}
	Q_{w}&=\,&\frac{\epsilon_{w}\pi}{8G_Nk_{\infty}}\left[\rho_s^2\sinh\left(2\gamma_w\right)+\alpha' \epsilon_{p}\epsilon_{w}\tanh \gamma_p\right]\,,\\[1mm]
	\label{eq:E}
	E&=\,&\frac{\pi \rho _s^2}{8G_N} \left[1+\cosh \left(2 \gamma _p\right)+\cosh \left(2 \gamma
	_w\right)+\frac{\alpha'}{4\rho_s^2} \left(-9+4 \epsilon_{p}\epsilon_{w}\tanh \gamma_p \tanh \gamma_w\right)\right]\,,\\[1mm]
	\label{eq:S}
	S&=\,& \frac{\pi^2\rho_s^3 \cosh \gamma_p \cosh \gamma_w}{2G_N}\,.
\end{eqnarray}

\subsubsection{Four-dimensional black holes}
\begin{eqnarray}
	\beta &= & 4 \pi \rho_s \cosh \gamma_p \cosh \gamma_w \,, \hspace{5mm}\Phi_p =  \epsilon_p k_\infty^{-1} \tanh \gamma_p \,, \hspace{5mm}\Phi_w = \epsilon_w k_\infty \tanh \gamma_w \,, \\[1mm]
	Q_{p}& = & \frac{\epsilon_p k_{\infty}\rho_s \sinh \left(2\gamma_p\right) }{8G_N} \left[1+\frac{\alpha'}{2\rho_s^2}\left(1 + \frac{\epsilon_p \epsilon_{w} \tanh \gamma_w}{4 \tanh \gamma_p }\right)\right]\,, \label{eq:Qp4d} \\[1mm]
	Q_{w}&= & \frac{\epsilon_{w} k_{\infty}^{-1} \rho_s \sinh \left(2\gamma_w\right)}{8G_N} \left[1+\frac{\alpha'}{2\rho_s^2}\left(1 + \frac{\epsilon_p \epsilon_{w} \tanh \gamma_p}{4 \tanh \gamma_w }\right)\right] \label{eq:Qw4d} \,,\\[1mm]
	E\;  &= & \; \;\frac{\rho _s \left(\cosh \left(2 \gamma _p\right)+\cosh \left(2 \gamma_w\right)+2\right)}{8G_N} \left[1+\frac{\alpha'}{2\rho_s^2}\left(\frac{\cosh (2\gamma_p)+\cosh (2\gamma_w)-2 }{\cosh (2\gamma_p)+\cosh (2\gamma_w)+ 2} \right.\right.\nonumber\\[1mm]
	&&\left.\left.+ \frac{\epsilon_p \epsilon_{w}\tanh \gamma_p \tanh \gamma_w}{4}\right)\right]\,, \label{eq:E4d} \\[1mm]
	S  &=&  \frac{\pi \rho_s^2 \cosh \gamma_p \cosh \gamma_w }{G_N} \left[1+ \frac{\alpha'}{2\rho_s^2}\left(1 + \frac{\epsilon_p \epsilon_{w}\tanh \gamma_p \tanh \gamma_w}{4} \right) \right]  \label{eq:S4d} \,.
\end{eqnarray}

\section{Thermodynamics from the Euclidean on-shell action}\label{sec:chargesfromI}

In the saddle-point approximation the Euclidean on-shell action of the black hole gives the dominant contribution to the grand-canonical partition function \cite{Gibbons:1976ue}. This leads to the so-called \textit{quantum statistical relation},
\begin{equation}\label{eq:qsr}
	I_{\infty}=\beta \,{\cal G}=\beta \left(E-\Phi_p \,Q_p-\Phi_w \,Q_w\right)-S\,,
\end{equation}
where $I_{\infty}$ is the renormalized Euclidean on-shell action and ${\cal G}$ is the grand-canonical potential,\footnote{Notice that the definition of ${\cal G}$ is compatible with the Smarr formulas we obtained in the previous chapter if it incorporates the mass and the term containing $\Phi_{\alpha'}$.} which is regarded as a function of the (inverse) temperature and the chemical potentials. Knowing ${\cal G}={\cal G}\left(\beta, \Phi_p, \Phi_w\right)$ suffices to extract all the thermodynamic quantities since the mass, charges and entropy can be obtained (assuming the first law of black hole mechanics) as follows:
\begin{equation}\label{eq:chargesfromI}
	Q_{p}=-\frac{\partial {\cal G}}{\partial {\Phi}_{p}}\,, \hspace{5mm}Q_{w}=-\frac{\partial {\cal G}}{\partial {\Phi}_{w}}\,,\hspace{5mm} S=-\frac{\partial {\cal G}}{\partial{\beta^{-1}}}\, , \hspace{5mm} E={\cal G}+\Phi_p \,Q_p+\Phi_w \,Q_w+\beta^{-1} S\, .
\end{equation}
As shown e.g. in \cite{Reall:2019sah, Bobev:2022bjm, Cassani:2022lrk}, this method to obtain the thermodynamics is particularly useful when dealing with higher-derivative corrections.

The purpose of this section is to evaluate the Euclidean on-shell action of the two-charge black holes at first order in $\alpha'$ and check that the thermodynamics that we get match the ones obtained in the previous section. For simplicity, we are going to evaluate the $(d+1)$-dimensional Euclidean effective action in the string frame, since its dimensional reduction on ${\mathbb S}^1_{y}$ gives rise to much more terms \cite{Baron:2017dvb, Elgood:2020xwu, Ortin:2020xdm, Eloy:2020dko, Liu:2023fqq}. Instead, the $(d+1)$-dimensional action coincides with the ten-dimensional one \eqref{heterotic-08} up to an overall factor which is absorbed in the $(d+1)$-dimensional Newton constant ${\hat G}_{N}$.  

Then, the heterotic Euclidean on-shell action $I$ for a manifold $M$ with boundary $\partial M$ is given by 
\begin{equation}\label{eq:Euclidean_action}
	I=-\frac{{\hat g}_s^2}{16\pi {\hat G}_N}\int_{M} \diff^{d+1}x \sqrt{|\hat g|} \, {\cal L}_{\rm {eff}}+\frac{{\hat g}_s^2}{8\pi {\hat G}_N}\int_{\partial M}\diff^{d}x \sqrt{|\hat h|}\, e^{-2\hat \phi}{\hat K}+\dots\,,
\end{equation}
where 
\begin{equation}
	{\cal L}_{\rm {eff}}=e^{-2\hat \phi}\left [\hat R-4\,\partial^{\hat\mu}{\hat \phi}\,\partial_{\hat\mu} \hat \phi+\frac{1}{2\cdot 3!}{\hat H}^2+\frac{\alpha'}{8}{\hat R}_{(-)}{}_{\hat \mu \hat \nu \hat a \hat b}{\hat R}_{(-)}{}^{\hat\mu \hat\nu \hat a \hat b}\right] \,
\end{equation}
is the effective Lagrangian of the heterotic superstring at first order in $\alpha'$ (see appendix~\ref{sec-HST}). The second term in \eqref{eq:Euclidean_action} is the standard Gibbons-Hawking-York (GHY) term written in the string frame: ${\hat h}_{\mu\nu}$ represents the metric induced at $\partial M$ and $\hat K$ is the trace of the extrinsic curvature. Finally, the dots indicate additional boundary terms associated to the higher-derivative corrections, which on general grounds are expected to give a vanishing contribution for asymptotically-flat solutions, \cite{Reall:2019sah} (hence, we shall ignore them from now on). As observed in \cite{Tseytlin:1988tv, Chen:2021dsw}, the bulk contribution reduces to a boundary term after using the equation of motion of the dilaton \eqref{eq:eq2-08}, which implies
\begin{equation}
	{\cal L}_{\rm{eff}}=-2{\hat \nabla}^2 e^{-2\hat \phi}\, .
\end{equation}
Therefore, we have that \eqref{eq:Euclidean_action} reduces to:
\begin{equation}\label{eq:I}
	I=\frac{{\hat g}^2_s}{8\pi {\hat G}_N}\int_{\partial M}\diff^{d}x \sqrt{|\hat h|}\, e^{-2\hat \phi}\left({\hat K}-2\,n^{\hat \mu}\, \partial_{\hat \mu}\hat \phi\right)\,,
\end{equation}
where $n^{\hat\mu}$ is the unit normal to the boundary. 

Here we are interested in asymptotically-flat black holes whose boundary ${\partial M}$ has the topology of ${\mathbb S}^1_\beta \times {\mathbb S}^{d-2}$ ($\times {\mathbb S}^1_y$). As it is well known, the GHY term diverges in the limit in which the radius of the ${\mathbb S}^{d-2}$ goes to infinity, just as in flat spacetime. In order to obtain a finite on-shell action, we follow the prescription of \cite{Gibbons:1976ue}. This amounts to first consider a regulated spacetime $M_R$, where $R$ is a radial cutoff. The regulated spacetime then corresponds to the region $\rho\le R$, and its boundary $\partial M_{R}$ is the hypersurface $\rho=R$. Second, we introduce an auxiliary configuration with flat metric ${\hat\delta}_{R}$ and constant dilaton ${\hat \phi}_{R}$ chosen so that the induced fields (metric and dilaton) at $\rho=R$ coincide with the induced metric and dilaton of the black hole solution, namely ${\hat \delta}_{R}|_{\rho=R}={\hat g}|_{\rho=R}$ and $\hat \phi_R=\hat \phi|_{\rho=R}$. Once we have $\hat \delta_R$ and $\hat \phi_R$, we substract the regulated action associated with the flat spacetime $I_R[\hat\delta_R, \hat \phi_R]$ to the one associated with the black hole $I_R[\hat g, \hat \phi]$  and only then take the $R\to \infty$ limit. Summarizing, the renormalized action $I_{\infty}$ is given by
\begin{equation}
	I_{\infty}=\lim_{R\to \infty}\left(I_R[\hat g, \hat \phi]-I_R[\hat\delta_R, \hat \phi_R]\right)\, ,
\end{equation}
and, making use of \eqref{eq:I}, we get
\begin{equation}\label{eq:I_infty}
	I_{\infty}=\lim_{R\to \infty}\left\{\frac{{\hat g}_s^2}{8\pi {\hat G}_N}\int_{\rho=R}\diff^{d}x \sqrt{|\hat h|}\, e^{-2\hat \phi}\left[\left({\hat K}-{\hat K}_{{\hat\delta}_ R}\right)-2\,n^{\hat \mu}\, \partial_{\hat \mu}\hat \phi\right]\right\}\,,
\end{equation}
where ${\hat K}_{{\hat\delta}_ R}$ is the trace of the extrinsic curvature associated to the metric ${\hat\delta}_{R}$. For the two-charge black holes we are interested in, the auxiliary flat solution $\{{\hat \delta}_R, \hat \phi_R\}$ is given by
\begin{equation}
	\begin{aligned}
		-\hat \delta_R=\,&\frac{f(R)}{\zp(R)  {\tilde f}_{w}(R)}\diff\tau^2+\zz(R)\left(\diff\rho^2+\rho^2 \diff\Omega^2_{(d-2)}\right)+k^2_{R}\left[\diff y+\beta_{p}k^{-1}_{\infty}\left(\zp(R)^{-1}-1\right)\diff t\right]^2\,,\\
		\hat \phi_R=\,&\hat \phi (R)\,,
	\end{aligned}
\end{equation}
where $k_{R}^2=k^2_{\infty}\frac{\zp (R)}{\zm (R)}$. Now we have all the ingredients to evaluate \eqref{eq:I_infty} using the corrected solutions found in the previous section. Let us do this for the five- and four-dimensional solutions separately.

\subsubsection{Five-dimensional black holes} 

Expressing the result in the grand-canonical ensemble, we get that the Euclidean on-shell action of the five-dimensional two-charge black holes is given by
\begin{equation}\label{eq:onshellaction5d}
	I_{\infty}=\frac{\pi^2 {\rho}_s^3\cosh \gamma_p\cosh\gamma_w}{4 G_N}\left[1-\frac{9\alpha'}{4{\rho}_s^2}-\frac{\alpha' \epsilon_{p}\epsilon_w\tanh\gamma_p\tanh\gamma_w}{{ \rho}^2_s}\right]\, ,
\end{equation}
and we recall that
\begin{equation}
	\beta=2\pi \rho_s\cosh \gamma_p\, \cosh \gamma_w\,,\hspace{5mm}\Phi_{p}=\,\frac{\epsilon_{p}\tanh \gamma_p}{k_{\infty}}\,,\hspace{5mm}\Phi_{w}=\,\epsilon_{w}k_{\infty}\tanh \gamma_w\,.
\end{equation}
It is a straightforward calculation to show that the corrected charges that follow from the on-shell action (using \eqref{eq:chargesfromI}) are in perfect agreement with the ones we computed in the previous section, namely with eqs.~\eqref{eq:Qp}, \eqref{eq:Qw}, \eqref{eq:E} and \eqref{eq:S}.

\subsubsection{Four-dimensional black holes} 

In the four-dimensional case the on-shell action in the grand-canonical ensemble takes the form
\begin{equation}\label{eq:onshellaction4d}
	I_{\infty} = \frac{\pi \rho_s^2 \cosh \gamma_p \cosh \gamma_w}{G_N}  \left[1- \frac{\alpha'}{2\rho_s^2}\left(1 + \frac{\epsilon_p \epsilon_{w}\tanh \gamma_p \tanh \gamma_w}{4} \right) \right]  \,,
\end{equation}
with the inverse temperature and the chemical potentials given by
\begin{equation}
	\beta= 4\pi \rho_s\cosh \gamma_p\, \cosh \gamma_w\,,\hspace{5mm}\Phi_{p}=\,\frac{\epsilon_p \tanh \gamma_p}{k_\infty}\,,\hspace{5mm}\Phi_{w}=\,\epsilon_{w}\,k_{\infty}\tanh \gamma_w\,.
\end{equation}
As before, the charges (\ref{eq:Qp4d}), (\ref{eq:Qw4d}), (\ref{eq:E4d}) and (\ref{eq:S4d}) are properly recovered from \eqref{eq:chargesfromI}.


\section{Two-charge black holes from Schwarzshchild-Tangherlini}\label{sec:2QsBHfromST}

As already mentioned, the corrections to the thermodynamics of two-charge black holes have been previously studied in \cite{Giveon:2009da, Chen:2021dsw}. The strategy of these references is to find the $\alpha'$ corrections by performing a set of $O(2, 2)$ transformations (boost with parameter $\delta_{w}$ plus T-duality along $y$, followed by another boost with parameter $\delta_{p}$) to the Schwarzschild-Tangherlini black hole, whose $\alpha'$ corrections had been already studied in \cite{Callan:1988hs}. The main difference between these two references is that \cite{Chen:2021dsw} just focuses on the thermodynamic properties while in \cite{Giveon:2009da} the full corrected solutions are obtained by means of this technique. This is technically more complicated than just obtaining the thermodynamics, as one has to take into account the explicit $\alpha'$ corrections to the $O(2, 2)$ transformations. This might be the reason why the $\alpha'$-corrected thermodynamics obtained in these references do not agree with one another.

The goal of this section is to show that our results for the $\alpha'$-corrected thermodynamics of heterotic two-charge black holes are in agreement with those of \cite{Chen:2021dsw}. To this aim, we find convenient to review here their calculation. A key observation is that the Euclidean on-shell action remains invariant after the $O(2,2)$ transformation. Therefore,
\begin{equation}\label{eq:I=tildeI}
	I_{\infty}(\beta, \Phi_p, \Phi_w; \phi_{\infty}, k_{\infty})={\tilde I}_{\infty}(\tilde\beta; {\tilde\phi}_{\infty}, {\tilde k}_{\infty})\, ,
\end{equation}
where, following the conventions of \cite{Chen:2021dsw}, we are using tildes for the quantities associated to the Schwarzschild-Tangherlini solution.

The right-hand side of \eqref{eq:I=tildeI} is obtained from the $\alpha'$ corrections to the Schwarzschild-Tangherlini solution \cite{Callan:1988hs}. Focusing just on the thermodynamic quantities, we have 
\begin{equation}
	{\tilde E}=\frac{d-2}{d-3}\frac{\gamma_d \,{{\tilde R}_{\beta}}^{d-3}}{8\pi {\tilde G}_{N}} \left(1-\frac{\epsilon_d \, \alpha'}{4 {\tilde R}_{\beta}^2}\right)\,, \hspace{5mm} {\tilde S}= \frac{\gamma_d \, {\tilde R}_{\beta}^{d-2}}{4{\tilde G}_{N}}\left(1-\frac{\sigma_d \, \alpha'}{4 {\tilde R}_{\beta}^2}\right)\,,
\end{equation}
where ${\tilde R}_{\beta}\equiv {\tilde\beta}/(2\pi)$ is the  radius of the thermal circle ${\mathbb S}^1_{\beta}$ and
\begin{equation}
	\gamma_d=\omega_{d-2} \left(\frac{d-3}{2}\right)^{d-2}\,, \hspace{5mm}\epsilon_d=\frac{2(d-4)(d-2)}{d-3}\,, \hspace{5mm}\sigma_d=\frac{2(d-5)(d-2)^2}{(d-3)^2}\, .
\end{equation}
Assuming the quantum statistical relation \eqref{eq:qsr}, we get that the Euclidean on-shell action of the Schwarzschild-Tangherlini black hole is
\begin{equation}\label{eq:I_ST}
	{\tilde I}_{\infty}={\tilde \beta} {\tilde E}- {\tilde S}= \frac{\gamma_d \, {\tilde R}_{\beta}^{d-2}}{4{\tilde G}_{N} (d-3)} \left[1-\frac{(d-2)^2\alpha'}{2(d-3){\tilde R}_{\beta}^{2}}\right]\,.
\end{equation}
Because of \eqref{eq:I=tildeI}, the right-hand side of  \eqref{eq:I_ST} computes the Euclidean on-shell action of the two-charge black holes as well. This is, however, meaningless at this stage, since we have not yet specified the expressions for $\beta$ and the chemical potentials $\Phi_{p}$, $\Phi_{w}$ in terms of ${\tilde R}_{\beta}$ and the parameters of the $O(2,2)$ transformations. Such expressions can be found in \cite{Chen:2021dsw}. Taking into account all the possibilities for the signs of the winding and momentum charges, we find
\begin{eqnarray}
	\label{eq:relation_beta}
	R_{\beta}&=\,& {\tilde R}_{\beta}\cosh \delta_p\, \cosh \delta_w \left(1-\frac{\alpha' \epsilon_{p}\epsilon_{w}\tanh \delta_p \tanh \delta_w}{2{\tilde R}_{\beta}^2}\right)\,,\\[1mm]
	\label{eq:relation_Phip}
	\Phi_{p}&=\,&\frac{\epsilon_{p}\tanh \delta_p}{k_{\infty}}\left(1-\frac{\alpha' \epsilon_{p}\epsilon_{w}\tanh \delta_w}{{\tilde R}_{\beta}^2\sinh \left(2\delta_p\right)}\right)\,,\\[1mm]
	\label{eq:relation_Phiw}
	\Phi_{w}&=\,&\epsilon_{w} k_{\infty}\tanh \delta_w\left(1-\frac{\alpha' \epsilon_{p}\epsilon_{w}\tanh \delta_p}{{\tilde R}_{\beta}^2\sinh \left(2\delta_w\right)}\right)\,,
\end{eqnarray}
where $\delta_{p, w}$ represent the parameters of the $O(2, 2)$ transformations. In addition to this, one must also bear in mind the relation between the moduli of the solutions. In particular, we need the relation between the asymptotic values of the $d$-dimensional dilaton $e^{\phi_{\infty}}=g_s$, which is the following \cite{Chen:2021dsw}
\begin{equation}\label{eq:relation_gs}
	g_s^2=\,{\tilde g}_s^2 \cosh \delta_p\, \cosh \delta_w \left(1-\frac{\alpha' \epsilon_{p}\epsilon_{w}\tanh \delta_p \tanh \delta_w}{2{\tilde R}_{\beta}^2}\right)\,.
\end{equation}
Taking into account that $G_N\propto g_s^2$, one gets that the Newton constants are related by
\begin{equation}
	G_N={\tilde G}_{N}\cosh \delta_p\, \cosh \delta_w \left(1-\frac{\alpha' \epsilon_{p}\epsilon_{w}\tanh \delta_p \tanh \delta_w}{2{\tilde R}_{\beta}^2}\right)\, .
\end{equation}
Using this in \eqref{eq:I_ST}, we obtain
\begin{equation}
	I_{\infty}={\tilde \beta} {\tilde E}- {\tilde S}= \frac{\gamma_d \, {\tilde R}_{\beta}^{d-2}\cosh \delta_p \cosh \delta_w}{4{G}_{N} (d-3)} \left[1-\frac{\alpha'}{2 {\tilde R}_{\beta}^2}\left(\frac{(d-2)^2}{(d-3)}+\epsilon_{p}\epsilon_w \tanh \delta_p \tanh \delta_w\right)\right]\,.
\end{equation}
This already specifies the thermodynamics. However, the parametrization we are using here differs from the one(s) used in the previous sections. It is not difficult to find that the relation between ${\tilde R}_{\beta}, \delta_p, \delta_w$ and the parameters $\rho_s, \gamma_p, \gamma_w$ used in the previous sections to express the thermodynamics in the grand-canonical ensemble is given by 
\begin{eqnarray}
	{\tilde R}_{\beta}&=&\frac{2\rho_s}{d-3} \left(1-\frac{\epsilon_{p}\epsilon_{w} (d-3)^2\alpha'\tanh \gamma_p \tanh \gamma_w}{8\rho_s^2}\right)\, ,\\[1mm]
	\delta_p&=&\gamma_p+\frac{\epsilon_{p}\epsilon_{w} (d-3)^2\alpha' \tanh \gamma_w}{8\rho_s^2}\,,\\[1mm]
	\delta_w&=&\gamma_w+\frac{\epsilon_{p}\epsilon_{w} (d-3)^2\alpha' \tanh \gamma_p}{8\rho_s^2}\,.
\end{eqnarray}
Making use of these relations, we can write the on-shell action of the two-charge black holes in the grand-canonical ensemble is
\begin{equation}
	I_{\infty}=\frac{\omega_{d-2} \,\rho_s^{d-2}\, \cosh \gamma_p \cosh \gamma_w}{4 (d-3)G_N}\left\{1-\frac{(d-3)\alpha'}{8\rho_s^2}\left[(d-2)^2+\epsilon_{p}\epsilon_{w} (d-3)^2\tanh\gamma_p\tanh \gamma_w\right]\right\}\, .
\end{equation}
This reduces to \eqref{eq:onshellaction5d} and to \eqref{eq:onshellaction4d} when setting $d=5$ and $d=4$, respectively. Given the grand-canonical potential ${{\cal G}=\beta^{-1}I_{\infty}}$, we can obtain the charges, entropy and mass through \eqref{eq:chargesfromI}, as already discussed.
Expressing them in the grand-canonical ensemble, we obtain the following expressions
\begin{eqnarray}
	\beta&=\,&\frac{4\pi \rho_s}{d-3}\cosh \gamma_p\, \cosh \gamma_w\,,\hspace{5mm}
	\Phi_{p}=\,\frac{\epsilon_p\tanh \gamma_p}{k_{\infty}}\,,\hspace{5mm}
	\Phi_{w}=\,{\epsilon}_{w}k_{\infty}\tanh \gamma_w\,,\\[1mm]
	Q_p&=& Q^{(0)}_{p}\left[1-\frac{(d-3)^2\alpha'}{16\rho_s^2}\left(\sigma_d-2\epsilon_p\epsilon_{w}(4-d+\text{coth}^2\gamma_p)\tanh\gamma_p\tanh \gamma_w\right)\right]\,,\\[1mm]
	Q_w&=& Q^{(0)}_{w}\left[1-\frac{(d-3)^2\alpha'}{16\rho_s^2}\left(\sigma_d-2\epsilon_p\epsilon_{w}(4-d+\text{coth}^2\gamma_w)\tanh\gamma_p\tanh \gamma_w\right)\right]\,,\\[1mm]
	S&=&S^{(0)}\left[1-\frac{(d-3)^2\alpha'}{16\rho_s^2}\left(\sigma_d+2(d-5)\epsilon_p\epsilon_{w}\tanh\gamma_p\tanh\gamma_w\right)\right]\, ,
\end{eqnarray}
where 
\begin{equation}
	\begin{aligned}
		Q^{(0)}_p=\,&\frac{(d-3)\epsilon_pk_{\infty}\omega_{d-2}\rho_s^{d-3}\sinh (2\gamma_p)}{32\pi G_N}\,, \hspace{5mm} Q^{(0)}_w=\,\frac{(d-3)\epsilon_{w}\omega_{d-2}\rho_s^{d-3}\sinh (2\gamma_w)}{32\pi G_N k_{\infty}}\,, \\[1mm] 
		S^{(0)}=\,&\frac{\gamma_{d-2}\rho_s^{d-2}\cosh \gamma_p \cosh \gamma_w}{32G_N}\, .
	\end{aligned}
\end{equation}
Instead of the mass we provide the expression for the grand-canonical potential ${\cal G}$, which is simpler
\begin{equation}
	{\cal G}=\frac{\omega_{d-2} \, \rho_s^{d-3}}{16\pi G_N} \left[1-\frac{(d-3)\alpha'}{8\rho_s^2}\left((d-2)^2+(d-3)^2\,\epsilon_p\epsilon_{w}\tanh\gamma_p\tanh\gamma_w\right)\right]\,.
\end{equation}
The mass $E$ follows then from the last of \eqref{eq:chargesfromI}. It is now straightforward to compare these expressions with the ones we obtained in sections~\ref{sec:BH_thermodynamics} and \ref{sec:chargesfromI} and see that they are in perfect agreement. Furthermore, we have also checked that they agree with the corrected thermodynamics given in the appendix of \cite{Chen:2021dsw}, after using the map between the two parametrizations, provided in \eqref{eq:relation_beta}, \eqref{eq:relation_Phip} and \eqref{eq:relation_Phiw}.

	\clearpage{\pagestyle{empty}\cleardoublepage}

	\part{Vacua Solutions} \label{part:FM}


\def\d{{\delta}}

\chapter{New instabilities for  non-supersymmetric  AdS$_4$ orientifold vacua}  \label{ch:5}
We consider massive type IIA orientifold compactifications of the form AdS$_4 \times X_6$, where $X_6$ admits a Calabi--Yau metric and is threaded by background fluxes. From a 4d viewpoint,  fluxes generate a  potential whose vacua have been classified, including one $\cN=1$ and three perturbatively stable $\cN=0$ branches. We reproduce this result from a 10d viewpoint, by solving the type IIA equations at the same level of \bk{detail} as previously done for the $\cN=1$ branch. All solutions exhibit localized sources and parametric scale separation.  We then analyze the non-perturbative stability of the $\cN=0$ branches. We consider new 4d membranes, obtained from wrapping D8-branes on $X_6$ or D6-branes on its divisors, threaded by non-diluted worldvolume fluxes. Using them we show that all branches are compatible with the Weak Gravity Conjecture for membranes. In fact, most vacua satisfy the sharpened  conjecture that predicts superextremal membranes in $\cN=0$ settings, except for a subset whose non-perturbative stability remains an open problem.

\section{Introduction}
\label{s:intro}

AdS vacua are a key sector of the string Landscape. On the one hand, stable vacua should have a dual holographic description that allows us to access their dynamics at strong coupling. On the other hand, they have been subject to recent scrutiny within the context of the Swampland Programme \cite{Vafa:2005ui,Brennan:2017rbf,Palti:2019pca,vanBeest:2021lhn,Grana:2021zvf}, where several proposals to describe their general properties have been made. Out of them, the most relevant one for the discussion of this chapter is the AdS Instability Conjecture \cite{Ooguri:2016pdq,Freivogel:2016qwc}, which states that all $\cN=0$ AdS$_d$ vacua are unstable, in which case their holographic description would not make much sense. In particular, in perturbatively stable vacua supported by $d$-form fluxes, the instability is expected to arise at the non-perturbative level, from one or several superextremal $(d-2)$-branes that nucleate and expand towards the AdS$_d$ boundary \cite{Maldacena:1998uz}. The existence of such branes is predicted by a sharpening of the Weak Gravity Conjecture (WGC), which states that the WGC inequality is only saturated in supersymmetric settings \cite{Ooguri:2016pdq}. 

All these statements are particularly meaningful in string constructions where the compactification scale is much smaller than the AdS length scale, as then the nucleation can be described by means of an EFT valid at intermediate scales. In this sense, the DGKT-CFI proposal \cite{DeWolfe:2005uu,Camara:2005dc}, in which a parametric separation of scales is achieved by moving in an infinite family of AdS$_4$ vacua, represents an interesting arena to test these ideas. A quite general construction realising this feature is based on massive type IIA string theory compactified on a Calabi--Yau orientifold geometry with O6-planes and D6-branes and threaded by background fluxes,\footnote{See \cite{Derendinger:2004jn,Villadoro:2005cu} for previous similar constructions in toroidal orbifold settings.} and it is typically referred to as DGKT-like vacua. While a holographic description of these vacua remains elusive and some of their features are quite counter-intuitive \cite{Lust:2019zwm}, the proposal has passed non-trivial tests at the gravity side, like the approximate 10d description provided in \cite{Junghans:2020acz,Marchesano:2020qvg}. 

A general classification of DGKT-like vacua can be done using a 4d EFT description, which includes an F-term potential generated by fluxes that fixes the Calabi--Yau moduli. Such an analysis was carried out in \cite{Marchesano:2019hfb}, where at least four branches of perturbatively stable vacua -- one supersymmetric and three non-supersymmetric -- were shown to exist. All of these branches contain an infinite number of vacua that is generated by a rescaling of internal fluxes, as in the supersymmetric case, and along which parametric scale separation is achieved. Remarkably, the mass spectrum found in \cite{Marchesano:2019hfb} for some of these branches has an amusing holographic interpretation \cite{Conlon:2021cjk,Apers:2022tfm}, while the remaining branches do not present this feature \cite{Quirant:2022fpn}. 

Given this setup, the purpose of this chapter is to gain further insight into the non-supersymmetric branches of DGKT-like vacua, and in particular \bk{on} their perturbative and non-perturbative stability, following up on previous work on this subject \cite{Junghans:2020acz,Marchesano:2021ycx,Casas:2022mnz}. As a first step, one would like to confirm the 4d result on perturbative stability, or in other words to verify that the F-term potential from where the moduli masses are derived is reliable. The effective F-term potential in massive type IIA orientifold compactifications used in \cite{Marchesano:2019hfb} is derived either by performing a direct Kaluza--Klein reduction over a Calabi--Yau geometry threaded by internal fluxes \cite{Grimm:2004ua,Grimm:2011dx,Kerstan:2011dy}, or through the formalism of 4d three-form potentials  \cite{Bielleman:2015ina,Carta:2016ynn,Farakos:2017jme,Herraez:2018vae,Bandos:2018gjp,Lanza:2019xxg}. If one obtains a 10d description for these vacua that displays scale separation and an approximate Calabi--Yau metric, then it means that the derivation of the potential is accurate up to the said degree of approximation. \bk{This was shown to be the case in \cite{Junghans:2020acz} via a general description of approximate solutions to 10d massive IIA equation that correspond to DGKT-like vacua.} The degree of accuracy is given by the 10d dilaton vev or equivalently by the inverse AdS$_4$ length in string units, which both become parametrically small as we advance in the infinite family of vacua. \bk{In this chapter we confirm this picture by reproducing the four 4d branches of vacua mentioned above directly from a 10d perspective. The 10d background describing all these 4d vacua is provided at the same degree of explicitness as given for the supersymmetric branch in  \cite{Marchesano:2020qvg}, using a combination of the results in \cite{Junghans:2020acz} and  \cite{Marchesano:2020qvg}.}

We then turn to analyze the non-perturbative stability of these vacua, by considering the charge $Q$ and tension $T$ of their 4d membranes, along the lines of \cite{Aharony:2008wz,Narayan:2010em,Marchesano:2021ycx,Casas:2022mnz}. We focus in particular on $D(2p+2)$-branes wrapping $2p$-cycles of $X_6$ which are those that can nucleate in the context of the 4d EFT \cite{Lanza:2019xxg,Lanza:2020qmt}. According to the sharpened WGC at a generic vacuum one should find at least two membranes with $Q>T$. One made up of a D4-brane wrapping a two-cycle $\Sigma \subset X_6$ or a bound state containing it, and another one made up of a D8-brane wrapping $X_6$. Both objects we analyzed in \cite{Marchesano:2021ycx} for \bk{one} branch of non-supersymmetric vacua, with special attention to the microscopic description of D8-branes as BIons. It was found that D4-branes satisfy $Q=T$ at the level of accuracy that we are working, while D8-branes satisfy $Q>T$ in simple configurations, due to a mixture of curvature corrections to their charge and tension and further corrections due to their BIonic nature. However, closer inspection showed that this last statement depends on the specific configuration of space-time filling D6-branes in a given vacuum, and that for some vacua the corrections to the D8-brane charge and tension tip the scales towards $Q<T$  \cite{Casas:2022mnz}. Therefore, it would seem that in such vacua not only the sharpened WGC fails to be true, but even the WGC for 4d membranes itself. 

As we will see, this apparent tension with the WGC is solved when one considers more exotic D-brane configurations. In particular, we look at those which are BPS  in supersymmetric DGKT vacua. Namely, we consider D8-branes wrapping $X_6$ and D6-branes wrapping a divisor ${\cal S} \subset X_6$, both threaded by non-diluted worldvolume fluxes in their internal dimensions (i.e., with worldvolume fluxes comparable to the K\"ahler two-form). One can check that at least one of these objects satisfies $Q\geq T$ in non-supersymmetric DGKT-like vacua. Since they couple to the same three-forms as D4-branes and D8-branes, they realize the WGC for 4d membranes. In fact, in most cases they correspond to superextremal 4d membranes, as predicted by the sharpened WGC. Only in one subclass of $\cN=0$ vacua all the relevant 4d membranes are extremal, namely in those $\cN=0$ vacua without space-time-filling D6-branes which, from the 4d viewpoint, are related to supersymmetric ones by an overall sign flip of the four-form flux. Quite amusingly, it is precisely such vacua which display integer conformal dimensions for their would-be holographic dual. Whether there is some meaning behind this coincidence or the marginality is an artefact of the accuracy of our description remains an open question for the future. 

The chapter is organized as follows. In section \ref{sec-type2A} we briefly review type IIA democratic formulation and some properties of its compactifications on 6-dimensional manifolds. In section \ref{s:branch} we review the main features of DGKT-like vacua and the four branches of solutions found  \cite{Marchesano:2019hfb}. In section \ref{s:10d} we discuss how to describe such 4d vacua from a 10d viewpoint, first using the smearing approximation and then with a more accurate 10d background with localized sources. In section \ref{s:membranes} we address the non-perturbative stability of these vacua by analysing the extremality of 4d membranes in the probe approximation. We leave our conclusions for section \ref{s:conclu} and several technicalities for the appendices. Appendix \ref{ap:10deom} analyzes in detail the 10d equations of motion and Bianchi identities for all branches of vacua. Appendix \ref{ap:DBI} deduces the D-brane DBI expressions by means of which we compute the corresponding 4d membrane tension. 

\section{Review of type IIA compactifications}\label{sec-type2A}

In this chapter we study some classes of vacua AdS$_4$ of the effective action of type IIA superstring theory compactified on a 6-dimensional compact manifold $X_6$. In all the cases considered we have vanishing fermionic fields. For the sake of self-consistency, we give a short description of the bosonic sector of type IIA effective action, of the fermions supersymmetry transformation and of the actions of certain localized sources. 

\subsection{The theory}

The bosonic degrees of freedom of 10-dimensional type IIA are organized into  the graviton $G_{M N}$, the KR 2-form $B_{MN}$, the dilaton $\phi$, the RR 1-form $C_1$ and the RR 3-form $C_3$. They combine into the 2-derivative effective action
\begin{equation} \label{eq:action2A}
	\begin{split}
		S_{\text{IIA}} = & \quad   \frac{1}{2\kappa_{10}^2}\int d^{10} x \sqrt{G} \left[e^{-2\phi} \left(R + 4 \, \partial_M \phi \,\partial^M \phi - \frac{1}{2} |H|^2\right) - \frac{1}{2}|G_2|^2 - \frac{1}{4} |G_4|^2 \right] \\[2mm]
		& \; -\frac{1}{4 \kappa_{10}^2} \int B \wedge d C_3 \wedge d C_3 \,,
	\end{split}
\end{equation}
where $|\cdot|^2$ indicates the inner product on forms, $2\kappa_{10}^2 = (2\pi)^7 \alpha'{}^4$ and the fieldstrengths $H$, $G_2$, $G_4$ are
\begin{equation}
	H = dB \,, \qquad G_2 = d C_1 \,, \qquad G_4 = d C_3 - C_1 \wedge H \,.
\end{equation}
In our conventions the string length $\ell_s$ is defined as $\ell_s = 2\pi \sqrt{\alpha'}$. We are using the mostly plus signature. The action ($\ref{eq:action2A}$) is not unique in the sense that admits a deformation trough a parameter $m$ called Romans mass. It can be interpreted as the background value of a rank 0 fieldstrength $G_0$. Including the Romans mass we obtain	\begin{equation} \label{eq:action2Amassive}
	S_{\text{IIA}}^{\text{massive}} = S_{\text{IIA}} - \frac{1}{4\kappa_{10}^2} \int d^{10}x \sqrt{-G} \, m^2 - \frac{1}{4 \kappa_{10}^2}\int \frac{1}{3} \, m \, dC_3 \wedge B^3 + \frac{1}{20} m^2 B^5 \,.
\end{equation}

\subsubsection{Action in the democratic formulation}
We can describe in a more convenient way type IIA with the democratic formulation. The idea of such formulation is to introduce into the action the duals of the RR fields as independent fields. In order to not double the degrees of freedom we have to impose a duality condition. The democratic action therefore is equivalent to the original one (\ref{eq:action2Amassive}) after we impose further constraints. For this reason is usually called pseudo-action. Its explicit form is \cite{Bergshoeff:2001pv}
\begin{equation} \label{eq:action2Adem}
	S_{\text{IIA}}^{\text{dem}} =     \frac{1}{2\kappa_{10}^2}\int d^{10} x \sqrt{G} \left[e^{-2\phi} \left(R + 4 \, \partial_M \phi \,\partial^M \phi - \frac{1}{2} |H|^2\right) - \frac{1}{4} \sum_p |G_p|^2 \right] \,,
\end{equation}
where the sum runs over $p = 0, 2,4,6,8,10$. If we introduce the polyforms 
\begin{subequations}
	\begin{align}
		& \mathbf{C} = C_1 + C_3 + C_5 + C_7 + C_9 \,, \\[2mm]
		& \mathbf{G} = G_0 + G_2 + G_4 + G_6 + G_8 + G_{10} \,,
	\end{align}
\end{subequations}
and the operator $d_H = d - H \wedge $ we can relate the fieldstrengths $G_p$ with the gauge vectors $C_p$ via\footnote{The expressions should be interpreted as the direct sum of independent expressions, one for each rank of the polyform.}
\begin{equation}
	\mathbf{G} = d_H \mathbf{C} + G_0 \, e^B \,.
\end{equation}
The duality condition we have to impose to avoid the doubling of the degrees of freedom is 
\begin{equation}
	\mathbf{G} = \star_{10} \lambda \left(\mathbf{G} \right) \,, 
\end{equation}
where $\star_{10}$ is the 10-dimensional Hodge star operator and $\lambda$ is the operator which reverses the order of the indexes of  a $p$-form. Explicitly, it acts on a $p$-form $\alpha$ as
\begin{equation}
	\lambda \left(\alpha\right) = (-1)^{p(p-1)/2} \alpha \,.
\end{equation}
The Bianchi identities for the $G_p$ and $H$ are then
\begin{equation}
	dH = 0 \,, \qquad d_H \mathbf{G} = 0 \,.
\end{equation} 
The Bianchi for $\mathbf{G}$ implies that if $\mathbf{C}$ and $B$ are not globally defined, $d \left(e^{-B} \mathbf{C}\right)$ might have an harmonic part. Calling such harmonic part $\mathbf{\bar{G}}$ we obtain the decomposition
\begin{equation}
	d \left(e^{-B} \mathbf{C}\right) = d \left(e^{-B} \mathbf{\bar{C}}\right) + \mathbf{\bar{G}} \,.
\end{equation}
Dropping the bar over the gauge potentials, we can write
\begin{equation}
	\mathbf{G} = d_H \mathbf{C} + \mathbf{\bar{G}} \, e^B \,.
\end{equation}
$\mathbf{\bar{G}}$ represents the quantized part of $\mathbf{G}$ in setups with no sources.

\subsubsection{Brane sources in type II theories}

If we admit D$p$-branes sources we need to take into account their effective action. For D$p$ branes it takes the form
\begin{equation} \label{eq:actionDp}
	S_{\text{D}p} = S_{\text{DBI}} + S_{\text{CS}} \,.
\end{equation}
$S_{\text{DBI}}$ encodes the interaction among the brane and the NS sector bosonic fields. $ S_{\text{CS}} $ encodes the interactions with the RR fields. They have the explicit form 
\begin{subequations}
	\begin{align}
		S_{\text{DBI}} \quad = &\quad  - \mu_{\text{D}p} \int_\mathcal{W} d^{p+1} \xi \, e^{-\phi} \sqrt{|\det(g_{\alpha\beta} - \mathcal{F}_{\alpha\beta})|} \,, \\[2mm]
		S_{\text{CS}} \quad = & \quad  \mu_{\text{D}p} \int_\mathcal{W} \mathbf{C} \wedge e^{-\mathcal{F}} \wedge \sqrt{\frac{\hat{A}(4\pi \alpha' R_T)}{\hat{A}(4\pi \alpha' R_N)}} \,,
	\end{align}
\end{subequations}
where $\mathcal{W}$ is the worldvolume of the D$p$ brane, $g_{\alpha\beta}$ is the pullback of the spacetime metric $g_{\mu\nu}$ on the worldvolume $\mathcal{W}$, $\phi$ is the 10-dimensional dilaton, $\mathcal{F}_{\alpha\beta}$ is the pullback of the combination
\begin{equation}
	\mathcal{F} = B + 2\pi\alpha' F \,,
\end{equation}
where $B$ is the KR field and $F$ is the fieldstrength of the worldvolume gauge vector living on the D$p$ brane. $\hat{A}$ is the so-called A-roof genus. It is essentially a polynomial of the Pontryagin classes built with the curvature 2-forms $R_{T/N}$, which are respectively, the curvature two forms of the pullback of the metric $g_{\mu\nu}$ on the tangent and normal bundle of $\mathcal{W}$. The presence of such a factor is fundamental for proper anomaly cancellations, but working at leading order in $\alpha'$ we can neglect it.\footnote{Notice that the next to leading order terms induce lower dimensional D-brane charge and tension. The overall effect is important for the tadpole cancellation when we deal with D-branes of internal dimension larger than three.} It has indeed the expansion 
\begin{equation}
	\sqrt{\frac{\hat{A}(4\pi \alpha' R_T)}{\hat{A}(4\pi \alpha' R_N)}} = 1 + \mathcal{O}(\alpha'{}^2) \,.
\end{equation}
Finally, $\mu_{\text{D}p}^{-1} = (2\pi)^p (\alpha')^{(p+1)/2}$ and it is related with the physical brane tension $T_{\text{D}p}$ by
\begin{equation}
	T_{\text{D}p} = \mu_{\text{D}p} g_s^{-1} = \frac{1}{ (2\pi)^p (\alpha')^{(p+1)/2} g_s} \,.
\end{equation}
For anti D$p$-branes the sign of $S_{\text{CS}}$ is flipped. In particular, the actual value of $\mu_{\text{D}p}$ is fixed by requiring that the 1-loop amplitude of an open string with endpoints on parallel D$p$-branes is equal to the tree-level amplitude of a closed string propagating between two parallel D$p$-branes.

\subsubsection{Orientifold planes in type II theories}

And orientifold quotient is a quotient of the fields of the theory with respect to a $\mathbb{Z}_2$ symmetry. The operator $\mathcal{O}$ implementing it has the generic form 
\begin{equation}
	\mathcal{O} = \Omega_p \mathcal{R}  (-1)^{\bar{F}}
\end{equation}
where $\Omega_p$ is the world-sheet parity reversal operator, $\mathcal{R}$ is an involution operator for bosons and satisfies $\mathcal{R}^2 = (-1)^{\bar{F} + F}$. The net effect of an orientifold projection is the elimination of all the states and fields which are not invariant under the $\mathbb{Z}_2$ action of $\mathcal{O}$. The fixed points of $\mathcal{R}$ define a geometric locus called orientifold plane O$p$, as DD boundary conditions do with D$p$-branes. However, the two extended objects have an important difference: O$p$ planes are not dynamical. Nonetheless, they are sources for the dynamical fields of the theory. The action which encodes the couplings of an O$p$-plane with the bosonic fields of type II superstring is
\begin{equation} \label{eq:acitionOp}
	S_{\text{O}p} = - \mu_{\text{O}p} \int_\mathcal{W} d^{p+1} \xi \, e^{-\phi} \sqrt{|\det g_{\alpha\beta}|} + \mu_{\text{O}p} \int_\mathcal{W} \mathbf{C} \wedge \sqrt{\frac{L(4\pi \alpha' R_T)}{L(4\pi \alpha' R_N)}} \,,
\end{equation}
where $L$ is the Hirzebruch L-polynomial. Again we can neglect them at leading order in $\alpha'$.\footnote{Notice that the next to leading order terms induce lower dimensional charge and tension. The overall effect is important for the tadpole cancellation when we deal with O-planes of internal dimension larger than three.}
\begin{equation}
	\sqrt{\frac{L(4\pi \alpha' R_T)}{L(4\pi \alpha' R_N)}}  = 1 + \mathcal{O}(\alpha'{}^2)\,.
\end{equation}
All the other terms have the same interpretation of those appearing in the D$p$-brane action. The actual amount of charge carried by an O$p$-planes depends on $\mu_{\text{O}p}$. It can be fixed counting the amount of D$p$-branes that must be introduced to cancel out the amplitudes anomalies which appears after the projection of part of the spectrum. In particular, this is done for O$9$ planes in type I obtaining $\mu_{O9}= - 32 \mu_{D9}$. Then, one computes the charge carried by O$p$ planes compactifying on $9-p$ directions and performing T-dualities. The tadpole cancellation is then achieved iff\footnote{T-duality in a longitudinal direction maps a D$p$ brane into a D$(p-1)$ brane and a O$p$-plane into a O$(p-1)$-plane. D and N boundary conditions are indeed exchanged. Defining the orientifold projection in the T-dual setup $\mathcal{O}'$  via $ \mathcal{O}' = T \mathcal{O} T$, we obtain that a longitudinal direction for $\mathcal{R}$ is mapped into a transverse direction for $\mathcal{R}'$. Starting from an O9 and 32 D9, the number of D$p$ branes is not affected by T-duality. The number of O$p$ it is. Every transversal compact direction has two fixed points with respect to $\mathcal{R}$. We have therefore a total of $2^{9-p}$ O$p$-planes.}
\begin{equation}
	\mu_{\text{O}p} = -2^{p-4}\mu_{\text{D}p} \,.
\end{equation}

\subsubsection{Equations of motion}

We want to extract now the equations of motion of type IIA coupled with local sources. We consider the democratic action (\ref{eq:action2Adem}) and the localized sources action $S_\textit{loc}$ of the form (\ref{eq:actionDp}), (\ref{eq:acitionOp}). We obtain 
\begin{subequations} \label{eq:eomIIA}
	\begin{align}
		0 & = d_{-H} \star \mathbf{G} -  8 \kappa_{10}^2 \frac{\delta S_\textit{loc}}{\delta \mathbf{C}}  \,, \\[4mm]
		0 & = d \left(e^{-2\phi} \star H \right) + \frac{1}{2} \star \mathbf{G} \wedge \mathbf{G} - 2 \kappa_{10}^2 \frac{\delta S_\textit{loc}}{\delta B}\,, \\[4mm]
		0 & = \nabla^2 \phi - (\partial \phi)^2 + \frac{1}{4}R - \frac{1}{48} H^2 - \frac{1}{4} \frac{k_{10}^2}{\sqrt{-g}} e^{2\phi} \frac{\delta S_{\text{loc}}}{\delta \phi} \,, \\[4mm]
		\begin{split}
			0 & =  R_{MN} + 2 \nabla_M \nabla_N \phi - \frac{1}{4} H_{M PQ} H_{N}{}^{PQ} - \frac{1}{4} e^{2\phi} F_{MP} F_N{}^P \\[1mm]
			& \quad - \kappa_{10}^2 e^{2\phi}\left(-\frac{2 \kappa_{10}^2 e^{2\phi}}{\sqrt{-g}} \frac{\delta S_{\text{loc}}}{\delta g^{MN}} + \frac{g_{MN} }{2 \sqrt{-g}} \frac{\delta S_{\text{loc}}}{\delta \phi}\right) \,.
		\end{split}
	\end{align}
\end{subequations}

\subsubsection{Supersymmetry transformations}

For vanishing fermions, the only non-trivial supersymmetry transformations are those of the gravitinos $\psi_M^i$ and dilations $\lambda^i$, with $i = 1,2$ and $M$ a 10-dimensional spacetime index. Combining the two Majorana--Weyl spinors into a single Majorana spinor we have \cite{Bergshoeff:2001pv}
\begin{subequations}\label{eq:susy}
	\begin{align}
		\delta_\epsilon \psi_M & = (\partial_M + \frac{1}{4}\slashed{\omega} + \frac{1}{8}\Gamma_{11} \slashed{H}_M ) \epsilon + \frac{1}{16} e^\phi \sum_{p} \frac{1}{(2p)!}\slashed{G}_{2p} \Gamma_M (\Gamma_{11})^p \epsilon \,, \\[2mm]
		\delta_\epsilon \lambda & =  (\slashed{\partial}\phi + \frac{1}{12}\slashed{H}\Gamma_{11})\epsilon + \frac{1}{8} e^{\phi} \sum_p \frac{5-2p}{(2p)!} \slashed{G}_{2p} (\Gamma_{11})^p \epsilon \,, 
	\end{align}
\end{subequations} 
where $\epsilon$ is the supersymmetry parameter, $\Gamma_M$ are  gamma matrices in a Majorana representation, $\omega$ is the spin connection and the slash represents the Clifford map which acts on a $p$-form $\alpha$ as $\slashed{\alpha} = \alpha_{\mu_1 \dots \mu_p} \Gamma^{\mu_1 \dots \mu_p}$.

\subsection{Compactifications}

We analyze now the possible compactifications of IIA which have a 4-dimensional maximally symmetric vacua. We briefly review the conditions imposed by the EOMs and supersymmetry. 

\subsubsection{The ansatz}
We are interested in compactifications of the theory on 
\begin{equation} \label{eq:ansatzCompact}
	\mathcal{M}_{10} = \mathcal{M}_4 \times X_6 \,,
\end{equation}
with $X_6$ a 6-dimensional compact manifold and $\mathcal{M}_4$ a maximally symmetric 4-dimensional vacua. If we allow for warping, the 10-dimensional line element decomposes as (we split the index $M$ into $(\mu, m)$)
\begin{equation}
	g_{MN} dx^M dx^N = e^{2A(y)} g_{\mu\nu} dx^\mu dx^\nu g_{mn} dy^m dy^n \,,
\end{equation}
where $g_{\mu\nu}$ is the metric of the 4-dimensional maximally symmetric space $\mathcal{M}_4$ and $g_{mn}$ is the metric of 6-dimensional compact manifold $X_6$. This in particular implies that we are breaking the Lorentz group as
\begin{equation}
	SO(1,9) \rightarrow SO(1,3) \times SO(6) \,.
\end{equation} 
The only fermions we may be not vanishing are the  Majorana--Weyl fermions generating supersymmetry transformations. They decompose as
\begin{equation}
	{\epsilon}_\pm = \sum_J \epsilon_{+ \, J} \otimes \eta_{\pm \, J} + \epsilon_{- \, J} \otimes \eta_{\mp \, J}
\end{equation}
where the label $\pm$ indicates the chirality, $\epsilon_{\pm \, J}$ and $\eta_{\pm \,, J}$ are Weyl spinors in $\mathcal{M}_4$ and $\mathcal{M}_6$ and they satisfy $(\epsilon_{\pm \, J})^c = \epsilon_{\mp \, J}$, $(\eta_{\pm \,, J})^c = \eta_{\mp \,, J}$. Finally, by consistency the RR fluxes must decompose as 
\begin{equation}
	\mathbf{G} = d \text{vol}_{\mathcal{M}_4} \wedge \mathbf{\tilde{G}} + \mathbf{\hat{G}} \,, \qquad \mathbf{\tilde{G}} = \star_6 \lambda(\mathbf{\hat{G}}) \,,
\end{equation}
where $\star_6$ is the Hodge star operator on $\mathcal{M}_6$,  $d \text{vol}_{\mathcal{M}_4}$ is the volume form of $\mathcal{M}_4$ and $\lambda$ is the operator which reverses the order of the indexes. 

\subsubsection{A first look at the EOMs}

In order to explore the space of allowed compactifications we need to solve the equations of motion (\ref{eq:eomIIA}) with the ansatz (\ref{eq:ansatzCompact}). A simple class of solutions is obtained in absence of sources, vanishing KR and RR fields, no Romans mass and constant dilaton and warping factor. The result is that the manifold $\mathcal{M}_{10}$ must be Ricci-flat, as well as $\mathcal{M}_4$ and $X_6$. Then, $\mathcal{M}_4$ must be Minkowski space and $X_6$ a Ricci-flat space.

If we turn on the fluxes, i.e. we have non-trivial, quantized KR and RR fields and non-vanishing Romans mass, the dilaton and the warping factor can not be constant anymore. One can prove in this setup that if $X_6$ is a closed, smooth manifold, and we consider positive tension sources then the Ricci curvature of  $\mathcal{M}_4$ can not be non-negative. We have therefore a no go theorem \cite{Maldacena:2000mw} excluding 4-dimensional de Sitter and Minkowski vacua. The presence of fluxes in general will backreact, driving away $X_6$ from the Ricci flatness condition. 

If we turn on sources but we but we treat them in the smearing approximation, one can compensate the effects of fluxes backreaction and one can verify that solutions with Ricci-flat $X_6$ still exist. Moreover, negative tension extended objects (O-planes) contribute to the sign of the external Ricci curvature in the opposite direction with respect to the fluxes (and the positive tension extended objects), allowing to evade the no go theorem \cite{Giddings:2001yu}. Examples of Minkowski vacua have been successfully built. From a phenomenological perspective it is particularly interesting to consider the insertion of extended sources because they reduce the amount of supersymmetry of the $\mathcal{M}_4$ vacua.

\subsubsection{Supesymmetric backgrounds}

We want to study the constraints imposed by the requirement that in $\mathcal{M}_4$ we have $\mathcal{N} = 2$ supersymmetry. This is equivalent to the requirement that exist $\epsilon^i_\pm$ and $\eta^i_\pm$ such that we can decompose the 2 supersymmetry generators of type IIA $\epsilon^i$ as 
\begin{subequations} \label{eq:spinDeco}
	\begin{align}
		{\epsilon}^1 = \epsilon_{+}^1 \otimes \eta_{+}^1 + \epsilon_{-}^1 \otimes \eta_{-}^1 \,, \\[2mm]
		{\epsilon}^2 = \epsilon_{+}^2 \otimes \eta_{-}^2 + \epsilon_{-}^2 \otimes \eta_{+}^2 \,,
	\end{align}
\end{subequations}
with $\epsilon^i_\pm$ and $\eta^i_\pm$ Weyl spinors and $\epsilon^i$ Majorana--Weyl spinor. The decomposition (\ref{eq:spinDeco}) is well defined provided that the $\eta_{+}^i$ are globally defined and nowhere vanishing. Such condition for a single spinor is equivalent to request that the structure group of the theory, i.e. the group of the transitions maps, is reduced to the stabilizer of $\eta_{+}^i$, which is $SU(3)$ \cite{Grana:2005ny}. If the $\eta_{+}^i$ are parallel, they are not independent and we conclude that $X_6$ must be a manifold with $SU(3)$ structure. If the  $\eta_{+}^i$ are not everywhere parallel we have the most general case of an $SU(3) \times SU(3)$ structure.

Let's start considering an $SU(3)$ structure and let's call $\eta_\pm $ the independent spinor components normalized in such a way ${\eta}_\pm^\dagger \eta_\pm = 1/2$. We can then build the 2 objects
\begin{equation}
	J_{mn}  = - 2 i {\eta}_+^\dagger \gamma_7 \gamma_{mn} \eta_+ \,, \qquad
	\Omega  = \Omega^+ + i \Omega^- 
\end{equation}
with 
\begin{equation}
	\Omega^+_{mnp} = - 2 i {\eta}_-^\dagger \gamma_{mnp} \eta_+ \,,  \qquad  \bar{\Omega}_{mnp} = - 2 i{\eta}_+^\dagger \gamma_{mnp} \eta_- \,.
\end{equation}
Rising one index of $J_{mn}$ we obtain an operator which squares to $-\delta_m^n$. We can identify therefore $J_{m}{}^n$ with an almost complex structure and classify forms as $(p,q)$-forms with respect to it. Given that $J$ and $\Omega$ satisfy
\begin{equation}
	J \wedge J \wedge J = \frac{3}{4}i \Omega \wedge \bar{\Omega} \,, \qquad J \wedge \Omega = 0 \,,
\end{equation}
we can conclude that $J$ is of type $(1,1)$ and $\Omega$ is of type $(3,0)$. In the absence of fluxes, the vanishing supersymmetry transformations (\ref{eq:susy}) is equivalent to  the constraint
\begin{equation}
	\nabla_m \eta_\pm = 0 \,.
\end{equation}
If $\eta$ is covariantly constant then also $J$ and $\Omega$ are covariantly constant. In particular, this implies that $X_6$ is a complex K\"ahler manifold with holonomy group contained into  $SU(3)$, the stabilizer of $\eta_{\pm}$. It can be shown that this is equivalent to require that the manifold is Calabi-Yau. In general, to solve the supersymmetry conditions (\ref{eq:susy}) $\eta_\pm$ need not to be covariantly constant. We can then classify the possible solutions in terms of the intrinsic torsion of $\eta_\pm$, i.e. in terms of the obstructions of being covariantly constant. The common way to classify such obstructions is in terms of the the differential of $\Omega$ and $J$. We have \cite{Grana:2005ny}
\begin{subequations}
	\begin{align}
		& d J = \frac{3}{4}i (W_1 \bar{\Omega} - \bar{W}_1 \Omega) + W_4\wedge J + W_3 \,, \\
		& d \Omega = W_1 J^2 + W_2 \wedge J + \bar{W}_5\wedge \Omega \,,
	\end{align}
\end{subequations}
with $W_1$ a zero form, $W_{4,5}$ a 1-form, $W_2$ a 2-form and $W_3$ a 3-form.  $W_1$ parameterize the $(3,0)$ and $(0,3)$ parts of $dJ$ and the non-primitive $(2,2)$ part of $d\Omega$, $W_3$ the real primitive $(1,2)_0 \oplus (2,1)_0$ part of $dJ$, $W_4$ the real non-primitive $(2,1) \oplus (1,2)$ part of $dJ$, $W_2$ the primitive $(2,2)_0$ part of $d\Omega$ and $\bar{W}_5$ the $(3,1)$ part of $d\Omega$. A special class of $SU(3)$ structures we are going to use later is that of the \textit{half-flat} manifolds. They satisfy
\begin{equation}
	d \Omega^- = 0 \,, \qquad d (J \wedge J ) = 0 \,.
\end{equation}

Let's consider now the case of $SU(3) \times SU(3)$ structures. We define the bispinors 
\begin{equation}
	\Phi_\pm = \eta_+^1 \otimes \eta^2_\pm{}^\dagger\,.
\end{equation}
If $SU(3)$ structures were characterized in terms of $J$ and $\Omega$, $SU(3) \times SU(3)$ structures are characterized in terms of three real functions $\rho$, $\psi$ and $\theta$, a complex 1-form $v$, a real 2-form $j$ and a complex two form $\omega$. Such quantities can be extracted from the bispinors $\Phi_\pm$ through the relations
\begin{equation} \label{eq:ansatzbispinor}
	\Phi_+ = \rho \,  e^{i\theta} \exp[-i J_\psi] \,, \qquad \Phi_- = \rho \, \Omega_\psi  \,,
\end{equation}
with 
\begin{equation}
	J_\psi = \frac{1}{\cos(\psi)}j + \frac{i}{2 \tan(\psi)^2 } v\wedge\bar{v} \,, \qquad \Omega_\psi = v \wedge \exp[i\omega_\psi]\,,
\end{equation} 
and
\begin{equation}
	\omega_\psi = \frac{1}{\sin(\psi) }\left(\text{Re}\omega + \frac{i}{\cos(\psi) } \text{Im}\omega \right) \,.
\end{equation}
$v$, $j$ and $\omega$  are the data which defines the $SU(2)$ structure with respect to which both spinors transforms as singlets. The relations that they must satisfy can be deduced by those of an $SU(3)$ structure defining
\begin{equation}
	j = J - \frac{i}{2}v\wedge\bar{v}\,, \qquad \omega = \frac{1}{2} \iota_v \Omega \,.
\end{equation}
$\psi$ measures the departure from the parallel spinor condition. Finally, $\theta$ and $\rho$ are a phase and a rescaling factor.	The constraints associated with the vanishing of the supersymmetry transformations with external AdS$_4$ space can be written as \cite{Marchesano:2020qvg}
\begin{subequations} \label{eq:susyeq}
	\begin{align}
		& d_H \Phi_+ = - 2 \mu e^{-A} \text{Re}\Phi_- \,, \\[2mm]
		& d_H (e^A \text{Im}\Phi_- ) = - 3 \mu \text{Im} \Phi_+ + e^{4A} \star_6 \lambda{G} \,.
	\end{align}
\end{subequations}
where we absorbed the mean value of $e^{-A}$ into $\mu$, i.e. $\ell_s \mu = \sqrt{-\Lambda/3}$, with $\Lambda$ the cosmological constant. Replacing the ansatz (\ref{eq:ansatzbispinor}) into (\ref{eq:susyeq}) we obtain some geometrical constraints, a priori of the value of the KR and RR fields
\begin{subequations}
	\begin{align}
		& \rho = e^{3A - \phi} \sin \theta \,, \\[2mm]
		& d (e^{3A - \phi} \cos\psi \sin \theta) = 0\,, \\[2mm]
		& \text{Re} v = \frac{e^A}{2 \mu \sin \theta}d \theta  \,, \\[2mm]
		& d\left(\frac{1}{\sin \theta} J_\psi\right) = 2 \mu e^{-A} \text{Im} (v \wedge \omega_\psi) \,.
	\end{align}
\end{subequations}
The fluxes then are then completely fixed and satisfy 
\begin{equation}
	H = dB \,, \qquad \mathbf{G} = e^B \wedge \mathbf{F} \,,
\end{equation}
with
\begin{subequations}
	\begin{align}
		& B = - \cot(\theta) J_\psi + \tan \psi \, \text{Im} \omega \,, \\[2mm]
		& F_0 = - J_\psi \cdot d(e^{-\phi} \cos\psi \text{Im} v) + 5 \mu e^{-A - \phi} \cos\psi \cos \theta \,, \\[2mm]
		\begin{split}
			& F_2 = F_0 \cot \theta J_\psi - J_\psi \cdot d \text{Re}(\cos\psi e^{-\phi} v \wedge \omega_\psi)  + \mu \cos\psi e^{-A-\phi} \big[(5+2\tan^2 \psi) \sin\theta J_\psi  \\[1mm]
			& \qquad + 2 \sin \theta \text{Re} v\wedge \text{Im} v - 2 \cos \theta \tan^2\psi \text{Im} \omega_\psi \big]\,, 
		\end{split}\\[2mm]
		& F_4 = F_0 \frac{J_\psi^2}{2 \sin^2 \theta} + d \left[\cos\psi e^{-\phi} \left(J_\psi \wedge \text{Im}v - \cot \theta \text{Re}(v\wedge \omega_\psi)\right)\right] \,, \\[2mm]
		& F_6 = -\frac{1}{\cos^2\psi} d\text{vol}_{X_6} \left(F_0 \frac{\cos\theta}{\sin^3 \theta} + 3 \frac{\mu \cos\psi e^{-\phi}}{\sin \theta}\right) \,.
	\end{align}
\end{subequations}
A special class of $SU(3)\times SU(3)$ structures is obtained for small $\psi$, namely the $SU(3)\times SU(3)$ structures which are small deformations of an $SU(3)$ structure. In such a limit, $\Omega_\psi$ and $J_\psi$ can be interpreted as deformations of the $SU(3)$ structure forms $\Omega$ and $J$. At leading order in $\psi$ expansion we have 
\begin{subequations}
	\begin{align}
		& J_\psi = J + \mathcal{O}(\psi) =  j + \frac{i}{2 \tan^2\psi} v \wedge \bar{v} + \dots \,, \\[2mm]
		& \Omega_\psi = \Omega + \mathcal{O}(\psi) =  \frac{i}{\tan \psi} v \wedge \omega + \dots \,.
	\end{align}
\end{subequations}

\subsubsection{More on Calabi-Yau manifolds}

We briefly review Calabi--Yau (CY) manifolds. A $2n$-dimensional CY can be defined as a complex, compact, K\"ahler manifolds with holonomy group contained in $SU(n)$. An equivalent definition is requiring the manifold to have vanishing first Chern class instead of the condition on the holonomy group. Without entering into the details of Chern classes, the relevant property CYs satisfy is that they are exactly the compact, K\"ahler manifolds which admits a ricci-flat metric. More precisely, Yau's theorem says that given a compact K\"ahler manifold with closed K\"ahler form $J$ and associated metric $g$, exist and it is unique a closed K\"ahler form $J'$ in the same cohomology class of $J$ whose associated metric $g'$ is Ricci-flat.

$J$ is not the unique covariantly constant object we have. Exists and it is unique (up to normalization) the covariantly constant holomorphic $(n,0)$ form. For us are particularly interesting the 6-dimensional CY. Their Hodge diamond diagram is
\[
\begin{tikzcd}[row sep=tiny, column sep=tiny]
	& & & 1 & & \\[-1mm]
	& & 0 & & 0 & & \\[-1mm]
	& 0 & & h^{1,1} & & 0 & \\[-1mm]
	1 & & h^{2,1} & & h^{2,1} & & 1 & &  \\[-1mm]
	& 0 & & h^{1,1} & & 0 & &  \\[-1mm]
	& & 0 & & 0 & & \\[-1mm]
	& & & 1 & &
\end{tikzcd}
\]

\section{Branches of AdS$_4$ Calabi--Yau orientifold vacua}
\label{s:branch}

Our current understanding of  AdS$_4$ Calabi--Yau orientifold vacua is based on type IIA string theory compactified on a Calabi--Yau three-fold $X_6$. To this background we apply an orientifold quotient generated by $\Omega_p (-1)^{F_L}{\cal R}$,\footnote{Here $\Omega_p$ is the worldsheet parity reversal operator and ${F_L}$ is the space-time fermion number for the left-movers.} with ${\cal R}$ an anti-holomorphic involution of $X_6$ acting as ${\cal R} J_{\rm CY}=-J_{\rm CY}$, ${\cal R}\Omega_{\rm CY} = - \overline{\Omega}_{\rm CY}$ on its K\"ahler two-form and holomorphic three-form, respectively. The fixed locus $\Pi_{\rm O6}$ of ${\cal R}$ is made of  3-cycles which satisfy 
\begin{equation}
	J_{CY} \big \vert_{\Pi_{\rm O6}} = 0 \,, \qquad \text{Re} \Omega_{\rm CY} \big\vert_{\Pi_{\rm O6}} = 0 \,.
\end{equation}
They are special Lagrangian 3-cycles \cite{Hitchin:1999fh} calibrated by $\text{Im}\Omega_{\rm CY}$.\footnote{This statement implies \cite{mclean1990deformations} that $\text{Im}\Omega_{CY}$ is the volume form of the special Lagrangian cycles and that its integral on 3-cycles in the same homology class is smaller or equal to the cycles volumes. Thus, a calibrated manifold is volume minimizing in its holomogy class.} The presence of O6-planes reduces the background supersymmetry to 4d $\CN=1$, and induces an RR tadpole that can be cancelled by a combination of D6-branes wrapping special Lagrangian three-cycles \cite{Blumenhagen:2005mu,Blumenhagen:2006ci,Marchesano:2007de,Ibanez:2012zz}, D8-branes wrapping coisotropic five-cycles \cite{Font:2006na}, and background fluxes including the Romans mass. If background fluxes are involved, one recovers a metric background of the form 
\begin{equation}\label{eq:warped-product}
	ds^2 = e^{2A}ds^2_{\mathrm{AdS}_4} + ds^2_{X_6}\, ,
\end{equation}
with $A$ a function on $X_6$. This may either correspond to a 4d $\CN=1$ or $\CN=0$ vacuum. 

If O6-planes and background D-branes are treated as localized sources, the warping function $A$ is non-constant. Similarly, we have a 10d dilaton $e^\phi$ varying over $X_6$ with an average value $g_s$, and a metric on $X_6$ which is no longer Calabi--Yau, but should instead be a deformation to a $SU(3)\times SU(3)$ structure metric. This picture is based on the results of \cite{Junghans:2020acz,Marchesano:2020qvg}, which provided explicit approximate solutions for the 10d equations of motion and Bianchi identities of massive type IIA supergravity. Their key ingredient is an expansion of the said equations in a small parameter, which in the case at hand can be taken to be either $g_s$ or $|\hat{\mu}| = \ell_s/R$, the AdS$_4$ scale in 10d string frame and in string length $\ell_s  =  2\pi \sqrt{\a'}$ units \cite{Saracco:2012wc}. The zeroth order of the expansion treats $\delta$-function sources like O6-planes and D6-branes as if they were smeared over $X_6$, yielding a particularly simple solution with constant warping and dilaton, and a Calabi--Yau metric. The localized nature of these sources is already manifest in the first non-trivial correction to this background, which also displays the said deformation away from the Calabi--Yau metric.

The advantage of the smearing zeroth-order approximation is that it gives a direct connection with the 4d effective approach to describe these vacua. In the 4d picture one considers the set of moduli present in a large-volume Calabi--Yau compactification without fluxes, and a scalar potential generated by flux quanta that stabilizes them at certain vevs. The 4d approach reveals an interesting vacua structure already in the case of toroidal orientifold compactifications \cite{Derendinger:2004jn,Villadoro:2005cu,DeWolfe:2005uu,Camara:2005dc}, and it can be generalized to arbitrary Calabi--Yau geometries thanks to the simple form of the scalar potential in the large-volume regime \cite{Herraez:2018vae,Escobar:2018tiu,Escobar:2018rna,Marchesano:2019hfb,Marchesano:2020uqz}. In the following we  review the results of \cite{Marchesano:2019hfb}, which obtained several branches of supersymmetric and non-supersymmetric vacua using the 4d approach on arbitrary Calabi--Yau orientifold geometries. 

Calabi--Yau orientifold vacua can be described by a set of relations between the Calabi--Yau metric forms $\Omega_{\rm CY}$ and $J_{\rm CY}$ and the  background fluxes. To describe the latter it is convenient to use the democratic formulation of type IIA supergravity \cite{Bergshoeff:2001pv}, in which all RR potentials are grouped in a polyform ${\bf C} = C_1 + C_3 + C_5 + C_7 + C_9$ and so are their gauge invariant field strengths
\be
{\bf G} \,=\, d_H{\bf C} + e^{B} \wedge {\bf \bar{G}} =  d{\rm vol}_4 \wedge \tilde{G} + \hat{G} \, .
\label{bfG}
\ee
Here $H$ is the three-form NS flux, $d_H \equiv (d - H \wedge)$ is the $H$-twisted differential and ${\bf \bar{G}}$ a formal sum of closed $p$-forms on $X_6$. The second equality is specific to the metric background \eqref{eq:warped-product}, with ${\rm vol}_4$ the AdS$_4$ volume form, $\tilde{G}$ and $\hat{G}$ only have internal indices and satisfy the relation $\tilde{G} = - \lambda ( \star_6 \hat{G})$, and where $\lambda$ is the operator that reverses the order of the indices of a $p$-form. 
The Bianchi identities for these field strengths read
\begin{equation}\label{IIABI}
	\ell_s^{2} \,  d (e^{-B} \wedge {\bf G} ) = - \sum_\a \lambda \left[\delta (\Pi_\alpha)\right] \wedge e^{\frac{\ell_s^2}{2\pi} F_\alpha} \, ,  \qquad d H = 0 \, ,
\end{equation} 
where $\Pi_\alpha$ hosts a D-brane source with a quantized worldvolume flux $F_\alpha$, and $\delta(\Pi_\alpha)$ is the bump $\delta$-function form with support on $\Pi_\alpha$ and indices transverse to it, such that $\ell_s^{p-9} \d(\Pi_\a)$ lies in the Poincar\'e dual class to $[\Pi_\a]$. O6-planes contribute as D6-branes but with minus four times their charge and $F_\alpha \equiv 0$. In the absence of localized sources, each $p$-form within ${\bf \bar{G}}$ is quantized, so one can define the internal RR flux quanta in terms of the following integer numbers
\begin{equation}
	m \, = \,  \ell_s G_0\, ,  \quad  m^a\, =\, \frac{1}{\ell_s^5} \int_{X_6} \bar{G}_2 \wedge \tilde \omega^a\, , \quad  e_a\, =\, - \frac{1}{\ell_s^5} \int_{X_6} \bar{G}_4 \wedge \omega_a \, , \quad e_0 \, =\, - \frac{1}{\ell_s^5} \int_{X_6} \bar{G}_6 \, ,
	\label{RRfluxes}
\end{equation}
with $\omega_a$, $\tilde \omega^a$ integral Calabi--Yau-harmonic two- and four-forms such that $\ell_s^{-6} \int_{X_6} \omega_a \wedge \tilde{\omega}^b = \delta_a^b$, in terms of which we can expand the K\"ahler form as
\be
J_{\rm CY} = t^a \omega_a\, , \qquad - J_{\rm CY} \wedge J_{\rm CY} = {\cal K}_a \tilde{\omega}^a\, . 
\ee
Here ${\cal K}_a \equiv {\cal K}_{abc} t^bt^c$, with ${\cal K}_{abc} = - \ell_s^{-6} \int_{X_6} \omega_a \wedge \omega_b \wedge \omega_c$ the Calabi--Yau triple intersection numbers and $-\frac{1}{6} J_{\rm CY}^3 = - \frac{i}{8} \Omega_{\rm CY} \wedge \bar{\Omega}_{\rm CY}$ its volume form.

Even in the presence of localized sources, \eqref{RRfluxes} are taken as integer flux quanta that together with the H-flux quanta enter the F-term scalar potential. The latter has a series of extrema that have been classified in \cite{Marchesano:2019hfb}. In the following we consider four of the branches of vacua found therein, dubbed as class {\bf S1}. They consist of one infinite family of supersymmetric vacua and three non-supersymmetric ones. Given the 4d moduli stabilization data, which in the conventions of this chapter is reviewed in \cite[Appendix A]{Marchesano:2021ycx}, one obtains that the background fluxes describing such vacua must obey the following relations:
\be
[ H ]  = 6A G_0 g_s  [\re \Omega_{\rm CY} ] \, , \qquad \frac{1}{\ell_s^6} \int_{X_6} {G}_2 \wedge \tilde{\omega}^a =  BG_0 t^a\, ,  \qquad -\frac{1}{\ell_s^6} \int_{X_6} \hat{G}_4  \wedge \omega_a  =  CG_0 {\cal K}_a  \, , 
\label{intflux}
\ee
together with $\hat{G}_6  =  0$. Here $A, B, C \in \pr$ are constants that index the different branches, see table \ref{vacuresul} for their specific values. The stabilization of Calabi--Yau moduli in terms of flux quanta follows from these relations and
\be
\hat{e}_a \equiv e_a - \oh \frac{\cK_{abc} m^bm^c}{m}  = \left( C - \oh B^2\right) m {\cal K}_a \, .
\label{hate}
\ee

An important feature of these vacua is that the quanta of $H$-flux and $G_0$ are constrained by the RR-flux Bianchi identities, that in the presence of O6-planes and D6-branes read
\be
dG_0 = 0\, , \qquad d G_2 = G_0 H - 4 \d_{\rm O6} +   N_\a \d_{\rm D6}^\a \, ,  \qquad d \hat{G}_4 = G_2 \wedge H\, , \qquad d\hat{G}_6 = 0\, ,
\label{BIG}
\ee
we have defined $\d_{\rm D6/O6}\equiv \ell_s^{-2}  \d(\Pi_{\rm D6/O6})$ and $N_\a$ is the number D6-branes wrapping the three-cycle $\Pi^{\rm D6}_\a$. This in particular implies that
\be
{\rm P.D.} \left[4\Pi_{\rm O6}- N_\a \Pi_{\rm D6}^\a\right] = m [\ell_s^{-2} H]  \implies mh +N = 4 \, ,
\label{tadpole}
\ee
where to arrive to the last equation we have taken the simplifying choice P.D.$[\ell_s^{-2}H] = h [\Pi_{\rm O6}] = h [\Pi_{\rm D6}^\a]$, $\forall \a$. In all branches $A>0$, so it follows from \eqref{intflux} and that all sources are calibrated by $\im\, \Omega_{\rm CY}$ that $0 < mh \leq 4$. The remaining flux quanta $e_a, m^a$ are however unconstrained by RR tadpole conditions, and so one can choose them freely to fix ${\cal K}_a$ arbitrarily large. As one does, it is driven to a region of larger Calabi--Yau internal volume ${\cal V}_{\rm CY}  =  \frac{1}{6} {\cal K}_{abc}t^at^bt^c \equiv  \frac{1}{6} {\cal K}$, weaker 10d string coupling $g_s$ and smaller AdS$_4$ curvature. The latter is given by
\be
\mu = G_0 g_s \frac{2}{3}\sqrt{C^2 + \frac{1}{8}B^2}\, ,
\label{mu}
\ee
again measured in the 10d string frame. 

\begin{table}[H]
	\begin{center}
		\scalebox{1}{%
			\begin{tabular}{| c || c | c | c | c |c |c|}
				\hline
				Branch & $A$  & $B$  & $C$  & $\mu$  & SUSY & pert. stable \\
				\hline \hline
				\textbf{A1-S1}$+$  &$\frac{1}{15}$ &  $0$  & $\frac{3}{10}$  & $\frac{1}{5} G_0 g_s$ & Yes & Yes  \\ \hline
				\textbf{A1-S1}$-$  &$\frac{1}{15}$  & $0$  & $-\frac{3}{10}$  & $\frac{1}{5} G_0 g_s$ & No & Yes \\ 
				\hline
				\textbf{A2-S1}$\pm$    & $\frac{1}{12}$  & $\pm\frac{1}{2}$  & $-\frac{1}{4}$  & $\frac{1}{\sqrt{24}} G_0 g_s$ 	& No & Yes \\ 
				\hline
		\end{tabular}}      
	\end{center}
	\caption{Different branches of {\bf S1} solutions found in  \cite{Marchesano:2019hfb}. \label{vacuresul}}
\end{table}

Table \ref{vacuresul} shows the four different branches of solutions found in  \cite{Marchesano:2019hfb} that correspond to the relations \eqref{intflux}, with the different values for  $A, B, C$. The branch \textbf{A1-S1}$+$ corresponds to the infinite family of supersymmetric solutions found in \cite{DeWolfe:2005uu}, while \textbf{A1-S1}$-$ represents non-supersymmetric vacua whose four-form flux harmonic piece has a sign flip compared to the supersymmetric case. Just like their supersymmetric cousins, these non-supersymmetric vacua have a simple, universal flux-induced mass spectrum absent of tachyons below the BF bound \cite{Marchesano:2019hfb}. Finally, the branches \textbf{A2-S1}$\pm$ correspond to non-supersymmetric vacua that have been less studied in the literature. While their mass spectrum is harder to analyze in general (see \cite{Quirant:2022fpn} for the case of toroidal geometries), one can show that the potential is positive semidefinite \cite{Marchesano:2019hfb}, and therefore they are perturbative stable as well.

Since they only differ by the value of the constants $A, B, C$, all these branches have the same parametric dependence on their Kaluza--Klein and AdS scales for larges values of $\hat{e}$. In particular they reproduce the scaling $m_{\rm KK} \sim \hat{e}^{1/2} \mu$ observed in  \cite{DeWolfe:2005uu} for the supersymmetric branch that leads to parametric scale separation. If this estimate of scales survives the 10d description of these vacua, it means that we can trust our 4d effective analysis, and in particular the perturbative stability obtained from it. In the next section we will address the 10d description of all these branches, extending the analysis of \cite{Marchesano:2020qvg,Marchesano:2021ycx}. We will see that from the smearing approximation one can rederive table \ref{vacuresul}, and then provide the first-order correction to this approximation, that describes localized sources. Since in principle this confirms the perturbative stability of such vacua, we turn to analyze their non-perturbative stability in section \ref{s:membranes}.


\section{10d uplift and localized sources}
\label{s:10d}

In this section we recover the 4d results reviewed above from a 10d viewpoint. As we will see, the four branches of {\bf S1} solutions can also be obtained by solving the equations of massive type IIA supergravity up to a certain order in a perturbative expansion, following \cite{Saracco:2012wc,Junghans:2020acz,Marchesano:2020qvg}. We first show that solving the equations at zeroth order, in which localized sources appear to be smeared, already reproduces the four different branches of table \ref{vacuresul}. We then proceed to show that the solution for each of these branches of vacua can be extended to the first order in the perturbative expansion, where space-time O6-planes and D6-branes are treated as localized sources  in the internal dimensions.

\subsection{Smearing approximation}

Let us first address the 10d equations of motion and Bianchi identities in the smearing approximation. Since we are not restricted to supersymmetric backgrounds, we will follow the general approach of \cite{Junghans:2020acz}. In such a formalism, after making a perturbative expansion of the 10d equations,  one obtains that the zeroth order 10d equations are described by a smearing approximation, which is defined by means of the following prescription:

\begin{itemize}
	
	\item The metric on $X_6$ is taken to be Calabi--Yau, the warp factor dilaton to be constant, and the background fluxes to have a harmonic $p$-form profile in this metric. This implies that the flux Ansatz \eqref{intflux} is approximated by the following, more specific flux background
	\be
	H   = 6A G_0 g_s  \re \Omega_{\rm CY}  \, , \qquad  {G}_2 =  BG_0 J_{\rm CY}\, ,  \qquad  \hat{G}_4   =  CG_0  J_{\rm CY} \wedge  J_{\rm CY}  \, , \qquad\hat{G}_6  =  0\, . 
	\label{intfluxsm}
	\ee
	
	\item The three-form bump $\delta$-functions that appear in the Bianchi identities \eqref{BIG} are replaced by harmonic representatives in the same homology class. Taking in addition the simplifying choice of eq.\eqref{tadpole} one obtains that the only non-trivial Bianchi identity at this level reads
	\be
	d G_2 = G_0 H - mh \delta_{\rm O6}^{\rm h} = 0\, ,
	\label{BIG2sm}
	\ee
	where $ \delta_{\rm O6}^{\rm h}$ is the harmonic piece of the three-form bump $\delta$-function $\delta_{\rm O6}$. 
	
	\item The $\delta$-like sources $\delta_\a^{(3)}$ that appear in the dilaton and Einstein equations are replaced by constant terms describing their zero mode in a Fourier expansion. Assuming that all three-cycles wrapped O6-plane and D6-brane are calibrated by $\Im \Om$, as we will do in the following, one can relate these localized sources with the three-form bump functions as 
	\be
	\delta^{(3)}_{\a} \equiv \star_6 (\im \Omega \wedge \delta(\Pi_{\a})) \simeq \star_{\rm CY} (\im \Omega_{\rm CY} \wedge \delta(\Pi_{\a})) \to \frac{{\cal V}_{\Pi_\a}}{{\cal V}_{\rm CY}}\, ,
	\label{deltasm}
	\ee
	where in the second step we have taken the Calabi--Yau metric approximation and in the third one we have replaced the $\delta$-function by its zero mode. Here ${\cal V}_{\Pi_\a}$ is the volume of the three-cycle $\Pi_\a$ measured in string units. 
	
\end{itemize}

Applying these prescriptions to the 10d massive type IIA supergravity equations, one obtains a set of constraints on the parameters $A, B, C \in \pr$. In particular, from the Bianchi identity \eqref{BIG2sm} one obtains
\be
\frac{mh}{\ell_s^{2}}\frac{{\cal V}_{\Pi_{\rm O6}}}{{\cal V}_{\rm CY}} = 24AG_0^2 g_s \, ,
\label{BIG2c}
\ee
where we have also made use of \eqref{deltasm}. Additionally, by plugging \eqref{intfluxsm} into the equations of motion for the background fluxes one obtains that the only non-trivial equation is 
\be
\label{fluxeomsm}
{G}_2 \wedge  \star_{\rm CY} \hat{G}_4  + {G}_0 \star_{\rm CY} {G}_2  = - G_0^2\,  J_{\rm CY} \wedge  J_{\rm CY}\, B\left(2C+ \oh\right)\, ,
\ee
see Appendix \ref{ap:10deom} for details. Solving this equation already constrains the parameters of our Ansatz, to either satisfy $B=0$ or $C=-1/4$. Notice that, in the language of \cite{Marchesano:2019hfb}, these two choices precisely correspond to the branches {\bf A1} and {\bf A2}, respectively.

Finally, one must apply the above prescription to the 10d Einstein and dilaton equations of motion. One obtains the following relations
\bes
\label{10dEinsteinsm}
\begin{align}
	\label{10ddilsm}
	\mu^2 & = \frac{G_0^2g_s^2}{72} \left( 144A^2 +3B^2 +36C^2-1\right)\, , \\
	\label{10dE1sm}
	\frac{mh}{\ell_s^{2}}\frac{{\cal V}_{\Pi_{\rm O6}}}{{\cal V}_{\rm CY}} & = \frac{G_0^2g_s}{3} \left( 576A^2 +3B^2 +36C^2-1\right)\, , \\
	\label{10dE2sm}
	\frac{mh}{\ell_s^{2}}\frac{{\cal V}_{\Pi_{\rm O6}}}{{\cal V}_{\rm CY}} & = \frac{G_0^2g_s}{6} \left( 1584A^2 +3B^2 +84C^2-5\right)\, . 
\end{align}
\ees
From \eqref{10dE1sm} and  \eqref{10dE2sm} one finds 
\be
144A^2-1 = B^2 -4C^2\, ,
\ee
which plugged into \eqref{10ddilsm} reproduces \eqref{mu}. Additionally, using \eqref{BIG2c} and \eqref{10dEinsteinsm} one obtains that
\be
72A = 3+7B^2+20C^2\, .
\ee
These last two equations and \eqref{fluxeomsm} completely determine the allowed values for the parameters of our flux Ansatz. For the branch {\bf A1} one recovers $A =1/15$ and $C=\pm3/10$, while for the branch {\bf A2} one finds $A=1/12$ and $B =\pm 1/2$, precisely reproducing the content of table \ref{vacuresul}.

\subsection{First-order corrections and localization}

Let us now proceed beyond the smearing approximation and solve the 10d equations at the next order in the $g_s$ expansion. For this we follow the same strategy as in \cite{Marchesano:2021ycx}, and combine the results of \cite{Junghans:2020acz} and \cite{Marchesano:2020qvg}. More precisely, we consider the same metric and dilaton background obtained in \cite{Marchesano:2020qvg} for the supersymmetric case, and then we apply the approach in \cite{Junghans:2020acz} to obtain the flux background that solves the 10d equations at the same order of approximation. 

In the first-order solution found in \cite{Marchesano:2020qvg}, the background corresponding to \eqref{eq:warped-product} is described by a $SU(3)\times SU(3)$ structure metric on $X_6$ and a varying dilaton and warp factor of the form
\begin{subequations}	
	\label{solutionsu3}
	\begin{align}
		J & = J_{\rm CY} + \cO(g_s^2) \, , \qquad   \qquad  \Omega  = \Omega_{\rm CY} + g_s k +  \cO(g_s^2)\, , \\
		e^{-A}  & = 1 + g_s \varphi + \cO(g_s^2) \, , \qquad e^{\phi}   = g_s \left(1 - 3  g_s \varphi\right) + \cO(g_s^3)\, ,
	\end{align}
\end{subequations}   
where $k$ is a (2,1) primitive current  and $\varphi$ a real function that satisfies $\int_{X_6} \varphi = 0$. These two quantities are obtained by solving the Bianchi identity \eqref{BIG} for $G_2$ at the given order of approximation in the $g_s$ expansion. Expressing the internal two-form flux as
\be
G_2 = G_2^{\rm h} + d^\dag_{\rm CY} K + \cO(g_s)\, ,
\ee
where $G_2^{\rm h}$ is given by the smeared profile in \eqref{intfluxsm}, and $K$ is three-form current satisfying
\begin{equation}
	\Delta_{\rm CY} K = G_0H +  \delta_{\rm O6+D6}    = 6A G_0^2 g_s  \re \Omega_{\rm CY} -  mh \delta_{\rm O6}  +  \cO(g_s^2)\, ,
	\label{eq: K equation}
\end{equation}
where we have defined $\Delta_{\rm CY} = d^\dag_{\rm CY} d + d d^\dag_{\rm CY}$ and used \eqref{tadpole} and \eqref{intfluxsm}. The harmonic piece of the RHS of this equation vanishes due to \eqref{tadpole}, or equivalently due to \eqref{BIG2sm}. Hence there is always a solution for $K$, which at this order of approximation is of the form
\be
K = \varphi \re \Omega_{\rm CY}  + \re k \, ,
\label{formK}
\ee
with $\varphi$ satisfying a Laplace equation with $\delta$-sources on top of the O6-planes and D6-branes, see \cite{Casas:2022mnz} for a detailed discussion and several explicit examples. 

Given the above metric, dilaton and two-form flux background one may look for the profiles of the remaining internal fluxes such that {\it i)} they reduce to the smeared values \eqref{intfluxsm} at the lowest order in the $g_s$ expansion and {\it ii)} they solve the 10d equations of massive type IIA supergravity at the next order in the same expansion. This exercise is carried out in Appendix \ref{ap:10deom}, with the following result
\begin{subequations}
	\label{solutionflux}
	\begin{align}
		H & =   6A G_0 g_s \left(\re \Omega_{\rm CY} + R g_s K \right) -\frac{S}{2}   d\re \left(\bar{v} \cdot \Omega_{\rm CY} \right) + \cO(g_s^{3}) \label{H3sol} \, , \\
		\label{G2sol}
		G_2 & =    BG_0 J_{\rm CY} - J_{\rm CY} \cdot d(4 \varphi \im \Omega_{\rm CY} - \star_{\rm CY} K) + \cO(g_s) \, , \\
		G_4 & =   G_0 J_{\rm CY} \wedge J_{\rm CY} \left(C  - 12A g_s \varphi \right)+  S J_{\rm CY} \wedge g_s^{-1} d \im v + \cO(g_s^2) \, , \\
		G_6 & = 0\, ,
	\end{align}
\end{subequations}   
where $R, S \in \pr$ and $v$ is a (1,0)-form determined by
\be \label{ansatzVF}
v  = g_s \p_{\rm CY} f_\star + \cO(g_s^3)\, , \qquad \text{with} \qquad \Delta_{\rm CY} f_\star  = - g_s 8 G_0 \varphi \, .
\ee 
This background has the same form as in the supersymmetric case, and only differs \bk{on} the values that the constants $A,B,C,R,S$ take, which are different for each branch. The value of the new constants $R$ and $S$ are in fact determined by those that already appear in the smearing approximation, as follows
\be
6AR = 12A + 2C -1\, , \qquad S = 6A + 2C\, ,
\ee
yielding the content of table \ref{vacuresulns}.
\begin{table}[H]
	\begin{center}
		\scalebox{1}{%
			\begin{tabular}{| c || 	c | c | c | c |c |c|c|c|}
				\hline
				Branch & $A$  & $B$  & $C$  & $R$ & $S$ & $\mu$  & SUSY & pert. stable \\
				\hline \hline
				\textbf{A1-S1}$+$  &$\frac{1}{15}$ &  $0$  & $\frac{3}{10}$  & $1$ & $1$ & $\frac{1}{5} G_0 g_s$ & Yes & Yes  \\ \hline
				\textbf{A1-S1}$-$  &$\frac{1}{15}$  & $0$  & $-\frac{3}{10}$  & $-2$ & $-\frac{1}{5}$ & $\frac{1}{5} G_0 g_s$ & No & Yes \\ 
				\hline
				\textbf{A2-S1}$\pm$   & $\frac{1}{12}$  & $\pm\frac{1}{2}$  & $-\frac{1}{4}$ &$-1$ & $0$ & $\frac{1}{\sqrt{24}} G_0 g_s$ 	& No & Yes \\ 
				\hline
		\end{tabular}}      
	\end{center}
	\caption{Different branches of {\bf S1} solutions found in  \cite{Marchesano:2019hfb}, beyond the smearing approximation. \label{vacuresulns}}
\end{table}

These results suggest that we can have a 10d description for each of the vacua in table \ref{vacuresul} in which the internal geometry is well approximated by the Calabi--Yau metric, which is more and more accurate for larger values of $\hat{e}$, hence smaller values of $g_s$. In such a regime, our 4d estimate for the Kaluza--Klein scale is accurate, and below it we can trust our 4d effective potential, including the values for the flux-induced moduli masses. As a result, our 10d backgrounds should be free of perturbative instabilities, including those which belong to non-supersymmetric branches. It however remains to analyze their non-perturbative decay channels, and in particular those mediated by nucleating 4d membranes, which we now turn to discuss.


\section{4d membranes and non-perturbative instabilities}
\label{s:membranes}

To detect non-perturbative instabilities of the vacuum triggered by membrane nucleation one may follow \cite{Aharony:2008wz,Narayan:2010em} and consider probe 4d membranes that extend along a hyperplane $z = z_0$ within the Poincar\'e patch of AdS$_4$
\be
ds^2_4 =e^{\frac{2z}{R}} (-dt^2 + d\vec{x}^2) + dz^2\, ,
\label{PPatch}
\ee
where $R= |\mu|^{-1}$ is the AdS length scale, $\vec{x} = (x^1, x^2)$, and all coordinates range over $\pr$. A membrane with non-trivial tension $T$ will naturally be dragged towards $z \to -\infty$, except if it couples as $-\int C_3$ to a background four-form flux with vev $Q$
\be
\langle F_4 \rangle = -\frac{3Q}{R} d{\rm vol}_4
\qquad \Longrightarrow \qquad \langle C_3 \rangle = Q\, e^{\frac{3z}{R}} dt \wedge dx^1 \wedge dx^2 \, .
\label{3form}
\ee
We can interpret $Q$ as the  membrane charge with respect to a normalized three-form potential. 
Whenever $Q=T$  the energy dependence on $z_0$  due to the membrane tension cancels out with the potential energy $-\int \langle C_3 \rangle$ due to its charge. Moving the membrane along the coordinate $z$ is then a flat direction, as expected for BPS membranes. In fact, as argued in \cite{Koerber:2007jb}, membranes of this sort with $Q=T$ and near the AdS$_4$ boundary $z_0 \to \infty$ capture the BPS bound of spherical membranes in global coordinates at asymptotically large radius. This is particularly relevant for the stability of the vacuum, since it is precisely the domain walls that correspond to spherical membranes near the AdS boundary that determine whether the non-perturbative decay of one vacuum to another with lower energy is favourable or not. In this sense, one may interpret a membrane with $Q=T$ as mediating a marginal decay as it happens between supersymmetric vacua, while one with $Q>T$ is likely to signal a non-perturbative instability of the vacuum.\footnote{This correspondence typically assumes a thin domain wall, which is not always a good approximation.} 

Interestingly, the Weak Gravity Conjecture applied to 4d membranes implies that at each vacuum there must be one membrane with $Q \geq T$, for each independent membrane charge. Moreover, the refinement made in \cite{Ooguri:2016pdq} proposes that this inequality is only saturated in supersymmetric vacua. In non-supersymmetric vacua there should be a membrane with $Q>T$ for each independent membrane charge, therefore signalling an instability. In this section we consider these proposals in the context of the AdS$_4$ orientifold vacua of section \ref{s:branch}, following the same strategy as in \cite{Marchesano:2021ycx}, namely computing $Q$ and $T$ via dimensional reduction of D-brane actions. As we will see, the key observation to satisfy the WGC for 4d membranes is to consider branes whose internal dimensions are threaded by non-diluted worldvolume fluxes. 

\subsection{4d membrane charges and their Weak Gravity Conjecture}

In order to check the WGC and its refinement for the DGKT-like vacua of section \ref{s:branch}, let us start by reviewing and extending the results of \cite{Marchesano:2021ycx}, which addressed this question for {\bf A1-S1} vacua. First of all, one should make precise the WGC statement, in the sense that one should describe the set of independent membrane charges in these vacua. Naively, one would associate the set of membrane charges with the lattice of fluxes, as described by the $H$-flux and RR flux quanta \eqref{RRfluxes}. 
However, some of the points in this lattice do not correspond to independent flux quanta, as they are related to each other by large gauge transformations involving periodic shifts of the axions \`a la axion monodromy, see e.g. \cite{Berasaluce-Gonzalez:2012awn,Marchesano:2014mla,Herraez:2018vae}. After such identifications one is left with a set of membranes with torsional charges, that are related to discrete three-form gauge symmetries  \cite{Berasaluce-Gonzalez:2012awn,Buratti:2020kda}. It is not clear if the WGC should apply to such torsional membrane charges, but in the following we will not consider them. Instead, we will focus on those 4d fluxes that do not couple to any axion. In general, one can describe their quanta by using the set of flux invariants defined in \cite{Marchesano:2020uqz}. In DGKT-like vacua, such invariants reduce to the $H$-flux quanta, the Romans' parameter $m$ and the flux combinations $\hat{e}_a$. Hence, in a given vacuum one should look for membranes that, as one crosses them towards $z\to -\infty$, they make one of these flux quanta jump and take us to a vacuum with lower energy, or equivalently with larger AdS$_4$ scale $|\mu|$. In practice this amounts to jumps that decrease $|\hat{e}_a|$ and/or increase $|m|$ or $|h|$. Notice that the last two are constrained by the tadpole condition \eqref{tadpole}, and so in some cases it is not possible to increase their value. In those cases only the membranes that change $|\hat{e}_a|$ should be considered. 

As in \cite{Marchesano:2021ycx}, we only consider those 4d membranes that arise from wrapping D$(2p+2)$-branes on $2p$-cycles of $X_6$. Such membranes couple to the dynamical fluxes of the 4d theory \cite{Lanza:2019xxg,Lanza:2020qmt}, which include the flux quanta $m$ and $\hat{e}_a$ and exclude the $H$-flux quanta. The charge of each of these membranes can be obtained by dimensionally reducing their Chern-Simons action, which couples to the appropriate component of the flux polyform $\tilde{G}$ defined in \eqref{bfG}. Similarly to \cite{Marchesano:2021ycx}, one finds that in the smearing approximation this is equivalent to use (\ref{3form}) with
\begin{equation}
	\langle F_4 \rangle = \frac{1}{\ell_s^{2p +3}} \left(\int_{2p} \tilde{G}_{2p} \right) {\rm vol}_4 \,.
\end{equation}
The different charges read
\be
Q_{\rm D2} = 0 \, , \qquad Q_{\rm D4} =  \frac{ C }{D} \eta q_{\rm D4} T_{\rm D4}  \, , \qquad Q_{\rm D6} =  \frac{B}{2D} \eta q_{\rm D6}   T_{\rm D6} \, , \qquad Q_{\rm D8} =  -\frac{\eta q_{\rm D8}}{2D} T_{\rm D8}\, ,
\label{QDGKT}
\ee
where we have assumed that a D4-branes wraps a holomorphic curve $\Sigma$, a D6-brane a divisor ${\cal S}$, and D8-branes the whole of $X_6$, so their tension in 4d Planck units is given by\footnote{We go to Planck units using the relation among the 4-dimensional Planck mass $M_P$, the 4-dimensional dilaton $\phi_4$ and the string length $\ell_s$ \begin{equation} \label{eq:planck}
	M_P	\ell_s^{-1} = \sqrt{2} e^{\phi_4} = \sqrt{2} e^{\phi}  {\cal V}_{\rm CY}^{-1} \,.
\end{equation}}
\be
T_{\rm D2} = 1 \, , \qquad T_{\rm D4} = e^{K/2}{\cal V}_\Sigma \, , \qquad T_{\rm D6} =  e^{K/2}{\cal V}_{\cal S} \, , \qquad T_{\rm D8} =  e^{K/2}{\cal V}_{\rm CY}\, ,
\label{TDGKT}
\ee
with ${\cal V}$  the volume of each cycle in string units and $K$ given by
\begin{equation}
	K =  4 \phi_4 - \log \left(\frac{4}{3} {\cal K} \right) \,.
\end{equation}
In the EFT perspective, $K$ is the K\"ahler potential (see \cite{Marchesano:2021ycx}). The orientation of the cycle, or equivalently if we consider a D-brane or an anti-D-brane, is encoded in $q_{{\rm D}(2p+2)} =\pm1$. Finally we have defined
\be
\eta = {\rm sign}\, m\, , \qquad D= \sqrt{C^2 + \frac{1}{8}B^2}\, .
\ee
It is easy to see that these results reproduce those in sections 3 and 4 of \cite{Marchesano:2021ycx}. In there the branches {\bf A1-S1} were considered, for which $B=0$ and so $Q_{\rm D4} = \eta \eta_C q_{\rm D4}  T_{\rm D4}$, with $\eta_C = {\rm sign}\, C$. One just needs to choose $ q_{\rm D4}$ such that $\eta \eta_C q_{\rm D4} = 1$ for the extremal condition $Q=T$ to be met. As expected,  this choice corresponds to the D4-branes that decrease the value of $|\hat{e}_a|$ \cite{Marchesano:2021ycx}. In particular, the case $m >0$ selects D4-branes for supersymmetric {\bf A1-S1}$+$ vacua and anti-D4-branes for the non-supersymmetric branch {\bf A1-S1}$-$. The opposite choice leads to $Q=-T$. 

As also pointed out in \cite{Marchesano:2021ycx}, the energetics of D8-branes is more involved that for the rest, because they have an excess of space-time-filling D6-branes ending on them and stretching along $z \in [z_0, \infty)$ for $\eta  q_{\rm D8} = 1$, and along  $z \in (-\infty, z_0]$ for $\eta  q_{\rm D8} = -1$. Their presence contributes to the forces acting on the D8-brane transverse position, so that it can be encoded in an effective D8-brane charge. Generalising the computations in  \cite{Marchesano:2021ycx} one finds that
\be
Q_{\rm D8}^{\rm eff} =  \frac{24A-1}{2D} \eta q_{\rm D8} T_{\rm D8}\, ,
\label{QD8eff}
\ee
which for {\bf A1-S1} reduces to $Q_{\rm D8}^{\rm eff} =  \eta q_{\rm D8} T_{\rm D8}$. Therefore, by taking $q_{\rm D8} = \eta$, which corresponds to a flux jump that increases $|m|$, one finds again a marginal membrane jump. 

To sum up, for {\bf A1-S1} vacua one finds that 4d membranes made up from both D4-branes and D8-branes satisfy $Q=T$, at least when computing these quantities in the smearing approximation. This is expected for {\bf A1-S1}$+$ vacua, which are supersymmetric, but would contradict the refinement of the WGC for the non-supersymmetric {\bf A1-S1}$-$ vacua. In order to check such a refinement one should then consider corrections to the 4d membrane charge and tension. Just like for the 10d background, such corrections can be expanded in increasing powers of $g_s$. For the case of D4-branes, one may look at corrections to $Q$ and $T$ that come from considering the more precise metric and flux backgrounds \eqref{solutionsu3} and \eqref{solutionflux}. It turns out that such corrections vanish for both classes of {\bf A1-S1} vacua, and so D4-branes wrapping (anti-)holomorphic two-cycles yield extremal 4d membranes also for {\bf A1-S1}$-$ vacua, at least at this level of the approximation. 

The story for D8-branes wrapping $X_6$ is slightly more involved \cite{Marchesano:2021ycx}. First, their DBI and CS actions are subject to curvature corrections encoded in the second Chern class of $X_6$, such that they can be understood as a bound state of a D8-brane and {\em minus} a D4-brane wrapping the Poincar\'e dual of $c_2(X_6)/24$. The term minus refers to the fact that these curvature corrections induce negative D4-brane and tension. This does not affect the relation $Q=T$ in supersymmetric vacua, but it changes it towards $Q>T$ for non-supersymmetric  {\bf A1-S1}$-$ vacua, due to the sign flip for the internal four-form flux $\hat{G}_4$. This provides a mechanism analogous to the one pointed out in \cite{Maldacena:1998uz}, where a D5-branes wraps the $K3$ in AdS$_3 \times S^3 \times K3$, and which drags the resulting membrane towards the AdS boundary. 

However, such curvature corrections appear at the same order in $g_s$ as the first corrections to the smearing approximation, and so both effects should be considered simultaneously. For D8-branes, corrections due to source localization appear in two different ways. On the one hand, due to considering their DBI+CS action in the more precise background \eqref{solutionsu3} and \eqref{solutionflux}. On the other hand, by realising that the space-time-filling D6-branes ending on them are also localized sources for their worldvolume flux $\cF = B + \frac{\ell_s^2}{2\pi} F$. This second effect results in a BIon profile along the D8-brane transverse direction $z$, that encodes the energy of the D8/D6-brane system. Taking all the localization effects into account one obtains a correction to the quantity $Q-T$ in  {\bf A1-S1}$-$ vacua of the form $2\Delta_{\rm D8}^{\rm Bion} \equiv - e^{K/2} \frac{1}{\ell_s^6} \int_{\rm X_6}  J_{\rm CY} \wedge \cF^2_{\rm BIon}$, where $\cF_{\rm BIon}$ is the piece of D8-brane worldvolume flux sourced by the  the D6-branes ending on it  and the $H$-flux \cite{Marchesano:2021ycx}. This quantity was computed in \cite{Casas:2022mnz} for several toroidal orbifold geometries, where it was compared to the D8-brane curvature corrections. It was found that $\Delta_{\rm D8}^{\rm Bion}$ can have both signs depending on the relative positions of the space-time-filling D6-branes in such vacua. In particular, it was found that in some instances adding both sets of corrections tips the scale towards $Q<T$, in apparent tension with the (unrefined) WGC for 4d membranes. 

Despite these negative results, in the following we will argue that the WGC for membranes is satisfied in the DGKT-like vacua of section \ref{s:branch}. To do so, we will consider more exotic bound states of D$(2p+2)$-branes, and in particular D8 and D6-branes with non-diluted worldvolume fluxes threading their internal dimensions. The corresponding 4d membranes will not only provide new decay channels for {\bf A1-S1}$-$ vacua, but also for the non-supersymmetric branches {\bf A2-S1}. Indeed, notice that for the latter $D = \sqrt{3/32}$, and so the ratio $Q/T$ for a 4d membrane obtained from wrapping a plain D$(2p+2)$-brane is given by an irrational number smaller than one. Again, it is via considering exotic bound states that one can achieve 4d membranes with $Q>T$. 

\subsection{Exotic bound states of membranes}

Let us consider new D-brane bound states that are candidates to yield 4d membranes with $Q>T$ in non-supersymmetric vacua. The main strategy will be to identify those bound states that yield $Q=T$ in the supersymmetric case, and analyze similar objects in the non-supersymmetric branches. As we will see, the mismatch between $Q$ and $T$ arises at level of the smearing approximation, so we may phrase most of our discussion in terms of the approximate Calabi--Yau geometry. As advanced, the bound states of interests correspond to D8 and D6-branes with non-diluted worldvolume fluxes in the internal dimensions. More precisely, in the smeared approximation they can be described by the following conditions
\bes
\label{exoticBPS}
\begin{align}
	\label{exoticD8}
	\text{D8-brane on $X_6$:} & \qquad \cF \wedge \cF = 3 J_{\rm CY} \wedge J_{\rm CY}\, ,\\
	\label{exoticD6}
	\text{$k$ D6-branes on ${\cal S}$:} & \qquad \cF \wedge \cF = J_{\rm CY} \wedge J_{\rm CY}|_{\cal S}\, .
\end{align}
\ees
Here  $\cF = B + \frac{\ell_s^2}{2\pi} F$ is the worldvolume flux\footnote{Recall that in the smearing approximation $\cF_{\rm Bion}=0$, so also for D8-branes $\cF$ is a closed two-form.} threading the internal dimensions of the D$(2p+2)$-brane, which is the whole $X_6$ in the case of D8-branes and a divisor ${\cal S}$ in the case of D6-branes. 

We dub these objects exotic bound states because, in the large volume regime, they carry a  large lower-dimensional D-brane charge, induced by a large flux $\cF$ \cite{Douglas:1995bn}.\footnote{One should not confuse the two notions of charge present in our discussion. D-brane charges refer to the couplings of D$(2p+2)$-branes to the RR $(2p+1)$-form potentials in 10d supergravity, in the absence of background fluxes. The charges in \eqref{QDGKT} correspond instead to the 4d membrane charges \eqref{3form} obtained via dimensional reduction of a Chern-Simons action in a particular 10d flux background that corresponds to a vacuum.} This makes them exotic from the  model building viewpoint, as parametrically large D-brane charges can be in conflict with RR tadpole conditions. 
In the case at hand, the large D-brane charges carried by these bound states translate into 4d membranes that induce large shifts for the flux quanta $m^a$, $e_a$ and $e_0$, which are not constrained by tadpole conditions. Therefore,  one must consider them as part of the spectrum of 4d membranes, and as such they may mediate decays in non-supersymmetric vacua. We will now analyze their properties in the different DGKT-like branches of section \ref{s:branch}.

\subsubsection*{Supersymmetric vacua}

Let us discuss the properties of the D-branes \eqref{exoticBPS} in supersymmetric vacua. In fact, it proves useful to first consider the case of type IIA Calabi--Yau orientifold compactifications to Minkowski, in the absence of background fluxes. In this context, \eqref{exoticBPS} are particular solutions to the MMMS equations \cite{Marino:1999af} and as such the corresponding D-branes are BPS objects. One can also detect the BPSness of such D-branes by 
analysing their DBI action, see Appendix \ref{ap:DBI}. After imposing \eqref{exoticBPS} the DBI action linearizes and its integrand reads\footnote{Curvature corrections will modify this expression as well as the BPS conditions \eqref{exoticBPS}, shifting $\cF \wedge \cF$ by   $c_2(X_6)/24$ for D8-branes and by $c_2({\cal S})/24$ for D6-branes. Because this effect is subleading in the large volume regime, and is comparable to corrections to the smearing approximation, it will be neglected in the following. }
\be
d {\rm DBI} = e^{i \theta} g_s^{-1} \left.e^{-(\cF + iJ_{\rm CY})} \right|_{2p}  
\label{calibration}
\ee
where $p=3$ for D8-branes and $p=2$ for D6-branes. We say that both objects are calibrated by $e^{-(\cF + iJ_{\rm CY})}$, with $e^{-i\theta}$ their calibration phase. In compactifications to Minkowski 4d membranes are BPS for any calibration phase, but only two membranes with the same calibration phase are mutually BPS. This is a relevant statement because, as we will show below, in $\CN=1$ AdS$_4$ type IIA vacua all 4d membranes that are BPS have the same calibration phase. We have already run into some BPS 4d membranes in supersymmetric DGKT-like vacua, like a D4-brane on a holomorphic curve $\Sigma$, which corresponds to $\theta = \pi/2$. Other D-branes with the same phase are
\bes
\label{iBPS}
\begin{align}
	\label{iD6}
	\text{(anti-)D6-brane on ${\cal S}$ with} & \quad \cF^2 = J_{\rm CY}^2|_{\cal S}\, , \\
	\label{iD8}
	\text{D8-brane on $X_6$ with} & \quad  \cF^2 \wedge J_{\rm CY} = c J^3_{\rm CY} \, , \ c \leq 0 \quad {\rm and}  \quad 3\cF \wedge J_{\rm CY}^2 = \cF^3 \, , \\
	\label{iaD8}
	\text{anti-D8-brane on $X_6$ with} & \quad \cF^2 = 3 J_{\rm CY}^2\, .
\end{align}
\ees
We have encountered instances of \eqref{iD8} in our previous discussion, like the case $c=0$ which corresponds to a D8-brane with $\cF = 0$. Other cases in which $0 > c \sim \cO({\cal V}_{\rm CY}^{-2/3})$ represent D8-branes with a worldvolume flux that is approximately primitive, and corresponds to a solution to the $\alpha'$-corrected Donalson-Uhlenbeck-Yau equations \cite{Douglas:2001hw}. Such objects can be seen as bound states of D8-branes and a few D6, D4 and D2-branes, and were also considered as 4d membranes in \cite{Marchesano:2021ycx}. However, they are not particularly interesting from the viewpoint of the WGC for 4d membranes in non-supersymmetric vacua. On the one hand they carry positive D4-brane charge, and in {\bf A1-S1}$-$ vacua this contributes towards $Q<T$. So in order to look for membranes with $Q \geq T$ it is better to set $c = 0$. On the other hand, they are quite unnatural in {\bf A2-S1} vacua, because the non-diluted B-field sets $c \sim \cO(1)$. In any event, we see that our reasoning selects two new candidates for 4d membranes satisfying the WGC, which are quite similar to \eqref{exoticBPS}.

Let us now show that all these objects fulfil the extremal condition $Q=T$ in supersymmetric AdS$_4$ vacua. For this, we consider the 10d type IIA  supersymmetry conditions \cite[eq.(2.13)]{Marchesano:2020qvg} in the smearing approximation\footnote{This is equivalent to consider (\ref{eq:susyeq}) and set $A = 0$, $\theta = 0$, $e^\phi = g_s$ and take the limit $\psi= 0$, assuming that $J_\psi$ and $\Omega_\psi$ become the forms of a Calabi--Yau.}
\bes
\label{susy}
\begin{align}
	\label{susyg}
	d_H \im \Omega_{\rm CY} & -  g_s \star_6 \left(G_0 - G_2 + \hat{G}_4 - \hat{G}_6 \right) + 3 \mu \im \left( e^{-iJ_{\rm CY}}\right) = 0\, , \\
	d_H e^{-iJ_{\rm CY}} & + 2\mu \re \Omega_{\rm CY} = 0\, .
	\label{susydw}
\end{align}
\ees
We may pull-back \eqref{susyg} on a $2p$-cycle of $X_6$ wrapped by a D$(2p+2)$-brane. Then, by multiplying the result by $e^{-\cF}$ and using \eqref{bfG} one obtains
\be
\frac{1}{3\mu} \left(g_s^{-1} d_H\Omega_{\rm CY} - e^{-\cF} \wedge \tilde{G}\right)_{2p} = - g_s^{-1}  \im \left( e^{-\cF-iJ_{\rm CY}}\right)_{2p}  = \sin \theta\, d {\rm DBI}\, ,
\label{10dQT}
\ee
where in the last equality we have used \eqref{calibration}. By switching off the worldvolume flux $\cF$, one can see that both sides of this equation are related to the 4d membrane charge and tension that appear in \eqref{QDGKT}. By introducing $\cF$ one generalizes this notion for bound states that arise from such worldvolume fluxes. Indeed, upon integration of the rhs of \eqref{10dQT} one recovers $\sin \theta$ times the 4d membrane tension. Similarly, the lhs of \eqref{10dQT} encodes the 4d membrane effective charge. Upon integration on an internal $2p$-cycle it gives $\eta Q$, where $\eta = {\rm sign}\, m$ and
\be
Q = \frac{\eta e^{K/2}}{\ell_s^{2p}}  \int_{2p}  e^{-\cF} \wedge {\bf Q} \, ,  \quad \text{with} \quad  {\bf Q} = \sum_{p}  \frac{q_p}{p!}J_{\rm CY}^p\, ,
\label{Qform}
\ee
and the coefficients $q_p$ correspond to charge-to-tension ratios $Q_{{\rm D}(2p+2)}/T_{{\rm D}(2p+2)}$. Namely,
\be
q_0 = 0\, , \qquad q_1 = \frac{C}{D}\, , \qquad q_2 = -\frac{B}{2D}\, ,  \qquad q_3 = -\frac{24A-1}{2D}\, . 
\ee
Note that $Q$ reproduces \eqref{QDGKT} and \eqref{QD8eff} for 4d membranes with $\cF=0$ in all branches of DGKT-like vacua, and it extends the definition of charge to their bound states. In the supersymmetric branch we have that ${\bf Q} =  \im e^{iJ_{\rm CY}}$, and \eqref{10dQT} translates into 
\be
Q = \eta \sin \theta\, T\, .
\ee
This illustrates our claim that, in supersymmetric AdS$_4$ vacua, all 4d membranes with $Q=T$ have the same calibration phase. In the case at hand they have $\theta = \pi/2$ for $m>0$, like D4-branes wrapping holomorphic curves $\Sigma$ and the D-branes in \eqref{iBPS}. For $m<0$ they must instead have $\theta=-\pi/2$, like anti-D4-branes on $\Sigma$ and the anti-D-branes version of \eqref{iBPS}. It is easy to convince oneself that our reasoning is more general that the specifics of DGKT-like vacua, and it ultimately boils down to the interpretation of the 10d supersymmetry equations as the existence of calibrations for D-branes wrapping internal cycles of a compact manifold \cite{Martucci:2005ht,Koerber:2007jb}.

It may seem surprising that anti-D8-branes with worldvolume fluxes in supersymmetric vacua with $m>0$ can be BPS and that their transverse position in the AdS$_4$ coordinate $z$ is a flat direction. In this case, the D8-brane tension and (effective) charge add up to drag them away from the AdS boundary. However, the worldvolume flux condition \eqref{iaD8} implies that they form a bound state with a very large number $N \sim 9T_{\rm D8}/T_{\rm D4}$ of D4-branes. Hence, even if D4-branes have smaller tension, their large number makes them weight nine times more than a D8-brane. The bound state is calibrated by $\frac{4}{3} J_{\rm CY}^3$, which given the opposite orientation compared to $X_6$ results in a 4d membrane tension  $T= 8 T_{\rm D8}$, while $Q= -T_{\rm D8} + NT_{\rm D4} =  8 T_{\rm D8}$. It thus happens that the tension gained by the bound state as compared to its constituents precisely cancels the factor of $2T_{\rm D8}$ that would drag away from the AdS boundary an anti-D8-brane with $\cF=0$. 

Before turning to non-supersymmetric vacua, let us comment on the actual existence of the D-branes \eqref{iD6} and \eqref{iaD8} in supersymmetric DGKT vacua. The question is non-trivial, because in such vacua the K\"ahler moduli and the B-field axions take discrete values as a function of the background flux quanta. So everything is fixed in these BPS equations except the piece of worldvolume flux given by $F$, which is also quantized. Hence, for arbitrary values of the complexified K\"ahler moduli one may not be able to find examples of such D-branes, which also illustrates the somewhat exotic nature of these objects.

Let us first consider the anti-D8-branes. Since they wrap the whole of $X_6$, the equation in \eqref{iaD8} is directly related to the stabilization of Calabi--Yau moduli. In particular, \eqref{hate} implies
\be
3 J_{\rm CY}^2 = -10 \left(\frac{e_a}{m} - \oh \frac{\CK_{abc}m^bm^c}{m^2}\right) \tilde{\omega}^a \, ,
\ee
while the stabilization of B-field axions implies that
\be
\cF =  \left(n^a -\frac{m^a}{m} \right) \omega_a\, ,
\ee
where $n^a \in \pz$. Putting both conditions together one finds that \eqref{iaD8} amounts to
\be
\CK_{abc} \left(m^2n^bn^c - 2n^bm^c\right)  = 10 m e_a - 6 \CK_{abc}m^bm^c\, ,  \quad \forall a\, .
\label{dioD8BPS}
\ee
Given some choice of flux quanta $m, m^a, e_a$, one should find appropriate values of $n^a$ solving these equations. While both sides of \eqref{dioD8BPS}  are integer, it is not always true that a solution to such quadratic Diophantine equations exist. In the particularly simple case where $m^a = 0$ they reduce to $m\CK_{abc} n^bn^c = 10 e_a$, which do not have a solution unless $10e_a$ is a multiple of $m$, $\forall a$. 

Similar equations can be derived for the case of D6-branes. Assuming $k$ D6-branes wrapped on a Nef divisor ${\cal S}_a$ dual to $\omega_a$, and a quantized worldvolume flux of the form $\frac{\ell_s^2}{2\pi} F = \frac{n^b}{k} \mathbbm{1}_{k} \omega_b|_{{\cal S}_a}$, the BPS condition \eqref{iD6} amounts to 
\be
\CK_{abc} \left(\frac{mn^bn^c}{k} - 2 n^bm^c \right)  = \frac{10k}{3} e_a - \frac{8k}{3m} \CK_{abc}m^bm^c \, ,
\label{dioD6BPS}
\ee
where $\CK_{abc}n^bn^c/k \in \IZ$ must be satisfied \cite{Rabadan:2001mt}. In this case we have a single Diophantine equation to solve, and we have more freedom, in the sense that given $m, m^a, e_a$ we may adjust the values of both $k$ and $n^a$ to find solutions. In particular, it seems that one must take $k$ proportional to $3m$ in order to find solutions for generic values of the flux quanta. In the particular case where $m^a = 0$, the equation reduces to $\CK_{abc} \frac{n^bn^c}{k} = \frac{10k}{3m}e_a$, which should have solution whenever $[e_a\tilde{\omega}^a]$ is dual to the intersection of two divisors.

\subsubsection*{A1-S1$-$ vacua}

Let us now turn to non-supersymmetric {\bf A1-S1}$-$ vacua. Recall that these vacua are defined by a sign flip of the internal four-form $\hat{G}_4$ with respect to the supersymmetric ones. In other words, $Q_{\rm D4}$ flips sign and (for $m>0$) D4-branes wrapping holomorphic curves satisfy $Q=-T$ from the 4d viewpoint, while anti-D4-branes satisfy $Q=T$. For this reason, 4d membranes corresponding to \eqref{iaD8} cannot satisfy $Q>T$ in {\bf A1-S1}$-$ vacua with $m >0$, since both of their constituents (anti-D8-brane and D4-branes) contribute with a negative charge. One may instead consider their anti-object, which is nothing but \eqref{exoticD8}. As we will see, the corresponding 4d membrane satisfies $Q>T$ and provides a decay channel for this class of vacua. 

Indeed, \eqref{exoticD8} can be roughly seen as a bound state of a D8-brane and $N \sim 9T_{\rm D8}/T_{\rm D4}$ anti-D4-branes. The charges of both constituents are positive whenever $m>0$, and add up to $Q = 10 T_{\rm D8}$. Indeed, for D8-branes in {\bf A1-S1}$-$ vacua we have that \eqref{Qform} reads
\be
Q =  \frac{\eta e^{K/2}}{\ell_s^{6}}  \int_{X_6}  e^{-\cF} \wedge \left( -  J_{\rm CY} - \frac{1}{6} J_{\rm CY}^3\right) = 10 \eta T_{\rm D8} \, ,
\label{QA1S1-}
\ee
where in the second equality we have applied \eqref{exoticD8}. The 4d membrane tension is equal to that of its anti-object \eqref{iaD8}, namely $T = 8 T_{\rm D8}$, as can also be checked by using the results of appendix \ref{ap:DBI}. Therefore we obtain that $Q-T = 2T_{\rm D8}$, and the membrane is superextremal. 

It remains to see whether these objects actually exist for a given vacuum. As in their supersymmetric case, their BPS condition translates into a quadratic Diophantine equation:
\be
\CK_{abc} \left(mn^b -m^b\right)\left(mn^c -m^c \right) = -10 m\hat{e}_a\, ,
\ee
or equivalently
\be
\CK_{abc} \left(m^2n^bn^c - 2n^bm^c\right)  = - 10 m e_a + 4 \CK_{abc}m^bm^c\, ,  \quad \forall a\, .
\label{dioD8A1-}
\ee
These equations look a bit different from the supersymmetric case, but note from \eqref{hate} that  ${\rm sign}\, (m\hat{e}_a) = \pm$ for {\bf A1-S1}$\pm$ vacua, so we are essentially solving the same equations. As before, we do not expect that for arbitrary choices of $m, m^a, e_a$ one can find $n^a \in \pz$ such that all these equations are satisfied. In that case, D8-branes with $\cF^2 = 3 J_{\rm CY}^2$ do not exists. However, one can still argue that D8-branes with worldvolume fluxes such that $Q>T$ do still exist. 

Indeed, let us consider that the quantized piece of the worldvolume flux is of the form
\be
\frac{\ell_s^2}{2\pi} F = n^a \omega_a \, , \qquad \text{with} \qquad n^a = \pm\sqrt{3} t^a + \frac{m^a}{m} + \eps^a\, . 
\label{quantansatz}
\ee
Here $\eps^a \in \pr$ are chosen to be the smallest possible numbers such that $n^a \in \pz$ and $\eps^a \hat{e}_a = 0$. Generically, this second condition sets $\eps^a$ to be of the order of the largest quotient between two $\hat{e}_a$'s, which we denote by $M$. It also implies that we can write the worldvolume flux as
\be
\cF = \pm \sqrt{3} J_{\rm CY} + \cF_{\rm p}\, ,
\ee
where $\cF_{\rm p} = \eps^a \omega_a$ is a primitive (1,1)-form, that is $\cF_{\rm p} \wedge J_{\rm CY}^2 =0$. Plugging this expression for the worldvolume flux into eq.\eqref{ap:DBID8}, one obtains that the D8-brane DBI density reads
\be
d{\rm DBI}_{\rm D8} = g_s^{-1} \sqrt{\left(8 -  ||\eps||^2 \right)^2 + \left(\sqrt{3} ||\eps||^2 + \cO(||\eps||^3) \right)^2} d{\rm vol}_{X_6}\, ,
\label{DBID8eps}
\ee
where we have defined
\be
||\eps|| = \frac{1}{2} \sqrt{\cF_{{\rm p}, ab} \cF_{\rm p}^{ab} }   \sim \cO\left(\frac{M}{{\cal V}_{\rm CY}^{1/3}}\right)\, .
\label{epsdef}
\ee
In the following we will assume that $||\eps|| \ll 1$, because it is not clear that otherwise we have scale separation, or even that the K\"ahler moduli are stabilized in the supergravity regime. Under this assumption one can expand \eqref{DBID8eps} and obtains that the tension reads
\be
T = \left(8 -  ||\eps||^2_0 + 2  ||\eps||^4_0 + \dots\right) T_{\rm D8}\, ,
\ee
where we have defined $||\eps||^n_0 \equiv \int_{X_6} ||\eps||^n/ {\cal V}_{\rm CY}$, and the dots represent higher order terms in $||\eps||$. Similarly, one may compute the membrane tension from \eqref{QA1S1-}, obtaining
\be
Q = \left(10 -  ||\eps||^2_0 \right) \eta T_{\rm D8}\, .
\ee
Therefore we find that for $m > 0$
\be
Q - T =  2\left(1 -  ||\eps||^4_0 + \dots\right) T_{\rm D8}\, ,
\ee
and the membrane is superextremal. For $m<0$, one instead needs to consider the anti-D8-brane \eqref{iaD8} to find the same result. 

Even if these 4d membranes may not be in the thin-wall approximation, one may apply the reasoning of \cite[section 5]{Marchesano:2021ycx} to argue that they represent a non-perturbative instability towards a vacuum with larger $|m|$ and smaller $|\hat{e}|$. Still they should not be considered in vacua where $|m|$ cannot be made larger due to the tadpole constraints, as for instance in models without space-time-filling D6-branes like in \cite{DeWolfe:2005uu}. In those cases, only membranes that vary $m^a$, $e_a$ and $e_0$ should be considered, like D4-branes and the D6-branes in \eqref{exoticD6}. For the latter, and under the same assumptions as in \eqref{dioD6BPS} their existence translates in the following  Diophantine equation
\be
\CK_{abc} \left(\frac{mn^bn^c}{k} - 2 n^bm^c \right)  = - \frac{10k}{3} e_a + \frac{2k}{3m} \CK_{abc}m^bm^c \, ,
\label{dioD6A1-}
\ee
which is again quite similar to that of the supersymmetric branch. As in there, we expect that one can choose appropriate values of $k$ and $n^a$ to find a solution. In the smearing approximation, we have that the tension of the corresponding 4d membrane is similar to its supersymmetric counterpart. Using eq.\eqref{ap:DBID6} and applying \eqref{exoticD6} one finds
\be
T = e^{K/2} \frac{1}{\ell_s^4} \left| \int_{\cal S} \cF \wedge J_{\rm CY}  \right| \, 
\ee
while its charge can be computed via \eqref{Qform} 
\be
Q =  e^{K/2} \frac{\eta }{\ell_s^{4}}  \int_{\cal S}  \cF \wedge  J_{\rm CY}\, .
\ee
The sign of the integral will depend of the sign of the projection of $\cF$ into $J_{\rm CY}|_{\cal S}$. For either sign and for each value of $\eta$ one can satisfy the extremal condition $Q=T$ by either considering a D6-branes or an anti-D6-brane satisfying \eqref{exoticD6}. 

We therefore only find superextremal 4d membranes when they arise from D8-branes. D4-branes and the D6-branes \eqref{exoticD6} are at best marginal. As this would contradict the WGC refinement proposed in \cite{Ooguri:2016pdq}, one may wonder if the equality $Q=T$ is an artefact of the smearing approximation. Following the same computations as in \cite[section 6]{Marchesano:2021ycx}, one can convince oneself that the D6-brane charge and tension do not vary when we consider them in the more precise background \eqref{solutionsu3} and \eqref{solutionflux}. Finally, one may add curvature corrections to the D6-brane action, which will modify its tension. However, the same corrections will also modify the worldvolume flux condition \eqref{exoticBPS}, in such a way that both effects cancel out. Therefore, at the level of approximation that we are working, we find that DGKT-like vacua in the {\bf A1-S1}$-$ branch without space-time-filling D6-branes are marginally stable. Whether further corrections tip the scale  towards $Q>T$ or not remains an open problem.

\subsubsection*{A2-S1 vacua}

Let us consider the last two branches of vacua, namely {\bf A2-S1}$\pm$, which can be discussed simultaneously. In this case, one can also show that D-brane bound states \eqref{exoticBPS} lead to 4d membranes with $Q>T$ whenever they exist. Discussing their existence is however more involved than in the \bk{\bf A1-S1}$\pm$ branches. Indeed, in the present vacua the worldvolume flux of, say, a D8-brane is of the form 
\be
\cF =  \left(n^a + Bt^a - \frac{m^a}{m} \right) \omega_a\, ,
\ee
and so when plugged into \eqref{exoticD8} there will be an explicit dependence on the K\"ahler moduli. As such, it is difficult to determine whether such an equation has a solution, unless the vevs of the K\"ahler moduli are known explicitly as a function of the background fluxes. 

\begin{figure}[H]
	\centering
	\includegraphics[width=0.65\linewidth]{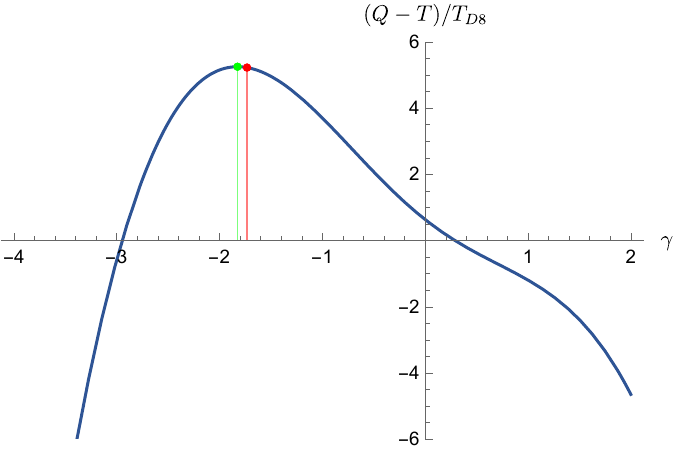}
	\caption{\label{plotD8} $Q-T$ for D8-branes in units of $T_{\rm D8}$ over $\gamma$ (blue) with $\eta = \eta_B = 1$ and $\epsilon = 0$. The dots correspond to the maximum of the curve $\gamma \simeq -1.82$ (green), and to $\gamma = - \sqrt{3}$ (red). $Q>T$ for the range $-2.95 \lesssim \gamma \lesssim 0.29$.}
\end{figure}

Nevertheless, one may still implement the approach previously used for D8-branes, and consider that the quantized piece of the worldvolume flux is of the form 
\be
\frac{\ell_s^2}{2\pi} F = n^a \omega_a \, , \qquad \text{with} \qquad n^a = \left(\gamma - B\right) t^a + \frac{m^a}{m} + \eps^a\, ,
\label{quantansatzA2}
\ee
with $\gamma \in \pr$ and $\eps^a$ satisfying the same constraints as in \eqref{quantansatz}. The corresponding worldvolume flux reads
\be
\cF = \gamma J_{\rm CY} + \cF_{\rm p}\, ,
\label{cfg}
\ee
and one may compute the 4d membrane tension and charge in terms of its parameters. As before, by plugging \eqref{cfg} into eq.\eqref{ap:DBID8} one obtains the following DBI density for D8-branes
\be
d{\rm DBI}_{\rm D8} = g_s^{-1}\sqrt{\left(3\g^2 - 1 -  ||\eps||^2 \right)^2 + \left(\g \left( \g^2 -3  - ||\eps||^2\right) + \cO(||\eps||^3) \right)^2} d{\rm vol}_{X_6}\, ,
\label{DBID8A2}
\ee
where $||\eps||$ is defined as in \eqref{epsdef}. From here one deduces that the corresponding 4d membrane tension reads
\be
T = \left[\left(1+\gamma^2\right)^{3/2} -  \left(\g^4-1\right) ||\eps||^2_0  + \dots\right] T_{\rm D8}\, .
\ee
The 4d membrane charge can be computed by plugging \eqref{cfg} into \eqref{Qform}:
\be
Q =  \frac{\eta e^{K/2}}{\ell_s^{6}}   \sqrt{\frac{2}{3}} \int_{X_6}  e^{-\cF} \wedge \left( - J_{\rm CY} - \eta_B \oh J_{\rm CY}^2 - \frac{1}{3} J_{\rm CY}^3\right) = \sqrt{\frac{2}{3}} \left( 3\g^2 - 3 \g \eta_B  +2 -  ||\eps||^2_0\right)  \eta T_{\rm D8} \, ,
\label{QA2S1}
\ee
where we have defined $\eta_B \equiv {\rm sign}\, B$.  From these expressions it is easy to see that $Q>T$ for $\gamma=-\eta_B \sqrt{3}$ and $||\eps||\ll1$, as claimed above. However, this value of $\gamma$ does not give the maximum possible value of $Q-T$. The actual value of the maximum and the range for which $Q-T$ is positive can be evaluated numerically (see figure \ref{plotD8}).

\begin{figure}[H]
	\centering
	\includegraphics[width=0.65\linewidth]{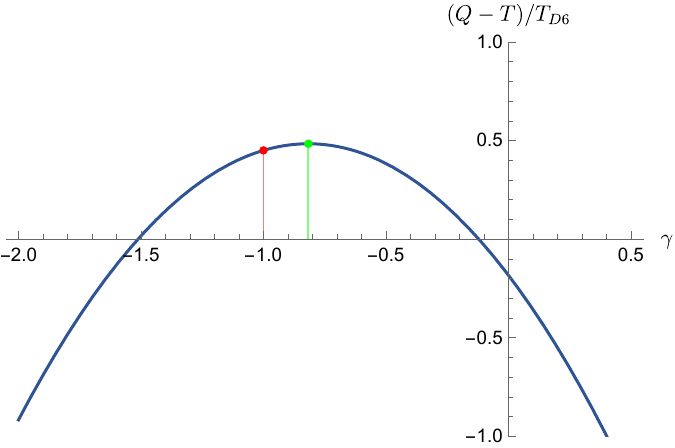}
	\caption{ \label{plotD6} $Q-T$ for D6-branes in units of $T_{D6}$ over $\gamma$ (blue) with $\eta = \eta_B = 1$ and $\epsilon = 0$ . The dots correspond to the maximum of the curve $\gamma = -\sqrt{2/3}$ (green) and to $\gamma = - 1$ (red). $Q>T$ for the range $-1.51 \lesssim \gamma \lesssim -0.12$.}
\end{figure}

One can implement the same strategy to analyze D6-branes with non-diluted worldvolume fluxes. This time, we take an Ansatz of the form \eqref{cfg} with $J_{\rm CY}$ representing the K\"ahler form pulled-back on the divisor ${\cal S}$, and $\cF_{\rm p}$ being a primitive (1,1)-form on the divisor, so that $\cF_{\rm p} \wedge J_{\rm CY} = 0$. We then encounter the following DBI density
\be
d{\rm DBI}_{\rm D6} = g_s^{-1} \sqrt{\left(\g^2 - 1 -  \varepsilon \right)^2 + 4\g^2}\, d{\rm vol}_{\cal S}\, ,
\label{DBID6A2}
\ee
where $\oh \cF_{\rm p} \wedge \cF_{\rm p} = \varepsilon\, d{\rm vol}_{\cal S}$. This leads to 
\be
T = \left(  1 + \g^2  -  \left(\g^2-1\right) \varepsilon_0  + \dots\right) T_{\rm D6}\, ,
\ee
with $\varepsilon_0 =  \int_{\cal S} \oh \cF_{\rm p} \wedge \cF_{\rm p}/ {\cal V}_{\cal S}$. The 4d membrane charge is again computed from \eqref{Qform}
\be
Q = \sqrt{\frac{2}{3}} \left(\eta_B-2\g \right) \eta T_{\rm D6}\, .
\ee
By choosing $\g = - \eta_B = - \eta$ one obtains that $Q>T$. Again, this is not the value of $\gamma$ that maximizes $Q-T$. The actual value of the maximum and the range for which $Q-T$ is positive can be evaluated numerically (see figure \ref{plotD6}).


\section{Discussion}
\label{s:conclu}

In this chapter we have analyzed the perturbative and non-perturbative stability of DGKT-like vacua, following up on previous similar work \cite{Marchesano:2021ycx,Casas:2022mnz}. The vacua that can be built from a given Calabi--Yau manifold organize themselves on different branches, one of which is supersymmetric and the rest is non-supersymmetric. Out of the latter, three share some key properties with the supersymmetric branch, like an infinite set of vacua indexed by internal fluxes, parametric scale separation as we move along this set, and perturbative stability for all of them. While there are  obvious differences between each of these branches, we have managed to give a unified treatment for all of them in terms of their stability. The final result is summarized in table \ref{t:final}.
\begin{table}[H]
	\begin{center}
		\scalebox{1}{%
			\begin{tabular}{| c || c | c | c | c |c |c|}
				\hline
				Branch & SUSY & pert. stable & sWGC D4 & sWGC D8 & non-pert. stable  \\
				\hline \hline
				\textbf{A1-S1}$+$   & Yes & Yes  & Yes  & Yes & Yes \\ \hline
				\textbf{A1-S1}$-$  & No & Yes & Marginal  & Yes & unclear if $N_{\rm D6} =0$ \\ 
				\hline
				\textbf{A2-S1}$\pm$   & No & Yes & Yes & Yes & No \\ 
				\hline
		\end{tabular}}      
	\end{center}
	\caption{Different branches of vacua, in terms of the sharpened WGC for membranes and their stability. \label{t:final}}
\end{table}

Perturbative stability can be addressed by providing a solution to the 10d equations of motion and Bianchi identities of massive type IIA that correspond to each of 4d vacua. The result is in section \ref{s:10d} and it follows the same approach of \cite{Junghans:2020acz,Marchesano:2020qvg}, by which one expands the 10d equations as a perturbative series in a small parameter (in our case $g_s$ or $\hat{\mu}$) and solves them up to next-to-leading order. The leading order corresponds to the so-called smearing approximation, which is the one used to derive the effective  potential used in \cite{Marchesano:2019hfb} to obtain a perturbatively stable vacuum for all the branches above. One can easily check that $\cO(g_s)$ corrections to the spectra derived in \cite{Marchesano:2019hfb} will not generate perturbative instabilities. 

The analysis of non-perturbative instabilities is more easily phrased in terms of the Weak Gravity Conjecture for 4d membranes, and more precisely via  the sharpening  proposed in \cite{Ooguri:2016pdq}. According to this more recent proposal,  there should be some extremal 4d membranes in  supersymmetric vacua, while non-supersymmetric vacua should contain superextremal membranes. There should be one of such objects per each independent membrane charge in our 4d EFT, which for the vacua of table \ref{t:final} translates into 4d membranes obtained from D4-branes and from D8-branes, or equivalently bound states that involve them. In terms of the definition of membrane charge $Q$ and tension $T$ given in the main text, in the supersymmetric branch \textbf{A1-S1}$+$ one should find membranes with $Q=T$, which is trivially satisfied by all D-branes that are (mutually) BPS in such a background. As for the non-supersymmetric branches, there should be at least one 4d membrane satisfying $Q>T$, separately for D4-branes and D8-branes. While this strict inequality is not aways realized by the most obvious choice of D8-branes \cite{Casas:2022mnz}, we have shown that by considering D8-branes threaded by non-diluted worldvolume fluxes one can construct 4d membranes that satisfy $Q>T$ in all $\cN=0$ branches, and therefore indicate an instability. Similarly, D6-branes with large internal worldvolume fluxes are bound states that involve D4-branes and which, in \textbf{A2-S1}$\pm$ vacua, correspond to membranes with $Q>T$, in line with the proposal in \cite{Ooguri:2016pdq}. The only case that escapes that proposal in the context of our analysis are D4-branes in \textbf{A1-S1}$-$ vacua, or any bound state without D8-brane charge. As already pointed out in \cite{Marchesano:2021ycx}, these objects are extremal even when considering first-order corrections to the smearing approximation. It could be that further corrections implement the inequality $Q>T$, but at the level of accuracy that we are working one should take some of the vacua in the \textbf{A1-S1}$-$ branch as marginally stable. In particular those vacua where the quantum of Romans mass $|m|$ cannot increase its value due to the tadpole constraint \eqref{tadpole}, like for instance when there are no space-time-filling D6-branes. 

In view of these results, it seems that a better understanding of DGKT-like vacua and their non-perturbative stability, as well as their connection with several Swampland criteria, seems like an interesting challenge for the future. We find particularly amusing that those non-supersymmetric vacua whose stability is still unclear at the current level of accuracy are those whose would-be holographic duals display integer conformal dimensions. Even if this coincidence does not seem to occur for 3d analogues \cite{Apers:2022zjx}, there could still something to be learnt if the same pattern is reproduced in further instances of 4d vacua. We hope that a more exhaustive analysis of the Landscape of AdS  vacua will shed some light into all these questions.


	\clearpage{\pagestyle{empty}\cleardoublepage} 
	
	
\def\d {{\rm d}}	

\chapter{Torsion in cohomology and dimensional reduction}  \label{ch:6}
Conventional wisdom dictates that $\IZ_N$ factors in the integral cohomology group $H^p(X_n, \IZ)$ of a compact manifold $X_n$ cannot be computed via smooth $p$-forms. We revisit this lore in light of the dimensional reduction of  string theory on $X_n$, endowed with a $G$-structure metric that leads to a supersymmetric EFT. If massive $p$-form eigenmodes of the Laplacian enter the EFT, then torsion cycles coupling to them will have a non-trivial smeared delta form, that is an EFT long-wavelength description of $p$-form currents of the $(n-p)$-cycles of $X_n$. We conjecture that, whenever torsion cycles are calibrated, their linking number can be computed via their smeared delta forms. From the EFT viewpoint, a torsion  factor in cohomology corresponds to a $\IZ_N$ gauge symmetry realised by a St\"uckelberg-like action, and calibrated torsion cycles to BPS objects that source the massive fields involved in it.

\section{Introduction}
\label{s:intro-07}

String theory compactifications provide a remarkable connection between the geometry of extra dimensions and the physics of Effective Field Theories (EFTs) \cite{Becker:2006dvp,Ibanez:2012zz,Blumenhagen:2013fgp,Baumann:2014nda,Tomasiello:2022dwe}. An early lesson that one obtains upon exploring this link is that the more an EFT quantity is protected against quantum corrections, the simpler is its description in geometric terms. Typical examples arise in the context of supergravity and supersymmetric gauge theories, where protection mechanisms involve both gauge invariance and renormalization effects constrained by supersymmetry. 

In several instances, discrete EFT data protected by gauge invariance are described in terms of the topology of the compact manifold $X_n$, while quantities protected by supersymmetry enjoy a simple description in terms of differential and/or algebraic geometry. A clear example of the second is BPS states or extended objects of the EFT, which can be obtained from, e.g., D-branes in type II compactifications at weak coupling. In that case, the BPSness condition requires that the D-brane extra dimensions wrap a $p$-cycle of $X_n$ that is calibrated, in the sense of \cite{Harvey:1982xk}. This condition not only has a neat differential geometric description for compactification manifolds $X_n$ with special holonomy, but it can be generalized whenever $X_n$ has a $G$-structure metric and a flux background that leads to a supersymmetric EFT \cite{Martucci:2005ht,Koerber:2005qi,Koerber:2010bx}. The central charge of the BPS object at tree-level is then determined by the integral over the $p$-cycle of the suitable $p$-form calibration, or generalizations that allow us to calibrate D-brane bound states. This picture also applies to space-time filling D-branes that are part of the background in type II orientifold compactifications, as well as to Euclidean D-branes that play the role of BPS instantons.  

An example of discrete EFT data with a topological higher-dimensional origin is the presence of discrete gauge symmetries. In the Abelian case, a $\IZ_N$ gauge symmetry of a $d$-dimensional EFT can be described by a Lagrangian coupling of the form \cite{Banks:2010zn}
\be
N  \, B_{d-2} \wedge F_2 ,
\label{BF}
\ee
where $B_{d-2}$ is a $(d-2)$-form of the EFT  dual to an axion $C_0$, and $F_2=dA_1$ is the field strength of the $U(1)$ boson gauged by $C_0$  \`a la St\"uckelberg. Finally, $N \in \IZ$ is the quantity that is described in terms of the topology of $X_n$. For instance, in type II orientifold compactifications, couplings of this form are specified by the homology classes of the $p$-cycles wrapped by space-time filling D-branes, which in turn determine the discrete gauge symmetries acting on the open string sector of the theory \cite{Berasaluce-Gonzalez:2011gos}. This case is particularly interesting because the discrete symmetry acts on the massless chiral spectrum of the EFT. However, it has the feature that the axion and gauge boson masses induced by \eqref{BF} are usually of the order of the string scale. This implies that the St\"uckelberg terms that complete  \eqref{BF} are not part of the EFT Lagrangian. 

A different setup where the coupling \eqref{BF} is realized is by threading the compact manifold $X_n$ with quantized background fluxes \cite{Berasaluce-Gonzalez:2012awn}. In this case, the coupling $N$ is determined by the flux quanta, or equivalently by an integral cohomology class in $X_n$. Here the interplay with the EFT cutoff is reversed with respect to the previous one. The St\"uckelberg-induced masses for axions and gauge bosons can lie below the EFT cutoff, but now the resulting discrete gauge symmetry acts on strings and particles that typically do not correspond to light states of the EFT. 

In this chapter, we are interested in yet another realization of discrete gauge symmetries, namely those that arise from torsion factors in the integral cohomology groups $H^p(X_n, \IZ)$. That such $\IZ_N$ factors correspond to $\IZ_N$ gauge symmetries can be seen in the AdS/CFT context by following the reasoning in \cite{Gukov:1998kn,Witten:1998wy},  applied to type II orientifold compactifications in \cite{Camara:2011jg}, and with subsequent work in similar setups in \cite{Grimm:2011tb,Berasaluce-Gonzalez:2012abm,Berasaluce-Gonzalez:2012awn,Mayrhofer:2014laa,Grimm:2015ona,Braun:2017oak}. As stressed in \cite{Berasaluce-Gonzalez:2012awn}, the realization of discrete gauge symmetries via torsion in cohomology is related to the setting with background fluxes by dualities such as mirror symmetry. This implies that the same EFT features should be realized, namely: {\it i)} St\"uckelberg couplings that are part of the EFT Lagrangian and {\it ii)}  charged objects that lie above the EFT cut-off. Indeed, as discussed in \cite{Camara:2011jg} such charged objects are given by D-branes wrapping torsion cycles of $X_n$, which from the EFT perspective look like particles and $(d-3)$-branes coupling to $A_1$ and $B_{d-2}$, respectively. 

From this simple observation, an apparent puzzle follows. If in this case \eqref{BF} and its St\"uckelberg  completion appears in the lower-dimensional EFT, is because torsion in cohomology is detected by the standard procedure of $p$-form dimensional reduction. This is rather counter-intuitive, in the sense that torsion cohomology groups are trivial in de Rham cohomology, or in other words their elements can only be represented by exact $p$-forms. Since the EFT data captured upon dimensional reduction typically involves integrals of $p$-forms over $p$-cycles, it is a priori not clear how torsion cohomology factors can translate into a St\"uckelberg  Lagrangian term in the EFT. This naive picture agrees with the standard lore that torsion in cohomology cannot be detected via smooth $p$-forms, and that one should resort to more advanced geometric techniques, like the computation of spectral sequences  \cite{Bott1982DifferentialFI} or to differential cohomology \cite{Apruzzi:2021nmk}.

This chapter addresses this puzzle and proposes a prescription to capture torsion in cohomology via the standard procedure of dimensional reduction. The basic idea is to use {\em smeared delta forms} to construct the integral basis in which ten-dimensional fields are expanded in order to perform the reduction. Here a delta $p$-form stands for the $p$-current $\delta_p(\Pi_{n-p})$ with legs transverse to an $(n-p)$-cycle $\Pi_{n-p} \subset X_n$, while its smeared version $\delta_p^{\rm sm}(\Pi_{n-p})$ corresponds to the projection into the light eigen-$p$-forms of the Laplacian. If one projects $\delta_p(\Pi_{n-p})$ to the zero-mode sector of the spectrum one simply obtains a harmonic $p$-form which is the de Rham Poincar\'e dual of $\Pi_{n-p}$, and torsion cycles are projected out. If however, one includes in the projection those non-vanishing eigenmodes that correspond to massive $p$-form fields entering the EFT, then torsion cycles can have a non-trivial smeared delta form, and translate into quantities of the EFT Lagrangian. 

More precisely, we propose that one should consider smeared delta forms of calibrated cycles in order to build the basis for the dimensional reduction. The physical intuition behind this proposal is that D-branes wrapping calibrated cycles correspond to BPS objects with a controlled backreaction, that one can use together with the picture developed in \cite{Goldberger:2001tn,Michel:2014lva,Polchinski:2015bea} to see their smeared delta function as an EFT long-wavelength description of the corresponding object.  This can then be used to extract information from the EFT, as done in \cite{Lanza:2020qmt,Lanza:2021udy,Lanza:2022zyg} in the context of 4d $\CN=1$ compactifications. In particular, D-branes wrapping calibrated torsion cycles can be seen as BPS operators that gather information on the massive sector of the EFT Lagrangian, like the kinetic terms of the fields that appear in \eqref{BF}. As a direct consequence of our proposal, the linking number between two calibrated torsion cycles can be computed using EFT data, or equivalently by defining a smeared version of the torsion linking number, as summarized in Conjecture \ref{conj:BPS}. 

The notion of calibrated torsion cycle or BPS operator with a $\IZ_N$ charge may seem puzzling. From a geometric viewpoint, calibrations in special 
holonomy manifolds are closed $p$-forms, and therefore they can never calibrate a torsion $p$-cycle. This obstruction is however absent in the more general set of manifolds with $G$-structure metrics, since there the exterior derivative of a calibration does not need to vanish, and one can indeed construct explicit examples with torsion $p$-cycles that are calibrated. From a physics viewpoint, due to the no-force condition between mutually BPS objects, one should always be able to stack an arbitrary number of them on top of each other without any binding energy. This fits naturally with a $\IZ$-valued charge, but not with a $\IZ_N$-valued one. To address this issue we construct examples of BPS objects with $\IZ_N$ charge, in the context of domain-wall solutions of type II string compactified on half-flat manifolds \cite{Gurrieri:2002wz}. We find that the process that annihilates $N$ BPS D-branes wrapped on a torsion cycle is indeed possible topologically, but not energetically favoured. As a result, it is energetically stable to stack an arbitrary number of such objects with $\IZ_N$ charge, as implied by the BPS condition. 

Expressing a delta form as a sum of eigenforms of the Laplacian is in general involved, as it requires knowledge of the massive $p$-form spectrum of a manifold. This difficulty is however less severe for three-dimensional  manifolds with isometries, a fact that we exploit to perform an explicit computation of a torsion linking number and its smeared version in twisted tori, in order to verify Conjecture \ref{conj:BPS}. While it seems challenging to extend such a computation to more general setups, one can provide physical evidence that our proposal should also be valid in SU(3)-structure manifolds. Indeed, using smeared delta forms of calibrated cycles as a basis for dimensional reduction fits perfectly with the framework developed in \cite{DAuria:2004kwe,Grana:2005ny,Kashani-Poor:2006ofe} to describe 4d $\CN=2$ gauged supergravities as EFTs of type II string compactifications and, in fact, one may argue that it is necessary for the consistency of the approach. Similar considerations can be drawn in the context of 4d $\CN=1$ type II orientifold vacua, where the BPS torsion objects are given by membranes ending on strings, and by space-time filling branes ending on membranes. 

In most of our examples it seems that an extension of Conjecture \ref{conj:BPS} is required. In such a generalization, the torsion linking number can be computed not only when elements of ${\rm Tor}H_{n-p}(X_n, \IZ)$ contain calibrated representatives, but also when they can be expressed as linear combinations of elements of $H_{n-p}(X_n, \IZ)$, all of them with calibrated representatives. This extension could in principle be applied to compute torsion factors in the cohomology of Calabi--Yau manifolds, whenever they contain light eigenforms of the Laplacian other than harmonic forms. One may even speculate that our approach could be useful to compute torsion linking numbers even in cases where torsion cycles cannot be related to calibrated submanifolds, by providing an estimate of the associated error in the linking number computation. In any event, our findings support that one may compute certain torsion topological invariants in terms of smeared or EFT data such as masses and kinetic terms, extending the dictionary between geometry and physics to the more subtle and unexplored sector that is torsion in cohomology. 

The rest of the chapter is organized as follows. In section \ref{s:proposal} we describe what is our proposal to compute the torsion linking numbers of a manifold via smeared delta forms, as well as an extension of such proposal. In section \ref{s:dimred} we motivate the proposal from a physics viewpoint, by interpreting torsion calibrated cycles as BPS objects of the EFT with a non-trivial backreaction. In section \ref{s:simple} we analyze our proposal in the context of domain-wall solutions of 4d $\CN=2$ EFTs obtained from compactifications of type IIA string theory on half-flat manifolds. The simplest examples of such manifolds are based on twisted three-tori, for which our conjecture can be verified explicitly using the techniques of section \ref{s:direct}. Section \ref{s:general} tests our proposal in the context of general SU(3)-structure compactifications of type IIA string theory, finding agreement with previous analysis in the literature and giving a more precise prescription to perform the dimensional reduction in this context. Section \ref{s:N=1} extends our general strategy to 4d $\CN=1$  type II orientifold vacua, and section \ref{s:nonBPS} contains some speculative remarks on how to perform a further extension to the case where no EFT BPS objects are available to detect torsion. We finally draw our conclusions in section \ref{s:conclu-07}. 

Several technical details have been relegated to the appendices. Appendix \ref{ap:NS5DW} analyzes a mirror dual setup to that of section \ref{s:simple} from a microscopic viewpoint, in order to classify the relevant set of BPS D-branes in both backgrounds. Appendix \ref{ap:spectra} analyzes the massive $p$-form spectrum for the case of the twisted three-torus, as a necessary step to perform the direct computation of the torsion linking number of section \ref{s:direct}.


\section{The proposal}
\label{s:proposal}

Let us consider a compact manifold $X_n$ of real dimension $n$, and a submanifold $\Pi_p \subset X_n$ which is a $p$-cycle. We can define a bump-delta $(n-p)$-current or distributional form $\delta (\Pi_p)$, such that
\be
\int_{X_n} \omega_p \wedge \delta (\Pi_p) = \int_{\Pi_p} \omega_p \, , 
\label{deltadef}
\ee
for any smooth $p$-form $\omega_p \in \Omega^p (X_n)$. If $X_n$ is endowed with a smooth metric $ds^2_{X^n}$ measured in string units $\ell_s = 2\pi \sqrt{\a'}$, one can solve the eigenvalue problem for $(n-p)$-forms
\be
\Delta b_{n-p}^i = \lambda_i^2 b_{n-p}^i \, , 
\ee
where $\Delta = d^\dag d + d d^\dag$ is the Laplace-de Rham operator, $\{ b_{n-p}^i\}_i$ is an orthonormal basis of eigenforms with respect to the Hodge product, and $\{\lambda_i^2\}_i$ the corresponding set of non-negative, dimensionless eigenvalues. Then one can expand the bump-delta $(n-p)$-form on such a basis
\be
\delta (\Pi_p) = \sum_i c_i  \, b^i_{n-p}\, , \qquad c_i = \int_{\Pi_p} \star  b^i_{n-p}\, ,
\label{deltaexp}
\ee
and from here define a {\em smeared} version of the delta-form, by keeping only those terms in the expansion that satisfy $\lambda_i < \lambda_{\rm max}$, for some choice of $\lambda_{\rm max}$. This  defines a smooth bump $(n-p)$-form localized within a tubular neighbourhood of radius $\ell_s/\lambda_{\rm max}$ around $\Pi_p$. Such a $(n-p)$-form can be identified with the Thom class of the normal bundle of $\Pi_p$, which is known to lie in the de Rham Poincar\'e dual to $[\Pi_p] \in H_p(X_n)$ \cite{Bott1982DifferentialFI}. Indeed, notice that all elements $b^{i}_{n-p}$ of the expansion \eqref{deltaexp} must be exact $(n-p)$-forms except those with vanishing eigenvalue, which must correspond to the harmonic representative of the Poincar\'e dual to $[\Pi_p]$. It follows that if $[\Pi_p]$ lies in a torsion class of $H_p(X_p, \IZ)$ then $\delta(\Pi_p)$ must be a sum of exact $(n-p)$-forms.

In string theory compactifications there is a natural choice of metric for $X_n$ that comes from solving the 10d supergravity equations of motion, as well as a natural choice of $\lambda_{\rm max}$ that one identifies with the compactification scale $m_{\rm KK}$. One can define $\ell_s m_{\rm KK}$ to be the typical spacing between positive eigenvalues $\lambda_i$, oftentimes estimated by the average radius Vol$(X_n)^{1/n}$. Physically, we understand $m_{\rm KK}$ as the energy scale below which we recover a $D$-dimensional EFT description, with $D=10-n$, that describes all eigenmodes with $\lambda_i \ll \ell_s m_{\rm KK}$  as $D$-dimensional fields. The standard practice in the string literature is to assume that only harmonic modes satisfy the requirement $\lambda_i \ll \ell_s m_{\rm KK}$, such that the procedure of dimensional reduction simply projects the spectrum of $p$-forms to the harmonic sector.\footnote{Alternatively, one may set the EFT cut-off $\Lambda_{\rm EFT}$ below any non-vanishing mode.} However, it has been shown that in certain compactification regimes, and in particular in six-dimensional manifolds with SU(3)-structure \cite{Gray1980TheSC,chiossi2002intrinsic,Hull:1986iu,Strominger:1986uh,LopesCardoso:2002vpf,Gauntlett:2003cy,Grana:2005jc,Koerber:2010bx}, one has a non-vanishing $p$-form eigenvalues well below the compactification scale. This will be the case of interest in this chapter, and henceforth our definition of smeared delta-form will correspond to the following:
\be
\delta^{\rm sm} (\Pi_p) = \sum_{\lambda_i \ll \ell_s m_{\rm KK}} c_i  \, b^{i}_{n-p}\, .
\label{deltasm-07}
\ee

Note that if $[\Pi_p] \in {\rm Tor} H_p (X_p, \IZ)$ then \eqref{deltasm-07} may contain no terms at all and, if it does, it will be a sum of exact $(n-p)$-forms. This reflects the difficulties in obtaining information from the torsion (co)homology classes from the viewpoint of the lower dimensional EFT, as integrals of \eqref{deltasm-07} over any $(n-p)$-cycle of $X_n$  simply vanish. There is however a well-defined topological invariant  for torsion homology classes, which is the torsion linking number. Given the torsion classes $[\Pi_p] \in {\rm Tor} H_p(X_n, \IZ)$ and $[\Pi_{n-p-1}] \in {\rm Tor} H_{n-p-1}(X_n, \IZ)$, one can define their linking number in terms of the bump delta-forms of two of their representatives as \cite{Camara:2011jg}  
\be
L( \Pi_{n-p-1},\Pi_p) = \int_{X_n}  d^{-1}  \delta(\Pi_{n-p-1}) \wedge \delta(\Pi_p)  \quad \mod \ 1 \, .
\ee
Following \cite{Horowitz:1989km}, we can rewrite this quantity as follows. Notice that $\{\lambda_i ^{-1}d\star b^i_{n-p}\}_i$ is an orthonormal basis of exact $(p+1)$-forms, so one can perform the expansion 
\be
\delta(\Pi_{n-p-1}) = \sum_i \frac{e_i}{\lambda_i} d\star b^i_{n-p} \, ,  \qquad e_i = \frac{(-1)^{n(n-p)}}{\lambda_i}\int_{\Pi_{n-p-1}}d^\dag  b^i_{n-p}\, ,
\label{deltaexp2}
\ee
from where one obtains
\be
L( \Pi_{n-p-1},\Pi_p) = \sum_i \frac{c_ie_i}{\lambda_i}   \quad \mod \ 1 \, .
\label{L}
\ee
One can now define a {\em smeared linking number}. From $[\Delta, d] = [\Delta, \star ] = 0$ it follows that $d\star  b_{n-p}^i$ has the same eigenvalue as $b_{n-p}^i$, and so the smeared version of \eqref{deltaexp2} corresponds to the same truncation as in \eqref{deltaexp}. Thus, it is natural to define the smeared analogue of \eqref{L} as
\be
L^{\rm sm}( \Pi_{n-p-1},\Pi_p) = \sum_{\lambda_i \ll \ell_s m_{\rm KK}} \frac{c_ie_i}{\lambda_i}   \quad \mod \ 1 \, .
\label{Lsm}
\ee

On the one hand, this quantity is not a topological invariant of $X_n$. Unlike for \eqref{L}, there is no reason for it to remain invariant under a continuous deformation of either of the representatives $\Pi_p$ or $ \Pi_{n-p-1}$. On the other hand, as we argue in section \ref{s:dimred}, whenever \eqref{Lsm} is non-vanishing the massive sector of the $D$-dimensional EFT  knows about the value of \eqref{L}, so there must be some way in which one can find about this topological invariant in terms of smeared data.

In the following we propose a solution to this puzzle, namely that one needs to focus on certain minimal-volume representatives within the torsion homology class. More precisely, we consider manifolds $X_n$ that contain calibration $p$-forms, and torsion $p$-cycles that are calibrated by them.  Calibration $p$-forms are standard objects in manifolds endowed with metrics of special holonomy \cite{Harvey:1982xk}. In that case they are closed $p$-forms, and therefore torsion $p$-cycles cannot be calibrated. However, using string theory one may generalize the notion of calibration to any Riemannian manifold $X_n$ that leads to a $D$-dimensional supersymmetric EFT with BPS objects \cite{Martucci:2005ht,Koerber:2005qi,Koerber:2010bx}. With this more general definition, which will be the one used in this chapter, calibrations may be non-closed $p$-forms that calibrate torsion $p$-cycles, as for instance happens in manifolds with $G$-structure metrics. An illustrative case for our discussion in the following sections  will be the case of six-dimensional manifolds with SU(3)-structure, whose metric is specified by the pair of calibrations $(J, \Omega)$, which can respectively calibrate two- and three-cycles that are torsion or even trivial in homology. In terms of this language, our proposal can be expressed as follows:

\begin{conjecture}
	
	A non-trivial smeared linking number between two calibrated torsion cycles equals their actual linking number. 
	
	\label{conj:BPS}
\end{conjecture}

From a physics viewpoint, D-branes wrapping calibrated cycles correspond to BPS objects of different dimensions in the lower-dimensional supersymmetric EFT. In this sense, Conjecture \ref{conj:BPS} can be understood as the equality between \eqref{L} and \eqref{Lsm} for the case of D-branes that wrap torsion cycles and that at the same time are mutually BPS, that is, they preserve some common supercharges in the lower-dimensional EFT.      Notice that equating \eqref{L} to \eqref{Lsm} implies a cancellation in the contribution of very massive modes to the torsion linking number. Physically this suggests that a protection mechanism against threshold corrections must be in place, which is indeed a characteristic feature of certain supersymmetric settings.

To make the proposal more precise, a number of comments are in order. First, some of the compactification manifolds that we will consider correspond to supersymmetric $D$-dimensional EFTs without vacua in the interior of their field space. Instead, they describe supersymmetric solutions that probe a family of metrics of $X_n$. For this reason, we require that the torsion representatives that are BPS/calibrated must be so in a region of the EFT field space, as opposed to in a single point. In particular, they must remain calibrated upon local deformations of the metric that either are moduli or involve energies below the compactification scale. Notice that, in general, calibrated $p$-cycles in fixed homology classes can cross walls of marginal or threshold stability when one deforms the metric of the compactification manifold, so this condition is a significant restriction in the definition of calibrated submanifolds, that we will dub strict calibration condition. In the string theory literature, examples of BPS objects with this property are the EFT strings and membranes defined in \cite{Lanza:2020qmt,Lanza:2021udy,Lanza:2022zyg}, so in this sense some the objects of study in this chapter can be thought of as their torsion analogues. 

Second, notice that Conjecture \ref{conj:BPS} implies that the smeared linking number should not vary upon infinitesimal deformations of the embedding of the torsion representatives that respect the calibration condition, which we will dub as BPS deformations. In the following sections we argue that this is indeed the case, by relating the coefficients $c_i$, $e_i$ with the volume of the respective $p$-cycles. Now, since we are interested in metrics with non-closed calibration $p$-forms, one may in principle encounter BPS deformations that vary the $p$-cycle volume. We have not found instances of this possibility for our more restrictive definition of calibrated cycle. However, in case that it occurred to apply Conjecture \ref{conj:BPS} one should consider the calibrated representative of $\Pi_p$ that locally minimizes its volume for a fixed metric in $X_n$. 

Finally, using the bilinearity of the linking number one may extend the conjecture to torsion $p$-cycles that are not calibrated by themselves, but that are linear combinations of calibrated cycles. For instance, let us consider a manifold $X_n$ with a $G$-structure metric and a pair of $p$-cycles $\Pi_p$ and $\Pi_p'$ calibrated by the same calibration, both in the strict sense, that correspond to the same class on $H_p(X_n, \IR)$, but such that $[\Pi_p^{\rm tor}] = [\Pi_p'] - [\Pi_p]$ is a non-trivial element of ${\rm Tor} H_p(X_n, \IZ)$. Then one may smear out both delta-forms separately, and define a smeared description of the torsion two-cycle as
\be
\delta^{\rm sm} (\Pi_p') - \delta^{\rm sm} (\Pi_p) \, .
\label{deldif}
\ee
By construction, this is an exact smooth $(n-p)$-form, from where one can extract the coefficients $c_i$ as in \eqref{deltasm-07}. A different realization of $\Pi_p^{\rm tor}$ in terms of BPS cycles, like for instance a representative $\Pi_p^{\rm tor}$ that is BPS by itself, may give rise to different coefficients $c_i$. However, the extension of the conjecture would imply that all these choices give rise to the same smeared linking number with a given BPS torsion $(n-p-1)$-cycle. Notice that with this extension one may not only compute torsion linking numbers via smeared data in $G$-structure manifolds with non-closed calibrations, but also in manifolds with metrics of special holonomy. 

To sum up, our proposal means that for manifolds endowed with certain metrics, one can compute some torsion invariants in terms of smeared/massive EFT data. One only needs {\it i)} the eigenforms of the Laplacian that correspond to their lowest eigenvalues and {\it ii)} the projection of torsion, strict-calibrated cycles into them.


\section{Localized sources and dimensional reduction}
\label{s:dimred}

The aim of this section is to motivate the content of Conjecture \ref{conj:BPS} from a physics viewpoint, by considering the effect of localized sources in compactifications of string theory. If these sources wrap torsion cycles in the compact dimensions and couple to the massive fields present in the lower-dimensional EFT, then by consistency of the low-energy description there should be terms in the EFT Lagrangian that know about their torsion linking number. The reason is that, in this case, the EFT contains localized objects charged under a discrete gauge symmetry (the torsion cohomology group) with a non-trivial backreaction at EFT wavelengths. The corresponding EFT Lagrangian has the form proposed in \cite{Camara:2011jg} (see also \cite{Grimm:2011tb,Berasaluce-Gonzalez:2012abm,Berasaluce-Gonzalez:2012awn,Mayrhofer:2014laa,Grimm:2015ona,Braun:2017oak}) to describe torsion in (co)homology from the viewpoint of dimensional reduction. However, this does not guarantee that one can compute the torsion linking number from smeared data. For this, one in addition needs that such localized sources appear as BPS objects of the EFT. 

\subsection{Localized sources in ten and four dimensions}

For concreteness, let us consider a static D4-brane in 10d, with worldvolume $\Sigma_5 = \IR \times \Sigma_4 \subset \IR^{1,9}$. Its backreaction sources a RR field strength $F_{4} = dC_{3}$, such that $dF_{4} = \delta_5(\Sigma_5)$  corresponds to the bump delta 5-form with support on $\Sigma_5$. On a 4-sphere $S^4$ surrounding this source, the pullback of $F_4$ is of the form $2\pi \Phi_{S^4}$, where $\Phi_{S^4}$ is such that $\int_{S^{4}} \Phi_{S^4} = 1$. Analogously to the Wu-Yang description of a 4d monopole, we need to describe the potential $C_{3}|_{S^4}$ as a connection, more precisely as the connection of a 2-gerbe on $S^4$, see e.g. \cite{Hitchin:1999fh}. 

Let us  recall how a probe D2-brane feels this background. In particular, let us consider the case where its worldvolume $\Sigma_{3}$ sweeps a 3-sphere $S^{3}$ at the equator of $S^{4}$. In analogy with the Wu-Yang monopole, the non-trivial pull-back of $F_{4}$ on $S^{4}$ has the effect that $C_{3}|_{S^{3}}$ is non-trivial in the cohomology of $S^3$. However, one can still globally write it as $d\lambda_{2}$, where $\lambda_{2}$ is not a globally well-defined smooth two-form, but nevertheless $e^{i\int_{\Pi_{2}} \lambda_{2}}$ is well-defined for any two-cycle $\Pi_{2}$ inside  $S^{3}$. This property amounts to saying that the wavefunction of the probe D2-brane is well-defined in the backreacted background of its magnetic dual. 

We now consider the particular 10d background $\IR^{1,3} \times X_6$, with $X_6$ a compact manifold with a given metric.  If the D4-brane wraps a 4-cycle $\Pi_4 \subset X_6$, then it will look like a point-like source in the 4d EFT, very much like a monopole. Its backreaction may be described by a 2-gerbe in the microscopic 10d picture, but its effective description at wavelengths larger than $1/m_{\rm KK}$  should correspond to a bundle similar to that of the Wu-Yang monopole. This is indeed the case whenever $[\Pi_4]$ is a non-trivial class in $H_4 (X_6,\IR)$. A probe D2-brane wrapping a two-cycle $\Pi_{2} \subset X_6$ with non-trivial transverse intersection $ \Pi_2 \cdot \Pi_{4} = Q$ looks, from the 4d viewpoint, like a (test) particle circling around the monopole-like source. The pull-back of $F_{4}$ on $S^2  \times \Pi_{2}$ with the two-sphere surrounding the source reads:
\be
2\pi \left( \Phi_{S^2} + d\chi_1\right) \wedge \delta_2(\Pi_4)|_{\Pi_{2}} = 2\pi Q \left( \Phi_{S^2} + d\chi_1\right) \wedge \left(\Phi_{\Pi_{2}} + d\tilde{\chi}_{1}\right)  ,
\label{pullF84d}
\ee
where the $\Phi$'s are volume forms normalized to unity, $\chi_1$ and $\tilde{\chi}_1$ are globally well-defined one-forms on $S^2$ and $\Pi_4$,  respectively, and $
\delta_2(\Pi_4)$ is the bump delta two-form of $\Pi_4$ in $X_6$. The result gives an integral of $2\pi Q$, that corresponds to the product of electric and magnetic charges. We can now restrict our attention to $\gamma \times \Pi_{2}$, where $\gamma$ is at the equator of $S^2$. The difference of connections $C_{3}$ on two patches overlapping over  this submanifold can be written as  
\be
C_{3} = d\lambda_{2}, \qquad \text{with} \quad \lambda_{2} = \lambda\, Q \left( \Phi_{\Pi_{2}} + d\tilde{\chi}_{1}\right) ,
\label{C3point}
\ee
and $\lambda$ a function $\lambda : \gamma \to S^1$ with a single winding. When we describe this system at energies well below the compactification scale, we simplify the internal profile for $\lambda_{2}$. In the effective description, one replaces $ \delta_2(\Pi_4)$ by a harmonic two-form $\omega_2^{\Pi_4}$ in the Poincar\'e dual class to $[\Pi_4]$, which is the lowest lying mode (the harmonic piece) of the Kaluza--Klein (KK) decomposition of $\delta_2(\Pi_4)$. Therefore, we write $ \lambda_{2} = \lambda Q \omega_2^{\Pi_4}$, as a more detailed profile would involve gauge transformations for massive $U(1)$'s that are beyond our 4d EFT description. 

If we now assume that $[\Pi_4] \in {\rm Tor}H_4(X_6, \IZ)$, then $\omega_2^{\Pi_4}$ vanishes, and the presence of a D4-brane wrapped on $\Pi_4$ remains undetected by any D2-brane wrapping a two-cycle $\Pi_2 \subset X_6$. Instead, as discussed in \cite{Camara:2011jg}, one needs to consider a D4-brane wrapping  a three-cycle $[\Pi_3] \in {\rm Tor}H_3(X_6, \IZ)$, which is perceived by the low-energy EFT as a 4d string. Let us for simplicity place this 4d string in $\IR^{1,1} \subset \IR^{1,3}$, and take $z = r e^{2\pi i \theta}$ to be the complex coordinate transverse to its worldsheet. This time one can provide a global description of the RR potential sourced by the D4-brane
\be
C_{3} = 2\pi d \left( \theta \rho_{2} \right) ,
\label{Cstringb}
\ee
where $\rho_2$ is a two-form on $X_6$ such that $d\rho_2 = \delta_3(\Pi_3)$ is the bump delta three-form of $\Pi_3$. This background is detected by a D2-brane with worldvolume $\g \times\Pi_2$, where $[\Pi_2] \in  {\rm Tor} H_2(X_6, \IZ)$ is a torsion class with linking number $L \in \mathbb{Q}$ with respect to $[\Pi_3]$, and $\gamma$ is a 4d worldline surrounding once the string location $\{z=0\}$. If we pull back \eqref{Cstringb} into the D2-brane worldvolume we obtain
\be
C_{3}|_{\g \times \Pi_{2}} = 2\pi L \left( d\phi + df(\phi) \right) \wedge \left(\Phi_{\Pi_{2}} + d\tilde{\chi}_{1}\right)  ,
\label{pullC74d}
\ee
where $\phi \in \IR/ \IZ$ parametrizes $\gamma$ and $f(\phi)$ is a periodic function in it, and we have used that $\int_{\Pi_{2}} \rho_{2} = L$. Therefore, we obtain that  $C_{3} = d\lambda_2$, with $\lambda_{2}$ of the form \eqref{C3point}, except for the replacement $Q \to L$. The fact that $L$ is not an integer number implies that the D2-brane picks a non-trivial phase $e^{2\pi i L}$ when circling around $\gamma$, which is a trait of 4d Aharanov-Bohm (AB) strings and signals the presence of a discrete gauge symmetry \cite{Banks:2010zn}. 

Let us describe the discrete gauge symmetry in terms of the gauge transformations involved in the backreacted D4-brane background. At the microscopic 10d level, these are of the form
\be
d \left( \lambda \rho_{2}\right) = d\lambda \wedge  \rho_{2} + \lambda \, \delta_3 (\Pi_{3}) ,
\label{truel}
\ee
with $\lambda$ well-defined on loops on $\IR^{1,3}$ but not on $\IR^{1,3}$ itself. It now remains to see what is the long-wavelength 4d EFT description of this transformation. As already discussed $\delta_3 (\Pi_{3})$ has no harmonic component, and the same can be assumed for $\rho_2$.\footnote{A priori nothing forbids $\rho_2$ to have a  harmonic piece, which would even be required if we impose that $\int_{\Pi_2} \rho_2 \in \IZ$ for any two-cycle $\Pi_2$ \cite{Marchesano:2014iea}. Following \cite{Marchesano:2014bia}, this piece would imply a non-trivial kinetic mixing between massive and massless $U(1)$'s of the compactification, which could then be removed by an  appropriate change of basis. To simplify the discussion, here we assume the absence of such a harmonic piece.} The question is then if $\delta_3 (\Pi_{3})$ has a non-trivial projection into the massive field content of the 4d EFT spectrum, or in other words if it has a non-trivial 4d smearing. If it does, the D4-brane backreaction should be seen by the 4d EFT, in the sense that it sources some of it fields, that pick a non-trivial profile involving wavelengths above $1/m_{\rm KK}$. So in the following we will assume that $\delta_3 (\Pi_{3})$ has a non-trivial 4d smearing, which is also necessary for Conjecture \ref{conj:BPS} to provide a non-trivial statement.

For simplicity let us assume that $X_6$ is such that ${\rm Tor}H_3(X_6,\IZ) =\IZ_N$. By the Universal Coefficient Theorem \cite{Bott1982DifferentialFI} and Poincar\'e duality this implies that ${\rm Tor}H_2(X_6,\IZ) =\IZ_N$, and also that $LN \in \IZ$. Let us in addition assume that there is a single exact eigen-three-form $b_3$ of the Laplacian with unit norm and a non-vanishing eigenvalue below the compactification scale. That is, we have a unique solution of the form $dd^{\dag} b_3 = \lambda_{\rm st}^2 b_3$, with $m_{\rm st} = \ell_s^{-1}\lambda_{\rm st} \ll m_{\rm KK}$. Then to obtain our 4d EFT via dimensional reduction we must consider the following set of $p$-forms
\be
b_3 , \qquad \star b_3, \qquad  \lambda_{\rm st}^{-1} d\star  b_3,  \qquad \lambda_{\rm st}^{-1} d^\dag b_3 ,
\label{massivepforms}
\ee
all of them with unit norm and the same eigenvalue, because they are associated with the same mass scale. The standard dimensional reduction procedure consists of expanding the 10d $p$-form potentials in a basis of harmonic forms plus the above, non-harmonic set. For instance, reducing the type IIA three-form $C_3$ to 4d with respect to the above non-harmonic sector gives 
\be
C_3 = 2\pi  \ell_s^{3}  \left(A_1 \wedge \om_2 +  C_0 \, \b_3 \right) ,
\label{C3sing}
\ee
where $A_1$ and $C_0$ describe a 1-form and a 0-form in 4d, respectively, and we have defined
\be
\b_3 = f\, b_3 , \qquad \om_2 =  \frac{1}{g \lambda_{\rm st}} d^\dag b_3 , \qquad \text{with} \quad f, g \in \IR .
\label{alom}
\ee
so $d\om_2 = N_{\rm eff} \b_3$ with $N_{\rm eff} = \frac{\lambda_{\rm st}}{fg}$. The reduction to 4d of the 10d kinetic term $\int F_4^2$ gives
\be
(2\pi  \hat{f})^2 \left( dC_0 - N_{\rm eff} A_1 \right)^2 + \frac{4\pi^2}{g^{2}} (dA_1)^2 , \qquad \hat{f} \coloneqq  f e^{\phi_4} M_{\rm P} = \frac{f e^{\phi}}{{\rm Vol}_{X_6}^{1/2}} M_{\rm P}  ,
\label{stuck}
\ee
namely a St\"uckelberg-like Lagrangian, where $e^{\phi_4}$ is the 4d and $e^{\phi}$ the 10d dilaton, Vol$_{X_6}$ is the volume of $X_6$ in string units, and $M_{\rm P}$ the 4d Planck mass. This is precisely the dimensional reduction scheme proposed in \cite{Camara:2011jg} to describe discrete gauge symmetries from torsion in cohomology, if one imposes the constraint $N_{\rm eff} = N$ and treats $C_0$ as an axion-like particle of unit periodicity $C_0 \sim C_0 +1$. In this case, the discrete gauge symmetry is generated by the shift
\be
2\pi C_0 \to 2\pi C_0 +\lambda , \qquad 2\pi A_1 \to  2\pi A_1 +  \frac{d\lambda}{N} ,
\label{disgauge}
\ee
with $\lambda \in 2\pi \IZ$. A particle with charge $NL$ under $A_1$ will pick up a phase $e^{2\pi i L}$ upon \eqref{disgauge}, for instance when circling a string of unit charge. This is how the 4d EFT reflects the linking number between torsion cycles on $X_6$, and in particular that ${\rm Tor} H_3(X_6,\IZ) \simeq $ ${\rm Tor} H_2(X_6,\IZ) =\IZ_N$. So while at this point we have not determined the parameters $f$ and $g$, consistency of the 4d EFT requires that they are constrained by $fg = \frac{\lambda_{\rm st}}{N}$. Therefore we have the relation
\be
\frac{1}{N} = \frac{fg}{\lambda_{\rm st}} = \frac{\hat{f}g}{m_{\rm st}} .
\label{Lphys}
\ee
Note that the expression in the middle resembles a smeared linking number, as defined in \eqref{L}, while the rhs corresponds to how the  EFT massive sector encodes this quantity.

The 4d effective Lagrangian \eqref{stuck} should be sufficient to give a long-wavelength description (more precisely in the range $(m_{\rm KK}^{-1}, m_{\rm st}^{-1}))$ of the backreaction of D4-branes wrapping torsion three-cycles of $X_6$. Recalling our 10d analysis, one may try to provide such a 4d description by directly smearing the 10d solution, that is by projecting the background \eqref{Cstringb} into the massive sector \eqref{massivepforms}. However, if one does so the gauge transformation \eqref{truel} translates into
\be
2\pi C_0 \to 2\pi C_0 + \frac{c}{f} \lambda , \qquad 2\pi A_1 \to  2\pi A_1 + \frac{c}{f}  \frac{d\lambda}{N} ,
\label{disgauge2}
\ee
where $\delta_3^{\rm sm}(\Pi_3) = c b_3$, and we have imposed that $N_{\rm eff} = N$. So only when $c=f$ we recover the expected gauge transformation \eqref{disgauge}. While this may seem surprising, it does not necessarily indicate any inconsistency of the 4d EFT. Instead, one may interpret it as the fact that smearing a 10d background is a classical procedure that may be subject to corrections, like quantum corrections associated to the fields above the compactification scale that one has truncated.   So in principle, it could be that these or other corrections modify the backreacted 4d background in such a way that the quotient $c/f$ disappears from \eqref{disgauge2}, and one recovers the gauge transformation \eqref{disgauge} consistent with \eqref{stuck}. If this was the case, \eqref{Lphys} should be interpreted as a smeared linking number after corrections have been taken into account. 

This proposal to solve the apparent inconsistency in \eqref{disgauge2} has the downside that it does not give a clear geometric prescription to compute the parameters $f$ and $g$ which, together with $\lambda_{\rm st}$, are the 4d EFT data that allow us to compute $N$. However, it gives us the guideline that one should try to consider D4-branes whose smeared backreaction does not suffer important corrections upon dimensional reduction. From a physics viewpoint, the best candidates to display this feature are D4-branes that preserve some supersymmetry of the background, namely BPS objects of the EFT, as the results of \cite{Blaback:2010sj} also suggest. In the next subsection we will argue why this is the right answer. 

Finally, it is instructive to perform the dimensional reduction of the RR potential $C_5$, dual to $C_3$ in 10d. An expansion in the relevant non-harmonic $p$-forms \eqref{massivepforms} gives
\be
C_{5} = 2\pi \ell_s^{5}\left( V_1 \wedge  \tilde{\omega}_4 +  B_2 \wedge \a_3 \right) , 
\label{C5sing}
\ee
where $V_1$ and $B_2$ are a 4d 1-form and 2-form in 4d, and we have defined
\be
\tilde{\omega}_4  = \frac{g}{\lambda_{\rm st}} d \star  b_3 , \qquad \a_3 =  f^{-1} \star  b_3  .
\label{tombe}
\ee
such that $\int_{X_6} \omega_2 \wedge \tilde{\omega}_4 = \int_{X_6} \a_3 \wedge \b_3 = 1$, as in \cite{Gurrieri:2002wz,DAuria:2004kwe,Grana:2005ny,Kashani-Poor:2006ofe}.  This is required for the fields $(V_1, B_2)$ to be quantized 4d duals to $(A_1, C_0)$. It also implies that $d\a_3 = N_{\rm eff} \tilde{\om}_4 = N \tilde{\om}_4$, so upon dimensional reduction one obtains 
\be
(2\pi g)^2 \left( dV_1 + N B_2  \right)^2 + \frac{4\pi^2}{\hat{f}^{2}}  (dB_2)^2 .
\label{stuckdual}
\ee
This 4d effective Lagrangian should describe the backreaction of D2-branes wrapping torsion two-cycles $\Pi_2$ of $X_6$, in the long-wavelength approximation. In these dual variables the discrete gauge symmetry reads
\be
2\pi V_1 \to 2\pi V_1 - \lambda_1 , \qquad 2\pi B_2 \to  2\pi B_2 +  \frac{d\lambda_1}{N} .
\label{disgaugedual}
\ee
A D2-brane wrapping a torsion two-cycle $\Pi_2$ such that $\delta_4^{\rm sm}(\Pi_2) = \frac{e}{\lambda_{\rm st}} d\star  b_3$ will not generate this shift via its backreaction, unless $e = g$. Again, one could interpret this mismatch as the result of non-trivial quantum corrections, and argue that the equality should hold for D2-branes wrapping calibrated two-cycles, as we proceed to argue.

\subsection{The supersymmetric case}
\label{ss:BPS}

One may summarize the reasoning of the previous subsection as follows. A D4-brane wrapping a torsion three-cycle $\Pi_3$ of $X_6$ becomes, upon compactification to 4d, an Aharanov-Bohm 4d string that realizes ${\rm Tor} H_3(X_6, \IZ) = \IZ_N$ as a discrete gauge symmetry. This object will be perceived by the 4d EFT if it couples to some massive $p$-form modes below the compactification scale $m_{\rm KK}$. In that case the backreaction has size $m_{\rm st} = \lambda_{\rm st}/\ell_s \ll m_{\rm KK}$, where $\lambda_{\rm st}$ is the eigenvalue of such massive eigenmodes, and there must be a term in the 4d EFT that describes such a backreacted solution at long wavelengths. This 4d Lagrangian term is \eqref{stuck}, with $N_{\rm eff} = \lambda_{\rm st}/fg = N$ encoding the topological information of the torsion homology group. Knowledge of the massive spectrum and of the parameters $f, g \in \IR$ thus allows us to compute torsion cohomology groups, and to represent them via smooth $p$-forms \eqref{alom} and \eqref{tombe} that are, from the 4d viewpoint, analogous to the harmonic representatives of de Rham cohomology groups. The parameters $f$ and $g$ are not determined from the 4d smearing of the backreaction of a D4-brane wrapping an arbitrary torsion three-cycle, since in general there can be significant corrections to the smeared background. Notice that these parameters are intrinsic of the 4d EFT, and so they only depend on the topology and metric of $X_6$. 

There is however a particular class of 4d strings for which quantum corrections should be under control, namely BPS fundamental strings of the EFT. We are particularly interested in D4-branes that correspond to the 4d EFT strings of \cite{Lanza:2020qmt,Lanza:2021udy,Lanza:2022zyg}, except that they source 4d axions with a mass $m_{\rm st}$ induced by a St\"uckelberg coupling. As stressed in \cite{Lanza:2020qmt}, near the string core and at wavelengths below $m_{\rm st}^{-1}$ one should be able to describe the 4d backreaction of these objects with a solution similar to that of standard EFT strings, implying that their tension is determined by the kinetic terms of the 4d EFT Lagrangian, and in particular by parameters like $f$. 

Geometrically, the BPS condition means that the torsion three-cycle $\Pi_3$ is calibrated by a complex three-form $\Omega$. This is not possible when $X_6$ is a Calabi--Yau, but it occurs in SU(3)-structure manifolds with a metric specified by $(J, \Omega)$ and a non-vanishing intrinsic torsion,\footnote{The two meanings of the word torsion should not be confused. By intrinsic torsion we mean the five torsion classes which enter in the description of manifolds with SU(3)-structure metrics, and which show up in the derivatives of the globally well-defined forms $\Omega$ and $J$ \cite{Gray1980TheSC,chiossi2002intrinsic,Hull:1986iu,Strominger:1986uh,LopesCardoso:2002vpf,Gauntlett:2003cy,Grana:2005jc,Koerber:2010bx}. In any other instance, the word torsion refers to torsion classes in (co)homology groups of $X_6$, and to their representatives.} which we will assume in the following. Notice that the calibration condition selects a specific representative within the torsion class $[\Pi_3] \in {\rm Tor} H_3(X_6, \IZ)$, that directly depends on the metric of $X_6$. Therefore, it is reasonable to assume that $f$, which also depends on the metric of $X_6$, can be computed from $\delta_3(\Pi_3)$ with $\Pi_3$ calibrated. More precisely, we will argue that $f$ can be computed from the smeared delta-form $\delta_3^{\rm sm}(\Pi_3)$.

To see this, let us assume that upon compactification of type IIA string theory on $X_6$ we recover a 4d EFT with $\CN= 2$ supersymmetry. One interesting framework to do so is when $X_6$ is an SU(3)-structure manifold with calibrations $(J,\Omega)$, as analyzed in \cite{Grana:2005ny,Kashani-Poor:2006ofe}.  Following their approach, we may expand $J$ and $\Omega$ in the set of harmonic two- and three-forms, respectively, plus the non-harmonic set \eqref{massivepforms}. Let us first consider $\Omega$ and assume an expansion of the form $\Omega = \Omega^{\rm harm} + i a \,  \a_3 +  b \, \b_3$, where $\Omega^{\rm harm}$ is a sum of harmonic three-forms and $a$, $b$ are real functions of the 4d fields. If we impose the condition $\star  \Omega = - i \Omega$ we find that the more precise form 
\be
\Omega = \Omega^{\rm harm} + {\rm Vol}_{X_6}^{1/2} \left(if \a_3 +  f^{-1} \b_3 \right) ,
\ee
where we have taken into account that $\int_{X_6} i \bar{\Omega} \wedge \Omega = 8 {\rm Vol}_{\rm X_6}$, and that $\a_3, \b_3$ are orthogonal to any harmonic form. Notice that here $f$ is not a fixed number, but depends on the choice of SU(3)-metric or, from the 4d viewpoint, on the vevs of the 4d scalar fields.

Next, we use that for a 4d BPS  string its tension is proportional to the  kinetic term of the axion to which it couples magnetically. In the  case at hand, the orthogonality of the massive modes implies that the string charge-to-mass ratio equals one, and so for a string inducing a single winding of $C_0$ around its core the tension is determined by the axion decay constant in \eqref{stuck} as $\hat{f}M_{\rm P}$. This quantity should correspond to the 4d string tension obtained from a D4-brane wrapped on a three-cycle $\Pi_3$ calibrated by $\Omega$, see \cite[section 6.4]{Lanza:2021udy}. We thus find\footnote{We change slightly the definition of the 4-dimensional Planck mass with respect to (\ref{eq:planck}). We use \begin{equation}
			M_P	\ell_s^{-1} = e^{\phi_4} =  e^{\phi}  \, {\rm Vol}_{X_6}^{-1} \,.
\end{equation} }
\be
\frac{\hat{f}}{M_{\rm P}}= \frac{e^{\phi}}{{\rm Vol}_{X_6}} \left| \int_{\Pi_3}  \Om \right| \implies  f = {\rm Vol}_{X_6}^{-1/2} \left| \int_{X_6}  \Om \wedge \delta_3(\Pi_3) \right| .
\label{fst}
\ee
Notice that in the second equation we can replace $\delta_3(\Pi_3) \to \delta_3^{\rm sm}(\Pi_3)$. Using that $\Pi_3$ is a torsion three-cycle and therefore $\delta_3(\Pi_3)$ is an exact three-form we finally obtain
\be
\delta_3^{\rm sm}(\Pi_3) = \b_3 .
\ee
That is, $f$ can be found from smearing the bump delta-form of a calibrated torsion three-cycle. 

Similarly, one may consider a D2-brane wrapping a BPS representative of $[\Pi_2] \in {\rm Tor} H_2 (X_6, \IZ)$, or in other words $\Pi_2$ is calibrated by $J$. A BPS particle of unit charge with respect to $A_1$ will have a mass $g M_{\rm P}$, so putting both statements together results in the equality
\be
g = {\rm Vol}_{X_6}^{-1/2} \left| \int_{X_6}  J \wedge \delta_4(\Pi_2) \right| .
\label{gpt}
\ee
Again, expanding $e^{iJ}$ in harmonic and non-harmonic forms and using the Hodge duality relations translates into the equality 
\be
\delta_4^{\rm sm}(\Pi_2) = \tilde{\om}_4 .
\ee
Equivalently, $g$ results from smearing the bump delta-form of a calibrated torsion two-cycle.

Notice that in this construction the torsion cycles $\Pi_3$ and $\Pi_2$ that lead to $f$ and $g$ have a minimal 4d charge and tension. Therefore we expect them to generate ${\rm Tor} H_3(X_6, \IZ)$ and ${\rm Tor} H_2(X_6, \IZ)$, respectively, and to have a linking number $1/N \mod 1$. When plugging the values of $f$ and $g$ into the smeared linking number one indeed finds that $L^{\rm sm} (\Pi_2, \Pi_3) =1 /N$, in agreement with Conjecture \ref{conj:BPS}. If instead $\Pi_2$ corresponds to a particle of charge $LN$, then repeating the same reasoning its smeared delta-form will have to be multiplied by $LN$, and we will recover a smeared linking number of $L$, again supporting the conjecture. 

An interesting point is that, when dealing with mutually BPS objects, one should be able to add up their tensions to compute the energy of the total system. Geometrically, this amounts to say that even if the topological charge of a calibrated torsion $p$-cycle or a sum of them lives in $\IZ_N$, its central charge lives in a lattice. This does not imply any contradiction with the $\IZ_N$ discrete system of the 4d EFT, provided that the process that reduces the number of BPS objects by $N$ has a non-vanishing energy which compensates for the loss of $N$ $p$-cycles. To illustrate this, let us consider the torsion three-cycle class $[\Pi_3]$ generating ${\rm Tor} H_3(X_6, \IZ)$ in a SU(3)-structure manifold. A set of $N$ D4-branes wrapping calibrated representatives $\Pi_{3,i}$ of this class looks like $N$ BPS strings in 4d. These can end on a 4d monopole, made up of a D4-brane wrapping a four-chain $\Sigma_4$ whose boundary is given by $\p \Sigma_4= \sum_i \Pi_{3,i}$ \cite{Camara:2011jg}. On the one hand, using Stokes' theorem one can relate the sum of string tensions with the integral of $d\Omega$ over $\Sigma_4$. On the other hand, the mass of the 4d monopole is proportional to the volume of $\Sigma_4$ which, if the monopole is BPS, is given by the integral of $\pm \oh J \wedge J$ over $\Sigma_4$. Therefore we find that the marginal stability of $N$ BPS AB strings implies
\be
\ell \left| \int_{X_6} d\Omega \wedge \delta_2(\Sigma_4)\right| + \left| \oh \int_{X_6} J \wedge J  \wedge \delta_2(\Sigma_4) \right| = {\rm const.} 
\label{BPSmono}
\ee
where $\ell$ is the length of the AB string in $\ell_s$ units.  Notice that this relation can only make sense if the monopole mass depends on $\ell$, which should then be a feature of  backgrounds with BPS AB strings and particles. We postpone a more precise explanation of this statement to the next section, where both quantities in \eqref{BPSmono} will be evaluated in a simple setup based on half-flat manifolds. For the time being, it is worth pointing out that the above reasoning leads to an interpretation of the non-closed two-form $\om_2$ in \eqref{alom}. Indeed, notice that in \eqref{BPSmono} we can replace $\delta_2(\Sigma_4) \to \delta_2^{\rm sm}(\Sigma_4)$ and that because the action of smearing commutes with the exterior derivative, $d \delta_2^{\rm sm}(\Sigma_4) = N \delta_3^{\rm sm}(\Pi_3) = N \b_3$. It is thus natural to guess that $\delta_2^{\rm sm}(\Sigma_4) = \om_2$ when $\Sigma_4$ is a calibrated four-chain, something that can be verified by noting that a 4d BPS monopole of unit charge has mass $g^{-1} M_{\rm P}$, and running a reasoning analogous to the previous ones. Similarly, one can deduce that $\delta^{\rm sm}_3(\Sigma_3) = \a_3$, where $\Sigma_3$ is a calibrated three-chain ending on $N$ calibrated torsion two-cycles. Therefore, one concludes that the set of harmonic plus non-harmonic forms in which one expands $J$, $\Omega$ and the RR potentials to obtain the 4d fields can be interpreted as smeared delta-forms of a basis of calibrated chains and cycles. Notice that this fits well with the notion that the set of forms $\{ \om_2, \a_3, \b_3, \tilde{\om}_4\}$ reflect quantization features of the 4d EFT, like axions of unit periodicity and $U(1)$ gauge symmetries. This quantization also implies that these $p$-forms generate a lattice just like quantized harmonic $p$-forms do, which seems to be in contradiction with the fact that these D-brane charges are torsion. However, as mentioned above when dealing with mutually BPS objects the mass/tensions are additive, which explains the lattice structure. Finally, while here we have considered a very simple case, it is reasonable to expect that this description of the reduction basis of $p$-forms extends to the general framework of SU(3)-structure manifold dimensional reduction analyzed in \cite{Grana:2005ny,Kashani-Poor:2006ofe}, as we will further discuss in section \ref{s:general}. 

It is also instructive to consider what happens when we slightly depart from a BPS embedding. In particular, let us take a D4-brane wrapping a calibrated torsion three-cycle $\Pi_3$ and perform a small deformation of its embedding, such that the torsion linking number with a calibrated torsion two-cycle $\Pi_2$  does not change
\be
L( \Pi_{2},\Pi_3) = \sum_i \frac{c_ie_i}{\lambda_i}  \, .
\label{Labs}
\ee
Notice that this quantity is not defined mod 1. Geometrically, this means that upon the deformation $\Pi_3$ does not cross $\Pi_2$. A simple deformation of this sort that changes the smeared linking number takes the form 
\be
c_{\rm KK} \to c_{\rm KK} - \eps , \qquad c_{\rm st} \to c_{\rm st} + \frac{\lambda_{\rm st}}{\lambda_{\rm KK}} \eps ,
\label{shiftKK}
\ee
where $c_{\rm KK}$ represents the coefficient of a mode above the compactification scale and $c_{\rm st}$ one below. We thus find that the naive gauge transformation \eqref{disgauge2} changes with a suppression factor of  $m_{\rm st}/m_{\rm KK}$ with respect to the BPS case. This is indeed the kind of suppression that one would expect from integrating out massive operators at the Kaluza-Klein scale, which supports the interpretation that the expected discrete gauge transformation \eqref{disgauge} could be restored for the non-BPS case, once that quantum corrections are taken into account. It would however be important to perform a more direct test of this proposal.

\subsection{Generalizations}

In our discussion so far we have focused in a type IIA setup, in which D4- and D2-branes look respectively like strings and particles in the 4d EFT. However, it is clear that the same reasoning can be applied to any other kind of string compactifications, as long as the 4d picture is similar. For instance, in type IIB compactified in a SU(3)-structure manifold, Aharanov-Bohm strings and particles would be realized by D3-branes wrapping torsion two- and three-cycles. There are other extended objects that can give rise to 4d AB strings and particles \cite{Camara:2011jg}, but in many instances they do not wrap calibrated cycles, and so the BPS property, which is an important ingredient of our logic, is missing.

Nevertheless, one may extend our reasoning in yet another direction, since there are other BPS objects in a 4d EFT that encode torsion in cohomology. Indeed, a key property of AB strings with $\IZ_N$ charge is that $N$ of them can end on a monopole, while $N$ AB particles can end on a 4d instanton \cite{Banks:2010zn}. As a general rule, $\IZ_N$ charges are detected in the 4d theory by $p$-branes ending on $(p-1)$-branes with $p = 0,1,2,3$, and in certain instances these $\IZ_N$ charges reflect torsion cohomology groups of the compactification manifold \cite{Berasaluce-Gonzalez:2012awn}. In our previous discussion we have focused on the cases $p=0$ and $p=1$, which are typically represented in 4d EFT language by the Lagrangians \eqref{stuck} and \eqref{stuckdual}, respectively, and are dual to each other. The case $p=2$ corresponds to 4d membranes ending on strings, and it is related to the following EFT Lagrangian piece \cite{Marchesano:2014mla}
\be
\left( dB_2 - N C_3\right)^2 ,
\label{p=2}
\ee
where $C_3$ is a three-form that couples to the membrane and $B_2$ is a two-form coupling to the string. In our previous type IIA setup, these objects would arise from D4-branes wrapping torsion two-cycles and the three-chain connecting them, respectively, and signal the presence of a non-trivial superpotential. The case $p=3$ describes 4d space-time filling branes ending on membranes, and the corresponding Lagrangian piece reads \cite{Lanza:2019xxg}
\be
\left( dD_3 - N A_4\right)^2 ,
\label{p=3}
\ee
where $A_4$ coupling to the space-time filling branes and $D_3$ to the membranes. In our type IIA setup these 4d objects arise from D6-branes wrapped on torsion three-cycles and on a four-chain linking them, respectively.\footnote{In most of the literature, these Lagrangians are shown to arise from compactifications with NS $H$-fluxes. In this case, $N$ represents an $H$-flux quantum and the feature of $p$-branes ending of $(p-1)$-branes has a microscopic description in terms of $d_H$ cohomology and its dual homology \cite{Evslin:2007ti}. By looking at concrete constructions, it is easy to convince oneself that such a setup is connected by mirror symmetry to the one that we are considering \cite{Tomasiello:2005bp,Marchesano:2006ns}.} 

The general philosophy of the previous subsection also applies to these St\"uckelberg-like couplings. That is, the 4d $p$-forms that appear in \eqref{p=2} and \eqref{p=3} should arise from expanding the 10d RR potentials on smeared delta-forms of calibrated torsion cycles of $X_6$. The resulting coefficients that multiply both expressions are the analogues of $f$ and $g$ in \eqref{stuck} and \eqref{stuckdual}, and so together with the relevant Laplace eigenvalue they determine $N$. Notice that the analogy is not straightforward, because  \eqref{p=2} and \eqref{p=3}  are not dual Lagrangians, which reflects the fact that D4-branes and D6-branes do not couple to dual 10d RR potentials. However one may consider a gauge instanton on the space-time filling D6-branes, which amounts to a D2-brane wrapping the same torsion three-cycle and coupling to a massive 4d axion $C_0'$ that arises from reducing $C_3$ on a coexact three-form like $\a_3$ in \eqref{tombe}. This EFT object is sensitive to the backreaction of D4-branes wrapping torsion two-cycles and the three-chain connecting them, and an analogy with the gauge transformations involving AB strings and particles can be drawn. The precise statement is that there exists a gauged $(-1)$-form symmetry that describes the discrete gauge symmetries of the EFT superpotential \cite{Hebecker:2017wsu,Heidenreich:2020pkc}. 

For the purposes of computing torsion in cohomology,  to consider these new terms in the Lagrangian may seem redundant, since in the type IIA constructions that we have discussed they are related to the same kind of torsion groups, namely ${\rm Tor} H_3(X_6, \IZ) \simeq$ ${\rm Tor} H_2(X_6, \IZ)$ and their linking number. However, an important difference is that the terms \eqref{p=2} and \eqref{p=3} appear in 4d $\CN=1$ string theory vacua, like in type II orientifold compactifications, while \eqref{stuck} and \eqref{stuckdual} typically appear in 4d $\CN=2$ compactifications without vacua, like the example considered in \cite{Gurrieri:2002wz}.  In this case the $\CN=2$ supersymmetry of the Lagrangian is realized off-shell, while solutions to the equations of motion at most preserve a fraction of this supersymmetry, like the domain-wall solution preserving four supercharges to be discussed in the next section. In practice this implies that the 10d background is not of the form $\IR^{1,3} \times X_6$, but instead a fibration of $X_6$ over a real line or a plane in $\IR^{1,3}$. Following the general philosophy of \cite{Gurrieri:2002wz,Grana:2005ny} we are entitled to carry out the usual procedure of dimensional reduction to 4d -- and therefore our discussion above -- as long as the variation of this fibration is very small compared to the compactification scale. The only additional thing that we need to take into account is that for Conjecture \ref{conj:BPS} to apply the objects like 4d strings and particles must be BPS with respect to the 4d solution, which is a stronger condition than being BPS in a would-be $\CN=2$ vacuum. In practice, this means that they must be calibrated also from the point of view of the fibration, as we will illustrate in the next section. 

The fact that AB strings and particles cannot be BPS in $\CN=1$ orientifold compactifications seems to clash with the proposal in \cite{Camara:2011jg}, in the sense that the basis of non-harmonic $p$-forms in which one expands the 10d RR potentials to obtain a St\"uckelberg Lagrangian \eqref{stuck} cannot come from smearing the delta-forms of calibrated cycles. Nevertheless, one can still make sense of such a basis of non-harmonic forms if one considers the extension of Conjecture \ref{conj:BPS} formulated around \eqref{deldif}. For instance, one could try to describe the torsion two- and three-cycles of a Calabi--Yau threefold as the difference of two calibrated cycles with equal volume, or some other combination of calibrated cycles. As long as there are some eigenmodes below the compactification scale that couple differently to these calibrated cycles, there will be non-harmonic $p$-forms that one builds from smearing their bump delta-forms. Finally, one should make sure that such harmonic forms have the appropriate parity under the orientifold action to lead to a St\"uckelberg term.


\section{A simple example}
\label{s:simple}

The simplest example of SU(3)-structure manifolds with torsion in cohomology are nilmanifolds or twisted tori, which in the context of type II string compactifications were initially considered in \cite{Gurrieri:2002wz,LopesCardoso:2002vpf,Kachru:2002sk}. Particularly interesting for our discussion is the setup of \cite{Gurrieri:2002wz}, in which the simplest kind of twisted torus is realized as a 4d domain-wall solution. In the following we will see how the objects defined in the previous sections, in particular torsion calibrated cycles and their smeared delta sources, are described in this case. 

\subsection{The 10d background}

Let us recall the main idea behind the construction in \cite{Gurrieri:2002wz}. One first considers a toroidal compactification of type IIB string theory to 4d with a backreacted NS5-brane wrapping a special Lagrangian three-cycle of ${\bf T}^6$ and extended along $\IR^{1,2} \subset \IR^{1,3}$ in the non-compact dimensions. The long-wavelength approximation  of this backreaction provides a domain-wall solution in 4d, which upon three T-dualities in ${\bf T}^6$ becomes a type IIA background with constant dilaton and a twisted six-torus $\tilde{\bf T}^6$ fibered over a non-compact direction. A simple generalization of this setup results in the following type IIA 10d string frame background:
\bea
\label{SU3ex}
ds^2 &= &ds^2_{\IR^{1,2}}+ \ell_s^2 V(d\xi)^2 + \ell_s^2 ds^2_{\tilde{\bf T}^6} , \\ 
ds^2_{\tilde{\bf T}^6}& = &(2\pi)^2 \left[\frac{R_1^2}{V_1}(\eta^1)^2+\frac{R_2^2}{V_2}(\eta^2)^2+\frac{R_3^2}{V_3}(\eta^3)^2+\frac{V R_4^2}{V_1}(\eta^4)^2+\frac{V R_5^2}{V_2}(\eta^5)^2+\frac{V R_6^2}{V_3}(\eta^6)^2 \right] ,
\label{SU3exb}
\eea
where $\xi$ is the 4d coordinate transverse to the domain-wall, $R_i$ are radii measured in string units, and $\eta^i$ are the left-invariant one-forms of the twisted six-torus, defined as
\begin{equation}
	\begin{array}{ll}
		\eta^1= d x^1 + M_1  x^6 d x^5   \,, & \qquad \eta^4= d x^4 \,, \\
		\eta^2= dx^2 + M_2 x^4 d x^6  \,, & \qquad \eta^5= d x^5 \,, \\
		\eta^3= d x^3 + M_3 x^5 d x^4   \,, & \qquad \eta^6=  d x^6 \, ,
	\end{array}
	\label{etas}
\end{equation}
with $M_i \in \mathbb{N}$. Finally, 
\be
V = V_1V_2V_3 , \qquad V_i = 1 - \zeta_i \xi , \qquad \zeta_i = \frac{M_i}{2\pi} \frac{R_i R_{i+3}}{R_4R_5R_6} .
\label{Vis}
\ee
To recover the case of \cite{Gurrieri:2002wz} one needs to take $M_i=M \in \mathbb{N}$ and $M_j= M_k=0$, with $i \neq j \neq k \neq i$. The solution applies to the range $(0,\xi_{\rm end})$, with $\xi_{\rm end} = {\rm min} \{\zeta_i^{-1}\}_i$, while for $\xi<0$ one should glue a direct product $\IR^{1,3} \times {\bf T}^6$, with torus radii $R_i$.\footnote{Our background differs slightly from the one  in \cite{Gurrieri:2002wz}, in the sense that therein the choice $V_i = \zeta_i \xi$ along the range $\xi \geq 0$ is taken, for a domain wall  at $\xi=0$. Both choices are compatible with the domain-wall analysis of \cite{Curio:2000sc,Behrndt:2001qa,Behrndt:2001mx}, but we find that our choice also reproduces the scalar flow features of $\oh$BPS domain walls in $\CN=1$ EFTs (see e.g.  \cite[section 4.3.2]{Lanza:2020qmt}) and is compatible with the presence of BPS AB strings as particles, as discussed below. \label{ft:dw}}

We refer to \cite{Marchesano:2006ns} for more details on the geometry and topology of this class of twisted six-tori. As in there, one can impose a $\IZ_2 \times \IZ_2$ orbifold projection that reduces the structure of the internal manifold to a genuine SU(3) structure, and which we will assume in the following. In the conventions $d{\rm vol}_{X_6} = -\frac{1}{6} J^3 = \frac{i}{8} \bar{\Omega} \wedge \Omega$, the SU(3)-structure calibrations $(J,\Omega)$ are given by
\bea
\label{Jex}
J & = & 4\pi^2 \left( t^1 \,\eta^1 \wedge \eta^4 +t^2\, \eta^2 \wedge \eta^5  + t^3 \,\eta^3 \wedge \eta^6 \right) , \\
\Om & = & {i} (2\pi)^3 V^{-1/2} R_1R_2R_3 \left( \eta^1 + i  \tau^1 \eta^4 \right) \wedge \left( \eta^2 + i \tau^2 \eta^5 \right) \left( \eta^3 + i \tau^3 \eta^6 \right) ,
\label{Omex}
\eea
with
\be
\label{defttau}
t^i = \frac{V^{1/2}}{V_i} R_iR_{i+3} , \qquad  \tau^i = V^{1/2} \frac{R_{i+3}}{R_i} .
\ee

The calibrated objects of this SU(3)-structure manifold are those $p$-chains whose volume is computed by integrating $\Omega$ or $e^{iJ}$. Recall, however, that we are interested in a particular kind of calibrated cycles. First, they need to be calibrated in a strict sense, meaning that upon varying the values of the $R_i$ they are still calibrated. Second, they need to be mutually BPS with the domain-wall source, in order to be actual BPS objects of the background \eqref{SU3ex}. This second criterion is more easily analyzed in the type IIB mirror background, as done in Appendix \ref{ap:NS5DW}. In our context, one finds the following BPS objects that are relevant to our discussion:

\begin{itemize}
	
	\item[-] A D4-brane wrapped on $\Pi_3^{\rm tor} = \{ x^4=x^5=x^6=0\}$ in $\tilde{\bf T}_6$ and extended along $\xi$.  
	
	\item[-] A D4-brane wrapped on $\Sigma_4^i = \{ x^i = x^{i+3} = 0\}$ in $\tilde{\bf T}_6$.
	
	\item[-] An Euclidean D2-brane on $\Pi_2^i = \Sigma_4^j \cap \Sigma_4^k$, with $i \neq j \neq k \neq i$ in $\tilde{\bf T}_6$ and extended along $\xi$.
	
	\item[-] An Euclidean D2-brane wrapped on $\Sigma_3 = \{ x^1=x^2=x^3=0\}$ in $\tilde{\bf T}_6$.
	
\end{itemize}
Notice that, when extending a D-brane along $\xi$, it does not make sense to do it beyond  $\xi_{\rm end}$, where the metric degenerates and we enter a strong coupling region. 

The submanifolds $(\Pi_3^{\rm tor}, \Sigma_4^i, \Pi_2^i, \Sigma_3)$ and others can be described via group theory techniques, by first writing the twisted six-torus as a coset $\tilde{\bf T}^6 = G/\Gamma$, with $G$ a Lie group of a 2-step nilpotent algebra and $\Gamma$ a co-compact lattice, and then exponentiating different set of generators of $G$, see \cite[Appendix A]{Marchesano:2006ns}. Using this framework and the results of \cite{Nomizu1954OnTC,Cenkl2000NILMANIFOLDSAA}, one can see that 
\be
{\rm Tor } H_3 (\tilde{\bf T}^6, \IZ)_{\IZ_2 \times \IZ_2} =  {\rm Tor} H_2 (\tilde{\bf T}^6, \IZ)_{\IZ_2 \times \IZ_2} = \IZ_M ,
\label{cohott6}
\ee
where $M = {\rm g.c.d.} (M_1, M_2, M_3)$, and the subindex represents those cycles invariant under the $\IZ_2 \times \IZ_2$ orbifold projection.\footnote{The generators of the $\IZ_2 \times \IZ_2$ orbifold group act on the left-invariant one forms as $\theta_1: (\eta^1, \eta^2, \eta^3, \eta^4, \eta^5, \eta^6) \mapsto  (\eta^1, -\eta^2, -\eta^3, \eta^4, -\eta^5, -\eta^6)$ and $\theta_2: (\eta^1, \eta^2, \eta^3, \eta^4, \eta^5, \eta^6) \mapsto  (-\eta^1, -\eta^2, -\eta^3, -\eta^4, -\eta^5, \eta^6)$ \cite{Marchesano:2006ns}. It is not obvious if the torsion cohomology of the orbifold quotient $\tilde{\bf T}^6/\IZ_2 \times \IZ_2$ corresponds to \eqref{cohott6} or if it has further elements. However, in case that some additional torsion cycles existed, one can show that they are not calibrated and they do not couple to any light eigenmode. Therefore one can ignore them for the purposes of this section. \label{ft:orbifold}}
The three-cycle $\Pi_3^{\rm tor}$ is the generator of ${\rm Tor } H_3 (\tilde{\bf T}^6, \IZ)_{\IZ_2 \times \IZ_2}$, while ${\rm Tor } H_2 (\tilde{\bf T}^6, \IZ)_{\IZ_2 \times \IZ_2}$ is generated by $\Pi_2^{\rm tor} = \sum_i (M_i/M) \Pi_2^i$. Additionally, $\Sigma_3$ is a three-chain with a boundary homotopic to $M \Pi_2^{\rm tor}$ and, if $M_i \neq 0$, $\Sigma_4^i$ is a four-chain with a boundary homotopic to $M_i \Pi_3^{\rm tor}$. All these $p$-chains are calibrated by either $\Omega$ or $e^{iJ}$, with a calibration phase that will depend on their orientation. The D-branes listed above are $\oh$BPS in the background \eqref{SU3ex}, which means that they preserve two supercharges out of the four supercharges preserved by the solution. The two supercharges that they preserve will depend on their orientation. For instance, a D4-brane wrapping $\Sigma_4^i$ looks like a $\oh$BPS monopole in 4d, and preserves two supercharges of the domain-wall solution. Reversing the orientation and wrapping the D4-brane on $-\Sigma_4^i$ corresponds to a 4d $\oh$BPS monopole with opposite charge and preserving the other two supercharges of the background, an object that we will refer to as anti-BPS monopole. Here we will not keep track of which objects preserve which supercharges, because a much more straightforward picture will arise when we interpret this system in terms of Hitchin flow equations. 

A D4-brane wrapping a chain $\Sigma_4^i$ with a boundary is not consistent by itself, as it develops a worldvolume anomaly, but one can make it consistent by attaching D4-branes wrapped on $\p \Sigma_4^i$. In the present setup, if the D4-brane wrapping $\Sigma_4^i$ is located at $\xi_0 \in (0, \xi_{\rm end})$, one can cure its worldvolume anomaly by wrapping $M_i$ D4-branes on $\Pi_3^{\rm tor}$, and connecting them to $\p \Sigma_4^i$. These $M_i$ D4-branes will look like 4d strings that extend along the coordinate $\xi$, and either end on an anti-monopole in a different location, or stretch up until the origin $\xi=0$. From the 4d perspective, in the first case we have a monopole-anti-monopole pair connected by $M_i$ AB strings, as expected for a 4d EFT with a Lagrangian of the form \eqref{stuckdual} and a monopole of charge $M_i/M$. In the second case, we have a 4d avatar of a Hanany-Witten brane creating effect \cite{Hanany:1996ie}, mirror dual to a D3-brane crossing the NS5-brane (the domain wall), with $M_i$ D1-branes stretching along both after the crossing. As stressed in \cite{Berasaluce-Gonzalez:2012awn}, this effects also signal the presence of a discrete gauge symmetry, encoded either in the Lagrangian \eqref{stuckdual} or its dual. Similarly, the worldvolume anomaly of an Euclidean D2-brane in $\Sigma_3$ can be cured by $M$ D2-branes wrapped on $\Pi_2^{\rm tor}$ and connected to $\p \Sigma_3$. From the 4d viewpoint this is perceived like $M$ AB Euclidean particles ending on an instanton \cite{Banks:2010zn}. Notice that in this case a configuration made of $\oh$BPS objects involves $M$ Euclidean AB particles extended along the coordinate $\xi$, that stretch either between the domain-wall source and the instanton or between an instanton-anti-instanton pair.

In terms of these 4d objects one can compute the quantities $f$ and $g$ that feature the discussion of section \ref{s:dimred}, by using \eqref{fst} and \eqref{gpt}. Since in our example there are many axions and gauge bosons, in order to isolate a pair of them in the Lagrangian, as in \eqref{stuck}, we must consider the particular case $M_i \neq 0$, while $M_j = M_k =0$ for $i\neq j \neq k \neq i$. One then finds 
\be
f = \left( \tau^1\tau^2 \tau^3 \right)^{-1/2} , \qquad g = \frac{1}{2\pi} \sqrt{\frac{t^i} {t^j t^k }} .
\label{fandg}
\ee
Additionally, via the direct computation of section \ref{s:direct} (see eq.\eqref{eq: U0T3}) or the results of Appendix \ref{ap:spectra}, one obtains that the smallest non-vanishing eigenvalue of $\tilde{\bf T}^6$ is
\be
\lambda_{\rm st} =  \zeta_i V_i^{-3/2} ,
\label{lamstex}
\ee
and so it follows that the first equality in \eqref{Lphys} is satisfied with $N=M_i$, even if all quantities depend on the coordinate $\xi$. 

One can also see that the unit-norm exact three-form eigenmode corresponding to \eqref{lamstex} is
\be
b_3 = f^{-1}  \eta^4 \wedge \eta^5 \wedge \eta^6  = f^{-1} \delta_3^{\rm sm} (\Pi_3^{\rm tor}),
\ee
where in the second equality we have again used the results of section \ref{s:direct}, cf. eq.\eqref{deltasmT6}. We thus find perfect agreement with the discussion of section \ref{s:dimred}, in which the definition of $f$ via a smeared delta bump-form coincides with the value in \eqref{fst}. A similar check can be made for $g$, and the combined result is such that Conjecture \ref{conj:BPS} is verified. In the following we will discuss how to extend this result to general $M_i \in \mathbb{N}$, using the 4d EFT description. 

\subsection{EFT description}
\label{ss:EFTdesc}

To obtain the 4d effective description of this system one may follow the approach in \cite{Gurrieri:2002wz}, or its extension to more general setups discussed in \cite{Grana:2005ny,Kashani-Poor:2006ofe}. One first defines the following basis of three-forms 
\begin{subequations}
	\label{alphabetas}
	\begin{align}
		\a_0 = \eta^1 \wedge \eta^2 \wedge \eta^3 , & \qquad \beta^0 = \eta^4 \wedge \eta^5 \wedge \eta^6 , \\
		\a_1 = \eta^4 \wedge \eta^2 \wedge \eta^3  , & \qquad \beta^1 = - \eta^1 \wedge \eta^5 \wedge \eta^6 , \\
		\a_2 = \eta^1 \wedge \eta^5 \wedge \eta^3  , & \qquad \beta^2 =  - \eta^4 \wedge \eta^2 \wedge \eta^6 , \\
		\a_3 = \eta^1 \wedge \eta^2 \wedge \eta^6  , & \qquad \beta^3 = - \eta^4 \wedge \eta^5 \wedge \eta^3 ,   
	\end{align}
\end{subequations}
and a basis of two- and four-forms
\begin{subequations}
	\label{omegas}
	\begin{align}
		\omega_1 = \eta^1 \wedge \eta^4 , \qquad \omega_2 = \eta^2 \wedge \eta^5 , \qquad \omega_3 = \eta^3 \wedge \eta^6 , \\
		\tilde{\omega}^1 = - \omega_2 \wedge \omega_3, \qquad \tilde{\omega}^2 = - \omega_3 \wedge \omega_1, \qquad  \tilde{\omega}^3 = - \omega_1 \wedge \omega_2 .
	\end{align}
\end{subequations}
This set of forms are those that are invariant under the $\IZ_2 \times \IZ_2$ projection that takes us to a genuine SU(3)-structure. Notice that they satisfy $\int_{\tilde{\bf T}^6} \a_i \wedge \beta^j = \int_{\tilde{\bf T}^6} \om_i \wedge \tilde{\om}^j = \delta_i^j$, with a specific normalization which is crucial for the discussion that follows, since we are going to expand both the calibrations $(J, \Omega)$ and the 10d RR fields in these forms, and the latter are going to define the 4d axion periodicities the global $U(1)$ gauge transformations. While for harmonic $p$-forms one has a clear prescription to define an integral basis, the same is not true for exact and co-exact elements of this set, which can be identified thanks to the relations
\be
d\om_i = -M_i \b^0 , \qquad d\a_0 = -M_i \tilde{\om}^i .
\ee
In the present setup such a normalization can be fixed by means of mirror symmetry, just as in \cite{Gurrieri:2002wz}. However, in the general setting of \cite{Grana:2005ny,Kashani-Poor:2006ofe} it is simply assumed as an input. In section \ref{s:general} we will argue that one can fix it by defining the set $\{\a_A, \b^B, \om_a, \tilde{\om}^b\}$ as smeared delta-forms. 

Following \cite{Grana:2005ny,Kashani-Poor:2006ofe} we expand the NS-NS sector in the above basis
\bea
\label{JcandOm}
J_c  &= &B+ iJ = 4\pi^2 (b^j + i t^j) \, \om_j , \\ \nonumber
\Om  &=  & {i} (2\pi)^3 \sqrt{\frac{t^1t^2t^3}{\tau^1\tau^2\tau^3}} \left( \a_0 + z^i \a_i - z^2 z^3 \b^1 - z^1 z^3 \b^2 - z^1 z^2 \b^3  + z^1z^2z^3 \b^0 \right), \quad  z^j = a^j + i \tau^j .
\eea
Similarly, one expands the RR potential $C_3$ as
\be
C_3 = 2\pi \ell_s^{3} \left[ A_1^i \wedge \om_i +  \theta^I \alpha_I + \tilde{\theta}_K \beta^K \right] ,
\ee
where $(\theta^I, \tilde{\theta}_I)$ with $I= (0,i)$ represent axions of unit periodicity. The dimensional reduction of this term gives 
\be
(2\pi)^2 \left[g_{ii} (dA_1^i)^2 + \hat{f}^{00} \left(d\tilde{\theta}_0 - M_i A_1^i\right)^2  + \hat{f}^{ii} (d\tilde{\theta}_i)^2 + \hat{f}_{II} (d{\theta}^I)^2  \right] ,
\label{StuckN=2gen}
\ee
plus a mass term for $\theta^0$. Here we have defined
\be
g_{ii} = (2\pi)^2\, \frac{t^j t^k}{t^i} , \qquad \hat{f}^{00} = e^{2\phi_4} (\tau^1\tau^2\tau^3)^{-1} M_{\rm P}^2 , \qquad \hat{f}^{ii} = e^{2\phi_4} \frac{\tau^i}{\tau^j\tau^k}  M_{\rm P}^2 ,
\ee
with $i \neq j \neq k \neq i$, and $\hat{f}_{II} = (\hat{f}^{II})^{-1} e^{4\phi_4}M_{\rm P}^4$, where $e^{\phi_4} = e^{\phi}/\sqrt{4\pi\,t^1t^2t^3}$ is the 4d dilaton. Notice that all these couplings depend on the domain-wall transverse coordinate $\xi$, while the axion vevs remain constant along it. The NS-NS sector of the compactification varies along $\xi$ via the non-trivial profile of the saxions $t^i, \tau^i$ along this coordinate, as captured by \eqref{defttau}, and in agreement with the results of \cite{Behrndt:2001mx,Mayer:2004sd}.

The lightest massive $p$-form mode has the following squared mass
\be
m_{\rm st}^2 = V^{-1} \left[\sum_i \frac{\zeta_i^2}{V_i^2} \right] e^{2\phi_4} M_{\rm P}^2 ,
\label{lamstexgen}
\ee
and so it is a priori not obvious how to compute the smeared linking number using \eqref{Lphys}. To do so, one must take into account that for generic $M_i$'s the torsion two-cycle $\Pi_2^{\rm tor} = \sum_i (M_i/M) \Pi_2^i$ is not a smooth calibrated cycle, but instead a linear combination of them. In this case, it is the extension of the conjecture made around \eqref{deldif} that should be applied. One obtains 
\be
\delta^{\rm sm}_4 (\Pi_2^{\rm tor}) =  \sum_i \frac{M_i}{M} \tilde{\om}^i ,
\ee
whose projection into the four-form eigenmode with the smallest non-vanishing eigenvalue gives
\be
g =   \frac{1}{2\pi}  \sqrt{ \sum_i \frac{M_i^{2}}{M^2} \frac{t^i} {t^j t^k }} .
\label{ggenM}
\ee
Defining $f = \sqrt{\hat{f}^{00}}$ one reproduces \eqref{Lphys} with $N=M = {\rm g.c.d.} (M_1, M_2, M_3)$, as expected. From a purely 4d EFT viewpoint, one can interpret \eqref{ggenM} as the gauge coupling of the linear combination of $U(1)$'s that develops a St\"uckelberg mass.

\subsection{Hitchin flow equations}
\label{ss:hitchin}

As already pointed out in \cite{Gurrieri:2002wz}, the background \eqref{SU3ex} can be understood geometrically as a fibration of a half-flat manifold $X_6$ over a real coordinate, that gives a seven-dimensional $G_2$-manifold $Y_7$. The general description of this kind of fibrations has been given in \cite{hitchin2001stable,chiossi2002intrinsic}, and are known as Hitchin flow equations. In the standard description, the real coordinate $z$ has a flat metric, and one constructs the $G_2$-structure forms
\bea
\varphi & = & dz \wedge J  - {\re \Om} , \\
\star \varphi & = & -dz \wedge {\im \Om} - \oh J \wedge J ,
\eea
where $J$ and $\Om$ are the $z$-dependent SU(3)-structure calibrations of $X_6$. Demanding that  $Y_7$ has $G_2$ holonomy amounts to impose that $\varphi$ is harmonic in $Y_7$. If we describe the 7d derivative as 
\be
d_7 = \p_z  dz \wedge  + \, d ,
\ee 
with $d$ the exterior derivative along the 6d fibre, this requirement reads
\bea
\label{dreom}
d {\im \Om} & = & \oh\p_z \left(J \wedge J\right) , \\
d J & = & - \p_z {\re \Om} .
\label{dj}
\eea
In our background the coordinate $\xi$ has a non-trivial metric, more precisely $dz = - V^{1/2}d\xi$, where the sign choice accounts for the difference in our background compared to \cite{Gurrieri:2002wz} (see footnote \ref{ft:dw}). The Hitchin flow equations then take the following form :
\bea
\label{dreomxi}
V^{1/2} d {\im \Om} & = & - \oh \p_\xi \left(J \wedge J\right) , \\
V^{1/2} d J & = &  \p_\xi {\re \Om} . 
\label{djxi}
\eea
Applied to the background \eqref{SU3ex} these equations reduce to
\be
\p_\xi V_i  =  - \zeta_i ,
\ee
which is clearly satisfied by \eqref{Vis}.

The Hitchin flow equations have a nice interpretation when it comes to D-branes on torsion cycles, that can be illustrated explicitly in the solution \eqref{SU3ex}. Let us consider $M$ D4-branes wrapped on $\Pi_3^{\rm tor}$ and extended along an interval $(0, \xi_0) \subset (0,\xi_{\rm end})$. At $\xi_0$  one places a D4-brane wrapping a four-chain $\Sigma_4$, such that its boundary coincides with the $M$ torsion three-cycles. From the 4d viewpoint, this represents a 4d monopole in which $M$ AB strings end, with their other end at the domain-wall source. Since both sets of D4-branes yielding the monopole and the AB strings are calibrated by $\star \varphi$, they must be mutually BPS, and satisfy the marginal stability condition \eqref{BPSmono}. Additionally, the total energy of the system must be given by its central charge, which is the integral of $\star \varphi$ over the full D4-brane worldvolume in the $G_2$ manifold $Y_7$, and it is easy to argue that this central charge must be independent of the monopole position $\xi_0$. 

Indeed, notice that shifting the value of $\xi_0$ corresponds to add $M$ AB strings extended along the interval $(\xi_0, \xi_0') \subset (0,\xi_{\rm end})$, with a D4-branes wrapping $\Sigma_4$ at each end, with opposite orientations. From the 4d viewpoint, this realizes a monopole-anti-monopole pair in which AB strings end. Since this object can annihilate by itself, one expects that its central charge vanishes. Microscopically, the whole object corresponds to a trivial four-cycle in the $G_2$ manifold $Y_7$, and so since $\star \varphi$ is closed its integral must vanish on it. So indeed the monopole-anti-monopole pair carries no central charge and changing the value of $\xi_0$ in the above BPS configuration should not change the energy of the system. In particular this energy should match that of a monopole placed at $\xi = \xi_{\rm end}$ which is equivalent to having $M$ AB strings, and to a monopole at $\xi =0$, which does not have any AB strings attached to it. 

This is indeed what the Hitchin flow equations are telling us, and in particular \eqref{dreom}. On the one hand, $\p_z J \wedge J$ represents the variation of the mass of BPS monopoles  when we move along $z$. On the other hand, ${\im \Omega}$ integrated along the torsion three-cycle measures the tension of a BPS 4d AB string, and by Stokes' theorem, this is equivalent to integrating ${d \im \Omega/M}$ over the four-chain $\Sigma_4$ linking $M$ of them. So what \eqref{dreom} is saying is that it is the monopoles in which $M$ AB strings can end the ones whose mass varies along the coordinate transverse to the domain wall. Moreover, there is a mass scale associated to the 4d string, which is its tension integrated along the interval $(0, \xi_0)$. For BPS objects, this energy increases with $\xi$ at the same rate as the monopole mass decreases, and that is why the total central charge and therefore the energy of the system stays constant. In our example \eqref{SU3ex} one can see that the factors of $V$ cancel for a 4d AB string, so the energy of $M_i$ BPS AB strings is given by $\ell_s^{-1} {\rm Vol}(\Pi_3^{\rm tor}) M_i \xi_0 \propto \xi_0 $. Additionally, the mass of the monopole in which such strings can end is given by $\ell_s^{-1} {\rm Vol}(\Sigma_4^i) \propto V_i|_{\xi_0}$. Therefore, it decreases linearly with $\xi_0$, precisely compensating the change in the energy of the AB strings.\footnote{One can engineer the BPS configuration of $M$ AB strings ending on a monopole by a Hanany-Witten brane-creation effect, as one can check using the mirror type IIB picture, see Appendix \ref{ap:NS5DW}. The interpretation is then that a Hanany-Witten effect does not change the energy of a BPS object. The mass of a monopole located at $\xi_0 \in (-\infty, 0)$ and at $\xi_0 \in (0,\xi_{\rm end})$ is the same, if in the second case we include the energy of the extended AB strings. That is, if at both sides we compute the energy or central charge of the gauge invariant operator.}

This example illustrates how \eqref{BPSmono} can be satisfied, and the expectation of subsection \ref{ss:BPS}, that one should be able to add up central charges of BPS objects in $\IZ$, even when their topological charge is $\IZ_N$. In the case at hand, $M$ D4-branes wrapping $\Pi_3^{\rm tor}$ can disappear by ending on a monopole, but a monopole nucleation process costs  energy, which is minimized for the case of BPS monopoles. The discussion above implies that this energy is at least that of $M$ BPS strings extended along the interval $(\xi_0, \xi_{\rm end})$. Therefore nucleating a monopole at $\xi_0$ is topologically possible, but not energetically favoured. In this sense, adding up an arbitrary number of  AB strings is well-defined in the BPS context, as well as considering a cone of 4d AB string charges.


\section{Direct computation}
\label{s:direct}

While the general arguments of section \ref{s:dimred} motivate Conjecture \ref{conj:BPS} for torsion cycles in SU(3)-structure manifolds, it is instructive to work out in detail how the conjecture is realized in explicit examples, by a direct comparison of the torsion linking number and its smeared version. In this section we perform such a comparison for the SU(3)-structure manifold of section \ref{s:simple}, more precisely for the twisted six-torus with a single metric flux. As we will see, the direct computation of the torsion linking number displays a series of cancellations between terms that is reminiscent of those that occur in the computation of topological indices, and that leaves the smeared torsion linking number \eqref{Lsm} as the only non-vanishing contribution. The reader not interested in these technical details may safely skip to the next section.

\subsubsection*{The setup}

Let us consider the twisted six-torus background in \eqref{SU3exb}, rewritten as
\be
ds^2_{\tilde{\bf T}^6} = (2\pi)^2 \sum_i \left(\frac{t^i}{\tau^i}(\eta^i)^2+t^i\tau^i(\eta^{i+3})^2 \right) ,
\label{SU3exc}
\ee
and with the definitions \eqref{defttau} and \eqref{etas}. In particular we consider $M_i \neq 0$ and $M_j = M_k =0$ with $i \neq j\neq k \neq i$. In this case, the metric background factorizes as $\tilde{\bf T}^6 = \tilde{\bf T}^3 \times {\bf T}^3$, and all the torsion cycles correspond to a direct product of a torsion one-cycle in $\tilde{\bf T}^3 \simeq \langle x^i, x^{j+3}, x^{k+3} \rangle$ and a non-trivial cycle in ${\bf T}^3 \simeq \langle x^j, x^k, x^{i+3}\rangle$. As a result, all torsion linking numbers of $\tilde{\bf T}^6$ stem from the torsion linking numbers between one-cycles in $\tilde{\bf T}^3$. Moreover, the calibration condition in $\tilde{\bf T}^6/\IZ_2 \times \IZ_2$ will translate into a subset of such torsion one-cycles. Therefore our strategy will be to verify Conjecture \ref{conj:BPS} for such a subset, then extend the result into calibrated two and three-cycles of $\tilde{\bf T}^6$, and finally check that the $\IZ_2 \times \IZ_2$ projection does not modify the statement. A necessary first step is to describe the set of massive $p$-form modes in $\tilde{\bf T}^3$, which one can accomplish using a general method for three-manifolds with isometries. 

\subsubsection*{Massive spectra of three-manifolds}

To describe the massive $p$-form spectrum of a twisted three-torus, one may use the method of  \cite{BenAchour:2015aah}, which applies to compact Riemannian three-dimensional manifolds $X_3$ with a continuous isometry.  Such a manifold admits a unit-norm Killing vector $\chi$, and we assume that its dual one-form satisfies 
\begin{equation}
	\star d \chi = \lambda_\chi \,\chi \,, \qquad \Delta_3 \chi = \lambda_\chi^2\, \chi \,, \qquad \chi^2 = 1  \,, \qquad  \lambda_\chi, \in \IR \, ,
\end{equation}
and that its integral curves are closed. Here $\star$ and $\Delta_3$ stand for the Hodge star operator and the Laplacian on $X_3$, respectively. Then, let $\{\phi_\a\}$ be an orthonormal basis of complex scalar eigenforms of the Laplacian such that\footnote{Notice that such 
	a basis always exists because $[\Delta_3, \mathcal{L}_\chi ] = 0$. }
\begin{equation} \label{eqbasisscalars}
	\Delta_3 \phi_\a = \sigma_\a^2 \phi_\a \,,\qquad \mathcal{L}_\chi \phi_\a = i \mu_\a \phi_\a \,, \qquad   \sigma_\a \in \mathbb{R} \,, \quad  \mu_\a \in \mathbb{R}\,.
\end{equation} 
Solving the second condition of (\ref{eqbasisscalars}) we can obtain the explicit dependence of  the $\{\phi_\a\}$ on the isometry coordinate $\th$ associated to $\chi$
\begin{equation}\label{eqphiexp}
	\phi_\a = e^{i \mu_\a \theta} K_\a \,, \qquad \th \sim \th + 2\pi r \, ,
\end{equation}
with $d\theta =\chi$ and $K_\a$ functions which do not depend on $\th$. Given that $\th$ parameterizes a closed integral curve of radius $r$ in a compact manifold, we obtain the quantization condition $\mu_\a r \in  \mathbb{Z} $\,. 

In this setup, it is possible to give a simple description of non-harmonic eigen-one-forms of the Laplacian, in terms of the Killing vector $\chi$ and the scalar eigenforms $\phi_\a$. We define 
\begin{equation}
	R_\a = d \phi_\a \,, \qquad S_\a = \star d (\phi_\a \chi ) \,, \qquad T_\a = \star d S_\a \,. \label{eq: BandC}
\end{equation}
It is easy to see that the set $R_\a$ forms a complete basis of exact eigen-one-forms. The set of co-exact one-forms $S_\a$ and $T_\a$ is closed under the action of the operator $ \star d$
\begin{equation}
	\star d S_\a = T_\a \,, \qquad \star d T_\a = \sigma_\a^2 S_\a +  \lambda_\chi T_\a\, ,
\end{equation}
from where one can find the following eigenforms of $\star d$
\begin{equation}
	U^\pm_\a = \left(\frac{1}{2}\pm\frac{ \lambda_\chi}{2 \sqrt{ \lambda_\chi^2 + 4 \sigma_\a^2}}\right) T_\a \pm \frac{\sigma_\a^2}{\sqrt{ \lambda_\chi^2 + 4 \sigma_\a^2}}S_\a \,.  \label{eq: Dpm}
\end{equation}
Therefore, since the action of the Laplacian $\Delta_3$ on co-closed forms amounts to $\star \, d \star d$, we obtain that the $U^\pm$ are eigenforms of the Laplace operator with eigenvalues
\begin{equation}
	(\lambda^\pm_\a)^2 = \sigma_\a^2 + \frac{ \lambda_\chi^2}{2} \pm \frac{ \lambda_\chi}{2}\sqrt{ \lambda_\chi^2 + 4 \sigma_\a^2 } \,.\label{eq: lambdapm}
\end{equation} 
Let us dub the constant eigenmode of the Laplacian as $\phi_0 = 1/\sqrt{V_3}$, with $V_3$ the volume of $X_3$. Then the eigenmode $U_0^-$ identically vanishes, while $U_0 \equiv U_0^+$ has eigenvalue $ \lambda_\chi^2$  with respect to $\Delta_3$ and takes form $U_0 =  \lambda_\chi^2 \, \chi \phi_0$. Moreover, the set of co-exact one-forms $U^\pm_\a $ are normalized to unity by multiplying them by the following factor 
\begin{equation}
	c_\a^{\pm} = \left[\frac{( \lambda_\chi^4+3 \lambda_\chi^2 \sigma_\a^2+\sigma_\a^4) - ( \lambda_\chi^2+\sigma_\a^2)\,\mu_\a^2}{2} \pm   \lambda_\chi \, \frac{( \lambda_\chi^4 + 5 \lambda_\chi^2 \sigma_\a^2 + 5\sigma_\a^4)-( \lambda_\chi^2+3\sigma_\a^2)\,\mu_\a^2}{2 \sqrt{ \lambda_\chi^2 + 4 \sigma_\a^2}}\right]^{-1/2}\,.
\end{equation}
In the following we will assume that the $U_\a^{\pm}$ have been normalized to unit norm. In particular, we have that $c_0 \equiv c_0^+ =  \lambda_\chi^{-2}$, and so $U_0 = \chi \phi_0$. 

The set $\{U_\a^\pm\}$ is part of the co-exact one-form eigenspectrum of $X_3$, but the above method does not guarantee that it is a complete set. In the particular case of $\tilde{\bf T}^3$ one can check that the whole co-exact spectrum is of this form, as verified in Appendix \ref{ap:spectra} by using the results of  \cite{Andriot:2018tmb}.

\subsubsection*{Computing the linking number}

Using that $\{U_0, U_\a^\pm\}$ is a complete basis of co-exact eigen-one-forms of $X_3$,\footnote{There may be more than one eigenform for a given eigenvalue, but this will not change our final result.} we can expand the bump delta two-form for a torsion one-cycle $\pi_1 \subset X_3$ as
\begin{equation}
	\delta^{(2)}(\pi_1) = K_0 \star U_0 + \sum_\a \left( K_\a^+  \star U^+_\a + K_\a^-  \star U^-_\a \right)\, , \label{delta2sm}
\end{equation} 
where
\be
K_0 = \int_{\pi_1} U_0, \qquad K_\a^\pm = \int_{\pi_1} U_\a^\pm \, .
\ee
In terms of these expressions, the linking number \eqref{L} between two torsion one-cycles reads
\begin{equation}
	L(\pi, \tilde{\pi}) = \frac{1}{ \lambda_\chi} K_0 \tilde{K}_0 + \sum_\a \left[ \frac{1}{\lambda_\a^+} K^+_\a \tilde{K}^+_\a + \frac{1}{\lambda_\a^-} K^-_\a \tilde{K}^-_\a \right]\, ,
	\label{LtT3}
\end{equation}
where the coefficients $\tilde{K}$ arise from integration over a different torsion one-cycle  $\tilde{\pi}_1$. 

We now impose the calibration condition. One can check that calibrated torsion two- and three-cycles in $\tilde{\bf T}^3 \times {\bf T}^3/\IZ_2 \times \IZ_2$ correspond to torsion one-cycles on $\tilde{\bf T}^3$ that are integral curves of $\chi$. For such one-cycles we have that $\int_{\pi_1} \alpha = \int_0^{2\pi r} \iota_\chi \alpha\, d\th$, for any one-form $\a$. Therefore 
\be
K_0 =  \int_0^{2\pi r} 	\iota_\chi  U_0 \, d\th = 2\pi r \phi_0 \, , \qquad K_\a^\pm =   \int_0^{2\pi r} 	\iota_\chi  U_\a^\pm \, d\th  \, .
\ee

From these expressions, one may compute each of the terms in the torsion linking number. Indeed, one first notices that
\begin{equation}
	\iota_\chi S_\a =   \lambda_\chi \phi_\a \,, \qquad \iota_\chi T_\a = ( \lambda_\chi^2 + \sigma_\a^2 -  \mu_\a^2) \, \phi_\a \,,
\end{equation}
which imply
\begin{equation}
	\iota_\chi U^\pm_\a = \pm \left\{\lambda^\pm_\a\bigg[ \lambda_\chi^2+\sigma_\a^2-\mu_\a^2\bigg]+\sigma_\a^2  \lambda_\chi\right\}\frac{c_\a^{\pm} }{\sqrt{ \lambda_\chi^2 + 4 \sigma_\a^2}} \phi_\a\,.
\end{equation}
As a result, the  massive eigenmodes with $\mu_\a \neq 0$ have a vanishing coefficient, since
\begin{equation}
	K^\pm_\a \propto \int_0^{2 \pi r}  \phi_\a \, d \th = 0  \,,
\end{equation}
where we have used \eqref{eqphiexp}. It remains to check the contribution of the modes with $\mu_\a = 0$ to \eqref{LtT3}. Recall that those modes with $\alpha\neq 0$ come in pairs, and one can check that they satisfy the following relation:
\begin{equation}
	\frac{1}{\lambda_\a^+} K^+_\a \tilde{K}^+_\a + \frac{1}{\lambda_\a^-} K^-_\a \tilde{K}^-_\a = \frac{\epsilon_\a}{ \lambda_\chi^2 + 4\sigma_\a^2} \int_0^{2 \pi r}  \phi_\a \, d\th \int_0^{2 \pi r}  \phi_\a \, d\th \,,
\end{equation}
where we have defined
\begin{equation}\label{eqdelta}
	\epsilon_\a \equiv \frac{(c_\a^{+})^2}{\lambda_\a^+} \left[\lambda_\a^+( \lambda_\chi^2+\sigma_\a^2) + \sigma_\a^2  \lambda_\chi\right]^2 + \frac{(c_\a^{-})^2}{\lambda^-_\a} \left[\lambda_\a^-( \lambda_\chi^2+\sigma_\a^2) + \sigma_\a^2  \lambda_\chi\right]^2 = 0 \,.
\end{equation}
That is, those massive eigenmodes with $\mu_\a =0$ have non-trivial coefficients $K_\a^\pm$, but for those contributing to the  bracket in \eqref{LtT3} there is a non-trivial cancellation by pairs, such that the sum cancels term by term. The surviving term in \eqref{LtT3} is the smeared linking number
\begin{equation}
	L^{\rm sm}(\pi_1,\tilde{\pi}_1) \equiv \frac{1}{ \lambda_\chi} K_0 \tilde{K}_0  =  \frac{4\pi^2r^2}{V_3  \lambda_\chi}\,.
	\label{Lsmexc}
\end{equation}
In a twisted three-torus with metric $ds^2_{\tilde{\bf T}^3} = (2\pi)^2 \left[(R_i\eta^i)^2 + (R_{j+3}\eta^{j+3})^2 + (R_{k+3}\eta^{k+3})^2 \right]$ and twist $d\eta^i = - N \eta^{j+3} \wedge \eta^{k+3}$ one obtains 
\be
U_0 =\frac{2\pi R_i}{\sqrt{V_3}}\,\eta^i ,\qquad\lambda_\chi = \frac{NR_i}{2\pi R_{j+3}R_{k+3}}\, , \qquad r^2 = R_i^2\, , \qquad V_3 = 8\pi^3 R_iR_{j+3}R_{k+3} \, .\label{eq: U0T3}
\ee
Therefore applying \eqref{Lsmexc} one recovers the result $L^{\rm sm}(\pi_1,\tilde{\pi}_1) = 1/N$, as expected.

\subsubsection*{Extension to $\tilde{\bf T}^6/\IZ_2 \times \IZ_2$}

Let us now see how the above computation extends to the SU(3)-structure manifold $\tilde{\bf T}^6/\IZ_2 \times \IZ_2$. We first consider the covering space  $\tilde{\bf T}^6 = \tilde{\bf T}^3 \times {\bf T}^3$ with metric \eqref{SU3exc}, where $\tilde{\bf T}^3$ is parametrized by the coordinates $\{x^i, x^{j+3}, x^{k+3}\}$ and $ {\bf T}^3$ by $\{x^{i+3}, x^j, x^k\}$,  with $i \neq j\neq k \neq i$. Given the factorization of the metric,  any eigenform of the Laplacian will be a wedge product of one in $\tilde{\bf T}^3$ and one in $ {\bf T}^3$. We are in particular interested in those eigenforms in which the bump delta-forms $\delta(\Pi_3^{\rm tor})$ and $\delta(\Pi_2^{\rm tor})$ are decomposed. It is easy to see that these fall in the subset
\bea
\left[ \star U_\a^\pm\right] & \wedge & \left(e^{2 \pi i n_{i+3} \, x^{i+3}}dx^{i+3}\right), \qquad n_{i+3} \in \IZ\, , \\
\left[ \star U_\a^\pm\right]  & \wedge & \left(e^{2 \pi (n_j  x^j+ n_k x^k)} dx^j \wedge dx^k \right), \qquad n_j, n_k \in \IZ\, ,
\eea
for  $\delta(\Pi_3^{\rm tor})$ and $\delta(\Pi_2^{\rm tor})$, respectively, where as above $\star$ stands for the Hodge star operator in $\tilde{\bf T}^3$. As a consequence,  the expansion of the smeared deltas $\delta^{\rm{sm}}(\Pi_3^{\rm tor})$ and $\delta^{\rm{sm}}(\Pi_2^{\rm tor})$ are given by $\star U_0\wedge dx^{i+3} $ and $\star U_0\wedge dx^{j}\wedge dx^k$, accordingly. That is, using the metric \eqref{SU3exc} one obtains
\be
\delta^{\rm{sm}}(\Pi_3^{\rm tor}) = \eta^{4} \wedge \eta^{5} \wedge \eta^{6}  
\qquad
\delta^{\rm{sm}}(\Pi_2^{\rm tor}) = -\eta^{j}\wedge \eta^{k}\wedge \eta^{j+3}\wedge \eta^{k+3}. \label{deltasmT6}
\ee
With regard to the complete expansion, it is easy to see that the wedge of one of these forms and its antiderivative will give a non-vanishing contribution only if $n_j = n_k = n_{i+3} =0$, that is if we select harmonic forms in ${\bf T}^3$. As a result, the computation of the linking number for calibrated cycles works precisely as outlined for $\tilde{\bf T}^3$, with the same vanishing coefficients and the same cancellations, and we end up again with the smeared torsion linking number \eqref{Lsmexc}. 

Let us now implement the $\IZ_2 \times \IZ_2$ orbifold projection, where each $\IZ_2$ generator $\th_1$ and $\th_2$ acts by flipping two coordinates on $\tilde{\bf T}^3$ and other two on $\tilde{\bf T}^3$, as follows from footnote \ref{ft:orbifold}. Since this is a product of two involutions, each acting on one submanifold, we can split the above exact eigenforms into even and odd under such involutions, and take (odd, odd) or (even, even) products, such that the result is invariant under the orbifold generators. While one could perform such an analysis explicitly, given our discussion above it is sufficient to show the action of these orbifold generators on the two-forms $\star U_\a^\pm$ only depends on the value of $\sigma_\a$ and $\mu_\a$, since then the orbifold projection will commute with relations that lead to the cancellations \eqref{eqdelta}, and they will also happen for orbifold-invariant massive modes. One can  show the assumption by using that $\th_1$ and $\th_2$ act as isometries when restricted to $\tilde{\bf T}^3$, as then they commute with $\Delta$, and ${\cal L}_\chi$. It then follows that they have a well-defined action on the basis of scalar wavefunctions $\{\phi_\a\}$, and act on the above set of co-exact one forms as
\be
\th_\a : S_\a \mapsto \nu_\a^{\sigma_\a, \mu_\a} S_\a   \, , \quad \th_\a : T_\a \mapsto \nu_\a^{\sigma_\a, \mu_\a} T_\a\  \implies \ \th_\a : U_\a^\pm  \mapsto \nu_\a^{\sigma_\a, \mu_\a}  U_\a^\pm\, .
\ee
That is, the orbifold group action on the massive modes of interest only depends on the value of $\sigma_\a$ and $\mu_\a$, as assumed. Finally, by construction, the orbifold projection leaves invariant the eigenmodes of $\tilde{\bf T}^3 \times {\bf T}^3$ that contribute to the smeared linking number.


\section{More general $\CN=2$ compactifications}
\label{s:general}

The extension of the setup in \cite{Gurrieri:2002wz} to more general type II string compactifications leading to 4d $\CN=2$ gauged supergravities has been performed in \cite{DAuria:2004kwe,Grana:2005ny,Kashani-Poor:2006ofe,Grana:2006hr}. In the following we focus on the framework developed in \cite{Grana:2005ny,Kashani-Poor:2006ofe}, which applies to SU(3)-structure manifolds. Such a framework relies on the existence of a set of smooth $p$-forms on an SU(3)-structure manifold  $X_6$:
\be
\{\om_a\} \in \Om^2(X_6)\, , \qquad \{\a_A, \b^B\} \in \Om^3(X_6)\, , \qquad \{\tilde{\om}^a\} \in \Om^4(X_6)\, , 
\ee
with $a=1, \dots, n_K$, $A,B = 1, \dots, n_{\rm c.s.}$, chosen such that 
\be
\int_{X_6} \om_a \wedge \tilde{\om}^b = \delta_a^b , \qquad \int_{X_6}  \a_A \wedge \b^B = \delta_A^B ,
\label{intrel}
\ee
and satisfying the relations 
\bes
\label{system}
\begin{align}
	\label{coma}
	d^\dag \om_a &= 0 , \\
	d\om_a &=  m_a{}^A \a_A + e_{aA} \b^A , \\
	d\a_A &= e_{aA} \tilde{\om}^a , \\
	d\b^B &=  -m_a{}^B \tilde{\om}^a , \\
	d\tilde{\om}^a & =  0 ,
	\label{doma}
\end{align}
\ees
with $m_a{}^A, e_{aA} \in \IZ$ such that $m_a{}^A e_{bA} = m_b{}^A e_{aA}$. Consistency of the dimensional reduction implies 
that the set is closed under the Hodge star operator:
\be
\tilde{\om}^a =  g^{ab} \star  \om_b , \qquad \star  \a_A = H_A^B \a_B + G_{AB}\b^B, \qquad \star  \b^A = F^{AB} \a_B - H^A_B \b^B ,
\label{hodgen}
\ee
mimicking the relations between harmonic forms in Calabi--Yau manifolds. 

Given this set of $p$-forms, one expands the SU(3)-structure calibrations in terms of them: 
\bea
\label{JcandOmgen}
J_c  &= &B+ iJ = 4\pi^2 (b^a + i t^a) \, \om_a , \\ \nonumber
\Om  &=  & Z^A\a_A - \cF_B\b^B ,
\eea
with $\cF_A = \p_A \cF$ the derivatives of the complex structure prepotential $\cF$. The 4d kinetic terms of the corresponding fields are governed by the same expressions as in the Calabi--Yau case, in terms of K\"ahler potentials  $K_\rho = - \log \int_{X_6} i \bar{\Om} \wedge \Om$ and $K_J = - \log \frac{4}{3}\int_{X_6} - J \wedge J \wedge J$ that correspond to Hitchin functionals  \cite{Grana:2005ny}. Finally, one should also expand the 10d RR potentials in this set of $p$-forms. In the case of type IIA compactifications such an expansion reads
\be
C_3 = 2\pi \ell_s^3  \left(A_1^a  \wedge \om_a + \tilde{C}_0^A \a_A + C_{0\, B} \b^B\right) ,
\label{C3gen}
\ee
leading to a set of axions and gauge vectors in 4d. The dual degrees of freedom are obtained from the expansion of $C_5$.

In the framework of \cite{Grana:2005ny,Kashani-Poor:2006ofe} there is no geometric interpretation for the set $\{\om_a, \a_A, \b^B, \tilde{\om}^b\}$, nor a clear prescription on how to build them from the light eigenmodes of the Laplacian. Notice that a key property of these $p$-forms is that they define the quantization features of the 4d EFT, either in terms of axion periodicities or $U(1)$ gauge transformations. As such, their definition should be connected to the presence of 4d EFT objects like strings, particles and instantons, which implement and detect global gauge transformations. We have already seen this connection in our discussion of section \ref{s:dimred}, in light of which one may propose to describe the set $\{\om_a, \a_A, \b^B, \tilde{\om}^b\}$ as smeared delta forms. 

Indeed, based on our previous discussion, it is natural to propose that the smooth $p$-forms $\{\om_a, \a_A, \b^B, \tilde{\om}^b\}$ correspond to smeared delta-forms $\delta^{\rm sm}_p(\Sigma_{6-p})$ of a set of strictly calibrated $(6-p)$-chains $\Sigma_{6-p} \subset X_6$, which encode the presence of BPS objects in the 4d EFT. More precisely, the closed four-forms $\tilde{\om}^b$ correspond to the smeared bump delta-forms $\delta^{\rm sm}_4(\Pi_2)$, where $[\Pi_2]$ belongs to the free part of $H_2(X_6, \IZ)$ if $\tilde{\om}^b$ is harmonic, and to ${\rm Tor} H_2(X_6, \IZ)$ if it is de Rham exact. Similarly, the subset of three-forms in $\{ \a_A, \b^B\}$ that are closed correspond to the smeared delta-forms $\delta^{\rm sm}_3(\Pi_3)$ of strictly calibrated three-cycles. The remaining set of smooth $p$-forms can be constructed by taking the anti-derivatives of the exact three- and four-forms and normalising them such that \eqref{intrel} is satisfied, which implies that the integers $m_a{}^A, e_{aA}$ encode the torsion linking numbers of $X_6$. Finally, as in our simple example above, one could also relate the non-closed two- and three-forms as the smeared version of delta-forms for calibrated four- and three-chains in $X_6$, whose boundary describes the torsional nature of some calibrated cycles. 

To make this picture more precise, let us consider the subcase $m_a{}^A =0$, which also resembles the setup considered in \cite{Camara:2011jg}. Then, the rank $r_{\bf e}$ of the matrix $e_{aA}$ should determine the number of harmonic two- and three forms of $X_6$ as $b_2(X_6) = n_K- r_{\bf e}$ and $b_3(X_6) = n_{\rm c.s.} - r_{\bf e}$. Clearly, the rank of $e_{aA}$ counts massive  eigenforms below the compactification scale, more precisely we should at least have $r_{\bf e}$ times a spectrum of the form \eqref{massivepforms}, as this is what we obtain from dimensionally reducing the RR sector of the theory. Indeed, let us consider the type IIA expansion \eqref{C3gen}, and for simplicity assume that in \eqref{hodgen} $H^A_B =0$, so that $G_{AB}F^{BC} = - \delta_A^C$. Then we find
\be
(2\pi)^2 \hat{F}^{AB} \left( dC_{0\,  A} - e_{aA} A_1^a\right) \left( dC_{0\, B} -e_{bB} A_1^b\right) + (2\pi)^2 g_{ab}\, dA_1^a \wedge dA_1^b\, , 
\label{disgaugen}
\ee
with $ \hat{F}^{AB} =   F^{AB} e^{2\phi_4} M_{\rm P}^2$, plus a mass term for $r_{\bf e}$ axions $\tilde{C}_0^A$.\footnote{These massive axions are more suitably described in terms of a 4d dual two-form $B_2$ involved in a gauging of the form \eqref{p=2}, see e.g. \cite{Marchesano:2014mla}.} The masses that one reads from such a mass term, the Lagrangian \eqref{disgaugen} and its dual reproduce the action of the Laplace operator on the set $\{\om_a, \a_A, \b^B, \tilde{\om}^b\}$ as expected \cite{Tomasiello:2005bp,Grana:2005ny,Kashani-Poor:2006ofe}. For instance, the action of the Laplacian on the closed forms $\b^A$ and $\tilde{\om}^a$ reads
\begin{eqnarray}
	\Delta \beta^A &= &F^{AB}e_{bB}g^{bc}e_{cC}\, \beta^C \, , \\
	\Delta \tilde{\om}^a &= & g^{ab} e_{bB}F^{BC}e_{cC} \, \tilde{\om}^c \, .
\end{eqnarray}
The diagonalization of these mass matrices gives us the set of massless and light $p$-form eigenmodes. Such a spectrum is by assumption complete, or otherwise the expansions \eqref{JcandOmgen} and \eqref{C3gen} would be missing light modes of the EFT.     Knowledge of these mass matrices and of the kinetic terms $C^{AB}$ and $g_{ab}$ leads to $e_{aA}$, in a generalization of the relation \eqref{Lphys}. As proposed in \cite{Camara:2011jg}, the matrix $e_{aA}$ is a sort of the inverse of the torsion linking numbers, and it encodes the torsion cohomology that is sensitive to the light EFT modes. This topological information is easier to extract if one performs a unimodular integral change of basis both in $\{\om_a, \tilde{\om}^b\}$ and in $\{\a_A, \b^B\}$ that take $e_{aA}$ to its Smith normal form
\be
e^{\rm Smith} =
\begin{pmatrix}
	k_1 \\ & k_2 \\ & & \ddots \\& & & k_{r_{\bf e}} & \dots & 0\\ & & & 0 &\dots & 0 \\  & & & \vdots &\ddots & \vdots \\ & & & 0 &\dots & 0
\end{pmatrix}
\label{eSmith}
\ee
with $k_i, k_i/k_{i+1} \in \IZ, \forall i$. In this basis the computation of the smeared linking number gives $\int_{X_6} d^{-1} \tilde{\om}^i \wedge \beta^j = k_{i}^{-1} \delta^{ij}$, for $i,j = 1, \dots, r_{\bf e}$, suggesting that the torsion cohomology groups are
\be
{\rm Tor }\, H^3(X_6, \IZ) \simeq {\rm Tor }\, H^4(X_6, \IZ) \simeq \IZ_{k_1} \times \dots \times \IZ_{k_m}\, ,
\label{torsmith}
\ee
where $|k_i| > 1$ for $i \leq m$ and $|k_i| = 1$ for $m< i \leq r_{\bf e}$. Those entries of \eqref{eSmith} with value $\pm1$ should correspond to calibrated $p$-cycles that are trivial in homology, but that nevertheless are detected by the 4d EFT because they couple to massive modes below the compactification scale.

The proposal that the closed $p$-forms within $\{\om_a, \a_A, \b^B, \tilde{\om}^b\}$ correspond to smeared bump delta-forms of calibrated cycles can be further motivated by considering the set of BPS objects in the 4d EFT. For instance, let us again consider type IIA with $m_a{}^A =0$ and look at the closed three-forms $\b^A$, which in the basis \eqref{eSmith} may either be harmonic or exact in de Rham cohomology. Each of these forms are related to an axion $C_{0\, A}$, and from the BPS completeness hypothesis \cite{Polchinski:2003bq}, or the EFT string completeness hypothesis \cite{Lanza:2021udy} applied to $\CN=2$ gauged supergravities, one expects a BPS string under which such an axion is magnetically charged. Then, the results of \cite{Grana:2005ny,Kashani-Poor:2006ofe} imply that $K_\rho = - \log \int_{X_6} i \bar{\Om} \wedge \Om$  describes the metric of the hypermultiplet moduli space, at least at the classical level. Because the axion kinetic terms only depend on $K_\rho$ and this has the same expression as in the ungauged case, the tension of a BPS string should have the same general expression as in a Calabi--Yau. That is, we have that
\be
\frac{{\cal T}^A}{M_{\rm P}^2} = \frac{e^{\phi}}{{\rm Vol}_{X_6}} \left| \int_{X_6} \Om \wedge \beta^A \right| \, .
\ee
Now, in this context BPS means that the D4-brane internal worldvolume $\Pi_3^A$ is calibrated by $\Omega$, and dimensionally reducing its DBI action one obtains
\be
\frac{{\cal T}^A}{M_{\rm P}^2} =  \frac{e^{\phi}}{{\rm Vol}_{X_6}}  \left| \int_{\Pi_3^A}  \Om \right| =  \frac{e^{\phi}}{{\rm Vol}_{X_6}}  \left| \int_{X_6}  \Om \wedge \delta(\Pi_3^A) \right| ,
\ee
which implies that $\b^A$ must be the smeared version of $\delta(\Pi_3^A)$. Note that this is a standard result when $\b^A$ is not exact in de Rham cohomology, since then $[\b^A]$ and $[\Pi_3^A]$ are related by standard Poincar\'e duality. Similarly, in 4d $\CN=2$ EFTs, the mass of charged BPS particles in Planck units is specified by their central charge. A key result of \cite{Grana:2005ny,Kashani-Poor:2006ofe} is that the kinetic terms of vector multiplet sector is encoded in the K\"ahler potential $K_J = - \log \frac{4}{3}\int_{X_6} - J \wedge J \wedge J$ also for gauged supergravities obtained from compactifications on SU(3)-structure manifolds. From here it follows that the central charges of BPS particles charged under the vector multiplets are precisely the periods of $B+iJ$, that is
\be
Z^a = e^{K_J/2} \int_{X_6} e^{B+iJ} \wedge \tilde{\om}^a .
\ee
As in the Calabi--Yau case, such BPS particles should arise from wrapping D2-branes on two-cycles $\Pi_2^a \subset X_6$  calibrated by $J$. By dimensionally reducing their DBI action we obtain
\be
\frac{m_a^2}{M_{\rm P}^2} = e^{K_J} \left| \int_{X_6} (B+i J)\wedge \delta (\Pi_2^a) \right|^2  ,
\ee
which implies that $\tilde{\om}^a$ should be the smeared version of $\delta(\Pi_2^a)$. Again, this is independent of whether $\tilde{\om}^a$ is de Rham exact or not. When $\tilde{\om}^a$ is exact, it has to be that $\Pi_2^a$ is either a torsion or a trivial class of $H_2(X_6, \IZ)$. Finally, the completeness hypothesis implies that there is a BPS particle per element of the basis $\{\tilde{\om}^a\}$, which again is a standard result in the Calabi--Yau case.


\section{Torsion D-branes in $\CN=1$ vacua}
\label{s:N=1}

Having discussed the physical meaning of the smeared torsion linking number in 4d $\CN=2$ settings, it is natural to wonder how Conjecture \ref{conj:BPS} can be physically realized in 4d $\CN=1$ string vacua. In the $\CN=2$ case, the realization was based on the existence of BPS AB particles and strings, a set of objects that will be essentially absent in 4d $\CN=1$ type II orientifold settings. Indeed, there are no BPS particles in $\CN=1$ vacua, and a 4d string that arises from a D$(p+1)$-brane wrapped on a $p$-cycle $\Pi_p \subset X_6$ can only be BPS if $\Pi_p$ is calibrated by a closed $p$-form.\footnote{The precise statement is that in 4d $\CN=1$ Minkowski vacua the calibration for 4d strings is $d_H$-closed  \cite{Martucci:2005ht,Koerber:2005qi,Koerber:2010bx}, an statement that also holds for the $\CN=0$ Minkowski vacua analyzed in \cite{Lust:2008zd}. In practice this implies that, even in compactifications with $H$-flux,  torsion cycles cannot be calibrated.} As such, torsion $p$-cycles cannot yield 4d BPS strings. The realisation of the smeared linking number must therefore be more subtle in this case. 

As already mentioned in section \ref{s:dimred}, one possibility is to invoke the extension of Conjecture \ref{conj:BPS} formulated around \eqref{deldif}, and look for torsion $p$-cycles that are not calibrated by themselves, but whose homology class can nevertheless be seen as a linear combination of calibrated $p$-cycles. This would in principle allow us to describe AB strings and particles and their associated smeared $p$-forms in 4d $\CN=1$ orientifold vacua, connecting our previous discussion with the setup of \cite{Camara:2011jg}.

The most natural realization of the smeared torsion linking form seems instead to involve the St\"uckelberg-like terms that involve 4d BPS objects in $\CN=1$ vacua, and that correspond to \eqref{p=2} and \eqref{p=3}. These two couplings represent 4d membranes ending on strings and space-time filling branes ending on membranes, and in the context of 4d $\CN=1$ supersymmetry they can be described in the language of three-form multiplets  \cite{Lanza:2019xxg}. Notice that in a type II compactification on a six-dimensional manifold $X_6$, a D$p$-brane that looks like a 4d particle and a D$(p+2)$-brane that looks like a 4d membrane can wrap the same $p$-cycle $\Pi_p \in X_6$, and the same holds for a D$(p+3)$-brane and an Euclidean  D$(p-1)$-brane instanton. So essentially we are trading the role of 4d particles and instantons for that of 4d membranes and space-time filling branes, in order to probe a similar set of torsion $p$-cycles with BPS objects. In practice, this implies that the topological information that was captured by \eqref{stuck} and \eqref{stuckdual} in the $\CN=2$ case, now is encoded in the couplings \eqref{p=2} and \eqref{p=3}.

To see how this works in practice, let us again focus on type IIA string theory on a compact SU(3)-structure manifold $X_6$, but now with an orientifold projection that introduces O6-planes. Assuming a 4d Minkowski vacuum leads to the following metric Ansatz
\be
ds^2 = e^{2A} ds^2_{\IR^{1,3}} + \ell_s^2 ds^2_{X_6} ,
\ee
with $A$ a warp factor that depends on the coordinates of $X_6$, whose SU(3)-structure metric satisfies the following equations
\be
d(3A-\phi) = H + idJ = 0 , \qquad d(e^{2A-\phi} {\re \Om}) =0 , \qquad \ell_s d( e^{4A-\phi} {\im \Om}) = - e^{4A} \star  F_2 ,
\label{minkIIA}
\ee
and $F_0 = F_4 =F_6 =0$, with $F_{2p}$ the gauge-invariant RR field strength. In this setup 4d strings made up of D4-branes wrapping three-cycles are calibrated by $\pm e^{2A-\phi}{\re \Om}$ which, as advanced, is a closed three-form. Therefore, there are no BPS strings of this sort that correspond to torsion homology classes. The same can be said for membranes, which are calibrated by $e^{3A-\phi}e^{B+iJ}$. 

The last equation in \eqref{minkIIA}, however, features a non-closed three-form that calibrates space-time filling D6-branes. As such, it can detect calibrated torsion three-cycles. That such three-cycles exist in certain SU(3)-structure vacua can be deduced from the results of \cite{Tomasiello:2005bp}, which imply that a D$p$-brane that is point-like in a Calabi--Yau manifold with $H$-flux is mapped by mirror symmetry to a D$(p+3)$-brane wrapping a torsion three-cycle $\Pi_3^{\rm tor}$. In our type IIA orientifold context, $\Pi_3^{\rm tor}$ will host a 4d BPS object if it is wrapped either by a D6-brane or by an Euclidean D2-brane. In practice, the DBI action of these objects is easier to analyze if, following \cite{Tomasiello:2007zq}, one trades the last equation in \eqref{minkIIA} by an equivalent one not involving the Hodge star operator. In the case at hand we find \cite{Marchesano:2014iea}
\be
\ell_s d( e^{-\phi} {\im \Om}) = - J \wedge F_2 ,
\label{IIAJF}
\ee
which already hints that part of the torsion data of $X_6$ is encoded in the RR flux $F_2$. 

In this context, it is illustrative to consider a simple example, like the twisted six-torus geometries analyzed in \cite{Marchesano:2006ns}. These correspond to an SU(3)-structure of the form \eqref{Jex} and \eqref{Omex}, with the simplification $V= V_i =1$, $\forall i$, and an orientifold action of the form ${\cal R}: (J,\Om) \mapsto {-}(J,\bar{\Om})$. The $p$-chains $\{\Pi_2^i, \Pi_3^{\rm tor}, \Sigma_3 , \Sigma_4^i \}$ play the same role in terms of torsion homology information as in section \ref{s:simple}, but from an EFT viewpoint they should be associated to either 4d membranes or space-time filling branes. For concreteness, let us consider a $\tilde{\bf T}^6$ with twisting $M_1 = -M_2 = N \in \mathbb{N}$ and $M_3=0$. This choice fixes the complexified K\"ahler moduli as $b^1+it^1 = b^2+it^2$ in the vacuum and leads to a background RR flux of the form
\be
F_2 = \ell_s K \left(\eta^1 \wedge \eta^4 -\eta^2 \wedge \eta^5\right) , \qquad K \in \mathbb{N}\, ,
\label{F2orism}
\ee
that solves \eqref{IIAJF} in the constant dilaton approximation, by setting $K t^1 = K t^2 = N e^{-\phi}R_1R_2R_3$. This sort of RR flux background is the one that appears in the type IIA orientifold flux literature, see e.g. \cite{Villadoro:2005cu,Camara:2005dc}, but it is important to realize that the expression is a consequence of the constant-dilaton/smeared approximation. Indeed, what occurs in this background is that there are eight O6-planes wrapped on $[\Pi_3^{\rm tor}]$. In a plain toroidal compactification one could cancel this charge by placing 32 D6-branes in the same three-cycle class of the covering space. In the twisted torus geometry, because $[\Pi_3^{\rm tor}]$ is $\IZ_N$-torsion, one only needs to place $32 - kN$ of such D6-branes, for some $k \in \IZ$, in order  to cancel the RR tadpole. The lack of D6-branes leads to a RR flux background that satisfies
\be
dF_2 \simeq -\ell_s k N \delta_3 (\Pi_3^{\rm tor}) ,
\label{F2oriunsm}
\ee
where for simplicity we have assumed all O6-planes and D6-branes on the same representative (otherwise one is led to more involved delta-source equations, like the ones solved in \cite{Casas:2022mnz}). Upon implementing the smearing approximation one obtains
\be
F_2 = - \ell_s k N\, d^{-1}\delta_3^{\rm sm} (\Pi_3^{\rm tor}) \, ,
\ee
which reproduces \eqref{F2orism} for $k =2K$, up to a harmonic form. The actual RR flux is, however, the one that solves \eqref{F2oriunsm}, since it is the only one that can satisfy Dirac's quantization condition, upon the appropriate choice of harmonic piece \cite{Marchesano:2014iea}. 

The couplings \eqref{p=2} and \eqref{p=3} are obtained upon dimensionally reducing the RR potentials
\bea
C_5 & = &  \ell_s^{5} 2\pi \left[ B_{2\, 0} \wedge \b^0 + B_2^i \wedge \a_i  + C_3^i \wedge \om_i\right]\, , \\
C_7 & = & \ell_s^{7} 2\pi \left[ D_{3\, i} \wedge \tilde{\om}^i + A_{4}^0 \wedge \a_0  + A_{4\, i}\wedge \b^i \right]\, ,
\eea
where we have taken into account the orientifold action, and expanded into $p$-forms that couple with unit charge to the calibrated $p$-chains $\{\Pi_2^i, \Pi_3^{\rm tor}, \Sigma_3 , \Sigma_4^i \}$.\footnote{For simplicity we are using the notation \eqref{alphabetas}, which results in an unusual convention in orientifold compactifications. The more standard one is obtained by interchanging the basis elements as $\a_i \leftrightarrow - \b^i$, cf. \cite[eq.(2.6)]{Camara:2005dc}.} One finds 
\be
(2\pi)^2 \left[ \hat{g}_{ii} (dC_3^i)^2 + \tilde{F}^{00}  \left(d B_{2\, 0} + M_i C_3^i \right)^2 + e^{-4\phi_4}M_P^{-4} \tilde{F}_{ii}  (dB_2^i)^2 \right]\, ,
\label{p=2ori}
\ee
and
\be
(2\pi)^2 \left[ \hat{g}^{ii} \left( d D_{3\, i} - M_i A_4^0 \right)^2 \right]  e^{-8\phi_4}M_P^{-8} \, ,
\label{p=3ori}
\ee
where we have assumed generic twists $M_i$ and have defined
\be
\hat{g}_{ii}=g_{ii} 
e^{-4\phi_4}M_P^{-4},\qquad
\tilde{F}^{AA} = F^{AA} 
e^{-2\phi_4} M_P^{-2}, 
\ee
and $\hat{g}^{ii}=1/\hat{g}_{ii}$, $\tilde{F}_{AA}=1/\tilde{F}^{AA}$. Note that $\hat{F}^{AB}$ defined below \eqref{disgaugen} satisfies $\hat{F}^{AB} = \tilde{F}^{AB} e^{4\phi_4}M_P^{4}$, and so \eqref{p=2ori} contains the same kind of information as \eqref{StuckN=2gen}. As a result, the computation of the smeared linking number works exactly as in section \ref{ss:EFTdesc}. The main difference is the expression of the smeared linking number in terms of 4d EFT quantities, which now involves the physical charges of membranes and strings ending on each other. More precisely in this $\CN=1$ setup, one finds that the relation \eqref{Lphys} is substituted by
\be
\frac{m_{\rm st}}{N} =  \sqrt{\frac{\hat{g}^{\a\a}}{\tilde{F}_{00}}} = e^{\phi_4} M_{\rm P} \sqrt{\frac{{g}^{\a\a}}{{F}_{00}}}\, ,
\ee
where $N=M = {\rm g.c.d.} (M_1, M_2, M_3)$ and $\hat{g}^{\a\a} = e^{4\phi_4}M_P^{4} {g}^{\a\a} = e^{4\phi_4}M_P^{4} \sum_i \frac{M_i^{2}}{M^2} g^{ii} $. Here $\hat{g}^{\a\a}$ represents the squared physical charge ${\cal Q}^2$ of a BPS membrane ending on a BPS string and $\tilde{F}_{00}$ the squared charge of such a string, as defined in \cite{Lanza:2020qmt}, see also \eqref{physQ}.


\section{Beyond the BPS case}
\label{s:nonBPS}

Conjecture \ref{conj:BPS} proposes a method to compute the linking number between two calibrated torsion cycles from smeared/EFT data. However, as already mentioned, in Calabi--Yau manifolds torsion $p$-cycles cannot be calibrated, or equivalently D-branes wrapped on them are not BPS objects of the EFT. The extension of the conjecture around \eqref{deldif} allows us to implement the same method whenever the torsion class $\Pi_p^{\rm tor}$ of interest is a linear combination of $p$-cycle homology classes with calibrated representatives. This more general setup could in principle occur in Calabi--Yau compactifications, and then the extended conjecture would imply that one can compute torsion in cohomology via smeared data, provided there exist massive eigenforms of the Laplacian below the compactification scale that couple to torsion $p$-cycles. Including such a set of light fields in the 4d EFT would presumably take us to a structure of the form \eqref{intrel} and \eqref{system}, in which giving a non-vanishing vev to a massive, light field deforms an SU(3)-holonomy metric to an SU(3)-structure one.  

Nevertheless, in general, one would expect that a torsion class in homology does not contain any calibrated representative, and neither can it be understood as a linear combination of homology classes with them. In that case, our discussion of section \ref{s:dimred} suggests that there should be non-trivial corrections  associated to this sector. More precisely, one would expect that in the EFT description the bump-delta form $\delta_{6-p} (\Pi_p^{\rm tor})$ can still be replaced by its smeared version, but only  up to a multiplying constant that could be interpreted as  wavefunction renormalization. That is, one does not simply project the delta into its lowest eigenmode component, but also has to multiply the result by some constant (or a field-dependent function) in order to correctly reproduce the 4d physics. Whenever this happens the smeared linking number and the exact linking number do not coincide, and one should rescale the smeared delta to make it so. 

It is hard to have an idea of the magnitude of this rescaling without an  example at hand where the computation can be carried out explicitly. The best we can do is to give an estimate for the error in the smeared linking number, as follows. In supersymmetric theories, the EFT kinetic terms obtained from truncating  massive modes at zero vev are exact up to corrections of ${\cal O}(\Lambda_{\rm EFT}/\Lambda_{\rm UV})$, see e.g. \cite{Buchmuller:2014vda}. In our case $\Lambda_{\rm EFT}$ corresponds to the mass $m_{\rm st}$ of the massive modes that we keep in our EFT, and $\Lambda_{\rm UV}$ to the compactification scale $m_{\rm KK}$. Notice that this is the suppression that we found in \eqref{shiftKK} when moving away from the minimal tension representative. 

If the torsion cycle is not calibrated, it means that the EFT data is not computing its minimal volume properly. In some cases, like in Calabi--Yau vacua the `expected' volume ${\cal V}_p^{\rm BPS}$, namely the integral of the appropriate calibration over $\Pi_p^{\rm tor}$,  vanishes. In general, the volume of a non-calibrated cycle should be larger than the integral of its would-be calibration over the given homology class. One can think of the mismatch between volumes as how much one needs to deform a would-be calibrated cycle to match the actual one. Finally, one usually converts differences of internal volumes to differences of field vevs via the physical 4d charge ${\cal Q}$ of the corresponding EFT object  \cite{Lanza:2021udy}, which here we define as 
\be
{\cal Q}^2(\Pi_p) = \sum_{\lambda_i \ll \ell_s m_{\rm KK}} c_i^2 ,
\label{physQ}
\ee
with $c_i$ the coefficients of the smeared delta-form of $\Pi_p$, as in \eqref{deltasm-07}.  

With these considerations in mind, let us consider the smeared linking number between a calibrated cycle $\Pi_{6-p-1}^{\rm tor}$ and a non-calibrated one $\Pi_p^{\rm tor}$. One may propose the following upper bound 
\be
\left| L- L^{\rm sm} \right| < \frac{{\cal V}
	_p - {\cal V}_{p}^{\rm BPS}}{{\cal Q}} \frac{m_{\rm st}}{m_{\rm KK}} ,
\ee
where ${\cal V}$ is the actual volume of $\Pi_p^{\rm tor}$, in string units. This upper bound estimates the error when computing the smeared linking number, with respect to the actual one $L$. If the bound is small, it still makes sense to compute \eqref{Lsm}, because it gives a good estimate of the actual linking number. That is, one may still use EFT data to characterize torsion in cohomology.


\section{Discussion}
\label{s:conclu-07}

In this chapter we have proposed a method to detect topological invariants of torsion cohomology groups via smooth $p$-forms. The proposal is based on what ${\rm Tor} H^p(X_n, \IZ)$ means when performing dimensional reduction of type II string theory on $X_n$ and obtaining a lower-dimensional EFT with a massive sector, and it can be summarized in two main points: 

\begin{itemize}
	
	\item[{\it i)}] If a D-brane wrapped on a torsion cycle $\Pi_p^{\rm tor}$ has a non-trivial backreaction at EFT wavelengths, it is because there are light massive eigenmodes of the Laplacian sourced by it. In geometric terms, this means that $\Pi_p$ has a non-trivial smeared delta form $\delta^{\rm sm}_{n-p} (\Pi_p)$, which is a necessary requirement to apply our approach.

	\item[{\it ii)}] Whenever $\Pi_p^{\rm tor}$ is calibrated, a D-brane wrapped on it is a BPS object of the theory whose smeared backreaction is protected from  dimensional reduction corrections. As a result one can compute the torsion linking numbers of $\Pi_p^{\rm tor}$ using its smeared delta form. 
\end{itemize}

This second statement, which is the content of Conjecture \ref{conj:BPS}, provides a method to detect the $\IZ_N$ factors in $H^p(X_n, \IZ)$. The method has a wider application if one assumes the extension of the conjecture made around \eqref{deldif}, and it would be really interesting to see if it can be applied to manifolds with special holonomy metrics. 

The use of smooth $p$-forms to compute torsion in cohomology may seem quite surprising because such $\IZ_N$ factors are projected out in de Rham cohomology groups. One should however keep in mind that in our approach we are starting with a set of objects that contain the information of singular homology groups, namely the bump delta forms $\delta_{n-p} (\Pi_p)$, and replacing them by a countable set of smooth forms $\delta_{n-p}^{\rm sm} (\Pi_p)$ that should remember part of the torsion data. As a possible analogy, one may consider a finite good cover of $X_n$ and its nerve $N$, which is a triangulation of $X_n$ \cite{Bott1982DifferentialFI}. We then consider the delta forms $\delta_{n-p} (\sigma_{p, \a})$ of the $p$-simplexes $\sigma_{p, \a}$ of $N$, with a small smearing (such that $1/\lambda_{\rm max}$ is below the spacings in $N$). This produces a lattice of smooth $p$-forms from which one can compute $H^p(X_n, \IZ)$ via singular cohomology. Our proposal can be thought of as a limit of this construction, in the sense that we perform a much more dramatic smearing, namely at wavelengths above ${\rm Vol}(X_n)^{1/n}$.  This more drastic coarse-graining is allowed geometrically because, if the assumptions behind Conjecture \ref{conj:BPS} are true, then there should be a $G$-structure manifold that one can construct by fibering $X_n$ over flat space, which is the EFT solution of a D-brane wrapped on $\Pi_p$. One could then use this non-compact, higher-dimensional manifold to compute topological information of $X_n$ via smeared data. 

It is also instructive to compare our approach with some of the discussion in the string theory literature, like the one carried in \cite{Tomasiello:2005bp} based on the classification theorems of Wall \cite{Wall:1966rcd} and \v Zubr \cite{Zhubr1978ClassificationOS}. As pointed out in \cite{Tomasiello:2005bp} these theorems classify six-manifolds up to diffeomorphisms and the classification data match the content of the massless sector of the compactification, discarding exact and co-exact $p$-forms. Our results are not in tension with this classification, because we need to endow $X_n$ with a $G$-structure metric in order to extract the torsion cohomology data. This choice of metric also specifies the set of calibrated cycles and the light spectrum of the Laplacian, so it is crucial in order to select those exact and co-exact forms that contain the torsion information. In this light, it would be interesting to see if the presence of torsion in cohomology restricts the choice of $G$-structure metrics on a manifold, or if one can always choose a $G$-structure metric where all the smeared delta forms of calibrated torsion cycles vanish. 

An important part of our analysis is based on constructing explicit examples of SU(3)-structure manifolds with calibrated torsion cycles. This allowed us to perform a direct comparison of the torsion linking number and its smeared version, where we observed a remarkable cancellation between terms in the eigenmode expansion of the delta form, that is reminiscent of the computations of topological indices. It would be very interesting to understand the meaning of this feature and if it is also realized in more involved setups, providing further evidence of Conjecture \ref{conj:BPS}. Our explicit constructions also provided concrete EFT descriptions of BPS configurations of branes ending on branes, like the 4d Aharanov-Bohm strings stretching between a domain wall and a monopole in $\CN=2$ gauged supergravities. Moreover, the properties of such objects resulted in a physical interpretation of the Hitchin flow equations, which could be useful to further understand the properties of these subtle objects. It is likely that this connection sheds light into the physics of $\CN=2$ gauged supergravities, like for instance when applied to the black hole supergravity solutions recently revisited in \cite{Angius:2023xtu}, and which share many properties with AB strings and particles. 

As a direct application of our proposal, we have revisited the dimensional reduction framework developed in \cite{Gurrieri:2002wz,DAuria:2004kwe,Grana:2005ny,Kashani-Poor:2006ofe} to furnish it with one of its main missing elements. That is, a geometric prescription to define the basis of $p$-forms in which the RR potentials and the calibration forms must be expanded. We have verified that our definition fits perfectly with the physical properties that these forms should have, and which define the periodicity properties of massive axions and gauge bosons of the 4d $\CN=2$ EFT. Such periodicities are crucial to define the global gauge transformations for massive $p$-form in more general setups. This applies in particular to  4d $\CN=1$ compactifications, where the St\"uckelberg-like couplings related to our method involve the gauging of three- and four-forms in 4d. 

To sum up, our findings seem to point out that torsion in cohomology could lead to specific, measurable physics in the massive sector of 4d EFTs obtained from string theory. It could be that exploiting this new link between geometry and physics could give us a new, more approachable understanding of the subtle objects that are torsion $p$-cycles.

	\clearpage{\pagestyle{empty}\cleardoublepage} 

	\part{Conclusions and Appendices}
	
	\chapter{Conclusions}

\section{English Version}

The goal of this thesis was to study some phenomenological aspects of string theory. We focused on the thermodynamics of the black hole solutions of the heterotic string theory (HST) effective action with $\alpha'$ corrections and on the compactifications of type II superstring theory. We successfully achieved our aim. The results of the thesis improved some well-known models, such as the Strominger--Vafa black hole and the DGKT-like CY orientifold vacuum. Now we have a description beyond the leading order, which includes certain quantum corrections. In the future, these examples can be used to study phenomena sensitive to these corrections. On top of that, we discovered and developed general approaches whose applications extend beyond string theory. We extended our understanding of black hole thermodynamics and Wald's formalism in theories with higher derivative corrections and non-trivial gauge transformations. We described how to include the effects of the torsional part of the homology group in compactifications. All these techniques open new possibilities to pinpoint constraints that a consistent EFT must satisfy, explore the string theory landscape and delve into our understanding of black holes from both the microscopic and macroscopic perspectives.

In part \ref{part:TO}, we described how to solve the equations of motion of HST at first order in $\alpha'$, obtaining explicit and analytical solutions representing 5-dimensional, 3-charge and 4-dimensional, 4-charge black holes. We did so in the supersymmetric, extremal non-supersymmetric, and non-extremal cases. Some of the extremal solutions have been further generalized to configurations describing multiple extremal black holes at equilibrium. One of the first results of the extended Wald's formalism is a modified Wald entropy formula for HST at first order in $\alpha'$ \cite{Elgood:2020nls}. We used our non-extremal black hole solutions to test such a proposal. We found that the entropy obtained using this formula (and not the one obtained applying Iyer and Wald's prescription) satisfies the thermodynamic relation $T^{-1} = \partial S/\partial M$. One of the main tools used within the extended Wald's formalism is the concept of gauge-covariant Lie derivative. We clarified its necessity by showing how it naturally emerges in the context of Kaluza--Klein dimensional reduction. We also explained how to define and obtain scalar charges in the first law of black hole thermodynamics using Wald's formalism. We did so in theories whose scalar sector kinetic term is that of a non-linear sigma model. As a byproduct, we showed that in these models, the scalar charges are determined by the value of the gauge charges and the electromagnetic potentials evaluated at the black hole horizon, proving that we do not have primary scalar hair. Using the extended Wald's formalism, we fully characterized the thermodynamics of the black hole solutions we obtained. We checked that for the 2-charge non-extremal configurations, the results match those extracted from the Euclidean on-shell action, and realize the predictions of \cite{Chen:2021dsw} based on duality arguments. In the extremal cases, we found that the higher derivative corrections introduce a non-positive correction to the mass. This is in agreement with the mild form of the weak gravity conjecture (WGC), which suggests that the state necessary for the decay might be the black hole itself in the presence of higher derivative corrections. This finding is also compatible with the sharpened version of the WGC because the mass of supersymmetric black holes does not receive corrections. Interestingly, we found that in the extremal cases, the entropy formula matches the microscopic predictions, which are supposed to be valid to all orders in $\alpha'$ \cite{Kraus:2005vz, Kraus:2006wn, Sen:2007qy}. Exploiting instead the absence of struts and singularities in the multi-center solutions, we have been able to conclude that first-order $\alpha'$ corrections do not spoil the cancellation of forces among black hole solutions in both the supersymmetric and non-supersymmetric cases. Moreover, charge conservation forbids fragmentation processes.
	
In part \ref{part:FM}, we studied the properties of a class of $\mathcal{M}_4\times X_6$ compactifications of type II theories. We considered a class of well-known AdS$_4$ orientifold vacua of massive type IIA. We showed that such vacua have an explicit 10-dimensional embedding, obtaining them as solutions of the 10-dimensional equations of motion in the smearing approximation. In this approximation, $X_6$ is a Calabi-Yau orientifold threaded by background fluxes. Such a description is unsatisfactory, especially because of the presence of the orientifold planes. We showed that if we consider a small deformation, i.e., we do not completely smear the sources, we can find a deformation of the CY metric of $X_6$ and of the other bosonic fields which still satisfy the EOMs. Such a result signals that a description in terms of localized sources exists and that the smearing approximation can be interpreted as the leading term of a perturbative expansion. Therefore, the smeared solutions are perturbatively stable. We analyzed the non-perturbative stability of the configurations by scanning for branes which can trigger a vacuum decay. We found that almost all the non-supersymmetric backgrounds admit superextremal branes and the supersymmetric ones admit only branes that at most saturate the WGC bound.\footnote{For the case A1-S1$-$, it is still unclear if we have non-perturbative stability and superextremal branes of dimension 4.} The results are in agreement with the predictions of the sharpened WGC and the AdS instability conjecture. Interestingly, in all these examples, we have scale separation. The supersymmetric case is a counterexample to the strong AdS distance conjecture. In the case of type II compactifications of the form $\mathcal{M}_4\times X_6$, with $X_6$ a compact manifold with non-trivial torsional (co)homology, we described a new mechanism to leak topological UV information in an effective theory. We proposed that whenever the torsion cycles of the manifold are calibrated by non-closed forms, the associated $\mathbb{Z}_p$ factors of the (co)homology group enter the EFT theory in the form of $\mathbb{Z}_p$ discrete gauge symmetries. This happens because for calibrated torsional cycles, the lowest, massive eigenmode of the Laplacian can encode all the topological information. More precisely, it can be used to determine the linking numbers of the torsional cycles. If such a mode is much smaller than the KK scale, the EFT can be sensitive to it. The existence of torsional cycles calibrated by non-closed forms is typical in the context of G-structure manifolds but does not apply to the CYs. As a byproduct, we explained how one can define the basis of forms used to expand the 10-dimensional fields in the compactification process. They are built by smearing the bump delta forms with support on the torsional cycles.

\section{Spanish Version}

El propósito de esta tesis era estudiar algunos aspectos fenomenológicos de la teoría de cuerdas. Nos centramos en la termodinámica de las soluciones de agujeros negros de la acción efectiva de la teoría de cuerdas heterótica (HST) con correcciones $\alpha'$ y en las compactificaciones de la teoría de supercuerdas de tipo II. Logramos nuestro objetivo. Los resultados de la tesis mejoran algunos modelos bien conocidos, como el agujero negro de Strominger--Vafa y el vacío \textit{orientifold} CY tipo DGKT. Ahora disponemos de una descripción valida más allá del primer orden, que incluye ciertas correcciones cuánticas. En el futuro, estos ejemplos podrán ser utilizados para estudiar fenómenos sensibles a dichas correcciones. Además, descubrimos y desarrollamos enfoques generales cuyas aplicaciones se extienden más allá de la teoría de cuerdas. Ampliamos nuestra comprensión de la termodinámica de agujeros negros y del formalismo de Wald en teorías con correcciones de derivadas superiores y transformaciones gauge no-triviales. Describimos cómo incluir los efectos de la parte torsional del grupo de homología en las compactificaciones. Todas estas técnicas abren nuevas posibilidades para identificar restricciones que una teoria de campos efectiva (EFT) consistente debe satisfacer, explorar el ``paisaj'' de la teoría de cuerdas y profundizar en nuestra comprensión de los agujeros negros desde perspectivas microscópicas y macroscópicas.

En la parte \ref{part:TO}, describimos cómo resolver las ecuaciones de movimiento de HST en primer orden en $\alpha'$, obteniendo soluciones explícitas y analíticas que representan agujeros negros en 5 dimensiones, con 3 cargas y en 4 dimensiones, con 4 cargas. Lo hemos hecho en los casos supersimétrico, no-supersimétrico extremo y no-extremo. Algunas de las soluciones extremas han sido generalizadas aún más para describir múltiples agujeros negros extremos en equilibrio. Uno de los primeros resultados del formalismo de Wald extendido es una fórmula de entropía de Wald modificada para la HST al primer orden en $\alpha'$ \cite{Elgood:2020nls}. Utilizamos nuestras soluciones de agujeros negros no-extremos para probar dicha propuesta. Encontramos que es la entropía obtenida usando esta fórmula (y no la obtenida aplicando la prescripción de Iyer y Wald) la que satisface la relación termodinámica $T^{-1} = \partial S/\partial M$. Una de las principales herramientas utilizadas en el formalismo de Wald extendido es el concepto de derivada de Lie covariante gauge. Demostramos la necesidad de usarla mostrando cómo surge naturalmente en el contexto de reducciones dimensionales de KK. También explicamos cómo definir y obtener las cargas escalares en la primera ley de la termodinámica de agujeros negros usando el formalismo de Wald. Lo hemos hecho en teorías cuyo término cinético del sector escalar es el de un modelo sigma no-lineal. Como subproducto, demostramos que en estos modelos, las cargas escalares están determinadas por el valor de las cargas y potenciales electromagnéticos evaluados en el horizonte del agujero negro, demostrando que los agujeros negros no tienen pelo escalar primario. Usando el formalismo de Wald extendido, caracterizamos completamente la termodinámica de las soluciones de agujeros negros que obtuvimos. Comprobamos que para las configuraciones no-extremas con 2 cargas, los resultados coinciden con los extraídos de la acción Euclidea \textit{on-shell}, y confirman las predicciones de \cite{Chen:2021dsw} basadas en argumentos de dualidad. En los casos extremos, encontramos que las correcciones de derivadas superiores introducen una corrección no positiva a la masa. Esto está de acuerdo con la versión suave de la conjetura de gravedad débil (WGC), que dice que el estado necesario para la desintegración podría ser el propio agujero negro en presencia de correcciones de derivadas superiores. Este hallazgo también es compatible con la versión refinada de la WGC porque la masa de los agujeros negros supersimétricos no recibe correcciones. Es interesante ver que en los casos extremos, la fórmula de la entropía coincide con las predicciones microscópicas, que se supone que son válidas a todos los órdenes en $\alpha'$ \cite{Kraus:2005vz, Kraus:2006wn, Sen:2007qy}. Aprovechando la ausencia de \textit{struts} y singularidades en las soluciones de múltiples centros, hemos demostrado que las correcciones a primer orden en $\alpha'$ no estropean la cancelación de fuerzas entre los agujeros negros en ambos casos supersimétricos y no supersimétricos. Además, la conservación de las cargas prohíbe los procesos de fragmentación.

En la parte \ref{part:FM}, estudiamos las propiedades de una clase de compactificaciones $\mathcal{M}_4\times X_6$ de teorías de cuerdas de tipo II. Consideramos una clase de vacíos AdS$_4$ en la teoría tipo IIA masiva de tipo \textit{orientifold} bien conocidos. Mostramos que tales vacíos tienen una immersíon explícita en 10 dimensiones, obteniéndolos como soluciones de las ecuaciones de movimiento de 10 dimensiones en la aproximación de \textit{smearing}. En esta aproximación, $X_6$ es un \textit{orientifold} de Calabi-Yau con flujos. Tal descripción es insatisfactoria, especialmente debido a la presencia de los planos \textit{orientifold}. Demostramos que si consideramos una pequeña deformación, es decir, no difuminamos completamente las fuentes, podemos encontrar una deformación de la métrica de CY de $X_6$ y de los otros campos bosónicos que aún satisfacen las ecuaciones de movimiento. Tal resultado indica que existe una descripción en términos de fuentes localizadas y que la aproximación de \textit{smearing} se puede interpretar como el primer término de una expansión perturbativa. Por lo tanto, las soluciones difuminadas son perturbativamente estables. Analizamos la estabilidad no-perturbativa de las configuraciones explorando para branas que puedan desencadenar una desintegracíon del vacío. Encontramos que casi todos los fondos no supersimétricos admiten branas superextremales y los supersimétricos admiten solo branas que como máximo saturan la WGC.\footnote{Para el caso A1-S1$-$, todavía no está claro si tenemos estabilidad no perturbativa y branas superextremales de dimensión 4.} Los resultados están en acuerdo con las predicciones de la WGC refinada y la conjetura de inestabilidad de AdS. Es interesante comprobar que, en todos estos ejemplos, tenemos separaciones de escalas. El caso supersimétrico es un contraejemplo de la versión fuerte de la conjetura de la distancia AdS. En el caso de compactificaciones de tipo II de la forma $\mathcal{M}_4\times X_6$, con $X_6$ una variedad compacta con (co)homología torsional no trivial, describimos un nuevo mecanismo para detectar información topológica UV en una teoría efectiva. Propusimos que siempre que los ciclos de torsión de la variedad estén calibrados por formas no cerradas, los factores $\mathbb{Z}_p$ asociados del grupo de (co)homología aparecen en la teoría EFT en forma de simetrías gauge discretas $\mathbb{Z}_p$. Esto sucede porque para ciclos de torsión calibrados, el autovector masivo más bajo del laplaciano puede codificar toda la información topológica. En concreto, puede usarse para determinar los números de enlace de los ciclos de torsión. Si dicho autovector es mucho más ligero que la escala de KK, la EFT puede ser sensible al mismo. La existencia de ciclos de torsión calibrados por formas no cerradas es típica en el contexto de variedades de compactificación con G-structura, pero no se aplica a las variedades CYs. Como subproducto, explicamos cómo se puede definir la base de formas utilizadas para expandir los campos de 10 dimensiones en el proceso de compactificación. Se construyen difuminando las formas delta con soporte en los ciclos de torsión.

	\clearpage{\pagestyle{empty}\cleardoublepage} 
	
	\appendix

	\chapter{Conventions}

This thesis is based on works which are using different conventions. We adapted them in such a way that we have just two different set of conventions: the one of \ref{part:TO} and the one of  \ref{part:FM}. Chapter \ref{ch:1} and appendices \ref{app:D} follow the conventions of \ref{part:FM}. Appendices \ref{app:B} and \ref{app:C} follow the conventions of \ref{part:TO}. We summarize briefly the differences.

\section{Conventions of part \ref{part:TO}}

We are following the conventions of \cite{Ortin:2015hya}

\subsubsection{Dimensional Reduction}

In the context of compactifications and dimensional reductions, we indicate with $\hat{\cdot}$ the higher dimensional fields. We indicate with greek letters the spacetime indexes. we split the higher dimensional spacetime indexes as $\hat{\mu} = (\mu, m)$, with $\mu$ and external, spacetime index and $m$ an interna,l compact manifold index. For flat indexes we use instead latin letters. We split them as $\hat{a} = (a, m)$. If we have to use both spacetime and flat indexes we indicate the internal curved indexes with a bar $\underline{m}$.

\subsubsection{Differential geometry conventions}
The Levi Civita connections $\nabla_\mu$ acts on vectors as
\begin{equation}
	\nabla_\mu \xi^\nu = \partial_\mu \xi^\nu - \Gamma^\nu_{\mu\rho} \xi^\rho \,,
\end{equation}
with 
\begin{equation}
	\Gamma^\nu_{\mu\rho} = \frac{1}{2} g^{\nu \alpha}\left(\partial_\mu g_{\rho\alpha} + \partial_\rho g_{\mu\alpha} - \partial_{\alpha}g_{\mu\rho} \right) \,.
\end{equation}
We define the Riemann curvature as
\begin{equation}
	R_{\mu \nu \sigma}{}^{\rho} = 2 \partial_{[\mu}\Gamma^\rho_{\nu] \sigma} + 2 \Gamma^\rho_{[\mu| \lambda} \Gamma^\lambda_{|\nu]\sigma} \,,
\end{equation}
which implies
\begin{equation}
	[\nabla_{\mu},\nabla_{\nu}]\xi^{\sigma}=R_{\mu\nu\rho}{}^{\sigma}\xi^{\rho}\,.
\end{equation}
Notice that this choice for the Riemann implies that the Ricci scalar of a sphere is negative
\begin{equation}
	R_{\text{S}^n} < 0 \,.
\end{equation}
The spin connection $\mathcal{D}$ acts on the vielbeins as
\begin{equation}
	\mathcal{D}e^{a}\equiv de^{a}-\omega^{a}{}_{b}\wedge e^{b}=0\,,
\end{equation}
and has curvature 2-form
\begin{equation}
	{R}{}^{{a}}{}_{{b}} = 
	d \omega{}^{{a}}{}_{{b}}
	- {\omega}{}^{{a}}{}_{{c}}
	\wedge  
	{\omega}{}^{{c}}{}_{{b}}\,.
\end{equation}
Given a $p$-form $\alpha$ on a $n$-dimensional Riemannian manifold $X_n$  with metric $g_{\mu\nu}$ 
\be
\alpha = \frac{1}{p!}\alpha_{\mu_1 \dots \mu_p} dx^{\mu_1} \wedge \cdots \wedge dx^{\mu_p} 
\ee
the Hodge dual is
\be
\star \, \alpha = \frac{1 }{p! (n-p)! \sqrt{|g|}} \epsilon^{\nu_1 \dots \nu_{n-p} \mu_{1} \dots \mu_p }  \, \alpha_{\mu_1 \dots \mu_p} \, g_{\nu_1 \rho_1} \dots g_{\nu_{n-p}\rho_{n-p}} \,  dx^{\rho_1} \wedge \cdots \wedge dx^{\rho_{n-p}} \,,
\ee
where $g = \text{det}(g_{\mu\nu})$ and $\epsilon$ is the Levi Civita symbol normalized as
\be 
\epsilon^{0\dots n-1} = 1 \,.
\ee
In our conventions we have
\be
\star^2 \, \alpha = s (-1)^{p(n-p)} \alpha \,, \qquad s = \text{sign}(g)\,.
\ee
The scalar product between forms is defined by 
\be
\alpha \cdot \beta =  \int_{X_n}   \alpha \wedge \star \, \beta = \frac{s (-1)^{p(n-p)}}{p!} \alpha_{\mu_1 \dots \mu_p} \, \beta^{\mu_1 \dots \mu_p} \,,
\ee
where the indexes of $\beta$ are raised with the inverse metric. The codifferential $d^\dagger$ over $p$-forms is defined as the adjoint of the differential with respect to the scalar product we introduced, i.e. it satisfies $ d^\dagger \alpha \cdot \beta   =  \alpha \cdot d \beta $.  In our conventions, the explicit expression of the codifferential acting on a $p$-form is
\be
d^\dagger \alpha = s (-1)^{n(p+1)+1} \star \, d \, \star \, \alpha \,.
\ee
The Laplace-de Rham operator is 
\be
\Delta = d d^\dagger + d^\dagger d \,.
\ee
On scalars, it is related with the Laplace-Beltrami operator $\nabla^2$  by
\be
\Delta = (-)^n \nabla^2 \,.
\ee

\subsubsection{Common bosonic sector effective action}

We use mostly minus signature $(+, - , \dots - )$. The action for the common bosonic sector in our conventions is 
\begin{equation}
	S
	=
	\frac{g_{s}^{2}}{16\pi G_{N}^{(10)}}
	\int d^{10}x\sqrt{|{g}|}\, 
	e^{-2{\phi}}\, 
	\left\{
	{R} 
	-4(\partial{\phi})^{2}
	+\tfrac{1}{12}{H}^2
	\right\}\,,
\end{equation}
the string coupling constant is defined as $g_s = \left< e^{\phi} \right>$. The relation between $\alpha'$ and $G_N^{(10)}$ is 
\begin{equation}
	G_N^{(10)} = 8 \pi^6 g_s^2 \alpha'{}^4 \,,
\end{equation}
and in our conventions the relation between $\alpha'$ and the string length $\ell_s$ is
\begin{equation}
	\ell_s = \sqrt{\alpha'} \,.
\end{equation}

\section{Conventions of part \ref{part:FM}}

We are following the conventions of \cite{Tomasiello:2022dwe} and \cite{Prieto:2024hkf}.

\subsubsection{Dimensional reduction}

In the context of dimensional reductions we use capital latin letters for 10-dimensional spacetime indexes. We split them as $ M = (\mu, m)$, indicating with greek letters the external spacetime indexes and latin letters the internal space indexes.

\subsubsection{Differential geometry conventions}

The Levi Civita connections $\nabla_\mu$ acts on vectors as
\begin{equation}
	\nabla_\mu \xi^\nu = \partial_\mu \xi^\nu - \Gamma^\nu_{\mu\rho} \xi^\rho \,,
\end{equation}
with 
\begin{equation}
	\Gamma^\nu_{\mu\rho} = \frac{1}{2} g^{\nu \alpha}\left(\partial_\mu g_{\rho\alpha} + \partial_\rho g_{\mu\alpha} - \partial_{\alpha}g_{\mu\rho} \right) \,.
\end{equation}
We define the Riemann curvature as
\begin{equation}
	R_{\mu \nu}{}^{\rho}{}_{\sigma} = 2 \partial_{[\mu}\Gamma^\rho_{\nu] \sigma} + 2 \Gamma^\rho_{[\mu| \lambda} \Gamma^\lambda_{\nu]\sigma} \,,
\end{equation}
which implies
\begin{equation}
	[\nabla_{\mu},\nabla_{\nu}]\xi^{\sigma}=R_{\mu\nu}{}^{\sigma}{}_{\rho} \,\xi^{\rho}\,.
\end{equation}
Notice that this choice for the Riemann implies that the Ricci scalar of a sphere is positive
\begin{equation}
	R_{\text{S}^n} > 0 \,.
\end{equation}
Given a $p$-form $\alpha$ on a $n$-dimensional Riemannian manifold $X_n$  with metric $g_{\mu\nu}$ 
\be
\alpha = \frac{1}{p!}\alpha_{\mu_1 \dots \mu_p} dx^{\mu_1} \wedge \cdots \wedge dx^{\mu_p} \,,
\ee
we use the conventions for the Hodge star operator
\be
\star \, \alpha = \frac{\sqrt{|g|} }{ p! (n-p)! } \epsilon_{\rho_1 \dots \rho_{n-p} \nu_{1} \dots \nu_p } g^{\nu_{1} \mu_{1}} \cdots g^{\nu_p \mu_p} \alpha_{\mu_1 \dots \mu_p} \, dx^{\rho_1} \wedge \cdots \wedge dx^{\rho_{n-p}} \,,
\ee
where $g = \text{det}(g_{\mu\nu})$ and $\epsilon$ is the Levi Civita symbol normalized as
\be 
\epsilon_{0\dots n-1} = 1 \,.
\ee
In our conventions we have
\be
\star^2 \, \alpha = s (-1)^{p(n-p)} \alpha \,, \qquad s = \text{sign}(g)\,.
\ee
The scalar product between forms is defined by 
\be
\alpha \cdot \beta =  \int_{X_n}  \star \, \alpha \wedge  \beta =  \frac{1}{p!}\alpha_{\mu_1 \dots \mu_p} \, \beta^{\mu_1 \dots \mu_p} \,,
\ee
where the indexes of $\beta$ are raised with the inverse metric. The codifferential $d^\dagger$ over $p$-forms is defined as the adjoint of the differential with respect to the scalar product we introduced, i.e. it satisfies $ d^\dagger \alpha \cdot \beta   =  \alpha \cdot d \beta$. In our conventions, the explicit expression of the codifferential acting on a $p$-form is
\be
d^\dagger \alpha = s (-1)^{np} \star \, d \, \star \, \alpha \,.
\ee
The Laplace-de Rham operator is 
\be
\Delta = d d^\dagger + d^\dagger d \,.
\ee
It is related with the Laplace-Beltrami operator $\nabla^2$ acting on scalars by
\be
\Delta = - \nabla^2  \,.
\ee

\subsubsection{Common bosonic sector action}

We use mostly minus signature $(+, - , \dots - )$.  The action for the common bosonic sector in our conventions is 
\begin{equation}
	S
	=
	\frac{1}{2\kappa_{10}^2}
	\int d^{10}x\sqrt{|{g}|}\, 
	e^{-2{\phi}}\, 
	\left\{
	{R} 
	+4(\partial{\phi})^{2}
	-\tfrac{1}{12}{H}^2
	\right\}\,,
\end{equation}
the string coupling constant is defined as $g_s = \left< e^{\phi} \right>$. The relation between $\alpha'$ and $\kappa_{10}$ is 
\begin{equation}
	2\kappa_{10}{}^2 = (2\pi)^7 \alpha'{}^4\,.
\end{equation}
and in our conventions the relation between $\alpha'$ and the string length $\ell_s$ is
\begin{equation}
	\ell_s = 2\pi\sqrt{\alpha'} \,.
\end{equation}
We use the following relation between 4d Planck units $M_p$ and string length $\ell_s$
\be
	\ell_s^2 	M_p ^2 = \eta \, { \text{Vol}_{X_6}} e^{-2\phi} = \eta \, e^{-2\phi_4} \,,  \qquad \eta = \begin{cases}
		1/2 \,, & \quad \text{chapter \ref{ch:5}} \,, \\
		1 \,, & \quad \text{chapter \ref{ch:6}} \,,
	\end{cases}
\ee
where $\phi$ is the 10d dilaton, $\text{Vol}_{X_6}$ is the volume of the internal manifold in string units and $\phi_4$ is the 4d dilaton. 

\subsubsection{Calabi--Yau conventions}

The Calabi-Yau structure of a threefold is completely characterized by two objects, an holomorphic $(3,0)$-form $\Omega_{CY}$ and a closed real kahler $(1,1)$-form $J_{CY}$. In our conventions they satisfy
\begin{equation}
	d\text{vol}_6 = -\frac{1}{6}J_{CY}\wedge J_{CY} \wedge J_{CY} \,, \quad \langle J_{CY},J_{CY} \rangle = 3 \,, \quad  * J_{CY} = -\frac{1}{2} J_{CY}\wedge J_{CY}\,, 
\end{equation}
which implies 
\begin{equation}
	\langle  J^2_{CY},J^2_{CY} \rangle  = 12 \,.
\end{equation}
The holomorphic three form is unique up to a constant rescaling. In our conventions we fix 
\begin{equation}
	\langle \Omega_{CY},\Omega_{CY} \rangle  = 8 \,,
\end{equation}
and hold
\begin{equation}
	\langle \text{Re}\Omega_{CY},\text{Re}\Omega_{CY} \rangle = 4\,, \quad \star  \Omega_{CY} = -i \Omega_{CY} \,, \quad d\text{vol}_6 = -\frac{i}{8}\Omega_{CY} \wedge \overline{\Omega}_{CY}\,.
\end{equation}

	\clearpage{\pagestyle{empty}\cleardoublepage} 
	
	\chapter{Details of black hole thermodynamics} \label{app:B}

\section{Motion in a Kaluza--Klein spacetime}
\label{eq:motioninKKspacetime}

To conclude, we study the geodesic motion of test particles in the
5-dimensional space, which is controlled by the equations

\begin{subequations}
	\label{eq:d5geodesicequations}
	\begin{align}
		\ddot{x}^{\hat{\mu}}
		+\hat{\Gamma}_{\hat{\nu}\hat{\rho}}{}^{\hat{\mu}}\dot{x}^{\hat{\nu}}
		\dot{x}^{\hat{\rho}}
		& =
		0\,,
		\\[2mm]
		\hat{g}_{\hat{\mu}\hat{\nu}}\dot{x}^{\hat{\mu}}\dot{x}^{\hat{\nu}}
		& =
		\alpha\,,
	\end{align}
\end{subequations}

\noindent
where $\alpha=0$ for massless particles and $\alpha=m^{2}$ for massive
particles. Rewriting these equations in terms of the 4-dimensional fields, we find
\begin{subequations}
	\label{eq:d4geodesicequations}
	\begin{align}
		\ddot{x}^{\mu}
		+\Gamma_{\nu\rho}{}^{\mu}\dot{x}^{\nu} \dot{x}^{\rho}
		-F_{\nu}{}^{\mu}\dot{x}^{\nu} k^{2}(\dot{z}+A_{\rho}\dot{x}^{\rho})
		-\tfrac{1}{2}\partial^{\mu}k^{2}(\dot{z}+A_{\rho}\dot{x}^{\rho})^{2}
		& =
		0\,,
		\\[2mm]
		g_{\mu\nu}\dot{x}^{\mu}\dot{x}^{\nu} -k^{2}(\dot{z}+A_{\rho}\dot{x}^{\rho})^{2}
		& =
		\alpha\,,
	\end{align}
\end{subequations}
plus the equation for $z(\xi)$. This equation is complicated but it is
entirely equivalent to the conservation of the momentum conjugate to $z$	
\begin{equation}
	P_{z}
	=
	-k^{2}(\dot{z}+A_{\rho}\dot{x}^{\rho})\,.
\end{equation}
Using this relation to eliminate $\dot{z}$ in
Eqs.~(\ref{eq:d4geodesicequations}), they take the form
\begin{subequations}
	\label{eq:d4geodesicequations2}
	\begin{align}
		\ddot{x}^{\mu}
		+\Gamma_{\nu\rho}{}^{\mu}\dot{x}^{\nu} \dot{x}^{\rho}
		& =
		P_{z}F^{\mu}{}_{\nu}\dot{x}^{\nu} 
		-\tfrac{1}{2}P_{z}^{2}\partial^{\mu}k^{-2}\,,
		\\[2mm]
		g_{\mu\nu}\dot{x}^{\mu}\dot{x}^{\nu} 
		& =
		\alpha +k^{-2}P_{z}^{2}\,,
	\end{align}
\end{subequations}
which are the equations of motion of a 4-dimensional particle with electric
charge $P_{z}$ and a spacetime-dependent effective mass squared
$\alpha+k^{-2}P_{z}^{2}$. The interaction with the scalar induces another	force term proportional to $P_{z}^{2}$. 	Eqs.~(\ref{eq:d5geodesicequations}) can be derived from the Polyakov-type action
\begin{equation}
	\label{eq:d5Polyakov}
	\hat{S}[e,x^{\hat{\mu}}]
	=
	-\tfrac{1}{2}\int d\xi
	\left\{e^{-1}\hat{g}_{\hat{\mu}\hat{\nu}}\dot{x}^{\hat{\mu}}\dot{x}^{\hat{\nu}}+e
	m^{2}\right\}\,.
\end{equation}
Rewriting  the action Eq.~(\ref{eq:d5Polyakov}) in terms of the 4-dimensional fields and performing a Legendre transformation to eliminate $z(\xi)$, one	arrives to 
\begin{equation}
	\label{eq:d4Polyakov}
	\hat{S}[e,x^{\mu}]
	=
	-\tfrac{1}{2}\int d\xi
	\left\{e^{-1}g_{\mu\nu}\dot{x}^{\mu}\dot{x}^{\nu}+e
	\left[m^{2}+k^{-2}P_{z}^{2}\right] +2P_{z} A_{\rho}\dot{x}^{\rho}\right\}\,.
\end{equation}
Eliminating $e$ replacing the (algebraic) solution to its equation of motion in the above action and using the fact that $P_{z}$ is constant, we get
\begin{equation}
	\label{eq:d4Nambu-Goto}
	\hat{S}[x^{\mu}]
	=
	-\int d\xi \sqrt{m^{2}+k^{-2}P_{z}^{2}}
	\sqrt{g_{\mu\nu}\dot{x}^{\mu}\dot{x}^{\nu}}
	-P_{z}\int d\xi A_{\rho}\dot{x}^{\rho}\,.
\end{equation}
The physical interpretation of this action is exactly the same as that of	Eqs.~(\ref{eq:d4geodesicequations2}), which, as expected, can be derived from Eq.~(\ref{eq:d4Polyakov}).	In the Einstein frame, the above action takes the form
\begin{equation}
	\label{eq:d4Nambu-Goto-Einsteinframe}
	\hat{S}[x^{\mu}]
	=
	-\int d\xi \sqrt{m^{2}(k/k_{\infty})^{-1} +k^{-3}q^{2}}
	\sqrt{g_{E\, \mu\nu}\dot{x}^{\mu}\dot{x}^{\nu}}
	-q\int d\xi A_{E\, \rho}\dot{x}^{\rho}\,,
\end{equation}
and it describes a particle of electric charge
\begin{equation}
	\label{eq:chargeversusmomentum}
	q
	=
	P_{z}k_{\infty}^{1/2}\,,
\end{equation}
and a position-dependent inertial mass that depends on the 5-dimensional mass and the charge and their couplings to the KK scalar.

\section{On the momentum maps at infinity} \label{sec-potentialsProve}

To show this we restrict to solutions corresponding to stationary and asymptotically flat BHs with no compact directions. We focus on theories with the structure of (\ref{eq:action-05}), but we work in generic dimensions $d$ and considering a $p+2$-form $F^\Sigma$.\footnote{The $p+2$-forms which admit a term with structure $F\wedge F$ are those which satisfy $2(p+2) = d$} In this setup we can expand the EOMs of $F^\Sigma$ in powers of the distance $r$ from the center of the BH. For large values of $r$, the metric will be the Minkowski one plus corrections suppressed by negative powers of $r$. The same applies for the scalars. Then, $F^\Sigma$ must satisfy asymptotically\footnote{We can have the second term only for $p+2$-forms whose rank is the self-dual one}
\begin{equation}
	d \left[I_{\Lambda\Sigma}(\phi_\infty) \star_{\text{mink}} F^\Sigma + R_{\Lambda \Sigma}(\phi_\infty) F^\Sigma  + \dots \right] = 0 \,.
\end{equation}
Combining this relation with the Bianchi identity we conclude that at leading order $F^\Sigma$ must be an harmonic of the Minkowski space. The same is true for $F_\Sigma$. In this configuration, the asymptotic surface $\Sigma_\infty^{d-2}$ is just a $d-2$-dimensional sphere $S^{d-2}_\infty$ which supports only two kind of harmonics: rank $d-2$ volume forms and scalar harmonics. The cases for which the integrals (\ref{eq:Integral}) and  (\ref{eq:Integral2}) do not vanish are, respectively, the $p=0$ and $p = d-4$ ones. The case $p=0$, $d=4$ is then the only one in which a $2$-form $F^\Sigma$ can have both the electrostatic and magnetostatic potentials non-vanishing at infinity. Let's study then the two cases. 

\subsubsection{Case 1: Electrostatic potential, $p =0$}
If $F^\Sigma$ is an harmonic form on Minkowski space we obtain that it must satisfy
\begin{equation}
	\Delta F^\Sigma = \nabla^\mu \nabla_\mu  F^\Sigma + \dots \,,
\end{equation}
where $\Delta$ is the Laplace--de Rahm operator. In Cartesian coordinates this equations is solved asking that the components $ F^\Sigma_{\mu\nu}$ are harmonic functions of Minkowski space. Restricting on time-independent harmonics, we obtain that the $ F^\Sigma_{\mu\nu}$ must be harmonics of the $d-1$-dimensional Euclidean space $\mathbb{E}^{d-1}$. This implies that in hyperspherical coordinates $(t,r,X^A)$ with $X^A = (\theta^a,\phi) $ and $\theta^a \in [0,\pi]$, $\phi \in [0,2\pi]$, $F_{\mu \nu}^\Sigma$ has form (we drop the index $\Sigma$)
\begin{equation}
	F_{\mu \nu} = \sum_k f^k_{\mu\nu}(r,X^A) H^k(r, X^A)
\end{equation}
where $H^k$ are harmonics of $\mathbb{E}^{d-1}$ and the $f^k_{\mu\nu}$ are coefficients one can easily determine performing the change of coordinates.\footnote{The index $k$ is not labeling the space of the harmonics. It labels the number of independent harmonics entering into a specific entries $F_{\mu \nu}^\Sigma$.} The main properties of the $f^k_{\mu\nu}$ are that they have a definite parity and a definite dependence in $r$. We have indeed
\begin{subequations}
	\begin{align}
		& f_{\mu\nu}^k(r,X^A) = r^q p_{\mu\nu}^k(X^A) \,, \qquad q = \# \text{indexes of angles} \,, \\[4mm]
		& p_{\mu\nu}^k(\text{P}X^A) = s_\mu s_\nu \, p_{\mu\nu}^k \,, \qquad s_\mu = \begin{cases}
			+1 \,, & \text{if} \, \,\mu = t,\theta^a \,, \\
			-1\ \,, & \text{if} \,\, \mu = r,\phi \,,
		\end{cases}
	\end{align}
\end{subequations}
where P is the parity operator that in hyperspherical coordinates acts as
\begin{equation}
	\text{P}: \qquad 	r  \rightarrow r \,, \qquad \theta^a \rightarrow \pi - \theta^a \,, \qquad \phi \rightarrow \phi + \pi \,.
\end{equation}
The harmonics of  $\mathbb{E}^{d-1}$ have also definite parity and power dependence
\begin{equation}
	\left(\frac{A}{r^{d-3+\ell}} + B r^\ell\right) Y^m_\ell (X^A) \,, \qquad \text{P}: Y^m_\ell \rightarrow (-1)^\ell Y^m_\ell \,,
\end{equation}
where  $Y^m_\ell (X^A)$ are the spherical harmonics for a $d-2$-dimensional sphere. $A$ and $B$ are coefficients are fixed by orthonormalization. We will only need harmonics with $B=0$. The electric charge of $F^\Sigma$ is finite provided that $F^\Sigma_{tr} \sim \mathcal{O}({r^{2-d}})$ which implies
\begin{equation}
	F_{tr} = \sum_{k,m} p_{tr}^k(X^A) \, c^{km}_{tr}\, \frac{Y_1^m(X^A)}{r^{d-2}} + \dots \,.
\end{equation}
where $c^{km}$ is a constant which is weighting the contribution of the different harmonics. Assuming that $F_{AB}^\Sigma \sim \mathcal{O}(r^0)$ we obtain 
\begin{equation}
	F_{AB} = \sum_{k,m} p_{AB}^k(X^A)\,c^{km}_{AB} \, \begin{cases} Y_1^m(X^A) + \dots \,, & \qquad d= 4 \,, \\
		Y_0^m(X^A) + \dots \,, &  \qquad d= 5 \,, \\
		\mathcal{O}(r^{5-d}) + \dots \,, & \qquad d > 5 \,. 
	\end{cases} 
\end{equation} 
The leading harmonics of $F_{iA}^\Sigma$ with $i = t,r$ are fixed by the previous choices and are
\begin{equation}
	F_{iA} = \sum_{k,m} p_{iA}^k(X^A) \, c^{km}_{iA} \, \begin{cases} Y_1^m(X^A) r^{-1} + \dots \,, & \qquad d= 4 \,, \\
		Y_0^m(X^A) r^{-1}  + \dots \,, &  \qquad d= 5 \,, \\
		\mathcal{O}(r^{4-d}) + \dots \,, & \qquad d > 5 \,. 
	\end{cases} 
\end{equation}
In particular, the $c^{km}_{iA}$ depend on $c^{km}_{AB}$ and $c^{km}_{tr}$.  If we further require that also $A_\mu^\Sigma \sim \mathcal{O}(r^0)$ we obtain that to $F^\Sigma_{rA} $ must go to zero faster than $1/r$, otherwise $ A^\Sigma_\mu$ develops a logarithmic divergence. In the $d=5$ case the only way one can achieve this is setting $c^{km}_{rA} = 0$, which imply that also the $c^{km}_{AB}$ are vanishing. In the $d=4$ we can have solution with non vanishing $c^{km}_{rA}$ and $c^{km}_{AB}$. Therefore, for the standard stationary BHs with horizon generated by Killing vector $k = \partial_t + \Omega \partial_\phi$ we have that asymptotically $\iota_k F^\Sigma$ is vanishing for $d>4$ and $P^\Sigma_k$ is closed. For $d=4$ we have instead
\begin{equation}
	\iota_k F^\Sigma =  \Omega \left( F_{\phi a} \, d\theta^a  \right) + \mathcal{O}(1/r^2)
\end{equation}
We obtain then the expansion
\begin{equation} \label{eq:expansionPk}
	P_k^\Sigma(\theta^a) = \Phi_\infty^\Sigma + \delta_{d,4} P_{\phi\, k}^\Sigma(\theta^a) + \dots 
\end{equation}
with
\begin{equation}
	d P_{\phi\, k}^\Sigma  = - F_{\phi a}^\Sigma \, d\theta^a \,.
\end{equation}
Notice that such a $P_{\phi\, k}^\Sigma$ must exist because the sphere does not support harmonic 1-forms. Therefore, $P_k^\Sigma(\theta^a)$ can have a coexact part $P_{\phi\, k}^\Sigma$  in $d=4$. However, such term does not contribute to the integral (\ref{eq:Integral}). It is simple to verify that all the components of $F^\Sigma$ are eigenfuctions of the parity operator P, i.e. transform with a definite sign. In particular, $F_{\phi a}$ is even under parity, which implies that $P_{\phi\, k}^\Sigma$ is odd
\begin{equation}
	P_{\phi\, k}^\Sigma(\text{P}\theta^a) = - P_{\phi\, k}^\Sigma(\theta^a)
\end{equation}	
The pullback of $F_\Sigma$ on $S_\infty^{d-2}$ also is an eigenform of the parity operator. Therefore, the pullbacks of $F_\Sigma$ and $F_\Sigma P_{\phi\, k}^\Sigma$ transform with opposite signs. Given that the volume form of $S_\infty^{d-2}$ is also an eigenform of the parity operator, we can conclude that if $q_\Sigma$ is not vanishing, the integral of  $F_\Sigma P_{\phi\, k}^\Sigma$ must vanish. We obtain
\begin{equation} 
	\int_{\Sigma^{d-2}} F_\Sigma \,  P_k^\Sigma  \,\,\,  \stackrel{\infty}{=}  \,\,\,  \Phi^\Sigma_{\infty} \, \, q_\Sigma\,.
\end{equation}
with $\Phi_\infty^\Sigma$ defined in (\ref{eq:expansionPk}).

\subsubsection{Case 2: Magnetostatic potential, $p = d-4$}

We consider the $d-2$ form $F^\Sigma$. The magnetic charge is well defined provided that the angular part of $F^\Sigma$ is at most constant in $r$ asymptotically. We have the expansion
\begin{equation}
	F_{A_1 \dots A_{d-2}} = \sum_{k,m} p^k_{A_1 \dots A_{d-2}}(X^A) \, c^{km}_{A_1 \dots A_{d-2}} \, Y_1^m (X^A)  + \dots \,.
\end{equation}
Imposing that the component $F_{trA_1 \dots A_{d-2}}$ is at most growing as $r^{-2}$ (in order to avoid logarithmic divergences in the gauge potential) we obtain
\begin{equation}
	F_{trA_1 \dots A_{d-4}} = \sum_{k,m} p^k_{trA_1 \dots A_{d-4}}(X^A) \, c^{km}_{trA_1 \dots A_{d-4}} \, \frac{Y_1^m (X^A)}{r^2}+ \dots \,.
\end{equation}
The harmonics of the remaining components are fixed to be
\begin{equation}
	F_{i A_1 \dots A_{d-3}} = \sum_{k,m} p^k_{iA_1 \dots A_{d-3}}(X^A) \, c^{km}_{iA_1 \dots A_{d-3}} \, \frac{Y_1^m(X^A)}{r}+ \dots
\end{equation}
Again the coefficients $c^{km}_{iA_1 \dots A_{d-3}}$ are combinations of $ c^{km}_{A_1 \dots A_{d-2}}$ and $c^{km}_{trA_1 \dots A_{d-4}}$ and they admit solutions with $F_{rA_1\dots A_{d-3}}^\Sigma \sim \mathcal{O}(r^{-2})$ and the same leading behavior for the other components of $F^\Sigma$. Notice that the expressions we found reduce to those of the previous section for $d=4$. 
For $d>4$ it is easy to verify that $\iota_k F_\Sigma$ vanishes asymptotically fast enough to make the momentum map $P_k{\,}_\Sigma$ closed and avoid logarithmic singularities in the dual gauge potential. Instead, for $d=4$,  $\iota_k F_\Sigma$ does not vanish and with our ansatz the dual gauge potential is singular. Again, we can restrict to solutions with non divergent for certain choices of the the $c^{km}_{iA}$ such that $F_{tA}^\Sigma \sim \mathcal{O}(r^{-2})$. We obtain 
\begin{equation}
	\iota_k F_\Sigma =\Omega F_\Sigma{\,}_{\phi a} \, d \theta^a  + \mathcal{O}(1/r^2) \,.
\end{equation}
Introducing $P_\Sigma^\phi{\,}_k$ such that
\begin{equation}
	d P_\Sigma^\phi{\,}_k = -  F_\Sigma{\,}_{\phi a} \, d \theta^a \,,
\end{equation}
we can expand asymptotically the momentum map $P_\Sigma{\,}_k$ as
\begin{equation} \label{eq:expansionPkdual}
	P_\Sigma{\,}_k(\theta^a) = \Phi_\infty{\,}_\Sigma + P_\Sigma^\phi{\,}_k(\theta^a) + \dots \,.
\end{equation}
It is simple to verify that $F_{AB}{\,}_{\Sigma}$ and $F_{AB}^\Sigma$ are even under parity. Therefore, $P_\Sigma^\phi{\,}_k$ must be odd
\begin{equation}
	P_\Sigma^\phi{\,}_k(\text{P}\theta^a) = - P_\Sigma^\phi{\,}_k(\theta^a) \,,
\end{equation}
and the integral of $F_{AB}^\Sigma \, P_\Sigma^\phi{\,}_k$ must vanish. Going back to $d\ge 4$, we got
\begin{equation}
	\int_{\Sigma^{d-2}} F^\Sigma \wedge P_k{}_{\,\Sigma}  \,\,\,  \stackrel{\infty}{=}   \,\,\,  \Phi_\Sigma{\,}_{\infty} \, \, q^\Sigma \,.
\end{equation}
with $\Phi_\infty^\Sigma$ defined in (\ref{eq:expansionPkdual}).

\section{Modified Einstein normalization} \label{sec-Einsteinnormalization}

If we dimensionally reduce the zeroth order HST effective action on a torus T$^n$ we obtain in the string frame
\begin{equation}
	\begin{split}
		S = & \; \frac{g_s^{(10-n)}{}^2}{16 \pi G_N^{(10-n)}} \int dx^{10-n} \sqrt{|g|}e^{-2\phi} \bigg\{R-4(\partial\phi)^2 - \frac{1}{8}\text{Tr}\left(\partial^a M \partial_a M^{-1}\right) \\
		& \quad   -\frac{1}{4} \mathcal{F}^T M^{-1} \cdot \mathcal{F} + \frac{1}{12} H^2 \bigg\} \,,
	\end{split}
\end{equation}
where we combined the KK vectors fieldstrengths $F^m = d A^m $ and the winding vectors fieldstrengths $G_m = d C_m$ in the vector $\mathcal{F}$ and the scalars $K_{mn}$ and $G_{mn}$ into the matrix of scalar fields $M$ 
\begin{equation}
	\mathcal{F} = \begin{pmatrix} F^m \\ G_m \end{pmatrix} \,, \qquad M^{-1} = \begin{pmatrix}
		G + K^T G^{-1} K & - K^T G^{-1} \\
		- G^{-1} K & G^{-1} 
	\end{pmatrix} \,.
\end{equation}
In order to go to the \textit{modified Einstein frame} we perform the rescaling (we introduce $d = 10-n$)
\begin{equation}
	g_{\mu\nu} = \Omega^2 g_E{}_{\mu\nu} \,, \qquad \Omega = e^{2 (\phi-\phi_\infty) /(d-2)} \,.
\end{equation}
In order to set the \textit{modified Einstein normalization} for the gauge fields we have to absorb the explicit dependencies on $g_s$
\begin{equation}
	\mathcal{F}_E = \mathcal{F} g_s^{2/(d-2)} \,, \qquad H_E = H g_s^{4/(d-2)} \,.
\end{equation}
The action becomes
\begin{equation}
	\begin{split}
		S = & \; \frac{1}{16 \pi G_N^{(d)}} \int dx^{d} \sqrt{|g_E|} \bigg\{R_E + \frac{4}{(d-2)}(\partial\phi)^2 - \frac{1}{8}\text{Tr}\left(\partial^a M \partial_a M^{-1}\right)  \\
		& \quad    - \frac{1}{4} e^{- \frac{4}{(d-2)} \phi} \, \mathcal{F}^T_E M^{-1} \cdot \mathcal{F}_E  + \frac{1}{12}  e^{-\frac{8}{(d-2)} \phi} H^2_E  \bigg\} \,.
	\end{split}
\end{equation}
Finally, notice that if $K_{mn}$ is vanishing and $G_{mn}$ is diagonal we can write the scalar sector as 
\begin{equation}
	\begin{split}
		\frac{1}{16 \pi G_N^{(d)}} \int dx^{d} \sqrt{|g_E|} \left[\frac{1}{2} g_{xy} \partial_\mu \varphi^x \partial^\mu \varphi^y \right]   \,,
	\end{split}
\end{equation}
with $(\varphi^x) = (\phi, \varphi^m )$ and $\varphi^m = \log(|G_{mm}|^{1/2})$. The non-vanishing components of the scalar metric $g_{xy}$ are 
\begin{equation}
	g_{\phi\phi} = \frac{8}{d-2} \,, \qquad g_{mm} = 2 \,.
\end{equation}
	\clearpage{\pagestyle{empty}\cleardoublepage} 
	
	\chapter{Details of black hole solutions constructions} \label{app:C}

\section{Relations among $10$-dimensional and $(10-n)$-dimensional fields} \label{sec-dictionary}
We start recalling the relations between the 10-dimensional fields (indicated with an hat) and the $(10-n)$-dimensional fields obtained via a $\text{T}^n$ dimensional reduction at first order in $\alpha'$. They are essentially those of \cite{Ortin:2020xdm}. Notice that we consistently truncated the YM fields. We decompose the indexes as $\hat{\mu} = (\mu, m)$, with $m = 1, \dots n$. We obtain
\begin{subequations}
	\begin{align}
		{g}_{\mu\nu} & = \hat{g}_{\mu\nu}-\hat{g}^{mn}\hat{g}_{\mu m}\hat{g}_{\nu n} \,,\\[2mm]
		A^n{}_\mu & = \hat{g}^{nm} \hat{g}_{m\mu}  \,, \\[2mm]
		G_{mn} & = - \hat{g}_{mn}  \,, \\[0mm]
		{ \phi} & = \hat{\phi} - \frac{1}{2}\log \det ( \hat{g}_{mn} ) \,, \\
		{B}^{(1)}{}_{\mu\nu} & = \hat{B}_{\mu\nu} + \hat{g}^{mn} \hat{g}_{m[\mu}\hat{B}_{\nu]n} - \frac{\alpha'}{4}\left(\hat{\Omega}_{(-)}^{(0)}{}_m{}^{\hat{a}}{}_{\hat{b}} \hat{\Omega}_{(-)}^{(0)}{}_{[\mu|}{}^{\hat{b}}{}_{\hat{a}}\right) \hat{g}^{mn} \hat{g}_{|\nu]n} \,, \\
		\begin{split}
			{C}^{(1)}{}_{m \mu}  &= \hat{B}_{\mu m} + \left[\hat{B}_{mn}- \frac{\alpha'}{4}\left(\hat{\Omega}_{(-)}^{(0)}{}_{m}{}^{\hat{a}}{}_{\hat{b}} \hat{\Omega}_{(-)}^{(0)}{}_{n}{}^{\hat{b}}{}_{\hat{a}}\right)\right] \hat{g}^{np}\hat{g}_{p \mu}  \\[2mm]
			\qquad & \qquad - \frac{\alpha'}{4} \left(\hat{\Omega}_{(-)}^{(0)}{}_m{}^{\hat{a}}{}_{\hat{b}} \hat{\Omega}_{(-)}^{(0)}{}_{\mu}{}^{\hat{b}}{}_{\hat{a}}\right) \,,
		\end{split} \\[2mm]
		{K}^{(1)}{}_{m n} & = \hat{B}_{mn} - \frac{\alpha'}{4}\left(\hat{\Omega}_{(-)}^{(0)}{}_{[m|}{}^{\hat{a}}{}_{\hat{b}} \hat{\Omega}_{(-)}^{(0)}{}_{|n]}{}^{\hat{b}}{}_{\hat{a}}\right)  \,. 
	\end{align} 
\end{subequations}
where $G_{mn}$ and $ {K}^{(1)}{}_{m n}$ are matrices of scalars, $A^n{} $ and $C^{(1)}_{m}$ are gauge vectors. $g_{\mu\nu}$, $B^{(1)}$ and $\phi$ are the lower dimensional metric, KR 2-form and dilaton. $g^{mn}$ is the inverse of $g_{mn}$.

In the setups considered in this work $K^{(1)}_{mn}=0$ and $G_{mn}$ is diagonal. In the most general case we have two independent $S^1$ circles parametrized by $w$ and $z$. Omitting the indexes of the trivial $\text{T}^{n-2}$ compactification, $G_{mn}$ has the explicit form
\begin{equation}
	G_{mn} = \begin{pmatrix}
		G_{ww} &  0 \\ 0 & G_{zz}  
	\end{pmatrix} \equiv  \begin{pmatrix}
		\ell^2 &  0 \\ 0 & k^2
	\end{pmatrix} \,.
\end{equation}
In order to study T-duality transformations along the $w$ and $z$ direction, following \cite{Elgood:2020xwu} it is useful to introduce the combinations 
\begin{subequations}\label{eqdefl1k1}
	\begin{align}
		& \ell^{(1)} = \ell\left[1 - \frac{\alpha'}{4}\left(\hat{\Omega}_{(-)}^{(0)}{}_{w}{}^{\hat{a}}{}_{\hat{b}} \hat{\Omega}_{(-)}^{(0)}{}_{w}{}^{\hat{b}}{}_{\hat{a}}\right)\hat{g}^{ww}\right] \,,  \\
		& k^{(1)} = k\left[1 - \frac{\alpha'}{4}\left(\hat{\Omega}_{(-)}^{(0)}{}_z{}^{\hat{a}}{}_{\hat{b}} \hat{\Omega}_{(-)}^{(0)}{}_{z}{}^{\hat{b}}{}_{\hat{a}}\right)\hat{g}^{zz}\right] \,.
	\end{align}
\end{subequations}

\section{Multi-center ansatz with $\beta_0 = \beta_{\mathcal{H}}$} \label{sec-curvature}

We consider the ansatz (\ref{eq:4dansatzLeading}) with generic functions $\mathcal{Z}_+$, $\mathcal{Z}_-$, $\mathcal{Z}_0$ of the 4-dimensional hyper-K\"ahler space with metric $d\sigma^2$. If the hyper-K\"ahler space is a 4-dimensional Gibbons--Hawking space we have explicitly
\begin{equation} \label{eq:ansatzsigma}
	d \sigma^2 = \mathcal{Z}_\mathcal{H} d\vec{x}_{(3)}^2 + \ell_{\infty }^2 \mathcal{Z}_\mathcal{H}^{-1} \bigg[dw + \ell_{\infty }^{-1}\beta_{\mathcal{H}} \, \chi \bigg]^2 \,,
\end{equation}
where $\vec{x}_{(3)}$ are coordinates of a flat 3-dimensional Euclidean space $\mathbb{E}^3$ and $\chi$ is a 1-form satisfying
\begin{equation} \label{eq:Constr}
	d \chi = \star_{(3)} d \mathcal{Z}_\mathcal{H}\,.
\end{equation}
Notice that (\ref{eq:Constr}) implies that $ \mathcal{Z}_\mathcal{H}$ is an harmonic function of $\mathbb{E}^3$ and $\chi$ is a 1-form of $\mathbb{E}^3$. In the limit  $\mathcal{Z}_{\mathcal{H}} = 1$, $\chi$ is closed and can be set to zero with a proper redefinition of $w$. We recover then $\mathbb{E}^4$
\begin{equation}
	d \sigma^2 =  d\vec{x}_{(4)}^2,
\end{equation}	
where $\vec{x}_{(4)}$ are the coordinates of the flat 4-dimensional Euclidean space. 

We start absorbing some of the $\beta_i$ factors of (\ref{eq:4dansatzLeading}) considering the map $z\rightarrow z \beta_+$ and $w \rightarrow w \beta_0$. We obtain 
\begin{subequations}
	\begin{align}
		d\hat{s}^{2}
		& =
		\frac{1}{\mathcal{Z}_{+}\mathcal{Z}_{-}}Wdt^{2}
		-\mathcal{Z}_{0} d\sigma^2
		\nonumber \\[1mm]
		& 
		-\frac{k_{\infty}^{2}\mathcal{Z}_{+}}{\mathcal{Z}_{-}}
		\left[dz+k_{\infty}^{-1}
		\left(\mathcal{Z}^{-1}_{+}-1\right)dt\right]^{2}
		-dy^{\tilde{m}}dy^{\tilde{m}} \,,
		\\[2mm]
		\hat{H}^{(0)}
		& = 
		\beta_+ \beta_- \, d\left[ k_{\infty}\left(\mathcal{Z}^{-1}_{-}-1\right)
		dt \wedge dz\right]
		+\star_\sigma d \mathcal{Z}_0 \,,
	\end{align}
\end{subequations}
with 
\begin{equation} 
	d \sigma^2 = \mathcal{Z}_\mathcal{H} d\vec{x}_{(3)}^2 + \ell_{\infty }^2 \mathcal{Z}_\mathcal{H}^{-1} \bigg[dw + \ell_{\infty }^{-1}\beta_0 \beta_{\mathcal{H}} \, \chi \bigg]^2 \,.
\end{equation}
Notice that the factor $\beta_0 \beta_{\mathcal{H}}$ cannot be absorbed by a change of orientation of $\mathbb{E}^3$. In the following we assume then
\begin{equation}
	\beta_0 \beta_{\mathcal{H}} = 1 \,, \qquad \beta_+ \beta_- = \frac{1 + 2\epsilon}{2}\,,
\end{equation}
with $\epsilon = 1,0$. In particular $\epsilon = 1$ correspond to the supersymmetric case. Notice that in the $\mathbb{E}^4$ limit the value of $\beta_0$ is irrelevant. We change coordinates introducing the coordinate $u = t - k_\infty z$. We then consider the vielbeins 
\begin{equation} \label{eq:vierbeinBasis}
	\hat{e}^{+}=\frac{du}{\mathcal{Z}_{-}}\,,
	\hspace{.5cm}
	\hat{e}^{-}=dt-\frac{\mathcal{Z}_{+}}{2}\,du\,,
	\hspace{.5cm}
	\hat{e}^{m}=\mathcal{Z}_{0}^{1/2}\,v^{m}\,,
\end{equation} 
such that $d\hat{s}^2 = 2 \hat{e}^+ \hat{e}^- - \hat{e}^m \hat{e}^n \delta_{mn}$ and $v^{m}$ is a vierbein of the 4-dimensional space with metric
$d\sigma^{2}=v^{m}v^{n}\delta_{mn}$. In the following we will use $\partial_m$ to indicate the partial derivative with respect to the flat indexes of the hyper-K\"ahler vielbeins $v^m$, i.e. we use
\begin{equation}
	\partial_m = v_m{}^{\underline{m}}\partial_{\underline{m}}\,.
\end{equation}
The components of the torsionful spin connection
\begin{equation}
	\hat{\Omega}_{(-)}{}_{\hat{a} \hat{b}}=\hat{\omega}_{\hat{a}\hat{b}}
	-\frac{1}{2}\hat{H}^{(0)}_{\hat{c} \hat{a} \hat{b}}\,\hat{e}^{\hat{c}}
\end{equation}
are given	by
\begin{subequations}
	\begin{align}
		\hat{\Omega}_{(-)\, +-}
		& =
		\frac{\epsilon}{\mathcal{Z}_{0}^{1/2}}\partial_{m}\log{\mathcal{Z}_{-}}
		\hat{e}^{m}\,,
		\\[2mm]
		\hat{\Omega}_{(-)\, -m}
		& =
		\frac{\epsilon}{\mathcal{Z}_{0}^{1/2}}\partial_{m}\log{\mathcal{Z}_{-}}
		\hat{e}^{+}\,,
		\\[2mm]
		\hat{\Omega}_{(-)\, +m}
		& =
		\frac{\mathcal{Z}_{-}}{2\mathcal{Z}_{0}^{1/2}}\partial_{m}\mathcal{Z}_{+} \hat{e}^{+}
		+\frac{1-\epsilon}{\mathcal{Z}_{0}^{1/2}}\partial_{m}\log{\mathcal{Z}_{-}}
		\hat{e}^{-}\,,
		\\[2mm]
		\hat{\Omega}_{(-)\, mn}
		& =
		\left[{\tilde{\omega}}_{pmn}+\mathbb{M}^{-}_{mnpq}\partial_{q}\log{\mathcal{Z}_{0}}\right]
		\frac{\hat {e}^{p}}{\mathcal{Z}_{0}^{1/2}}\,,
	\end{align}
\end{subequations} 

\noindent
where $\tilde{\omega}_{mn}$ is the hyper-K\"ahler spin connection\footnote{It
	is defined through $dv^{m}+\tilde{\omega}^{m}{}_{n}\wedge v^{n}=0$.} and
$\mathbb{M}^{-}_{mnpq}=\delta_{m[p} \delta_{q]n}-\frac{1}{2}\epsilon_{mnpq}$
are the anti-self-dual $\mathfrak{so}(3)$ subalgebra of of $\mathfrak{so}(4)$.
The non-vanishing components of the curvature 2-form are

\begin{subequations}
	\label{eq:curvature2form}
	\begin{align} 
		\hat{R}_{(-)\,-m} & =
		\frac{\hat{e}^{n}\wedge\hat{e}^{+}}{\mathcal{Z}_{0}}
		\epsilon\left[\nabla_{n} \partial_{m} \log \mathcal{Z}_{-}
		-\frac{1}{2}\partial_{m} \log \mathcal{Z}_{-}\partial_{n}\log{\mathcal{Z}_{0}}
		-{\mathbb
			M}^{-}_{pmnq}\partial_{p}\log{\mathcal{Z}_{-}}\partial_{q}\log{\mathcal{Z}_{0}}\right]\,,
		\\[2mm]
		\hat{R}_{(-)}{}_{+m}
		& =
		\frac{\mathcal{Z}_{-}\hat{e}^{n}\wedge
			\hat{e}^{+}}{2\mathcal{Z}_{0}}
		\left[\nabla_{n} \partial_{m}
		\mathcal{Z}_{+}-\frac{1}{2}\partial_{n}\log \mathcal{Z}_{0} \partial_{m}
		\mathcal{Z}_{+}+(\epsilon-1)\partial_{m}\log \mathcal{Z}_{-} \partial_{n}
		\mathcal{Z}_{0}\right.
		\nonumber \\[1mm]
		&\hspace{.5cm}
		\left.
		-\epsilon\partial_{m}
		\mathcal{Z}_{+}\partial_{n}\log{\mathcal{Z}_{-}}\right]
		+(1-\epsilon)\frac{e^{n}\wedge e^{-}}{\mathcal{Z}_{0}}
		\left[\nabla_{n}\partial_{m} \log{\mathcal{Z}_{-}}\right.
		\nonumber \\[1mm]
		& \hspace{.5cm}
		\left.
		-\frac{1}{2}\partial_{m}\log{\mathcal{Z}_{-}}
		\partial_{n} \log{\mathcal{Z}_{0}}
		-\mathbb{M}^{-}_{pmnq}\partial_{p}\log{\mathcal{Z}_{-}}\partial_{q}\log{\mathcal{Z}_{0}}\right]\,,\\[2mm]
		\hat{R}_{(-)}{}_{mn}
		& =
		\tilde{R}_{mn}+\tilde{F}_{mn}\,,
	\end{align}
\end{subequations} 

\noindent
where

\begin{subequations}
	\begin{align} 
		\tilde{R}_{mn}
		& =
		d\tilde{\omega}_{mn}+\tilde{\omega}_{mp}\wedge \tilde{\omega}{}_{pn}\,,
		\\[2mm]
		\tilde{F}_{mn}
		&=
		d\tilde{A}_{mn}+\tilde{A}_{mp}\wedge \tilde{A}{}_{pn}\,,
	\end{align}
\end{subequations} 

\noindent
and

\begin{equation}
	\tilde{A}_{mn}=\mathbb{M}^{-}_{mnpq}\partial_{q}\log{\mathcal{Z}_{0}} \,v^{p}\,.
\end{equation} 
When computing the $mn$ components of the curvature 2-form,
Eq.~(\ref{eq:curvature2form}) it is crucial that the spin connection
$\tilde{\omega}_{mn}$ and the connection $\tilde{A}_{mn}$ satisfy opposite
self-duality relations,

\begin{equation}
	\tilde{\omega}_{mn}
	=
	+\frac{1}{2}\epsilon_{mnpq}\,\tilde{\omega}_{pq}\,,
	\hspace{1cm}
	\tilde{A}_{mn}
	=
	-\frac{1}{2}\epsilon_{mnpq}\tilde{A}_{pq}\,,
\end{equation} 

\noindent
\textit{i.e.}, each of these connections belongs to one of the two orthogonal
subspaces, $\mathfrak{so}_{\pm}(3)$, in which
$\mathfrak{so}(4)=\mathfrak{so}_{+}(3)\oplus\mathfrak{so}_{-}(3)$ splits \cite{Chimento:2018kop, Ortin:2021win}.

\section{T-duality constraints}\label{sec-tduality}

The $0^{\rm th}$-order solution, understood as a family, is invariant under the $0^{\rm th}$-order Buscher T~duality transformations \cite{Buscher:1987sk,Buscher:1987qj}: one member of the family is transformed into another with different values of the parameters. 
Explicitly we have 
\begin{subequations} \label{eqmapTdual}
	\begin{align}
		&T_z: \qquad 	q_+ \leftrightarrow q_- \,, \qquad \beta_+ \leftrightarrow \beta_- \,, \qquad k_\infty \leftrightarrow 1/k_\infty \,,\\
		&T_w: \qquad 	q_\mathcal{H}\leftrightarrow q_0 \,,  \qquad \beta_{\mathcal{H}} \leftrightarrow \beta_0 \,, \qquad \ell_\infty \leftrightarrow 1/\ell_\infty  \,,
	\end{align}
\end{subequations}
where $T_z$ is the operator implementing T-duality along the circle $S^1_z$ and $T_w$ is the operator implementing T-duality along the circle $S^1_w$.
The physical interpretation of the transformations of the parameters generated by $T_z$ is the expected one (see later for the relation among the $q_i$ and the actual physical charges): momentum and winding are interchanged and the radius of the compact dimension is inverted. The same holds for $T_w$, but now the quantities exchanged are the charges associated to the NS5 branes and the Kaluza--Klein monopoles. In order to better  appreciate this, we can study T-duality in the dimensionally-reduced solution. Despite the fact that we are dimensionally reducing 10-dimensional HST on a 6-dimensional torus, the compactification is trivial in 4 directions and the remaining ones, parameterized by $z$ and $w$, are simply 2 independent 1-dimensional reductions. Therefore, T-duality for each circle takes the form of a discrete symmetry transformation of the action \cite{Bergshoeff:1994dg} that, essentially, interchanges two vector fields and inverts a scalar. 
At $0^{\rm th}$ order, these transformations are given by
\begin{subequations}\label{eq:Tdualityfirstorder}
	\begin{align}
		&T_z: \qquad 	C_z \leftrightarrow A^z \,, \qquad k \leftrightarrow 1/k \,,\\
		&T_w: \qquad 	C_w \leftrightarrow A^w \,, \qquad \ell \leftrightarrow 1/\ell  \,,
	\end{align}
\end{subequations}
and the remaining ones must transform as follow
\begin{equation}\label{eqtdual1}
	T_{z,w}:\qquad	ds^2_{E} \leftrightarrow ds^2_{E} \,, \qquad e^{-2\phi}\leftrightarrow e^{-2\phi} \,,
\end{equation}
It is reasonable to expect that the T~duality invariance of the family of
solutions is preserved at first order in $\alpha'$ when the
$\alpha'$-corrected Buscher T~duality transformations derived in
Refs.~\cite{Bergshoeff:1995cg,Elgood:2020xwu} are used because we are just
dealing with a more accurate description of exactly the same physical
system. At first order in $\alpha'$ the trivial transformations (\ref{eqtdual1}) have the same form. The non-trivial ones (\ref{eq:Tdualityfirstorder}) are slightly modified 
\begin{subequations}\label{eqtdual2}
	\begin{align}
		&T_z: \qquad 	C^{(1)}_z \leftrightarrow A^z \,, \qquad k \leftrightarrow 1/k^{(1)} \,,\\
		&T_w: \qquad 	C^{(1)}_w \leftrightarrow A^w \,, \qquad \ell \leftrightarrow 1/\ell^{(1)}  \,.
	\end{align}
\end{subequations}
The main difference with the $0^{\rm th}$-order ones is that the relation between the vectors $C^{(1)}_m$ and the higher-dimensional fields is modified by	$\alpha'$ corrections (cfr. equation \ref{eqvecC4d}). The scalar combinations $k^{(1)}$ and $\ell^{(1)}$ contain $\alpha'$ corrections as well (cfr. equation (\ref{eq:scalarcombination})). Then, the first-order solution can
only be self-dual if the $\alpha'$ corrections one finds for $A^m$, $k$ and $\ell$ are	related to the explicit $\alpha'$ corrections of $C^{(1)}_m$, $k^{(1)}$ and $\ell^{(1)}$ in a very specific way. Thus, the expected T~duality invariance of the $\alpha'$-corrected solution can be used to simplify the problem of finding the corrections and also to test them.

We follow the same logic of \cite{Cano:2021nzo, Zatti:2023oiq}.  
It has been verified in several solutions \cite{Cano:2021nzo,Ortin:2021win,Cano:2022tmn, Zatti:2023oiq} that whenever the $\alpha'$ corrections do not modify the asymptotic values of the fields and the form of the poles of the functions relevant for the definitions of the physical charges (in this case, $\delta \mathcal{Z}_{h0}$, $\delta\mathcal{Z}_{h-}$ and $\delta\mathcal{Z}_+$), the zeroth order map between parameters (\ref{eqmapTdual}) is valid at first order too. In order to obtain non-trivial constraints we assume that this is true in this case too.\footnote{Notice that this assumption cannot generate any loophole in the final result, which is building the BH solutions. The solution is a solution independently of the path we followed to obtain it. Anyway, one can verify that for our choice of coordinates and mechanism to fix the integration constants, the solutions we get fulfill this property.}

We impose now that our ansatz (\ref{ansatz4d}) satisfies the transformation rules (\ref{eqtdual1}) and (\ref{eqtdual2}). Imposing the invariance of the Einstein metric and the dilaton we obtain the invariance of the string frame metric. From the invariance of the string frame metric we obtain
\begin{equation}\label{eqtdualds}
	T_m : \quad \frac{W_{tt}}{\mathcal{Z}_+\mathcal{Z}_-} \leftrightarrow  \frac{W_{tt}}{\mathcal{Z}_+\mathcal{Z}_-} \,, \quad  \mathcal{Z}_0\mathcal{Z}_{\mathcal{H}} \leftrightarrow  \mathcal{Z}_0\mathcal{Z}_{\mathcal{H}} \,, \quad W_{rr} \leftrightarrow W_{rr}\,.
\end{equation}
Combing these relations with the invariance of the dilaton we further get
\begin{equation}\label{eqtdualdila}
	T_m: \quad \frac{c_\phi}{r^2 \mathcal{Z}_{h-}'} W_{tt} \left(\frac{\mathcal{Z}_{h-}}{\mathcal{Z}_{-}}\right)^2\leftrightarrow   \frac{c_\phi}{r^2 \mathcal{Z}_{h-}'} W_{tt} \left(\frac{\mathcal{Z}_{h-}}{\mathcal{Z}_{-}}\right)^2 \,.
\end{equation}
From the transformation property of $A^z$  we obtain 
\begin{equation} \label{eqtdualzp}
	T_z[\mathcal{Z}_+] = \mathcal{Z}_{h-}\left(1-\alpha' \frac{\Delta_C}{\beta_-}\right)\,,
\end{equation}
where we used the transformation properties of $\beta_\pm$ of equation (\ref{eqmapTdual}). Using the invariance of $\Delta_C$ under $T_z$ we can invert the relation and obtain 
\begin{equation}\label{eqtdualzmh}
	T_z[\mathcal{Z}_{h-}] = \mathcal{Z}_{+}\left(1+\alpha' \frac{\Delta_C}{\beta_-}\right)\,.
\end{equation}
Combining the transformation property of $k$ with the relation (\ref{eqtdualzp}) we get
\begin{equation}\label{eqtdualzm}
	T_z [\mathcal{Z}_-] = \mathcal{Z}_+ \frac{\mathcal{Z}_{h-}}{\mathcal{Z}_-}\left(1 +2\alpha'\Delta_k-\alpha'\frac{\Delta_C}{\beta_-}\right)\,.
\end{equation}
Combining equations (\ref{eqtdualds}), (\ref{eqtdualzp}) and (\ref{eqtdualzm}) we obtain the transformation properties of $W_{tt}$
\begin{equation}\label{eqtdualwtt}
	T_z[W_{tt}] = W_{tt}\left(\frac{\mathcal{Z}_{h-}}{\mathcal{Z}_-}\right)^2\left(1+2\alpha' \Delta_C - 2\alpha' \frac{\Delta_C}{\beta_-}\right)\,.
\end{equation}
Replacing (\ref{eqtdualzmh}), (\ref{eqtdualzm}) and (\ref{eqtdualwtt}) into (\ref{eqtdualdila}) we can obtain a new expression for $\mathcal{Z}_-$ 
\begin{equation} \label{eqtdualzmbis}
	\mathcal{Z}_- = \mathcal{Z}_{h-}\left(1-\alpha' \frac{\Delta_C}{\beta_+}+\alpha' \Delta_k\right) \sqrt{\frac{c_\phi T_z[\mathcal{Z}_{h-}']}{T_z[c_\phi]\mathcal{Z}_{h-}'}} \,.
\end{equation}
The transformation property of $A^w$ is trivial and does not produce any constraint. Finally, the transformation property of $\ell$ combined with the invariance of the metric produces
\begin{subequations}
	\begin{align}
		& T_w [\mathcal{Z}_0] = \mathcal{Z}_{\mathcal{H}} (1-\alpha' \Delta_\ell) \,, \\[2mm]
		& T_w [\mathcal{Z}_{\mathcal{H}}] = \mathcal{Z}_0 (1 + \alpha' \Delta_\ell) \,.
	\end{align}
\end{subequations}

\section{Solving the EOMs of the 2-charge black hole}\label{app:correctedsol}
The purpose of this appendix is to explain in more detail the procedure we have followed to solve the $\alpha'$-corrected equations motion \eqref{eq:eq1-08}, \eqref{eq:eq2-08} and \eqref{eq:eq3-08}.

\subsubsection{The ansatz} 

Let us begin motivating the ansatz we have used for the metric $\hat g_{\mu\nu}$ and two-form $\hat B$. This is given in \eqref{eq:ansatz_metric} and \eqref{eq:ansatz_B}, which we repeat here for convenience:
\begin{eqnarray}
	\diff{\hat s}^2&=&\frac{f}{\zp  {\tilde f}_{w}}\diff t^2-\zz\left(f^{-1}{\diff}\rho^2+\rho^2 {\diff}\Omega^2_{(d-2)}\right)-k^2_{\infty}\frac{\zp}{ {\tilde f}_{w}}\left[{\diff}y+\beta_{p}k^{-1}_{\infty}\left(\zp^{-1}-1\right){\diff}t\right]^2\,,\\[1mm]
	{\hat B}&=&\beta_w k_{\infty}\left({f}_{w}^{-1}-1\right) {\diff}t \wedge {\diff}y\,.
\end{eqnarray}
For the dilaton we only assume a dependence on just the radial coordinate $\rho$.

The ``recipe'' followed to fix the ansatz is essentially to keep the same field components active as in the two-derivative solution. Spherical symmetry reduces the number of independent components of the metric to four: ${\hat g}_{tt}, {\hat g}_{\rho \rho}, {\hat g}_{yy}$ and ${\hat g}_{ty}$. These are in one-to-one correspondence with the functions $f, f_p, {\tilde f}_{w}$ and $g$. The reason to choose this particular parametrization is that we expect the form of these functions will be simpler, just by experience with the two-derivative ones. The last function to be considered is $f_{w}$, which is associated to the only non-vanishing component of the two-form $\hat B$. Since we are going to treat the $\alpha'$ corrections in a perturbative fashion, the form of these functions must be:
\begin{equation}
	\begin{aligned}
		{f}_{p}=\,&1+\frac{q_p}{\rho^{d-3}}+\alpha' {\delta f_p}\, , \hspace{5mm} {\tilde f}_{w}=\,1+\frac{q_w}{\rho^{d-3}}+\alpha' {\delta {\tilde f}_{w}}\, ,\hspace{5mm}\zz=\,1+\alpha' {\delta \zz}\, ,\\[1mm]
		{f}=&\,1-\frac{\rho_s^2}{\rho^{d-3}}+\alpha' {\delta f}\, ,\hspace{5mm} {f}_{w}=\,1+\frac{q_w}{\rho^{d-3}}+\alpha' {\delta {f}_{w}}\, ,
	\end{aligned}
\end{equation}
so that the two-derivative solution is properly recovered in the  $\alpha'\to 0$ limit. 

When plugging the above ansatz in the corrected equations of motion, one gets a coupled system of second-order differential equations for the unknown functions. In what follows we describe the procedure that we have followed in order to solve it, focusing on the five-dimensional case.

\subsubsection{The equation of motion of $\hat B$} 

The expression for the dilaton can be found by solving the equation of motion of the two-form $\hat B$. There is just one independent component which is not trivially satisfied, and it yields the following equation
\begin{equation}
	\frac{f''_w}{f'_w}-\frac{2f'_w}{f_w}+\frac{g'}{g}+\frac{{\tilde f}'_w}{{\tilde f}_w}+\frac{d-2}{\rho}-2\hat \phi'=0\,,
\end{equation}
where primes denote derivatives with respect to $\rho$. It is solved by 
\begin{equation}
	e^{-2\hat{\phi} }= - \frac{(d-3) c_{\hat{\phi}}}{\rho^{d-2} f'_w} \left(\frac{f_w}{\tilde{f}_w}\right)^2 \tilde{f}_w \, g^{-(d-3)/{2}} \,,
\end{equation}
where $c_{\hat \phi}$ is an appropriate integration constant which we are going to fix imposing that the asymptotic value of the dilaton (string coupling) is not renormalized. Notice that the the Bianchi identity is identically satisfied.

\subsubsection{Einstein and dilaton equations}

As explained in the main text, the strategy to solve the corrected Einstein and dilaton equations is to expand the unknown functions $\Psi=\{\delta f, \delta f_p, \delta f_w, \delta {\tilde f}_w, \delta g\}$ in a series in $1/\rho^2$, 
\begin{equation}\label{eq:asymptotic_exp}
	\Psi=\frac{a_{\Psi}}{\rho^2}+\sum_{n>1}^N \frac{b^{(n)}_{\Psi}}{\rho^{2n}}\,,
\end{equation}
and then we solve \eqref{eq:eq1-08} and \eqref{eq:eq2-08} order by order in $1/\rho^2$. This leads to a set of algebraic equations that determine the values of the coefficients $b_{\Psi}^{(n)}$ in terms of $a_{\Psi}$ and of the parameters of the two-derivative solution, $\rho_s, q_p, q_w$. Two out of the five integration constants $a_{\Psi}$ can be fixed right away. These are $a_{f_p}$ and $a_{f_w}$, which are both set to zero by imposing that the charges of the black hole do not receive $\alpha'$ corrections. We can then set $c_{\hat{\phi}} = q_w e^{-2 \hat{\phi}_\infty} $. The next step is to find the generating functions that produce the asymptotic expansions \eqref{eq:asymptotic_exp}. This is done with the help of \texttt{Mathematica}.\footnote{For this purpose, one has to compute the solution up to sufficiently high order in $1/\rho^2$. }
Finally, we must fix the three remaining integration constants $a_g$, $a_{f}$ and $a_{\tilde f_w}$. Since we want to express the solution in the microcanonical ensemble, we must fix one of these constants (let us say, $a_g$) by imposing the mass does not receive $\alpha'$ corrections.  The resulting solution turns out to be singular at the horizon for arbitrary values of the two remaining integration constants, $a_f$ and $a_{\tilde f_w}$.\footnote{In particular, the dilaton and the Kaluza-Klein scalar diverge when $\rho \to \rho_H$.} Demanding regularity  imposes two conditions which fix both $a_f$ and $a_{\tilde f_w}$, leaving us with the solution reported in section~\ref{sec:alpha_corrected_BHs}.

\section{The coset space SU$(2)/$U$(1)$}
\label{app:coset}

The su$(2)$ algebra 

\begin{equation}
	[T_{i},T_{j}] = - \epsilon_{ijk}T_{k}\,,
	\hspace{1cm}
	i,j,k=1,2,3\,,
\end{equation}

\noindent
can be split into horizontal ($P_{a}$) and vertical ($M$) components
with the commutation relations

\begin{equation}
	\label{eq:su2algebrasplit}
	[P_{a},P_{b}] = - \epsilon_{ab}M\,,
	\hspace{1cm}
	[M,P_{a}] = - \epsilon_{ab}P_{b}\,,
	\hspace{1cm}
	a,b=1,2\,.
\end{equation}

A SU$(2)/$U$(1)$ corset representative, $u$ can be constructed by
exponentiation

\begin{equation}
	u= e^{x^{1}P_{1}}  e^{x^{2}P_{2}}\,,
\end{equation}

\noindent
where $x^{a}$ are two coordinates unrelated to those of chapter \ref{ch:3}.

The left-invariant Maurer-Cartan 1-form is

\begin{equation}
	\label{eq:MC1-form}
	\begin{aligned}
		V
		& = -u^{-1}du
		\\
		& \\
		& = (-\cos{x^{2}}dx^{1})P_{1} +(-dx^{2})P_{2}
		+\sin{x^{2}}dx^{1}M
		\\
		& \\
		& \equiv
		v^{a}P_{a}+\vartheta M\,.
	\end{aligned}
\end{equation}

The horizontal components $v^{a}$ will be used as Zweibeins while the vertical
component $\vartheta$ will play the role of U$(1)$ connection.  The
Maurer-Cartan equations are

\begin{equation}
	dV-V\wedge V=0\,,
	\,\,\,\,\,
	\Rightarrow
	\,\,\,\,\,
	\left\{
	\begin{array}{rcl}
		dv^{a}& = &  -\epsilon^{ab}v^{b}\wedge \vartheta\,,
		\\
		& & \\
		d\vartheta
		& = &
		-\tfrac{1}{2}\epsilon_{ab}v^{a}\wedge v^{b}\,.
	\end{array}
	\right.
\end{equation}

The coordinates $x^{a}$ are related to the standard $\theta,\varphi$ by

\begin{equation}
	x^{2} = \pi/2-\theta\,,
	\hspace{1cm}
	x^{1} =-\varphi\,,
\end{equation}

\noindent
and, in terms of these coordinates

\begin{equation}
	\label{eq:componentsMC1-form}
	v^{1} = \sin{\theta}d\varphi\,,
	\hspace{.5cm}
	v^{2} = d\theta\,,
	\hspace{.5cm}
	\vartheta = -\cos{\theta}d\varphi\,.
\end{equation}

Using the invariant metric $\delta_{ab}$, we get

\begin{subequations}
	\begin{align}
		ds^{2}
		& =
		\delta_{ab}v^{a}\otimes v^{b} = d\Omega^{2}{}_{(2)}\,,
		\\
		& \nonumber \\
		\omega_{(2)}
		& =
		-\tfrac{1}{2}\epsilon_{ab}v^{a}\wedge v^{b}
		=
		\sin{\theta}d\theta\wedge d\varphi\,,
		\\
		& \nonumber \\
		d\vartheta
		& =
		\omega_{(2)}\,.
	\end{align}
\end{subequations}
	\clearpage{\pagestyle{empty}\cleardoublepage} 
	
	\chapter{Details of type IIA compactifications} \label{app:D}

\section{Explicit solution of the EOMs}
\label{ap:10deom}

In this appendix we analyze and solve the 10d equations of motion (EOMs) and Bianchi identities of massive type IIA supergravity, with a compactification Ansatz of the form \eqref{eq:warped-product}, using the conventions of \cite{Tomasiello:2022dwe}. 
The Bianchi identities \ref{IIABI} for the polyform $\mathbf{G}$ encode both the EOMs and the Bianchi identities of the RR internal fluxes $\hat{G}$. Exploiting (\ref{bfG}) we obtain the Bianchi identities
\begin{subequations}\label{App10dEOMbianchi}
	\begin{align}
		& d{G}_0 = 0\,, \label{BiG0}\\
		& d {G}_2 = {G}_0 \wedge H - 4 \d_{\rm O6} +   N_\a \d_{\rm D6}^\a \, , \label{BiG2}\\
		& d \hat{G}_4 = {G}_2 \wedge H\, ,\label{BiG4} \\
		& d\hat{G}_6 = 0\, , \label{BiG6}
	\end{align}
\end{subequations}
and dualising the relations obtained for the external fluxes trough $\tilde{G} = -\lambda(\star_6 \hat{G})$ and the Ansatz \ref{eq:warped-product} we obtain the EOMs
\begin{subequations}
	\begin{align}
		& d(e^{4A}\star_{6}{G}_2) + e^{4A} H\wedge \star_{6}\hat{G}_4 = 0  \,, \label{eomG2} \\
		& 	d(e^{4A}\star_{6}\hat{G}_4) + e^{4A} H\wedge \star_{6}\hat{G}_6 = 0 \,, \label{eomG4}\\
		& 	d(e^{4A}\star_{6}\hat{G}_6) = 0 \label{eomG6} \,.
	\end{align}
\end{subequations}
The Bianchi identity and the EOM for the NSNS flux are instead 
\begin{subequations}
	\begin{align}
		& dH = 0 \,, \label{BiH} \\
		& d(e^{-2\phi+4A}\star_{6}H) + e^{4A}\star_{6} \hat{G}_6 \wedge \hat{G}_4 + e^{4A}\star_{6}\hat{G}_4 \wedge {G}_2 + e^{4A}\star_{6} {G}_2 \wedge {G}_0 = 0 \,.  \label{eomH} 
	\end{align}
\end{subequations}
The dilaton and the Einstein EOMs in our conventions are finally (see \cite{Junghans:2020acz})
\allowdisplaybreaks
\begin{subequations}
	\begin{align}
		\begin{split}
			0 = & \; 12 \frac{\tau^2}{\omega^2} + 12 \frac{\tau^2}{\omega^2}\left(\partial \omega \right)^2 + 4 \frac{\tau^2}{\omega} \nabla^2 \omega + 12 \frac{\tau}{\omega}\left(\partial \omega \right)\left(\partial \tau \right) + \tau \nabla^2 \tau 
			+ \left(\partial \tau \right)^2  - \frac{1}{2}\tau^2 \left|H\right|^2 \\ & \; - \sum_{q=0}^6 \frac{q-1}{4}|\hat{G}_q|^2  + \frac{\tau}{4}\sum \delta_i^{(3)} \,,
		\end{split} \\
		\begin{split}
			0 = & \; -\tau^2 R_{mn} + 4 \frac{\tau^2}{\omega} \nabla_m \partial_n \omega + \frac{\tau}{\omega}g_{mn}\left(\partial \omega\right)\left(\partial \tau \right) + \frac{1}{4}g_{mn}\tau \nabla^2 \tau + \frac{1}{4}g_{mn}\left(\partial \tau \right)^2 \\
			& \; + 2 \tau \nabla_m \partial_n \tau - 2\left(\partial_m \tau \right)\left(\partial_n \tau \right) 
			+ \frac{1}{2}\tau^2 \left(|H|_{mn}^2-\frac{1}{4}g_{mn}|H|^2\right) \\ & \; + \frac{1}{2}\sum_{q=0}^6\left(|\hat{G}_q|^2_{mn}- \frac{q-1}{8}g_{mn}|\hat{G}_q|^2\right) + \frac{\tau}{2}\sum_i \left(\Pi_{i,mn}-\frac{7}{8}g_{mn}\right)\delta_i^{(3)} \,,
		\end{split} \\
		\begin{split}
			0 = & \; -8 \nabla^2 \,\tau - 24 \frac{\tau}{\omega^2}-\frac{32}{\omega}\left(\partial \omega\right)\left(\partial \tau\right) - 24\frac{\tau}{\omega^2}\left(\partial \omega\right)^2 - 16 \frac{\tau}{\omega}\nabla^2 \omega + 2 \tau R_{mn}g^{mn} \\
			& \;  - \tau|H|^2 + \sum_i \delta^{(3)}_i\,,
		\end{split}
	\end{align}
\end{subequations}
with
\begin{equation}
	\Pi_{i,mn} = -\frac{2}{\sqrt{g_{\pi_i}}}\frac{\delta \sqrt{g_{\pi_i}}}{\delta g^{mn}} \,, \qquad |F_p|^2_{mn} = \frac{\delta |F_p|^2}{\delta g^{mn}}\,, \qquad \tau = e^{-\phi} \,, \qquad \omega = R e^A\,,
\end{equation}
with $R$ the AdS$_4$ radius, which implies that $\langle e^A \rangle = 1$. We finally have the delta-function sources
\begin{equation}
	\sum_i  \delta^{(3)}_i = \star_{\rm CY} \bigg[ \text{Im} \Omega _{\rm CY}\wedge \left(4 \d_{\rm O6} -   N_\a \d_{\rm D6}^\a\right) \bigg] \,.
\end{equation}

\subsubsection{Smearing approximation}

The smearing approximation assumes the dilaton and the warp factor to be constant, and the internal metric to be Calabi--Yau. With the Ansatz (\ref{intfluxsm}) and the tadpole cancellation condition (\ref{tadpole}) all fluxes are harmonic and the only non-trivial Bianchi identity is that of ${G}_2$ 
\begin{equation}
	d{G}_2 =  6 A  g_s {G}_0^2 \, \text{Re}(\Omega_{\rm CY}) - \frac{mh}{\ell_s^2} \delta_{O6} = 0  \,,
\end{equation}
which provides the following constraint
\begin{equation}\label{const1}
	24 A g_s {G}_0^2  = \frac{mh}{\ell_s^2} \star_{\rm CY} \left[ \text{Im}_{\rm CY}\wedge \delta_{O6} \right] = \frac{mh}{\ell_s^2}  \frac{{\cal V}_{\Pi_{O6}}}{{\cal V}_{\rm CY}}\, ,
\end{equation}
where in the last step we approximated the bump delta functions trough their constant Fourier modes, according to (\ref{deltasm}). Under the same assumptions, the only non-trivial flux EOM is the one of $H$ 
\begin{equation}\label{const2}
	d\star_{\rm CY}H = - 2 G_0^2 J_{\rm CY}^2 \, B  \left(C + \frac{1}{4}\right) = 0\,.
\end{equation}
Finally, the EOM of the dilaton and the Einstein equations take the form
\begin{subequations} 
	\begin{align}
		0 = & \; 12 \mu^2  \frac{e^{-2\phi}}{e^{2A}}-\frac{1}{2}e^{-2\phi}|H|^2+\frac{1}{4}|{G}_0|^2 - \frac{1}{4}|{G}_2|^2 - \frac{3}{4}|\hat{G}_4|^2 - \frac{5}{4}|\hat{G}_6|^2+ \frac{1}{4}e^{-\phi} \sum_i \delta^{(3)}_i \,, \label{EOMeindil1}\\
		\begin{split}
			0 = & \; \frac{1}{2}e^{-2\phi}\left(|H|^2_{mn} - \frac{1}{4}g_{mn}|H|^2\right) +\frac{1}{2}\left(|{G}_2|^2_{mn} - \frac{1}{8}g_{mn}|{G}_2|^2\right) \\
			& +\frac{1}{2}\left(|\hat{G}_4|^2_{mn} - \frac{3}{8}g_{mn}|\hat{G}_4|^2\right) +\frac{1}{2}\left(|\hat{G}_6|^2_{mn} - \frac{5}{8}g_{mn}|\hat{G}_6|^2\right) \\
			& +\frac{1}{16}g_{mn}|{G}_0|^2 + \frac{1}{2}e^{-\phi} \sum \left(\Pi_{i,mn}-\frac{7}{8}g_{mn}\right) \delta^{(3)}_i \,, \label{EOMeindil2}
		\end{split} \\
		0 = & -24 \mu^2 \frac{e^{-\phi}}{e^{2A}} - e^{-\phi}|H|^2 + \sum \delta^{(3)}_i \label{EOMeindil3} \,,
	\end{align}
\end{subequations}
where we have introduced $\mu^2 = 1/R^2$, as in the main text. Here one should make the replacement $e^\phi \to \left<e^\phi\right> = g_s$ and $e^A \to  \left<e^{A}\right> = 1$, and smear the delta functions as follows 
\begin{equation}
	\qquad \sum_i  \delta^{(3)}_i = \frac{mh}{\ell_s^2} \star_{\rm CY} \left[ \text{Im}_{\rm CY}\wedge \delta_{O6} \right] = \frac{mh}{\ell_s^2}  \frac{{\cal V}_{\Pi_{O6}}}{{\cal V}_{\rm CY}}\,.
\end{equation}
It is easy to verify that the smearing Ansatz (\ref{intfluxsm}) implies in our conventions 
\begin{equation}
	\begin{split} \label{fieldstr}
		& |H|^2 = 144 A^2 g_s^2 {G}_0^2 \,, \quad |{G}_2|^2 = 3 B^2 {G}_0^2 \,, \quad |\hat{G}_4|^2 = 12 C^2 {G}_0^2 \,, \quad |\hat{G}_6|^2 = 0 \,, \\
		&  |H|^2_{mn} = \frac{1}{2}g_{mn}|H|^2 \,, \quad |{G}_2|^2_{mn} = \frac{1}{3}g_{mn}|{G}_2|^2  \,, \quad |\hat{G}_4|^2_{mn} = \frac{2}{3}g_{mn}|\hat{G}_4|^2  \,, \quad |\hat{G}_6|^2_{mn} = 0 \,.
	\end{split}
\end{equation}
Combining (\ref{EOMeindil1}) and (\ref{EOMeindil3}) and exploiting (\ref{fieldstr}) we obtain
\begin{subequations} \label{App10dEOMeq01}
	\begin{align}
		\mu^2 & = \frac{{G}_0^2g_s^2}{72} \left( 144A^2 +3B^2 +36C^2-1\right)\, , \\
		\frac{mh}{\ell_s^{2}}\frac{{\cal V}_{\Pi_{\rm O6}}}{{\cal V}_{\rm CY}} & = \frac{{G}_0^2g_s}{3} \left( 576A^2 +3B^2 +36C^2-1\right)\, \label{const3} .
	\end{align}
\end{subequations}
Replacing (\ref{App10dEOMeq01}) into (\ref{EOMeindil2}) an taking the trace we find  
\begin{equation}
	\frac{mh}{\ell_s^{2}}\frac{{\cal V}_{\Pi_{\rm O6}}}{{\cal V}_{\rm CY}} = \frac{{G}_0^2g_s}{6} \left( 1584A^2 +3B^2 +84C^2-5\right)\,. \label{const4}
\end{equation}
Summarising, equations \ref{const1}, \ref{const2}, \ref{const3} and \ref{const4} tell us that a vacua must satisfy 
\begin{align}
	& 24 A = \frac{1}{3} \left( 576A^2 +3B^2 +36C^2-1\right) = \frac{1}{6} \left( 1584A^2 +3B^2 +84C^2-5\right) \,, \\
	& B\left(C+\frac{1}{4}\right) = 0 \,.
\end{align} 

\subsubsection{First-order corrections}

Away from the smearing approximation the dilaton and the warping factor are no more constant. Inspired by the results of \cite{Marchesano:2020qvg}, we assume the following Ansatz for the warp factor and the dilaton
\begin{equation}\label{ansatzDW}
	e^{-A}  = 1 + g_s \varphi + \cO(g_s^2) \, , \qquad e^{\phi}   = g_s \left(1 - 3  g_s \varphi\right) + \cO(g_s^3)\, ,
\end{equation} 
with $\varphi$ a real function. The metric $g_{mn}$ is no more Calabi--Yau, but we assume that its departure from the Calabi--Yau conditions can be described by a series of $\cO(g_s^n)$ corrections. At the level of the first correction there still exist a three-form $\Omega$ and a two-form $J$ such that 
\begin{equation}
	\star_6\, \text{Re}(\Omega) = \text{Im}(\Omega) + \mathcal{O}(g_s^2) \label{Tromega}\,, \qquad \star_6 \, J = -\frac{1}{2}\, J^2 + \mathcal{O}(g_s^2)\,,
\end{equation}
with $\star_6$ the corrected Hodge star operator. Inspired again by \cite{Marchesano:2020qvg} we assume moreover that $\Omega$ and $J$ satisfy
\begin{subequations} 
	\begin{align}
		& \text{Re}(\Omega) = \text{Re}(\Omega_{\rm CY})(1-g_s \varphi) + g_s K + \mathcal{O}(g_s^2) \,, \label{Reomega}\\
		& \text{Im}(\Omega) = \text{Im}(\Omega_{\rm CY}) (1+g_s\varphi) - g_s \star_{\rm CY} K + \mathcal{O}(g_s^2) \,,  \label{Imomega}\\
		& J = J_{\rm CY} + \mathcal{O}(g_s^2)\,, \label{eq017}
	\end{align}
\end{subequations}
with $K$ a current three-form such that 
\begin{equation}\label{kdef}
	\Delta_{\rm CY} K =  6 A g_s {G}_0^2\text{Re}(\Omega_{\rm CY})- \frac{mh}{\ell_s^2} \delta_{O6} + \mathcal{O}(g_s^2)\,,
\end{equation}
where $\Delta_{\rm CY}$ is the Laplace operator associated to the uncorrected Calabi--Yau metric (notice that such current can be easily build, see the construction in \cite{Hitchin:1999fh}). 

We will show now explicitly that the fluxes (\ref{solutionflux}) satisfy the EOMs and the Bianchi identities for certain values of the real constants $S$ and $R$ therein. We start discussing the Bianchi identities. According to Hodge theory, a form admits a unique decomposition in exact, harmonic and co-exact components.\footnote{Notice that such a decomposition depends on the metric. We define it with respect to the Calabi--Yau one.} The Ansatz (\ref{solutionflux}) tell us that neither $\hat{G}_6$ nor $H$ have a co-exact part and takes $G_0$ as a constant, therefore (\ref{BiG0}), (\ref{BiG6}) and (\ref{BiH}) are automatically satisfied. According to \cite{Marchesano:2020qvg}, the most general $K$ that satisfies (\ref{kdef}) is a closed three-form current which can be written as $K = \tilde{\varphi} \text{Re}(\Omega_{\rm CY})+c \,\text{Im}(\Omega_{\rm CY})+ \text{Re}(k)$ with $k$ a (2,1) primitive current. The requirement of $K$ being closed implies that $k_{2,1}$ must satisfy  (see \cite{Casas:2022mnz} for more details on the constraints that $\tilde{\varphi}$ and $k$ must satisfy)
\begin{subequations}\label{kconditions}
	\begin{equation} 
		k_{2,1} = \partial k_{1,1} + \bar{\partial} k_{2,0} + k_{2,1}^{h}\,,\qquad
		\partial k_{2,1} = \bar{\partial} \tilde{\varphi}\, \Omega_{\rm CY} \,, \qquad
		\text{Re}(\bar{\partial}k_{2,1}) = 0 \,.
	\end{equation}
\end{subequations}    
If we set $c=0$ and we require that $\varphi = \tilde{\varphi}$ it is straightforward to obtain
\begin{equation} \label{coDK1}
	d^\dagger_{\rm CY} K = \star_{\rm CY} d \left[2\varphi \text{Im}(\Omega_{\rm CY})\right] - V_{1,1}\,,
\end{equation} 
with $V_{1,1} = \star_{\rm CY} \left( \partial \bar{\partial} k_{1,1}\right)$ a primitive (1,1)-form. Exploiting the K\"ahler identity $\left[d^c, J \cdot \right] = d^\dagger$ with $d^c = -i(\partial - \bar{\partial})$ it is possible to remove $V_{1,1}$ from (\ref{coDK1}) and we obtain
\begin{equation}
	d^\dagger_{\rm CY} K = -J\cdot d \left[4\varphi \text{Im}(\Omega_{\rm CY})-\star_{\rm CY}K\right]\,,
\end{equation}
which implies that the Bianchi identity (\ref{BiG2}) is satisfied at first order in $g_s'$, we have indeed
\begin{equation}
	d{G}_2 = dd^\dagger_{\rm CY} K = \Delta_{\rm CY}K = {G}_0\wedge H - 4 \d_{\rm O6} +   N_\a \d_{\rm D6}^\a + \mathcal{O}(g_s^2)\,.
\end{equation}
Differentiating $\hat{G}_4$ we obtain instead
\begin{equation}
	\begin{split}
		d\hat{G}_4  & = 12 A g_s {G}_0 \left\{\star_{\rm CY} \left[d\varphi \wedge \text{Im}(\Omega_{\rm CY})\right]\wedge\text{Re}(\Omega_{\rm CY})\right\} \\ & = d^\dagger_{\rm CY}K \wedge H + \mathcal{O}(g_s^2) = {G}_2 \wedge H + \mathcal{O}(g_s^2)\, .
	\end{split}
\end{equation} 
In the first step we used the identity (see \cite[prop. 1.2.31]{Huybrechts:2005} adapted to the conventions of \cite{Tomasiello:2022dwe})
\begin{equation} \label{IdentityP}
	\star_{\rm CY} \left[J^s \wedge \alpha \right] = (-1)^{\frac{k(k+3)}{2}+1}\, \frac{s!}{(3-k-s)!}J^{3-k-s} \wedge I\left( \alpha\right)\,,
\end{equation}
where $\alpha$ is a primitive $k$-form and $I$ is the operator 
\begin{equation}
	I = \sum_{p,q = 0}^3 i^{p-q} \, \Pi^{p,q} \,,
\end{equation}  
to obtain $J^2_{\rm CY} \wedge d\varphi = -2 \star_{\rm CY} d^c \varphi$. In the second step we used equation (\ref{coDK1}). 

Let us now discuss the fluxes EOMs. The EOM of $\hat{G}_6$ (\ref{eomG6}) is trivially satisfied. Exploiting that $\hat{G}_6 = 0$ the EOM of $\hat{G}_4$ (\ref{eomG4}) becomes
\begin{equation}
	d \left[e^{4A} \star_{6} \hat{G}_4 \right] = 0\,.
\end{equation}  
Exploiting the fact that the exact and co-exact part of $G_4$ are of order $\mathcal{O}(g_s)$ and that $g_s \star_6 = g_s \star_{\rm CY} + \mathcal{O}(g_s^2)$ it further reduces to 
\begin{equation} \label{G4eq1}
	4\left(2C+6A\right) g_s  {G}_0 \, d \varphi \wedge J_{\rm CY} + d \star_{\rm CY} d \left[ S g_s^{-1} J_{\rm CY} \wedge \text{Im}(v)\right] = 0 + \mathcal{O}(g_s^2)\,.
\end{equation} 
Using the identity (\ref{IdentityP}) it is straightforward to prove that $ \star_{\rm CY} \left( J_{\rm CY} \wedge \text{Im}(v) \right) $ is a closed form. Exploiting the definition of $v$ and $f_*$ (\ref{ansatzVF}) the last term of (\ref{G4eq1}) takes the form
\begin{equation}
	d \star_{\rm CY} d \left[ \frac{S}{g_s}  J_{\rm CY} \wedge \text{Im}(v)\right] = \frac{S}{2} \star_{\rm CY}^{-1}\left(J_{\rm CY} \wedge d_c \Delta_{\rm CY} f_* \right) = -4 S g_s {G}_0 \, d \varphi \wedge J_{\rm CY} \,.
\end{equation}
The EOM of $\hat{G}_4$ is then simply
\begin{equation}\label{cond2}
	4\left(2C+6A-S\right) g_s  {G}_0 \, d \varphi \wedge J_{\rm CY} = 0 + \mathcal{O}(g_s^2)\,.
\end{equation}
The EOM of ${G}_2$ (\ref{eomG2}) reduces to 
\begin{equation}
	d \star_{\rm CY} {G}_2 = 0 + \mathcal{O}(g_s)\, ,
\end{equation}
which is trivially satisfied because ${G}_2$ does not have an exact part. The EOM of $H$ becomes
\begin{equation}\label{eomHstep}
	d(g_s^{-2}(1+2g_s \varphi)\star_6 H) + \star_{\rm CY}\hat{G}_4 \wedge{G}_2 + \star_{\rm CY} {G}_2 \wedge {G}_0 = 0 + \mathcal{O}(g_s) \,, 
\end{equation}
which we can evaluate term by term. Combining the ansatz (\ref{Reomega}) and (\ref{Imomega}) with the transformation property (\ref{Tromega}) we obtain
\begin{equation} 
	\star_6 \text{Re}(\Omega_{\rm CY}) = \text{Im}(\Omega_{\rm CY})(1+2\varphi g_s) - 2 g_s \star_{\rm CY} K + \mathcal{O}(g_s^2)\,,
\end{equation}
and the first term of (\ref{eomHstep}) becomes
\begin{equation}
	\begin{split}
		(1) \, = &\;  24 \, A \, {G}_0 d\varphi \wedge \text{Im}(\Omega_{\rm CY}) - 6 (2A - AR) {G}_0 \star_{\rm CY} d^\dagger_{\rm CY} K \\ & -\frac{S}{2g_s^2} d\star_{\rm CY}   d\re \left(\bar{v} \cdot \Omega_{\rm CY} \right) + \mathcal{O}(g_s)\,,
	\end{split}
\end{equation}
where there latter term further reduce to
\begin{equation}
	-\frac{S}{2g_s^2} d\star_{\rm CY}   d\re \left(\bar{v} \cdot \Omega_{\rm CY} \right) = - 4 {G}_0 S d\varphi \wedge \text{Im}(\Omega_{\rm CY}) \,.
\end{equation}
The second term of (\ref{eomHstep}) becomes
\begin{equation}
	(2) = - 2 \, C \, {G}_0 \left(B\, {G}_0 J_{\rm CY}^2  - 4 d\varphi \wedge \text{Im}(\Omega_{\rm CY}) + \star_{\rm CY} d^\dagger_{\rm CY}K \right) + \mathcal{O}(g_s) \,,
\end{equation}
where we used the K\"ahler identity $\left[J\cdot, J\wedge\right] = H$ with $H$ the operator that on $k$-forms act as $H \alpha = (3-k)\alpha$. The third term of (\ref{eomHstep}) becomes simply
\begin{equation} 
	(3) = {G}_0 \star_{\rm CY} d^\dagger_{\rm CY} K_3 - \frac{1}{2} B {G}_0 J_{\rm CY}^2 + \mathcal{O}(g_s) \,.
\end{equation}
Replacing everything in the EOM of $H$, equation (\ref{eomH}) finally becomes
\begin{equation}\label{cond1}
	\begin{split}
		& \left(24 A + 8 C - 4 S\right) {G}_0 \, d\varphi \wedge \text{Im}(\Omega_{\rm CY}) + {G}_0 (1-2C - 12 A + 6A R) \star_{\rm CY} d^\dagger_{\rm CY} K \\  & - 2 G_0^2 \, B \left( C + \frac{1}{4}\right) = 0 + \mathcal{O}(g_s) \,.
	\end{split}
\end{equation}  

Summarising, equations (\ref{cond2}) and (\ref{cond1}) tell us that a vacua of the form (\ref{solutionflux}) exist provided that $S$ and $R$ satisfy
\begin{equation}
	S = 2C+6A\,, \qquad 6AR = 12A + 2C - 1\,. 
\end{equation} 


\section{DBI computation}
\label{ap:DBI}

In this appendix we derive the expressions for the DBI action that are used in section \ref{s:membranes} to compute the tension of D8-branes and D6-branes with internal worldvolume fluxes. 

Let us start by considering a D8-brane wrapping the whole of a Calabi--Yau manifold $X_6$. Ignoring curvature corrections, the contribution to the DBI coming from the internal dimensions involves the square root of 
\begin{equation}
	\label{eq:dbii}
	\operatorname{det}\left(g_{a b}-\mathcal{F}_{a b}\right)=\operatorname{det} g \operatorname{det}(\mathbb{I}+A)=\operatorname{det} g\left(1-\frac{t_{2}}{2}+\frac{t_{2}^{2}}{8}-\frac{t_{4}}{4}+\operatorname{det} A\right)\, ,
\end{equation}
where we have used the Cayley–Hamilton theorem and introduced the definitions $A\equiv -g^{-1}\mathcal{F}$ and $t_k\equiv \text{Tr}\left(A^k\right)$. Assuming that $\cF$ is a (1,1)-form it follows that
\begin{align}
	-\frac{1}{2} t_{2} &=\left(\frac{1}{2} \mathcal{F} \wedge J \wedge J\right)^{2} +\left(\mathcal{F} \wedge \mathcal{F} \wedge J\right) \cdot d\text{vol}_{X_6}\, , \\
	\left(\frac{t_{2}^{2}}{8}-\frac{t_{4}}{4}\right) &=\left[\left(\frac{1}{2} \mathcal{F} \wedge \mathcal{F} \wedge J\right)^{2} -2\left(\frac{1}{2} \mathcal{F} \wedge J \wedge J\right) \cdot\left(\frac{1}{6} \mathcal{F} \wedge \mathcal{F} \wedge \mathcal{F}\right)\right] \, ,\\
	\operatorname{det} A &=\frac{1}{36}\left(\mathcal{F}\wedge\mathcal{F}\wedge\mathcal{F}\right)^2 \, ,
\end{align} 
where $d\text{vol}_{X_6}=-\frac{1}{6}J\wedge J\wedge J$ and the product means contraction of two six-form with the metric. Putting everything together,  the integrand of the  DBI action  for a D8 wrapping the whole of $X_6$  and with a $\left(1,1\right)$ worldvolume flux on it  can be written as
\begin{equation}
	d{\rm DBI}_{\rm D8} = g_s^{-1}
	\sqrt{\left(\frac{1}{6} J \wedge J \wedge J-\frac{1}{2} \mathcal{F} \wedge \mathcal{F} \wedge J\right)^{2}+\left(\frac{1}{6} \mathcal{F} \wedge \mathcal{F} \wedge \mathcal{F}-\frac{1}{2} \mathcal{F} \wedge J \wedge J\right)^{2}} d\text{vol}_{X_6}\, .
	\label{ap:DBID8}
\end{equation}	
Therefore, whenever $\mathcal{F}\wedge\mathcal{F}=3 J\wedge J$ we obtain a perfect square, signalling that we have a BPS configuration. This is just a  particular solution of the MMMS equations \cite{Marino:1999af}, which in our conventions read
\begin{align}
	\frac{1}{6} \mathcal{F} \wedge \mathcal{F} \wedge \mathcal{F}-\frac{1}{2} \mathcal{F} \wedge J \wedge J
	=\tan \theta\,  \left(\frac{1}{6} J \wedge J \wedge J - \frac{1}{2} \mathcal{F} \wedge \mathcal{F} \wedge J\right)\, ,
\end{align}
with $\theta$ defined as in \eqref{calibration}.

We can apply the same reasoning considering to a D6-brane wrapping an internal 4-cycle $\mathcal{S}$ of $X_6$. In this case, the determinant that appears in the DBI action can be expressed as
\begin{equation}
	\label{eq:dbii6}
	\operatorname{det}\left(g_{a b}-\mathcal{F}_{a b}\right)=\operatorname{det} g \operatorname{det}(\mathbb{I}+A)=\operatorname{det} g\left(1-\frac{t_{2}}{2}+\operatorname{det} A\right)\, ,
\end{equation}
Assuming that $\cF$ is a (1,1)-form and denoting by $J$ the pull-back of $J_{\rm CY}$ on ${\cal S}$ we have that 
\begin{align}
	-\frac{1}{2} t_{2} &=\left( \mathcal{F} \wedge J\right)^{2}  +\left(\mathcal{F} \wedge \mathcal{F}\right) \cdot d\text{vol}_{\mathcal{S}}\, , \\
	\operatorname{det} A &=\left(\frac{1}{2}\mathcal{F}\wedge\mathcal{F}\right)^2 \, ,
\end{align}
where $d\text{vol}_{\mathcal{S}}=-\frac{1}{2}J\wedge J$. Taking into account all these terms,  the internal part of the  DBI action  for a D6 wrapping a four-cycle $\mathcal{S}$ and with  a $\left(1,1\right)$ internal worldvolume flux on it reads
\begin{align}
	d{\rm DBI}_{\rm D6} = g_s^{-1} \sqrt{\left(-\frac{1}{2} J \wedge J+\frac{1}{2} \mathcal{F} \wedge \mathcal{F}\right)^{2}+(J \wedge \mathcal{F})^{2}}\;d\text{vol}_{\mathcal{S}}\, .
	\label{ap:DBID6}
\end{align}
We see that for $J\wedge J = \cF \wedge \cF$ the interior of the square root becomes a perfect square. Accordingly, we recover again a solution of the MMMS equations \cite{Marino:1999af}, which for the case at hand read 
\begin{align}
	\tan^{-1}\theta  \left(J \wedge \mathcal{F}\right)= \frac{1}{2} J \wedge J-\frac{1}{2} \mathcal{F} \wedge \mathcal{F}\, .
\end{align}

\section{NS5-branes and domain wall solutions}
\label{ap:NS5DW}

In  \cite{Gurrieri:2002wz} a 4d domain-wall solution based on a backreacted N5-brane is considered, in order to arrive at a background of the form \eqref{SU3ex} via mirror symmetry. In this appendix we review the NS5-brane setup, and describe a set of type IIB D-branes that are mutually BPS with the NS5-brane. Such D-branes have a well-defined interpretation from the 4d domain-wall viewpoint, and are in one-to-one correspondence with the type IIA D-brane objects discussed in section \ref{s:simple}. 

Following  \cite{Gurrieri:2002wz}, let us consider a toroidal compactification of type IIB string theory to 4d. Let us call the three complex coordinates of ${\bf T}^6 = ({\bf T}^2)_1 \times ({\bf T}^2)_2 \times ({\bf T}^2)_3$ as $dz^j = dx^j + i dx^{j+3}$, $j=1,2,3$. Then one places $M$ NS5-branes along the three-cycle $(1,0)_1(1,0)_2(0,1)_3$ (that is, the coordinates $\{x^1, x^2, x^6\}$) and spanning two spatial dimensions in $\IR^{1,3}$. One then backreacts such NS5-branes and smears the solution down to 4d. In this approximation the harmonic function $V$ that describes an NS5-brane backreaction becomes a linear function of its transverse coordinate in 4d, therefore $V = 1- \zeta \xi$ with $\zeta$ a real constant.\footnote{See footnote \ref{ft:dw} for an explanation of different choice of $V$ compared to \cite{Gurrieri:2002wz}.} One obtains the background
\bea
\label{Hex}
ds^2 &= &ds^2_{\IR^{1,2}}+ \ell_s^2 V(d\xi)^2 + \ell_s^2 ds^2_{{\bf T}^6}, \\
ds^2_{{\bf T}^6} & = & (2\pi)^2 \left[ (r_1 dx^1)^2+(r_2 dx^2)^2+V(r_3 dx^3)^2+V( r_4 dx^4)^2+V(r_5 dx^5)^2+(r_6 dx^6)^2 \right] , \\
H & = & - M dx^4 \wedge dx^5 \wedge dx^3, \\
e^{2\phi} & = & e^{2\phi_0} V .
\eea
with $M \in \mathbb{N}$. Upon three T-dualities along the coordinates $\{x^1, x^2, x^3\}$ one is led to a type IIA background with constant dilaton, no $H$-flux, and an internal metric of the form \eqref{SU3exb}, more precisely with $M_3=M$ and $R_j = 1/r_j$ for $j=1,2,3$. 

This smeared solution is interpreted as the long-wavelength description of the $M$ NS5-branes'  backreaction, more precisely as the 4d domain-wall solution that is perceived at wavelengths much larger than the compactification radii. As such, a D-brane that is BPS in the microscopic background should also be so in its long-wavelength approximation. In practice, this means that if we consider D-branes that are mutually BPS with the NS5-branes sourcing the solution, they should correspond to BPS objects in the 4d domain-wall background. To match out discussion with that of section \ref{s:simple}, we will consider type IIB D-branes whose embedding survives the $\IZ_2 \times \IZ_2$ orbifold projection with generators $\theta_1: (x^1, x^2, x^3, x^4, x^5, x^6) \mapsto  (x^1, -x^2, -x^3, x^4, -x^5, -x^6)$ and $\theta_2: (x^1, x^2, x^3, x^4, x^5, x^6) \mapsto  (-x^1, -x^2, -x^3, -x^4, -x^5, x^6)$.

For instance, D1 and D5-branes are mutually supersymmetric with respect to an NS5-brane if the system has $2+4k$ ND directions.\footnote{In an abuse of language, we are borrowing the nomenclature used for configurations of pairs of D-branes.} We may then consider

\begin{itemize}
	
	\item[-] A D5-brane wrapping  $(T^2)_i \times (T^2)_j$ and extended along $\xi$  $\Longrightarrow$ 6 ND directions. 
	
	\item[-] A D1-brane extended along $\xi$ $\Longrightarrow$ 6 ND directions. 
	
	\item[-] An Euclidean D5-brane wrapped on ${\bf T}^6$ $\Longrightarrow$ 6 ND directions.
	
\end{itemize}
Additionally, a D3-brane is mutually BPS with an NS5-brane if the system has $4k$ ND directions. So for instance we can have:

\begin{itemize}
	
	\item[-] A D3-brane wrapping  $(0,1)_i(0,1)_j(1,0)_k$ and point-like along $\xi$  $\Longrightarrow$ 4 or 8 ND directions. 
	
	\item[-] An Euclidean D3-brane on  $(0,1)_i(1,0)_j(1,0)_k$ and along $\xi$  $\Longrightarrow$ 4 or 8 ND directions. 
	
\end{itemize}
All these objects are mutually BPS with the NS5-brane sources. If their orientation is reversed, they will still be BPS in the NS5-brane background, but preserving a disjoint set of supercharges. 

Upon three T-dualities along $\{x^1, x^2, x^3\}$, these D-branes are mapped to some of the type IIA D-branes discussed in section \ref{s:simple}. More precisely,  the D1-brane becomes a D4-brane wrapped on $\Pi_3^{\rm tor}$ and the Euclidean D5-brane is mapped to the Euclidean D2-brane wrapping the three-chain $\Sigma_3$. Similarly, the D3-branes become the D4-branes on the four-chains $\Sigma_4^i$ and the Euclidean D3-branes become the Euclidean D2-branes wrapped on $\Pi_2^i$. The configurations of AB strings and particles ending on monopoles and instantons described in section \ref{s:simple} have a clear microscopic origin in this dual type IIB setup. For instance, a D3-brane wrapped on $(0,1)_1(0,1)_2(1,0)_3$ intersects the NS5-brane at a point in ${\bf T}^6$ and is pointlike in 4d, by a Hanany-Witten effect (see e.g. \cite[Appendix B]{Berasaluce-Gonzalez:2012awn}) it must have $M$ D1-branes stretched along $\xi$ between the NS5-branes and its 4d location. The same occurs in the smeared description \eqref{Hex}, where the Freed-Witten anomaly induced by the $H$-flux on such a D3-brane is cured by the same stack of $M$ D1-branes.

\section{Massive $p$-from spectra in twisted tori}
\label{ap:spectra}

The aim of this appendix is to provide the explicit $p$-form eigenforms and eigenvalues of the twisted tori $\tilde{\bf T}^3$, whose $p$-form  spectrum was discussed in section \ref{s:direct} based on the method of  \cite{BenAchour:2015aah}. Furthermore, we compare the results with the analyzes carried out in \cite{Andriot:2018tmb}, to show that the said spectrum is complete.

Consider the three-dimensional twisted torus  $\tilde{\bf T}^3$ with metric 
\be
ds^2_{\tilde{\bf T}^3} = (2\pi)^2 \left[(R_i\eta^i)^2 + (R_{j+3}\eta^{j+3})^2 + (R_{k+3}\eta^{k+3})^2 \right] \,,
\ee
and twist $d\eta^{i} = - N \eta^{j+3}\wedge \eta^{k+3}$. We choose angular coordinates $x^a\in[0,1]$ such that the $\eta^a$ are parametrized as 
\be
\eta^{i} = dx^{i} + N x^{k+3} dx^{j+3},\hspace{1em}\eta^{j+3} = dx^{j+3},\hspace{1em}\eta^{{k+3}}=dx^{{k+3}},
\ee
with $N=2\pi\lambda_{\chi} R_{k+3} R_{j+3}/R_{i}$ and  $\lambda_{\chi}\in \mathbb{R}$.

Then, we take the one-form Killing vector $\chi=2\pi R_{i} \eta^{i}/\sqrt{V}$, which satisfies the assumed properties presented in section \ref{s:direct}:
\begin{equation}
	\star d \chi = \lambda_\chi \,\chi \,, \qquad \Delta_3 \chi = \lambda_\chi^2\, \chi \,, \qquad \chi^2 = 1  \,, \qquad  \lambda_\chi, \in \IR \, ,
\end{equation}
where $V=8\pi^3R_iR_{j+3}R_{k+3}$ is the volume of $\tilde{\bf T}^3$.

Let $\{\phi_i\}$ represent the basis of complex scalar eigenforms of the Laplacian. The explicit shape of these eigenforms is \cite{Andriot:2018tmb}

\bea
&\phi_{p,q} = \frac{e^{2\pi i q x^{j+3}}e^{2\pi i p x^{k+3}}}{\sqrt{V}},\label{eq: phipq}\\
& \phi_{k,l,n} = \sqrt{\frac{2\pi R_{j+3}}{|N|V}}\frac{1}{\sqrt{2^n n!\sqrt{\pi}}} e^{2\pi i k(x^{i} + N x^{k+3} x^{j+3})}e^{2\pi l x^{k+3}} \sum_{m\in\mathbb{Z}}e^{2\pi i k mx^{k+3}}\Phi_n^\lambda(\omega_m),\label{eq: phikln}
\eea
with eigenvalues
\bea
& \sigma^2_{p,q}= \frac{q^2}{R^2_{j+3}}+\frac{p^2}{R^2_{k+3}}  ,\\
&\sigma^2_{k,l,n} = \frac{k^2}{R_{i}^2} + (2 n +1)k \,\frac{|\lambda_{\chi}|}{R_{i}},
\eea
where we have defined $\lambda=k\lambda_\chi /R_{i}$, $\omega_m= 2\pi R_{j+3}(x^{j+3} + \frac{m}{N}+\frac{l}{kN})$ and $\Phi_{n}^{\lambda}(z)=|\lambda|^{1/4}\Phi_n(|\lambda|^{1/2}z)$ with $\Phi_n$ being the Hermite polynomial. The integers have the following ranges
\be
p,q \in \mathbb{Z}^2,\,\,\,k\in \mathbb{Z}/\{0\},\,\,\, n\in \mathbb{N},\,\, \,l=0,1,\dots,|k|-1.
\ee

Moreover, the exact one-forms are given by the complete set $\{d\phi_i\}$. Based on the approach presented in section \ref{s:direct}, we observe that the sets of co-exact one-forms, denoted by $S_i$ and $T_i$ and defined in equation \eqref{eq: BandC}, exhibit closure properties when subjected to the $\star d$ operator. \newline 
Substituting the scalar eigenforms \eqref{eq: phipq} and \eqref{eq: phikln} into the sets $S_i$ and $T_i$, we obtain the eigenforms $U_{i}^{\pm}$ defined in \eqref{eq: Dpm}. Also, the corresponding eigenvalues $\lambda^{\pm}_i$ can be obtained from \eqref{eq: lambdapm}.

Using the first tower of scalar eigenforms, $\phi_{p,q}$, we acquire the following results:
\vspace{1em}

$\text{For}\,\, p,q \in \mathbb{Z}^2/\{0,0\}:$
\bea
&U^{\pm}_{p,q} = \frac{2\pi\,\phi_{p,q}}{\sqrt{(\lambda^{\pm}_{p,q})^2 +\sigma_{p,q}^2 }} \left(   \frac{(\lambda^{\pm}_{p,q})^2-\sigma_{p,q}^2 }{i \lambda_{\chi}}\,\,R_{i} \,\eta^{i}- \frac{R_{j+3} }{R_{k+3}}\,p\,\eta^{j+3}+\frac{R_{k+3}}{R_{j+3}}\,q\,\eta^{k+3}\right),\nonumber\\
&(\lambda^\pm_{p,q})^2 = \frac{q^2}{R^2_{j+3}}+\frac{p^2}{R^2_{k+3}}  + \frac{ \lambda_\chi^2}{2} \pm \frac{ \lambda_\chi}{2}\sqrt{ \lambda_\chi^2 +\frac{4\,q^2}{R^2_{j+3}}+\frac{4\,p^2}{R^2_{k+3}}   } \,. 
\eea

For $p=q=0$:
\bea
& U^{-}=0, \hspace{1em} U^{+}_{0,0} = \frac{2\pi R_{i}}{\sqrt{V}}\,\eta^{i},\\
&\lambda^{+}_{0,0} = \lambda_{\chi}^2.
\eea

Comparing with the analysis carried out in \cite{Andriot:2018tmb} one can see that the above results reproduce exactly half of the co-exact one-form spectrum (see \cite{Andriot:2018tmb}, eq. (2.34) and Table 1).

Finally, for the second type of eigen-scalars, $\phi_{k,l,n}$, due to its very intricate expression (see eq. (2.59) from \cite{Andriot:2018tmb} for details), let us focus exclusively on the resulting eigenvalues and their degeneracy. Substituting again in \eqref{eq: lambdapm} we obtain 

\be
(\lambda_{k,l,n}^{\pm})^2 = \frac{k^2}{R_{i}^2} + \frac{(2n +1)k |\lambda_{\chi}|}{R_{i}} + \frac{\lambda_\chi^2}{2} \pm \frac{\lambda_\chi}{2}\sqrt{\lambda_\chi^2 + \frac{4k^2}{R_{i}^2} + \frac{4(2n +1)k |\lambda_{\chi}|}{R_{i}}},\,\,\, n\in\mathbb{N}^*. \label{eq: lambdakln}
\ee

Therefore, we observe an exact correspondence with the remaining spectrum. It is important to note that we have shifted $n$ to $n+1$ in \eqref{eq: lambdakln}. This is because the solution \eqref{eq: lambdakln} with $n=0$ is in fact trivial using our method, and the matching with \cite{Andriot:2018tmb} occurs at our $n\geq1$.

	\clearpage{\pagestyle{empty}\cleardoublepage}


	\addcontentsline{toc}{chapter}{\bibname}

	\bibliographystyle{Bibliography/JHEP2015}
	\bibliography{Bibliography/referencesZatti}

\providecommand{\href}[2]{#2}\begingroup\raggedright\begin{thebibliography}{100}

\bibitem{Cano:2021nzo}
P.~A. Cano, T.~Ort\'\i{}n, A.~Ruip\'erez and M.~Zatti,
  \emph{{Non-supersymmetric black holes with \ensuremath{\alpha}'
  corrections}}, \href{https://doi.org/10.1007/JHEP03(2022)103}{\emph{JHEP}
  {\bfseries 03} (2022) 103}
  [\href{https://arxiv.org/abs/2111.15579}{{\ttfamily 2111.15579}}].

\bibitem{Ortin:2021win}
T.~Ort\'\i{}n, A.~Ruip\'erez and M.~Zatti, \emph{{Extremal stringy black holes
  in equilibrium at first order in $\alpha'$}},
  \href{https://doi.org/10.21468/SciPostPhysCore.6.4.072}{\emph{SciPost Phys.
  Core} {\bfseries 6} (2023) 072}
  [\href{https://arxiv.org/abs/2112.12764}{{\ttfamily 2112.12764}}].

\bibitem{Marchesano:2022rpr}
F.~Marchesano, J.~Quirant and M.~Zatti, \emph{{New instabilities for
  non-supersymmetric AdS$_{4}$ orientifold vacua}},
  \href{https://doi.org/10.1007/JHEP10(2022)026}{\emph{JHEP} {\bfseries 10}
  (2022) 026} [\href{https://arxiv.org/abs/2207.14285}{{\ttfamily
  2207.14285}}].

\bibitem{Cano:2022tmn}
P.~A. Cano, T.~Ort\'\i{}n, A.~Ruip\'erez and M.~Zatti, \emph{{Non-extremal,
  \ensuremath{\alpha}'-corrected black holes in 5-dimensional heterotic
  superstring theory}},
  \href{https://doi.org/10.1007/JHEP12(2022)150}{\emph{JHEP} {\bfseries 12}
  (2022) 150} [\href{https://arxiv.org/abs/2210.01861}{{\ttfamily
  2210.01861}}].

\bibitem{Ballesteros:2023iqb}
R.~Ballesteros, C.~G\'omez-Fayr\'en, T.~Ort\'\i{}n and M.~Zatti, \emph{{On
  scalar charges and black hole thermodynamics}},
  \href{https://doi.org/10.1007/JHEP05(2023)158}{\emph{JHEP} {\bfseries 05}
  (2023) 158} [\href{https://arxiv.org/abs/2302.11630}{{\ttfamily
  2302.11630}}].

\bibitem{Gomez-Fayren:2023wxk}
C.~Gomez-Fayren, P.~Meessen, T.~Ortin and M.~Zatti, \emph{{Wald entropy in
  Kaluza-Klein black holes}},
  \href{https://doi.org/10.1007/JHEP08(2023)039}{\emph{JHEP} {\bfseries 08}
  (2023) 039} [\href{https://arxiv.org/abs/2305.01742}{{\ttfamily
  2305.01742}}].

\bibitem{Casas:2023wlo}
G.~F. Casas, F.~Marchesano and M.~Zatti, \emph{{Torsion in cohomology and
  dimensional reduction}},
  \href{https://doi.org/10.1007/JHEP09(2023)061}{\emph{JHEP} {\bfseries 09}
  (2023) 061} [\href{https://arxiv.org/abs/2306.14959}{{\ttfamily
  2306.14959}}].

\bibitem{Zatti:2023oiq}
M.~Zatti, \emph{{$\alpha'$ corrections to 4-dimensional non-extremal stringy
  black holes}}, \href{https://doi.org/10.1007/JHEP11(2023)185}{\emph{JHEP}
  {\bfseries 11} (2023) 185}
  [\href{https://arxiv.org/abs/2308.12879}{{\ttfamily 2308.12879}}].

\bibitem{Massai:2023cis}
S.~Massai, A.~Ruip\'erez and M.~Zatti, \emph{{Revisiting $\alpha'$ corrections
  to heterotic two-charge black holes}},
  \href{https://doi.org/10.1007/JHEP04(2024)150}{\emph{JHEP} {\bfseries 04}
  (2024) 150} [\href{https://arxiv.org/abs/2311.03308}{{\ttfamily
  2311.03308}}].

\bibitem{Fernandez-Melgarejo:2023kwk}
J.~J. Fernandez-Melgarejo, G.~Giorgi, C.~Gomez-Fayren, T.~Ortin and M.~Zatti,
  \emph{{Democratic actions with scalar fields: symmetric sigma models,
  supergravity actions and the effective theory of the type IIB superstring}},
  \href{https://arxiv.org/abs/2401.00549}{{\ttfamily 2401.00549}}.

\bibitem{Gomez-Fayren:2024cpl}
C.~G\'omez-Fayr\'en, T.~Ort\'\i{}n and M.~Zatti, \emph{{Gravitational
  higher-form symmetries and the origin of hidden symmetries in Kaluza-Klein
  compactifications}},
  \href{https://doi.org/10.21468/SciPostPhysCore.8.1.010}{\emph{SciPost Phys.
  Core} {\bfseries 8} (2025) 010}
  [\href{https://arxiv.org/abs/2405.16706}{{\ttfamily 2405.16706}}].

\bibitem{Ortin:2015hya}
T.~Ortin, \emph{{Gravity and Strings}}, Cambridge Monographs on Mathematical
  Physics. Cambridge University Press, 2nd ed.~ed., 7, 2015,
  \href{https://doi.org/10.1017/CBO9781139019750}{10.1017/CBO9781139019750}.

\bibitem{Blumenhagen:2013fgp}
R.~Blumenhagen, D.~L\"ust and S.~Theisen, \emph{{Basic concepts of string
  theory}}, Theoretical and Mathematical Physics. Springer, Heidelberg,
  Germany, 2013,
  \href{https://doi.org/10.1007/978-3-642-29497-6}{10.1007/978-3-642-29497-6}.

\bibitem{Ibanez:2012zz}
L.~E. Ibanez and A.~M. Uranga, \emph{{String theory and particle physics: An
  introduction to string phenomenology}}. Cambridge University Press, 2, 2012.

\bibitem{Hull1987}
C.~M. Hull, \emph{Lectures on Non-Linear Sigma-Models and Strings},
  pp.~77--168.
\newblock Springer US, 1987.

\bibitem{Green:2012oqa}
M.~B. Green, J.~H. Schwarz and E.~Witten, \emph{{Superstring Theory Vol. 1}:
  {25th Anniversary Edition}}, Cambridge Monographs on Mathematical Physics.
  Cambridge University Press, 11, 2012,
  \href{https://doi.org/10.1017/CBO9781139248563}{10.1017/CBO9781139248563}.

\bibitem{Green:2012pqa}
M.~B. Green, J.~H. Schwarz and E.~Witten, \emph{{Superstring Theory Vol. 2}:
  {25th Anniversary Edition}}, Cambridge Monographs on Mathematical Physics.
  Cambridge University Press, 11, 2012,
  \href{https://doi.org/10.1017/CBO9781139248570}{10.1017/CBO9781139248570}.

\bibitem{Polchinski:1998rq}
J.~Polchinski, \emph{{String theory. Vol. 1: An introduction to the bosonic
  string}}, Cambridge Monographs on Mathematical Physics. Cambridge University
  Press, 12, 2007,
  \href{https://doi.org/10.1017/CBO9780511816079}{10.1017/CBO9780511816079}.

\bibitem{Polchinski:1998rr}
J.~Polchinski, \emph{{String theory. Vol. 2: Superstring theory and beyond}},
  Cambridge Monographs on Mathematical Physics. Cambridge University Press, 12,
  2007,
  \href{https://doi.org/10.1017/CBO9780511618123}{10.1017/CBO9780511618123}.

\bibitem{Buscher:1987sk}
T.~H. Buscher, \emph{{A Symmetry of the String Background Field Equations}},
  \href{https://doi.org/10.1016/0370-2693(87)90769-6}{\emph{Phys. Lett. B}
  {\bfseries 194} (1987) 59}.

\bibitem{Buscher:1987qj}
T.~H. Buscher, \emph{{Path Integral Derivation of Quantum Duality in Nonlinear
  Sigma Models}},
  \href{https://doi.org/10.1016/0370-2693(88)90602-8}{\emph{Phys. Lett. B}
  {\bfseries 201} (1988) 466}.

\bibitem{Palti:2019pca}
E.~Palti, \emph{{The Swampland: Introduction and Review}},
  \href{https://doi.org/10.1002/prop.201900037}{\emph{Fortsch. Phys.}
  {\bfseries 67} (2019) 1900037}
  [\href{https://arxiv.org/abs/1903.06239}{{\ttfamily 1903.06239}}].

\bibitem{vanBeest:2021lhn}
M.~van Beest, J.~Calder\'on-Infante, D.~Mirfendereski and I.~Valenzuela,
  \emph{{Lectures on the Swampland Program in String Compactifications}},
  \href{https://doi.org/10.1016/j.physrep.2022.09.002}{\emph{Phys. Rept.}
  {\bfseries 989} (2022) 1} [\href{https://arxiv.org/abs/2102.01111}{{\ttfamily
  2102.01111}}].

\bibitem{Harlow:2018jwu}
D.~Harlow and H.~Ooguri, \emph{{Constraints on Symmetries from Holography}},
  \href{https://doi.org/10.1103/PhysRevLett.122.191601}{\emph{Phys. Rev. Lett.}
  {\bfseries 122} (2019) 191601}
  [\href{https://arxiv.org/abs/1810.05337}{{\ttfamily 1810.05337}}].

\bibitem{Susskind:1995da}
L.~Susskind, \emph{{Trouble for remnants}},
  \href{https://arxiv.org/abs/hep-th/9501106}{{\ttfamily hep-th/9501106}}.

\bibitem{Arkani-Hamed:2006emk}
N.~Arkani-Hamed, L.~Motl, A.~Nicolis and C.~Vafa, \emph{{The String landscape,
  black holes and gravity as the weakest force}},
  \href{https://doi.org/10.1088/1126-6708/2007/06/060}{\emph{JHEP} {\bfseries
  06} (2007) 060} [\href{https://arxiv.org/abs/hep-th/0601001}{{\ttfamily
  hep-th/0601001}}].

\bibitem{Banks:1992is}
T.~Banks, M.~O'Loughlin and A.~Strominger, \emph{{Black hole remnants and the
  information puzzle}},
  \href{https://doi.org/10.1103/PhysRevD.47.4476}{\emph{Phys. Rev. D}
  {\bfseries 47} (1993) 4476}
  [\href{https://arxiv.org/abs/hep-th/9211030}{{\ttfamily hep-th/9211030}}].

\bibitem{Giddings:1993km}
S.~B. Giddings, \emph{{Constraints on black hole remnants}},
  \href{https://doi.org/10.1103/PhysRevD.49.947}{\emph{Phys. Rev. D} {\bfseries
  49} (1994) 947} [\href{https://arxiv.org/abs/hep-th/9304027}{{\ttfamily
  hep-th/9304027}}].

\bibitem{Ooguri:2006in}
H.~Ooguri and C.~Vafa, \emph{{On the Geometry of the String Landscape and the
  Swampland}},
  \href{https://doi.org/10.1016/j.nuclphysb.2006.10.033}{\emph{Nucl. Phys. B}
  {\bfseries 766} (2007) 21}
  [\href{https://arxiv.org/abs/hep-th/0605264}{{\ttfamily hep-th/0605264}}].

\bibitem{Calderon-Infante:2023ler}
J.~Calder\'on-Infante, A.~Castellano, A.~Herr\'aez and L.~E. Ib\'a\~nez,
  \emph{{Entropy bounds and the species scale distance conjecture}},
  \href{https://doi.org/10.1007/JHEP01(2024)039}{\emph{JHEP} {\bfseries 01}
  (2024) 039} [\href{https://arxiv.org/abs/2306.16450}{{\ttfamily
  2306.16450}}].

\bibitem{vandeHeisteeg:2023ubh}
D.~van~de Heisteeg, C.~Vafa and M.~Wiesner, \emph{{Bounds on Species Scale and
  the Distance Conjecture}},
  \href{https://doi.org/10.1002/prop.202300143}{\emph{Fortsch. Phys.}
  {\bfseries 71} (2023) 2300143}
  [\href{https://arxiv.org/abs/2303.13580}{{\ttfamily 2303.13580}}].

\bibitem{Basile:2023blg}
I.~Basile, D.~L\"ust and C.~Montella, \emph{{Shedding black hole light on the
  emergent string conjecture}},
  \href{https://arxiv.org/abs/2311.12113}{{\ttfamily 2311.12113}}.

\bibitem{Ooguri:2016pdq}
H.~Ooguri and C.~Vafa, \emph{{Non-supersymmetric AdS and the Swampland}},
  \href{https://doi.org/10.4310/ATMP.2017.v21.n7.a8}{\emph{Adv. Theor. Math.
  Phys.} {\bfseries 21} (2017) 1787}
  [\href{https://arxiv.org/abs/1610.01533}{{\ttfamily 1610.01533}}].

\bibitem{Lust:2019zwm}
D.~L\"ust, E.~Palti and C.~Vafa, \emph{{AdS and the Swampland}},
  \href{https://doi.org/10.1016/j.physletb.2019.134867}{\emph{Phys. Lett. B}
  {\bfseries 797} (2019) 134867}
  [\href{https://arxiv.org/abs/1906.05225}{{\ttfamily 1906.05225}}].

\bibitem{Coudarchet:2023mfs}
T.~Coudarchet, \emph{{Hiding the extra dimensions: A review on scale separation
  in string theory}},
  \href{https://doi.org/10.1016/j.physrep.2024.02.003}{\emph{Phys. Rept.}
  {\bfseries 1064} (2024) 1}
  [\href{https://arxiv.org/abs/2311.12105}{{\ttfamily 2311.12105}}].

\bibitem{Penrose:1968ar}
R.~Penrose, \emph{{Structure of space-time}},  in \emph{{Battelle Rencontres}},
  pp.~121--235, 1968.

\bibitem{Hawking:1971vc}
S.~W. Hawking, \emph{{Black holes in general relativity}},
  \href{https://doi.org/10.1007/BF01877517}{\emph{Commun. Math. Phys.}
  {\bfseries 25} (1972) 152}.

\bibitem{Hawking:1973uf}
S.~W. Hawking and G.~F.~R. Ellis, \emph{{The Large Scale Structure of
  Space-Time}}, Cambridge Monographs on Mathematical Physics. Cambridge
  University Press, 2, 2023,
  \href{https://doi.org/10.1017/9781009253161}{10.1017/9781009253161}.

\bibitem{Schwarzschild:1916uq}
K.~Schwarzschild, \emph{{On the gravitational field of a mass point according
  to Einstein's theory}}, {\emph{Sitzungsber. Preuss. Akad. Wiss. Berlin (Math.
  Phys. )} {\bfseries 1916} (1916) 189}
  [\href{https://arxiv.org/abs/physics/9905030}{{\ttfamily physics/9905030}}].

\bibitem{Birkhoff:1923}
G.~D. {Birkhoff} and R.~E. {Langer}, \emph{{Relativity and modern physics}}.
  Harvard University Press, 1923.

\bibitem{Israel:1967wq}
W.~Israel, \emph{{Event horizons in static vacuum space-times}},
  \href{https://doi.org/10.1103/PhysRev.164.1776}{\emph{Phys. Rev.} {\bfseries
  164} (1967) 1776}.

\bibitem{Kerr:1963ud}
R.~P. Kerr, \emph{{Gravitational field of a spinning mass as an example of
  algebraically special metrics}},
  \href{https://doi.org/10.1103/PhysRevLett.11.237}{\emph{Phys. Rev. Lett.}
  {\bfseries 11} (1963) 237}.

\bibitem{Carter:1971zc}
B.~Carter, \emph{{Axisymmetric Black Hole Has Only Two Degrees of Freedom}},
  \href{https://doi.org/10.1103/PhysRevLett.26.331}{\emph{Phys. Rev. Lett.}
  {\bfseries 26} (1971) 331}.

\bibitem{Robinson:1975bv}
D.~C. Robinson, \emph{{Uniqueness of the Kerr black hole}},
  \href{https://doi.org/10.1103/PhysRevLett.34.905}{\emph{Phys. Rev. Lett.}
  {\bfseries 34} (1975) 905}.

\bibitem{Israel:1967za}
W.~Israel, \emph{{Event horizons in static electrovac space-times}},
  \href{https://doi.org/10.1007/BF01645859}{\emph{Commun. Math. Phys.}
  {\bfseries 8} (1968) 245}.

\bibitem{Reissner:1916cle}
H.~Reissner, \emph{{\"Uber die Eigengravitation des elektrischen Feldes nach
  der Einsteinschen Theorie}},
  \href{https://doi.org/10.1002/andp.19163550905}{\emph{Annalen Phys.}
  {\bfseries 355} (1916) 106}.

\bibitem{Nordstrom:1918}
G.~{Nordstr{\"o}m}, \emph{{On the Energy of the Gravitation field in Einstein's
  Theory}}, {\emph{Koninklijke Nederlandse Akademie van Wetenschappen
  Proceedings Series B Physical Sciences} {\bfseries 20} (1918) 1238}.

\bibitem{Papaetrou:1947ib}
A.~Papaetrou, \emph{{A Static solution of the equations of the gravitational
  field for an arbitrary charge distribution}}, {\emph{Proc. Roy. Irish Acad.
  A} {\bfseries 51} (1947) 191}.

\bibitem{Ruffini:1971bza}
R.~Ruffini and J.~A. Wheeler, \emph{{Introducing the black hole}},
  \href{https://doi.org/10.1063/1.3022513}{\emph{Phys. Today} {\bfseries 24}
  (1971) 30}.

\bibitem{Bardeen:1973gs}
J.~M. Bardeen, B.~Carter and S.~W. Hawking, \emph{{The Four laws of black hole
  mechanics}}, \href{https://doi.org/10.1007/BF01645742}{\emph{Commun. Math.
  Phys.} {\bfseries 31} (1973) 161}.

\bibitem{Hawking:1972hy}
S.~W. Hawking and J.~B. Hartle, \emph{{Energy and angular momentum flow into a
  black hole}}, \href{https://doi.org/10.1007/BF01645515}{\emph{Commun. Math.
  Phys.} {\bfseries 27} (1972) 283}.

\bibitem{Sudarsky:1992ty}
D.~Sudarsky and R.~M. Wald, \emph{{Extrema of mass, stationarity, and
  staticity, and solutions to the Einstein Yang-Mills equations}},
  \href{https://doi.org/10.1103/PhysRevD.46.1453}{\emph{Phys. Rev. D}
  {\bfseries 46} (1992) 1453}.

\bibitem{Hawking:1971tu}
S.~W. Hawking, \emph{{Gravitational radiation from colliding black holes}},
  \href{https://doi.org/10.1103/PhysRevLett.26.1344}{\emph{Phys. Rev. Lett.}
  {\bfseries 26} (1971) 1344}.

\bibitem{Israel:1986gqz}
W.~Israel, \emph{{Third Law of Black-Hole Dynamics: A Formulation and Proof}},
  \href{https://doi.org/10.1103/PhysRevLett.57.397}{\emph{Phys. Rev. Lett.}
  {\bfseries 57} (1986) 397}.

\bibitem{Kehle:2022uvc}
C.~Kehle and R.~Unger, \emph{{Gravitational collapse to extremal black holes
  and the third law of black hole thermodynamics}},
  \href{https://arxiv.org/abs/2211.15742}{{\ttfamily 2211.15742}}.

\bibitem{Bekenstein:1973ur}
J.~D. Bekenstein, \emph{{Black holes and entropy}},
  \href{https://doi.org/10.1103/PhysRevD.7.2333}{\emph{Phys. Rev. D} {\bfseries
  7} (1973) 2333}.

\bibitem{Hawking:1974rv}
S.~W. Hawking, \emph{{Black hole explosions}},
  \href{https://doi.org/10.1038/248030a0}{\emph{Nature} {\bfseries 248} (1974)
  30}.

\bibitem{Hawking:1975vcx}
S.~W. Hawking, \emph{{Particle Creation by Black Holes}},
  \href{https://doi.org/10.1007/BF02345020}{\emph{Commun. Math. Phys.}
  {\bfseries 43} (1975) 199}.

\bibitem{Emparan:2008eg}
R.~Emparan and H.~S. Reall, \emph{{Black Holes in Higher Dimensions}},
  \href{https://doi.org/10.12942/lrr-2008-6}{\emph{Living Rev. Rel.} {\bfseries
  11} (2008) 6} [\href{https://arxiv.org/abs/0801.3471}{{\ttfamily
  0801.3471}}].

\bibitem{Horowitz:2012nnc}
G.~T. Horowitz, ed., \emph{{Black holes in higher dimensions}}. Cambridge Univ.
  Pr., Cambridge, UK, 2012.

\bibitem{Wald:1993nt}
R.~M. Wald, \emph{{Black hole entropy is the Noether charge}},
  \href{https://doi.org/10.1103/PhysRevD.48.R3427}{\emph{Phys. Rev. D}
  {\bfseries 48} (1993) R3427}
  [\href{https://arxiv.org/abs/gr-qc/9307038}{{\ttfamily gr-qc/9307038}}].

\bibitem{Iyer:1994ys}
V.~Iyer and R.~M. Wald, \emph{{Some properties of Noether charge and a proposal
  for dynamical black hole entropy}},
  \href{https://doi.org/10.1103/PhysRevD.50.846}{\emph{Phys. Rev. D} {\bfseries
  50} (1994) 846} [\href{https://arxiv.org/abs/gr-qc/9403028}{{\ttfamily
  gr-qc/9403028}}].

\bibitem{Wall:2015raa}
A.~C. Wall, \emph{{A Second Law for Higher Curvature Gravity}},
  \href{https://doi.org/10.1142/S0218271815440149}{\emph{Int. J. Mod. Phys. D}
  {\bfseries 24} (2015) 1544014}
  [\href{https://arxiv.org/abs/1504.08040}{{\ttfamily 1504.08040}}].

\bibitem{Hollands:2022fkn}
S.~Hollands, A.~D. Kov\'acs and H.~S. Reall, \emph{{The second law of black
  hole mechanics in effective field theory}},
  \href{https://doi.org/10.1007/JHEP08(2022)258}{\emph{JHEP} {\bfseries 08}
  (2022) 258} [\href{https://arxiv.org/abs/2205.15341}{{\ttfamily
  2205.15341}}].

\bibitem{Jacobson:1993vj}
T.~Jacobson, G.~Kang and R.~C. Myers, \emph{{On black hole entropy}},
  \href{https://doi.org/10.1103/PhysRevD.49.6587}{\emph{Phys. Rev. D}
  {\bfseries 49} (1994) 6587}
  [\href{https://arxiv.org/abs/gr-qc/9312023}{{\ttfamily gr-qc/9312023}}].

\bibitem{Komar:1958wp}
A.~Komar, \emph{{Covariant conservation laws in general relativity}},
  \href{https://doi.org/10.1103/PhysRev.113.934}{\emph{Phys. Rev.} {\bfseries
  113} (1959) 934}.

\bibitem{PereniguezRodriguez:2022eal}
D.~Pere\~n\'\i{}guez Rodr\'\i{}guez, \emph{{Classical and Stringy Properties of
  Black Holes}}, Ph.D. thesis, U. Autonoma, Madrid (main), Madrid, Autonoma U.,
  9, 2022.

\bibitem{Elgood:2020svt}
Z.~Elgood, P.~Meessen and T.~Ort\'\i{}n, \emph{{The first law of black hole
  mechanics in the Einstein-Maxwell theory revisited}},
  \href{https://doi.org/10.1007/JHEP09(2020)026}{\emph{JHEP} {\bfseries 09}
  (2020) 026} [\href{https://arxiv.org/abs/2006.02792}{{\ttfamily
  2006.02792}}].

\bibitem{Elgood:2020mdx}
Z.~Elgood, D.~Mitsios, T.~Ort\'\i{}n and D.~Pere\~n\'\i{}guez, \emph{{The first
  law of heterotic stringy black hole mechanics at zeroth order in
  \ensuremath{\alpha}'}},
  \href{https://doi.org/10.1007/JHEP07(2021)007}{\emph{JHEP} {\bfseries 07}
  (2021) 007} [\href{https://arxiv.org/abs/2012.13323}{{\ttfamily
  2012.13323}}].

\bibitem{Elgood:2020nls}
Z.~Elgood, T.~Ort\'\i{}n and D.~Pere\~n\'\i{}guez, \emph{{The first law and
  Wald entropy formula of heterotic stringy black holes at first order in
  $\alpha'$}}, \href{https://doi.org/10.1007/JHEP05(2021)110}{\emph{JHEP}
  {\bfseries 05} (2021) 110}
  [\href{https://arxiv.org/abs/2012.14892}{{\ttfamily 2012.14892}}].

\bibitem{Meessen:2022hcg}
P.~Meessen, D.~Mitsios and T.~Ort\'\i{}n, \emph{{Black hole chemistry, the
  cosmological constant and the embedding tensor}},
  \href{https://doi.org/10.1007/JHEP12(2022)155}{\emph{JHEP} {\bfseries 12}
  (2022) 155} [\href{https://arxiv.org/abs/2203.13588}{{\ttfamily
  2203.13588}}].

\bibitem{Mitsios:2021zrn}
D.~Mitsios, T.~Ort\'\i{}n and D.~Pere\~n\'\i{}guez, \emph{{Komar integral and
  Smarr formula for axion-dilaton black holes versus S duality}},
  \href{https://doi.org/10.1007/JHEP08(2021)019}{\emph{JHEP} {\bfseries 08}
  (2021) 019} [\href{https://arxiv.org/abs/2106.07495}{{\ttfamily
  2106.07495}}].

\bibitem{Ortin:2022uxa}
T.~Ortin and D.~Pere\~niguez, \emph{{Magnetic charges and Wald entropy}},
  \href{https://doi.org/10.1007/JHEP11(2022)081}{\emph{JHEP} {\bfseries 11}
  (2022) 081} [\href{https://arxiv.org/abs/2207.12008}{{\ttfamily
  2207.12008}}].

\bibitem{Bandos:2023zbs}
I.~Bandos and T.~Ortin, \emph{{Noether-Wald charge in supergravity: the
  fermionic contribution}},  \href{https://arxiv.org/abs/2305.10617}{{\ttfamily
  2305.10617}}.

\bibitem{Ballestaros:2023ipa}
R.~Ballesteros and T.~Ort\'\i{}n, \emph{{Hairy black holes, scalar charges and
  extended thermodynamics}},
  \href{https://arxiv.org/abs/2308.04994}{{\ttfamily 2308.04994}}.

\bibitem{Gibbons:1996af}
G.~W. Gibbons, R.~Kallosh and B.~Kol, \emph{{Moduli, scalar charges, and the
  first law of black hole thermodynamics}},
  \href{https://doi.org/10.1103/PhysRevLett.77.4992}{\emph{Phys. Rev. Lett.}
  {\bfseries 77} (1996) 4992}
  [\href{https://arxiv.org/abs/hep-th/9607108}{{\ttfamily hep-th/9607108}}].

\bibitem{Prabhu:2015vua}
K.~Prabhu, \emph{{The First Law of Black Hole Mechanics for Fields with
  Internal Gauge Freedom}},
  \href{https://doi.org/10.1088/1361-6382/aa536b}{\emph{Class. Quant. Grav.}
  {\bfseries 34} (2017) 035011}
  [\href{https://arxiv.org/abs/1511.00388}{{\ttfamily 1511.00388}}].

\bibitem{Ortin:2002qb}
T.~Ortin, \emph{{A Note on Lie-Lorentz derivatives}},
  \href{https://doi.org/10.1088/0264-9381/19/15/101}{\emph{Class. Quant. Grav.}
  {\bfseries 19} (2002) L143}
  [\href{https://arxiv.org/abs/hep-th/0206159}{{\ttfamily hep-th/0206159}}].

\bibitem{kn:Lich}
A.~Lichnerowicz, \emph{Spineurs harmoniques}, {\emph{C. R. Acad. Sci. Paris}
  {\bfseries 257} (1963) 7}.

\bibitem{kn:Kos}
Y.~Kosmann, \emph{D\'eriv\'ees de lie des spineurs}, {\emph{C. R. Acad. Sci.
  Paris S\'er. A} {\bfseries 262} (1966) A289}.

\bibitem{kn:Kos2}
Y.~Kosmann, \emph{D\'eriv\'ees de lie des spineurs},
  \href{https://doi.org/10.1007/BF02428822}{\emph{Annali Mat. Pura Appl.}
  {\bfseries 91} (1972) 317}.

\bibitem{Hurley:1994cfa}
D.~J. Hurley and M.~A. Vandyck, \emph{{On the concepts of Lie and covariant
  derivatives of spinors. Part 1}},
  \href{https://doi.org/10.1088/0305-4470/27/13/030}{\emph{J. Phys. A}
  {\bfseries 27} (1994) 4569}.

\bibitem{Vandyck:1988ei}
M.~A.~J. Vandyck, \emph{{On the problem of space-time symmetries in the theory
  of supergravity}}, \href{https://doi.org/10.1007/BF00759185}{\emph{Gen. Rel.
  Grav.} {\bfseries 20} (1988) 261}.

\bibitem{Vandyck:1988gc}
M.~A. Vandyck, \emph{{On the problem of space-time symmetries in the theory of
  supergravity 2: N=2 supergravity and spinorial lie derivatives}},
  \href{https://doi.org/10.1007/BF00760090}{\emph{Gen. Rel. Grav.} {\bfseries
  20} (1988) 905}.

\bibitem{Barnich:2001jy}
G.~Barnich and F.~Brandt, \emph{{Covariant theory of asymptotic symmetries,
  conservation laws and central charges}},
  \href{https://doi.org/10.1016/S0550-3213(02)00251-1}{\emph{Nucl. Phys. B}
  {\bfseries 633} (2002) 3}
  [\href{https://arxiv.org/abs/hep-th/0111246}{{\ttfamily hep-th/0111246}}].

\bibitem{Scherk:1979zr}
J.~Scherk and J.~H. Schwarz, \emph{{How to Get Masses from Extra Dimensions}},
  \href{https://doi.org/10.1016/0550-3213(79)90592-3}{\emph{Nucl. Phys. B}
  {\bfseries 153} (1979) 61}.

\bibitem{Heusler:1993cj}
M.~Heusler and N.~Straumann, \emph{{The First law of black hole physics for a
  class of nonlinear matter models}},
  \href{https://doi.org/10.1088/0264-9381/10/7/008}{\emph{Class. Quant. Grav.}
  {\bfseries 10} (1993) 1299}.

\bibitem{Coleman:1991ku}
S.~R. Coleman, J.~Preskill and F.~Wilczek, \emph{{Quantum hair on black
  holes}}, \href{https://doi.org/10.1016/0550-3213(92)90008-Y}{\emph{Nucl.
  Phys. B} {\bfseries 378} (1992) 175}
  [\href{https://arxiv.org/abs/hep-th/9201059}{{\ttfamily hep-th/9201059}}].

\bibitem{Janis:1968zz}
A.~I. Janis, E.~T. Newman and J.~Winicour, \emph{{Reality of the Schwarzschild
  Singularity}}, \href{https://doi.org/10.1103/PhysRevLett.20.878}{\emph{Phys.
  Rev. Lett.} {\bfseries 20} (1968) 878}.

\bibitem{Agnese:1994zx}
A.~G. Agnese and M.~La~Camera, \emph{{General spherically symmetric solutions
  in charged dilaton gravity}},
  \href{https://doi.org/10.1103/PhysRevD.49.2126}{\emph{Phys. Rev. D}
  {\bfseries 49} (1994) 2126}.

\bibitem{Gaillard:1981rj}
M.~K. Gaillard and B.~Zumino, \emph{{Duality Rotations for Interacting
  Fields}}, \href{https://doi.org/10.1016/0550-3213(81)90527-7}{\emph{Nucl.
  Phys. B} {\bfseries 193} (1981) 221}.

\bibitem{Pacilio:2018gom}
C.~Pacilio, \emph{{Scalar charge of black holes in Einstein-Maxwell-dilaton
  theory}}, \href{https://doi.org/10.1103/PhysRevD.98.064055}{\emph{Phys. Rev.
  D} {\bfseries 98} (2018) 064055}
  [\href{https://arxiv.org/abs/1806.10238}{{\ttfamily 1806.10238}}].

\bibitem{Bandos:2016smv}
I.~A. Bandos and T.~Ortin, \emph{{On the dualization of scalars into (d
  \ensuremath{-} 2)-forms in supergravity. Momentum maps, R-symmetry and gauged
  supergravity}}, \href{https://doi.org/10.1007/JHEP08(2016)135}{\emph{JHEP}
  {\bfseries 08} (2016) 135}
  [\href{https://arxiv.org/abs/1605.05559}{{\ttfamily 1605.05559}}].

\bibitem{Astefanesei:2018vga}
D.~Astefanesei, R.~Ballesteros, D.~Choque and R.~Rojas, \emph{{Scalar charges
  and the first law of black hole thermodynamics}},
  \href{https://doi.org/10.1016/j.physletb.2018.05.005}{\emph{Phys. Lett. B}
  {\bfseries 782} (2018) 47}
  [\href{https://arxiv.org/abs/1803.11317}{{\ttfamily 1803.11317}}].

\bibitem{Lee:1990nz}
J.~Lee and R.~M. Wald, \emph{{Local symmetries and constraints}},
  \href{https://doi.org/10.1063/1.528801}{\emph{J. Math. Phys.} {\bfseries 31}
  (1990) 725}.

\bibitem{Strominger:1996sh}
A.~Strominger and C.~Vafa, \emph{{Microscopic origin of the Bekenstein-Hawking
  entropy}}, \href{https://doi.org/10.1016/0370-2693(96)00345-0}{\emph{Phys.
  Lett. B} {\bfseries 379} (1996) 99}
  [\href{https://arxiv.org/abs/hep-th/9601029}{{\ttfamily hep-th/9601029}}].

\bibitem{Horowitz:1996ay}
G.~T. Horowitz, J.~M. Maldacena and A.~Strominger, \emph{{Nonextremal black
  hole microstates and U duality}},
  \href{https://doi.org/10.1016/0370-2693(96)00738-1}{\emph{Phys. Lett. B}
  {\bfseries 383} (1996) 151}
  [\href{https://arxiv.org/abs/hep-th/9603109}{{\ttfamily hep-th/9603109}}].

\bibitem{Sen:2007qy}
A.~Sen, \emph{{Black Hole Entropy Function, Attractors and Precision Counting
  of Microstates}}, \href{https://doi.org/10.1007/s10714-008-0626-4}{\emph{Gen.
  Rel. Grav.} {\bfseries 40} (2008) 2249}
  [\href{https://arxiv.org/abs/0708.1270}{{\ttfamily 0708.1270}}].

\bibitem{Castro:2007hc}
A.~Castro, J.~L. Davis, P.~Kraus and F.~Larsen, \emph{{5D Black Holes and
  Strings with Higher Derivatives}},
  \href{https://doi.org/10.1088/1126-6708/2007/06/007}{\emph{JHEP} {\bfseries
  06} (2007) 007} [\href{https://arxiv.org/abs/hep-th/0703087}{{\ttfamily
  hep-th/0703087}}].

\bibitem{Castro:2007ci}
A.~Castro, J.~L. Davis, P.~Kraus and F.~Larsen, \emph{{Precision Entropy of
  Spinning Black Holes}},
  \href{https://doi.org/10.1088/1126-6708/2007/09/003}{\emph{JHEP} {\bfseries
  09} (2007) 003} [\href{https://arxiv.org/abs/0705.1847}{{\ttfamily
  0705.1847}}].

\bibitem{Castro:2008ne}
A.~Castro, J.~L. Davis, P.~Kraus and F.~Larsen, \emph{{String Theory Effects on
  Five-Dimensional Black Hole Physics}},
  \href{https://doi.org/10.1142/S0217751X08039724}{\emph{Int. J. Mod. Phys. A}
  {\bfseries 23} (2008) 613} [\href{https://arxiv.org/abs/0801.1863}{{\ttfamily
  0801.1863}}].

\bibitem{DominisPrester:2008ynb}
P.~Dominis~Prester and T.~Terzic, \emph{{$\alpha'$-exact entropies for BPS and
  non-BPS extremal dyonic black holes in heterotic string theory from
  ten-dimensional supersymmetry}},
  \href{https://doi.org/10.1088/1126-6708/2008/12/088}{\emph{JHEP} {\bfseries
  12} (2008) 088} [\href{https://arxiv.org/abs/0809.4954}{{\ttfamily
  0809.4954}}].

\bibitem{Castro:2008ys}
A.~Castro and S.~Murthy, \emph{{Corrections to the statistical entropy of five
  dimensional black holes}},
  \href{https://doi.org/10.1088/1126-6708/2009/06/024}{\emph{JHEP} {\bfseries
  06} (2009) 024} [\href{https://arxiv.org/abs/0807.0237}{{\ttfamily
  0807.0237}}].

\bibitem{Kutasov:1998zh}
D.~Kutasov, F.~Larsen and R.~G. Leigh, \emph{{String theory in magnetic
  monopole backgrounds}},
  \href{https://doi.org/10.1016/S0550-3213(99)00144-3}{\emph{Nucl. Phys. B}
  {\bfseries 550} (1999) 183}
  [\href{https://arxiv.org/abs/hep-th/9812027}{{\ttfamily hep-th/9812027}}].

\bibitem{Kraus:2005vz}
P.~Kraus and F.~Larsen, \emph{{Microscopic black hole entropy in theories with
  higher derivatives}},
  \href{https://doi.org/10.1088/1126-6708/2005/09/034}{\emph{JHEP} {\bfseries
  09} (2005) 034} [\href{https://arxiv.org/abs/hep-th/0506176}{{\ttfamily
  hep-th/0506176}}].

\bibitem{Cano:2018qev}
P.~A. Cano, P.~Meessen, T.~Ort\'\i{}n and P.~F. Ram\'\i{}rez,
  \emph{{$\alpha'$-corrected black holes in String Theory}},
  \href{https://doi.org/10.1007/JHEP05(2018)110}{\emph{JHEP} {\bfseries 05}
  (2018) 110} [\href{https://arxiv.org/abs/1803.01919}{{\ttfamily
  1803.01919}}].

\bibitem{Kastor:2008xb}
D.~Kastor, \emph{{Komar Integrals in Higher (and Lower) Derivative Gravity}},
  \href{https://doi.org/10.1088/0264-9381/25/17/175007}{\emph{Class. Quant.
  Grav.} {\bfseries 25} (2008) 175007}
  [\href{https://arxiv.org/abs/0804.1832}{{\ttfamily 0804.1832}}].

\bibitem{Ortin:2021ade}
T.~Ort\'\i{}n, \emph{{Komar integrals for theories of higher order in the
  Riemann curvature and black-hole chemistry}},
  \href{https://doi.org/10.1007/JHEP08(2021)023}{\emph{JHEP} {\bfseries 08}
  (2021) 023} [\href{https://arxiv.org/abs/2104.10717}{{\ttfamily
  2104.10717}}].

\bibitem{Cano:2023dyg}
P.~A. Cano and M.~David, \emph{{The extremal Kerr entropy in higher-derivative
  gravities}}, \href{https://doi.org/10.1007/JHEP05(2023)219}{\emph{JHEP}
  {\bfseries 05} (2023) 219}
  [\href{https://arxiv.org/abs/2303.13286}{{\ttfamily 2303.13286}}].

\bibitem{Chimento:2018kop}
S.~Chimento, P.~Meessen, T.~Ortin, P.~F. Ramirez and A.~Ruiperez, \emph{{On a
  family of $\alpha'$-corrected solutions of the Heterotic Superstring
  effective action}},
  \href{https://doi.org/10.1007/JHEP07(2018)080}{\emph{JHEP} {\bfseries 07}
  (2018) 080} [\href{https://arxiv.org/abs/1803.04463}{{\ttfamily
  1803.04463}}].

\bibitem{Cano:2018brq}
P.~A. Cano, S.~Chimento, P.~Meessen, T.~Ort\'\i{}n, P.~F. Ram\'\i{}rez and
  A.~Ruip\'erez, \emph{{Beyond the near-horizon limit: Stringy corrections to
  Heterotic Black Holes}},
  \href{https://doi.org/10.1007/JHEP02(2019)192}{\emph{JHEP} {\bfseries 02}
  (2019) 192} [\href{https://arxiv.org/abs/1808.03651}{{\ttfamily
  1808.03651}}].

\bibitem{Cano:2019ycn}
P.~A. Cano, S.~Chimento, R.~Linares, T.~Ort\'\i{}n and P.~F. Ram\'\i{}rez,
  \emph{{$\alpha'$ corrections of Reissner-Nordstr\"om black holes}},
  \href{https://doi.org/10.1007/JHEP02(2020)031}{\emph{JHEP} {\bfseries 02}
  (2020) 031} [\href{https://arxiv.org/abs/1910.14324}{{\ttfamily
  1910.14324}}].

\bibitem{Cano:2021rey}
P.~A. Cano and A.~Ruip\'erez, \emph{{String gravity in D=4}},
  \href{https://doi.org/10.1103/PhysRevD.105.044022}{\emph{Phys. Rev. D}
  {\bfseries 105} (2022) 044022}
  [\href{https://arxiv.org/abs/2111.04750}{{\ttfamily 2111.04750}}].

\bibitem{Campbell:1991kz}
B.~A. Campbell, N.~Kaloper and K.~A. Olive, \emph{{Classical hair for
  Kerr-Newman black holes in string gravity}},
  \href{https://doi.org/10.1016/0370-2693(92)91452-F}{\emph{Phys. Lett. B}
  {\bfseries 285} (1992) 199}.

\bibitem{Natsuume:1994hd}
M.~Natsuume, \emph{{Higher order correction to the GHS string black hole}},
  \href{https://doi.org/10.1103/PhysRevD.50.3949}{\emph{Phys. Rev. D}
  {\bfseries 50} (1994) 3949}
  [\href{https://arxiv.org/abs/hep-th/9406079}{{\ttfamily hep-th/9406079}}].

\bibitem{Giveon:2009da}
A.~Giveon, D.~Gorbonos and M.~Stern, \emph{{Fundamental Strings and Higher
  Derivative Corrections to d-Dimensional Black Holes}},
  \href{https://doi.org/10.1007/JHEP02(2010)012}{\emph{JHEP} {\bfseries 02}
  (2010) 012} [\href{https://arxiv.org/abs/0909.5264}{{\ttfamily 0909.5264}}].

\bibitem{Reall:2019sah}
H.~S. Reall and J.~E. Santos, \emph{{Higher derivative corrections to Kerr
  black hole thermodynamics}},
  \href{https://doi.org/10.1007/JHEP04(2019)021}{\emph{JHEP} {\bfseries 04}
  (2019) 021} [\href{https://arxiv.org/abs/1901.11535}{{\ttfamily
  1901.11535}}].

\bibitem{Ma:2023qqj}
L.~Ma, Y.~Pang and H.~Lu, \emph{{Higher derivative contributions to black hole
  thermodynamics at NNLO}},
  \href{https://doi.org/10.1007/JHEP06(2023)087}{\emph{JHEP} {\bfseries 06}
  (2023) 087} [\href{https://arxiv.org/abs/2304.08527}{{\ttfamily
  2304.08527}}].

\bibitem{Bergshoeff:1989de}
E.~A. Bergshoeff and M.~de~Roo, \emph{{The Quartic Effective Action of the
  Heterotic String and Supersymmetry}},
  \href{https://doi.org/10.1016/0550-3213(89)90336-2}{\emph{Nucl. Phys. B}
  {\bfseries 328} (1989) 439}.

\bibitem{Bergshoeff:1994dg}
E.~Bergshoeff, I.~Entrop and R.~Kallosh, \emph{{Exact duality in string
  effective action}},
  \href{https://doi.org/10.1103/PhysRevD.49.6663}{\emph{Phys. Rev. D}
  {\bfseries 49} (1994) 6663}
  [\href{https://arxiv.org/abs/hep-th/9401025}{{\ttfamily hep-th/9401025}}].

\bibitem{Bergshoeff:1995as}
E.~Bergshoeff, C.~M. Hull and T.~Ortin, \emph{{Duality in the type II
  superstring effective action}},
  \href{https://doi.org/10.1016/0550-3213(95)00367-2}{\emph{Nucl. Phys. B}
  {\bfseries 451} (1995) 547}
  [\href{https://arxiv.org/abs/hep-th/9504081}{{\ttfamily hep-th/9504081}}].

\bibitem{Meessen:1998qm}
P.~Meessen and T.~Ortin, \emph{{An Sl(2,Z) multiplet of nine-dimensional type
  II supergravity theories}},
  \href{https://doi.org/10.1016/S0550-3213(98)00780-9}{\emph{Nucl. Phys. B}
  {\bfseries 541} (1999) 195}
  [\href{https://arxiv.org/abs/hep-th/9806120}{{\ttfamily hep-th/9806120}}].

\bibitem{Bergshoeff:1995cg}
E.~Bergshoeff, B.~Janssen and T.~Ortin, \emph{{Solution generating
  transformations and the string effective action}},
  \href{https://doi.org/10.1088/0264-9381/13/3/002}{\emph{Class. Quant. Grav.}
  {\bfseries 13} (1996) 321}
  [\href{https://arxiv.org/abs/hep-th/9506156}{{\ttfamily hep-th/9506156}}].

\bibitem{Elgood:2020xwu}
Z.~Elgood and T.~Ortin, \emph{{T duality and Wald entropy formula in the
  Heterotic Superstring effective action at first-order in
  \ensuremath{\alpha}'}},
  \href{https://doi.org/10.1007/JHEP10(2020)097}{\emph{JHEP} {\bfseries 10}
  (2020) 097} [\href{https://arxiv.org/abs/2005.11272}{{\ttfamily
  2005.11272}}].

\bibitem{Sahoo:2006pm}
B.~Sahoo and A.~Sen, \emph{{alpha-prime - corrections to extremal dyonic black
  holes in heterotic string theory}},
  \href{https://doi.org/10.1088/1126-6708/2007/01/010}{\emph{JHEP} {\bfseries
  01} (2007) 010} [\href{https://arxiv.org/abs/hep-th/0608182}{{\ttfamily
  hep-th/0608182}}].

\bibitem{Faedo:2019xii}
F.~Faedo and P.~F. Ramirez, \emph{{Exact charges from heterotic black holes}},
  \href{https://doi.org/10.1007/JHEP10(2019)033}{\emph{JHEP} {\bfseries 10}
  (2019) 033} [\href{https://arxiv.org/abs/1906.12287}{{\ttfamily
  1906.12287}}].

\bibitem{Kastor:2009wy}
D.~Kastor, S.~Ray and J.~Traschen, \emph{{Enthalpy and the Mechanics of AdS
  Black Holes}},
  \href{https://doi.org/10.1088/0264-9381/26/19/195011}{\emph{Class. Quant.
  Grav.} {\bfseries 26} (2009) 195011}
  [\href{https://arxiv.org/abs/0904.2765}{{\ttfamily 0904.2765}}].

\bibitem{Gross:1986mw}
D.~J. Gross and J.~H. Sloan, \emph{{The Quartic Effective Action for the
  Heterotic String}},
  \href{https://doi.org/10.1016/0550-3213(87)90465-2}{\emph{Nucl. Phys. B}
  {\bfseries 291} (1987) 41}.

\bibitem{Metsaev:1987zx}
R.~R. Metsaev and A.~A. Tseytlin, \emph{{Order alpha-prime (Two Loop)
  Equivalence of the String Equations of Motion and the Sigma Model Weyl
  Invariance Conditions: Dependence on the Dilaton and the Antisymmetric
  Tensor}}, \href{https://doi.org/10.1016/0550-3213(87)90077-0}{\emph{Nucl.
  Phys. B} {\bfseries 293} (1987) 385}.

\bibitem{Chemissany:2007he}
W.~A. Chemissany, M.~de~Roo and S.~Panda, \emph{{alpha'-Corrections to
  Heterotic Superstring Effective Action Revisited}},
  \href{https://doi.org/10.1088/1126-6708/2007/08/037}{\emph{JHEP} {\bfseries
  08} (2007) 037} [\href{https://arxiv.org/abs/0706.3636}{{\ttfamily
  0706.3636}}].

\bibitem{Fontanella:2019avn}
A.~Fontanella and T.~Ort\'\i{}n, \emph{{On the supersymmetric solutions of the
  Heterotic Superstring effective action}},
  \href{https://doi.org/10.1007/JHEP10(2021)130}{\emph{JHEP} {\bfseries 06}
  (2020) 106} [\href{https://arxiv.org/abs/1910.08496}{{\ttfamily
  1910.08496}}].

\bibitem{Maldacena:1996ky}
J.~M. Maldacena, \emph{{Black holes in string theory}}, Ph.D. thesis, Princeton
  U., 1996.
\newblock \href{https://arxiv.org/abs/hep-th/9607235}{{\ttfamily
  hep-th/9607235}}.

\bibitem{Alonso-Alberca:2002wsh}
N.~Alonso-Alberca, E.~Lozano-Tellechea and T.~Ortin, \emph{{Geometric
  construction of Killing spinors and supersymmetry algebras in homogeneous
  space-times}},
  \href{https://doi.org/10.1088/0264-9381/19/23/309}{\emph{Class. Quant. Grav.}
  {\bfseries 19} (2002) 6009}
  [\href{https://arxiv.org/abs/hep-th/0208158}{{\ttfamily hep-th/0208158}}].

\bibitem{Gibbons:1976ue}
G.~W. Gibbons and S.~W. Hawking, \emph{{Action Integrals and Partition
  Functions in Quantum Gravity}},
  \href{https://doi.org/10.1103/PhysRevD.15.2752}{\emph{Phys. Rev. D}
  {\bfseries 15} (1977) 2752}.

\bibitem{Sahoo:2006vz}
B.~Sahoo and A.~Sen, \emph{{BTZ black hole with Chern-Simons and higher
  derivative terms}},
  \href{https://doi.org/10.1088/1126-6708/2006/07/008}{\emph{JHEP} {\bfseries
  07} (2006) 008} [\href{https://arxiv.org/abs/hep-th/0601228}{{\ttfamily
  hep-th/0601228}}].

\bibitem{Ortin:2020xdm}
T.~Ortin, \emph{{O(n, n) invariance and Wald entropy formula in the Heterotic
  Superstring effective action at first order in $\alpha'$}},
  \href{https://doi.org/10.1007/JHEP01(2021)187}{\emph{JHEP} {\bfseries 01}
  (2021) 187} [\href{https://arxiv.org/abs/2005.14618}{{\ttfamily
  2005.14618}}].

\bibitem{Ma:2022nwq}
L.~Ma, Y.~Pang and H.~Lu, \emph{{Improved Wald formalism and first law of
  dyonic black strings with mixed Chern-Simons terms}},
  \href{https://doi.org/10.1007/JHEP10(2022)142}{\emph{JHEP} {\bfseries 10}
  (2022) 142} [\href{https://arxiv.org/abs/2202.08290}{{\ttfamily
  2202.08290}}].

\bibitem{Tachikawa:2006sz}
Y.~Tachikawa, \emph{{Black hole entropy in the presence of Chern-Simons
  terms}}, \href{https://doi.org/10.1088/0264-9381/24/3/014}{\emph{Class.
  Quant. Grav.} {\bfseries 24} (2007) 737}
  [\href{https://arxiv.org/abs/hep-th/0611141}{{\ttfamily hep-th/0611141}}].

\bibitem{Ortin:2024emt}
T.~Ort\'\i{}n and M.~Zatti, \emph{{On the thermodynamics of the black holes of
  the Cano-Ruip\'erez 4-dimensional string effective action}},
  \href{https://arxiv.org/abs/2411.10417}{{\ttfamily 2411.10417}}.

\bibitem{Duff:1994an}
M.~J. Duff, R.~R. Khuri and J.~X. Lu, \emph{{String solitons}},
  \href{https://doi.org/10.1016/0370-1573(95)00002-X}{\emph{Phys. Rept.}
  {\bfseries 259} (1995) 213}
  [\href{https://arxiv.org/abs/hep-th/9412184}{{\ttfamily hep-th/9412184}}].

\bibitem{Cheung:2018cwt}
C.~Cheung, J.~Liu and G.~N. Remmen, \emph{{Proof of the Weak Gravity Conjecture
  from Black Hole Entropy}},
  \href{https://doi.org/10.1007/JHEP10(2018)004}{\emph{JHEP} {\bfseries 10}
  (2018) 004} [\href{https://arxiv.org/abs/1801.08546}{{\ttfamily
  1801.08546}}].

\bibitem{Hamada:2018dde}
Y.~Hamada, T.~Noumi and G.~Shiu, \emph{{Weak Gravity Conjecture from Unitarity
  and Causality}},
  \href{https://doi.org/10.1103/PhysRevLett.123.051601}{\emph{Phys. Rev. Lett.}
  {\bfseries 123} (2019) 051601}
  [\href{https://arxiv.org/abs/1810.03637}{{\ttfamily 1810.03637}}].

\bibitem{Bellazzini:2019xts}
B.~Bellazzini, M.~Lewandowski and J.~Serra, \emph{{Positivity of Amplitudes,
  Weak Gravity Conjecture, and Modified Gravity}},
  \href{https://doi.org/10.1103/PhysRevLett.123.251103}{\emph{Phys. Rev. Lett.}
  {\bfseries 123} (2019) 251103}
  [\href{https://arxiv.org/abs/1902.03250}{{\ttfamily 1902.03250}}].

\bibitem{Charles:2019qqt}
A.~M. Charles, \emph{{The Weak Gravity Conjecture, RG Flows, and
  Supersymmetry}},  \href{https://arxiv.org/abs/1906.07734}{{\ttfamily
  1906.07734}}.

\bibitem{Loges:2019jzs}
G.~J. Loges, T.~Noumi and G.~Shiu, \emph{{Thermodynamics of 4D Dilatonic Black
  Holes and the Weak Gravity Conjecture}},
  \href{https://doi.org/10.1103/PhysRevD.102.046010}{\emph{Phys. Rev. D}
  {\bfseries 102} (2020) 046010}
  [\href{https://arxiv.org/abs/1909.01352}{{\ttfamily 1909.01352}}].

\bibitem{Cano:2019oma}
P.~A. Cano, T.~Ort\'\i{}n and P.~F. Ramirez, \emph{{On the extremality bound of
  stringy black holes}},
  \href{https://doi.org/10.1007/JHEP02(2020)175}{\emph{JHEP} {\bfseries 02}
  (2020) 175} [\href{https://arxiv.org/abs/1909.08530}{{\ttfamily
  1909.08530}}].

\bibitem{Andriolo:2020lul}
S.~Andriolo, T.-C. Huang, T.~Noumi, H.~Ooguri and G.~Shiu, \emph{{Duality and
  axionic weak gravity}},
  \href{https://doi.org/10.1103/PhysRevD.102.046008}{\emph{Phys. Rev. D}
  {\bfseries 102} (2020) 046008}
  [\href{https://arxiv.org/abs/2004.13721}{{\ttfamily 2004.13721}}].

\bibitem{Loges:2020trf}
G.~J. Loges, T.~Noumi and G.~Shiu, \emph{{Duality and Supersymmetry Constraints
  on the Weak Gravity Conjecture}},
  \href{https://doi.org/10.1007/JHEP11(2020)008}{\emph{JHEP} {\bfseries 11}
  (2020) 008} [\href{https://arxiv.org/abs/2006.06696}{{\ttfamily
  2006.06696}}].

\bibitem{Cano:2020qhy}
P.~A. Cano and A.~Murcia, \emph{{Electromagnetic Quasitopological Gravities}},
  \href{https://doi.org/10.1007/JHEP10(2020)125}{\emph{JHEP} {\bfseries 10}
  (2020) 125} [\href{https://arxiv.org/abs/2007.04331}{{\ttfamily
  2007.04331}}].

\bibitem{Cano:2021tfs}
P.~A. Cano and A.~Murcia, \emph{{Duality-invariant extensions of
  Einstein-Maxwell theory}},
  \href{https://doi.org/10.1007/JHEP08(2021)042}{\emph{JHEP} {\bfseries 08}
  (2021) 042} [\href{https://arxiv.org/abs/2104.07674}{{\ttfamily
  2104.07674}}].

\bibitem{Arkani-Hamed:2021ajd}
N.~Arkani-Hamed, Y.-t. Huang, J.-Y. Liu and G.~N. Remmen, \emph{{Causality,
  unitarity, and the weak gravity conjecture}},
  \href{https://doi.org/10.1007/JHEP03(2022)083}{\emph{JHEP} {\bfseries 03}
  (2022) 083} [\href{https://arxiv.org/abs/2109.13937}{{\ttfamily
  2109.13937}}].

\bibitem{Jones:2019nev}
C.~R.~T. Jones and B.~McPeak, \emph{{The Black Hole Weak Gravity Conjecture
  with Multiple Charges}},
  \href{https://doi.org/10.1007/JHEP06(2020)140}{\emph{JHEP} {\bfseries 06}
  (2020) 140} [\href{https://arxiv.org/abs/1908.10452}{{\ttfamily
  1908.10452}}].

\bibitem{Israel-Kahn}
W.~Israel and K.~A. Khan, \emph{Collinear particles and bondi dipoles in
  general relativity}, {\emph{Il Nuovo Cimento (1955-1965)} {\bfseries 33}
  (1964) 331}.

\bibitem{Costa:2000kf}
M.~S. Costa and M.~J. Perry, \emph{{Interacting black holes}},
  \href{https://doi.org/10.1016/S0550-3213(00)00577-0}{\emph{Nucl. Phys. B}
  {\bfseries 591} (2000) 469}
  [\href{https://arxiv.org/abs/hep-th/0008106}{{\ttfamily hep-th/0008106}}].

\bibitem{Brill:1963yv}
D.~R. Brill and R.~W. Lindquist, \emph{{Interaction energy in
  geometrostatics}}, \href{https://doi.org/10.1103/PhysRev.131.471}{\emph{Phys.
  Rev.} {\bfseries 131} (1963) 471}.

\bibitem{Hartle:1972ya}
J.~B. Hartle and S.~W. Hawking, \emph{{Solutions of the Einstein-Maxwell
  equations with many black holes}},
  \href{https://doi.org/10.1007/BF01645696}{\emph{Commun. Math. Phys.}
  {\bfseries 26} (1972) 87}.

\bibitem{Meessen:2017rwm}
P.~Meessen, T.~Ort\'\i{}n and P.~F. Ram\'\i{}rez, \emph{{Dyonic black holes at
  arbitrary locations}},
  \href{https://doi.org/10.1007/JHEP10(2017)066}{\emph{JHEP} {\bfseries 10}
  (2017) 066} [\href{https://arxiv.org/abs/1707.03846}{{\ttfamily
  1707.03846}}].

\bibitem{Marolf:2000cb}
D.~Marolf, \emph{{Chern-Simons terms and the three notions of charge}},  in
  \emph{{International Conference on Quantization, Gauge Theory, and Strings:
  Conference Dedicated to the Memory of Professor Efim Fradkin}}, pp.~312--320,
  6, 2000, \href{https://arxiv.org/abs/hep-th/0006117}{{\ttfamily
  hep-th/0006117}}.

\bibitem{kn:ORZ}
T.~Ort\'{\i}n, A.~Ruip\'erez and M.~Zatti, ``work in progress.''.

\bibitem{Dabholkar:1989jt}
A.~Dabholkar and J.~A. Harvey, \emph{{Nonrenormalization of the Superstring
  Tension}}, \href{https://doi.org/10.1103/PhysRevLett.63.478}{\emph{Phys. Rev.
  Lett.} {\bfseries 63} (1989) 478}.

\bibitem{Dabholkar:1990yf}
A.~Dabholkar, G.~W. Gibbons, J.~A. Harvey and F.~Ruiz~Ruiz, \emph{{Superstrings
  and Solitons}},
  \href{https://doi.org/10.1016/0550-3213(90)90157-9}{\emph{Nucl. Phys. B}
  {\bfseries 340} (1990) 33}.

\bibitem{Sen:1994eb}
A.~Sen, \emph{{Black hole solutions in heterotic string theory on a torus}},
  \href{https://doi.org/10.1016/0550-3213(95)00063-X}{\emph{Nucl. Phys. B}
  {\bfseries 440} (1995) 421}
  [\href{https://arxiv.org/abs/hep-th/9411187}{{\ttfamily hep-th/9411187}}].

\bibitem{Cvetic:1995uj}
M.~Cvetic and D.~Youm, \emph{{Dyonic BPS saturated black holes of heterotic
  string on a six torus}},
  \href{https://doi.org/10.1103/PhysRevD.53.R584}{\emph{Phys. Rev. D}
  {\bfseries 53} (1996) 584}
  [\href{https://arxiv.org/abs/hep-th/9507090}{{\ttfamily hep-th/9507090}}].

\bibitem{Dabholkar:1995nc}
A.~Dabholkar, J.~P. Gauntlett, J.~A. Harvey and D.~Waldram, \emph{{Strings as
  solitons and black holes as strings}},
  \href{https://doi.org/10.1016/0550-3213(96)00266-0}{\emph{Nucl. Phys. B}
  {\bfseries 474} (1996) 85}
  [\href{https://arxiv.org/abs/hep-th/9511053}{{\ttfamily hep-th/9511053}}].

\bibitem{Callan:1995hn}
C.~G. Callan, J.~M. Maldacena and A.~W. Peet, \emph{{Extremal black holes as
  fundamental strings}},
  \href{https://doi.org/10.1016/0550-3213(96)00315-X}{\emph{Nucl. Phys. B}
  {\bfseries 475} (1996) 645}
  [\href{https://arxiv.org/abs/hep-th/9510134}{{\ttfamily hep-th/9510134}}].

\bibitem{Sen:1995in}
A.~Sen, \emph{{Extremal black holes and elementary string states}},
  \href{https://doi.org/10.1142/S0217732395002234}{\emph{Mod. Phys. Lett. A}
  {\bfseries 10} (1995) 2081}
  [\href{https://arxiv.org/abs/hep-th/9504147}{{\ttfamily hep-th/9504147}}].

\bibitem{Dabholkar:2004yr}
A.~Dabholkar, \emph{{Exact counting of black hole microstates}},
  \href{https://doi.org/10.1103/PhysRevLett.94.241301}{\emph{Phys. Rev. Lett.}
  {\bfseries 94} (2005) 241301}
  [\href{https://arxiv.org/abs/hep-th/0409148}{{\ttfamily hep-th/0409148}}].

\bibitem{Dabholkar:2004dq}
A.~Dabholkar, R.~Kallosh and A.~Maloney, \emph{{A Stringy cloak for a classical
  singularity}},
  \href{https://doi.org/10.1088/1126-6708/2004/12/059}{\emph{JHEP} {\bfseries
  12} (2004) 059} [\href{https://arxiv.org/abs/hep-th/0410076}{{\ttfamily
  hep-th/0410076}}].

\bibitem{Cano:2018hut}
P.~A. Cano, P.~F. Ram\'\i{}rez and A.~Ruip\'erez, \emph{{The small black hole
  illusion}}, \href{https://doi.org/10.1007/JHEP03(2020)115}{\emph{JHEP}
  {\bfseries 03} (2020) 115}
  [\href{https://arxiv.org/abs/1808.10449}{{\ttfamily 1808.10449}}].

\bibitem{Ruiperez:2020qda}
A.~Ruip\'erez, \emph{{Higher-derivative corrections to small black rings}},
  \href{https://doi.org/10.1088/1361-6382/abff9b}{\emph{Class. Quant. Grav.}
  {\bfseries 38} (2021) 145011}
  [\href{https://arxiv.org/abs/2003.02269}{{\ttfamily 2003.02269}}].

\bibitem{Cano:2021dyy}
P.~A. Cano, A.~Murcia, P.~F. Ram\'\i{}rez and A.~Ruip\'erez, \emph{{On small
  black holes, KK monopoles and solitonic 5-branes}},
  \href{https://doi.org/10.1007/JHEP05(2021)272}{\emph{JHEP} {\bfseries 05}
  (2021) 272} [\href{https://arxiv.org/abs/2102.04476}{{\ttfamily
  2102.04476}}].

\bibitem{Susskind:1993ws}
L.~Susskind, \emph{{Some speculations about black hole entropy in string
  theory}},  \href{https://arxiv.org/abs/hep-th/9309145}{{\ttfamily
  hep-th/9309145}}.

\bibitem{Horowitz:1996nw}
G.~T. Horowitz and J.~Polchinski, \emph{{A Correspondence principle for black
  holes and strings}},
  \href{https://doi.org/10.1103/PhysRevD.55.6189}{\emph{Phys. Rev. D}
  {\bfseries 55} (1997) 6189}
  [\href{https://arxiv.org/abs/hep-th/9612146}{{\ttfamily hep-th/9612146}}].

\bibitem{Horowitz:1997jc}
G.~T. Horowitz and J.~Polchinski, \emph{{Selfgravitating fundamental strings}},
  \href{https://doi.org/10.1103/PhysRevD.57.2557}{\emph{Phys. Rev. D}
  {\bfseries 57} (1998) 2557}
  [\href{https://arxiv.org/abs/hep-th/9707170}{{\ttfamily hep-th/9707170}}].

\bibitem{Damour:1999aw}
T.~Damour and G.~Veneziano, \emph{{Selfgravitating fundamental strings and
  black holes}},
  \href{https://doi.org/10.1016/S0550-3213(99)00596-9}{\emph{Nucl. Phys. B}
  {\bfseries 568} (2000) 93}
  [\href{https://arxiv.org/abs/hep-th/9907030}{{\ttfamily hep-th/9907030}}].

\bibitem{Chen:2021emg}
Y.~Chen and J.~Maldacena, \emph{{String scale black holes at large D}},
  \href{https://doi.org/10.1007/JHEP01(2022)095}{\emph{JHEP} {\bfseries 01}
  (2022) 095} [\href{https://arxiv.org/abs/2106.02169}{{\ttfamily
  2106.02169}}].

\bibitem{Chen:2021dsw}
Y.~Chen, J.~Maldacena and E.~Witten, \emph{{On the black hole/string
  transition}}, \href{https://doi.org/10.1007/JHEP01(2023)103}{\emph{JHEP}
  {\bfseries 01} (2023) 103}
  [\href{https://arxiv.org/abs/2109.08563}{{\ttfamily 2109.08563}}].

\bibitem{Brustein:2021cza}
R.~Brustein and Y.~Zigdon, \emph{{Black hole entropy sourced by string winding
  condensate}}, \href{https://doi.org/10.1007/JHEP10(2021)219}{\emph{JHEP}
  {\bfseries 10} (2021) 219}
  [\href{https://arxiv.org/abs/2107.09001}{{\ttfamily 2107.09001}}].

\bibitem{Matsuo:2022kvx}
Y.~Matsuo, \emph{{Fluid model of a black hole-string transition}},
  \href{https://doi.org/10.1103/PhysRevD.107.126003}{\emph{Phys. Rev. D}
  {\bfseries 107} (2023) 126003}
  [\href{https://arxiv.org/abs/2205.15976}{{\ttfamily 2205.15976}}].

\bibitem{Balthazar:2022hno}
B.~Balthazar, J.~Chu and D.~Kutasov, \emph{{On Small Black Holes in String
  Theory}},  \href{https://arxiv.org/abs/2210.12033}{{\ttfamily 2210.12033}}.

\bibitem{Ceplak:2023afb}
N.~\v{C}eplak, R.~Emparan, A.~Puhm and M.~Toma\v{s}evi\'c, \emph{{The
  correspondence between rotating black holes and fundamental strings}},
  \href{https://arxiv.org/abs/2307.03573}{{\ttfamily 2307.03573}}.

\bibitem{Mathur:2018tib}
S.~D. Mathur and D.~Turton, \emph{{The fuzzball nature of two-charge black hole
  microstates}},
  \href{https://doi.org/10.1016/j.nuclphysb.2019.114684}{\emph{Nucl. Phys. B}
  {\bfseries 945} (2019) 114684}
  [\href{https://arxiv.org/abs/1811.09647}{{\ttfamily 1811.09647}}].

\bibitem{Callan:1988hs}
C.~G. Callan, Jr., R.~C. Myers and M.~J. Perry, \emph{{Black Holes in String
  Theory}}, \href{https://doi.org/10.1016/0550-3213(89)90172-7}{\emph{Nucl.
  Phys. B} {\bfseries 311} (1989) 673}.

\bibitem{Kaloper:1997ux}
N.~Kaloper and K.~A. Meissner, \emph{{Duality beyond the first loop}},
  \href{https://doi.org/10.1103/PhysRevD.56.7940}{\emph{Phys. Rev. D}
  {\bfseries 56} (1997) 7940}
  [\href{https://arxiv.org/abs/hep-th/9705193}{{\ttfamily hep-th/9705193}}].

\bibitem{Bedoya:2014pma}
O.~A. Bedoya, D.~Marques and C.~Nunez, \emph{{Heterotic $\alpha$'-corrections
  in Double Field Theory}},
  \href{https://doi.org/10.1007/JHEP12(2014)074}{\emph{JHEP} {\bfseries 12}
  (2014) 074} [\href{https://arxiv.org/abs/1407.0365}{{\ttfamily 1407.0365}}].

\bibitem{Eloy:2020dko}
C.~Eloy, O.~Hohm and H.~Samtleben, \emph{{Duality Invariance and Higher
  Derivatives}}, \href{https://doi.org/10.1103/PhysRevD.101.126018}{\emph{Phys.
  Rev. D} {\bfseries 101} (2020) 126018}
  [\href{https://arxiv.org/abs/2004.13140}{{\ttfamily 2004.13140}}].

\bibitem{RuiperezVicente:2020qfw}
A.~Ruip\'erez~Vicente, \emph{{Black holes in string theory with
  higher-derivative corrections}}, Ph.D. thesis, U. Autonoma, Madrid (main),
  2020.

\bibitem{Bobev:2022bjm}
N.~Bobev, V.~Dimitrov, V.~Reys and A.~Vekemans, \emph{{Higher derivative
  corrections and AdS5 black holes}},
  \href{https://doi.org/10.1103/PhysRevD.106.L121903}{\emph{Phys. Rev. D}
  {\bfseries 106} (2022) L121903}
  [\href{https://arxiv.org/abs/2207.10671}{{\ttfamily 2207.10671}}].

\bibitem{Cassani:2022lrk}
D.~Cassani, A.~Ruip\'erez and E.~Turetta, \emph{{Corrections to AdS$_{5}$ black
  hole thermodynamics from higher-derivative supergravity}},
  \href{https://doi.org/10.1007/JHEP11(2022)059}{\emph{JHEP} {\bfseries 11}
  (2022) 059} [\href{https://arxiv.org/abs/2208.01007}{{\ttfamily
  2208.01007}}].

\bibitem{Baron:2017dvb}
W.~H. Baron, J.~J. Fernandez-Melgarejo, D.~Marques and C.~Nunez, \emph{{The Odd
  story of \ensuremath{\alpha}'-corrections}},
  \href{https://doi.org/10.1007/JHEP04(2017)078}{\emph{JHEP} {\bfseries 04}
  (2017) 078} [\href{https://arxiv.org/abs/1702.05489}{{\ttfamily
  1702.05489}}].

\bibitem{Liu:2023fqq}
J.~T. Liu and R.~J. Saskowski, \emph{{Consistent truncations in higher
  derivative supergravity}},
  \href{https://arxiv.org/abs/2307.12420}{{\ttfamily 2307.12420}}.

\bibitem{Tseytlin:1988tv}
A.~A. Tseytlin, \emph{{Mobius Infinity Subtraction and Effective Action in
  $\sigma$ Model Approach to Closed String Theory}},
  \href{https://doi.org/10.1016/0370-2693(88)90421-2}{\emph{Phys. Lett. B}
  {\bfseries 208} (1988) 221}.

\bibitem{Vafa:2005ui}
C.~Vafa, \emph{{The String landscape and the swampland}},
  \href{https://arxiv.org/abs/hep-th/0509212}{{\ttfamily hep-th/0509212}}.

\bibitem{Brennan:2017rbf}
T.~D. Brennan, F.~Carta and C.~Vafa, \emph{{The String Landscape, the
  Swampland, and the Missing Corner}},
  \href{https://doi.org/10.22323/1.305.0015}{\emph{PoS} {\bfseries TASI2017}
  (2017) 015} [\href{https://arxiv.org/abs/1711.00864}{{\ttfamily
  1711.00864}}].

\bibitem{Grana:2021zvf}
M.~Gra\~na and A.~Herr\'aez, \emph{{The Swampland Conjectures: A Bridge from
  Quantum Gravity to Particle Physics}},
  \href{https://doi.org/10.3390/universe7080273}{\emph{Universe} {\bfseries 7}
  (2021) 273} [\href{https://arxiv.org/abs/2107.00087}{{\ttfamily
  2107.00087}}].

\bibitem{Freivogel:2016qwc}
B.~Freivogel and M.~Kleban, \emph{{Vacua Morghulis}},
  \href{https://arxiv.org/abs/1610.04564}{{\ttfamily 1610.04564}}.

\bibitem{Maldacena:1998uz}
J.~M. Maldacena, J.~Michelson and A.~Strominger, \emph{{Anti-de Sitter
  fragmentation}},
  \href{https://doi.org/10.1088/1126-6708/1999/02/011}{\emph{JHEP} {\bfseries
  02} (1999) 011} [\href{https://arxiv.org/abs/hep-th/9812073}{{\ttfamily
  hep-th/9812073}}].

\bibitem{DeWolfe:2005uu}
O.~DeWolfe, A.~Giryavets, S.~Kachru and W.~Taylor, \emph{{Type IIA moduli
  stabilization}},
  \href{https://doi.org/10.1088/1126-6708/2005/07/066}{\emph{JHEP} {\bfseries
  07} (2005) 066} [\href{https://arxiv.org/abs/hep-th/0505160}{{\ttfamily
  hep-th/0505160}}].

\bibitem{Camara:2005dc}
P.~G. Camara, A.~Font and L.~E. Ibanez, \emph{{Fluxes, moduli fixing and
  MSSM-like vacua in a simple IIA orientifold}},
  \href{https://doi.org/10.1088/1126-6708/2005/09/013}{\emph{JHEP} {\bfseries
  09} (2005) 013} [\href{https://arxiv.org/abs/hep-th/0506066}{{\ttfamily
  hep-th/0506066}}].

\bibitem{Derendinger:2004jn}
J.-P. Derendinger, C.~Kounnas, P.~M. Petropoulos and F.~Zwirner,
  \emph{{Superpotentials in IIA compactifications with general fluxes}},
  \href{https://doi.org/10.1016/j.nuclphysb.2005.02.038}{\emph{Nucl. Phys. B}
  {\bfseries 715} (2005) 211}
  [\href{https://arxiv.org/abs/hep-th/0411276}{{\ttfamily hep-th/0411276}}].

\bibitem{Villadoro:2005cu}
G.~Villadoro and F.~Zwirner, \emph{{N=1 effective potential from dual type-IIA
  D6/O6 orientifolds with general fluxes}},
  \href{https://doi.org/10.1088/1126-6708/2005/06/047}{\emph{JHEP} {\bfseries
  06} (2005) 047} [\href{https://arxiv.org/abs/hep-th/0503169}{{\ttfamily
  hep-th/0503169}}].

\bibitem{Junghans:2020acz}
D.~Junghans, \emph{{O-Plane Backreaction and Scale Separation in Type IIA Flux
  Vacua}}, \href{https://doi.org/10.1002/prop.202000040}{\emph{Fortsch. Phys.}
  {\bfseries 68} (2020) 2000040}
  [\href{https://arxiv.org/abs/2003.06274}{{\ttfamily 2003.06274}}].

\bibitem{Marchesano:2020qvg}
F.~Marchesano, E.~Palti, J.~Quirant and A.~Tomasiello, \emph{{On supersymmetric
  AdS$_{4}$ orientifold vacua}},
  \href{https://doi.org/10.1007/JHEP08(2020)087}{\emph{JHEP} {\bfseries 08}
  (2020) 087} [\href{https://arxiv.org/abs/2003.13578}{{\ttfamily
  2003.13578}}].

\bibitem{Marchesano:2019hfb}
F.~Marchesano and J.~Quirant, \emph{{A Landscape of AdS Flux Vacua}},
  \href{https://doi.org/10.1007/JHEP12(2019)110}{\emph{JHEP} {\bfseries 12}
  (2019) 110} [\href{https://arxiv.org/abs/1908.11386}{{\ttfamily
  1908.11386}}].

\bibitem{Conlon:2021cjk}
J.~P. Conlon, S.~Ning and F.~Revello, \emph{{Exploring the holographic
  Swampland}}, \href{https://doi.org/10.1007/JHEP04(2022)117}{\emph{JHEP}
  {\bfseries 04} (2022) 117}
  [\href{https://arxiv.org/abs/2110.06245}{{\ttfamily 2110.06245}}].

\bibitem{Apers:2022tfm}
F.~Apers, J.~P. Conlon, S.~Ning and F.~Revello, \emph{{Integer conformal
  dimensions for type IIa flux vacua}},
  \href{https://doi.org/10.1103/PhysRevD.105.106029}{\emph{Phys. Rev. D}
  {\bfseries 105} (2022) 106029}
  [\href{https://arxiv.org/abs/2202.09330}{{\ttfamily 2202.09330}}].

\bibitem{Quirant:2022fpn}
J.~Quirant, \emph{{Noninteger conformal dimensions for type IIA flux vacua}},
  \href{https://doi.org/10.1103/PhysRevD.106.066017}{\emph{Phys. Rev. D}
  {\bfseries 106} (2022) 066017}
  [\href{https://arxiv.org/abs/2204.00014}{{\ttfamily 2204.00014}}].

\bibitem{Marchesano:2021ycx}
F.~Marchesano, D.~Prieto and J.~Quirant, \emph{{BIonic membranes and AdS
  instabilities}}, \href{https://doi.org/10.1007/JHEP07(2022)118}{\emph{JHEP}
  {\bfseries 07} (2022) 118}
  [\href{https://arxiv.org/abs/2110.11370}{{\ttfamily 2110.11370}}].

\bibitem{Casas:2022mnz}
G.~F. Casas, F.~Marchesano and D.~Prieto, \emph{{Membranes in AdS$_{4}$
  orientifold vacua and their Weak Gravity Conjecture}},
  \href{https://doi.org/10.1007/JHEP09(2022)034}{\emph{JHEP} {\bfseries 09}
  (2022) 034} [\href{https://arxiv.org/abs/2204.11892}{{\ttfamily
  2204.11892}}].

\bibitem{Grimm:2004ua}
T.~W. Grimm and J.~Louis, \emph{{The Effective action of type IIA Calabi-Yau
  orientifolds}},
  \href{https://doi.org/10.1016/j.nuclphysb.2005.04.007}{\emph{Nucl. Phys. B}
  {\bfseries 718} (2005) 153}
  [\href{https://arxiv.org/abs/hep-th/0412277}{{\ttfamily hep-th/0412277}}].

\bibitem{Grimm:2011dx}
T.~W. Grimm and D.~Vieira~Lopes, \emph{{The N=1 effective actions of D-branes
  in Type IIA and IIB orientifolds}},
  \href{https://doi.org/10.1016/j.nuclphysb.2011.10.019}{\emph{Nucl. Phys. B}
  {\bfseries 855} (2012) 639}
  [\href{https://arxiv.org/abs/1104.2328}{{\ttfamily 1104.2328}}].

\bibitem{Kerstan:2011dy}
M.~Kerstan and T.~Weigand, \emph{{The Effective action of D6-branes in N=1 type
  IIA orientifolds}},
  \href{https://doi.org/10.1007/JHEP06(2011)105}{\emph{JHEP} {\bfseries 06}
  (2011) 105} [\href{https://arxiv.org/abs/1104.2329}{{\ttfamily 1104.2329}}].

\bibitem{Bielleman:2015ina}
S.~Bielleman, L.~E. Ibanez and I.~Valenzuela, \emph{{Minkowski 3-forms, Flux
  String Vacua, Axion Stability and Naturalness}},
  \href{https://doi.org/10.1007/JHEP12(2015)119}{\emph{JHEP} {\bfseries 12}
  (2015) 119} [\href{https://arxiv.org/abs/1507.06793}{{\ttfamily
  1507.06793}}].

\bibitem{Carta:2016ynn}
F.~Carta, F.~Marchesano, W.~Staessens and G.~Zoccarato, \emph{{Open string
  multi-branched and K\"ahler potentials}},
  \href{https://doi.org/10.1007/JHEP09(2016)062}{\emph{JHEP} {\bfseries 09}
  (2016) 062} [\href{https://arxiv.org/abs/1606.00508}{{\ttfamily
  1606.00508}}].

\bibitem{Farakos:2017jme}
F.~Farakos, S.~Lanza, L.~Martucci and D.~Sorokin, \emph{{Three-forms in
  Supergravity and Flux Compactifications}},
  \href{https://doi.org/10.1140/epjc/s10052-017-5185-y}{\emph{Eur. Phys. J. C}
  {\bfseries 77} (2017) 602}
  [\href{https://arxiv.org/abs/1706.09422}{{\ttfamily 1706.09422}}].

\bibitem{Herraez:2018vae}
A.~Herraez, L.~E. Ibanez, F.~Marchesano and G.~Zoccarato, \emph{{The Type IIA
  Flux Potential, 4-forms and Freed-Witten anomalies}},
  \href{https://doi.org/10.1007/JHEP09(2018)018}{\emph{JHEP} {\bfseries 09}
  (2018) 018} [\href{https://arxiv.org/abs/1802.05771}{{\ttfamily
  1802.05771}}].

\bibitem{Bandos:2018gjp}
I.~Bandos, F.~Farakos, S.~Lanza, L.~Martucci and D.~Sorokin,
  \emph{{Three-forms, dualities and membranes in four-dimensional
  supergravity}}, \href{https://doi.org/10.1007/JHEP07(2018)028}{\emph{JHEP}
  {\bfseries 07} (2018) 028}
  [\href{https://arxiv.org/abs/1803.01405}{{\ttfamily 1803.01405}}].

\bibitem{Lanza:2019xxg}
S.~Lanza, F.~Marchesano, L.~Martucci and D.~Sorokin, \emph{{How many fluxes fit
  in an EFT?}}, \href{https://doi.org/10.1007/JHEP10(2019)110}{\emph{JHEP}
  {\bfseries 10} (2019) 110}
  [\href{https://arxiv.org/abs/1907.11256}{{\ttfamily 1907.11256}}].

\bibitem{Aharony:2008wz}
O.~Aharony, Y.~E. Antebi and M.~Berkooz, \emph{{On the Conformal Field Theory
  Duals of type IIA AdS(4) Flux Compactifications}},
  \href{https://doi.org/10.1088/1126-6708/2008/02/093}{\emph{JHEP} {\bfseries
  02} (2008) 093} [\href{https://arxiv.org/abs/0801.3326}{{\ttfamily
  0801.3326}}].

\bibitem{Narayan:2010em}
P.~Narayan and S.~P. Trivedi, \emph{{On The Stability Of Non-Supersymmetric AdS
  Vacua}}, \href{https://doi.org/10.1007/JHEP07(2010)089}{\emph{JHEP}
  {\bfseries 07} (2010) 089} [\href{https://arxiv.org/abs/1002.4498}{{\ttfamily
  1002.4498}}].

\bibitem{Lanza:2020qmt}
S.~Lanza, F.~Marchesano, L.~Martucci and I.~Valenzuela, \emph{{Swampland
  Conjectures for Strings and Membranes}},
  \href{https://doi.org/10.1007/JHEP02(2021)006}{\emph{JHEP} {\bfseries 02}
  (2021) 006} [\href{https://arxiv.org/abs/2006.15154}{{\ttfamily
  2006.15154}}].

\bibitem{Bergshoeff:2001pv}
E.~Bergshoeff, R.~Kallosh, T.~Ortin, D.~Roest and A.~Van~Proeyen, \emph{{New
  formulations of D = 10 supersymmetry and D8 - O8 domain walls}},
  \href{https://doi.org/10.1088/0264-9381/18/17/303}{\emph{Class. Quant. Grav.}
  {\bfseries 18} (2001) 3359}
  [\href{https://arxiv.org/abs/hep-th/0103233}{{\ttfamily hep-th/0103233}}].

\bibitem{Maldacena:2000mw}
J.~M. Maldacena and C.~Nunez, \emph{{Supergravity description of field theories
  on curved manifolds and a no go theorem}},
  \href{https://doi.org/10.1142/S0217751X01003937}{\emph{Int. J. Mod. Phys. A}
  {\bfseries 16} (2001) 822}
  [\href{https://arxiv.org/abs/hep-th/0007018}{{\ttfamily hep-th/0007018}}].

\bibitem{Giddings:2001yu}
S.~B. Giddings, S.~Kachru and J.~Polchinski, \emph{{Hierarchies from fluxes in
  string compactifications}},
  \href{https://doi.org/10.1103/PhysRevD.66.106006}{\emph{Phys. Rev. D}
  {\bfseries 66} (2002) 106006}
  [\href{https://arxiv.org/abs/hep-th/0105097}{{\ttfamily hep-th/0105097}}].

\bibitem{Grana:2005ny}
M.~Grana, J.~Louis and D.~Waldram, \emph{{Hitchin functionals in N=2
  supergravity}},
  \href{https://doi.org/10.1088/1126-6708/2006/01/008}{\emph{JHEP} {\bfseries
  01} (2006) 008} [\href{https://arxiv.org/abs/hep-th/0505264}{{\ttfamily
  hep-th/0505264}}].

\bibitem{Hitchin:1999fh}
N.~J. Hitchin, \emph{{Lectures on special Lagrangian submanifolds}},
  {\emph{AMS/IP Stud. Adv. Math.} {\bfseries 23} (2001) 151}
  [\href{https://arxiv.org/abs/math/9907034}{{\ttfamily math/9907034}}].

\bibitem{mclean1990deformations}
R.~McLean, \emph{Deformations and Moduli of Calibrated Submanifolds}. Duke
  University, 1990.

\bibitem{Blumenhagen:2005mu}
R.~Blumenhagen, M.~Cvetic, P.~Langacker and G.~Shiu, \emph{{Toward realistic
  intersecting D-brane models}},
  \href{https://doi.org/10.1146/annurev.nucl.55.090704.151541}{\emph{Ann. Rev.
  Nucl. Part. Sci.} {\bfseries 55} (2005) 71}
  [\href{https://arxiv.org/abs/hep-th/0502005}{{\ttfamily hep-th/0502005}}].

\bibitem{Blumenhagen:2006ci}
R.~Blumenhagen, B.~Kors, D.~Lust and S.~Stieberger, \emph{{Four-dimensional
  String Compactifications with D-Branes, Orientifolds and Fluxes}},
  \href{https://doi.org/10.1016/j.physrep.2007.04.003}{\emph{Phys. Rept.}
  {\bfseries 445} (2007) 1}
  [\href{https://arxiv.org/abs/hep-th/0610327}{{\ttfamily hep-th/0610327}}].

\bibitem{Marchesano:2007de}
F.~Marchesano, \emph{{Progress in D-brane model building}},
  \href{https://doi.org/10.1002/prop.200610381}{\emph{Fortsch. Phys.}
  {\bfseries 55} (2007) 491}
  [\href{https://arxiv.org/abs/hep-th/0702094}{{\ttfamily hep-th/0702094}}].

\bibitem{Font:2006na}
A.~Font, L.~E. Ibanez and F.~Marchesano, \emph{{Coisotropic D8-branes and
  model-building}},
  \href{https://doi.org/10.1088/1126-6708/2006/09/080}{\emph{JHEP} {\bfseries
  09} (2006) 080} [\href{https://arxiv.org/abs/hep-th/0607219}{{\ttfamily
  hep-th/0607219}}].

\bibitem{Saracco:2012wc}
F.~Saracco and A.~Tomasiello, \emph{{Localized O6-plane solutions with Romans
  mass}}, \href{https://doi.org/10.1007/JHEP07(2012)077}{\emph{JHEP} {\bfseries
  07} (2012) 077} [\href{https://arxiv.org/abs/1201.5378}{{\ttfamily
  1201.5378}}].

\bibitem{Escobar:2018tiu}
D.~Escobar, F.~Marchesano and W.~Staessens, \emph{{Type IIA Flux Vacua with
  Mobile D6-branes}},
  \href{https://doi.org/10.1007/JHEP01(2019)096}{\emph{JHEP} {\bfseries 01}
  (2019) 096} [\href{https://arxiv.org/abs/1811.09282}{{\ttfamily
  1811.09282}}].

\bibitem{Escobar:2018rna}
D.~Escobar, F.~Marchesano and W.~Staessens, \emph{{Type IIA flux vacua and
  $\alpha'$-corrections}},
  \href{https://doi.org/10.1007/JHEP06(2019)129}{\emph{JHEP} {\bfseries 06}
  (2019) 129} [\href{https://arxiv.org/abs/1812.08735}{{\ttfamily
  1812.08735}}].

\bibitem{Marchesano:2020uqz}
F.~Marchesano, D.~Prieto, J.~Quirant and P.~Shukla, \emph{{Systematics of Type
  IIA moduli stabilisation}},
  \href{https://doi.org/10.1007/JHEP11(2020)113}{\emph{JHEP} {\bfseries 11}
  (2020) 113} [\href{https://arxiv.org/abs/2007.00672}{{\ttfamily
  2007.00672}}].

\bibitem{Koerber:2007jb}
P.~Koerber and L.~Martucci, \emph{{D-branes on AdS flux compactifications}},
  \href{https://doi.org/10.1088/1126-6708/2008/01/047}{\emph{JHEP} {\bfseries
  01} (2008) 047} [\href{https://arxiv.org/abs/0710.5530}{{\ttfamily
  0710.5530}}].

\bibitem{Berasaluce-Gonzalez:2012awn}
M.~Berasaluce-Gonzalez, P.~G. Camara, F.~Marchesano and A.~M. Uranga, \emph{{Zp
  charged branes in flux compactifications}},
  \href{https://doi.org/10.1007/JHEP04(2013)138}{\emph{JHEP} {\bfseries 04}
  (2013) 138} [\href{https://arxiv.org/abs/1211.5317}{{\ttfamily 1211.5317}}].

\bibitem{Marchesano:2014mla}
F.~Marchesano, G.~Shiu and A.~M. Uranga, \emph{{F-term Axion Monodromy
  Inflation}}, \href{https://doi.org/10.1007/JHEP09(2014)184}{\emph{JHEP}
  {\bfseries 09} (2014) 184} [\href{https://arxiv.org/abs/1404.3040}{{\ttfamily
  1404.3040}}].

\bibitem{Buratti:2020kda}
G.~Buratti, J.~Calderon, A.~Mininno and A.~M. Uranga, \emph{{Discrete
  Symmetries, Weak Coupling Conjecture and Scale Separation in AdS Vacua}},
  \href{https://doi.org/10.1007/JHEP06(2020)083}{\emph{JHEP} {\bfseries 06}
  (2020) 083} [\href{https://arxiv.org/abs/2003.09740}{{\ttfamily
  2003.09740}}].

\bibitem{Douglas:1995bn}
M.~R. Douglas, \emph{{Branes within branes}}, {\emph{NATO Sci. Ser. C}
  {\bfseries 520} (1999) 267}
  [\href{https://arxiv.org/abs/hep-th/9512077}{{\ttfamily hep-th/9512077}}].

\bibitem{Marino:1999af}
M.~Marino, R.~Minasian, G.~W. Moore and A.~Strominger, \emph{{Nonlinear
  instantons from supersymmetric p-branes}},
  \href{https://doi.org/10.1088/1126-6708/2000/01/005}{\emph{JHEP} {\bfseries
  01} (2000) 005} [\href{https://arxiv.org/abs/hep-th/9911206}{{\ttfamily
  hep-th/9911206}}].

\bibitem{Douglas:2001hw}
M.~R. Douglas, \emph{{D-branes and N=1 supersymmetry}},  in \emph{{Strings
  2001: International Conference}}, 5, 2001,
  \href{https://arxiv.org/abs/hep-th/0105014}{{\ttfamily hep-th/0105014}}.

\bibitem{Martucci:2005ht}
L.~Martucci and P.~Smyth, \emph{{Supersymmetric D-branes and calibrations on
  general N=1 backgrounds}},
  \href{https://doi.org/10.1088/1126-6708/2005/11/048}{\emph{JHEP} {\bfseries
  11} (2005) 048} [\href{https://arxiv.org/abs/hep-th/0507099}{{\ttfamily
  hep-th/0507099}}].

\bibitem{Rabadan:2001mt}
R.~Rabadan, \emph{{Branes at angles, torons, stability and supersymmetry}},
  \href{https://doi.org/10.1016/S0550-3213(01)00560-0}{\emph{Nucl. Phys. B}
  {\bfseries 620} (2002) 152}
  [\href{https://arxiv.org/abs/hep-th/0107036}{{\ttfamily hep-th/0107036}}].

\bibitem{Apers:2022zjx}
F.~Apers, M.~Montero, T.~Van~Riet and T.~Wrase, \emph{{Comments on classical
  AdS flux vacua with scale separation}},
  \href{https://doi.org/10.1007/JHEP05(2022)167}{\emph{JHEP} {\bfseries 05}
  (2022) 167} [\href{https://arxiv.org/abs/2202.00682}{{\ttfamily
  2202.00682}}].

\bibitem{Becker:2006dvp}
K.~Becker, M.~Becker and J.~H. Schwarz, \emph{{String theory and M-theory: A
  modern introduction}}. Cambridge University Press, 12, 2006,
  \href{https://doi.org/10.1017/CBO9780511816086}{10.1017/CBO9780511816086}.

\bibitem{Baumann:2014nda}
D.~Baumann and L.~McAllister, \emph{{Inflation and String Theory}}, Cambridge
  Monographs on Mathematical Physics. Cambridge University Press, 5, 2015,
  \href{https://doi.org/10.1017/CBO9781316105733}{10.1017/CBO9781316105733},
  [\href{https://arxiv.org/abs/1404.2601}{{\ttfamily 1404.2601}}].

\bibitem{Tomasiello:2022dwe}
A.~Tomasiello, \emph{{Geometry of String Theory Compactifications}}. Cambridge
  University Press, 1, 2022,
  \href{https://doi.org/10.1017/9781108635745}{10.1017/9781108635745}.

\bibitem{Harvey:1982xk}
R.~Harvey and H.~B. Lawson, Jr., \emph{{Calibrated geometries}},
  \href{https://doi.org/10.1007/BF02392726}{\emph{Acta Math.} {\bfseries 148}
  (1982) 47}.

\bibitem{Koerber:2005qi}
P.~Koerber, \emph{{Stable D-branes, calibrations and generalized Calabi-Yau
  geometry}}, \href{https://doi.org/10.1088/1126-6708/2005/08/099}{\emph{JHEP}
  {\bfseries 08} (2005) 099}
  [\href{https://arxiv.org/abs/hep-th/0506154}{{\ttfamily hep-th/0506154}}].

\bibitem{Koerber:2010bx}
P.~Koerber, \emph{{Lectures on Generalized Complex Geometry for Physicists}},
  \href{https://doi.org/10.1002/prop.201000083}{\emph{Fortsch. Phys.}
  {\bfseries 59} (2011) 169} [\href{https://arxiv.org/abs/1006.1536}{{\ttfamily
  1006.1536}}].

\bibitem{Banks:2010zn}
T.~Banks and N.~Seiberg, \emph{{Symmetries and Strings in Field Theory and
  Gravity}}, \href{https://doi.org/10.1103/PhysRevD.83.084019}{\emph{Phys. Rev.
  D} {\bfseries 83} (2011) 084019}
  [\href{https://arxiv.org/abs/1011.5120}{{\ttfamily 1011.5120}}].

\bibitem{Berasaluce-Gonzalez:2011gos}
M.~Berasaluce-Gonzalez, L.~E. Ibanez, P.~Soler and A.~M. Uranga,
  \emph{{Discrete gauge symmetries in D-brane models}},
  \href{https://doi.org/10.1007/JHEP12(2011)113}{\emph{JHEP} {\bfseries 12}
  (2011) 113} [\href{https://arxiv.org/abs/1106.4169}{{\ttfamily 1106.4169}}].

\bibitem{Gukov:1998kn}
S.~Gukov, M.~Rangamani and E.~Witten, \emph{{Dibaryons, strings and branes in
  AdS orbifold models}},
  \href{https://doi.org/10.1088/1126-6708/1998/12/025}{\emph{JHEP} {\bfseries
  12} (1998) 025} [\href{https://arxiv.org/abs/hep-th/9811048}{{\ttfamily
  hep-th/9811048}}].

\bibitem{Witten:1998wy}
E.~Witten, \emph{{AdS / CFT correspondence and topological field theory}},
  \href{https://doi.org/10.1088/1126-6708/1998/12/012}{\emph{JHEP} {\bfseries
  12} (1998) 012} [\href{https://arxiv.org/abs/hep-th/9812012}{{\ttfamily
  hep-th/9812012}}].

\bibitem{Camara:2011jg}
P.~G. Camara, L.~E. Ibanez and F.~Marchesano, \emph{{RR photons}},
  \href{https://doi.org/10.1007/JHEP09(2011)110}{\emph{JHEP} {\bfseries 09}
  (2011) 110} [\href{https://arxiv.org/abs/1106.0060}{{\ttfamily 1106.0060}}].

\bibitem{Grimm:2011tb}
T.~W. Grimm, M.~Kerstan, E.~Palti and T.~Weigand, \emph{{Massive Abelian Gauge
  Symmetries and Fluxes in F-theory}},
  \href{https://doi.org/10.1007/JHEP12(2011)004}{\emph{JHEP} {\bfseries 12}
  (2011) 004} [\href{https://arxiv.org/abs/1107.3842}{{\ttfamily 1107.3842}}].

\bibitem{Berasaluce-Gonzalez:2012abm}
M.~Berasaluce-Gonzalez, P.~G. Camara, F.~Marchesano, D.~Regalado and A.~M.
  Uranga, \emph{{Non-Abelian discrete gauge symmetries in 4d string models}},
  \href{https://doi.org/10.1007/JHEP09(2012)059}{\emph{JHEP} {\bfseries 09}
  (2012) 059} [\href{https://arxiv.org/abs/1206.2383}{{\ttfamily 1206.2383}}].

\bibitem{Mayrhofer:2014laa}
C.~Mayrhofer, E.~Palti, O.~Till and T.~Weigand, \emph{{On Discrete Symmetries
  and Torsion Homology in F-Theory}},
  \href{https://doi.org/10.1007/JHEP06(2015)029}{\emph{JHEP} {\bfseries 06}
  (2015) 029} [\href{https://arxiv.org/abs/1410.7814}{{\ttfamily 1410.7814}}].

\bibitem{Grimm:2015ona}
T.~W. Grimm, T.~G. Pugh and D.~Regalado, \emph{{Non-Abelian discrete gauge
  symmetries in F-theory}},
  \href{https://doi.org/10.1007/JHEP02(2016)066}{\emph{JHEP} {\bfseries 02}
  (2016) 066} [\href{https://arxiv.org/abs/1504.06272}{{\ttfamily
  1504.06272}}].

\bibitem{Braun:2017oak}
V.~Braun, M.~Cvetic, R.~Donagi and M.~Poretschkin, \emph{{Type II String Theory
  on Calabi-Yau Manifolds with Torsion and Non-Abelian Discrete Gauge
  Symmetries}}, \href{https://doi.org/10.1007/JHEP07(2017)129}{\emph{JHEP}
  {\bfseries 07} (2017) 129}
  [\href{https://arxiv.org/abs/1702.08071}{{\ttfamily 1702.08071}}].

\bibitem{Bott1982DifferentialFI}
R.~Bott and L.~W. Tu, \emph{Differential forms in algebraic topology},  in
  \emph{Graduate texts in mathematics}, 1982.

\bibitem{Apruzzi:2021nmk}
F.~Apruzzi, F.~Bonetti, I.~n.~G. Etxebarria, S.~S. Hosseini and
  S.~Schafer-Nameki, \emph{{Symmetry TFTs from String Theory}},
  \href{https://arxiv.org/abs/2112.02092}{{\ttfamily 2112.02092}}.

\bibitem{Goldberger:2001tn}
W.~D. Goldberger and M.~B. Wise, \emph{{Renormalization group flows for brane
  couplings}}, \href{https://doi.org/10.1103/PhysRevD.65.025011}{\emph{Phys.
  Rev. D} {\bfseries 65} (2002) 025011}
  [\href{https://arxiv.org/abs/hep-th/0104170}{{\ttfamily hep-th/0104170}}].

\bibitem{Michel:2014lva}
B.~Michel, E.~Mintun, J.~Polchinski, A.~Puhm and P.~Saad, \emph{{Remarks on
  brane and antibrane dynamics}},
  \href{https://doi.org/10.1007/JHEP09(2015)021}{\emph{JHEP} {\bfseries 09}
  (2015) 021} [\href{https://arxiv.org/abs/1412.5702}{{\ttfamily 1412.5702}}].

\bibitem{Polchinski:2015bea}
J.~Polchinski, \emph{{Brane/antibrane dynamics and KKLT stability}},
  \href{https://arxiv.org/abs/1509.05710}{{\ttfamily 1509.05710}}.

\bibitem{Lanza:2021udy}
S.~Lanza, F.~Marchesano, L.~Martucci and I.~Valenzuela, \emph{{The EFT stringy
  viewpoint on large distances}},
  \href{https://doi.org/10.1007/JHEP09(2021)197}{\emph{JHEP} {\bfseries 09}
  (2021) 197} [\href{https://arxiv.org/abs/2104.05726}{{\ttfamily
  2104.05726}}].

\bibitem{Lanza:2022zyg}
S.~Lanza, F.~Marchesano, L.~Martucci and I.~Valenzuela, \emph{{Large Field
  Distances from EFT strings}},
  \href{https://doi.org/10.22323/1.406.0169}{\emph{PoS} {\bfseries CORFU2021}
  (2022) 169} [\href{https://arxiv.org/abs/2205.04532}{{\ttfamily
  2205.04532}}].

\bibitem{Gurrieri:2002wz}
S.~Gurrieri, J.~Louis, A.~Micu and D.~Waldram, \emph{{Mirror symmetry in
  generalized Calabi-Yau compactifications}},
  \href{https://doi.org/10.1016/S0550-3213(03)00045-2}{\emph{Nucl. Phys. B}
  {\bfseries 654} (2003) 61}
  [\href{https://arxiv.org/abs/hep-th/0211102}{{\ttfamily hep-th/0211102}}].

\bibitem{DAuria:2004kwe}
R.~D'Auria, S.~Ferrara, M.~Trigiante and S.~Vaula, \emph{{Gauging the
  Heisenberg algebra of special quaternionic manifolds}},
  \href{https://doi.org/10.1016/j.physletb.2005.01.084}{\emph{Phys. Lett. B}
  {\bfseries 610} (2005) 147}
  [\href{https://arxiv.org/abs/hep-th/0410290}{{\ttfamily hep-th/0410290}}].

\bibitem{Kashani-Poor:2006ofe}
A.-K. Kashani-Poor and R.~Minasian, \emph{{Towards reduction of type II
  theories on SU(3) structure manifolds}},
  \href{https://doi.org/10.1088/1126-6708/2007/03/109}{\emph{JHEP} {\bfseries
  03} (2007) 109} [\href{https://arxiv.org/abs/hep-th/0611106}{{\ttfamily
  hep-th/0611106}}].

\bibitem{Gray1980TheSC}
A.~Gray and L.~M. Hervella, \emph{The sixteen classes of almost hermitian
  manifolds and their linear invariants}, {\emph{Annali di Matematica Pura ed
  Applicata} {\bfseries 123} (1980) 35}.

\bibitem{chiossi2002intrinsic}
S.~Chiossi and S.~Salamon, \emph{The intrinsic torsion of su(3) and $g_2$
  structures},  2002.

\bibitem{Hull:1986iu}
C.~M. Hull, \emph{{Superstring compactifications with torsion and space-time
  supersymmetry}},  in \emph{{First Torino Meeting on Superunification and
  Extra Dimensions}}, 3, 1986.

\bibitem{Strominger:1986uh}
A.~Strominger, \emph{{Superstrings with Torsion}},
  \href{https://doi.org/10.1016/0550-3213(86)90286-5}{\emph{Nucl. Phys. B}
  {\bfseries 274} (1986) 253}.

\bibitem{LopesCardoso:2002vpf}
G.~Lopes~Cardoso, G.~Curio, G.~Dall'Agata, D.~Lust, P.~Manousselis and
  G.~Zoupanos, \emph{{NonKahler string backgrounds and their five torsion
  classes}}, \href{https://doi.org/10.1016/S0550-3213(03)00049-X}{\emph{Nucl.
  Phys. B} {\bfseries 652} (2003) 5}
  [\href{https://arxiv.org/abs/hep-th/0211118}{{\ttfamily hep-th/0211118}}].

\bibitem{Gauntlett:2003cy}
J.~P. Gauntlett, D.~Martelli and D.~Waldram, \emph{{Superstrings with intrinsic
  torsion}}, \href{https://doi.org/10.1103/PhysRevD.69.086002}{\emph{Phys. Rev.
  D} {\bfseries 69} (2004) 086002}
  [\href{https://arxiv.org/abs/hep-th/0302158}{{\ttfamily hep-th/0302158}}].

\bibitem{Grana:2005jc}
M.~Grana, \emph{{Flux compactifications in string theory: A Comprehensive
  review}}, \href{https://doi.org/10.1016/j.physrep.2005.10.008}{\emph{Phys.
  Rept.} {\bfseries 423} (2006) 91}
  [\href{https://arxiv.org/abs/hep-th/0509003}{{\ttfamily hep-th/0509003}}].

\bibitem{Horowitz:1989km}
G.~T. Horowitz and M.~Srednicki, \emph{{A Quantum Field Theoretic Description
  of Linking Numbers and Their Generalization}},
  \href{https://doi.org/10.1007/BF02099875}{\emph{Commun. Math. Phys.}
  {\bfseries 130} (1990) 83}.

\bibitem{Marchesano:2014iea}
F.~Marchesano, D.~Regalado and G.~Zoccarato, \emph{{On D-brane moduli
  stabilisation}}, \href{https://doi.org/10.1007/JHEP11(2014)097}{\emph{JHEP}
  {\bfseries 11} (2014) 097} [\href{https://arxiv.org/abs/1410.0209}{{\ttfamily
  1410.0209}}].

\bibitem{Marchesano:2014bia}
F.~Marchesano, D.~Regalado and G.~Zoccarato, \emph{{U(1) mixing and D-brane
  linear equivalence}},
  \href{https://doi.org/10.1007/JHEP08(2014)157}{\emph{JHEP} {\bfseries 08}
  (2014) 157} [\href{https://arxiv.org/abs/1406.2729}{{\ttfamily 1406.2729}}].

\bibitem{Blaback:2010sj}
J.~Blaback, U.~H. Danielsson, D.~Junghans, T.~Van~Riet, T.~Wrase and
  M.~Zagermann, \emph{{Smeared versus localised sources in flux
  compactifications}},
  \href{https://doi.org/10.1007/JHEP12(2010)043}{\emph{JHEP} {\bfseries 12}
  (2010) 043} [\href{https://arxiv.org/abs/1009.1877}{{\ttfamily 1009.1877}}].

\bibitem{Evslin:2007ti}
J.~Evslin and L.~Martucci, \emph{{D-brane networks in flux vacua, generalized
  cycles and calibrations}},
  \href{https://doi.org/10.1088/1126-6708/2007/07/040}{\emph{JHEP} {\bfseries
  07} (2007) 040} [\href{https://arxiv.org/abs/hep-th/0703129}{{\ttfamily
  hep-th/0703129}}].

\bibitem{Tomasiello:2005bp}
A.~Tomasiello, \emph{{Topological mirror symmetry with fluxes}},
  \href{https://doi.org/10.1088/1126-6708/2005/06/067}{\emph{JHEP} {\bfseries
  06} (2005) 067} [\href{https://arxiv.org/abs/hep-th/0502148}{{\ttfamily
  hep-th/0502148}}].

\bibitem{Marchesano:2006ns}
F.~Marchesano, \emph{{D6-branes and torsion}},
  \href{https://doi.org/10.1088/1126-6708/2006/05/019}{\emph{JHEP} {\bfseries
  05} (2006) 019} [\href{https://arxiv.org/abs/hep-th/0603210}{{\ttfamily
  hep-th/0603210}}].

\bibitem{Hebecker:2017wsu}
A.~Hebecker, P.~Henkenjohann and L.~T. Witkowski, \emph{{What is the Magnetic
  Weak Gravity Conjecture for Axions?}},
  \href{https://doi.org/10.1002/prop.201700011}{\emph{Fortsch. Phys.}
  {\bfseries 65} (2017) 1700011}
  [\href{https://arxiv.org/abs/1701.06553}{{\ttfamily 1701.06553}}].

\bibitem{Heidenreich:2020pkc}
B.~Heidenreich, J.~McNamara, M.~Montero, M.~Reece, T.~Rudelius and
  I.~Valenzuela, \emph{{Chern-Weil global symmetries and how quantum gravity
  avoids them}}, \href{https://doi.org/10.1007/JHEP11(2021)053}{\emph{JHEP}
  {\bfseries 11} (2021) 053}
  [\href{https://arxiv.org/abs/2012.00009}{{\ttfamily 2012.00009}}].

\bibitem{Kachru:2002sk}
S.~Kachru, M.~B. Schulz, P.~K. Tripathy and S.~P. Trivedi, \emph{{New
  supersymmetric string compactifications}},
  \href{https://doi.org/10.1088/1126-6708/2003/03/061}{\emph{JHEP} {\bfseries
  03} (2003) 061} [\href{https://arxiv.org/abs/hep-th/0211182}{{\ttfamily
  hep-th/0211182}}].

\bibitem{Curio:2000sc}
G.~Curio, A.~Klemm, D.~Lust and S.~Theisen, \emph{{On the vacuum structure of
  type II string compactifications on Calabi-Yau spaces with H fluxes}},
  \href{https://doi.org/10.1016/S0550-3213(01)00285-1}{\emph{Nucl. Phys. B}
  {\bfseries 609} (2001) 3}
  [\href{https://arxiv.org/abs/hep-th/0012213}{{\ttfamily hep-th/0012213}}].

\bibitem{Behrndt:2001qa}
K.~Behrndt, S.~Gukov and M.~Shmakova, \emph{{Domain walls, black holes, and
  supersymmetric quantum mechanics}},
  \href{https://doi.org/10.1016/S0550-3213(01)00052-9}{\emph{Nucl. Phys. B}
  {\bfseries 601} (2001) 49}
  [\href{https://arxiv.org/abs/hep-th/0101119}{{\ttfamily hep-th/0101119}}].

\bibitem{Behrndt:2001mx}
K.~Behrndt, G.~Lopes~Cardoso and D.~Lust, \emph{{Curved BPS domain wall
  solutions in four-dimensional N=2 supergravity}},
  \href{https://doi.org/10.1016/S0550-3213(01)00193-6}{\emph{Nucl. Phys. B}
  {\bfseries 607} (2001) 391}
  [\href{https://arxiv.org/abs/hep-th/0102128}{{\ttfamily hep-th/0102128}}].

\bibitem{Nomizu1954OnTC}
K.~Nomizu, \emph{On the cohomology of compact homogeneous spaces of nilpotent
  lie groups}, {\emph{Annals of Mathematics} {\bfseries 59} (1954) 531}.

\bibitem{Cenkl2000NILMANIFOLDSAA}
B.~Cenkl and R.~D. Porter, \emph{Nilmanifolds and associated lie algebras over
  the integers}, {\emph{Pacific Journal of Mathematics} {\bfseries 193} (2000)
  5}.

\bibitem{Hanany:1996ie}
A.~Hanany and E.~Witten, \emph{{Type IIB superstrings, BPS monopoles, and
  three-dimensional gauge dynamics}},
  \href{https://doi.org/10.1016/S0550-3213(97)00157-0}{\emph{Nucl. Phys. B}
  {\bfseries 492} (1997) 152}
  [\href{https://arxiv.org/abs/hep-th/9611230}{{\ttfamily hep-th/9611230}}].

\bibitem{Mayer:2004sd}
C.~Mayer and T.~Mohaupt, \emph{{Domain walls, Hitchin's flow equations and
  G(2)-manifolds}},
  \href{https://doi.org/10.1088/0264-9381/22/2/010}{\emph{Class. Quant. Grav.}
  {\bfseries 22} (2005) 379}
  [\href{https://arxiv.org/abs/hep-th/0407198}{{\ttfamily hep-th/0407198}}].

\bibitem{hitchin2001stable}
N.~Hitchin, \emph{Stable forms and special metrics},  2001.

\bibitem{BenAchour:2015aah}
J.~Ben~Achour, E.~Huguet, J.~Queva and J.~Renaud, \emph{{Explicit vector
  spherical harmonics on the 3-sphere}},
  \href{https://doi.org/10.1063/1.4940134}{\emph{J. Math. Phys.} {\bfseries 57}
  (2016) 023504} [\href{https://arxiv.org/abs/1505.03426}{{\ttfamily
  1505.03426}}].

\bibitem{Andriot:2018tmb}
D.~Andriot and D.~Tsimpis, \emph{{Laplacian spectrum on a nilmanifold,
  truncations and effective theories}},
  \href{https://doi.org/10.1007/JHEP09(2018)096}{\emph{JHEP} {\bfseries 09}
  (2018) 096} [\href{https://arxiv.org/abs/1806.05156}{{\ttfamily
  1806.05156}}].

\bibitem{Grana:2006hr}
M.~Grana, J.~Louis and D.~Waldram, \emph{{SU(3) x SU(3) compactification and
  mirror duals of magnetic fluxes}},
  \href{https://doi.org/10.1088/1126-6708/2007/04/101}{\emph{JHEP} {\bfseries
  04} (2007) 101} [\href{https://arxiv.org/abs/hep-th/0612237}{{\ttfamily
  hep-th/0612237}}].

\bibitem{Polchinski:2003bq}
J.~Polchinski, \emph{{Monopoles, duality, and string theory}},
  \href{https://doi.org/10.1142/S0217751X0401866X}{\emph{Int. J. Mod. Phys. A}
  {\bfseries 19S1} (2004) 145}
  [\href{https://arxiv.org/abs/hep-th/0304042}{{\ttfamily hep-th/0304042}}].

\bibitem{Lust:2008zd}
D.~Lust, F.~Marchesano, L.~Martucci and D.~Tsimpis, \emph{{Generalized
  non-supersymmetric flux vacua}},
  \href{https://doi.org/10.1088/1126-6708/2008/11/021}{\emph{JHEP} {\bfseries
  11} (2008) 021} [\href{https://arxiv.org/abs/0807.4540}{{\ttfamily
  0807.4540}}].

\bibitem{Tomasiello:2007zq}
A.~Tomasiello, \emph{{Reformulating supersymmetry with a generalized Dolbeault
  operator}}, \href{https://doi.org/10.1088/1126-6708/2008/02/010}{\emph{JHEP}
  {\bfseries 02} (2008) 010} [\href{https://arxiv.org/abs/0704.2613}{{\ttfamily
  0704.2613}}].

\bibitem{Buchmuller:2014vda}
W.~Buchmuller, C.~Wieck and M.~W. Winkler, \emph{{Supersymmetric Moduli
  Stabilization and High-Scale Inflation}},
  \href{https://doi.org/10.1016/j.physletb.2014.07.024}{\emph{Phys. Lett. B}
  {\bfseries 736} (2014) 237}
  [\href{https://arxiv.org/abs/1404.2275}{{\ttfamily 1404.2275}}].

\bibitem{Wall:1966rcd}
C.~T.~C. Wall, \emph{{Classification problems in differential topology. V}},
  \href{https://doi.org/10.1007/BF01389738}{\emph{Invent. Math.} {\bfseries 1}
  (1966) 355}.

\bibitem{Zhubr1978ClassificationOS}
A.~V. Zhubr, \emph{Classification of simply connected six-dimensional spin
  manifolds}, {\emph{Journal of Soviet Mathematics} {\bfseries 10} (1978) 451}.

\bibitem{Angius:2023xtu}
R.~Angius, J.~Huertas and A.~M. Uranga, \emph{{Small black hole explosions}},
  \href{https://doi.org/10.1007/JHEP06(2023)070}{\emph{JHEP} {\bfseries 06}
  (2023) 070} [\href{https://arxiv.org/abs/2303.15903}{{\ttfamily
  2303.15903}}].

\bibitem{Kraus:2006wn}
P.~Kraus, \emph{{Lectures on black holes and the AdS(3) / CFT(2)
  correspondence}}, {\emph{Lect. Notes Phys.} {\bfseries 755} (2008) 193}
  [\href{https://arxiv.org/abs/hep-th/0609074}{{\ttfamily hep-th/0609074}}].

\bibitem{Prieto:2024hkf}
D.~Prieto, \emph{{Moduli Stabilization and Stability in Type II/F-theory flux
  compactifications}},  phd thesis, UAM, 1, 2024.

\bibitem{Huybrechts:2005}
D.~Huybrechts, \emph{Complex Geometry: An Introduction}. Springer Berlin
  Heidelberg, 2005, \href{https://doi.org/10.1007/b137952}{10.1007/b137952}.

\end{thebibliography}\endgroup

\end{document}